\documentclass[sn-vancouver,Numbered]{sn-jnl}

\usepackage[fontsize=10.5pt]{fontsize}

\usepackage{geometry}
\usepackage{graphicx}%
\usepackage{multirow}%
\usepackage{amsmath,amssymb,amsfonts}%
\usepackage{amsthm}%
\usepackage{mathrsfs}%
\usepackage[title]{appendix}%
\usepackage{xcolor}%
\usepackage{textcomp}%
\usepackage{manyfoot}%
\usepackage{booktabs}%
\usepackage{algorithm}%
\usepackage{algorithmicx}%
\usepackage{algpseudocode}%
\usepackage{listings}%

\usepackage{amssymb}
\usepackage{comment}
\usepackage{tgschola}
\usepackage[mathscr]{euscript}
\usepackage{color}

\usepackage{mathrsfs}
\usepackage{lipsum}
\usepackage{hyperref}
\usepackage{xcolor}
\definecolor{xll}{rgb}{0,0,0}
\hypersetup{
  colorlinks=true,
  citecolor=brown,
  linkcolor=magenta,
  urlcolor=magenta,
}

\usepackage{enumitem}

\usepackage{amsmath,amsfonts,amssymb}
\usepackage{mathtools}

\usepackage{multicol}
\usepackage[T1]{fontenc}
\usepackage{epstopdf}
\usepackage{empheq}

\usepackage{titlesec}
\renewcommand{\thesection}{\Roman{section}}
\usepackage{pxfonts}

\usepackage[condensed,math]{anttor}
\usepackage[T1]{fontenc}

\titleformat{\section}{\large\bfseries}{\thesection}{.5em}{}
\titleformat{\subsection}{\color{teal}\normalsize\bfseries}{\thesubsection}{.5em}{}

\DeclareMathAlphabet{\mathcal}{OMS}{cmsy}{m}{n}
\renewcommand{\d}{\mathrm{d}}
\renewcommand{\v}[1]{\mathbf{#1}}
\renewcommand{\bf}[1]{\mathbf{#1}}
\renewcommand{\rm}[1]{\mathrm{#1}}

\usepackage[symbol]{footmisc}
\numberwithin{equation}{section}

\usepackage{chngcntr}
\counterwithout{equation}{section}
\renewcommand\theequation{\arabic{equation}}
\renewcommand\thefigure{\arabic{figure}}
\usepackage{wrapfig}

\def\thesection{\arabic{section}\hspace*{0.cm}}

\usepackage{titletoc}
\titlecontents{section}
              [.25cm]
              {\normalsize\sc\thecontentslabel}
              {\normalsize\makebox[.25cm][l]{}}
              {}
              {\normalsize\titlerule*[.5pc]{$\cdot$}\contentspage
              \hspace*{.5cm}}

\titlecontents{subsection}
              [1.cm]
              {\footnotesize\thecontentslabel}
              {\footnotesize\makebox[0.5cm][l]{}}
              {}
              {\footnotesize\titlerule*[.5pc]{$\cdot$}\contentspage
              \hspace*{.5cm}}

\usepackage{indentfirst}
\newcommand{\x}{\mathrm{X}}
\newcommand{\y}{\mathrm{Y}}
\newcommand{\epsc}{\varepsilon_{\rm{c}}}
\newcommand{\Pc}{P_{\rm{c}}}
\newcommand{\hP}{\widehat{P}}
\newcommand{\hr}{\widehat{r}}
\newcommand{\hM}{\widehat{M}}
\newcommand{\heps}{\widehat{\varepsilon}}
\newcommand{\ep}{\overline{\varepsilon}}

\usepackage{setspace}

\renewcommand\theequation{\arabic{section}.\arabic{equation}}

\usepackage[font=footnotesize]{caption} 

\usepackage{tabularx}

\usepackage{scalerel}

\usepackage{tabularray}

\begin{document}

\title[xxxxxx]{\Large \bfseries Novel Scalings of Neutron Star Properties from
 Analyzing Dimensionless Tolman--Oppenheimer--Volkoff Equations}


\author*[1]{\fnm{Bao-Jun} \sur{Cai}}\email{bjcai87@gmail.com}

\author*[2]{\fnm{Bao-An} \sur{Li}}\email{Bao-An.Li@tamuc.edu}


\affil*[1]{\orgdiv{Quantum Machine Learning Laboratory}, \orgname{Shadow Creator Inc.}, \orgaddress{\street{}\city{Pudong New District}, \postcode{201208}, \state{Shanghai}, \country{People's Republic of China}}}

\affil*[2]{\orgdiv{Department of Physics and Astronomy}, \orgname{East Texas A\&M University}, \orgaddress{\street{}\city{Commerce}, \postcode{75429-3011}, \state{Texas}, \country{USA}}}


\abstract{The Tolman--Oppenheimer--Volkoff (TOV) equations govern the radial evolution of pressure and energy density in static neutron stars (NSs) in hydrodynamical equilibrium. Using the reduced pressure and energy density with respect to the NS central energy density, the original TOV equations can be recast into dimensionless forms. While the traditionally used integral approach for solving the original TOV equations require an input nuclear Equation of State (EOS), the dimensionless TOV equations can be anatomized by using the reduced pressure and energy density as polynomials of the reduced radial coordinate without using any input nuclear EOS. It has been shown in several of our recent works that interesting and novel perspectives 
about NS core EOS can be extracted directly from NS observables by using the latter approach.
Our approach is based on \underline{i}ntrinsic and  \underline{p}erturbative \underline{a}nalyses of the \underline{d}imensionless (IPAD) TOV equations (IPAD-TOV).
In this review article, we first discuss the length and energy density scales of NSs as well as the dimensionless TOV equations for scaled variables and their perturbative solutions near NS cores. We then review several new insights into NS physics gained from solving perturbatively the scaled TOV equations. Whenever appropriate, comparisons with the traditional approach from solving the original TOV equations will be made.
In particular, we first show that the nonlinearity of the TOV equations basically excludes a linear EOS for dense matter in NS cores. We then show that perturbative analyses of the scaled TOV equations enable us to reveal novel scalings of the NS mass, radius and the compactness with certain combinations of the NS central pressure and energy density. Thus, observational data on either mass, radius or compactness can be used to constrain directly the core EOS of NS matter independent of the still very uncertain nuclear EOS models. As examples, the EOS of the densest visible matter in our Universe before the most massive neutron stars collapse into black holes (BHs) as well as the central EOS of a canonical or a 2.1 solar mass NS are extracted without using any nuclear EOS model. In addition, we show that causality in NSs sets an upper bound of about 0.374 for the ratio of pressure over energy density and correspondingly a lower limit for trace anomaly in supra-dense matter. We also demonstrate that the strong-field gravity plays a fundamental role in extruding a peak in the density/radius profile of the speed of sound squared (SSS) in massive NS cores independent of the nuclear EOS. Finally, some future perspectives of NS research using the new approach reviewed here by solving perturbatively the dimensionless TOV equations are outlined.}




\maketitle

\begin{minipage}{16.cm}
\renewcommand{\contentsname}{Content}
\setcounter{tocdepth}{2}
{
\begin{spacing}{1.35}
\tableofcontents
\end{spacing}
}
\end{minipage}

\newpage

\centerline{\large Notations of key quantities used in this review (under units of $c=G=1$)}

\renewcommand*\tablename{\footnotesize TAB.}
\begin{table}[h!]
\renewcommand{\arraystretch}{1.2}
\centerline{\normalsize
\begin{tabular}{c||c|c} 
\hline
{\sc symbol}&{\sc meaning}&{\sc equations}\\\hline\hline
$\varepsilon$&energy density of NS matter&\\\hline
$\varepsilon_0\approx150\,\rm{MeV}/\rm{fm}^3$& energy density at nuclear saturation density&\\\hline
$\varepsilon_{\rm{c}}$& central energy density of NS matter&\\\hline
$\widehat{\varepsilon}\equiv\varepsilon/\varepsilon_{\rm{c}}$&reduced energy density with respect to $\varepsilon_{\rm{c}}$&\\\hline
$\overline{\varepsilon}\equiv\varepsilon/\varepsilon_0$& reduced energy density with respect to $\varepsilon_0$&\\\hline
$\overline{\varepsilon}_{\rm{c}}\equiv \varepsilon_{\rm{c}}/\varepsilon_0\equiv\rm{Y}$& reduced central energy density with respect to $\varepsilon_0$&\\\hline
$\mu\equiv\widehat{\varepsilon}-\widehat{\varepsilon}_{\rm{c}}\equiv\widehat{\varepsilon}-1$&dimensionless energy density based on $\widehat{\varepsilon}_{\rm{c}}\equiv 1$&Eq.\,(\ref{RE-small1})\\\hline
$W=Q\equiv(4\pi\varepsilon_{\rm{c}})^{-1/2}$&mass/length scale&Eq.\,(\ref{RE-WQ})\\\hline
\shortstack{~~\\$\rho=\rho_{\rm{n}}+\rho_{\rm{p}}$\\~~}&\shortstack{baryon number density \\of neutron (n) and proton (p)} & \\\hline
$\rho_0\equiv\rho_{\rm{sat}}\approx0.16\,\rm{fm}^3$&nuclear saturation density&\\\hline
$\rho_{\rm{c}}$&central baryon number density&\\\hline
$\delta=(\rho_{\rm{n}}-\rho_{\rm{p}})/\rho$& isospin asymmetry of neutron-rich matter&\\\hline
\shortstack{~~\\$E(\rho,\delta)$\\~~}& \shortstack{equation of state (EOS) of asymmetric nuclear\\ matter of isospin asymmetry $\delta$}&\\\hline
$\Sigma\equiv M_{\odot}/\rm{km}\approx1.477$&constant related to solar mass&\\\hline
$P$&pressure of NS matter&\\\hline
$P_{\rm{c}}$&central pressure of NS matter&\\\hline
$\widehat{P}\equiv P/\varepsilon_{\rm{c}}$&reduced pressure with respect to $\varepsilon_{\rm{c}}$&\\\hline
$\phi\equiv P/\varepsilon\equiv\widehat{P}/\widehat{\varepsilon}$& ratio of pressure over energy density&Eq.\,(\ref{def-phi})\\\hline
$\phi_{\rm{c}}\equiv P_{\rm{c}}/\varepsilon_{\rm{c}}\equiv\widehat{P}_{\rm{c}}\equiv\x$& central ratio of pressure over energy density& Eq.\,(\ref{RE-small2})\\\hline
$\Delta\equiv 1/3-\phi$& dimensionless trace anomaly&\\\hline
$\Delta_{\rm{c}}\equiv1/3-\x$& dimensionless trace anomaly at NS center&\\\hline
$s^2\equiv \d P/\d \varepsilon$&speed of sound squared (SSS)&Eq.\,(\ref{def-s2})\\\hline
$s_{\rm{c}}^2$&central speed of sound squared
&\\\hline
$\gamma\equiv \d\ln P/\d\ln \varepsilon=s^2/\phi$& polytropic index&Eq.\,(\ref{def_gamma})\\\hline
$\gamma_{\rm{c}}\equiv s_{\rm{c}}^2/\phi_{\rm{c}}$&central polytropic index&\\\hline
$M_{\rm{NS}}$& masses of generally stable NSs&\\\hline
\shortstack{~~\\$M_{\rm{NS}}^{\max}\equiv M_{\rm{TOV}}$\\ ~~\\~~}&\shortstack{NS maximum mass supported  by\\ a given EOS at TOV configuration\\ where the mass-radius curve peaks}&\\\hline
$\widehat{M}_{\rm{NS}}\equiv \widehat{M}\equiv M_{\rm{NS}}/W$&reduced NS mass with respect to $W$&\\\hline
$R$&radii of generally stable NSs&\\\hline
$R_{\max}\equiv R_{\rm{TOV}}$&radii of NSs at TOV configuration&\\\hline
$\widehat{r}\equiv r/Q$&reduced distance from NS center with respect to $Q$&\\\hline
$\widehat{R}\equiv R/Q$& reduced NS radius with respect to $Q$&\\\hline
\shortstack{~~\\$\Psi\equiv 2\d\ln M_{\rm{NS}}/\d\ln\varepsilon_{\rm{c}}$\\~~}&\shortstack{logarithmic derivative of NS mass \\with respect to central energy density}&Eq.\,(\ref{def-Psi})\\\hline
$\xi\equiv M_{\rm{NS}}/R$& NS compactness coefficient&Eq.\,(\ref{def-compactness})\\\hline
$\xi_{\max}\equiv M_{\rm{NS}}^{\max}/R_{\max}
    \equiv M_{\rm{TOV}}/R_{\rm{TOV}}$& compactness of NSs at TOV configuration\\\hline
\end{tabular}}
        \caption{Notations for main quantities used in this review, related equations (if applied) are shown in the third column.}\label{tab_notations}        
\end{table}

\newpage
\section{Introduction}\label{SEC_1}

The Nature and Equation of State (EOS) of superdense matter in neutron stars (NSs)\,\cite{Walecka1974,Collins1975,Chin1977,Freedman1977-1,Freedman1977-2,Freedman1977-3,Baluni1978,Wiringa1988,Akmal1998} have long been among the most important unsolved questions in nuclear astrophysics\,\cite{Migdal1978,Morley1979,Shuryak1980,Bailin1984,Lattimer2001,Dan02,Steiner2005,Alford2008,LCK08,Watts2016,Ozel2016,Oertel2017,Vidana2018,Bur2021,Dri2021,Lov2022,Sor2024,Kumar2024,Baym2018,Bai2019,Ors2019,Li2019,Dex2021,Lattimer:2021emm}.
The EOS for cold matter refers to the functional relationship between pressure $P$ and energy density $\varepsilon$ of the system under consideration, namely $P=P(\varepsilon)$. 
In this review, we adopt units in which $c=G=1$.
Closely related to the EOS is the speed of sound squared (SSS) defined as\,\cite{Landau1987}
\begin{equation}\label{def-s2}
\boxed{
    s^2\equiv c^2\frac{\d P}{\d\varepsilon}=
    \frac{\d P}{\d\varepsilon}.}
\end{equation}
It is a measure of the stiffness of the EOS.
Another important quantity for a NS is its compactness:
\begin{equation}\label{def-compactness}
\boxed{
    \xi\equiv\frac{GM_{\rm{NS}}}{Rc^2}=\frac{M_{\rm{NS}}}{R},}
\end{equation}
here $M_{\rm{NS}}$ and $R$ are the NS mass and radius, respectively.
The third dimensionless quantity is the ratio of pressure over energy density:
\begin{equation}\label{def-phi}
\boxed{
    \phi\equiv P/\varepsilon.}
\end{equation}
From the $\phi$ and $s^2$ defined above, the polytropic index
can be define as 
\begin{equation}\label{def_gamma}
\boxed{
\gamma\equiv\frac{\d\ln P}{\d\ln\varepsilon}=\frac{s^2}{P/\varepsilon}=\frac{s^2}{\phi},}
\end{equation}
which is also a dimensionless quantity.
Studying these quantities, their relationships and the roles they play in determining properties of NSs have been among the major objectives of many research in nuclear astrophysics in recent decades. In particular, finding the EOS of densest visible matter existing in our Universe is an ultimate goal of astrophysics in the era of high-precision multi-messenger astronomy\,\cite{sathyaprakash2019}.
However, despite of much effort using various data especially those thanks to the observational progresses made since the discovery of GW170817\,\cite{Abbott2017,Abbott2018} and the recent NASA's NICER (Neutron Star Interior Composition Explorer) mass-radius measurements for PSR J0740+6620\,\cite{Fon21,Riley21,Miller21,Salmi22,Ditt24,Salmi24},  PSR J0030+0451\,\cite{Riley19,Miller19,Vin24} and PSR J0437-4715\,\cite{Choud24,Reard24}, as well as various nuclear EOS models and new data analysis tools over the last few decades, information about the NS core EOS remains ambiguous and quite elusive\,\cite{Rai16,Rai17,Rai18,Rai21,Raithel2023,De2018,Tews2018,Lim2018,Lim2019,Drischler2020,Han2021,Radice2018,Most2018,Fatt2018,Bose2018,Baym2019,Huang2022,Biswas2022,Weih2020,XieLi2020,Most2019,Baus2019,Baus2020,Montana2019,Malfatti2019,Greif2019,Fuku2020,LiAng2020,LiAng2021,Kap2021,Drischler2021PRC,Drischler2022PRC,Kojo2022,Fuji22HDL,Leg21,Mam2021,Breschi2022,Malik2022,Perego2022,Blacker2023PRD,Blacker2024PRD,Han2023,Som2023,Brandes2023,Brandes2023-a,Fuji2023,ZhangLi2023a,ZhangLi2023b,Alam2024,Christian2024PRD,Chu2024PRD,HuangChun2024,Issifu2024,Kom24PRD,Malik2024PRD,Patkos2024,Rather2023PRD,Rather2024,Tsang2024NA}.
Reviewing our recent contributions to the global efforts of unraveling the nature and EOS of NS matter based on observational data is the main goal of this article. We have summarized recently the upper bound on $\phi$ due to the strong-field gravity in General Relativity (GR) in a short review\,\cite{CL24-c}. Here we aim at a more comprehensive review of our unique contributions using a novel approach in analyzing the TOV equations in the context of existing work on NS physics by many others in nuclear astrophysics. 

\renewcommand*\figurename{\footnotesize FIG.}
\begin{figure}[h!]
\centering
\includegraphics[width=16.cm]{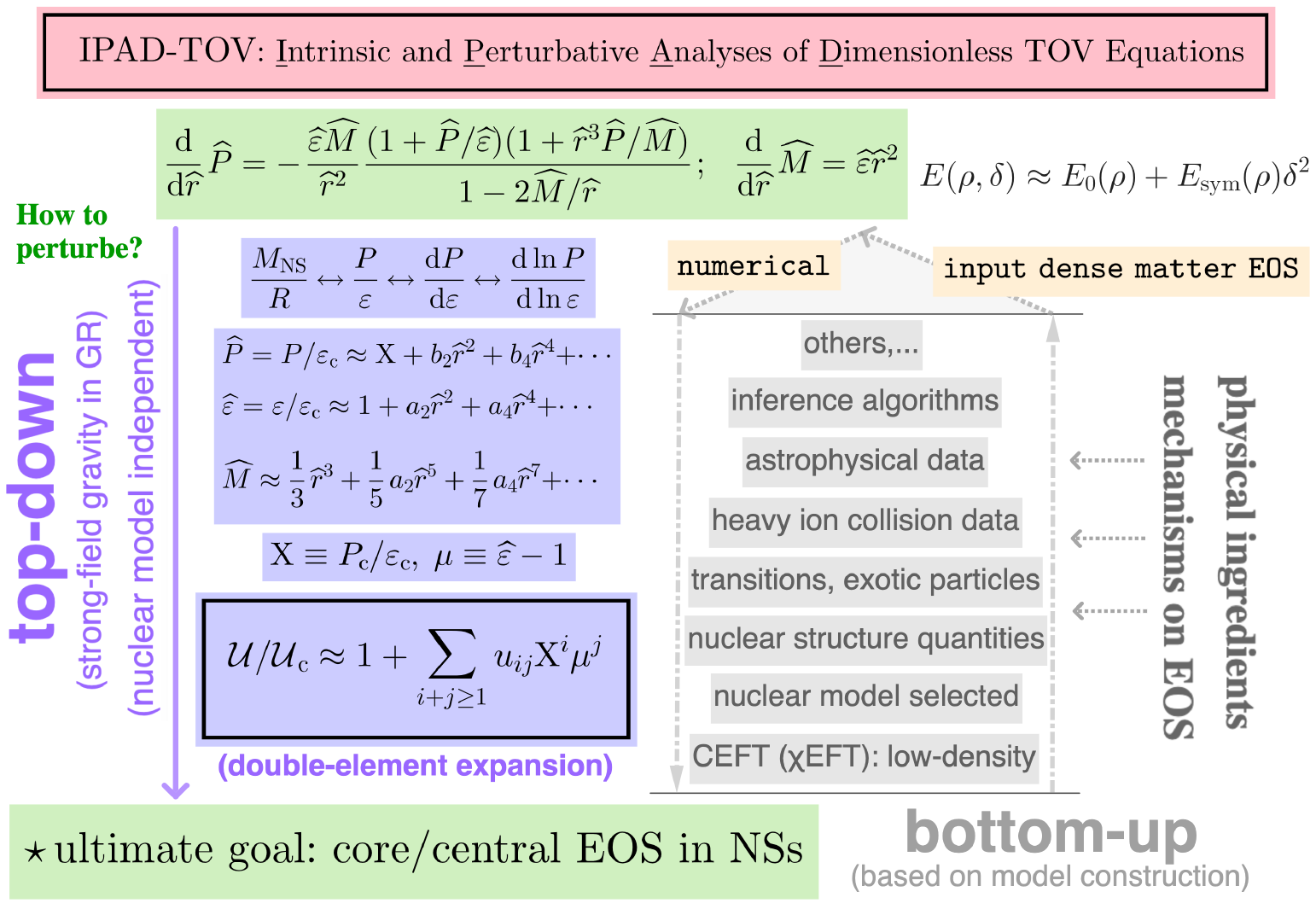}
\caption{(Color Online). 
Comparison between the two approaches for extracting the EOS from solving the TOV equations: (right) in the conventional (bottom-up) approach, one first constructs a model EOS for NSs from low to high densities using different inputs/constraints at the corresponding density regions as indicated by the arrows. In our IPAD-TOV (top-down) approach (left), the central/core EOS in NSs is straightforwardly extracted from the TOV equations themselves without using any model EOS. See the text for more details on the notations.
}\label{fig_TB}
\end{figure}
The basic equations for describing (spherical static) NSs are the Tolman--Oppenheimer--Volkoff (TOV) equations\,\cite{TOV39-1,TOV39-2,Misner1973} (given near the top of FIG.\,\ref{fig_TB}), obtained from GR under hydrodynamic equilibrium (see SECTION \ref{SEC_2} for an essential and brief introduction); any NS EOS investigation relies unavoidably on solving and analyzing the TOV equations.
In the conventional {\color{xll}(bottom-up)} approach, one first constructs/builds an appropriate NS matter EOS from low to high densities using different inputs/constraints in the corresponding density regions, then puts it into the TOV equations to obtain a sequence $M(R)$ of NS mass $M$ versus its radius $R$. Presently, the large uncertainties mainly come from the NS EOS-model construction step since many different mechanisms, models and constraints exist and they often can explain all existing observations equally well. Thus, comparing the predicted $M(R)$ with observational data in such a way may introduce spurious effects and still can not distinguish different or exclude some input NS EOSs. Moreover, in this approach although the EOS up to about $1.5$ times the saturation density $\rho_0$ of symmetric nuclear matter (SNM) could be determined/constrained quite well by both reliable nuclear theories and/or experiments in terrestrial nuclear laboratories, they generally have little impact on NS mass. On the other hand, the ingredients largely affecting the NS masses at high densities are poorly known and still have very large uncertainties. The necessary steps and constrains on the nuclear EOS often considered presently in solving the TOV equations in the traditional approach are listed on the right hand side of FIG.\,\ref{fig_TB}. 

As we shall discuss in great detail, another way of solving the TOV equations is to first recast them into dimensionless (scaled) forms by using the reduced pressure and energy density with respect to the NS central energy density Refs.\,\cite{CLZ23-a,CLZ23-b,CL24-a,CL24-b,CL24-c}.
The scaled TOV equations can by solved perturbatively near NS centers by expanding the reduced pressure and energy density as polynomials of the reduced radial coordinate without using any input nuclear EOS. Since our approach is based on \underline{i}ntrinsic and \underline{p}erturbative \underline{a}nalyses of the \underline{d}imensionless (IPAD) TOV equations (IPAD-TOV),\footnote{English meaning of Apple's iPad from Cambridge Dictionary: a brand name for a tablet (aka, small computer) that is controlled by touch rather than having a keyboard. In our IPAD-TOV approach, properties of supra-dense NS matter governed by the TOV equations are probed perturbatively without using a specific input model EOS.} we refer this novel method as 
the {\color{xll}top-down approach to differentiate it from the traditional bottom-up one.} Interesting new features about the NS matter EOS, e.g., the $s^2$, $\xi$ and $\phi$ in NS cores, can be directly extracted from NS observational data without using any model EOS. Fundamentally, this is made possible by the fact that the TOV equations inherently couple the pressure, energy density and the mass, i.e., the EOS is implicitly encapsulated in them. Therefore, extracting information about the NS EOS does not necessarily have to rely on any specific input NS EOS model as long as enough and accurate NS $M(R)$ data is available. 
For a comparison with the traditional approach, shown in the left side of FIG.\,\ref{fig_TB} are key steps in our {\color{xll}top-down approach.} 

\begin{figure}[h!]
\centering
\includegraphics[width=14.cm]{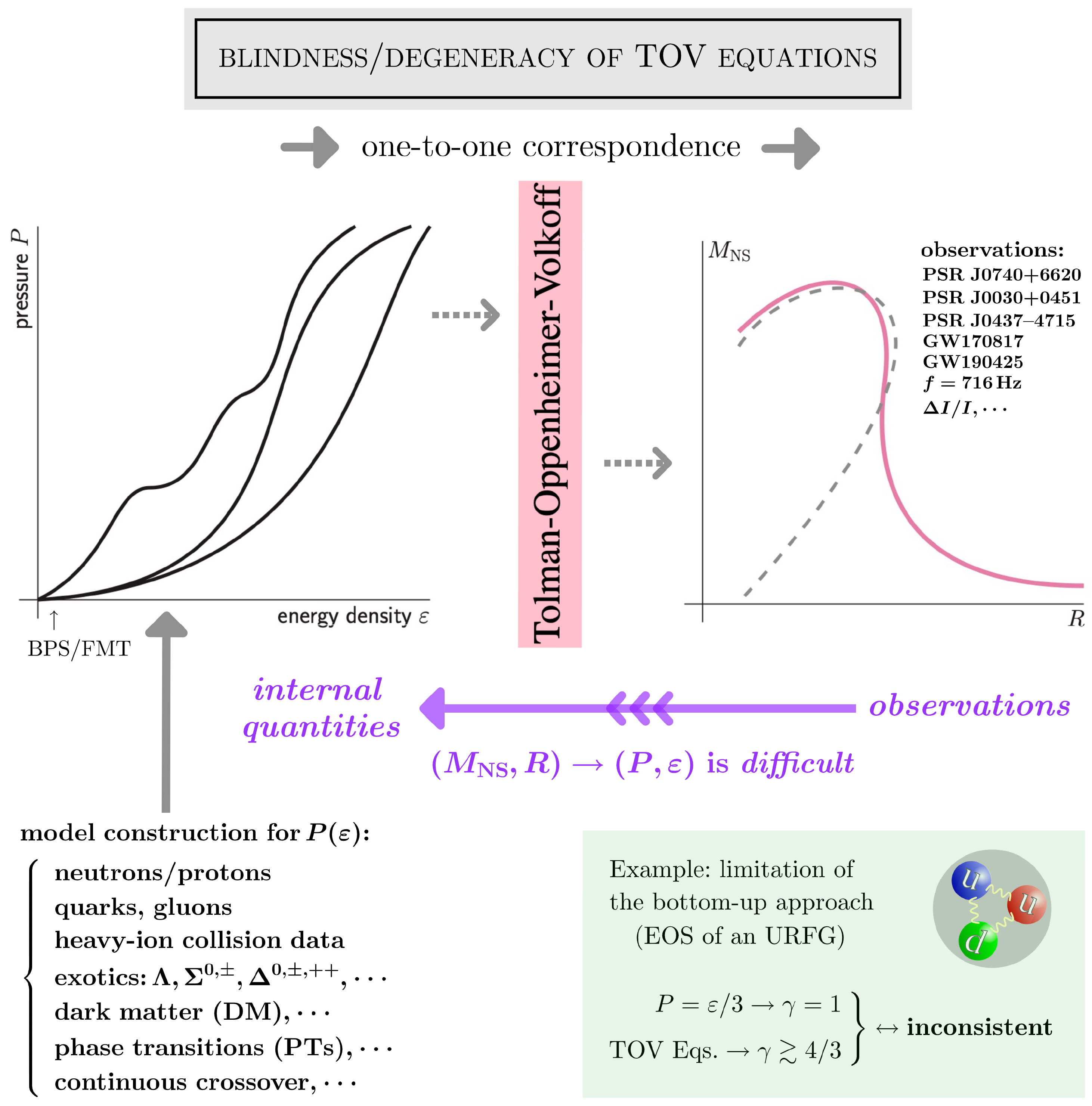}
\caption{(Color Online). 
The TOV equations are blind in the sense that regardless of the composition of the NSs, a unique M-R relation is determined as long as the same EOS $P(\varepsilon)$ is constructed and there are multiple ways to do so. Shown near the bottom are features of a linear EOS of the form $P=\varepsilon/3$ for an ultra-relativistic Fermi gas in NS cores.
Such EOS is excluded by analyzing the scaled TOV equations without any prior knowledge about its core EOS. See the text for details.
}\label{fig_TOVBD}
\end{figure}

In essence, the traditional approach is a forward-modeling of NS properties given an EOS model while our new approach is a backward inference of the NS EOS directly from a given set of NS observational data. Our approach is also fundamentally different from the Bayesian statistical inference of NS EOS from observational data. In the standard Bayesian inference, the forward-modeling is a necessary step to evaluate the likelihood for a given EOS to reproduce the observational data in each step of the Markov Chain Monte Carlo (MCMC) sampling. On the contrary, in principle our approach does not need any NS EOS model as a middle agent but the observational data alone. 
Thus, the knowledge on NS EOS extracted in our new approach can be used as an unbiased reference in 
comparing predictions of various NS EOS models based on nuclear many-body theories. As we shall demonstrate, some of the novel scalings of NS properties we derive are highly accurate up to high-orders of the polynomial expansions without using any nuclear EOS models. Nevertheless, to evaluate the accuracy and/or determine the truncation order for analyzing some NS properties, we do use existing information on NS EOS from various models in the literature. Moreover, we also use a large set of microscopic and/or phenomenological nuclear EOSs available in the literature and $10^5$ randomly generated nuclear EOSs in a meta-model within the traditional approach to verify quantitatively the novel scalings revealed from our new approach. 
Thus, in this sense and context, most of the novel scaling properties of NS properties revealed in our analyses are largely instead of absolutely independent of nuclear EOS models. For this reason, we shall try to distinguish the NS matter EOS ($P=P(\varepsilon)$) determined by the TOV equations themselves and the EOS from nuclear EOS-models in the following discussions when it is necessary and possible.

As mentioned earlier, it has been very challenging to extract information about the nature and EOS of supra-dense matter inside NSs. A critical reason for this is 
the blindness or degeneracy 
of the TOV equations about the composition of NSs. Namely, 
regardless how/what the energy density is made of, as long as the same EOS  $P(\varepsilon)$ is given in the traditional approach, a unique $M(R)$ sequence is determined. 
Similarly, in the {\color{xll}top-down approach},
even with the $P(\varepsilon)$ and its characteristics extracted directly from the $M(R)$ observations, there is still no explicit information about the unique NS composition underlying the  $P(\varepsilon)$ extracted
\,\cite{CLZ23-a,CLZ23-b,CL24-a,CL24-b,CL24-c}. Illustrated in FIG.\,\ref{fig_TOVBD} are some key points about the blindness of TOV equations. 
This intrinsic feature of the TOV equations is independent of the techniques people may use to solve them. As well documented in the literature, various combinations of different mechanisms or models including the formations of various new particles, such as hyperons, baryon resonances and possible phase transitions to quarks and gluons, can lead to the same NS EOS $P(\varepsilon)$. The resulting mass-radius (M-R) curve thus cannot distinguish the composition of NSs with the same EOS unless one looks into observables from microscopic processes happening inside NSs.

Both the conventional approach for solving the TOV equations and our perturbative analysis of its central solutions (SECTION \ref{SEC_2}) have their own advantages and limitations.
For example, our {\color{xll}top-down approach} is expected to work well for extracting the NS core EOS due to its perturbative nature. While for studying low-density properties of NSs such as those in the crust, the conventional approach using nuclear EOS models as an input is necessary and useful. Nevertheless, main features of the central EOSs from
the two approaches have to match.  Therefore studies of some NS properties using both approaches may be beneficial. They may provide complementary information leading to a deeper understanding of superdense matter in strong gravitational fields.
For example, {\color{xll}if one feeds an inappropriate EOS into the TOV equations in the traditional approach, then the results would be misleading.} As we shall discuss in details below, without making any prior assumption about NS EOS, analyses of the scaled TOV equations themselves can basically exclude some core EOSs. In particular, a linear EOS of the form $P=\varepsilon/3$ for an ultra-relativistic Fermi gas in NS cores is clearly excluded. As a baseline, this excluded EOS is indicated at the bottom of FIG.\,\ref{fig_TOVBD}.

Stimulated by the exciting progresses achieved recently in NS observations and the strong curiosity to explore the NS EOS under extreme gravity by many people in the field, we first gather below a few relevant questions to start our discussions in this review:
\begin{enumerate}[label=(\alph*)]
\item How NS mass $M_{\rm{NS}}$ and radius $R$ depend on the NS matter EOS? How can this dependence be revealed by the general-relativistic stellar structure equations themselves in an EOS-model independent manner?

\item What is the EOS of the densest visible matter existing in our Universe before it collapses into black holes (BHs)? Can it be accessed/constrained directly using certain astrophysical data such as observed NS radii and/or masses without using any input EOS model?
\item Is a linear EOS in the form of $P=\zeta\varepsilon+\rm{const.}$ (with $\zeta$ being a constant) basically consistent with the TOV equations especially near NS centers?
Equivalently, can the dense matter in NS cores have a constant speed of sound (CSS)? Notice that the speed of sound squared (SSS) $s^2$ is defined in Eq.\,(\ref{def-s2}). 

\item What is the relation between NS compactness $\xi=GM_{\rm{NS}}/Rc^2=M_{\rm{NS}}/R$ (Eq.\,(\ref{def-compactness})) involving two {\color{xll}macroscopic} observables  $M_{\rm{NS}}$ and $R$ and the ratio $\phi=P/\varepsilon$ (Eq.\,(\ref{def-phi})) from two NS {\color{xll} internal} quantities $P$ and $\varepsilon$?
Using units of $c=G=1$, both $\xi$ and $\phi$ are dimensionless.

\item What is the relation between NS compactness presently defined as $\xi=M_{\rm{NS}}/R$ and the NS stiffness quantified by $s^2$?
Are there other quantities affecting the NS compactness besides its mass/radius ratio?

\item Is the upper limit for $\phi=P/\varepsilon$ as $\phi\leq1$ from the Principle of Causality of Special Relativity sufficient considering the dense matter in NS cores?
If it is insufficient,  how and to what extend this upper limit could be improved?
Since NS contains the densest visible matter in our Universe, this upper limit holds universally for all stable matter.

\item The existing causality boundaries based on various theories and/or assumptions are generally high above the NS maximum masses predicted by most EOS models. Is there a causal limit set by the TOV equations themselves using only NS observational data? A related question is whether a massive NS can have a small radius about 10\,km based on the TOV equations alone without using any EOS model.

\item Closely related to the last question, what is the upper limit of NS compactness? What are the ultimate energy density and/or pressure allowed in most massive NSs?

\item Can we estimate the maximum central baryon density in massive NSs from their radii observed?

\item What is the SSS in the core of a canonical NS with mass about 1.4$M_{\odot}$ (here $M_{\odot}$ is the solar mass)? Is the QCD conformal bound for SSS as $s^2\leq1/3$ violated in any NS of different masses?

\item If the SSS $s^2$ is upper bounded to a lower value different from 1, what is its corresponding impact on radii of NSs with masses about $2M_{\odot}$?
Similarly, can we upper bound $s^2$ under certain assumptions?

\item Does a sharp phase transition signaled by a sudden vanishing of $s^2$ occur in NS cores?
If not, is a continuous crossover characterized by a smooth reduction of $s^2$ allowed?
Equivalently, does the NS contain a soft core?

\item Is there a peak in the density or radial profile of $s^2$ in NSs? If yes, what is its physical origin? Where is its radial location and what is its size (enhancement with respect to the QCD conformal limit)? Can the currently available NS data invariably generate a peaked $s^2$ structure by solving the TOV equations without using any EOS-model?

\item Continued with the last question and similarly, is there a peak in the density or radial profiles of $s^2$ in Newtonian stars and how can we understand the results in comparison with NSs in GR?

\item Can the dense matter in NS cores be conformal or nearly conformal adopting certain empirical criterion (e.g., in terms of the trace anomaly $\Delta=1/3-\phi$ or the measure $\Theta=[\Delta^2+(s^2-\phi)^2]^{1/2}$ that would vanish at the QCD conformal limit)?

\item What is the physical origin of the existence of a maximum mass for stable NSs? How can we extract/estimate this limit from the TOV equations without using any EOS model?

\item Putting aside tentatively the various EOS model predictions, can we understand generally the empirical evidence from observations (e.g., canonical NSs and massive NSs have similar radii about 12-13\,km) for the ``vertical'' shape ($\d M_{\rm{NS}}/\d R\to\infty$ or $\d R/\d M_{\rm{NS}}\approx0$) of the NS M-R curve for $M_{\rm{NS}}/M_{\odot}\approx1.4\mbox{-}2.2$ based on the TOV equations alone?

\item If there exists a peak in the trace anomaly $\Delta=1/3-\phi$ around a reduced energy density $\overline{\varepsilon}_{\ell}\equiv\varepsilon_\ell/\varepsilon_0$, is there a corresponding peak around $\overline{\varepsilon}_{\ell}^{\ast}$ in the SSS $s^2$ profile? If so, 
how are these two positions $\overline{\varepsilon}_{\ell}$ and $\overline{\varepsilon}_{\ell}^{\ast}$ related? Answers to these questions may have relevance for determining the SSS in finite-temperature QCD matter (in the crossover region), here $\varepsilon_0\approx150\,\rm{MeV}/\rm{fm}^3$ is the energy density at $\rho_0$. 
\end{enumerate}

Obviously the above questions are not isolated from each other. Of course, there are also many other interesting and important questions under intense investigations by experts in the NS science community. Within our limited knowledge in the field, we try to answer the above questions in a unified framework in the context of existing literature. We acknowledge {\it a prior} that our opinions might be biased unintentionally and there are certainly issues that we touched on but can not fully address. We shall try to identify these issues for future investigations. 

This review is mostly based on our earlier work in Refs.\,\cite{CLZ23-a,CLZ23-b,CL24-a,CL24-b,CL24-c} with more details and some new additions. When necessary we compare our results with contributions by others using different approaches in addressing the same issues. The rest of this review is organized as follows: SECTION \ref{SEC_2} gives an essential introduction to the perturbative treatment/analysis of the dimensionless TOV equations; in SECTION \ref{SEC_3} we review briefly the status of dense matter EOS in NS cores and the related $s^2$ profile; SECTION \ref{SEC_4} is devoted to the central EOS obtained via the mass, radius and compactness scalings; SECTION \ref{SEC_7} investigates in details the SSS in NSs including a possible origin of the peaked structure in $s^2$ profiles; we then  discuss in SECTION \ref{SEC_56} the gravitational (lower) bound for the trace anomaly $\Delta=1/3-\phi$ and the related ratio $\phi=P/\varepsilon$, which were reviewed in some details in Ref.\,\cite{CL24-c}; a new causality boundary for NS M-R curve together with its implications, and the implications of a positive $\Delta$ are also given in this section. Finally, the conclusions and  caveats of work as well as a few perspectives for stimulating future studies along this line are given in SECTION \ref{SEC_8} .

\setcounter{equation}{0}
\section{Scaled Variables in TOV Equations and Their Perturbative Treatments}\label{SEC_2}

This section introduces some basic ingredients of the method on extracting the core EOS of NS matter. Subsection \ref{sub_DTOV} gives the dimensionless TOV equations, based on which the radial dependence of $\heps(\hr)$, $\hP(\hr)$ and $\hM(\hr)$ is discussed in Subsection \ref{sub_Char}. We also discuss in Subsection \ref{sub_Char} the characteristic scales in the dimensionless TOV equations and the double-element expansion using two small quantities $\x$ and $\mu$ (or equivalently using $\x$ and $\widehat{r}$). Finally in Subsection \ref{sub_DEP123} we give the basic results on the perturbative expansions of $\heps(\hr)$, $\hP(\hr)$ and $\hM(\hr)$ near NS centers ($\hr\approx0$), the relation of our results to the Lane-Emden equation for studying Newtonian stars is also given in this subsection.

\subsection{The dimensionless TOV equations}\label{sub_DTOV}

As described in SECTION \ref{SEC_1},  the TOV equations describe the radial evolution of pressure $P(r)$ and mass $M(r)$ of a NS under static hydrodynamic equilibrium conditions\,\cite{TOV39-1,TOV39-2,Misner1973}.
Specifically, they are originally written as (adopting $c=1$)
\begin{align}
\frac{\d P}{\d r}=-\frac{GM\varepsilon}{r^2}\left(1+\frac{P}{\varepsilon}\right)\left(1+\frac{4\pi r^3P}{M}\right)
\left(1-\frac{2GM}{r}\right)^{-1},~~
\frac{\d M}{\d r}=4\pi r^2\varepsilon,
\end{align}
here the mass $M=M(r)$, pressure $P=P(r)$ and energy density $\varepsilon=\varepsilon(r)$ are functions of the distance $r$ from NS center. The TOV equations are obviously nonlinear. They are traditionally solved numerically by selecting a central pressure to start the integration towards the surface of a NS of radius $R$ where the pressure $P(R)=0$ for a given input EOS $P(\varepsilon)$.

The central energy density $\epsc$ is an especially important quantity. It is straightforwardly connected to the central pressure $\Pc$ via the EOS $\Pc=P(\epsc)$.
Using $\epsc$, we can construct respectively a mass scale $W$ and a length scale $Q$ as:
\begin{equation}\label{RE-WQ}
W=\frac{1}{G}\frac{1}{\sqrt{4\pi G\varepsilon_{\rm{c}}}}
=\frac{1}{\sqrt{4\pi\varepsilon_{\rm{c}}}}
,~~Q=\frac{1}{\sqrt{4\pi G\epsc}}=\frac{1}{\sqrt{4\pi\varepsilon_{\rm{c}}}},
\end{equation}
the second steps in the above relations are taken with $G=1$.

The above TOV equations could then be recast in the following dimensionless (scaled) forms\,\cite{CLZ23-a,CLZ23-b,CL24-a,CL24-b,CL24-c},
\begin{equation}\label{TOV-ds}
\boxed{
\frac{\d\widehat{P}}{\d\widehat{r}}
=-\frac{\widehat{\varepsilon}\widehat{M}}{\widehat{r}^2}
\frac{(1+{\widehat{P}}/{\widehat{\varepsilon}})
(1+{\widehat{r}^3\widehat{P}}/{\widehat{M}})}{1-
{2\widehat{M}}/{\widehat{r}}},~~\frac{\d\widehat{M}}{\d\widehat{r}}=\widehat{r}^2\widehat{\varepsilon},}
\end{equation}
where $\widehat{P}=P/\epsc$, $\heps=\varepsilon/\varepsilon_{\rm{c}}$, $\widehat{r}=r/Q$ and $\hM=M/W$.
The terms on the right hand side of the pressure evolution equation can be classified as the following\,\cite{Sil2004,CL24-a},
\begin{enumerate}[label=(\alph*)]
\item The front factor $-{\widehat{\varepsilon}\widehat{M}}/{\widehat{r}^2}$ is for Newtonian stars\,\cite{Chan10-a} under hydrostatic equilibrium conditions, and the evolution equations becomes at this limit as,
\begin{equation}\label{RE-NewtonPre}
\frac{\d\hP}{\d\hr}=-\frac{\hM\heps}{\hr^2},~~\frac{\d\hM}{\d\hr}=\hr^2\heps.
\end{equation}
\item The two terms in the numerator represent special relativity (SR) corrections and the ratio $\widehat{P}/\widehat{\varepsilon}=\phi$ is a pure matter effect (due to the absence of $\widehat{r}$), and it is zero if $\widehat{P}=0$ is taken. The $\widehat{r}^3\widehat{P}/\widehat{M}$ is the coupling between matter (characterized by $\widehat{P}$) and geometry (by $\widehat{r}^3/\widehat{M}$), which also vanishes if $\widehat{P}=0$. 
\item The denominator $1-2\widehat{M}/\widehat{r}$ is a General Relativity (GR) effect; it remains even when $\widehat{P}$ vanishes on the surface of NS. The factor $2\widehat{M}/\widehat{R}$ can be sizable for massive and compact NSs, making the GR factor $(1-2\widehat{M}/\widehat{R})^{-1}$ in the TOV equations large, here $\widehat{R}$ is the reduced radius of a NS defined via the termination condition
\begin{equation}\label{RE-term}
P(R)=0\leftrightarrow\hP(\widehat{R})=0.
\end{equation}
Similarly, in terms of the reduced variables the NS mass is given by
\begin{equation}\label{Mrr}
M_{\rm{NS}}=\widehat{M}_{\rm{NS}}W,~~\mbox{with}~~\widehat{M}_{\rm{NS}}\equiv \widehat{M}(\widehat{R})=\int_0^{\widehat{R}}\d\hr\hr^2\widehat{\varepsilon}(\hr).
\end{equation}
\end{enumerate}

Before discussing the scales of variables in the dimensionless TOV equations and their corresponding perturbative treatments, we may infer important properties of NSs directly from the equations of (\ref{TOV-ds}) without solving them numerically. In particular, we would like to first investigate if the TOV equations themselves can put fundamental restrictions on (1) the core EOS of NS matter and (2) the radial-dependence of the relevant quantities in NSs. Summarized in the following of this subsection and the next subsection are our results.

Firstly, we can analytically solve the dimensionless TOV equations (\ref{TOV-ds}) for the linear EOS $P=\zeta\varepsilon$ or equivalently $\hP=\zeta\heps$, here $\zeta$ is a constant.
The results are given by,
\begin{align}\label{hPeps-linear}
\hP(\hr)=\frac{1}{\hr^2}\frac{2\zeta^2}{1+6\zeta+\zeta^2},~~\heps(\hr)=\frac{1}{\hr^2}\frac{2\zeta}{1+6\zeta+\zeta^2},~~\hM(\hr)=\frac{2\zeta\hr}{1+6\zeta+\zeta^2}.
\end{align}
Obviously, both the reduced pressure and energy density diverge at the NS center $\hr=0$. On the other hand, the mass enclosed even in a very small sphere with radius $\widehat{r}$ near the center is finite, and this is because the singularity in $\heps$ with respect to $\hr$ is removed by the volume integration $\widehat{M}
\sim\int\heps\d\v{x}\sim\int\heps \hr^2\d\hr\sim\hr$. {\color{xll}Thus, overall a linear EOS in the form of $P=\zeta\varepsilon$ is fundamentally inconsistent with the nature of TOV equations themselves describing NSs at hydrodynamical equilibrium, at least near the NS center.}
Consider an ultra-relativistic Fermi gas (URFG) as an example, which could be used to approximately describe quark matter at extremely high densities above about $40\rho_0$\,\cite{Bjorken83,Kur10}, its EOS is given by
\begin{equation}\label{EOS-URFG}
P=\varepsilon/3.
\end{equation}
According to (\ref{hPeps-linear}), we have $\hP=1/14\hr^2$ and $\heps=3/14\hr^2$, both approaching $\infty$ for $\widehat{r}=0$.
This means although the approximate conformal symmetry of quark matter may be realized theoretically at these very high densities\,\cite{Ann23},  the latter could not be used to describe the dense matter in NS cores. Of course, to our best knowledge, there is no evidence indicating that densities close to $40\rho_0$ can be realized in any NS as we know it presently. 

While the expressions in (\ref{hPeps-linear}) show the diverging behavior of the reduced pressure and energy density near $\hr=0$, globally, the M-R relation of a NS is given by the last one of (\ref{hPeps-linear}) as\,\cite{Lightman1975},
\begin{equation}\label{MR-UST}
\hM_{\rm{NS}}=3\widehat{R}/14\leftrightarrow M_{\rm{NS}}={3R}/{14G},
\end{equation}
with $\widehat{R}$ the reduced NS radius with respect to $Q$.
Numerically, this gives
\begin{equation}\label{MR-URFG}
R/\rm{km}\approx6.9M_{\rm{NS}}/M_{\odot},~~\mbox{for an URFG}.
\end{equation}
Considering the PSR J0740+6620 with a mass about $2.08M_{\odot}$\,\cite{Fon21},  Eq.\,(\ref{MR-URFG}) gives $R\approx 14.3\,\rm{km}$, which is merely consistent with the 68\% upper boundary of the observational radius of this NS\,\cite{Salmi22,Salmi24,Riley21,Miller21,Ditt24}.

Next, let's examine if a constant shift by $\widehat{\Phi}$ of $\hP=\zeta\heps$ can remove the unphysical singularity in the radial profile of pressure. The generalized linear EOS can be written as 
\begin{equation}\label{tk-0}
\hP=\zeta\heps+\widehat{\Phi},~~\mbox{where }\widehat{\Phi}=\hP_{\rm{c}}-\zeta\heps_{\rm{c}}=\hP_{\rm{c}}-\zeta.
\end{equation}
Because of the basic requirement $0\leq \widehat{P}_{\rm{c}}\leq1$, we have $-\zeta\leq\widehat{\Phi}\leq1-\zeta$. Thus, the magnitude of the constant $\widehat{\Phi}$ is smaller than 1, e.g., for $\zeta=1/3$, we have $-1/3\leq\widehat{\Phi}\leq2/3$.
The smallness of $\widehat{\Phi}$ implies we can develop relevant expansions in terms of it.
{\color{xll} We also notice that the linear EOS of Eq.\,(\ref{tk-0}) predicts a constant-speed-of-sound (CSS), i.e., $s^2=\rm{const.}=\zeta$.}
In the following, we provide two pieces of evidence on the inconsistency between the TOV equations and the generalized linear EOS of the form (\ref{tk-0}). In this subsection, we show that the TOV equations with (\ref{tk-0}) still generate singularities for $\hP(\hr)$ and $\heps(\hr)$ near $\hr\approx0$. 
 Then, in the following sections we show that the central SSS $s_{\rm{c}}^2$ is not a constant (see Eq.\,(\ref{sc2-TOV}) or Eq.\,(\ref{sc2-GG})). Since only a linear EOS of the form (\ref{tk-0}) would give a constant $s_{\rm{c}}^2$, therefore this form of the EOS is completely excluded.

For example, as $|\widehat{\Phi}|$ is generally smaller than 1, we can obtain straightforwardly to linear order in $\widehat{\Phi}$:
\begin{align}
\hP(\hr)\approx&\frac{1}{\hr^2}\frac{2\zeta^2}{1+6\zeta+\zeta^2}+\frac{2(1+2\zeta)\widehat{\Phi}}{(1+3\zeta)(2+\zeta)}
,\label{kt-1}\\
\heps(\hr)\approx&\frac{1}{\hr^2}\frac{2\zeta}{1+6\zeta+\zeta^2}-\frac{3(1+\zeta)\widehat{\Phi}}{(1+3\zeta)(2+\zeta)},\label{kt-2}\\
\hM(\hr)\approx&\frac{2\zeta\hr}{1+6\zeta+\zeta^2}-\frac{(1+\zeta)\widehat{\Phi}\hr^3}{(1+3\zeta)(2+\zeta)},\label{kt-3}
\end{align}
here each term on the right hand side is given in (\ref{hPeps-linear}).
For $\zeta=1/3$, the corrections to pressure, energy density and mass are $5\widehat{\Phi}/7$, $-6\widehat{\Phi}/7$ and $-2\widehat{\Phi}\hr^3/7$, respectively; so $\widehat{P}(\hr)\approx1/14\hr^2+4\widehat{\Phi}/7$, $\heps(\hr)\approx3/14\hr^2-6\widehat{\Phi}/7$ and $\hM(\hr)\approx3\hr/14-2\widehat{\Phi}\hr^3/7$.
Thus, it is clear that the finite-constant $\widehat{\Phi}$ can not remove the singularities of $\hP(\hr)$ and $\heps(\hr)$. This implies that the linear EOS (\ref{tk-0}) is also basically inconsistent with the nature of TOV equations.
Artificially taking $\zeta=0$, the above expressions reduce to $\hP=\widehat{\Phi}$,  $\heps=-3\widehat{\Phi}/2$ and $\hM=-\widehat{\Phi}\hr^3/2$. The latter two are obviously unphysical if $\widehat{\Phi}$ is positive (though the relation $\d\hM/\d\hr=\heps \hr^2$ still holds); the $\zeta=0$ in this case also means the term $\zeta\heps$ in $\hP$ is not allowed.

One can show the above linear approximations are exact if $\widehat{\Phi}=1$, i.e.,
\begin{equation}\label{ext-1}
\boxed{
\mbox{exact solution for linear EOS $\hP=\heps+\widehat{\Phi}$:}~~
\hP(\hr)=\frac{1}{4\hr^2}+\frac{\widehat{\Phi}}{2},~~\heps(\hr)=\frac{1}{4\hr^2}-\frac{\widehat{\Phi}}{2},
~~\hM(\hr)=\frac{\hr}{4}-\frac{\widehat{\Phi}\hr^3}{6}.}
\end{equation}
Obviously, both the pressure and energy density are singular at $\hr=0$.
The ratio $\phi=P/\varepsilon=\hP/\heps$ is approximated as $\phi\approx1+4\widehat{\Phi}\hr^2$ which is generally smaller than 1 since $\widehat{\Phi}<0$.
Moreover, the compactness is:
\begin{equation}
\xi=
{\hM}/{\widehat{R}}\approx4^{-1}\left(1-{32\hM^2\widehat{\Phi}}/{3}\right)\approx4^{-1}\left(1-{2\widehat{R}^2\widehat{\Phi}}/{3}\right),
\end{equation}
i.e., a small negative $\widehat{\Phi}$ reduces $\widehat{R}$ and therefore increases the compactness $\xi$.  Physically, this is because a negative $\widehat{\Phi}$ reduces the pressure and makes the attractive gravity more apparent compared with $\widehat{\Phi}=0$.

\begin{figure}[h!]
\centering
\includegraphics[width=9.cm]{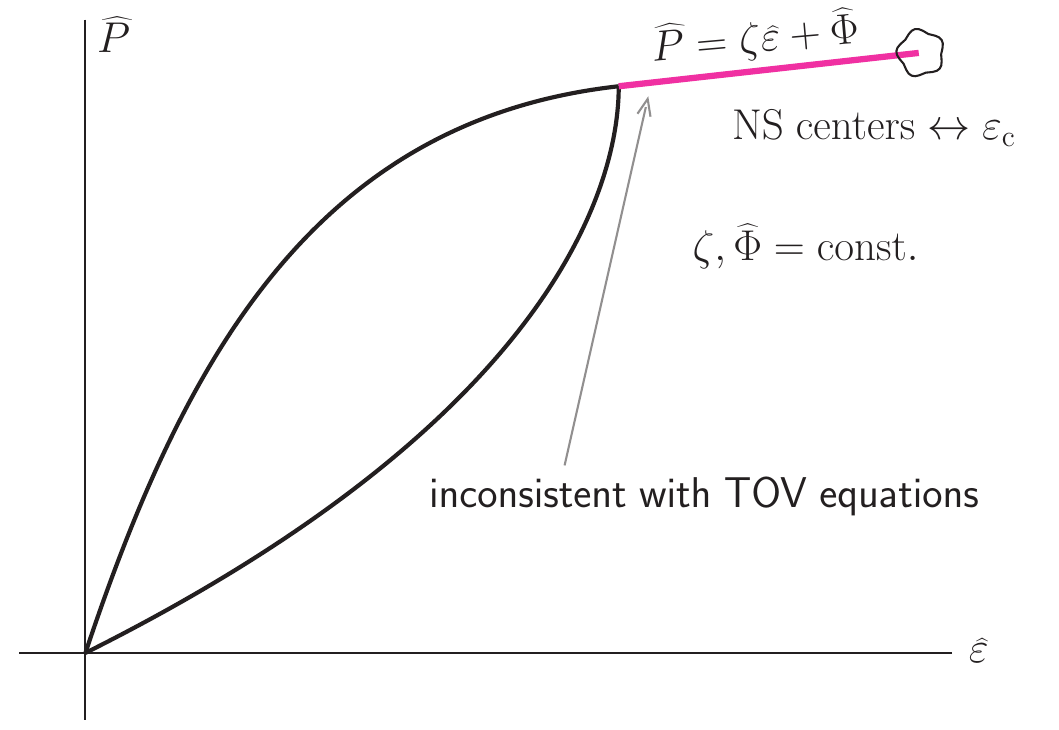}
\caption{(Color Online). The linear EOS $\hP=\zeta\heps+\widehat{\Phi}$ near NS centers is inconsistent with the TOV equations.}\label{fig_s2const}
\end{figure}

Without technical difficulties, we may find that the profiles for $\heps$ and $\hP$ still have singularities when higher order terms in $\widehat{\Phi}$ are included.
{\color{xll}Consequently, since the linear EOSs $\hP=\zeta\heps$ and its extension $\hP=\zeta\heps+\widehat{\Phi}$ (with $\zeta$ and $\widehat{\Phi}$ being constants) are basically inconsistent for describing the dense matter (especially) near NS centers, as sketched in FIG.\,\ref{fig_s2const},  and the causality requirement on the SSS as $s^2\leq1$ is equivalent to $P/\varepsilon\leq1$ only for the linear EOS $\hP=\zeta\heps$, the upper limit for $P/\varepsilon$ should be refined to be smaller than 1\,\cite{CL24-c}.}
This is because the EOS in NS cores could significantly be nonlinear, we discuss this and related issues in more details in SECTION \ref{SEC_56}.
Our discussions above also demonstrate {\color{xll}that a constant-sound-speed model is excluded by the TOV equations for describing NS cores.} This finding should necessarily be taken into account in building NS EOS models.

\subsection{Characteristic scales in the TOV equations and the double-element expansions of scaled variables}\label{sub_Char}

Next, we investigate the properties of $\hP$, $\heps$ and $\hM$ under the transformation $\hr\to-\hr$\,\cite{CL24-a}.
Actually, we have found only the even order terms in $\hr$ are allowed in $\hP$ and $\heps$ from discussions in the previous subsection, though the linear EOS $\hP=\phi\heps+\widehat{\Phi}$ leads to unphysical solutions.
The reduced NS mass $\widehat{M}$ as a function of radial distance $\widehat{r}$ is:
\begin{align}\label{ok-1}
\widehat{M}(\widehat{r})=\int_0^{\widehat{r}}\d xx^2\widehat{\varepsilon}(x),
\end{align}
here $x$ is an integration variable.
Under the coordinate transformation $
x\to-x$,  we have $
\d x\to-\d x$, $x^2\to x^2$ and $\heps(x)\to\heps(-x)$,
then the mass $\widehat{M}(\widehat{r})$ transforms as,
\begin{equation}\label{ok-2}
\widehat{M}(\widehat{r})\to
-\int_0^{-\widehat{r}}\d xx^2\widehat{\varepsilon}(-x).
\end{equation}
On the other hand,  we have straightforwardly from Eq.\,(\ref{ok-1}) by inverting the self-variable $\hr$ as,
\begin{equation}\label{ok-3}
\widehat{M}(-\widehat{r})=\int_0^{-\widehat{r}}\d xx^2\widehat{\varepsilon}(x).
\end{equation}
Similarly, starting from the pressure evolution equation, we shall obtain
\begin{align}
\widehat{P}(\widehat{r})=-&\int_0^{-\widehat{r}}\d x\frac{\widehat{\varepsilon}(x)\widehat{M}(x)}{x^2}
\frac{[1+\widehat{P}(-x)/\widehat{\varepsilon}(x)][1+x^3\widehat{P}(-x)/\widehat{M}(x)]}{1-2\widehat{M}(x)/x},\label{kq-1}\\
\widehat{P}(-\widehat{r})=-&\int_0^{-\widehat{r}}\d x\frac{\widehat{\varepsilon}(x)\widehat{M}(x)}{x^2}
\frac{[1+\widehat{P}(x)/\widehat{\varepsilon}(x)][1+x^3\widehat{P}(x)/\widehat{M}(x)]}{1-2\widehat{M}(x)/x},\label{kq-2}
\end{align}
by changing the integration variable $x$ and the self-variable $\hr$, respectively.

In order that both Eq.\,(\ref{ok-2}) and Eq.\,(\ref{ok-3}) hold, only two possibilities exist: (a)
$\widehat{\varepsilon}(-x)=\widehat{\varepsilon}(x)$ and $\widehat{M}(-x)=-\widehat{M}(x)$,  or (b)
 $\widehat{\varepsilon}(-x)=-\widehat{\varepsilon}(x)$ and $\widehat{M}(-x)=\widehat{M}(x)$.
Since we have the physical requirement that $\widehat{\varepsilon}(0)=1$ at $\widehat{r}=0$, only the possibility (a) above is allowed as the option (b) would lead to $\widehat{\varepsilon}(0)=0$ that is unphysical. This means that $
\widehat{\varepsilon}(\widehat{r})=\widehat{\varepsilon}(-\widehat{r})$ and $\widehat{M}(\widehat{r})=-\widehat{M}(-\widehat{r})$.
Then in order to make (\ref{kq-1}) and (\ref{kq-2}) be consistent with each other, only one possibility for the $\hr$-dependence of $\hP$ remains, $
\widehat{P}(\widehat{r})=\widehat{P}(-\widehat{r})$.
These analyses show that $\widehat{\varepsilon}$ and $\widehat{P}$ are even functions of $\widehat{r}$ and $\widehat{M}$ is an odd function of $\widehat{r}$, though physically $\widehat{r}$ is non-negative.
In fact, the structure of $\hM$ as a function of $\hr$ could be seen immediately from the mass evolution equation, i.e., the evenness of $\heps$ implies that $\hM$ is an odd function of $\hr$ and consequently $\hM(0)=0$.
Therefore, we have the following expansions for $\widehat{\varepsilon}$,  $\widehat{P}$ and $\widehat{M}$ near $\widehat{r}=0$:
\begin{empheq}[box=\fbox]{align}
\heps(\hr)\approx&1+a_2\hr^2+a_4\hr^4+a_6\hr^6+\cdots,\label{ee-heps}\\
\hP(\hr)\approx&\x+b_2\hr^2+b_4\hr^4+b_6\hr^6+\cdots,\label{ee-hP}\\
\hM(\hr)\approx &\frac{1}{3}\hr^3+\frac{1}{5}a_2\hr^5+\frac{1}{7}a_4\hr^7+\frac{1}{9}a_6\hr^9+\cdots,\label{ee-hM}
\end{empheq}
here $\x$ is the ratio of central pressure over energy density 
\begin{equation}\label{RE-small2}
\boxed{
\x\equiv\widehat{P}_{\rm{c}}\equiv P_{\rm{c}}/\varepsilon_{\rm{c}}.}
\end{equation}
In this review,
we may use $\x$ to denote the ratio $P_{\rm{c}}/\varepsilon_{\rm{c}}$ as much as possible to avoid notation confusion (captions in some figures may still use $\widehat{P}_{\rm{c}}$).
As a direct consequence, we find that $s^2(\hr)=s^2(-\hr)$, i.e., there would be no odd terms in $\hr$ in the expansion of $s^2$.
We notice that in Ref.\,\cite{Ecker2022}, the authors parametrized their SSS as a function of $\hr$ by fitting the inferred results of $s^2$ for different values of NS mass. However, their $s^2$ parameterization contains odd terms in $\hr$.

For a typical NS central energy density of
$
\varepsilon_{\rm{c}}=500\,\rm{MeV}/\rm{fm}^3$,
we obtain $W\approx Q\approx 11\,\rm{km}$\,\cite{CLZ23-b} according to (\ref{RE-WQ}) using $\rm{MeV}/\rm{fm}^3\approx1.32\times10^{-6}\,\rm{km}^{-2}$ in units of $c=G=1$.
Considering a $2M_{\odot}$ NS with a radius about $R\approx 12\,\rm{km}$, we have
\begin{equation}\label{RE-WQnum}
\widehat{M}_{\rm{NS}}=M_{\rm{NS}}/W\approx0.18,~~
\widehat{R}=R/Q\approx1.1,
\end{equation}
here $M_{\rm{NS}}\equiv M(R)$ is the NS mass.
The $\widehat{R}\sim\mathcal{O}(1)$ means that the expansions of relevant quantities over $\hr$ are safe within a wide range of radial distance if the coefficients $\{a_k\}$ and $\{b_k\}$ are normal,   although we mainly focus on the small-$\hr$ expansion, i.e., quantities near the NS center.
On the other hand,  the two estimates of (\ref{RE-WQnum}) together indicate that $a_2$ should be negative, which is a natural requirement as $\heps$ is a monotonically decreasing function of $\hr$ when going out from the center.
We will show that $a_2<0$ in the next subsection by perturbatively solving the dimensionless TOV equations.
The smallness of $\hr$ near NS center is equivalent to the smallness of the relative energy density 
\begin{equation}\label{RE-small1}
\boxed{
\mu\equiv \heps-\widehat{\varepsilon}_{\rm{c}}=\heps-1<0.}
\end{equation}
As discussed in the previous subsection, the nonlinearity of the EOS in NS cores implies that the ratio $\phi=P/\varepsilon$ is smaller than 1\,\cite{CL24-c}, and in particular we have for its central value $\x<1$ for the $\x$ defined in Eq.\,(\ref{RE-small2}).
Combining (\ref{RE-small1}) and (\ref{RE-small2}) enables us to develop a general scheme for perturbative double expansions of a NS quantity $\mathcal{U}$ over $\mu$ (or equivalently over $\hr$) and $\x$\,\cite{CL24-c}:
\begin{equation}\label{double-exp}
\boxed{
\mathcal{U}/\mathcal{U}_{\rm{c}}\approx1+\sum_{i+j\geq1}u_{ij}\x^i\mu^j,}
\end{equation}
here $\mathcal{U}_{\rm{c}}$ is the corresponding quantity at the center.
Knowing the coefficients $\{u_{ij}\}$ enables us to reconstruct the $\mathcal{U}/\mathcal{U}_{\rm{c}}$.
This double-element perturbative expansion becomes eventually exact as $\mu\to0$ or $\hr\to0$, providing us a unique and reliable approach to study the EOS of NS cores.
In particular, we will give the central EOS in SECTION \ref{SEC_4},  e.g., the radial dependence of $P/P_{\rm{c}}$ in Subsection \ref{sub_DEPc}, by working out its expansion over $\mu$ and $\x$.

Here we want to emphasize that $\x$ is an important dimensionless quantity for NSs. It combines the central pressure and energy density and the relation of them is determined by the EOS (or equivalently, the $\x$ defines the EOS). So the $\x$ characterizes NS {\color{xll}microscopic} properties. Another important dimensionless quantity for NSs is the compactness $\xi\equiv M_{\rm{NS}}/R$, which on the other hand characterizes obviously some NS {\color{xll}macroscopic} properties. Based on an order-of-magnitude estimate, we expect as both $\xi$ and $\x$ are smaller than 1 that,
\begin{equation}\label{rel-xi-x}
\xi=\tau(\x)\approx \tau_1\x+\tau_2\x^2+\cdots,
\end{equation}
where the dimensionless coefficient $\tau_i\sim\mathcal{O}(1)$. Naturally, the TOV equations give us such macroscopic-microscopic connection.
In addition, the SSS $s^2$ of Eq.\,(\ref{def-s2}) is also dimensionless. Using a similar dimension analysis and order-of-magnitude estimate, we may figure out important features of $s^2$ based soly on the TOV equations without using any nuclear EOS model. This analysis will be given in Subsection \ref{sub_s2ORDER}.

\subsection{Perturbative expansions of energy density, pressure and mass functions near NS centers}\label{sub_DEP123}

\begin{figure}[h!]
\centering
\includegraphics[height=7.cm]{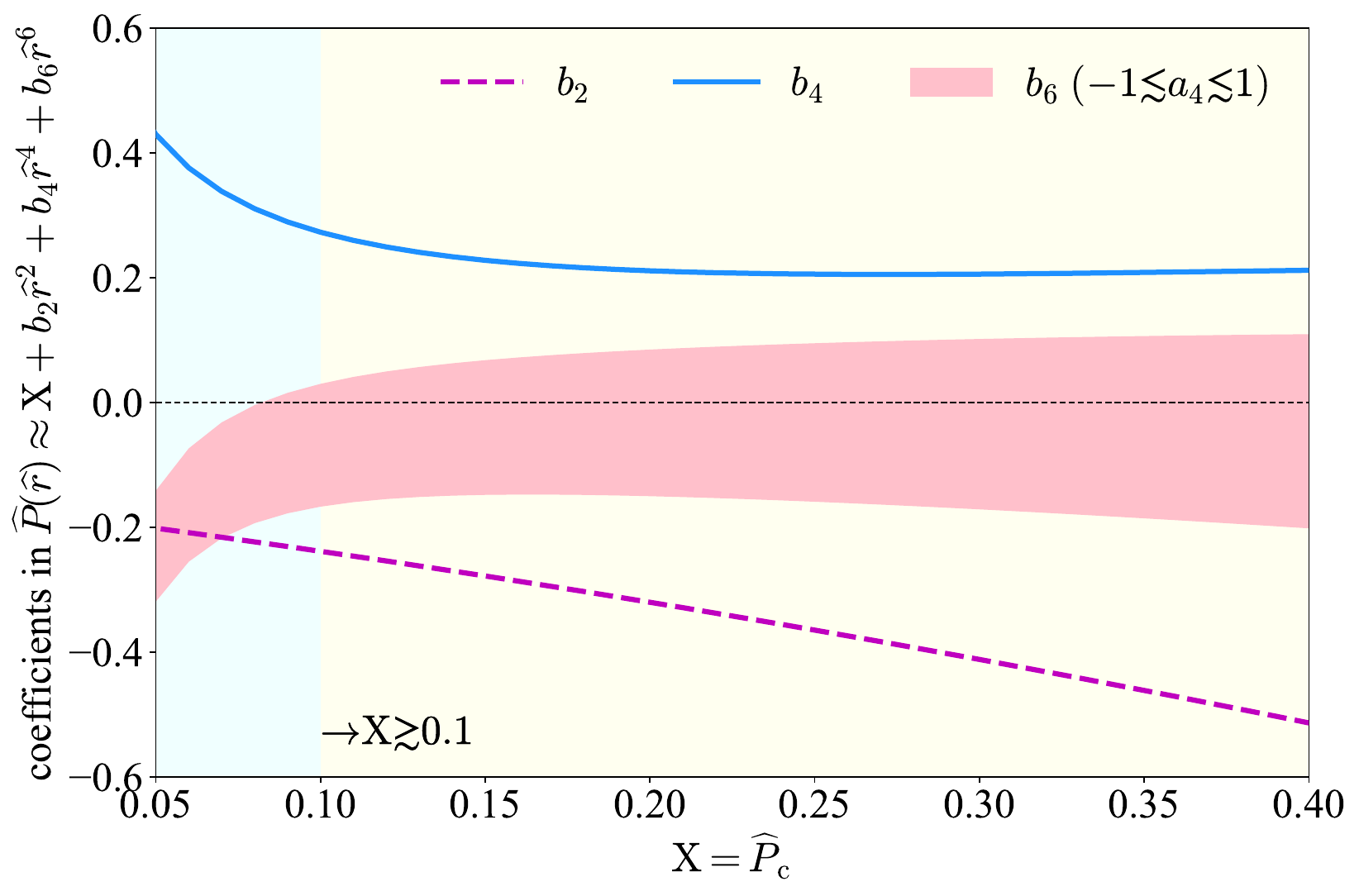}
\caption{(Color Online).  The $\x$-dependence of coefficients $b_2$, $b_4$ and $b_6$.
}\label{fig_b2b4b6}
\end{figure}

The relationships between the coefficients $\{a_k\}$ and $\{b_k\}$ could be determined by the pressure evolution in the TOV equations.
The results are\,\cite{CLZ23-a}
\begin{empheq}[box=\fbox]{align}
b_2=&-\frac{1}{6}\left(1+4\x+3\x^2\right),\label{ee-b2}\\
b_4=&-\frac{2a_2}{15}+\left(\frac{1}{12}-\frac{3}{10}a_2\right)\x+\frac{1}{3}\x^2+\frac{1}{4}\x^3,\label{ee-b4}\\
b_6=&-\frac{1}{216}-\frac{a_2^2}{30}-\frac{a_2}{54}-\frac{5a_4}{63}+\left(\frac{a_2}{45}-\frac{4a_4}{21}-\frac{1}{54}\right)\x+\left(\frac{2a_2}{15}-\frac{1}{18}\right)\x^2-\frac{1}{6}\x^3-\frac{1}{8}\x^4,\label{ee-b6}
\end{empheq}
etc., and all the odd terms of $\{b_k\}$ and $\{a_k\}$ are zero.
The coefficient $a_2$ can be expressed in terms of $b_2$ via the SSS:
\begin{equation}\label{s2_r_exp}
s^2=\frac{\d\widehat{P}}{\d\widehat{\varepsilon}}=\frac{\d\widehat{P}}{\d\widehat{r}}\cdot\frac{\d\widehat{r}}{\d\widehat{\varepsilon}}=\frac{b_2+2b_4\widehat{r}^2+\cdots}{a_2+2a_4\widehat{r}^2+\cdots}.
\end{equation}
Evaluating it at $\hr=0$ gives $s_{\rm{c}}^2=b_2/a_2$,  or inversely 
\begin{equation}\label{def-a2x}
a_2=b_2/s_{\rm{c}}^2.
\end{equation}
Since $s_{\rm{c}}^2>0$ and $b_2<0$, we find $
a_2<0$, i.e., the energy density is a monotonically decreasing function of $\hr$ near $\hr\approx0$.
The coefficient $b_4$ could be rewritten as\,\cite{CL24-a}
\begin{equation}
\label{ee-b4-another}
\boxed{
b_4=-\frac{1}{2}b_2\left(\x+\frac{4+9\x}{15s_{\rm{c}}^2}\right),}
\end{equation}
which is definitely positive.
On the other hand, the sign of $a_4$ is undetermined. It is very relevant for the onset of a peaked behavior in $s^2$ as a function of $\hr$. We shall discuss this issue in details in SECTION \ref{SEC_7}. 
We find since $|b_2|\lesssim\mathcal{O}(1)$ and $s_{\rm{c}}^2\lesssim\mathcal{O}(1)$ near NS centers that $b_4\lesssim\mathcal{O}(1)$. In fact, the magnitudes of all these expansion coefficients are expected to be $\lesssim \mathcal{O}(1)$ as shown in FIG.\,\ref{fig_b2b4b6} for their $\x$-dependence, here the expression for $s_{\rm{c}}^2$ is given in Eq.\,(\ref{sc2-TOV}).
Notice that the $\x$ in realistic NSs is generally greater than about 0.1.
The smallness of $b_6$ adopting $-1\lesssim a_4\lesssim1$ (which is quite close to zero as we shall show) implies that we can well predict the $\hr$-dependence of $\hP$ near $\hr\approx0$. The result is expected to be (nearly) independent of the EOS models, see Subsection \ref{sub_DEPc} for more detailed discussion and the related FIG.\,\ref{fig_s2_prep_r}.

Next, we estimate the matter and geometry corrections in the dimensionless TOV equations of (\ref{TOV-ds}). Using the $s_{\rm{c}}^2$ of Eq.\,(\ref{sc2-GG}), we obtain around $\widehat{r}\approx0$:
\begin{align}
    \hP/\heps\approx&\x-\frac{1}{6}\frac{1+\Psi}{4+\Psi}\left[1
    +\frac{7+\Psi}{4+\Psi}\cdot4\x+\frac{\Psi^2+14\Psi+88}{(4+\Psi)^2}\cdot3\x^2
    \right]\hr^2,\\
    \hr^3\hP/\hM\approx&3\x-\frac{1}{10}\frac{11+5\Psi}{4+\Psi}\left[
    1+\frac{5\Psi^2+40\Psi+53}{5\Psi^2+31\Psi+44}\cdot4\x+\frac{5\Psi^3+69\Psi^2+402\Psi+392}{(11+5\Psi)(4+\Psi)^2}\cdot3\x^2
    \right]\hr^2,\\
    2\hM/\hr\approx&\frac{2}{3}\hr^2\left[1-\frac{3}{10\x}\frac{1}{4+\Psi}
    \left[1+\frac{3}{4+\Psi}\cdot4\x-\frac{\Psi^2+18\Psi+8}{(4+\Psi)^2}\cdot3\x^2\right]\hr^2
    \right],
\end{align}
where $\Psi\equiv 2\d\ln M_{\rm{NS}}/\d\ln\varepsilon_{\rm{c}}\geq0$ is defined in Eq.\,(\ref{def-Psi}).
If $\Psi=0$ is taken, then corresponding to the classifications given below Eq.\,(\ref{TOV-ds}), the matter, matter-geometry coupling and GR geometry term is, respectively,
\begin{align}
    \hP/\heps\approx&\x-\frac{1}{24}\left(1+7\x+\frac{33}{2}\x^2\right)\hr^2,\\
    \hr^3\hP/\hM\approx&3\x-\frac{11}{40}\left(1+\frac{53}{11}\x+\frac{147}{22}\x^2\right)\hr^2,\\
     2\hM/\hr\approx&\frac{2}{3}\hr^2\left[1-\frac{3}{40\x}\left(1+3\x-\frac{3}{2}\x^2\right)\hr^2\right].
\end{align}
The ratio $\hP/\heps$ is discussed/reviewed in details in our recent work\,\cite{CL24-c} and would be briefly mentioned in SECTION \ref{SEC_56}.
These expressions are in the general expansion form of (\ref{double-exp}), from them we can find that: (a) near the NS centers both $\hP/\heps$ and $\hr^3\hP/\hM$ could be sizable (compared to ``1''). In particular, the matter-geometry coupling $\hr^3\hP/\hM$ dominates over the matter correction $\hP/\heps$ by a factor of 3,  and (b) for finite $\hr$ values the GR term $2\hM/\hr$ becomes sizable.
These corrections together make the GR structure equations for NSs different from their Newtonian counterparts. They do have important impact on the SSS in NSs compared to Newtonian stars. We shall discuss these issues in some more details in SECTION \ref{SEC_7}.

In the Newtonian limit $\x\to0$, the coefficient $b_2$ is $-1/6$. This result can be obtained by expanding the solution of the Lane--Emden equation\,\cite{Chan10-a}. The latter is given by
\begin{equation}
\frac{1}{\omega^2}\frac{\d}{\d\omega}\left(\omega^2\frac{\d\theta}{\d\omega}\right)+\theta^n=0,
\end{equation}
where $\omega$ is the dimensionless radius related to our $\hr=r/Q$ by
\begin{equation}
\omega=\frac{\hr}{\sqrt{(n+1)\x}};
\end{equation}
the index $n$ appears in the polytropic EOS via 
\begin{equation}
P=K\varepsilon^{1+1/n}=K\varepsilon_{\rm{c}}^{1+1/n}\theta^{n+1}=P_{\rm{c}}\theta^{n+1},
\end{equation} so $P/P_{\rm{c}}=\theta^{n+1}$.
The boundary conditions for $\theta$ are $\theta(0)=1$ and $\theta'(0)=0$ with the derivative taken with respect to $\omega$.
For example, for the EOS with $n=0$, one has $\omega=\hr/\sqrt{\x}$ , $\theta(\omega)=1-\omega^2/6$ and therefore $P/P_{\rm{c}}=\theta(\omega)=1-\omega^2/6=1-\hr^2/6\x$. For $n=1$, we have $\omega=\hr/\sqrt{2\x}$ and the Lane-Emden equation has the solution $\theta(\omega)=\sin\omega/\omega$ and so $P/P_{\rm{c}}=\theta^2(\omega)\approx1-\omega^2/3+(2/45)\omega^4+\cdots\approx1-\hr^2/6\x+\hr^4/90\x^2+\cdots$. For $n=5$, we have $\omega=\hr/\sqrt{6\x}$ and the solution of the Lane--Emden equation is $\theta(\omega)=(1+\omega^2/3)^{-1/2}$ and therefore $P/P_{\rm{c}}\approx1-\omega^2+2\omega^4/3+\cdots
=1-\hr^2/6\x+\hr^4/54\x^2+\cdots$.
Generally, one has\,\cite{Chan10-a},
\begin{equation}
\theta(\omega)=1-\frac{1}{6}\omega^2+\frac{n}{120}\omega^4+\cdots,
\end{equation}
and therefore
\begin{equation}\label{LE-1}
P/P_{\rm{c}}=\theta^{n+1}(\omega)\approx1-\frac{1}{6\x}\hr^2+\frac{n}{n+1}\frac{1}{45\x^2}\hr^4+\cdots.
\end{equation}
Recalling the expansion (\ref{ee-hP}), we have
\begin{equation}
P/P_{\rm{c}}\approx1+\frac{b_2}{\x}\hr^2+\frac{b_4}{\x}\hr^4+\cdots
\approx1-\frac{1}{6\x}\hr^2+\frac{1}{45\x s_{\rm{c}}^2}\hr^4,
\end{equation}
where the second approximation is taken under the limit $\x=\widehat{P}_{\rm{c}}\to0$.
Using  the $s_{\rm{c}}^2$ of Eq.\,(\ref{sc2-GG}), we could further write $P/P_{\rm{c}}$ as
\begin{equation}\label{LE-2}
P/P_{\rm{c}}\approx1-\frac{1}{6\x}\hr^2+\frac{3}{4+\Psi}\frac{1}{45\x^2}\hr^4
\xrightarrow[]{\Psi=0}1-\frac{1}{6\x}\hr^2+\frac{1}{60\x^2}\hr^4,
\end{equation}
where $\Psi\geq0$, see Eq.\,(\ref{def-Psi}).
Comparing these expressions, we find our expression for the pressure reproduces the solution of Lane--Emden equation for a general index $n$ at order $\omega^2$ (or $\hr^2$). However, the higher-order terms (starting from $\omega^4$ or $\hr^4$) depend on the index $n$.
Comparing (\ref{LE-1}) and (\ref{LE-2}), we have 
\begin{equation}\label{n-Psi}
\boxed{
n=\frac{3}{1+\Psi};}
\end{equation}
thus $n\leq3$ or equivalently $1+n^{-1}\geq4/3$, indicating $(P/\varepsilon)^{-1}(\d P/\d\varepsilon)\geq4/3$ for the  polytropic EOS.
We may encounter several times that the index $s^2/(P/\varepsilon)$ should be larger than 4/3 in NSs in this review.
For NSs with $\Psi\gtrsim2$ (roughly corresponding to $M_{\rm{NS}}\lesssim2M_{\odot}$), see FIG.\,\ref{fig_k-fac}, we then have $n\lesssim1$.

Finally, we can construct our EOS by expanding the pressure $\widehat{P}$ over energy density $\widehat{\varepsilon}$ as\,\cite{CLZ23-a}
\begin{equation}\label{Peps-dk}
\widehat{P}=\sum_{k=1}d_k\widehat{\varepsilon}^k\approx d_1\widehat{\varepsilon}+d_2\widehat{\varepsilon}^2+d_3\widehat{\varepsilon}^3+\cdots,
\end{equation}
from which we obtain two sum rules for the coefficients $\{d_k\}$,
\begin{align}\label{Peps-sumrule}
\x=\sum_{k=1}d_k,~~
s_{\rm{c}}^2=\left.\frac{\d \widehat{P}}{\d\widehat{\varepsilon}}\right|_{\widehat{r}=0\leftrightarrow
\widehat{\varepsilon}_{\rm{c}}=1}=\sum_{k=1}kd_k.
\end{align}
These sum rules are useful for studying the $\hr$-dependence of $s^2$, to be discussed in details in Subsection \ref{sub_s2_1st}.

\setcounter{equation}{0}
\section{Brief Review of NS Matter EOS and Density Dependence of its Speed of Sound}\label{SEC_3}

This section is not meant to be a comprehensive review of all existing work on constraining the EOS in the literature. Instead, 
we briefly review some existing constraints on the supra-dense neutron-rich matter EOS in NSs and point out a few remaining major issues most relevant to the topic of this article. In particular, Subsection \ref{sub_REV_EOS} is devoted to a short summary of observational constraints on the EOS of superdense neutron-rich matter in NSs since GW170817\,\cite{Abbott2017,Abbott2018}.
In Subsection \ref{HDEsym}, we comment on the main findings about the high-density behavior of nuclear symmetry energy that has been widely recognized as the most uncertain term in the EOS of supra-dense neutron-rich nucleonic matter.
In Subsection \ref{sub_REV_s2} we outline our current understanding on $s^2$ and point out several major issues that our new approach for solving the scaled TOV equations may help address at least to some extent. 

\subsection{Observational constraints on the EOS of superdense neutron-rich matter}\label{sub_REV_EOS}
\begin{figure}[h!]
\centering
\includegraphics[height=8.5cm]{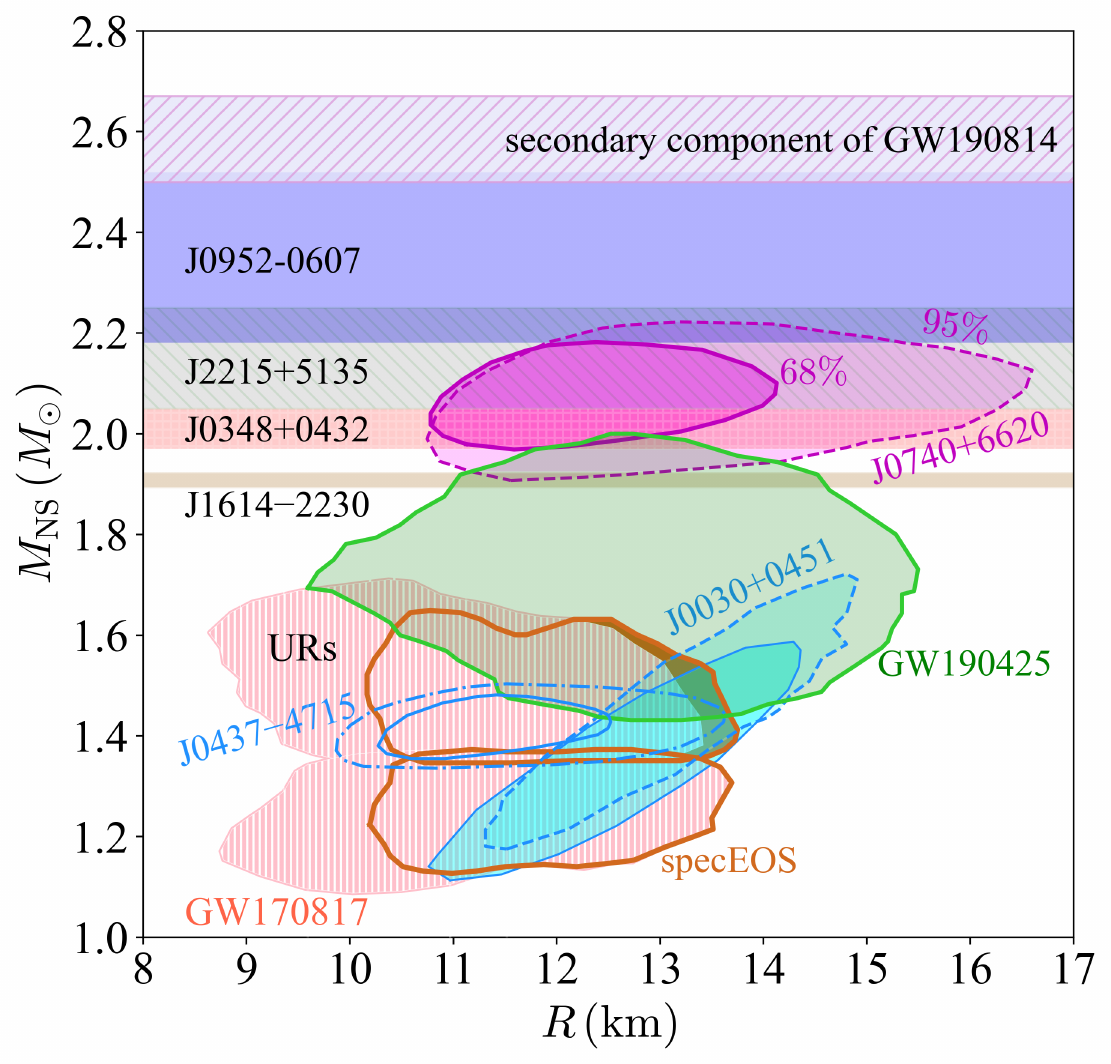}
\caption{(Color Online). A (partial) summary on NSs observed, these include the three GW events GW170817\,\cite{Abbott2017,Abbott2018}, GW190425\,\cite{Abbott2020-a} and GW190814\,\cite{Abbott2020}, the NICER mass-radius joint observations for PSR J0740+6620\,\cite{Riley21}, PSR J0030+0451\,\cite{Riley19} and PSR J0437-4715\,\cite{Choud24} (shown with the 68\% and 95\% confidence levels, respectively), the joint X-ray and optical study of redback pulsar PSR J2215+5135\,\cite{Sul24}, the black widow pulsar PSR 0952-0607, the mass of PSR J1614-2230\,\cite{Dem10,Arz18} using Shapiro delay and mass of PSR J0348+0432\,\cite{Ant13} via its spectroscopy. See the text for details and Ref.\,\cite{Tan2022-b} for a similar plot.}\label{fig_NSMR-REV}
\end{figure}

We first summarize in FIG.\,\ref{fig_NSMR-REV} several recently observed NSs (or the potential candidates), a similar plot was given in Ref.\,\cite{Tan2022-b}; here:
\begin{enumerate}[label=(\alph*)]
\item NICER's simultaneous mass-radius observations using the X-ray data from NASA's mission for three pulsars, namely PSR J0740+6620 with a mass about $2.08_{-0.07}^{+0.07}M_{\odot}$\,\cite{Fon21,Riley21,Miller21,Salmi22,Salmi24,Ditt24}, PSR J0030+0451 with a mass about $1.4M_{\odot}$\,\cite{Riley19,Miller19,Vin24} and PSR J0437-4715 with a mass about $1.418_{-0.037}^{+0.037}M_{\odot}$\,\cite{Choud24} (see also Ref.\,\cite{Reard24}).
\item A joint X-ray and optical (U-band) study of the massive redback pulsar PSR J2215+5135 with its mass about $2.15_{-0.10}^{+0.10}M_{\odot}$\,\cite{Sul24}.
\item Two GW events, i.e., GW170817\,\cite{Abbott2017,Abbott2018}and GW190425\,\cite{Abbott2020-a} give us constraints on the mass and radius under certain assumptions in their theoretical modelings, as well as the information of tidal deformability $\Lambda$; specifically, we have $R_1\approx R_2\approx11.9_{-1.4}^{+1.4}\,\rm{km}$ and $1.36\lesssim M_{\rm{NS}}^{(1)}/M_{\odot}\lesssim1.58$ and $1.18\lesssim M_{\rm{NS}}^{(2)}/M_{\odot}\lesssim1.36$ for GW170817 and GW190425\,\cite{Abbott2020-a} has $13.1\,\rm{km}\lesssim R_1\lesssim14.9\,\rm{km}$ with $1.6\lesssim M_{\rm{NS}}^{(1)}/M_{\odot}\lesssim1.9$ and $13.3\,\rm{km}\lesssim R_2\lesssim14.9\,\rm{km}$ with $1.5\lesssim M_{\rm{NS}}^{(2)}/M_{\odot}\lesssim1.7$ using the low-spin prior\,\cite{Abbott2021}.
\item Mass of PSR J1614-2230 about $1.908_{-0.016}^{+0.016}M_{\odot}$ using the technique of Shapiro delay\,\cite{Dem10,Arz18}.
\item Mass of  PSR J0348+0432 about $2.01_{-0.04}^{+0.04}M_{\odot}$ using its spectroscopy\,\cite{Ant13}.
\end{enumerate}

Among these NSs, the NICER results for PSR J0740+6620 and PSR J0030+0451 are very valuable since both the masses and radii are available. In particular, the former is extremely useful as it is among the most massive NSs observed so far.  
For example, Ref.\,\cite{Riley21} showed that either a very soft or a very stiff EOS is fundamentally excluded at a 68\% confidence level considering the joint mass-radius observation for PSR J0740+6620. Similar inference results on the NS matter EOS were also given by Ref.\,\cite{Miller21} via three models (polytrope, spectral and GP). Without surprise, incorporating PSR J0740+6620 essentially increases the pressure (compared with the case in which only PSR J0030+0451 is used for the inference), i.e., a stiffer EOS is necessary. 
In Ref.\,\cite{Ruther2024}, the M-R posterior distributions for the “Baseline” and “New” scenarios using the PP and CS models were given. In their studies, the constraints from N$^3$LO chiral effective field theory (CEFT or $\chi$EFT) band up to $1.5\rho_0$ and $1.1\rho_0$ was used.
The ``Baseline'' scenario uses observation information of Ref.\,\cite{Salmi22} for PSR J0740+6620, that of Ref.\,\cite{Vin24} for PSR J0030+0451; while the ``New'' scenario uses the results of Ref.\,\cite{Salmi24} for PSR J0740+6620, those of Ref.\,\cite{Vin24} for PSR J0030+0451 (including background constraints) and the information of Ref.\,\cite{Choud24} for the newly announced observation of PSR J0437+4715. Ref.\,\cite{Ruther2024} found that both the updated NICER's mass/radius measurements and the new CEFT inputs can provide effective constraints on the NS M-R relation.
On the other hand, we may point out that although GW190425 puts a weak constraint on the inference of NS radii, it may have effective influence on extracting the trace anomaly\,\cite{CL24-b}, which is also useful for constraining the NS matter EOS.

Recently, an observation of the black widow pulsar PSR J0952-0607 with a mass about $2.35_{-0.17}^{+0.17}M_{\odot}$ was announced \,\cite{Romani22}. It also has the the second fastest known spin rate (about 707\,Hz) among all the pulsars observed so far. Similarly, the secondary component of GW190814 was found to have a mass about $2.59^{+0.08}_{-0.09}M_{\odot}$\,\cite{Abbott2020}.
The PSR J0952-0607 and the secondary component of GW190814 are potential NS candidates. However, there exist contentious debates on both the nature and the formation mechanisms of these massive objects. This is especially obvious for the minor component of GW190814, see, e.g., Refs.\,\cite{Most2020MNRAS,ZhangLi2020ApJ,Tews2021,Nathanail2021,Godzieba2021,Bombaci2021,Fat2020PRC,Sed2020PRD,Dexh2021PRC,Lopes2022,Cao2022PRD}.

\begin{figure}[h!]
\centering
\includegraphics[height=6.5cm]{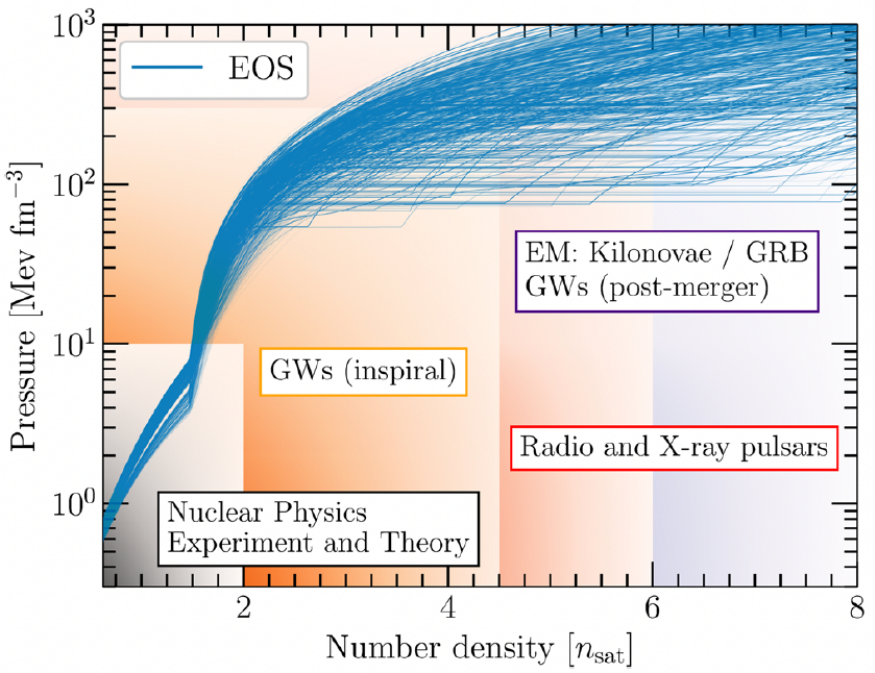}\quad
\includegraphics[height=7.cm]{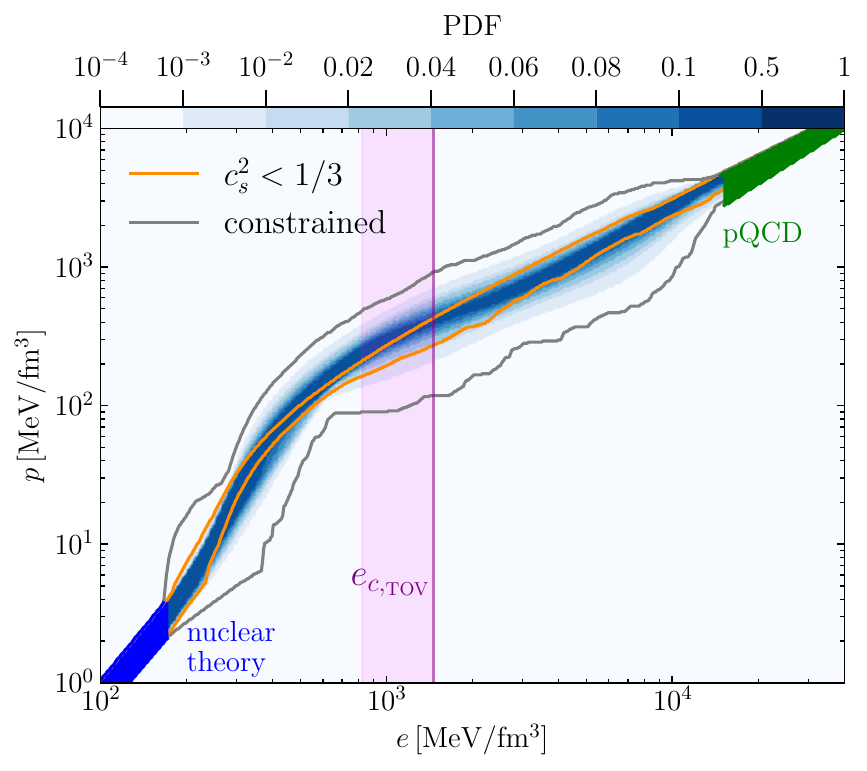}
\caption{(Color Online).  Left panel: overview of the constraints on dense matter EOS from various approaches/methods/theories for low to high densities. Figure taken from Ref.\,\cite{Pang2023}.
Right panel: a contemporary summary of constraints on supra-dense NS matter EOS, the EOS being relevant to realistic NSs (indicated by the vertical solid line) is still largely uncertain. Figure taken from Ref.\,\cite{Altiparmak2022}.
}\label{fig_RN-EOS-over}
\end{figure}

The EOS of NS matter from low to high densities could be constructed by using different methods/theories.
The left panel of FIG.\,\ref{fig_RN-EOS-over} gives an overview of this process\,\cite{Pang2023}:  (a) at relatively low densities $\lesssim2\rho_0$ with $\rho_0\approx0.16\,\rm{fm}^{-3}$ being the nuclear saturation density,  CEFT could provide reliable restrictions\,\cite{Lynn16,Holt17,Drischler19PRL,Leon20,Huth21,Keller23} for the EOS of nucleonic matter with an arbitrary proton fraction $x_{\rm{p}}$, although the uncertainty becomes gradually larger at supra-saturation densities\,\cite{Drischler2022PRC}; (b) available heavy-ion collision data\,\cite{Tsang12,Russ16PRC,Lynch22}
constrain the EOS of neutron-rich hadronic matter within a similar and/or somewhat larger density region compared to (a) above, while nuclear structure observables, e.g., neutron-skin thickness\,\cite{Adh21} and the electric dipole polarizability $\alpha_{\rm{D}}$\,\cite{Zhang15alpha} of heavy nuclei provide useful information on the EOS of neutron-rich nucleonic matter at sub-saturation densities; (c) GW signals emitted during the inspiral of a binary NS (BNS) contain new information about NS matter EOS at densities around $2\rho_0$, e.g., the GW170817\,\cite{Abbott2017,Abbott2018} puts effective constraints on the tidal deformability which in turn should limit the dense matter EOS around such densities; (d) heavy and massive NSs like PSR J0740+6620 (X-ray observation)\,\cite{Fon21,Riley21,Miller21,Salmi22,Salmi24,Ditt24}, PSR J0348+0432 (spectroscopy)\,\cite{Ant13}, PSR J1614-2230 (Shapiro delay)\,\cite{Dem10} and PSR J2215+5135 (X-ray+optical)\,\cite{Sul24} can potentially probe dense matter EOS at even larger densities up to about $8\rho_0$. They may also affect the extraction of the maximum mass of stable NSs; (e) at extremely (asymptotically) large densities $\gtrsim40\rho_0$ the EOS could practically be calculated by using the perturbative QCD\,\cite{Bjorken83,Kur10}, providing an asymptotical boundary condition for NS matter EOS\,\cite{Kur10,Kurk14,Kurk16,Gorda18,Gorda21,Vuorinen2024}.

Many interesting issues regarding NS matter EOS exist. For example, an approximate conformal symmetry of quark matter has been predicted at very large densities\,\cite{Bjorken83,Kur10}. Incorporating the pQCD effects may soften the NS matter EOS\,\cite{Gorda2023}, because the pressure at a given energy density is reduced if the conformal symmetry is realized\,\cite{Gorda2023}.
However, due to the extremely high density required to realize such conformal symmetry of quark matter, the application of pQCD predictions in NSs has been questioned repeatedly in the literature, see, e.g., Ref.\,\cite{Zhou2024}. On the other hand, 
low-density crust EOS\,\cite{BPS71} at about $\rho\lesssim\rho_0/2$ has little impact on NS mass. Nevertheless, its uncertain size may have appreciable influence on predicting accurately NS radii. Moreover, it is important for understanding possible pasta phases predicted to exist in the crust and some interesting astrophysical phenomena, e.g., NS glitches and oscillations, see, e.g., Refs.\,\cite{Newton:2021vyd,Sotani:2024mlb,Shchechilin:2024kjv}.

A contemporary summary of constraints on supra-dense NS matter EOS\,\cite{Altiparmak2022} is shown in the right panel of FIG.\,\ref{fig_RN-EOS-over}.
{\color{xll} Although the NS matter EOS at densities $\lesssim(1\mbox{-}2)\rho_0$\,\cite{Drischler2022PRC} and very large densities $\gtrsim40\rho_0$\,\cite{Gorda2023,Altiparmak2022} is well constrained by the CEFT and pQCD, respectively, the EOS being relevant to realistic NSs is still largely uncertain (indicated by the vertical solid line). Massive NSs provide a unique opportunity to further restrict the supra-dense EOS at densities about $2\rho_0\lesssim\rho\lesssim8\rho_0$; and the cores of these NSs contain the densest stable matter visible in our Universe. } 

\subsection{Relevance, constraints and longstanding issues of high-density nuclear symmetry energy in NS matter}\label{HDEsym}
As we mentioned in SECTION \ref{SEC_1}, almost all the existing works adopt certain NS matter EOS models (either phenomenological or microscopic such as energy density functionals or mean-field theories). In the standard approach, they are put into the TOV equations or other related frameworks to predict NS properties in forward-modelings or to infer the probability distributions of EOS model parameters in Bayesian analyses from the observational data. The results are often EOS model dependent and therefore have sizable uncertainties. Before describing further the analysis of NS matter EOS using the perturbative treatments of the TOV equations (in several following sections), it is necessary here to discuss briefly the most important cause of the very uncertain EOS of supra-dense neutron-rich nucleonic matter.

Among the many possible origins, the isospin-dependent part of the asymmetric nuclear matter (ANM) EOS, namely the nuclear symmetry energy especially its high-density behavior occupies an important position. At the nucleonic level, 
the EOS of ANM can be constructed using the energy $E(\rho,\delta)$ per nucleon in an infinite nuclear matter with total density $\rho=\rho_{\rm{n}}+\rho_{\rm{p}}$ and isospin asymmetry $\delta=(\rho_{\rm{n}}-\rho_{\rm{p}})/\rho$ where $\rho_{\rm{n}}$ and $\rho_{\rm{p}}$ are densities of neutrons and protons, respectively. The EOS of symmetric nuclear matter $E_0(\rho)\equiv E(\rho,0)$ around the saturation density $\rho_0$ can be expanded as\,\cite{CaiLi2021PRC-Aux}
\begin{equation}\label{fE0}
E_0(\rho)\approx E_0(\rho_0)+\frac{1}{2}K_0\chi^2+\frac{1}{6}J_0\chi^3+\mathcal{O}(\chi^4),~~\chi\equiv\frac{\rho-\rho_0}{3\rho_0},
\end{equation}
defining the coefficients $K_0$---incompressibility of SNM, $J_0$---skewness coefficient of SNM, $E_0(\rho_0)$ is the nucleon binding energy in SNM.
Similarly, the symmetry energy $E_{\rm{sym}}(\rho)$ can be expanded as
\begin{equation}\label{fEsym}
E_{\rm{sym}}(\rho)\equiv\left.\frac{1}{2}\frac{\partial^2E(\rho,\delta)}{\partial\delta^2}\right|_{\delta=0}
\approx S+L\chi+\frac{1}{2}K_{\rm{sym}}\chi^2+\frac{1}{6}J_{\rm{sym}}\chi^3+\mathcal{O}(\chi^4),
\end{equation}
here $S\equiv E_{\rm{sym}}(\rho_0)$, $L$, $K_{\rm{sym}}$ and $J_{\rm{sym}}$ are respectively the magnitude, slope, curvature and skewness of the symmetry energy.
The even higher order terms in $\delta$ is relatively smaller, and is often neglected,  leading to the well-known parabolic approximation $E(\rho,\delta)\approx E_0(\rho)+E_{\rm{sym}}(\rho)\delta^2$ for the EOS of ANM.
Once the $E(\rho,\delta)$ is known, the pressure of ANM is obtained by the basic thermodynamic relation,
\begin{equation}
P(\rho,\delta)=\rho^2\frac{\partial E(\rho,\delta)}{\partial\rho}.
\end{equation}

\begin{figure}[h!]
\centering
\includegraphics[height=12.cm]{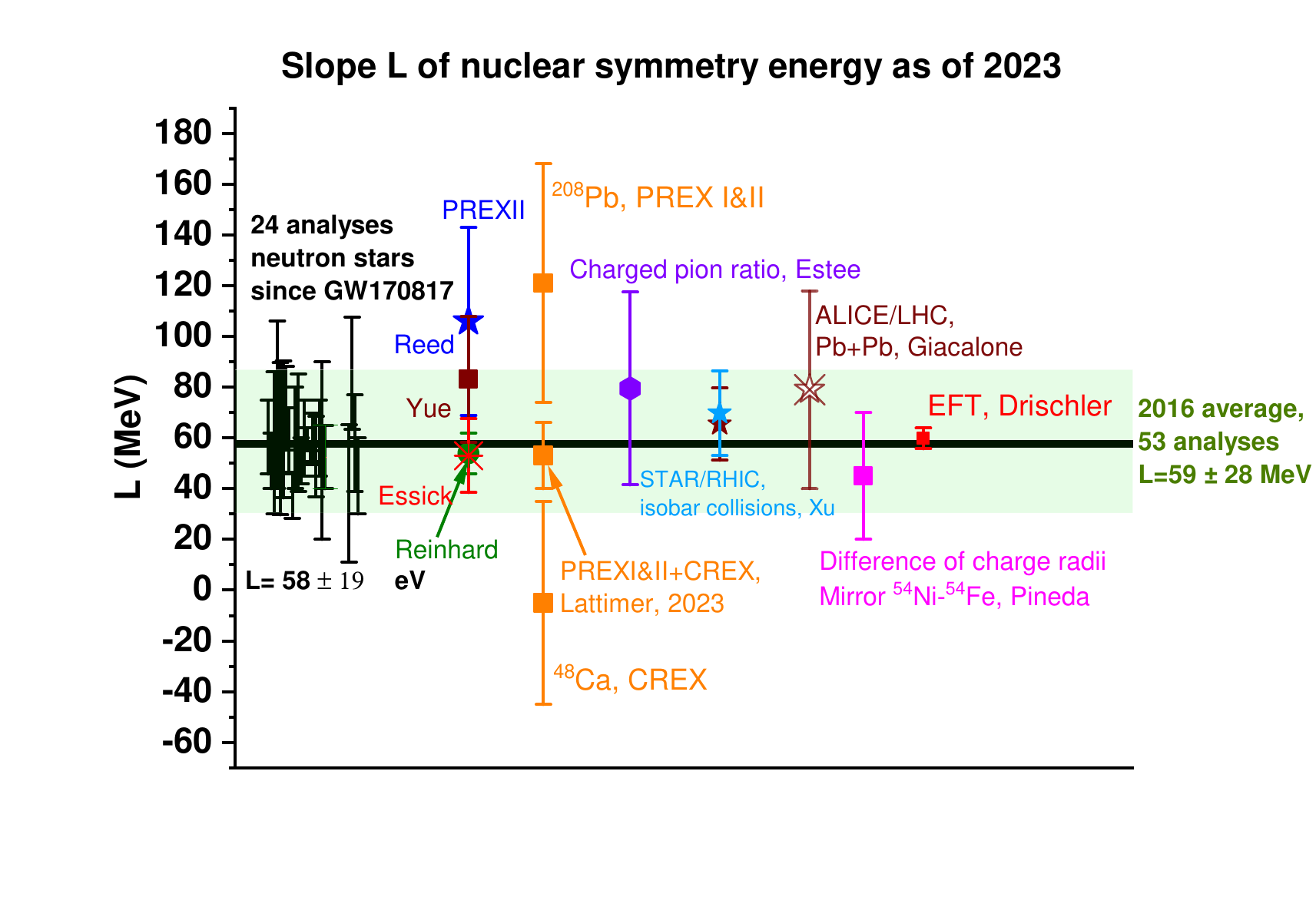}
\vspace{-1.5cm}
\caption{(Color Online). Constraints on the slope parameter $L$ of symmetry energy up to year 2023, via analyses of
several recent terrestrial experiments and NS observables since GW170817 in comparison with earlier systematics and the chiral EFT prediction. Starting from the left are the $L$ values from (1) 24 independent analyses of neutron star observables carried out by various groups between 2017 and 2021, they give an average $L\approx58\pm 19$\,MeV (thick horizontal black line)\,\cite{LCXZ2021}; (2) the original analysis of the PREX-II data\,\cite{Adh21} by Reed et al.\,\cite{Reed:2021nqk} and 3 independent analyses of PREX-II data together with different combinations of terrestrial and/or astrophysical data by Reinhard et al.\,\cite{Reinhard:2021utv}, Essick et al.\,\cite{Essick2021} and Yue et al.\,\cite{Yue:2021yfx}, respectively; (3) liquid drop model analyses using separately the PREX-I and II data together, CREX data only, and the combination of all data assuming they are equally reliable by Lattimer\,\cite{Lattimer:2023rpe}; (4) charged pion ration in heavy-ion reactions at RIKEN by Estee et al.\,\cite{SpiRIT:2021gtq}; (5) the ratio of average transverse momentum (sky blue  star) and the ratio of charged particle multiplicities (black star) in isobar collisions  $^{96}$Zr+$^{96}$Zr and $^{96}$Ru+$^{96}$Ru) from STAR/RHIC experiments analyzed by Xu et al.\,\cite{Xu:2022ikx}; (6) using neutron-skin thickness of $^{208}$Pb inferred by Giacalone et al. from $^{208}$Pb + $^{208}$Pb collisions measured by the ALICE/LHC Collaboration\,\cite{Giacalone:2023cet},
(7) the difference of charge radii of the mirror pair $^{54}$Ni-$^{54}$Fe by Pineda et al.\,\cite{mirror}; (8) the chiral EFT prediction by Dirschler et al.\,\cite{Drischler19PRL}. The horizontal band covering $L\approx59\pm 28$\,MeV is the 2016 average of 53 earlier analyses of various data mostly from terrestrial nuclear experiments\,\cite{Li:2013ola,Oertel2017}.
Figure modified from a figure in Ref.\,\cite{ZhangLi2023a}.}\label{fig_RN-L}
\end{figure}

\begin{figure}[h!]
\centering
\includegraphics[height=7.cm]{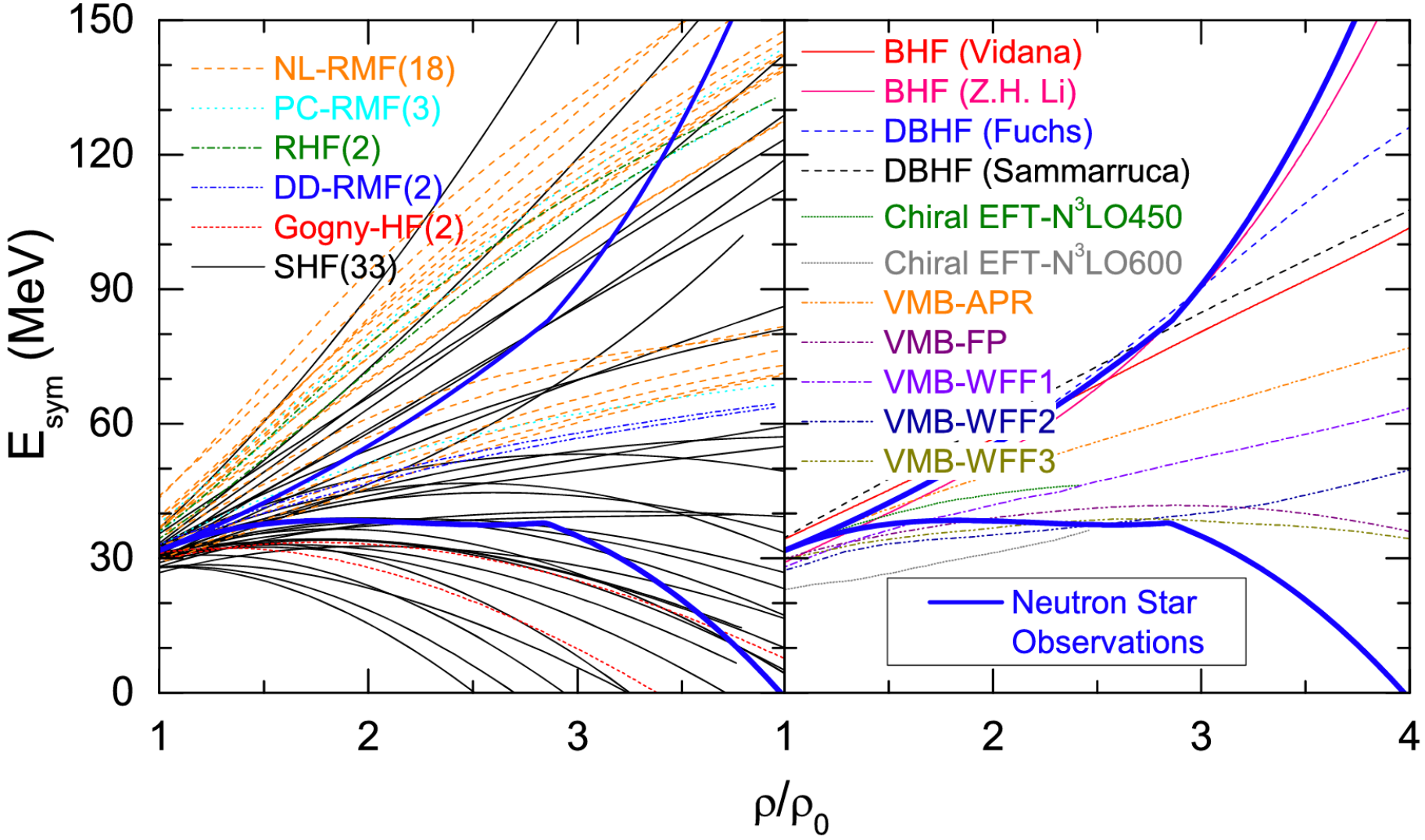}
\caption{(Color Online). Nuclear symmetry energy $E_{\rm{sym}}(\rho)$ at supra-saturation densities
predicted by different models in comparison with the upper and lower 68\% confidence boundaries (thick blue curves) from directly inverting several NS observables available since GW170817 by brute force in the high-density EOS parameter space\,\cite{ZhangLi2020ApJ}. Left: 60 examples of energy density functionals and/or phenomenological models. Right: 11 examples of {\it ab initio} and/or microscopic nuclear many-body theories.
Figure taken from Ref.\,\cite{ZhangLi2020ApJ} with some inputs from Ref.\,\cite{Chen:2017och}.}\label{fig_Esymrho}
\end{figure}

Thanks to the hard work of especially high-energy heavy-ion reaction community over the last few decades, the (density dependence of) EOS of cold SNM is relatively well constrained up to about $4\rho_0$, see, e.g., Refs.\,\cite{Sor2024,Tsang2024NA}.
However, the nuclear symmetry energy is relatively well determined only within a narrow region around the saturation density $\rho_0$. Specifically, the magnitude $S$ of symmetry energy has been well determined
to be about $S=31.6\pm 2.7$ MeV \cite{Li:2013ola} or $S=31.7\pm 3.2$ MeV \cite{Oertel2017} from surveys of large number of analyses of both nuclear and astrophysical data, as well as $S=31.7\pm 1.1$ MeV by chiral EFT prediction \cite{Drischler19PRL}. Unfortunately, its density dependence especially at supra-saturation densities is still very poorly understood. For example, FIG.\,\ref{fig_RN-L} is a recent update on the systematics of the slope parameter $L$ from (1) analyses of several recent terrestrial nuclear experiments listed in its caption, and (2) 24 independent analyses of NS observables since GW170817 (giving an average of $L\approx58\pm 19\,\rm{MeV}$\,\cite{LCXZ2021}) in comparison with (1) its earlier systematics ($L\approx59\pm 28$\,MeV based on 53 independent analyses of various nuclear and astrophysical data available before 2016\,\cite{Oertel2017}) as well as (2) the chiral EFT prediction of $L\approx59.8\pm 4.1\, \rm{MeV}$\,\cite{Drischler19PRL}. Clearly, the uncertainty of $L$ is still large. Not surprisingly, the high-order parameters $K_{\rm{sym}}$ and $J_{\rm{sym}}$ are even more poorly constrained, given the large uncertainty window of $E_{\rm{sym}}(\rho)$ at suprasaturation densities shown in FIG.\,\ref{fig_Esymrho}. 

\begin{figure}[h!]
\centering
\includegraphics[height=7.5cm]{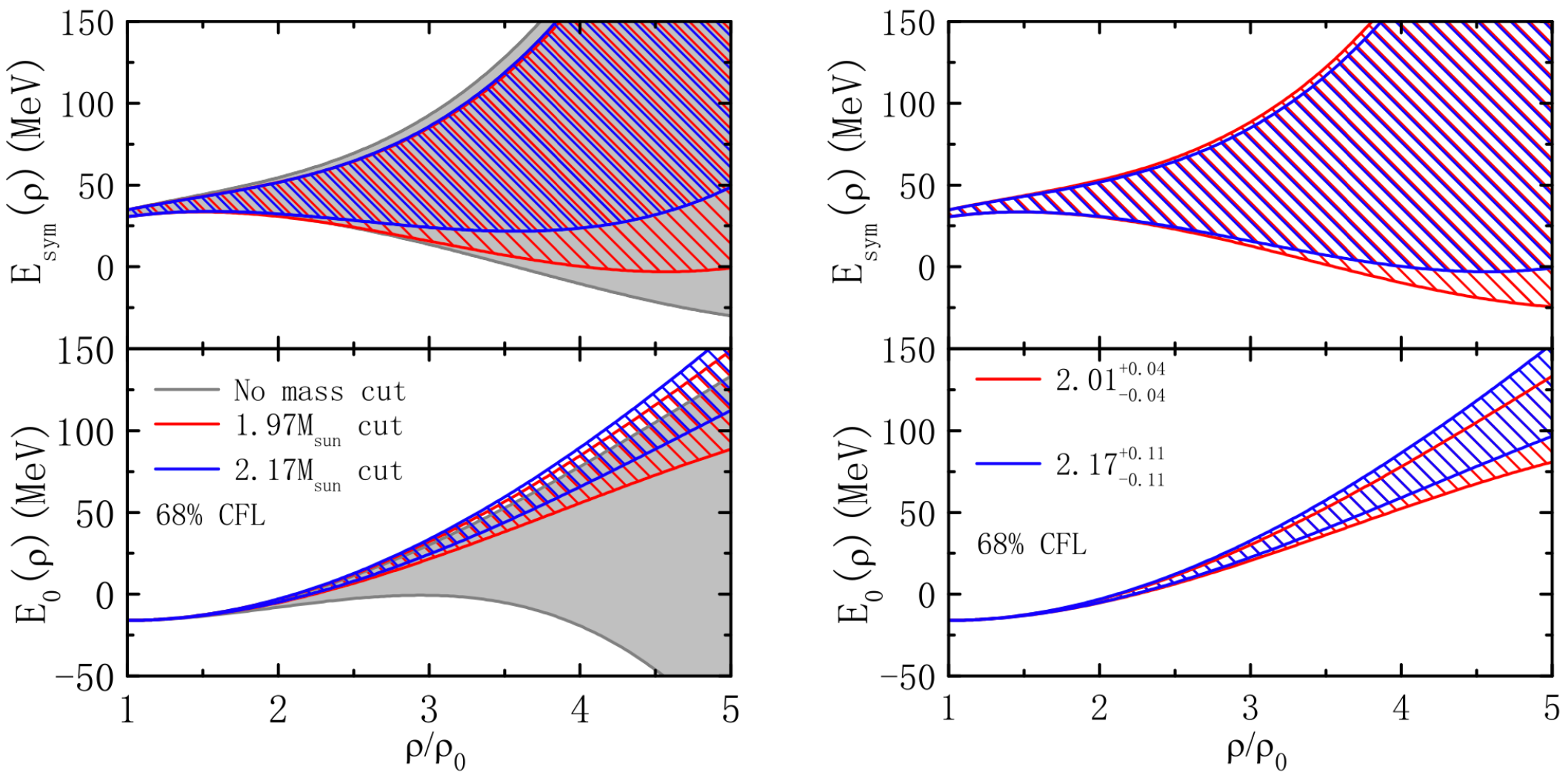}
\caption{(Color Online). 68\% posterior confidence boundaries of the SNM EOS
$E_0(\rho)$ (lower panels) and nuclear symmetry energy $E_{\rm{sym}}(\rho)$ (upper panels) inferred from a comprehensive Bayesian 
analysis\,\cite{Xie2019} using combined data of 
$R_{1.4}$ (radii of canonical NSs) from GW170817 by LIGO/VIRGO and low mass X-ray binaries by Chandra-XMM-Newton Collaborations with different minimum $M_{\rm{TOV}}$ values. Figure taken from Ref.\,\cite{Xie2019}.
}\label{fig_XieLi19ApJEsym}
\end{figure}

Shown in FIG.\,\ref{fig_Esymrho} are comparisons of the high-density $E_{\rm{sym}}(\rho)$ predicted by different models in comparison with the upper and lower 68\% confidence boundaries (thick blue curves) from directly inverting several NS observables available since GW170817. Here, clearly demonstrated is the fact that different approaches including non-relativistic energy density functionals, relativistic mean-field models, microscopic theories using realistic potentials or {\it ab initio} forces may predict very different high-density behaviors of $E_{\rm{sym}}(\rho)$.
This situation is partially because extracting symmetry energy straightforwardly from both astrophysical observations and/or terrestrial experimental data is very difficult and often rather model-dependent. Moreover, it is not always clear what observables are sensitive to the high-density behavior of nuclear symmetry energy in either astrophysics or nuclear physics. For example, shown with the thick blue curves in FIG.\,\ref{fig_Esymrho} are the upper and lower boundaries of $E_{\rm{sym}}(\rho)$ from directly inverting several NS observables available since GW170817 by brute force in the high-density EOS parameter space\,\cite{ZhangLi2020ApJ}.
There is a large opening window between them at supra-saturation densities above about $2\rho_0$. This is mostly because the radii and/or tidal deformations of canonical NSs observed relatively accurately so far are not so sensitive to the pressure significantly above $2\rho_0$, see, e.g., discussions in Refs.\,\cite{LCXZ2021,ZhangLi2023a}, as NS radii are determined by the condition $P(R)=0.$ The latter can be heavily masked by the remaining uncertainties of nuclear EOS at low densities. Thus, establishing a direction connection between the NS radius R and its core EOS, made possible as we shall demonstrate in detail by solving perturbatively the scaled TOV equations, will be extremely invaluable for narrowing down the uncertainty range of high-density nuclear symmetry energy. 

As another example, shown in FIG.\,\ref{fig_XieLi19ApJEsym} are the 68\% posterior confidence boundaries of the SNM EOS
$E_0(\rho)$ and nuclear symmetry energy $E_{\rm{sym}}(\rho)$ inferred from a comprehensive Bayesian 
analysis\,\cite{Xie2019} using combined data of 
$R_{1.4}$ from GW170817 by LIGO/VIRGO and low mass X-ray binaries by Chandra-XMM-Newton Collaborations. It is seen that using different minimum $M_{\rm{TOV}}$ in the analysis affects significant the lower boundary of $E_0(\rho)$ but has little effect on $E_{\rm{sym}}(\rho)$ at supra-saturation densities. Overall, while the high-density SNM EOS is relatively well constrained once the minimum $M_{\rm{TOV}}$ is considered, the available NS radius and mass data used in the Bayesian analyses do not constrain much the symmetry energy $E_{\rm{sym}}(\rho)$ at high densities above about $2\rho_0$. We refer the interested readers to Ref.\,\cite{LCXZ2021} for a review on the progress in constraining nuclear symmetry energy using NS observables since GW170817.

\subsection{Existing constraints and critical issues on the speed of sound in NSs}\label{sub_REV_s2}

\begin{figure}[h!]
\centering
\includegraphics[height=5.5cm]{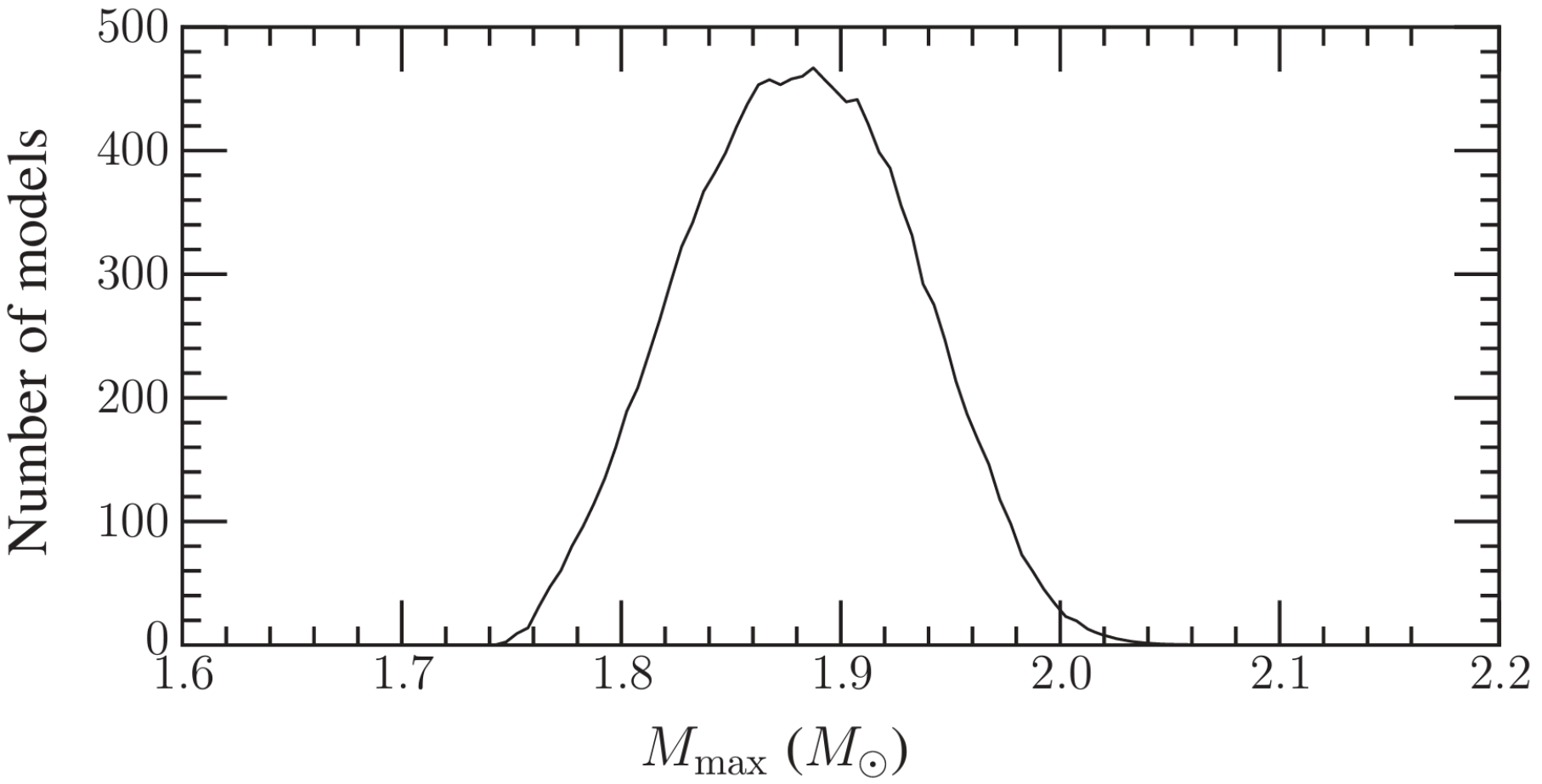}
\caption{Histogram of the number of models as a function of the
maximum/threshold mass supported, in these models the criteria $s^2<1/3$ is fulfilled. Figure taken from Ref.\,\cite{Bed15}.}\label{fig_RN-s2-BS}
\end{figure}

We now review briefly the current constraints on the speed of sound squared (SSS) $s^2$ in NS matter. The ultimate limit of $s^2$ was discussed in 1960's by Zel'dovich using a model with a vector-field\,\cite{Zeldovich61}. It was shown $s^2\to1$ and $P\to\varepsilon$ are the fundamental limits.
This issue was re-addressed in Ref.\,\cite{Bed15} in 2015 by studying the possible tension between $s^2<1/3$ and the observed massive NSs. They found that the number of models (in their simulation) consistent with low-density EOS and $s^2<1/3$ abruptly decreases even to disappear for $M_{\rm{NS}}\gtrsim2M_{\odot}$, see FIG.\,\ref{fig_RN-s2-BS}, implying $s^2$ should unavoidably be larger than 1/3 somewhere considering massive NSs. Since the quark matter at very large densities $\gtrsim40\rho_0$\,\cite{Bjorken83} has the approximate conformal symmetry and its EOS approaches that of an URFG with $P\approx\varepsilon/3$, its $s^2$ naturally approaches $1/3$ at these densities. One thus expects that the $s^2$ as a function of density (or energy density) may develop a bump structure with $s^2>1/3$ and then tend to $1/3$ asymptotically at extremely large densities.
{\color{xll}This expectation raises two important questions: (a) is this peaked structure in the density/radius profile of $s^2$ realized anywhere (e.g., near the center or some other places) within NSs? (b) what is the physical origin of such peak if it happens in NSs?}

Stimulated strongly by the findings in Ref.\,\cite{Bed15}, many interesting works, see, e.g., Refs.  \,\cite{Tews2018,Baym2019,Cas19,McLerran2019,Dua20,Ferr20,Mot20,Malf20,Zhao2020,Min21,Stone21,Otto20,Mott21,Jak21,Sen21,Kap2021,Tan2022-a,Tan2022-b,Altiparmak2022,Drischler2022PRC,Huang2022,Kojo2022,Ecker2022,Ecker23,Fuji22,Fuji2023,Han2023,Mar23,Mar24,Pro2024,Som2023,Liu23,Brandes2023,Brandes2023-a,Tak23,Fan24,Jim24,Roy2024,Gholami2024,Pegios2024PRD,Silva2024PRD} have been carried out to investigate the issues mentioned above.
For instance, in the quarkyonic picture\,\cite{McLerran2019,Pang2023,Fuji2024b,Fujimoto2024PRLQY}, a special arrangement of nucleons and quarks in a combined Fermi sphere forces the appearance of a peak in $s^2$ at NS densities. At high densities\,\cite{Pang24}, the baryon Fermi momentum is large and the degree of freedom (dof) deep within the Fermi sphere is Pauli blocked. Thus, creating a particle-hole excitation from deep within the Fermi sea requires large amounts of energy and momentum. Therefore, one can treat such excitation as weakly interacting. Due to the asymptotical freedom of quark matter at very high densities\,\cite{Bjorken83}, one naturally expects the existence of quark matter at high densities where quarks behave as nearly free particles. Consequently, one may treat the dof deep within the Fermi sphere as quarks. On the other hand, the dof near the Fermi surface can be excited with low energy and momentum transfers, allowing it to be treated as nucleons arising from quark correlations\,\cite{Pang24}. The baryon density $\rho$ in this model increases less rapidly in the quarkyonic phase. The suppression of the susceptibility
$\d\rho/\d\mu$ with $\mu\equiv\d\varepsilon/\d\rho$ leads to a rapid increase in
SSS\,\cite{McLerran2019} as $s^2=\rho\d\mu/\mu\d\rho$, generating a peak in $s^2$, see Refs.\,\cite{McLerran2019,Pang24,Fuji2024b} for more detailed analyses of the quarkyonic model and its applications in addressing problems associated with NS matter EOS.

\begin{figure}[h!]
\centering
\includegraphics[height=7.cm]{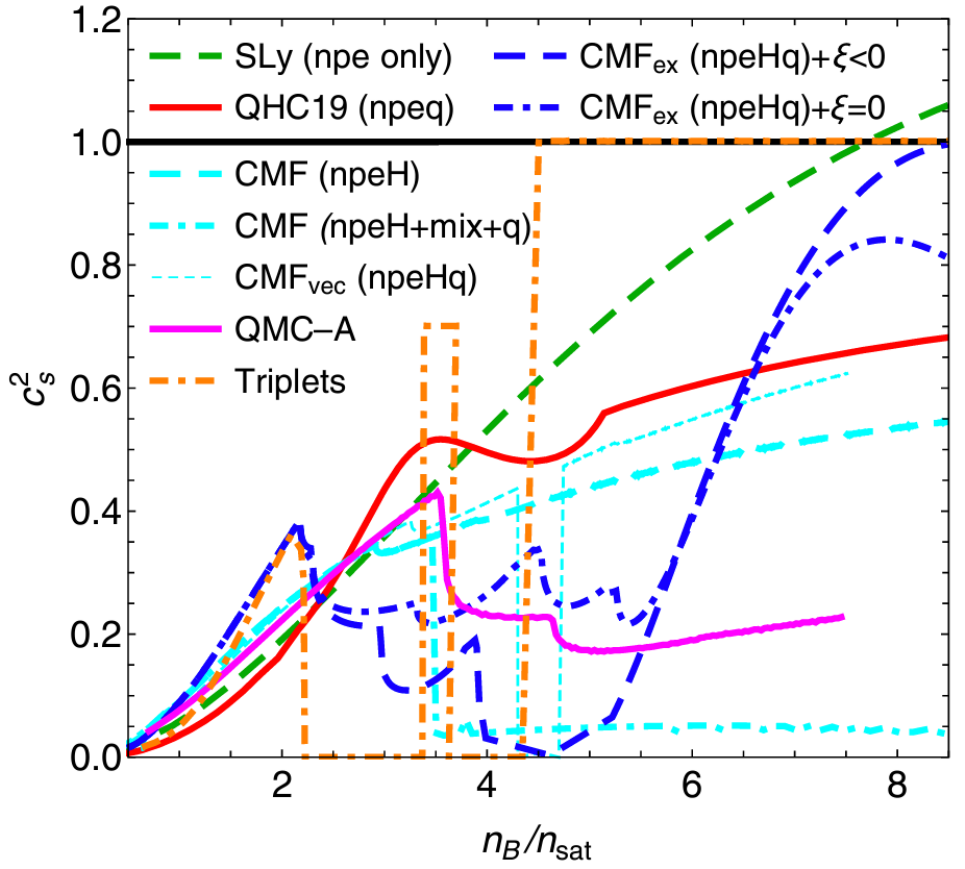}\quad
\includegraphics[height=6.cm]{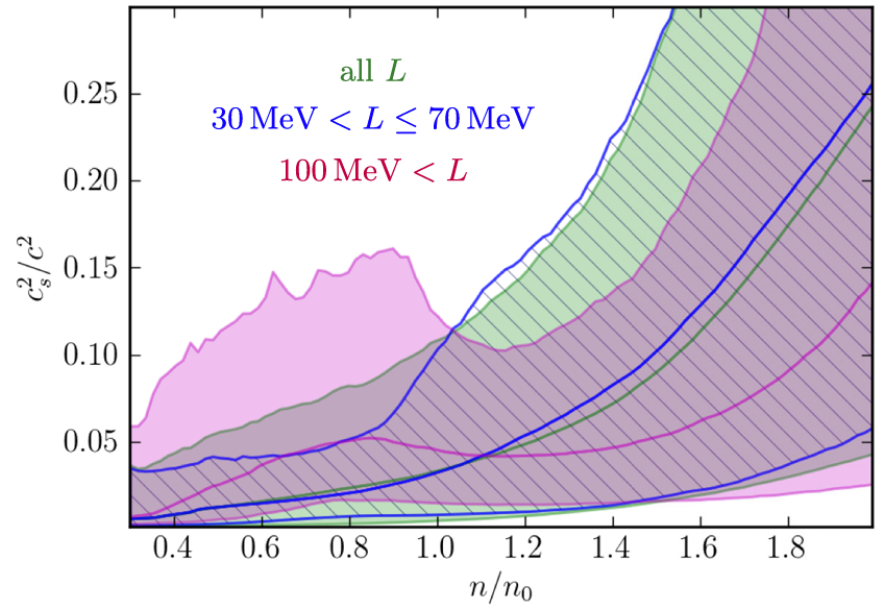} 
\caption{(Color Online). Left panel: Density-dependence of $s^2$ obtained using several phenomenological models, the basic conclusion is that $s^2$ has non-monotonic structures at some densities.
Figure taken from Ref.\,\cite{Tan2022-b}.
Right panel: median and 90\% 1-dimensional symmetric posterior
credible regions for $s^2\equiv c_{\rm{s}}^2$
at densities smaller than 2 times saturation density. Figure taken from Ref.\,\cite{Essick2021}.}\label{fig_RN-s2-summary}.
\end{figure}

Shown in the left panel of FIG.\,\ref{fig_RN-s2-summary} are examples of the predicted $s^2$ as a function of density obtained via several phenomenological models\,\cite{Tan2022-b}. One interesting feature of these results is that the $s^2$ is a non-monotonic function of density (the existence of plateaus, kinks, bumps).
However, the origin of these structures is still largely unknown and has significant model dependence.
For example,  the approximate conformal symmetry of quark matter and its possible realization at extremely high densities in massive NS cores\,\cite{Ann18,Ann20,Ann23,Gorda2023,Gorda21} may induce a peak in $s^2$, indicating possibly the occurrence of a sharp phase transition or a continuous crossover signaled by a smooth reduction of $s^2$.
Similarly,  Ref.\,\cite{ZhangLi2023a} showed recently that a purely nucleonic matter EOS model may also generate a peak in $s^2$ in dense neutron-rich matter accessible in massive NSs and/or relativistic heavy-ion collisions depending on the high-density behavior of nuclear symmetry energy. It is also interesting to note the work of Ref.\,\cite{Cao23} which demonstrated that the $s^2$ in self-bound quark stars made purely of absolutely stable deconfined quark matter may not show the peaked behavior. 
Furthermore, a very recent Bayesian analysis\,\cite{Mro23} of X-ray measurements and GW observations of NSs\,\cite{Abbott2017,Abbott2018,Abbott2020} incorporating the pQCD predictions\,\cite{Gorda21} shows that the peaked behavior in $s^2$ is consistent with but not required by these astrophysical data and pQCD predictions.
A similar inference on the behavior of $s^2$ as a function of energy density incorporating the presently existing data also implies that $s^2$ may have a weak/wide peak, even including the newly announced massive black widow pulsar PSR J0952-0607\,\cite{Romani22}. 
The peak in $s^2$ may also emerge in a hadron-quark hybrid model 
with excluded volume effects of baryons and chiral dynamics\,\cite{Kouno2024PRD}.

Both qualitatively and quantitatively, there are interesting discrepancies among the reported findings about the behavior of $s^2$. For illustrating the diversity of predictions, we mention a few of them below. For example, Ref.\,\cite{Brandes2023-a} found a peak in $s^2$ at about $630\,\rm{MeV}/\rm{fm}^3$,  which is very close to the central energy density about $670\,\rm{MeV}/\rm{fm}^3$.
Including updated 
measurement derived by fitting models of X-ray emission from the NS surface to NICER
data accumulated through 21 April 2022\,\cite{Ditt24} has little impact on the $s^2$ profile; specifically, the analysis still does not indicate an apparent peak in the density profile of $s^2$.
On the other hand, in a very recent study\,\cite{Kom24PRD} using a large model-agnostic ensemble via Gaussian processes conditioned with state-of-the-art astrophysical and theoretical inputs, the peaked structure in $s^2$ was found to be stable when first-order phase transitions (FOPTs) are included in the inference.

Although there exist large uncertainties about the high-density behavior of $s^2$, the monotonicity of $s^2$ at $\rho\lesssim2\rho_0$ (at zero temperature) is relatively well established. This conclusion benefited significantly from CEFT calculations conditioned on the relevant astrophysical data as shown in the right panel of FIG.\,\ref{fig_RN-s2-summary} (taken from Ref.\,\cite{Essick2021}). It is seen that $s^2$ is monotonically increasing with density for all slope parameter $L$ of the symmetry energy.
Similar and consistent result on $s^2$ at densities $\lesssim2\rho_0$ was also given in Refs.\,\cite{Tews2018,Keller23}. In particlar, Ref.\,\cite{Keller23} found that the finite-temperature may induce a local minimum in $s^2$ at low densities.
Interestingly, we can find from the right panel of FIG.\,\ref{fig_RN-s2-summary} that if one assumes $L > 100\,\rm{MeV}$ a local maximum in $s^2$ just below the saturation density may emerge (however the uncertainties are quite large to make definite conclusions).
This is because the EOSs that are stiff at low densities (corresponding to large
$L$) probed by neutron-skin experiments need to be softened at densities above $\rho_0$ to be consistent with the astrophysical data from GW170817\,\cite{Abbott2017,Abbott2018}. It is useful to note that a number of studies on $s^2$ used the findings of the PREX-II experiment \,\cite{Adh21} as a justification for using $L\gtrsim100\,\rm{MeV}$. However, as already indicated by the results shown in FIG.\,\ref{fig_RN-L}, such choice of a very large $L$ is not fully justified if one considers all published results as equally reliable within their error bars reported. In fact, a large $L$ is not really necessary to account for the result from the PREX-II experiment\,\cite{Adh21} based on several studies in the literature. For instance, a very recent study of Ref.\,\cite{Yue24} found that a $L\approx55\,\rm{MeV}$ can describe adequately both the PREX-II\,\cite{Adh21} and the CREX\,\cite{Adh22} data simultaneously if a strong iso-vector spin-orbit interaction is considered. Similarly, if one treats the results of PREX-I, PREX-II and CREX experiments as equally reliable within the reported experimental error bars, a unified analysis of them within an extended liquid droplet model leads to a value of $L\approx53\pm 13\,\rm{MeV}$\,\cite{Lattimer:2023rpe}. The values of $L$ from these two independent analyses of PREX and CREX data are consistent with the established global systematics of $L$ based on over 90 independent studies of various data of astrophysical observations and terrestrial nuclear experiments as shown in FIG.\,\ref{fig_RN-L}.

Interestingly, a peaked $s^2$ may also occur in heavy-ion collisions (HICs)\,\cite{Kutt23,Olii23,Soren21PRL,Yao24}.
For example, Ref.\,\cite{Kutt23} gave a Bayesian inference of $s^2$ using HIC data of mean transverse kinetic energy and elliptic flow of protons in the beam energy range $\sqrt{s_{\rm{NN}}}\approx2\mbox{-}10\,\rm{GeV}$. In their studies, the peak in $s^2$ profile is located at about $4\rho_0$. This peaked behavior is quite similar to that inferred from NS data.
However, if only 13 HIC data points were used in the analysis the peak in $s^2$ disappears within the relevant energy density region. This may indicate {\color{xll} that it is too early to make any robust conclusion on the possible bump structures in $s^2$ from analyzing the HIC data.}
We shall discuss more on $s^2$ in Subsection \ref{sub_DenseQCD}.

\begin{figure}[h!]
\centering
\includegraphics[width=9.0cm]{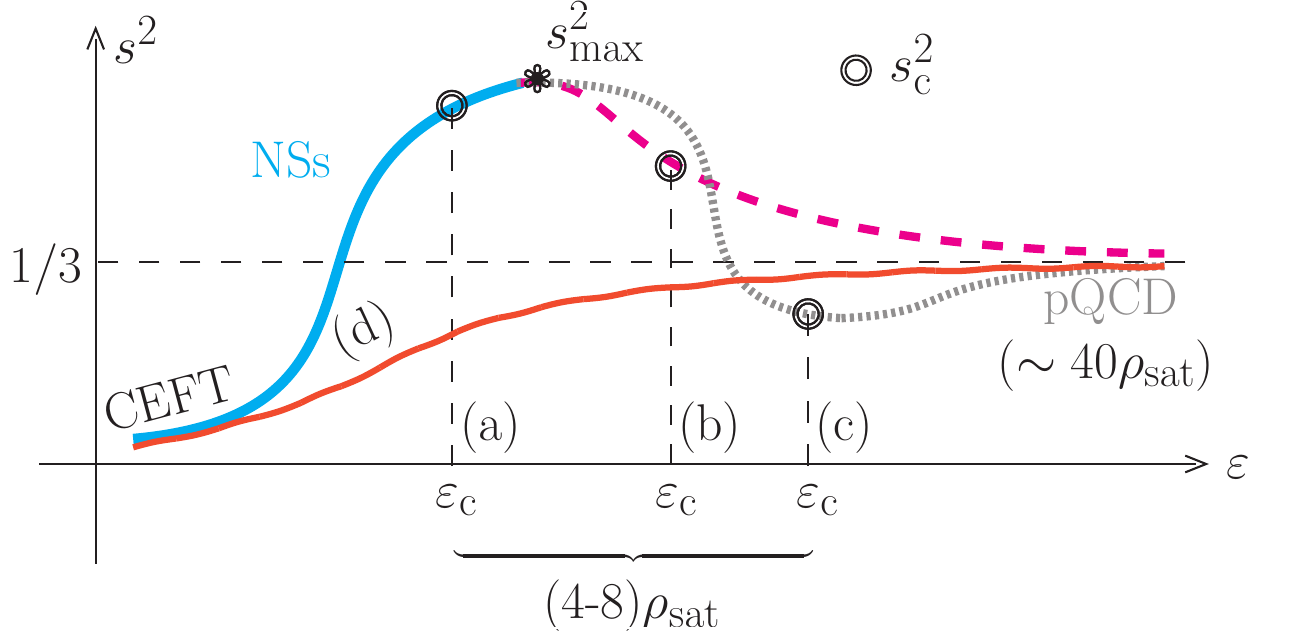}
\caption{(Color Online). Anatomy of patterns of extrapolating the $s^2$ in the core of NSs with $\rho_{\rm{c}}\approx\rho_{\max}\approx(4\mbox{-}8)\rho_{\rm{0}}$ to high densities $\sim40\rho_{\rm{0}}$ (pQCD),  here the double-circle on each line denotes its $s_{\rm{c}}^2$ and $s_{\max}^2$ is the global maximum value of the SSS; $\rho_0\equiv\rho_{\rm{sat}}$.
Figure taken from Ref.\,\cite{CL24-a}.
}\label{fig_s2pQCDkk}
\end{figure}

Shown in FIG.\,\ref{fig_s2pQCDkk} is an anatomy of different patterns of extrapolating the $s^2$ in NSs with their maximum densities $\lesssim8\rho_0$ to high densities $\sim40\rho_0$ (pQCD region). Here the double-circle on each line denotes its central value and $s_{\max}^2$ is the global maximum value of the SSS\,\cite{CL24-a}; a similar plot was given in Ref.\,\cite{Tews2018}.
The $s^2$ in patterns (a) and (d) shows a monotonic behavior and the difference lies in whether $s_{\rm{c}}^2$ is greater (smaller) than $1/3$; the $s^2$ in patterns (b) and (c) shows a peak at densities smaller than the central density of NSs where the $s_{\rm{c}}^2$ in pattern (b) (pattern (c)) is larger (smaller) than $1/3$.
Both pattern (b) and (c) indicate a continuous crossover behavior near the NS center.
Other possible nontrivial features in $s^2$ (like plateau, spike or bump, etc.) are not sketched in FIG.\,\ref{fig_s2pQCDkk}. We recommend Ref.\,\cite{Tan2022-b} and Ref.\,\cite{Mro23} for more detailed discussions on these issues.

{\color{xll}
To summarize this section, we emphasize the following points on the dense NS matter EOS and its speed of sound: (a) the EOS at $\rho\lesssim\rho_0$ is well constrained using CEFT; (b) the inference/restriction on NS matter EOS becomes eventually uncertain as density increases from $\rho\gtrsim\rho_0$ to $\rho\lesssim8\rho_0$, due to the lack/difficulties of first-principle calculations, model EOSs are often adopted and this unavoidably introduce model-dependence into the final constraints; (c) although pQCD at asymptotically large densities predicts a quite concise EOS (approximately an URFG), its relevance to NS matter EOS and impact on NS observations need more detailed investigations; (d) the SSS $s^2$ is larger than 1/3 ( that of an URFG) in massive NSs and has a peaked structure at some density/energy density when the asymptotic limit of the EOS as $P\to\varepsilon/3$ at extremely large densities is imposed; (e) however, such a peaked structure may or may not be realized within NSs since the maximum density of NSs is far smaller than the pQCD density; and (f) the physical origin of a peak in $s^2$ profile in NSs has significant model-dependence and is therefore still quite elusive.}
Improving the understanding of these issues is a major science driver of many ongoing research in nuclear astrophysics. It is also one of our major motivations for writing the current review.

{\color{xll} We also emphasize here that efforts to constrain the EOS of supra-dense neutron-rich matter near NS cores using the relevant NS observational data (especially NS radius) within the conventional framework are unavoidably EOS-model dependent and also suffer from the remaining uncertainties of low-density EOS. On the other hand, as we shall discuss in the next section, NS radius and mass scaling relations from dissecting the scaled TOV equations themselves provide unique insights directly into the NS core EOS without using any nuclear EOS model.}

\setcounter{equation}{0}
\section{Scalings of NS Mass, Radius and Compactness with its Central EOS}\label{SEC_4}

In this section,  we use the perturbative analysis of the TOV equations of SECTION \ref{SEC_2} as a tool to study the central EOS of NSs. We first give an intuitive argument in Subsection \ref{sub_SG} on the mass-, radius-scalings based on dimensional analysis and order-of-magnitude estimate using the excellent self-gravitating and quantum degenerate nature of NSs, following a perturbative analysis directly from the TOV equations. We then give in Subsection \ref{sub_Densest} the EOS of the densest visible matter existing in our Universe, namely the EOS of the NS cores at the maximum-mass configuration $M_{\rm{TOV}}$. 
Two examples (applications) of the method are given in the sequent subsections. In Subsection \ref{sub_rhoc}, we shall estimate the central baryon density $\rho_{\rm{c}}$ realized in NSs, and by taking the double-element expansion based on $\mu$ of (\ref{RE-small1}) and $\x$ of (\ref{RE-small2}) we give in Subsection \ref{sub_DEPc} the result for $P/P_{\rm{c}}$ and compare it with the state-of-the-art constraint available in the literature. Go beyond the NSs at TOV configuration, we study 
the central EOSs in canonical NSs and a NS with $M_{\rm{NS}}\approx2M_{\odot}$ in Subsection \ref{sub_1420}. 
The key quantity $\Psi=2\d\ln M_{\rm{NS}}/\d\ln\varepsilon_{\rm{c}}$ for describing generally stable NSs is given in Subsection \ref{sub_PsiVert}.
In Subsection \ref{sub_E1_count}, we discuss the counter-intuitive feature of the NS mass scaling established in the previous few subsections, then in Subsection \ref{sub_MTOV} we work out the details on the estimates of the maximum masses for generally stable NSs along the M-R curve and those at the TOV configuration.

In this section, we focus on the scalings of NS mass, radius and compactness as well as several related issues. In Section \ref{SEC_7} and Section \ref{SEC_56}, we shall shift our focus to the speed of sound squared and the trace anomaly, respectively.

\subsection{Self-gravitating and quantum degenerate properties of dense NS matter}\label{sub_SG}

Since NSs are self-gravitating systems\,\cite{Lan32,Rho74,Hartle77,Hartle78PR,Lind84, Lind92,Kalogera96,Lattimer05,Lattimer10,Shapiro1983}, one expects that a larger energy density $\varepsilon$ induces a smaller radius $R$.
By temporally neglecting the general relativistic effects\,\cite{Chan10-a}, we  have $\d P/\d r=-GM\varepsilon/r^2$ for Newtonian stars and therefore $P\sim-GM\varepsilon/r\sim Gr^2\varepsilon^2$ where $\d M/\d r=4\pi r^2\varepsilon$ or $M\sim r^3\varepsilon$ is used.
Consequently,
\begin{equation}\label{Isca-1}
r\sim(P/\varepsilon)^{1/2}/\sqrt{G\varepsilon}
\sim1/\sqrt{G\varepsilon},
\end{equation} 
the second relation follows because $P$ and $\varepsilon$ have the same dimension ($c=1$).
The factor $(P/\varepsilon)^{1/2}$ in (\ref{Isca-1}) can also be obtained as follows: NSs are supported mainly by the neutron degenerate pressure (at zero temperature), a larger pressure $P$ may lead to a larger radius $R$, i.e., 
\begin{equation}\label{Isca-2}
R\sim P^{\sigma},
\end{equation} where $\sigma>0$.
In order to infer the value of $\sigma$, we notice that $P\sim\int M\varepsilon/r^2\d r$ as a function of $r$ is even (see Subsection \ref{sub_Char}) and thus $P\sim\rm{const}.+Br^2+\cdots$ (being equivalent to $\widehat{P}\approx\x+b_2\widehat{r}^2+\cdots$), from which one obtains $\sigma=1/2$ and $R\sim P^{1/2}+\cdots$.
The absence of a linear term ($\propto r$) in the expansion of $P$ over $r$ could also be understood by the boundary condition $\d P/\d r=0$ at $r=0$ (pressure cannot have a cusp-like singularity).

By combining the self-gravitating and quantum-degenerate nature of NSs, we have $
R\sim (P/\varepsilon)^{1/2}/\sqrt{G\varepsilon}+\cdots$, since $P/\varepsilon$ is the relevant dimensionless quantity in combining $P$ and $\varepsilon$. The NS M-R relation is conventionally obtained by integrating the TOV equations starting from a given central energy density $\varepsilon_{\rm{c}}$\,\cite{CaiLi2016,LiF2022,CaiChen12,CaiChen17,CaiLi15},  so the above scaling could also be written in the form of $
R\sim \x^{1/2}/\sqrt{G\varepsilon_{\rm{c}}}+\cdots\sim\x^{1/2}/\sqrt{\varepsilon_{\rm{c}}}+\cdots$,  where $\x\equiv\widehat{P}_{\rm{c}}\equiv P_{\rm{c}}/\varepsilon_{\rm{c}}$.
Similar arguments give for the mass as 
$
M_{\rm{NS}}\sim R^3\varepsilon_{\rm{c}}\sim\x^{3/2}/\sqrt{\varepsilon_{\rm{c}}}+\cdots$.
Going one step further by putting back the general relativistic effects, NS radius $R$ in GR should scale as
\begin{equation}\label{Isca-4}
R\sim(\x^{1/2}/\sqrt{\varepsilon_{\rm{c}}})\cdot \vartheta(\x),\end{equation} 
with $\vartheta(\x)$ the general-relativistic correction.
Without doing detailed calculations, one can immediately infer that
$\vartheta(\x)<1$, since the stronger gravity in GR than Newtonian's should effectively reduce the NS radius.

We can reveal the specific form of $\vartheta(\x)$ by considering the pressure expansion of Eq.\,(\ref{ee-hP}) to order $\widehat{r}^2$; using the expression for the coefficient $b_2$ of Eq.\,(\ref{ee-b2}) and the basic definition of NS radius of Eq.\,(\ref{RE-term}), one has\,\cite{CLZ23-a}
\begin{equation}\label{def-hR}
\widehat{R}=\left(\frac{6\x}{1+3\x^2+4\x}\right)^{1/2}\sim\left(\frac{\x}{1+3\x^2+4\x}\right)^{1/2},~~\mbox{from}~~\widehat{P}(\widehat{R})=0,
\end{equation}
and so,
\begin{equation}\label{def-vartheta}
\boxed{
\vartheta(\x)=\left(\frac{1}{1+3\x^2+4\x}\right)^{1/2}.}
\end{equation}
After obtaining the GR factor $\vartheta(\x)$, we then obtain the radius-scaling\,\cite{CLZ23-a,CLZ23-b,CL24-a,CL24-b,CL24-c}
\begin{empheq}[box=\fbox]{align}
R=\widehat{R}Q=\left(\frac{3}{2\pi G}\right)^{1/2}\nu_{\rm{c}}\sim \nu_{\rm{c}},~~\mbox{with}~~\nu_{\rm{c}}\equiv
\frac{\x^{1/2}}{\sqrt{\varepsilon_{\rm{c}}}}\left(\frac{1}{1+3\x^2+4\x}\right)^{1/2},\label{gk-radius}
\end{empheq}
as well as the mass-scaling\,\cite{CLZ23-a,CLZ23-b,CL24-a,CL24-b,CL24-c}
\begin{empheq}[box=\fbox]{align}
M_{\rm{NS}}\approx\frac{1}{3}\widehat{R}^3W=\left(\frac{6}{\pi G^3}\right)^{1/2}\Gamma_{\rm{c}}\sim\Gamma_{\rm{c}},~~\mbox{with}~~\Gamma_{\rm{c}}
\equiv\frac{\x^{3/2}}{\sqrt{\varepsilon_{\rm{c}}}}\left(\frac{1}{1+3\x^2+4\x}\right)^{3/2},\label{gk-mass}
\end{empheq}
Dividing (\ref{gk-mass}) and (\ref{gk-radius}) gives the scaling for NS compactness\,\label{CL24-b}
\begin{equation}\label{gk-comp}
\boxed{
\xi=\frac{M_{\rm{NS}}}{R}
\approx\frac{2\Pi_{\rm{c}}}{G}\sim\Pi_{\rm{c}},~~\mbox{with}~~
\Pi_{\rm{c}}=\frac{\x}{1+3\x^2+4\x}.}
\end{equation}
Comparing it with Eq.\,(\ref{rel-xi-x}), we find $\tau_1\approx2$ and all $\tau_i\sim\mathcal{O}(1)$.
The above scaling (\ref{gk-comp}) implies that NS compactness directly probes its core EOS
$\x$, in the sense that $\Pi_{\rm{c}}$ is an increasing function of $\x$.
This means a NS with larger $\x$ is more compact than that with a smaller $\x$; i.e., $\x$ plays the role of the compactness.
Otherwise, $\x$ and $\xi$ are basically two different quantities, e.g., Specially Relativity requires $\x\leq1$ while General Relativity limits $\xi\leq1/2$.

\begin{figure}[h!]
\centering
\includegraphics[height=7.5cm]{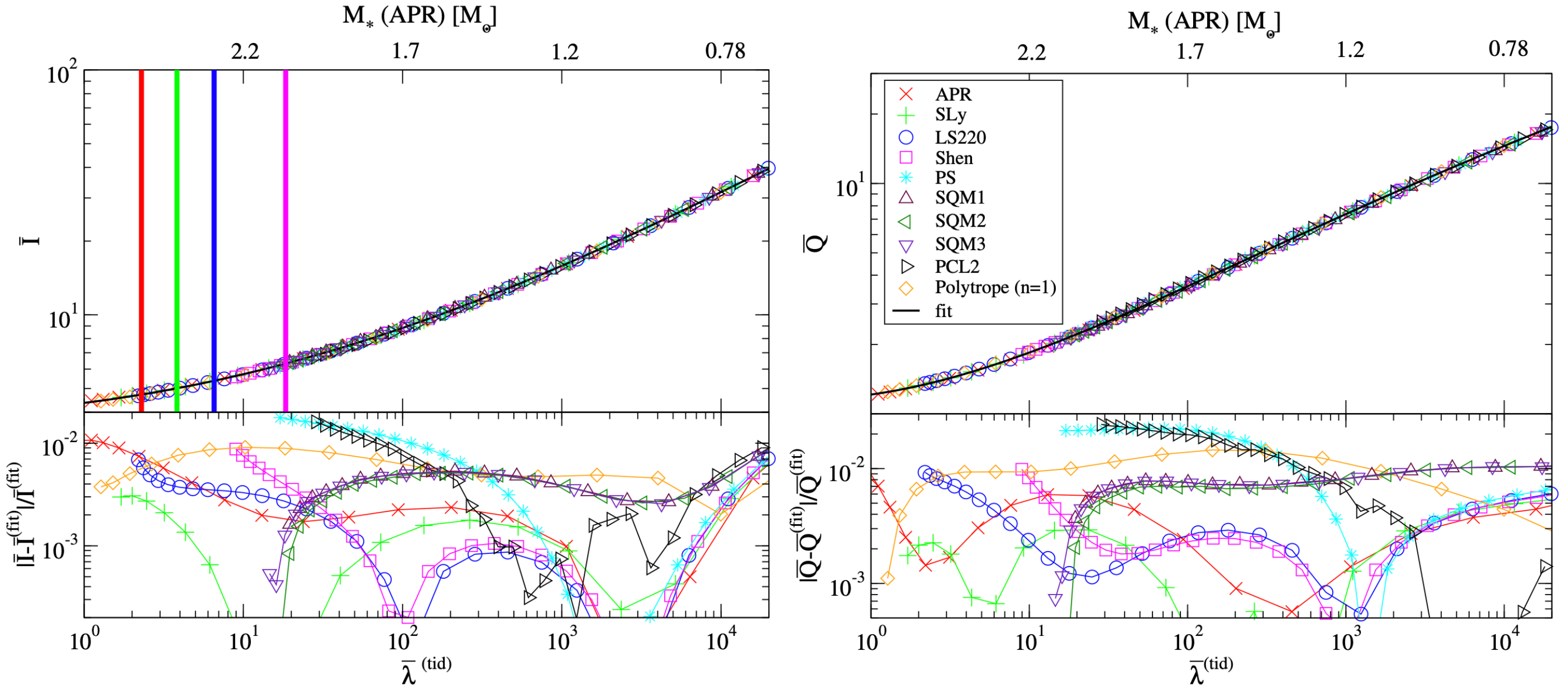}
\caption{(Color Online). The $I$-Love (left panel) and Love-${Q}$ relations for various EOSs. Figure taken from Ref.\,\cite{Yagi13}.
}\label{fig_Yagi}
\end{figure}

The above scalings relate directly the radii and/or masses of NSs with their central EOS ${P}_{\rm{c}}(\varepsilon_{\rm{c}})=P(\varepsilon_{\rm{c}})$. They are direct consequences of the TOV equations themselves without assuming any particular structure, composition and/or EOS for NSs. Once the $\nu_{\rm{c}}$ or $\Gamma_{\rm{c}}$ is constrained within a certain range by NS observational data, it can be used to determine the NS central EOS. We notice that the radius scaling may be slightly  affected by the still uncertain low-density EOS especially through the NS crust although its thickness accounts for only about 10\% of the whole radius\,\cite{BPS71,Iida1997,XuJ}.

In the expressions for $\vartheta(\x)$, $\Gamma_{\rm{c}}$, $\nu_{\rm{c}}$ and $\Pi_{\rm{c}}$, the term $3\x^2+4\x$ is the GR correction, which is far smaller than 1 if the Newtonian limit with $\x\to0$ is taken.
Therefore, we have
\begin{equation}\label{Newtion-SC}
\boxed{
\mbox{Newtonian stars:}~~\vartheta\approx1,~~M_{\rm{NS}}\sim\frac{{\x}^{3/2}}{\sqrt{\varepsilon_{\rm{c}}}}\sim P_{\rm{c}}^{3/2}\varepsilon_{\rm{c}}^{-2},~~R\sim\frac{{\x}^{1/2}}{\sqrt{\varepsilon_{\rm{c}}}}\sim P_{\rm{c}}^{1/2}\varepsilon_{\rm{c}}^{-1},~~\xi\approx2\x.}
\end{equation}
The GR correction bends the compactness $\xi$ at large ratio of central pressure over energy density.
Consequently, we have for Newtonian stars:
\begin{equation}\label{Pc-M2R4}
    P_{\rm{c}}\sim{M_{\rm{NS}}^2}/{R^4}.
\end{equation}
This relation can also be obtained by dimension analysis on the Newtonian stellar equations.

Extensive studies on scaling relations among NS observables exist in the literature, see, e.g. Refs.\,\cite{Rho74,Hartle77,Hartle78PR,Kalogera96,Lattimer05,Tsui05,Yagi13,Wen19,JiangYagi20}.
These EOS-independent universal scalings are mostly for NS bulk properties, such as the moment of inertia $I$, tidal Love number $k_2$, (spin-induced) quadrupole moment ${Q}$, compactness $\xi$, and frequencies ($f$ or $\omega$) of various oscillation modes of NSs. 
The moment of inertia (using $c=1$) is defined as\,\cite{Hartle1967},
\begin{equation}\label{def_MI}
\boxed{
I=-\frac{2}{3G}\int_0^R\d rr^3\omega(r)\left(\frac{\d }{\d r}j(r)\right)
=\frac{8\pi}{3}\int_0^R\d rr^4\left[\varepsilon(r)+P(r)\right]\exp\left[\lambda(r)\right]j(r)\omega(r),}
\end{equation}
here $j(r)=\exp[-2^{-1}(\nu(r)+\lambda(r))]$ and $\omega(r)$ is the rotational drag; $\nu(r)$ and $\lambda(r)$ are two metric functions\,\cite{Hartle1967}.
The rotational drag satisfies a differential equation\,\cite{Hartle1967},
\begin{equation}
\boxed{
\frac{\d}{\d r}\left(r^4j(r)\frac{\d}{\d r}\omega(r)\right)+4r^3\omega(r)\frac{\d}{\d r}j(r)=0.}
\end{equation}
Consequently, $
I=[{R^4}/{6G}][{\d \omega}/{\d r}]_{r=R}$.
Knowing one or more observables, these scalings (e.g., $I$-Love-${Q}$, $I$-Love-$\xi$ or $M\omega$-$\xi$) enable the finding of other observables or bulk properties that have not been or hard to be measured; FIG.\,\ref{fig_Yagi} shows the celebrated $I$-Love-${Q}$ relations for NSs\,\cite{Yagi13}.
They are completely different from the scalings of NS radius and mass separately as functions of the reduced core pressure over energy density we derived above.
Among the earlier mass scalings closest to ours is the one showing\,\cite{Hartle78PR,Kalogera96,Lattimer05} 
\begin{equation}\label{HKL}
\boxed{
M_{\rm{NS}}^{\max}\sim D_{\rm{M}}\varepsilon_{\rm{c}}^{-1/2},}
\end{equation} with the coefficient $D_{\rm{M}}$ estimated empirically using either sometimes simplified or certain selected microscopic dense matter EOSs, often leading to sizable EOS model-dependence or lack solid/clear physical origins.
Another fundamental difference is that our scalings link directly the macroscopic observables with microscopic core variables  (different combinations of the core pressure and energy density via $\nu_{\rm{c}}$ and $\Gamma_{\rm{c}}$). This feature enables us to extract the core EOS ${P}_{\rm{c}}(\varepsilon_{\rm{c}})=P(\varepsilon_{\rm{c}})$ once the NS mass or radius is constrained observationally, instead of just the individual value of $\varepsilon_{\rm{c}}$ or the pressure around $(1\mbox{-}2)\rho_0$ in the previous studies of NS scaling observables. 

\begin{figure}[h!]
\centering
\includegraphics[width=6.5cm]{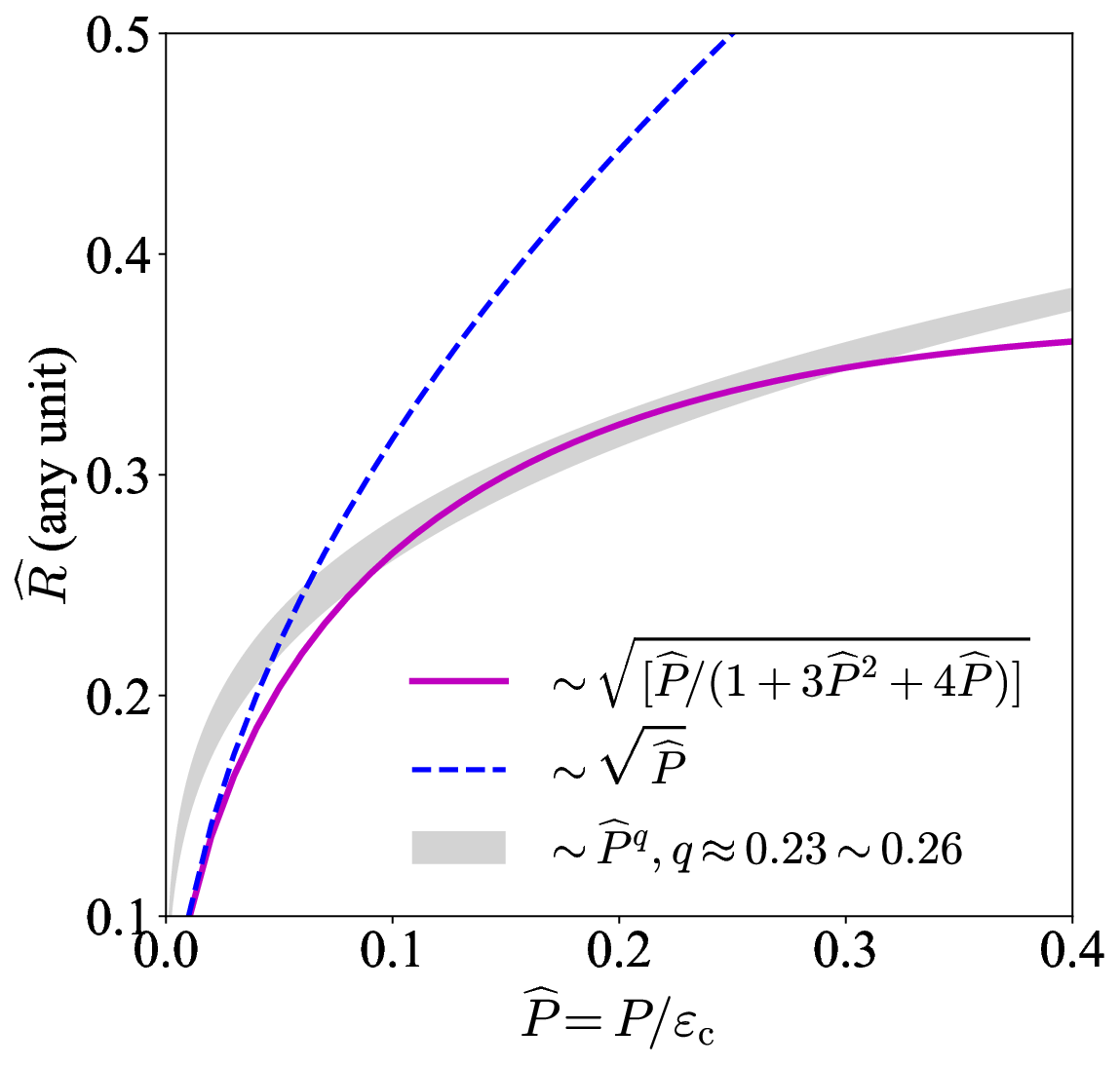}\quad
\includegraphics[width=10.cm]{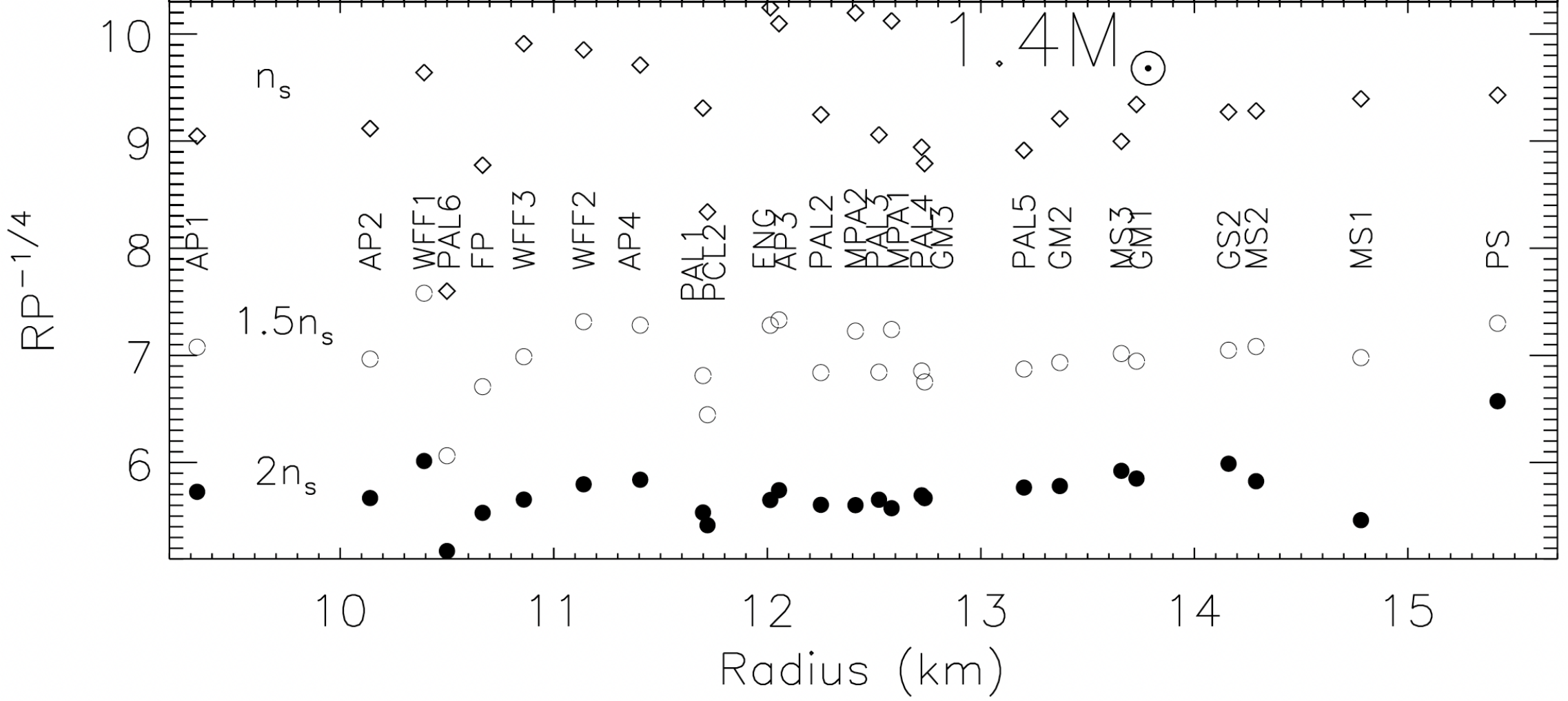}
\caption{(Color Online). Left panel: comparisons of NS radius scalings ($\widehat{R}$ vs reduced pressure $\widehat{P}$) for both general-relativistic (magenta curve) NSs using our Eq.\,(\ref{def-hR}) and Newtonian (blue) stars with the empirical power law $R\sim P^{q}$ (gray band) from Lattimer and Prakash\,\cite{Lattimer2001}, here $q\approx0.23\mbox{-}0.26$. Figure taken from Ref.\,\cite{CLZ23-b}.
Right panel: the empirical radius scaling $R\sim P^{0.25}$ obtained for canonical NSs from using two dozens of NS matter EOSs by Lattimer and Prakash. Figure taken from Ref.\,\cite{Lattimer2001}. }\label{fig_RSca}
\end{figure}

Very interestingly, an empirical power law for NS radii was found earlier by Lattimer and Prakash as $R\sim D_{\rm{R}}P^{q}$ with $q\approx0.23\mbox{$\sim$}0.26$\,\cite{Lattimer2001}, where $P$ is the pressure at about $(1\mbox{-}2)\rho_0$, see the right panel of FIG.\,\ref{fig_RSca} for the scaling $R\sim P^{0.25}$ for canonical NSs using two dozens of NS matter EOSs existing at that time\,\cite{Lattimer2001}. A comparison between our radius scaling in Eq.\,(\ref{def-hR}) and theirs is shown in the left panel of FIG.\,\ref{fig_RSca}. The narrow width of the gray band indicates the very weak EOS model-dependence involved in their empirical power law. Firstly, it is clearly seen that the Newtonian prediction $\widehat{R}\sim \widehat{P}^{1/2}$ (by neglecting the GR correction $3\widehat{P}^2+4\widehat{P}$) indicated by the blue dashed line deviates significantly from the empirical power law $\widehat{P}^q$\,\cite{Lattimer2001}. On the other hand, our full scaling $\widehat{R}\sim[\widehat{P}/(1+3\widehat{P}^2+4\widehat{P})]^{1/2}$ (magenta curve) is rather consistent with the empirical scaling by Lattimer and Prakash, especially for $0.1\lesssim\widehat{P}\lesssim0.3$ (which is the most relevant region for $P/\varepsilon$ in NS cores). Since our scalings are directly from analyzing the TOV equations themselves without using any model EOS, they provide independently a fundamental physics basis for the NS empirical radius power law by Lattimer and Prakash.

\subsection{EOS of the densest visible matter existing in our Universe}\label{sub_Densest}

The maximum-mass configuration (or the TOV configuration) along the NS M-R curve is a special point\,\cite{CL24-c}.
Consider a typical NS M-R curve near the TOV configuration from right to left, the radius $R$ (mass $M_{\rm{NS}}$) eventually decreases (increases), the compactness $\xi=M_{\rm{NS}}/R$ correspondingly increases and reaches its maximum value at the TOV configuration.
When going to the left along the M-R curve even further, the stars becomes unstable and then may collapse into black holes (BHs).
So the NS at the TOV configuration is denser than its surroundings and the cores of such NSs contain the critically stable densest matter visible in the Universe.
Mathematically, we describe the TOV configuration by\,\cite{CL24-c},
\begin{equation}\label{MTOV-cr}
\left.\frac{\d M_{\rm{NS}}}{\d\varepsilon_{\rm{c}}}\right|_{M_{\rm{NS}}=M_{\rm{NS}}^{\max}=M_{\rm{TOV}}}
=0,~~\left.\frac{\d ^2M_{\rm{NS}}}{\d\varepsilon_{\rm{c}}^2}\right|_{M_{\rm{NS}}=M_{\rm{NS}}^{\max}=M_{\rm{TOV}}}
<0.
\end{equation}
Using the NS mass scaling of Eq.\,(\ref{gk-mass}), we can obtain an expression for the central SSS\cite{CLZ23-b,CL24-a,CL24-c},
\begin{equation}\label{sc2-GG}
\boxed{\mbox{for stable NSs along the M-R curve:}~~
s_{\rm{c}}^2={\x}\left(1+\frac{1+\Psi}{3}\frac{1+3{\x}^2+4{\x}}{1-3{\x}^2}\right),}
\end{equation}
where\,\cite{CL24-a,CL24-c} 
\begin{equation}\label{def-Psi}
\boxed{
\Psi=\frac{2\varepsilon_{\rm{c}}}{M_{\rm{NS}}}\frac{\d M_{\rm{NS}}}{\d\varepsilon_{\rm{c}}}
=2\frac{\d\ln M_{\rm{NS}}}{\d\ln\varepsilon_{\rm{c}}}\geq0.}
\end{equation}
We study the factor $\Psi$ in Subsection \ref{sub_PsiVert}.
For the TOV configuration where $\Psi=0$, we have\,\cite{CL24-a,CL24-c}
\begin{equation}\label{sc2-TOV}
\boxed{
\mbox{for NSs at the TOV configuration:}~~
s^2_{\rm{c}}=\x\left(1+\frac{1}{3}\frac{1+3\x^2+4\x}{1-3\x^2}\right).}
\end{equation}
Using the Principle of Causality of Special Relativity, we immediately obtain\,\cite{CLZ23-a,CL24-c}
\begin{equation}\label{Xupper}
\boxed{
s_{\rm{c}}^2\leq1\leftrightarrow\x\lesssim0.374
\equiv\x_+.}
\end{equation}
Physically, although the causality condition requires apparently ${\x}\leq1$, the GR nature of strong-field gravity makes the 
EOS of superdense NS core matter highly nonlinear as indicated by the nonlinear dependence of $s_{\rm{c}}^2$ on ${\x}$. Consequently, the maximum speed of sound reachable in massive NSs is much smaller than what is allowed by causality with a linear EOS\,\cite{CL24-c}.
The upper bound on $\x$ and $\phi=P/\varepsilon$ is recently reviewed in Ref.\,\cite{CL24-c}, we will in SECTION \ref{SEC_56} briefly outline the main results.
After obtaining the $s_{\rm{c}}^2$ and using the expression for $R$ of Eq.\,(\ref{gk-radius}), we can calculate the derivative of NS radius $R$ with respective to $\varepsilon_{\rm{c}}$, the result is:
\begin{empheq}[box=\fbox]{align}\label{fgk-1}
\frac{\d R}{\d\varepsilon_{\rm{c}}}=&\frac{\d R}{\d M_{\rm{NS}}}\frac{\d M_{\rm{NS}}}{\d\varepsilon_{\rm{c}}}
=\frac{\d}{\d\varepsilon_{\rm{c}}}\left[\left(\frac{3}{2\pi G}\right)^{1/2}\nu_{\rm{c}}\right]
=\left(\frac{R}{\varepsilon_{\rm{c}}}\right)
\cdot\left(\frac{\Psi}{6}-\frac{1}{3}\right).
\end{empheq}
This means if $\Psi$ is around 2, the dependence of $R$ on $\varepsilon_{\rm{c}}$ would be weak.
Moreover, since $\d M_{\rm{NS}}/\d\varepsilon_{\rm{c}}>0$ for stable NSs, the signs of $\d R/\d\varepsilon_{\rm{c}}$ and that of $\d R/\d M_{\rm{NS}}$ are the same.
Assuming $\Psi\approx\rm{const.}$ for a given NS mass, Eq.\,(\ref{fgk-1}) further gives
\begin{equation}\label{fgk-dd}
\boxed{
    R\sim\varepsilon_{\rm{c}}^{\Psi/6-1/3}.}
\end{equation}
For NSs at the TOV configuration, $\Psi=0$, Eq.\,(\ref{fgk-1}) gives $\d R/\d\varepsilon_{\rm{c}}<0$\,\cite{CLZ23-a} and Eq.\,(\ref{fgk-dd}) leads to $R\sim\varepsilon_{\rm{c}}^{-1/3}$, i.e., as $\varepsilon_{\rm{c}}$ increases, the radius $R$ decreases (self-gravitating property), as expected.
We discuss the importance of $\Psi$ further in Subsection \ref{sub_s2canon} when we estimate the central SSS in a canonical NS.

At this point, we can express $M_{\rm{NS}}$ in terms of $s_{\rm{c}}^2$ by inverting the relation (\ref{sc2-GG}) to obtain $\x=\x(s_{\rm{c}}^2,\Psi)$ and inserting the latter into (\ref{gk-mass}), the result is
\begin{align}
    M_{\rm{NS}}\sim&\frac{1}{\sqrt{\varepsilon_{\rm{c}}}}\left(\frac{\x(s_{\rm{c}}^2,\Psi)}{1+3\x^2(s_{\rm{c}}^2,\Psi)+4\x(s_{\rm{c}}^2,\Psi)}\right)^{3/2}\equiv\frac{\mathcal{M}(s_{\rm{c}}^2,\Psi)}{\sqrt{\varepsilon_{\rm{c}}}}\notag\\
    =&\frac{1}{\sqrt{\varepsilon_{\rm{c}}}}
    \cdot\frac{3\sqrt{3}s_{\rm{c}}^3}{(4+\Psi)^{3/2}}
    \left[1-18\frac{5+2\Psi}{(4+\Psi)^2}s_{\rm{c}}^2
    +\frac{81}{2}\frac{148+126\Psi+29\Psi^2}{(4+\Psi)^4}s_{\rm{c}}^4+\cdots
    \right]\notag\\
    \to&\frac{1}{\sqrt{\varepsilon_{\rm{c}}}}\frac{3\sqrt{3}s_{\rm{c}}^3}{8}\left(1-\frac{45s_{\rm{c}}^2}{8}+\frac{2997s_{\rm{c}}^4}{128}+\cdots\right),\label{kl-1}
\end{align}
where the last line follows by taking $\Psi=0$.
The function $\mathcal{M}(s_{\rm{c}}^2,\Psi)$ monotonically increases with $s_{\rm{c}}^2$ (and takes its maximum value about $\mathcal{M}(1,0)\approx0.046$), therefore for a given $\varepsilon_{\rm{c}}$, a larger $s_{\rm{c}}^2$ (stiffer EOS) generates a larger NS mass, which is a well-known result.

\begin{figure}[h!]
\centering
\includegraphics[width=11.cm]{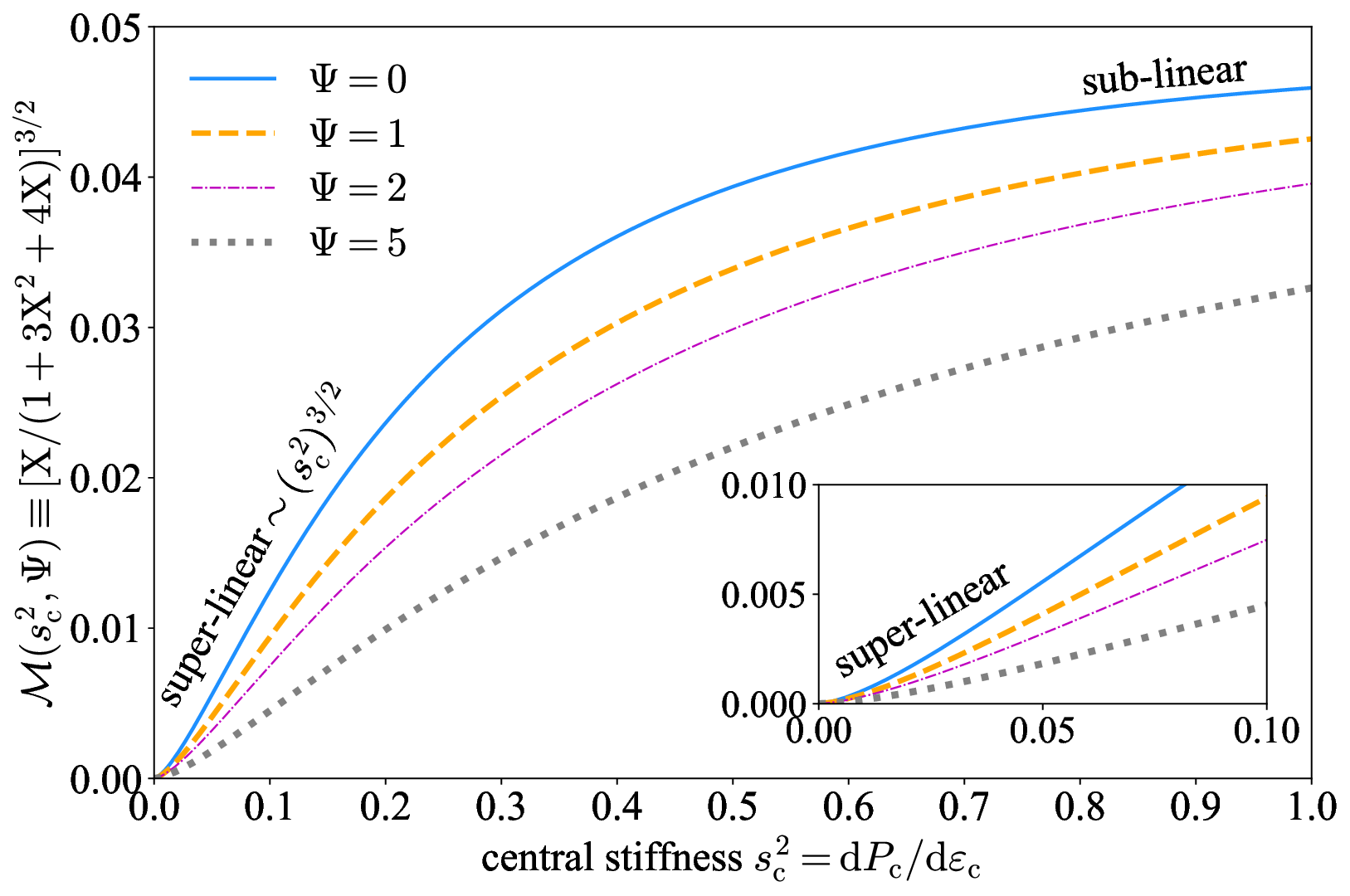}
\caption{(Color Online). The dependence of the function $\mathcal{M}(s_{\rm{c}}^2,\Psi)$ on stiffness $s_{\rm{c}}^2$.
For small stiffness $s_{\rm{c}}^2$, the growth rate of $\mathcal{M}(s_{\rm{c}}^2,\Psi)$ is super-linear while for $s_{\rm{c}}\approx1$, the rate is sub-linear.
}\label{fig_Msc2}
\end{figure}

Eq.\,(\ref{kl-1}) implies that the growth rate of $M_{\rm{NS}}$ over $s_{\rm{c}}^2$ (for a given $\varepsilon_{\rm{c}}$) is super-linear for small stiffness $s_{\rm{c}}^2\approx0$, i.e., $\mathcal{M}(s_{\rm{c}}^2,\Psi)\sim (s_{\rm{c}}^2)^{3/2}$.
On the other hand, we can show that the growth rate of $M_{\rm{NS}}$ is sub-linear for $s_{\rm{c}}^2\lesssim1$.
By introducing $\varphi=s_{\rm{c}}^2-1\lesssim0$, we can obtain from Eq.\,(\ref{sc2-GG}) that
\begin{equation}\label{kl-2}
    \x\approx\x_+(\Psi)+L_1\varphi+L_2\varphi^2+\mathcal{O}(\varphi^3),
\end{equation}
here $\x_+(\Psi)$ is the upper bound on $\x_+$ determined by $s_{\rm{c}}^2\leq1$, e.g., $\x_+(0)\approx0.374$ of (\ref{Xupper}).
Without doing detailed derivation one can infer that $L_1<0$ since $\varphi<0$ and $\x_+(\Psi)$ is the upper bound for $\x$.
Specifically, 
\begin{align}
    L_1=&\frac{[1-3\x_+^2(\Psi)]\x_+(\Psi)[1+\x_+(\Psi)][1+3\x_+(\Psi)]}{1+8\x_+(\Psi)+8\x_+^2(\Psi)-12\x_+^3(\Psi)-21\x_+^4(\Psi)},\\
    L_2=&-\frac{2[1-3\x_+^2(\Psi)]\x_+^2(\Psi)[1-\x_+(\Psi)][1+\x_+(\Psi)]^2[1+3\x_+(\Psi)]^2[2+9\x_+(\Psi)+18\x_+^2(\Psi)+9\x_+^3(\Psi)]}{[1+8\x_+(\Psi)+8\x_+^2(\Psi)-12\x_+^3(\Psi)-21\x_+^4(\Psi)]^3},
\end{align}
For all $\x_+(\Psi)\leq\x_+(0)\approx0.374$, $L_1>0$ ($L_2<0$) monotonically increases (decreases) with $\x_+(\Psi)$, e.g., $L_1\approx0.155$, $L_2\approx-0.107$ for $\x_+(0)$ and so $\x\approx0.374+0.155\varphi-0.107\varphi^2$.
Putting the $\x$ of (\ref{kl-2}) into $\mathcal{M}(s_{\rm{c}}^2,\Psi)=\mathcal{M}(\varphi+1,\Psi)$ gives
\begin{equation}
\mathcal{M}(s_{\rm{c}}^2,\Psi)/\mathcal{M}(1,\Psi)\approx 1+T_1\varphi+T_2\varphi^2+\mathcal{O}(\varphi^3),
~~
    \mathcal{M}(1,\Psi)=\left(\frac{\x_+(\Psi)}{1+3\x_+^2(\Psi)+4\x_+(\Psi)}\right)^{3/2},
\end{equation}
where,
\begin{align}
    T_1=&\frac{3}{2}\frac{[1-3\x_+^2(\Psi)]^2}{1+8\x_+(\Psi)+8\x_+^2(\Psi)-12\x_+^3(\Psi)-21\x_+^4(\Psi)},\\
    T_2=&\frac{3}{8}\frac{[1-3\x_+^2(\Psi)]^2}{[1+8\x_+(\Psi)+8\x_+^2(\Psi)-12\x_+^3(\Psi)-21\x_+^4(\Psi)]^3}
    \times    \Big[1-24\x_+(\Psi)-282\x_+^2(\Psi)-820\x_+^3(\Psi)\notag\\
&\hspace{2cm}-504\x_+^4(\Psi)+1344\x_+^5(\Psi)+1746\x_+^6(\Psi)-324\x_+^7(\Psi)-945\x_+^8(\Psi)\Big].
\end{align}
Then for all $\x_+(\Psi)$ allowed, $T_1$ is greater than zero; while $T_2>0$ only for $\x_+(\Psi)\lesssim0.03$ or equivalently $\Psi\gtrsim85$. For all realistic NSs, the $\Psi$ factor is roughly smaller than 10, see FIG.\,\ref{fig_k-fac}, so we can treat $T_2<0$.
A negative $T_2$ shows that the growth rate of $M_{\rm{NS}}$ when $s_{\rm{c}}^2\approx1$ is sub-linear.
See FIG.\,\ref{fig_Msc2} for the dependence of $\mathcal{M}(s_{\rm{c}}^2,\Psi)$ on stiffness $s_{\rm{c}}^2$; the super- or sub-linear feature of the growth of $\mathcal{M}$ over $s_{\rm{c}}^2$ implies that $M_{\rm{NS}}$ increases much faster (slower) when $s_{\rm{c}}^2$ is small (large), e.g., we have $\mathcal{M}(0.5,0)/\mathcal{M}(0.25,0)\approx1.42$ and $\mathcal{M}(1.0,0)/\mathcal{M}(0.5,0)\approx1.17$, when the $s_{\rm{c}}^2$ is doubled.
Another feature of FIG.\,\ref{fig_Msc2} is that the curve of $\mathcal{M}(s_{\rm{c}}^2,\Psi)$ becomes more flat for smaller $\Psi$, indicating the stiffness $s_{\rm{c}}^2$ has a more obvious effect on stable NSs away from the TOV configuration (so the $\Psi$ is large). For example, we have $\mathcal{M}(0.5,1)/\mathcal{M}(0.25,1)\approx1.52$ besides $\mathcal{M}(0.5,0)/\mathcal{M}(0.25,0)\approx1.42$ given previously.
The nonlinearity of the growth rate of $M_{\rm{NS}}$ over $s_{\rm{c}}^2$ (for a given $\varepsilon_{\rm{c}}$) near some other SSS (say, $s_{\rm{c}}^2\approx0.5$) could be analyzed similarly and would not be discussed further.

The second condition of the TOV configuration (\ref{MTOV-cr}) can similarly induce useful information/constraint on the EOS.
After some long but straightforward derivations, we obtain
\begin{equation}
\frac{\d^2M_{\rm{NS}}}{\d\varepsilon_{\rm{c}}^2}
\sim\left(1-\frac{s_{\rm{c}}^2}{{\x}}\right)\left[
\left(\frac{s_{\rm{c}}^2}{{\x}}-\frac{\d s_{\rm{c}}^2}{\d{\x}}\right)
+{\x}\left(1-\frac{s_{\rm{c}}^2}{{\x}}\right)\frac{12{\x}^2+12{\x}+4}{9{\x}^4+12{\x}^3-4{\x}-1}\right].\label{ss-5}
\end{equation}
Now, one can not directly use the expression for $s_{\rm{c}}^2=\d P_{\rm{c}}/\d\varepsilon_{\rm{c}}$ given by  (\ref{sc2-TOV}) to evaluate the derivative $\d s_{\rm{c}}^2/\d{\x}$ since (\ref{sc2-TOV}) is obtained via the condition $\d M_{\rm{NS}}/\d\varepsilon_{\rm{c}}=0$, namely if (\ref{sc2-TOV}) is put in (\ref{ss-5}) the expression is identically zero.
Instead, demanding generally $\d^2M_{\rm{NS}}/\d\varepsilon_{\rm{c}}^2<0$ leads us to,
\begin{equation}\label{io-7}
\left.\frac{\d s_{\rm{c}}^2}{\d{\x}}\right|_{M_{\rm{NS}}^{\max}}<
\sigma_{\rm{c}}^2\equiv\frac{\d}{\d{\x}}\left[{\x}\left(1
+\frac{1}{3}\frac{1+3{\x}^2+4{\x}}{1-3{\x}^2}
\right)\right]=
\frac{2}{3}\frac{9{\x}^4-3{\x}^2+4{\x}+2}{(3{\x}^2-1)^2},
\end{equation}
since the in-front factor $1-s_{\rm{c}}^2/{\x}$ in (\ref{ss-5}) is negative.
We may use this criterion in SECTION \ref{SEC_7} (Subsection \ref{sub_s2_1st}) when discussing the radial dependence of the SSS.

\begin{figure}[h!]
\centering
\includegraphics[height=7.cm]{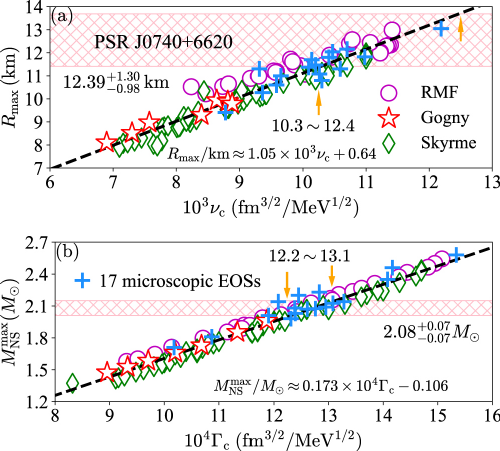}\qquad
\includegraphics[height=7.cm]{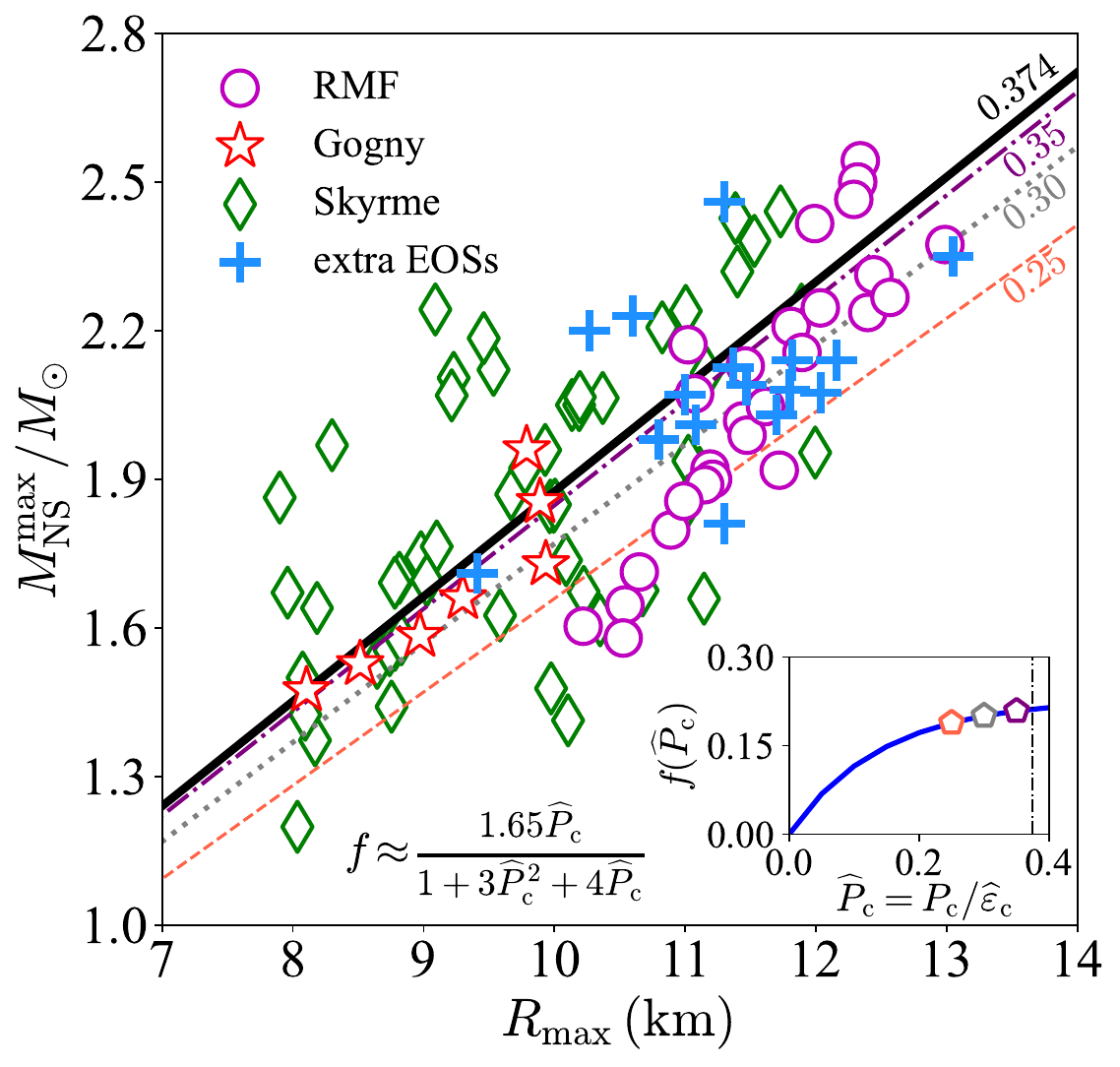}
\caption{(Color Online). 
Panel (a): the $R_{\max}$-$\nu_{\rm{c}}$ correlation
using 104 EOS samples (colored symbols), see Ref.\,\cite{CLZ23-a} for more detailed descriptions on these EOSs,  the constraints on the mass\,\cite{Fon21} and radius\,\cite{Riley21} of PSR J0740+6620 are shown by the pink hatched bands. Panel (b): similar as the left panel but fot $M_{\rm{NS}}^{\max}$-$\Gamma_{\rm{c}}$.
 correlations with 104 EOS samples (colored symbols), 
The two orange arrows and captions nearby in each panel indicate the corresponding ranges for $\nu_{\rm{c}}$ and $\Gamma_{\rm{c}}$ defined in Eq.\,(\ref{gk-radius}) and Eq.\,(\ref{gk-mass}), respectively.
Figures taken from Ref.\,\cite{CLZ23-a}.
Right panel: scatters of $M_{\rm{NS}}^{\max}$ versus $R_{\max}$ with the corresponding 104 EOSs, the captions near each lines are the fiducial values for ${\x}$.
Figure taken from Ref.\,\cite{CLZ23-b}.
}\label{fig_MmaxS}
\end{figure}

We show in FIG.\,\ref{fig_MmaxS} the $R_{\max}$-$\nu_{\rm{c}}$ (panel (a)) and the $M_{\rm{NS}}^{\max}$-$\Gamma_{\rm{c}}$ (panel (b)) correlations by using 87 phenomenological and 17 extra microscopic NS EOSs with and/or without considering hadron-quark phase transitions and hyperons by solving numerically the original TOV equations, see Ref.\,\cite{CLZ23-a} for more details on these EOS samples.
The observed strong linear correlations demonstrate vividly that the $R_{\max}$-$\nu_{\rm{c}}$ and $M_{\rm{NS}}^{\max}$-$\Gamma_{\rm{c}}$ scalings are nearly universal. 
Since the TOV configuration is very near the mass threshold before NSs collapse into BHs, and it is well known that certain properties of BHs are universal and only depend on quantities like mass, charge and angular momentum, it is not surprising that the predicted mass and radius scalings at the TOV configuration hold very well with the diverse set of EOSs. It is also particularly interesting to notice that EOSs allowing phase transitions and/or hyperon formations confirm consistently the same scalings first revealed from our analyses of the scaled TOV equations without using any EOS. By performing linear fits of the results obtained from using these EOS samples in solving the TOV equations in the traditional approach, the quantified scaling relations are determined to be\,\cite{CLZ23-a,CLZ23-b,CL24-a,CL24-c}
 \begin{equation}\label{Rmax-n}
 \boxed{
R_{\rm{max}}/\rm{km}
\approx A_{\rm{R}}^{\max}\nu_{\rm{c}}+B_{\rm{R}}^{\max}
\approx 1.05_{-0.03}^{+0.03}\times 10^3 \left(\frac{\nu_{\rm{c}}}{\rm{fm}^{3/2}/\rm{MeV}^{1/2}}\right)+0.64_{-0.25}^{+0.25},}
\end{equation} with its Pearson's coefficient about 0.958 and 
\begin{equation}\label{Mmax-G}
\boxed{
M_{\rm{NS}}^{\rm{max}}/M_{\odot} 
\approx A_{\rm{M}}^{\max}\Gamma_{\rm{c}}+B_{\rm{M}}^{\max}
\approx 1.73_{-0.03}^{+0.03}\times 10^3  \left(\frac{\Gamma_{\rm{c}}}{\rm{fm}^{3/2}/\rm{MeV}^{1/2}}\right)-0.106_{-0.035}^{+0.035},}
\end{equation} with its Pearson's coefficient about 0.986, respectively, here $\nu_{\rm{c}}$ and $\Gamma_{\rm{c}}$ are measured in $\rm{fm}^{3/2}/\rm{MeV}^{1/2}$.
In addition, the standard errors (ste's) for the radius and mass fittings are about 0.031 and 0.003 for these EOS samples.
In FIG.\,\ref{fig_MmaxS}, the condition $M^{\max}_{\rm{NS}}\gtrsim1.2M_{\odot}$ used is necessary to mitigate influences of uncertainties in modeling the crust EOS\,\cite{BPS71,Iida1997,XuJ} for low-mass NSs. For the heavier NSs studied here, it is reassuring to see that although the above 104 EOSs predicted quite different crust properties, they all fall closely around the same scaling lines consistently, especially for the $M_{\rm{NS}}^{\max}$-$\Gamma_{\rm{c}}$ relation.

Using the numerical forms of Eqs.\,(\ref{Rmax-n}) and (\ref{Mmax-G}), we similarly have,
\begin{equation}\label{rel-1}
\frac{M_{\rm{NS}}^{\max}}{M_{\odot}}\approx\frac{1.65{\x}}{1+3{\x}^2+4{\x}}\left(\frac{R_{\max}}{\rm{km}}-0.64\right)-0.106.
\end{equation}
The in-front coefficient $f\approx1.65{\x}/(1+3{\x}^2+4{\x})$ is a slow-varying function of ${\x}$, and it essentially explains the conventional quasi-linear correlation between $M_{\rm{NS}}^{\max}/M_{\odot}$ and $R_{\max}$ from model calculations.
In the right panel of FIG.\,\ref{fig_MmaxS}, the scatters of $M_{\rm{NS}}^{\max}$ versus $R_{\max}$ using the above 104 EOSs\,\cite{CLZ23-a} are shown.
The fitting lines of Eq.\,(\ref{rel-1}) adopt four fiducial values for ${\x}$ (captioned near the lines with the same color), from which one finds that for ${\x}\gtrsim0.3$ the expression (\ref{rel-1}) could reasonably describe the EOS samples.
However, obvious dispersions can be seen for these scatters.
In addition, certain EOSs are unfavored by the causal limit ${\x}\lesssim0.374$ (solid black line, corresponding to $s_{\rm{c}}^2\leq1$ discussed above).
We shall discuss the causality boundary for NSs in details in SECTION \ref{SEC_56}.

\renewcommand*\tablename{\footnotesize TAB.}

\begin{table}[h!]
\renewcommand{\arraystretch}{1.5}
\centerline{\normalsize
\begin{tabular}{c|c|c|c|c|c} 
\hline
Ref.& radius\,(km) &$10^3\nu_{\rm{c}}$&$\varepsilon_{\rm{c}}$&$P_{\rm{c}}$&$s_{\rm{c}}^2$\\\hline\hline
       \cite{Riley21} &$12.39_{-0.98}^{+1.30}$&$11.2_{-0.9}^{+1.2}$&$901_{-287}^{+214}$&$218_{-125}^{+93}$&$0.45_{-0.18}^{+0.14}$\\\hline
    \cite{Miller21}  &$13.7_{-1.5}^{+2.6}$&$12.4_{-1.4}^{+2.5}$&$656_{-339}^{+187}$&$124_{-99}^{+53}$&$0.32_{-0.14}^{+0.08}$\\\hline
    \cite{Salmi22} &$12.90_{-0.97}^{+1.25}$&$11.7_{-0.9}^{+1.2}$&$794_{-235}^{+181}$&$173_{-89}^{+69}$&$0.39_{-0.13}^{+0.09}$\\\hline
    \cite{Salmi24}&$12.49_{-0.88}^{+1.28}$&$11.3_{-0.9}^{+1.3}$&$879_{-312}^{+208}$&$208_{-140}^{+94}$&$0.44_{-0.21}^{+0.14}$\\\hline
    \cite{Ditt24}&$12.76_{-1.02}^{+1.49}$&$11.5_{-1.2}^{+1.8}$&$822_{-383}^{+255}$&$184_{-157}^{+105}$&$0.40_{-0.22}^{+0.15}$\\\hline
    \end{tabular}}
        \caption{Most probable values for the central energy density $\varepsilon_{\rm{c}}$ (fourth column), central pressure $P_{\rm{c}}$ (fifth column) and the central SSS $s_{\rm{c}}^2$ (sixth column) for PSR J0740+6620, here its mass about $2.08_{-0.07}^{+0.07}M_{\odot}$ and so $12.2\lesssim{10^4\Gamma_{\rm{c}}}/[{\rm{fm}^{3/2}/\rm{MeV}^{1/2}]}
\lesssim13.1$ are used for five cases for the radius\,\cite{Riley21,Miller21,Salmi22,Salmi24,Ditt24}; $10^3\nu_{\rm{c}}$ is measured in unit $\rm{fm}^{3/2}/\rm{MeV}^{1/2}$, $\varepsilon_{\rm{c}}$ and $P_{\rm{c}}$ are measured in $\rm{MeV}/\rm{fm}^3$.
Table taken from Ref.\,\cite{CLZ23-a} with modifications.}\label{sstab_aa}        
\end{table}
Given a $\Gamma_{\rm{c}}$ (mass observation) or $\nu_{\rm{c}}$ (radius observation), either the mass scaling or the radius scaling can lead to a functional relation between $P_{\rm{c}}$ and $\varepsilon_{\rm{c}}$, namely the central EOS:
\begin{align}
P_{\rm{c}}(\varepsilon_{\rm{c}})\approx& f_{\rm{M}}^{2/3}\varepsilon_{\rm{c}}^{4/3}\cdot\left(1+4f_{\rm{M}}^{2/3}\varepsilon_{\rm{c}}^{1/3}+19f_{\rm{M}}^{4/3}\varepsilon_{\rm{c}}^{2/3}+
100f_{\rm{M}}^2\varepsilon_{\rm{c}}+\cdots\right),\label{P1-pert}\\
P_{\rm{c}}(\varepsilon_{\rm{c}})\approx&f_{\rm{R}}^2\varepsilon_{\rm{c}}^2\cdot
\left(1+4f_{\rm{R}}^2\varepsilon_{\rm{c}}+19f_{\rm{R}}^4\varepsilon_{\rm{c}}^2+100f_{\rm{R}}^6\varepsilon_{\rm{c}}^3+\cdots\right)\label{P2-pert},
\end{align}
where $
{f_{\rm{M}}}/[{\rm{fm}^{3/2}/\rm{MeV}^{1/2}}]=({M_{\rm{NS}}^{\max}/M_{\odot}+0.106})/{1730}$ and $
{f_{\rm{R}}}/[{\rm{fm}^{3/2}/\rm{MeV}^{1/2}}]=({R_{\max}/\rm{km}-0.64})/{1050}$.

With the scalings given above, if the mass and radius of a massive NS are observed simultaneously, then one can determine individually its central pressure and central energy density. As an example, for one of the most massive NSs observed/confirmed so far, e.g., PSR J0740+6620, it is reasonable to use the scallings at the TOV configuration. The extracted central properties shown in TAB.\,\ref{sstab_aa} for PSR J0740+6620 are indeed very interesting. In particular, Eq.\,(\ref{P1-pert}) implies that a heavier NS has a smaller upper $\varepsilon_{\rm{c}}$ (as the ratio $P_{\rm{c}}/\varepsilon_{\rm{c}}$ is bounded from above) and therefore is easier to collapse into a BH\,\cite{Hawking1973}. The observational radius about $12.39_{-0.98}^{+1.30}\,\rm{km}$\,\cite{Riley21} at 68\% confidence level and mass about $2.08_{-0.07}^{+0.07}M_{\odot}$\,\cite{Fon21} of PSR J0740+6620 are also shown in the left panels of FIG.\,\ref{fig_MmaxS}.
The strong linear correlations of $M_{\rm{NS}}^{\max}$-$\Gamma_{\rm{c}}$ and $R_{\max}$-$\nu_{\rm{c}}$ then enable us to read out the corresponding values for $\Gamma_{\rm{c}}$ and $\nu_{\rm{c}}$.
Their intersections with the two scaling lines indicate that $10.3\lesssim10^{3}\nu_{\rm{c}}/[\rm{fm}^{3/2}/\rm{MeV}^{1/2}]\lesssim12.4$ and  $12.2\lesssim{10^4\Gamma_{\rm{c}}}/[{\rm{fm}^{3/2}/\rm{MeV}^{1/2}]}
\lesssim13.1$, respectively, in PSR J0740+6620 (indicated by the orange arrows). 
For a comparison, we notice that the hadron-quark hybrid EOS ALF2 predicts its $M_{\rm{NS}}^{\max}\approx2.09M_{\odot}$\,\cite{AFL2} that is very close to the mass of PSR J0740+6620. It also predicts $P_{\rm{c}}\approx310\,\rm{MeV}/\rm{fm}^{3}$ and $\varepsilon_{\rm{c}}\approx1100\,\rm{MeV}/\rm{fm}^3$,   therefore $\Gamma_{\rm{c}}\approx12.4\times10^{-4}\,\rm{MeV}^{1/2}/\rm{fm}^{3/2}$, being consistent with the $\Gamma_{\rm{c}}$ extracted for PSR J0740+6620 from the mass scaling shown in FIG.\,\ref{fig_MmaxS}.
Different observational constraints on the radius may lead to different ranges for $\nu_{\rm{c}}$. We found that the radius constraint of $13.7_{-1.5}^{+2.6}\,\rm{km}$\,\cite{Miller21} gives $11.0\lesssim10^{3}\nu_{\rm{c}}/[\rm{fm}^{3/2}/\rm{MeV}^{1/2}]\lesssim14.9$; and the recent analysis including NICER's background estimates\,\cite{Salmi22} found a radius of $12.90_{-0.97}^{+1.25}\,\rm{km}$, leading to $10.8\lesssim10^{3}\nu_{\rm{c}}/[\rm{fm}^{3/2}/\rm{MeV}^{1/2}]\lesssim12.9$. 
See the third column of TAB.\,\ref{sstab_aa} for these values for $\nu_{\rm{c}}$.
If we use the very recently updated constraints on the radius as a result of using more observational data accumulated\,\cite{Salmi24,Ditt24}, the $\nu_{\rm{c}}$ should be updated correspondingly, see the last two rows of TAB.\,\ref{sstab_aa}.
In the following discussions, we mainly use the radius of Ref.\,\cite{Riley21}.
Having obtained the $\Gamma_{\rm{c}}$ and $\nu_{\rm{c}}$, the factor $\Pi_{\rm{c}}=\Gamma_{\rm{c}}/\nu_{\rm{c}}$ defined in Eq.\,(\ref{gk-comp}) could be obtained as
\begin{equation}
\boxed{
\mbox{PSR J0740+6620:}~~{\x}\approx0.24_{-0.07}^{+0.05},~~\mbox{under}~~R_{\max}\approx12.39_{-0.98}^{+1.30}\,\rm{km}.}
\end{equation}

From (\ref{P1-pert}), we obtain then
\begin{equation}\label{gammac_M}
\gamma_{\rm{c}}^{(\rm{M})}=\frac{s_{\rm{c}}^2}{\x}
\approx \frac{4}{3}\left(1+f_{\rm{M}}^{2/3}\varepsilon_{\rm{c}}^{1/3}+\frac{11}{2}f_{\rm{M}}^{4/3}\varepsilon_{\rm{c}}^{3/3}+34f_{\rm{M}}^{2}\varepsilon_{\rm{c}}+\cdots\right),~~f_{\rm{M}}\approx\left(\frac{M_{\rm{NS}}^{\max}+0.106}{1730}\right)\rm{fm}^{3/2}/\rm{MeV}^{1/2},
\end{equation}
under the basic definition for the polytropic index of Eq.\,(\ref{def_gamma}).
Similarly, we have from (\ref{P2-pert}) that
\begin{equation}\label{gammac_R}
\gamma_{\rm{c}}^{(\rm{R})}\approx2\left(1+2f_{\rm{R}}^2\varepsilon_{\rm{c}}+11f_{\rm{R}}^4\varepsilon_{\rm{c}}^2+68f_{\rm{R}}^6\varepsilon_{\rm{c}}^3+\cdots\right),~~f_{\rm{R}}\approx\left(\frac{M_{\rm{NS}}^{\max}-0.64}{1050}\right)\rm{fm}^{3/2}/\rm{MeV}^{1/2}.
\end{equation}
Eqs.\,(\ref{gammac_M}) and (\ref{gammac_R}) indicate that the $\gamma_{\rm{c}}$ is greater than 4/3.
For example, taking $M_{\rm{NS}}^{\max}\approx2M_{\odot}$, $R_{\max}\approx12\,\rm{km}$ and $\varepsilon_{\rm{c}}\approx800\,\rm{MeV}/\rm{fm}^3$ gives the first-order corrections as $f_{\rm{M}}^{2/3}\varepsilon_{\rm{c}}^{1/3}\approx0.106$ and $f_{\rm{R}}^2\varepsilon_{\rm{c}}\approx0.094$, respectively.
We discuss in Subsection \ref{sub_conformal} the polytropic index $\gamma_{\rm{c}}$ further using the expression for $s_{\rm{c}}^2$ of Eq.\,(\ref{sc2-TOV}) obtained via the condition $\d M_{\rm{NS}}^{\max}/\d\varepsilon_{\rm{c}}=0$, see information of TAB.\,\ref{sstab}.

\begin{figure}[h!]
\centering
\includegraphics[width=8.3cm]{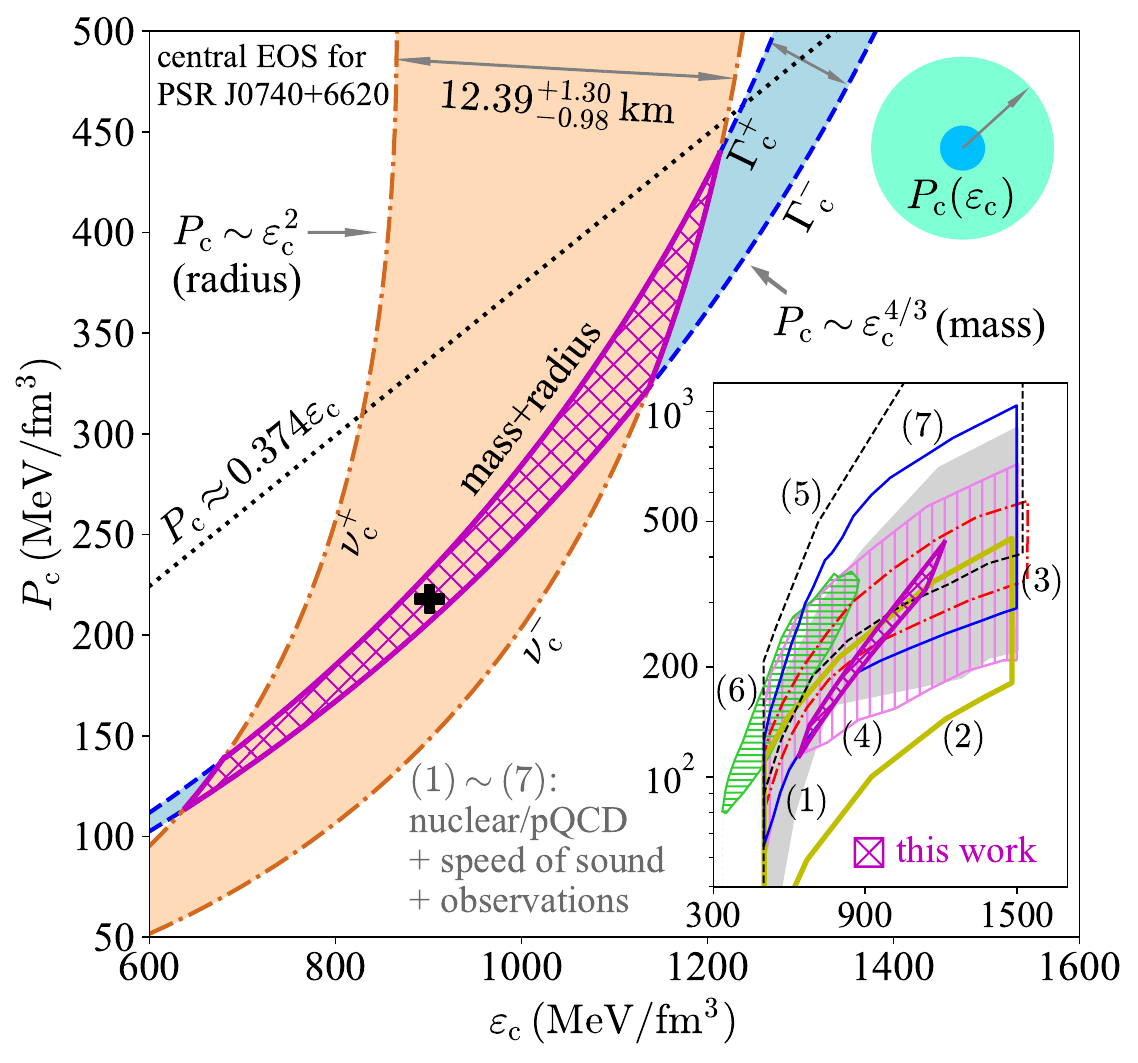}\quad
\includegraphics[width=8.3cm]{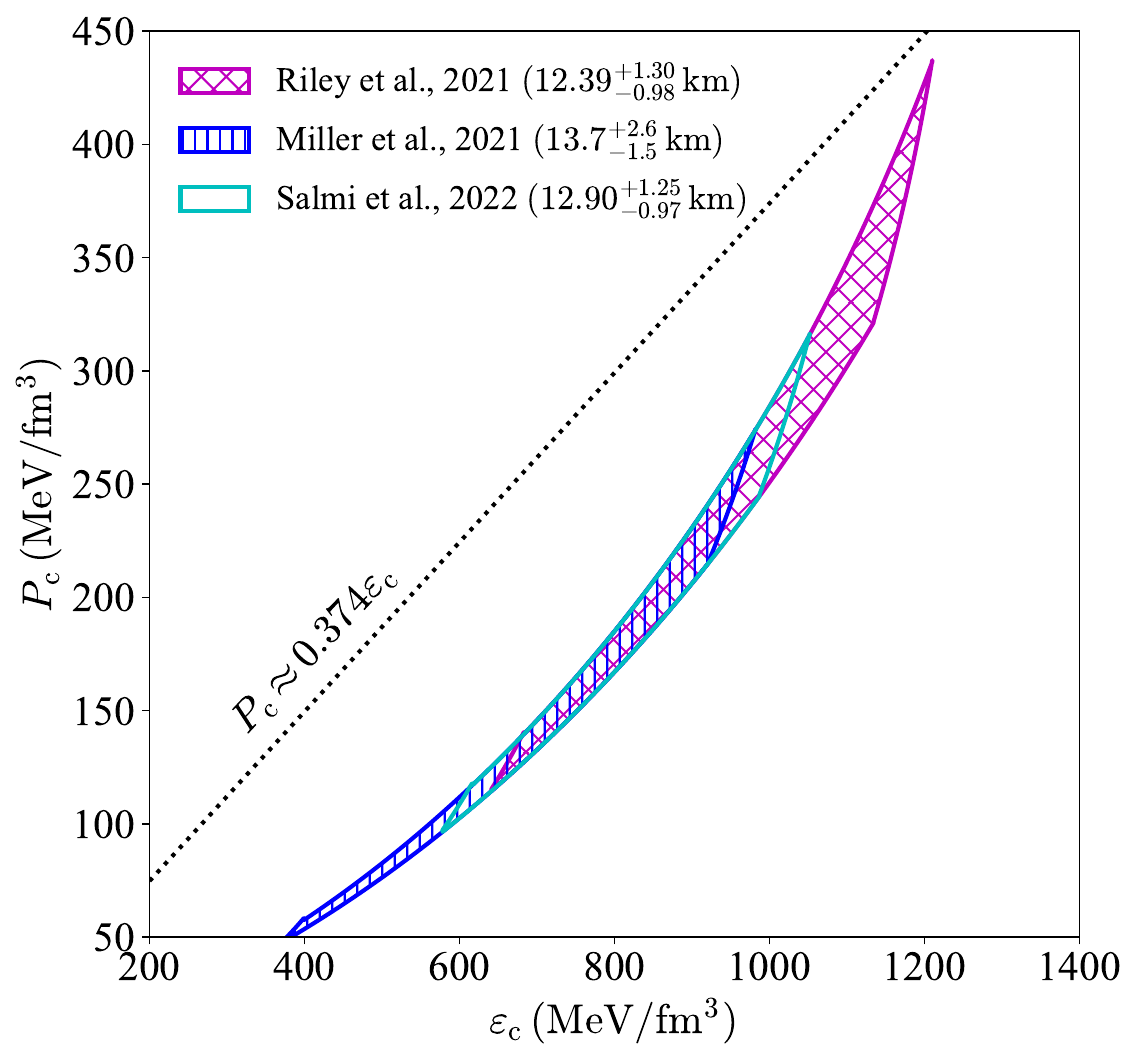}
\caption{(Color Online). Left panel: constraints on the central EOS $P_{\rm{c}}(\varepsilon_{\rm{c}})$ independently from the NICER measurements on the mass of $2.08_{-0.07}^{+0.07}M_{\odot}$\,\cite{Fon21} (blue band) and the radius of $12.39_{-0.98}^{+1.30}\,\rm{km}$\,\cite{Riley21} (chocolate band) for PSR J0740+6620, respectively.
The inset plots our central EOS (magenta crossed band) and predictions/constraints from a few other approaches (the energy density varies from 500 to 1500$\,\rm{MeV}/\rm{fm}^3$).
Right panel: constraining bands for the central EOS of PSR J0740+6620 using different radii\,\cite{Riley21,Miller21,Salmi22}.
Figures taken from Ref.\,\cite{CLZ23-a}.
}\label{fig_eP-PSR740+6620}
\end{figure}

The $\Gamma_{\rm{c}}$ and $\nu_{\rm{c}}$ extracted from NICER's joint mass-radius observation for PSR J0740+6620 provide two independent constraints on the NS central EOS as shown in the left panel of FIG.\,\ref{fig_eP-PSR740+6620} by the blue and chocolate bands, respectively.
Since the uncertainty of the radius is currently about three times larger than that of the mass measurement,  the resulting constraining band of the central EOS from the radius measurement is much wider. 
This means the central EOS is essentially given by the mass measurement via $\Gamma_{\rm{c}}$ while the radius effectively sets the upper- and lower-limits for $\varepsilon_{\rm{c}}$. 
In the overlapping area of the two constraining bands,  the most probable NS central EOS can then be written as\,\cite{CLZ23-a}
\begin{equation}
\mbox{PSR J0740+6620:}~~P_{\rm{c}}(\varepsilon_{\rm{c}})\approx0.012\varepsilon_{\rm{c}}^{4/3}\cdot\left(1+0.047\varepsilon_{\rm{c}}^{1/3}+0.0026\varepsilon_{\rm{c}}^{2/3}+0.00016\varepsilon_{\rm{c}}+\cdots\right),
\end{equation}
with roughly $
640\,\rm{MeV}/\rm{fm}^3\lesssim\varepsilon_{\rm{c}}\lesssim1210\,\rm{MeV}/\rm{fm}^3$.
The inset of the left panel of FIG.\,\ref{fig_eP-PSR740+6620} compares our constraints on the central EOS (shown as magenta crossed band) with predictions on the dense NS matter EOS from several other approaches, these include (1) combination of low-density nuclear,  high-density pQCD theories and the tidal deformability from GW event GW170817\,\cite{Abbott2018} shown as the grey shaded band\,\cite{Ann18}; (2) pQCD with constraint of the SSS at very high densities shown as the yellow solid band\,\cite{Kom2022PRL}; (3) similar physical inputs as (1) combining Gaussian-process (GP) inference algorithm using the red dash-dotted band\,\cite{Gorda2023}; (4) the construction of EOS ensembles consistent with nuclear theories,  pQCD and astronomical observations and continuous SSS shown by the pink vertical hatched band\,\cite{Altiparmak2022}; (5) a nonparametric inference via GP adopting similar astronomical observations by the black dashed band  together with (6) the prediction on the central pressure-density of PSR J0740+6620 shown by green horizontal hatched band\,\cite{Leg21}; and (7) the constraint from an earlier model analysis of the NICER data by the blue solid band\,\cite{Miller21}.  

Our constraining band on the central NS EOS directly from the NICER data using the mass and radius scalings without using any specific input EOS model is mostly consistent with but much narrower than the NS EOS bands from the previous studies. For example, we have $\varepsilon_{\rm{c}}\approx901_{-287}^{+214}\,\rm{MeV}/\rm{fm}^3$ and $P_{\rm{c}}\approx218_{-125}^{+93}\,\rm{MeV}/\rm{fm}^3$ (black ``+'' in FIG.\,\ref{fig_eP-PSR740+6620}), respectively. 
This value of central energy density is consistent with $P_{\rm{c}}\approx312_{-169}^{+226}\,\rm{MeV}/\rm{fm}^3$ and $\varepsilon_{\rm{c}}\approx904_{-327}^{+329}\,\rm{MeV}/\rm{fm}^3$ and $\x\approx0.35$ for a $2.1M_{\odot}$ NS from a recent Bayesian inference\,\cite{Brandes2023}.
Notice that the inferred radius of that work\,\cite{Brandes2023} is about 11.6\,km, which is somewhat smaller than the observational radius for PSR J0740+6620, therefore leading to a much larger $\x$ (the matter is denser).
Including the black widow pulsar PSR J0952-0607 in the same inference framework, Ref.\,\cite{Brandes2023-a} then found $R\approx12.3\,\rm{km}$ as well as $P_{\rm{c}}\approx186\,\rm{MeV}/\rm{fm}^3$ and $\varepsilon_{\rm{c}}\approx628\,\rm{MeV}/\rm{fm}^3$ for a $2.1M_{\odot}$ NS and so $\x\approx0.30$, this is still larger than the $\x$ we extracted for PSR J0740+6620.
Also notice that our constraining band has obvious deviation from the one in Ref.\,\cite{Leg21}. One possible reason is that the inferred central baryon density in the latter work is about $3.6\rho_0$, which is much smaller than ours about $6\rho_0$, see Subsection \ref{sub_rhoc} for discussion/estimate on it.
If different radius constraints\,\cite{Miller21,Salmi22,Riley21} were used in the inference, we may obtain slightly different constraining band for the central EOS.
The result is shown in the right panel of FIG.\,\ref{fig_eP-PSR740+6620} from which one can easily find the constraining bands for $P_{\rm{c}}(\varepsilon_{\rm{c}})$ in all three cases are narrow, and the uncertainties on $P_{\rm{c}}(\varepsilon_{\rm{c}})$ mainly reflect the observational uncertainty on the radius of PSR J0740+6620.
See TAB.\,\ref{sstab_aa} for the extracted $P_{\rm{c}}$ and $\varepsilon_{\rm{c}}$ for these cases.

\begin{figure}[h!]
\centering
\includegraphics[width=16.cm]{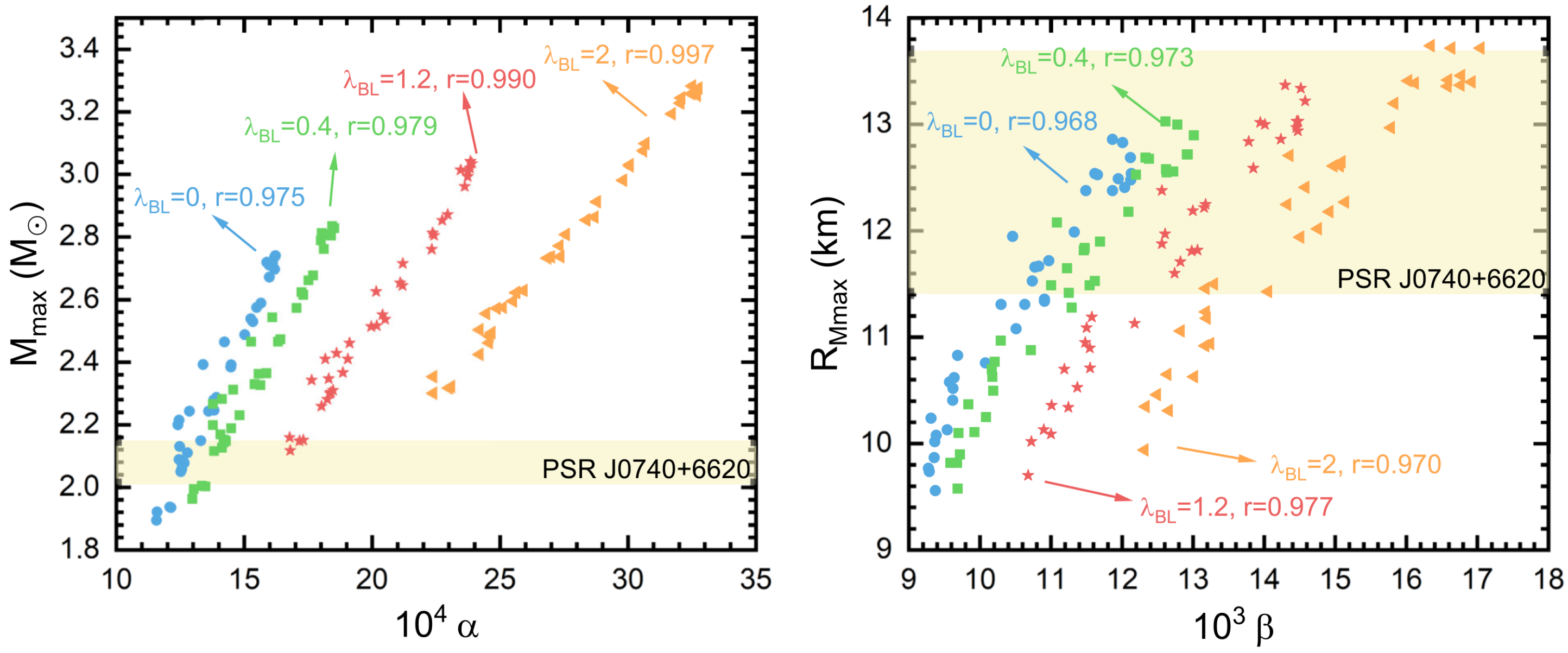}
\caption{(Color Online). Mass and radius scalings considering anisotropic pressure effects. Figure taken from Ref.\,\cite{YangWen2024}.
}\label{fig_YW24}
\end{figure}

Recently, the above NS mass and radius scalings were generalized by considering anisotropic pressure effects in Ref.\,\cite{YangWen2024}. In their analysis, there exists a tangential pressure $P_{\rm{t}}$ besides the conventional radial pressure $P$; the relation between them is
\begin{equation}
\delta P=    P_{\rm{t}}-P=\frac{\lambda_{\rm{BL}}}{3}\frac{(\varepsilon+3P)(\varepsilon+P)r^3}{r-2M},
\end{equation}
here $\lambda_{\rm{BL}}$ is a parameter characterizing the pressure anisotropy.
A new term $2\delta P/r$ should be added to the right hand side of the $\d P/\d r$ in the TOV equations.
Using a similar method as ours, the authors of Ref.\,\cite{YangWen2024} derived the following scalings
\begin{equation}
    M_{\rm{NS}}^{\max}\sim\alpha\equiv\Gamma_{\rm{c}}\left(1-\frac{\lambda_{\rm{BL}}}{2\pi}\right)^{-3/2},~~R\sim\beta\equiv\nu_{\rm{c}}\left(1-\frac{\lambda_{\rm{BL}}}{2\pi}\right)^{-1/2},
\end{equation}
where $\Gamma_{\rm{c}}$ and $\nu_{\rm{c}}$ are the same ones defined in Eqs.\,(\ref{gk-mass}) and (\ref{gk-radius}), respectively.
Their numerical verifications of the scalings are shown in FIG.\,\ref{fig_YW24} using different values for $\lambda_{\rm{BL}}$ between 0 and 2.
They also derived the expression for $s_{\rm{c}}^2$, which involves the parameter $\lambda_{\rm{BL}}$. In this case, the causality condition $s_{\rm{c}}^2\leq1$ may induce a different upper bound for $\x$.

\begin{figure}[h!]
\centering
\includegraphics[height=8.cm]{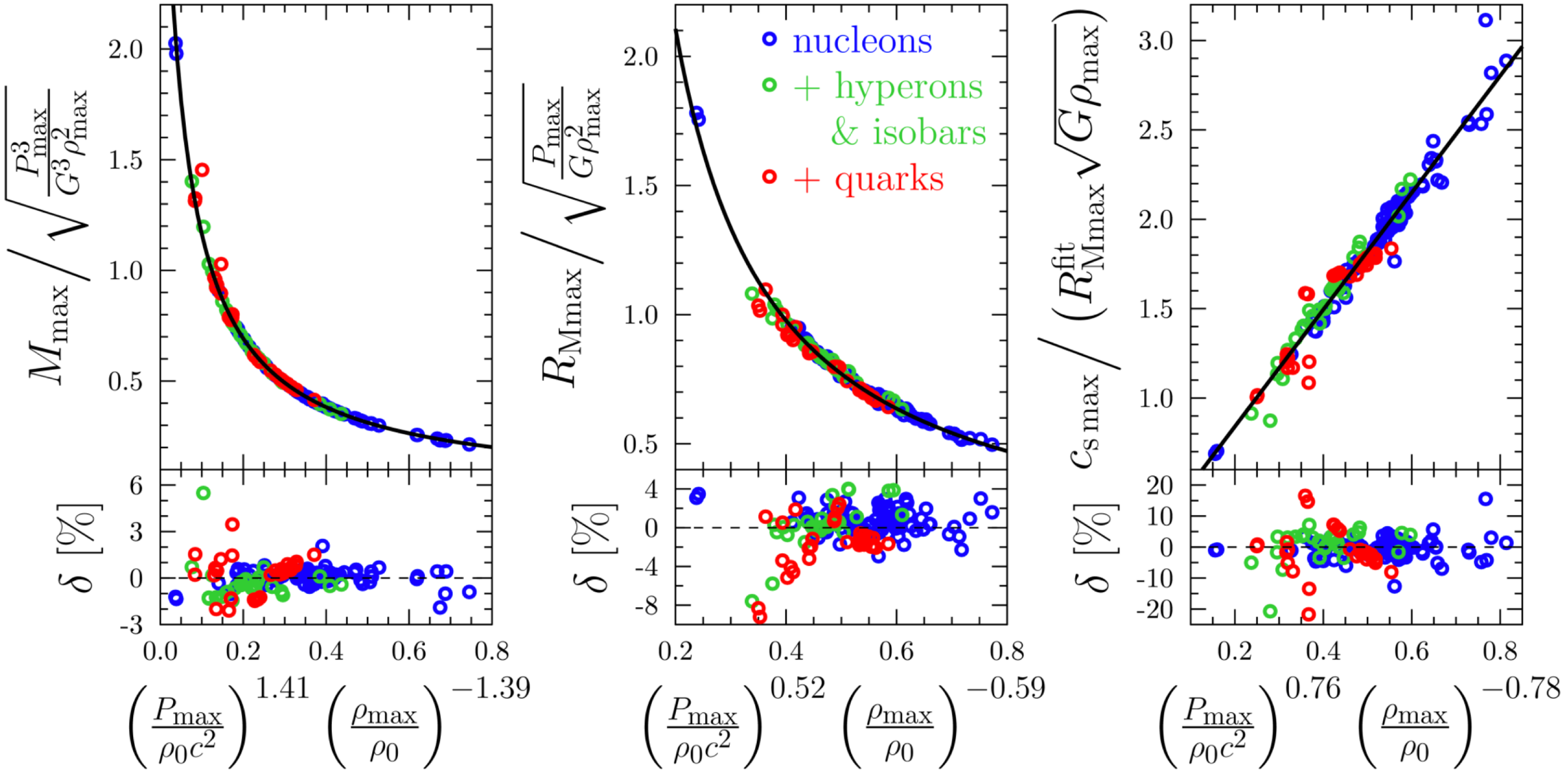}
\caption{(Color Online).  Scalings of the TOV mass $M_{\rm{TOV}}=M_{\rm{NS}}^{\max}$,  the corresponding radius $R_{\max}$ as well as the speed of sound with certain combinations of NS central quantities. Figure taken from Ref.\,\cite{Ofeng2023}.
}\label{fig_Ofeng}
\end{figure}

The TOV configuration is very important/useful for exploring properties of superdense matter, as we pointed out in the beginning of this subsection. Besides our scalings for NS mass and radius given above, another recent work in Ref.\,\cite{Ofeng2023} found that the TOV mass $M_{\rm{NS}}^{\max}=M_{\rm{TOV}}$ and radius $R_{\max}$ scale empirically as\,\cite{Ofeng2023},
\begin{equation}
M_{\rm{NS}}^{\max}\approx\frac{a_{\rm{M}}\varepsilon_{\rm{c}}^{-1/2}\x^{3/2}}{b_{\rm{M}}+c_{\rm{M}}\x^{p_{\rm{M}}}\varepsilon_{\rm{c}}^{q_{\rm{M}}}},~~
R_{\max}\approx\frac{a_{\rm{R}}\varepsilon_{\rm{c}}^{-1/2}\x^{1/2}}{b_{\rm{R}}+c_{\rm{R}}\x^{p_{\rm{R}}}\varepsilon_{\rm{c}}^{q_{\rm{R}}}},~~\x=P_{\rm{c}}/\varepsilon_{\rm{c}},
\end{equation}
here $a_{\rm{M,R}}$, $b_{\rm{M,R}}$, $c_{\rm{M,R}}$, $p_{\rm{M,R}}$ and $q_{\rm{M,R}}$ are fitting parameters\,\cite{Ofeng2023}. Their numerical verifications of the scalings using a large and diverse set of model EOSs are shown in FIG.\,\ref{fig_Ofeng}. It is seen that the relative uncertainties of their fittings are generally smaller than $8\%$, though the physical origin and meaning of their fitting parameters are still somewhat elusive.
See also the very recent work in Ref.\,\cite{Ofeng24}.

The above analysis/constraint on the central EOS for PSR J0740+6620 is based on the assumption that it is at the TOV configuration.
In Subsection \ref{sub_1420} we shall study the mass and radius scalings for NSs with arbitrary masses and extract the central EOS for two specific cases: a canonical NS with a mass of $M_{\rm{NS}}=1.4M_{\odot}$ and a $2.1M_{\odot}$ NS (e.g., PSR J0740+6620).

Very recently, Sun and Lattimer revisited the correspondence between NS M-R relation and the underlying nuclear EOS with the goal of inferring the latter more accurately and efficiently from observations compared to currently known approaches for solving the longstanding NS inverse structure problem\,\cite{Sun2024}. They used two approaches: (1) the perturbative solution of the dimensionless TOV equations and (2) fitting the energy density, pressure and the speed of sound as powers of NS mass and radius. In their perturbative analysis, they truncated the dimensionless TOV equations at the first leading-order in $\hr^2$, i.e., $\hP(\hr)\approx\x+b_2\hr^2$ and $\heps(\hr)\approx1+a_2\hr^2$ where $b_2$ is given by Eq.\,(\ref{ee-b2}), then set simultaneously the pressure and energy density to vanish at NS radius $\widehat{R}$.
This gives $\x+b_2\widehat{R}^2=0$ and $1+a_2\widehat{R}^2=0$, thus $\widehat{R}^2=-\x /b_2$ and $a_2=b_2/\x$. The resulting expression for $\widehat{R}$ is the same as Eq.\,(\ref{def-hR}), while that for $a_2$ is different from $a_2=b_2/s_{\rm{c}}^2$ using $s_{\rm{c}}^2$ of Eq.\,(\ref{sc2-TOV}) or Eq.\,(\ref{sc2-GG}).
Consequently, they obtained a different scaling form for the SSS (compared with Eqs.\,(\ref{sc2-TOV}) and (\ref{sc2-GG}) both having high-order terms in $\x$)\,\cite{Sun2024}:
\begin{equation}\label{s2-SL}
\boxed{
\mbox{Sun and Lattimer:}~~
    s^2={\d\hP}/{\d\heps}={b_2}/{a_2}=\x.}
\end{equation}
It is interesting to notice that the above expression holds for arbitrary radial distance $\hr$ (including $\hr=0$ at NS center), leading to a polytropic index of $\gamma=s^2/\phi=\x/\phi$ that equals to 1 at the center (i.e., $\gamma_{\rm{c}}=1$), because $s_{\rm{c}}^2=\x$. Moreover, since $s^2$ in Eq.\,(\ref{s2-SL}) is independent of $\hr$, it is equivalent to having an average SSS of $\x$ that is exactly what Eq.\,(\ref{def-AVERS2}) (in Subsection \ref{sub_conjecture}) defining the average SSS in NSs requires even for SSS varying with $\hr$. 

Going beyond the leading-order analysis given in Ref.\,\cite{Sun2024} by including more terms in expanding the pressure and energy density may essentially change the expressions for $\widehat{R}$ and SSS.
For example, considering the next-to-leading order correction
by keeping the $b_4$- and $a_4$-terms, neglecting the even higher order terms and requiring the pressure and energy density to vanish simultaneously on the surface at the $\widehat{R}^4$ order, i.e., 
$\x+b_2\widehat{R}^2+b_4\widehat{R}^4=0$ and $1+a_2\widehat{R}^2+a_4\widehat{R}^4=0$, we obtain the expression below for the central SSS and its difference $s_{\rm{c}}^2-\x$ from the leading-order one in Eq.\,(\ref{s2-SL}):
\begin{equation}
s_{\rm{c}}^2=\frac{b_2}{a_2}=\frac{\x+b_4\widehat{R}^4}{1+a_4\widehat{R}^4},
~~s_{\rm{c}}^2-\x=\frac{(b_4-a_4\x)\widehat{R}^4}{1+a_4\widehat{R}^4}.
\end{equation}
Using the same approximation, the SSS on the surface of radius $\widehat{R}$ is given by:
\begin{equation}\label{s2-surf}
s_{\rm{surf}}^2=\frac{b_2+2b_4\widehat{R}^2}{a_2+2a_4\widehat{R}^2}=\frac{\x-b_4\widehat{R}^4}{1-a_4\widehat{R}^4}.
\end{equation}
Consequently,
\begin{equation}
s_{\rm{c}}^2-s_{\rm{surf}}^2
=\frac{\x+b_4\widehat{R}^4}{1+a_4\widehat{R}^4}-\frac{\x-b_4\widehat{R}^4}{1-a_4\widehat{R}^4}
=\frac{2(b_4-a_4\x)\widehat{R}^4}{(1-a_4\widehat{R}^4)(1+a_4\widehat{R}^4)}
=\frac{2(s_{\rm{c}}^2-\x)}{1-a_4\widehat{R}^4}.
\end{equation}
Since $\d\hM/\d\hr>0$ and $\d\heps/\d\hr<0$, applying them at radius $\widehat{R}$ gives $a_4\widehat{R}^4<1$. Therefore $1-a_4\widehat{R}^4>0$.
Under the (reasonable) assumption  $s_{\rm{c}}^2>s_{\rm{surf}}^2$, one then obtains $s_{\rm{c}}^2>\x$ or equivalently $\gamma_{\rm{c}}\equiv s_{\rm{c}}^2/\x>1$, indicating qualitative effects of the next-to-leading order correction on the SSS.
Moreover, this analysis shows again that the NS central EOS could not be conformal signaled by $\gamma_{\rm{c}}\approx1$.

\begin{figure}[h!]
\centering
\includegraphics[height=6.5cm]{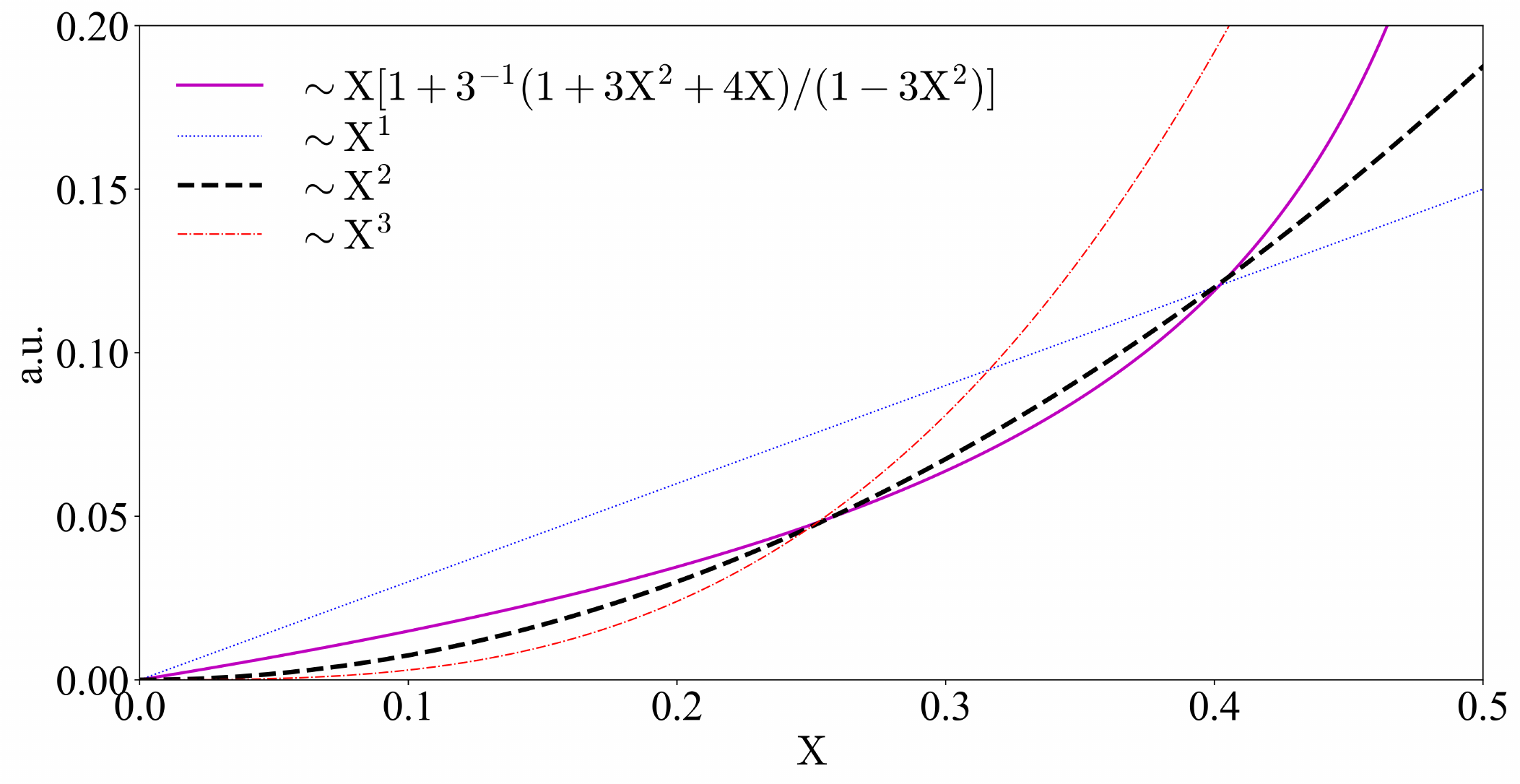}
\caption{(Color Online). The central SSS of Eq.\,(\ref{sc2-TOV}) can be mapped onto an effective power-law form $\x^{k}$ with ${k}\approx2$ (black dashed line). For typical NSs, $\x\gtrsim0.1$.
}\label{fig_SSD3}
\end{figure}
 
In the power-law fitting scheme, Sun and Lattimer suggested that\,\cite{Sun2024}
\begin{equation}
\boxed{
    \mathcal{O}=a[{\mathcal{O}}]\left(\frac{M_{\rm{NS}}^{\max}}{M_{\odot}}\right)^{b[{\mathcal{O}}]}\left(\frac{R_{\max}}{10\,\rm{km}}\right)^{c[{\mathcal{O}}]},}
\end{equation}
for a certain quantity $\mathcal{O}$ (e.g., energy density, pressure and the speed of sound). They found that their power-law fits for the energy density, pressure and speed of sound are accurate to about 1.5\%, 4.5\%, and 3.6\%, respectively, with respect to the exact solutions of TOV equations using the traditional approach. At the TOV configuration, the power-law exponents for fitting the central pressure are found to be $b[P_{\rm{c}}]\approx2.8$ and $c[P_{\rm{c}}]\approx-5.3$, respectively. Interestingly, an extra factor about $M_{\rm{NS}}^{\max}/R_{\max}$ emerges compared with Eq.\,(\ref{Pc-M2R4}), i.e., $P_{\rm{c}}\sim M_{\rm{NS}}^{\max,3}/R_{\max}^5\approx(M_{\rm{NS}}^{\max,2}/R_{\max}^4)(M_{\rm{NS}}^{\max}/R_{\max})\sim \varepsilon_{\rm{c}}\Pi_{\rm{c}}^2$.
Additionally, by fitting the central SSS $b[s_{\rm{c}}]\approx1.9$ and $c[s_{\rm{c}}]\approx-2.4$ were found\,\cite{Sun2024}, implying $s_{\rm{c}}\sim(M_{\rm{NS}}^{\max}/R_{\max})^2$. In fact, these findings can be well explained quantitatively by the mass and radius scalings of Eqs.\,(\ref{gk-mass}) and (\ref{gk-radius}) as well as the expression for $s_{\rm{c}}^2$ of Eq.\,(\ref{sc2-TOV}).
As shown in the left panel of FIG.\,\ref{fig_RSca}, the function $\x/(1+3\x^2+4\x)$ could be mapped onto an effective power-law form $\x^{q}$ with $q\approx1/2$.
Similarly, the central SSS for TOV NSs of Eq.\,(\ref{sc2-TOV}) can be mapped onto the following power-law form:
\begin{equation}\label{sc2EFF}
    s_{\rm{c}}^2=\x\left(1+\frac{1}{3}\frac{1+3\x^2+4\x}{1-3\x^2}\right)\sim\x^{{k}},~~\rm{with}~~{k}\approx2,
\end{equation}
as shown in FIG.\,\ref{fig_SSD3}.
With these effective forms, we then obtain $
R_{\max}
\sim P_{\rm{c}}^{q/2}
\varepsilon_{\rm{c}}^{-q/2-1/2}$ from Eq.\,(\ref{gk-radius}) and $M_{\rm{NS}}^{\max}\sim P_{\rm{c}}^{3q/2}\varepsilon_{\rm{c}}^{-3q/2-1/2}$ from Eq.\,(\ref{gk-mass}), so
\begin{equation}
\boxed{
P_{\rm{c}}
\sim \left(M_{\rm{NS}}^{\max}\right)^{1+q^{-1}}R_{\max}^{-3-q^{-1}}\sim M_{\rm{NS}}^{\max,3}/R_{\max}^{5},~~\rm{with}~~q\approx1/2.}
\end{equation}
Dividing $M_{\rm{NS}}^{\max}$ and $R_{\max}$ leads to ${M_{\rm{NS}}^{\max}}/{R_{\max}}\sim({P_{\rm{c}}}/{\varepsilon_{\rm{c}}})^{q}=\x^{q}$, or inversely $\x\sim \left({M_{\rm{NS}}^{\max}}/{R_{\max}}\right)^{q^{-1}}\approx\left({M_{\rm{NS}}^{\max}}/{R_{\max}}\right)^{2}$ (using $q\approx1/2$).
Then Eq.\,(\ref{sc2EFF}) gives:
\begin{equation}
\boxed{
s_{\rm{c}}\sim
\x^{{k}/2}\sim\left({M_{\rm{NS}}^{\max}}/{R_{\max}}\right)^{{k}/2q}\approx(M_{\rm{NS}}^{\max}/R_{\max})^2,~~\mbox{with}~~q\approx1/2,~{k}\approx2.}
\end{equation}
Interestingly, if one adopts Eq.\,(\ref{s2-SL}) for the central SSS as $s_{\rm{c}}^2=\x$ and considering that $M_{\rm{NS}}^{\max}/R_{\max}=\widehat{M}_{\rm{NS}}^{\max}/\widehat{R}_{\max}\approx2\widehat{R}_{\max}^2/15\sim \widehat{R}^2_{\max}\sim\x/(1+3\x^2+4\x)$\,\cite{Sun2024} which still maps effectively onto $\x^{q}$ with $q\approx1/2$, then we obtain $s_{\rm{c}}\sim \x^{1/2}\sim(M_{\rm{NS}}^{\max}/R_{\max})$; it is a factor of $M_{\rm{NS}}^{\max}/R_{\max}$ different from their power-law fittings mentioned above\,\cite{Sun2024}.
Finally, for Newtonian stars (with $\x\to0$),  $q\approx1$ (which follows directly from Eq.\,(\ref{gk-mass})), so $M_{\rm{NS}}^{\max}/R_{\max}\sim\x$; and $s_{\rm{c}}^2\approx 4\x/3\sim \x$ (indicating ${k}\approx1$). Consequently, one obtains   $s_{\rm{c}}\sim\x^{1/2}\sim(M_{\rm{NS}}^{\max}/R_{\max})^{1/2}$. Thus, the leading-order approximation for SSS in Eq.\,(\ref{s2-SL}) is more appropriate for light NSs or Newtonian stars.

\subsection{Estimating the central baryon density of NSs at the TOV configuration}\label{sub_rhoc}

\begin{figure}[h!]
\centering
\includegraphics[height=7.1cm]{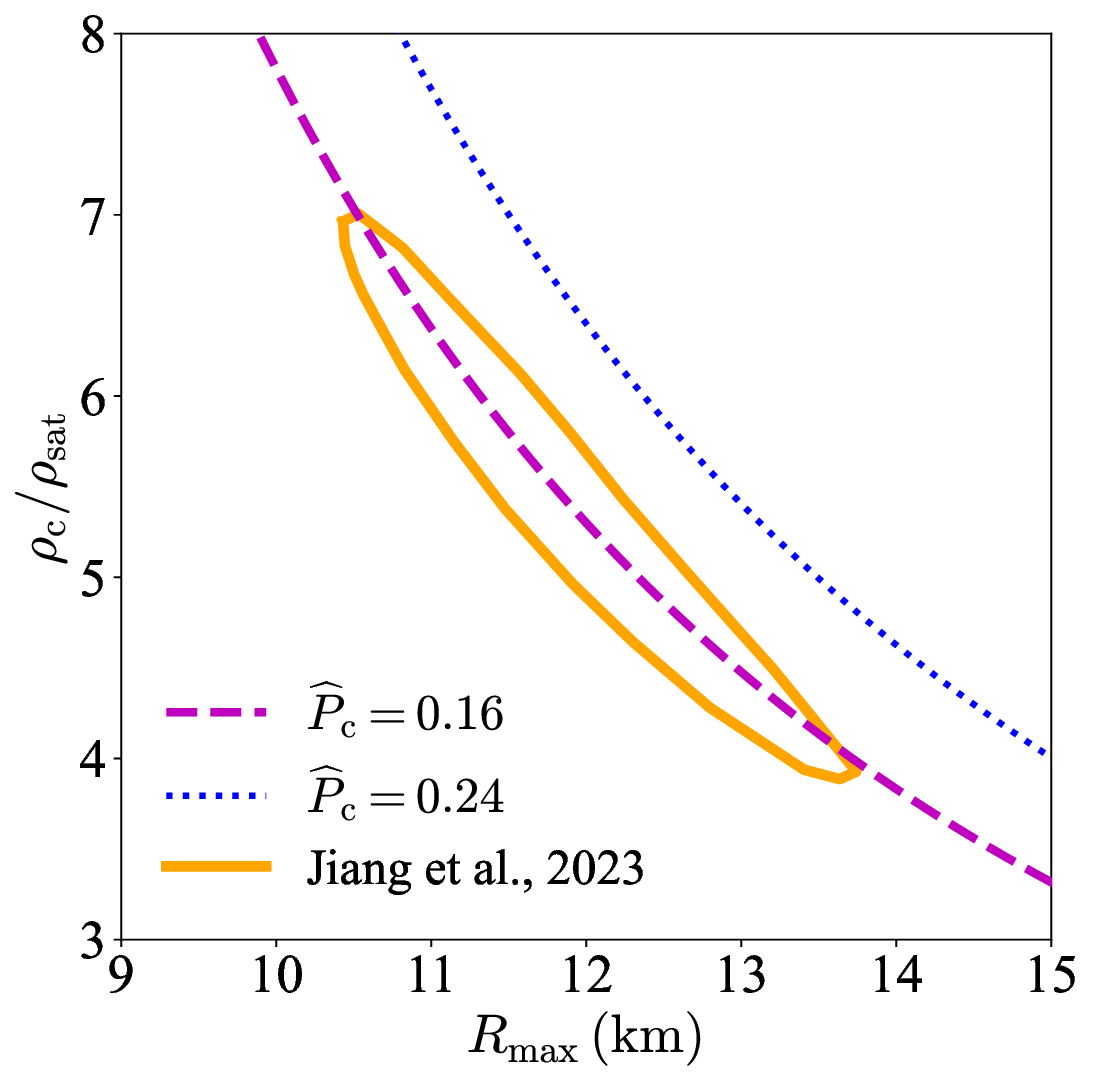}\quad
\includegraphics[height=7.cm]{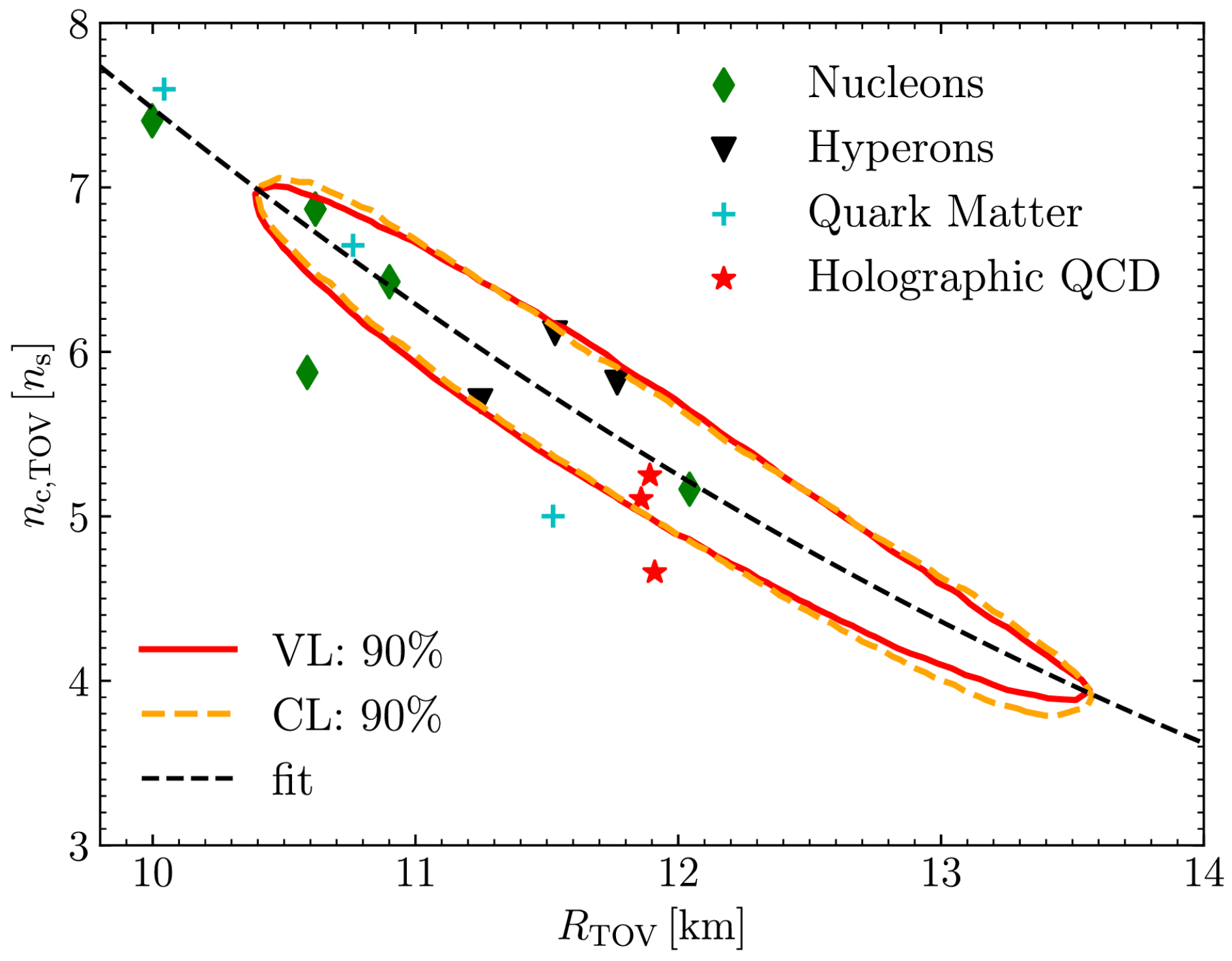}
\caption{(Color Online). Left panel: correlation between $\rho_{\rm{c}}/\rho_{\rm{sat}}$ and radius $R_{\max}$ adopting two reference values for ${\x}$ (0.16 and 0.24).
Figure taken from Ref.\,\cite{CLZ23-b}.
Right panel: an empirical prediction on $\rho_{\rm{c}}/\rho_{\rm{sat}}$-$R_{\max}\equiv R_{\rm{TOV}}$ used in Ref.\,\cite{Jiang2023ApJ}. Figure taken from Ref.\,\cite{Jiang2023ApJ}.
}\label{fig_Rrhoc}
\end{figure}

In this subsection, we use our mass and radius scalings to estimate the central baryon density $\rho_{\rm{c}}$ of NSs at the TOV configuration.
In order to estimate the baryon density $\rho$, a relation between $\varepsilon$ and $\rho$ is necessary. An effective approximation is $\varepsilon\approx M_{\rm{N}}\rho$ with $M_{\rm{N}}\approx939\,\rm{MeV}$ the mass of a static nucleon.
Then by using the scaling of Eq.\,(\ref{Rmax-n}) we can derive a relation between NS central baryon density $\rho_{\rm{c}}$ (in terms of $\rho_{\rm{sat}}=\rho_0\approx0.16\,\rm{fm}^{-3}$) and the radius $R_{\max}$. The result is,
\begin{equation}\label{Rrhoc}
\frac{\rho_{\rm{c}}}{\rho_{\rm{sat}}}
\approx\frac{7.35\times10^3{\x}}{1+3{\x}^2+4{\x}}\left(\frac{R_{\max}}{\rm{km}}-0.64\right)^{-2}.
\end{equation}
We show in FIG.\,\ref{fig_Rrhoc} the relation between this $\rho_{\rm{c}}/\rho_{\rm{sat}}$ and the radius $R_{\max}$ under two values of ${\x}$, namely ${\x}=0.16$ and ${\x}=0.24$.
For a comparison,  an algorithmic/empirical prediction from Ref.\,\cite{Jiang2023ApJ} is also shown (orange solid contour),  with the original figure being shown in the right panel.
Specifically, Ref.\,\cite{Jiang2023ApJ} parameterized $\rho_{\rm{c}}/\rho_{\rm{sat}}$ for the NSs at TOV configuration via a parabolic approximation, 
\begin{equation}
\frac{\rho_{\rm{c}}}{\rho_{\rm{sat}}}\approx \overline{d}_0\left[1-\left(\frac{R_{\max}}{10\,\rm{km}}\right)\right]+\overline{d}_1\left(\frac{R_{\max}}{10\,\rm{km}}\right)^2,
\end{equation}
where $\overline{d}_0\approx27.6$ and $\overline{d}_1\approx7.5$ are two fitting parameters.
 It is seen that the overall correlation between $\rho_{\rm{c}}/\rho_{\rm{sat}}$ and $R_{\max}$ from our analyses here is consistent with that from Ref.\,\cite{Jiang2023ApJ}.

Interestingly, we see from Eq.\,(\ref{Rrhoc}) that the scaling of $\rho_{\rm{c}}/\rho_{\rm{sat}}$ with $R_{\max}$ is roughly 
\begin{equation}
\rho_{\rm{c}}/\rho_{\rm{sat}}\sim R_{\max}^{-2}\cdot\left[1+\mbox{corrections of }R_{\max}^{-1}\right],
\end{equation}
once a ${\x}$ is specified.
The correlation shown in FIG.\,\ref{fig_Rrhoc} may help us to estimate the maximum central (baryon) density when future radius measurements of very massive NSs are available\,\cite{Tang2024}.
Since the correlation (\ref{Rrhoc}) is obtained under the approximation $\varepsilon_{\rm{c}}\approx M_{\rm{N}}\rho_{\rm{c}}$ and principally one needs to use $E(\rho_{\rm{c}})+M_{\rm{N}}$ to replace $M_{\rm{N}}$ where $E(\rho_{\rm{c}})$ is the mechanical energy per nucleon at $\rho_{\rm{c}}$ has its own uncertainties.
Doing so will shift down the lines in FIG.\,\ref{fig_Rrhoc} as $E(\rho_{\rm{c}})>0$ at $\rho_{\rm{c}}$. 
Similarly, using the mass scaling of Eq.\,(\ref{Mmax-G}), we may obtain another estimate for $\rho_{\rm{c}}/\rho_{\rm{sat}}$ (still under $\varepsilon\approx\rho M_{\rm{N}}$)
\begin{equation}
{\rho_{\rm{c}}}/{\rho_{\rm{sat}}}\approx2\times10^4
\left(\frac{{\x}}{1+3{\x}^2+4{\x}}\right)^3
\left(\frac{M_{\rm{NS}}^{\max}}{M_{\odot}}+0.106\right)^{-2}\sim M_{\rm{NS}}^{\max,-2}\cdot\left[1+\mbox{corrections of }M_{\rm{NS}}^{\max,-1}\right].
\end{equation}
It indicates that a heavier NS has a relatively smaller central baryon density with the same central EOS ${\x}$.

Based on the basic thermodynamic relation $\rho\partial\varepsilon/\partial\rho=P+\varepsilon$, one can obtain the perturbative expansion $\rho$ using the coefficients $a_k$ and $b_k$, see Eqs.\,(\ref{ee-heps}) and (\ref{ee-hP}).
The result is\,\cite{CLZ23-b},
\begin{equation}
\boxed{
\widehat{\rho}\equiv
\rho/\rho_{\rm{c}}\approx1+\left(\frac{b_2/s_{\rm{c}}^2}{1+{\x}}\right)\widehat{r}^2+\frac{1}{1+{\x}}\left(a_4-\frac{{b_2^2}/{2s_{\rm{c}}^2}}{1+{\x}}\right)\widehat{r}^4
,\label{pk-4}}
\end{equation}
without any surprise, $\rho<\rho_{\rm{c}}$ for finite $\widehat{r}\neq0$, i.e., baryon density like the energy density is a monotonically decreasing function of $\hr$.
Combing (\ref{pk-4}) and $\varepsilon\approx 1+a_2\hr^2=1+s_{\rm{c}}^{-2}b_2\hr^2$ gives to order $\mu^2$ and ${\x}^2$,
\begin{align}\label{kkk}
\rho/\rho_{\rm{c}}
\approx
\widehat{\varepsilon}-\mu\left(1+\frac{4}{3}\mu\right){\x}\left(1-{\x}\right),
\end{align} here $\mu=\widehat{\varepsilon}-1<0$, see the definition of Eq.\,(\ref{RE-small1}).
Therefore, we have
$\widehat{\rho}\approx\widehat{\varepsilon}$ to leading-order near the center and $\widehat{\rho}\gtrsim\widehat{\varepsilon}$ considering finite ${\x}$ (notice that the finite-${\x}$ of Eq.\,(\ref{kkk}) is very small).
Eq.\,(\ref{kkk}) is an example of the double-element perturbative expansion based on $\mu$ and $\x$, we study in Subsection \ref{sub_DEPc} the central EOS using the double-element expansion.

\begin{figure}[h!]
\centering
\includegraphics[width=7.cm]{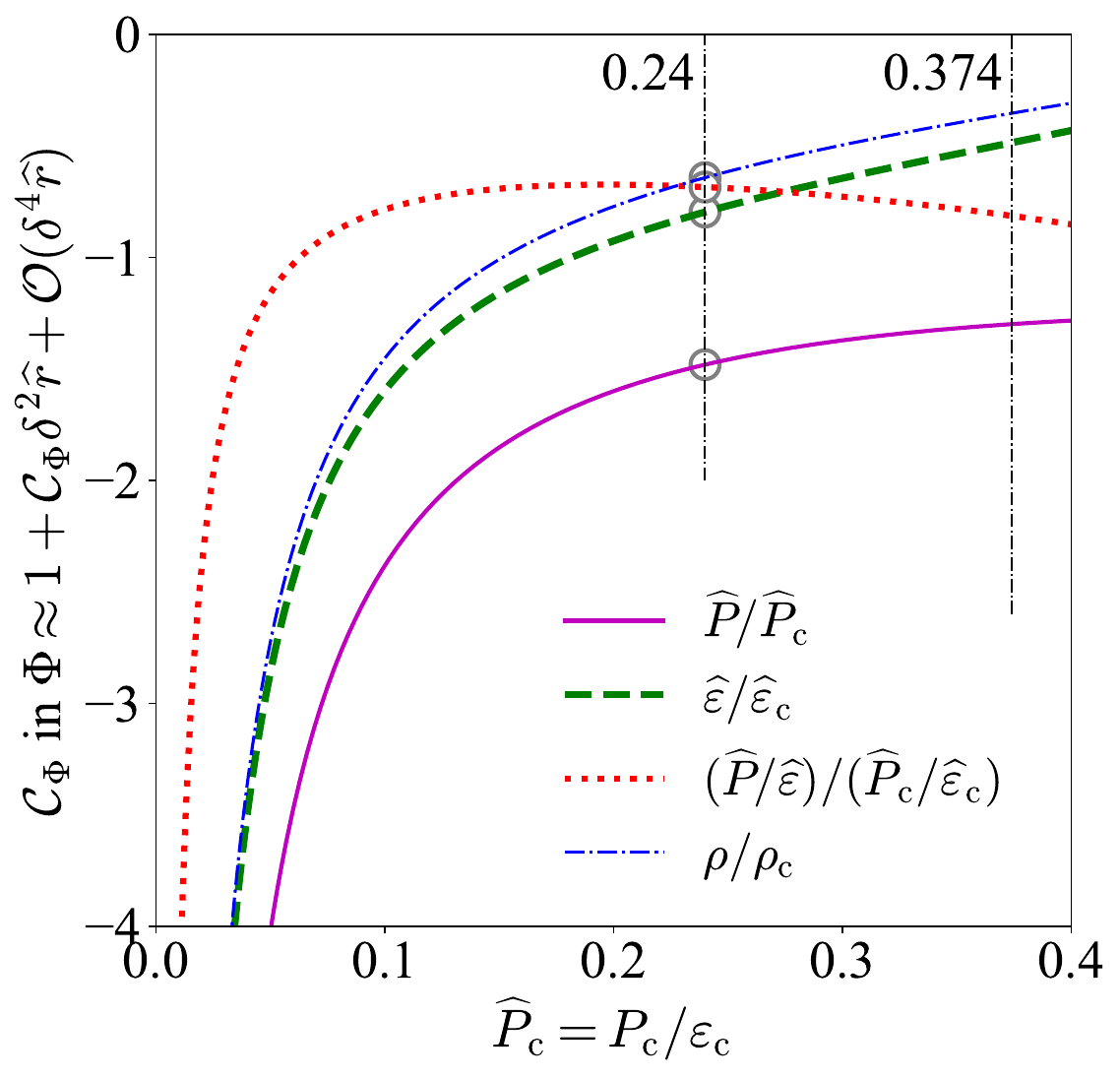}
\caption{(Color Online). The ${\x}$-dependence of coefficients $\mathcal{C}_{\Phi}$ appeared in $\Phi\approx 1+\mathcal{C}_{\Phi}\widehat{r}^2+\mathcal{O}(\widehat{r}^4)$ for $\Phi=\widehat{P}/{\x}$, $\widehat{\varepsilon}/\widehat{\varepsilon}_{\rm{c}}$, $\phi/\x=(\widehat{P}/\widehat{\varepsilon})/({\x}/\widehat{\varepsilon}_{\rm{c}})$ and $\rho/\rho_{\rm{c}}$. Figure taken from Ref.\,\cite{CLZ23-b}.
}\label{fig_eps_P_rho}
\end{figure}

At this point, it is necessary to figure out the tendency of $\hP$, $\heps$, $\phi$ and $\widehat{\rho}=\rho/\rho_{\rm{c}}$ for finite $\hr$.
Specifically, we have
\begin{equation}\label{all_dec1}
\hP/{\x}\approx1+\frac{b_2}{{\x}}\hr^2,
~~
\heps/\heps_{\rm{c}}\approx1+\frac{b_2}{s_{\rm{c}}^2}\hr^2,
\end{equation}
and
\begin{equation}\label{all_dec2}
\phi/{\x}\approx1+b_2\left(\frac{1}{{\x}}-\frac{1}{s_{\rm{c}}^2}\right)\hr^2
=1+\frac{b_2}{\x}\left[1-\left(1+\frac{1}{3}\frac{1+3\x^2+4\x}{1-3\x^2}\right)^{-1}\right]\hr^2,
\end{equation}
as well as (\ref{pk-4}) for $\widehat{\rho}$; here $b_2<0$ is given by Eq.\,(\ref{ee-b2}).
To illustrate the $\hr$-dependence of several quantities all together, in FIG.\,\ref{fig_eps_P_rho}, we show the ${\x}$-dependence of the coefficient $\mathcal{C}_{\Phi}$ appeared in the general approximation $\Phi\approx 1+\mathcal{C}_{\Phi}\widehat{r}^2+\mathcal{O}(\widehat{r}^4)$ for $\Phi=\widehat{P}/{\x}$, $\widehat{\varepsilon}/\widehat{\varepsilon}_{\rm{c}}$, $\phi/\x=(\widehat{P}/\widehat{\varepsilon})/({\x}/\widehat{\varepsilon}_{\rm{c}})$ and $\rho/\rho_{\rm{c}}$, respectively, for the $M_{\rm{NS}}^{\max}$ configuration.
By taking ${\x}\approx0.24$ for PSR J0740+6620\,\cite{CLZ23-a}, we have numerically $\widehat{P}/{\x}\approx 1-1.5\widehat{r}^2$, $\widehat{\varepsilon}/\widehat{\varepsilon}_{\rm{c}}\approx1-0.8\widehat{r}^2$, $(\widehat{P}/\widehat{\varepsilon})/({\x}/\widehat{\varepsilon}_{\rm{c}})\approx1-0.7\widehat{r}^2$, and $\rho/\rho_{\rm{c}}\approx1-0.6\widehat{r}^2$, here $\widehat{\varepsilon}_{\rm{c}}=\varepsilon_{\rm{c}}/\varepsilon_{\rm{c}}=1$.
It is clearly shown that all these $\mathcal{C}_{\Phi}$'s are definitely negative for $0\leq{\x}\lesssim0.374$. On the other hand, the coefficient $\mathcal{C}_{\Phi}$ for $\Phi=s^2/s_{\rm{c}}^2$ may be either positive or negative as we shall demonstrate later (FIG.\,\ref{fig_NEG} and FIG.\,\ref{fig_AD}).

The radial expansion (\ref{pk-4}) is very useful and we could give an order of magnitude estimate for the coefficient $a_4$ appearing in the expansion of $\widehat{\varepsilon}\approx1+a_2\widehat{r}^2+a_4\widehat{r}^4$\,\cite{CLZ23-b}.
The starting point is that both $\widehat{\varepsilon}$ and $\widehat{\rho}=\rho/\rho_{\rm{c}}$ are decreasing functions of radial distance from the center.
We may use the following elementary result: If $y(x)=1+Ax+Bx^2$ ($A<0$) is a decreasing function defined for positive $x$, then the first-order derivative $y'(x)$ should be negative, and consequently $
B\leq-{A}/{2x_{\max}}$,
with $x_{\max}$ the maximum value of $x$ the $y$ could take.
Applying this criterion to $\widehat{\varepsilon}\approx1+a_2\widehat{r}^2+a_4\widehat{r}^4$ and  (\ref{pk-4}) and treating them as functions of $\widehat{r}^2$ gives us two inequalities,
\begin{align}
a_4\leq&-\frac{b_2}{2s_{\rm{c}}^2}\frac{1}{\widehat{R}^2},~~\rm{and}~~
a_4\leq\frac{b_2}{2s_{\rm{c}}^2}\left(\frac{b_2}{1+{\x}}-\frac{1}{\widehat{R}^2}\right),\label{est-1}
\end{align}
respectively, where $\widehat{R}=R/Q\approx1$ (see (\ref{RE-WQnum})) and $a_2=b_2/s_{\rm{c}}^2$ is used here.
By taking ${\x}\approx0.24$, inequalities of (\ref{est-1}) give then $a_4\lesssim0.40$ and $a_4\lesssim0.51$, respectively.
Similarly, if ${\x}\approx0.16$ is adopted, then we have $a_4\lesssim0.55$ and $a_4\lesssim0.69$.
Therefore, $a_4\sim\mathcal{O}(1)$; and there is no surprise on that since $\heps\sim\mathcal{O}(1)$ near the NS center and $\hr\leq\widehat{R}\sim\mathcal{O}(1)$,  so all the expansion coefficients $a_k$ are natural on the order of $\mathcal{O}(1)$.
The sign and magnitude of $a_4$ is very relevant for the emergence of a peak in $s^2$, we will discuss it in more details in SECTION \ref{SEC_7}.
On the other hand, both $\widehat{\varepsilon}$ and $\widehat{\rho}$ are obviously decreasing functions of $\widehat{r}$ if $a_4<0$.

\subsection{Double-element expansion for pressure of NS matter}\label{sub_DEPc}

\begin{figure}[h!]
\centering
\includegraphics[width=8.5cm]{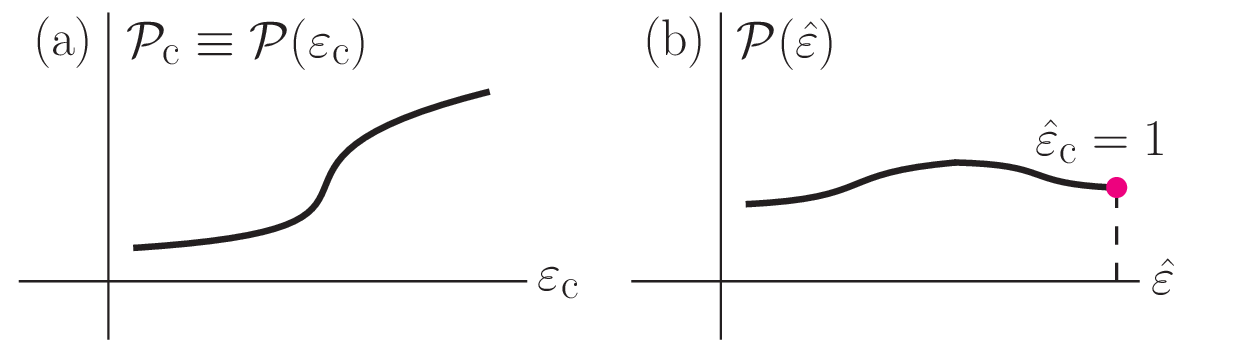}
\caption{Sketch for two types of dependence of certain physical quantities on energy density.
Figures taken from Ref.\,\cite{CLZ23-a}.
}\label{fig_Pec}
\end{figure}

In this review, we may encounter two types of dependence of a quantity $\mathcal{P}$ on the energy density, one is the dependence of the (central) $\mathcal{P}_{\rm{c}}\equiv\mathcal{P}(\varepsilon_{\rm{c}})$ on $\varepsilon_{\rm{c}}$ (panel (a) of FIG.\,\ref{fig_Pec}) or equivalently on ${\x}$, 
e.g.,  the dependence of central SSS (\ref{sc2-TOV}) or (\ref{sc2-GG}) on $\x$.
The other is the dependence of $\mathcal{P}$ on the reduced energy density $\widehat{\varepsilon}=\varepsilon/\varepsilon_{\rm{c}}$,  as sketched in panel (b) of FIG.\,\ref{fig_Pec}.
The latter actually encapsulates the radial-dependence of the quantity, i.e., a finite $\widehat{r}$ corresponds to a finite $\widehat{\varepsilon}<1$.
We may write explicitly the subscripts to avoid potential confusions.

\begin{figure}[h!]
\centering
\includegraphics[height=7.cm]{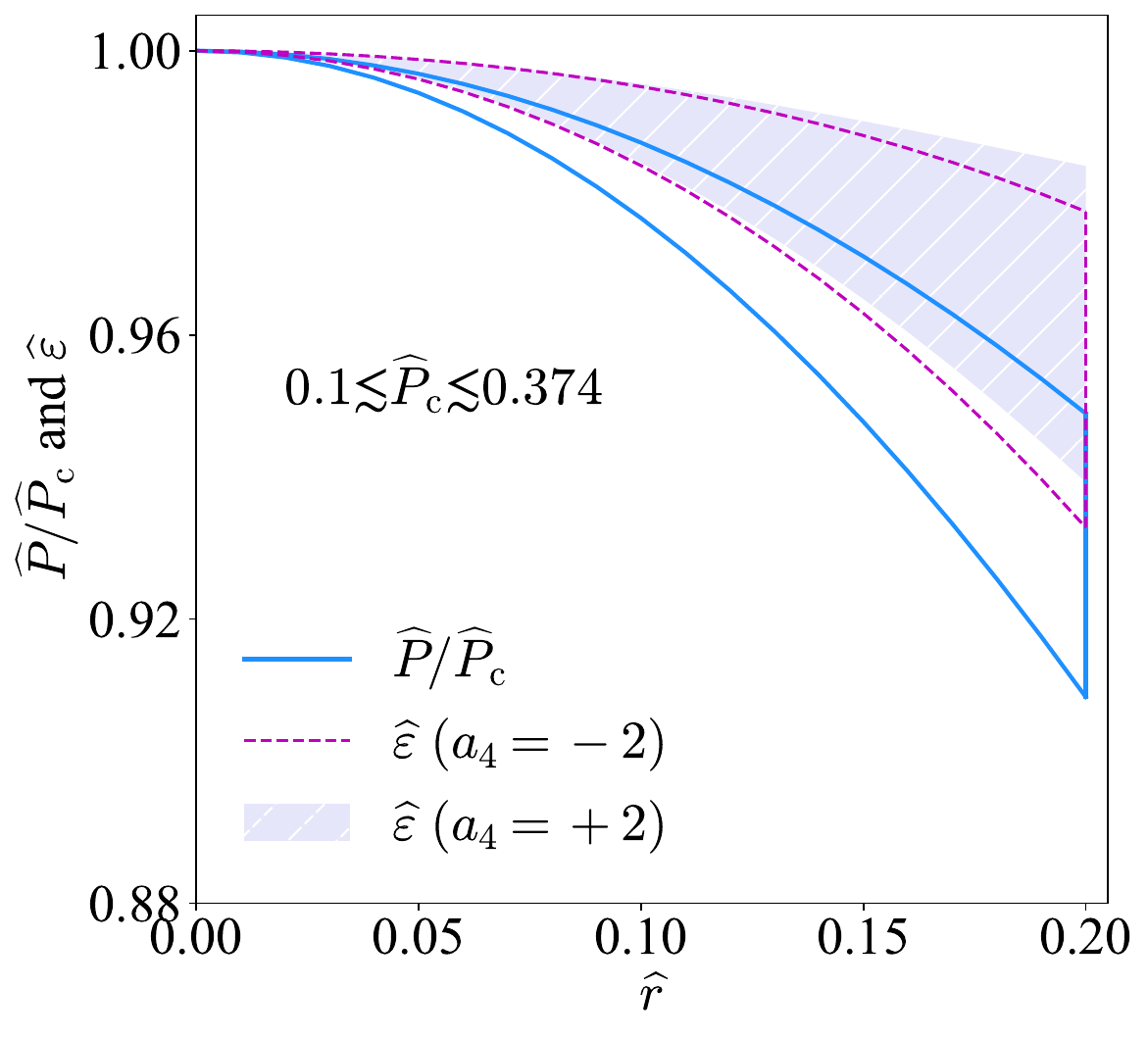}
\caption{(Color Online).  Radial dependence of $\widehat{\varepsilon}$ and $\widehat{P}/{\x}$ to $\widehat{r}\approx0.20$ (about one-fifth of the reduced radius $\widehat{R}$), here $0.1\lesssim{\x}\lesssim0.374$ as a typical range for NSs is adopted.
Figure taken from Ref.\,\cite{CL24-a}.
}\label{fig_s2_prep_r}
\end{figure}

The expansion coefficients of $\widehat{P}$ given in Eq.\,(\ref{ee-hP}) and $\widehat{\varepsilon}$ given in Eq.\,(\ref{ee-heps}) over $\widehat{r}$ are heuristic as they contain fundamentally the intrinsic information on the NS structure starting from the center.
For example,  the coefficients $a_2=b_2/s_{\rm{c}}^2$, $b_2$ of Eq.\,(\ref{ee-b2}) and $b_4$ of Eq.\,(\ref{ee-b4}) are all independent of the matter EOS while the coefficient $a_4$
has certain randomness (owing to the uncertainties of the EOS).
In the appendix of Ref.\,\cite{CL24-a}, we give an example on how the coefficient $a_4$ is related to the dense matter EOS and its maximum size is found to be around 1 based on our current best knowledge about nuclear EOS. {\color{xll}Therefore, at places with small $\widehat{r}$ (i.e., near the NS center) where higher-order coefficients have weak impact,  the resulted EOS is expected to be nearly model-independent, and so has the predictive power.}
Considering typical NSs with $\widehat{R}\approx1$, see an order-of-magnitude estimate of Eq.\,(\ref{RE-WQnum}), the form of the energy density $\widehat{\varepsilon}\approx 1+a_2\widehat{r}^2+a_4\widehat{r}^4+\cdots$ indicates the magnitude of $a_2$ and $a_4$, etc., is $\sim\mathcal{O}(1)$.
See also the discussion/estimate given in (\ref{est-1}).
Consequently, the contribution from the term $a_4\widehat{r}^4$ to $\widehat{\varepsilon}$, e.g.,  for $\widehat{r}\approx0.20$ (one-fifth of the reduced radius $\widehat{R}$), is $\lesssim0.4\%$ (compared with 1) using $-2\lesssim a_4\lesssim2$ (which is a rather large magnitude for $a_4$ compared to the best estimate).
Although $a_4$ has tiny impact on the $\widehat{\varepsilon}$, it may become relevant for the peaked behavior in $s^2$, and this is the issue of SECTION \ref{SEC_7}.
Similarly,  the $b_4$-term contributes $\lesssim0.5\%$ to $\widehat{P}/{\x}$ for $\widehat{r}\approx0.20$ and ${\x}\gtrsim0.1$.
In FIG.\,\ref{fig_s2_prep_r}, we show the radial dependence of $\widehat{\varepsilon}$ and $\widehat{P}/{\x}$ from the NS center to $\widehat{r}\approx0.20$.
Naturally,  the $b_6$- and $a_6$-terms have even smaller contributions to $\widehat{P}/{\x}$ and $\widehat{\varepsilon}$, respectively.

\begin{figure}[h!]
\centering
\includegraphics[height=7.cm]{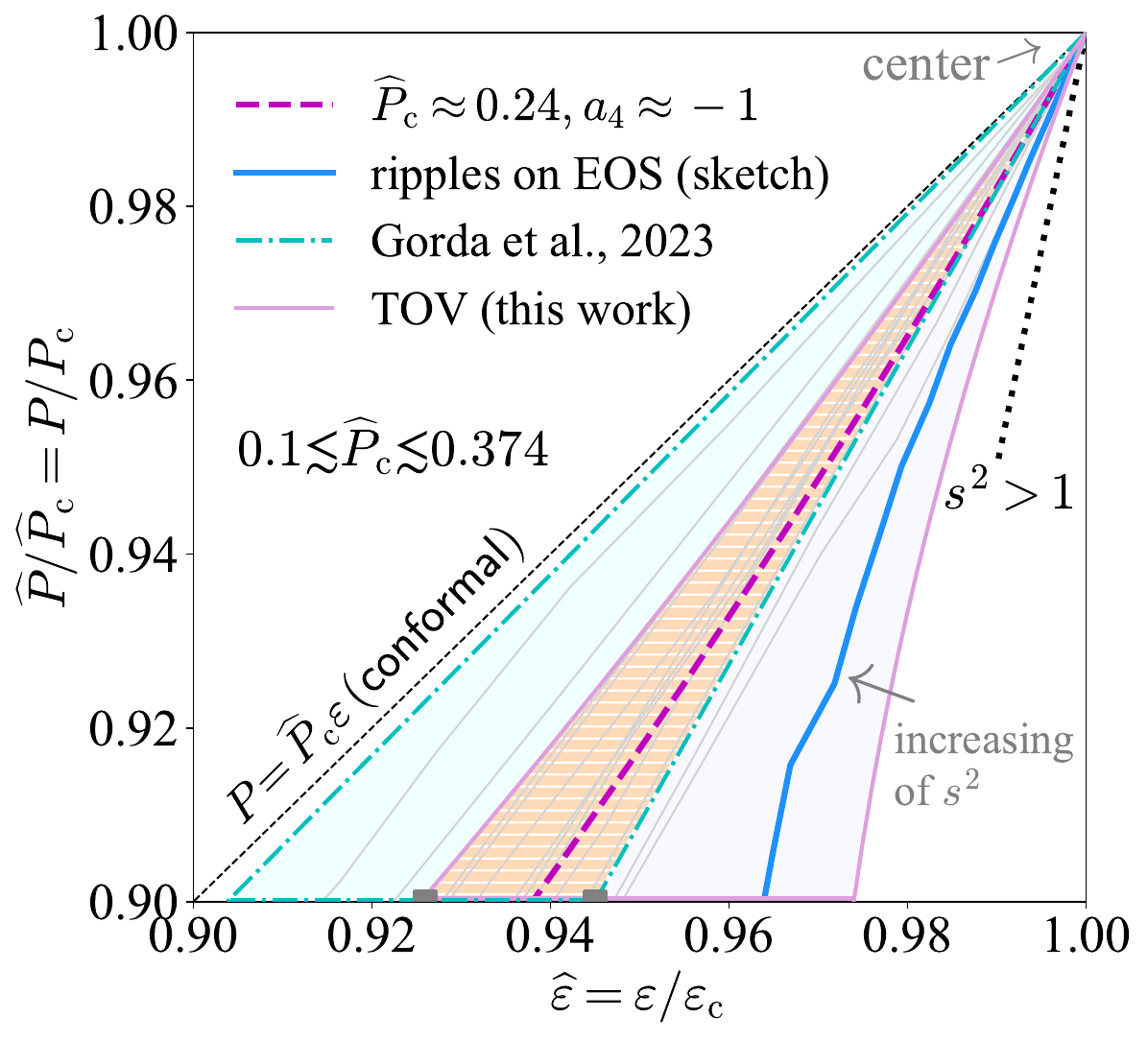}\qquad
\includegraphics[height=7.cm]{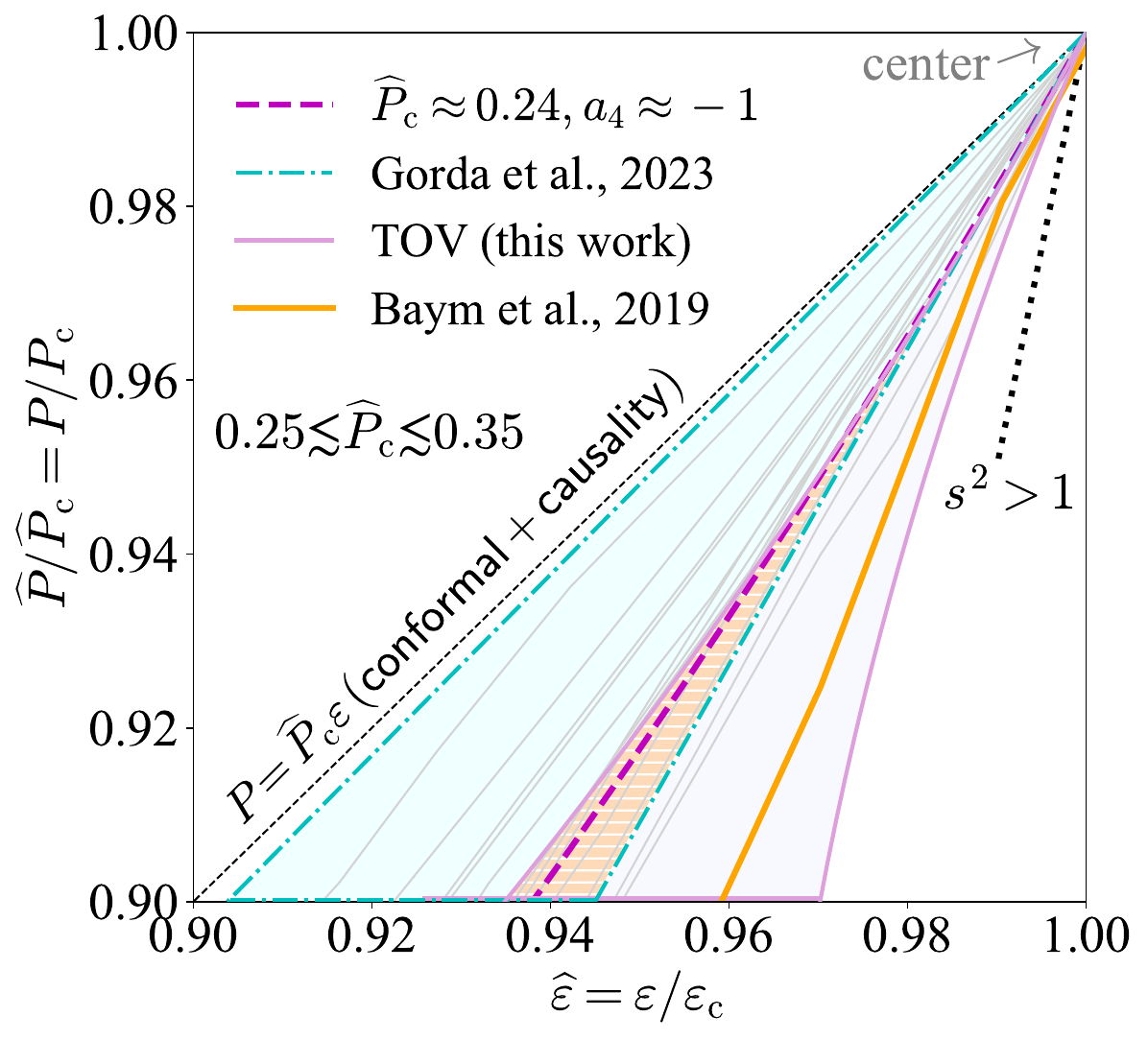}
\caption{(Color Online).  Left panel: core EOS of NSs by expanding $\widehat{P}$ and $\widehat{\varepsilon}$ over $\widehat{r}$ to order $\widehat{r}^4$ where $0.1\lesssim{\x}\lesssim0.374$ and $-2\lesssim a_4\lesssim2$ (plum band). 
The EOS incorporating the pQCD effects at densities $\gtrsim40\rho_{\rm{sat}}$ is shown by the dashed-dotted cyan band\,\cite{Gorda2023}.
An instance with ${\x}\approx0.24$ and $a_4\approx-1$  (dashed magenta) and a few empirical EOSs are shown (solid grey).
An increasing of $s^2$ indicates a peak occurring there.
Figure taken from Ref.\,\cite{CL24-a}. Right panel: same as the left panel but adopting $0.25\lesssim\x\lesssim0.35$, the continuous crossover EOS from Ref.\,\cite{Baym2019} (their set-B)  is shown by the orange line.
}\label{fig_s2_prep}
\end{figure}

We can work out the central EOS, e.g.,  at the order $\mu^3{\x}^3$\,\cite{CL24-a},
\begin{empheq}[box=\fbox]{align}\label{zf-3}
\widehat{P}/{\x}=P/P_{\rm{c}}\approx&1+\frac{4}{3}\mu+\frac{16}{15}\mu^2+\frac{4}{15}\mu^3
+\left(\frac{4}{3}-\frac{4}{5}\mu-\frac{268}{135}\mu^2\right)\mu{\x}\notag\\
&+\left[2+\left(\frac{28}{3}-\frac{256a_4}{3}\right)\mu+\left(\frac{370}{27}-\frac{30208}{315}a_4\right)\mu^2
\right]\mu{\x}^2\notag\\
&+\left[4+\left(\frac{1280a_4}{3}-\frac{262}{15}\right)\mu
+\left(\frac{68608}{105}a_4+\frac{2048}{3}a_6
-\frac{1496}{27}\right)\mu^2
\right]\mu{\x}^3+\mathcal{O}\left({\x}^4,\mu^4\right),
\end{empheq}
here the expansions of $b_2$, $b_4$ and $b_6$ of Eqs.\,(\ref{ee-b2}), (\ref{ee-b4}) and (\ref{ee-b6}) over ${\x}$ are also executed.
As an order-of-magnitude estimate, we have $|\mu^3{\x}^3|\sim10^{-5}$ for ${\x}\sim0.2$ and $\mu\sim-0.1$, implying $a_6$ has small effect on the core EOS near $\widehat{r}=0$, namely $|2048a_6\mu^3{\x}^3/3|\lesssim1\%$ by considering $a_6\sim\mathcal{O}(1)$.
In the left panel of FIG.\,\ref{fig_s2_prep}, we show the core EOS $\widehat{P}$-$\widehat{\varepsilon}$ near $\widehat{r}=0$ by considering the expansion of $\widehat{P}$ and $\widehat{\varepsilon}$ over $\widehat{r}$ to order $\widehat{r}^4$ using $0.1\lesssim{\x}\lesssim0.374$ and $-2\lesssim a_4\lesssim2$,  an example with ${\x}\approx0.24$ and $a_4\approx-1$ is given.
Notice that the plum band of the left panel of FIG.\,\ref{fig_s2_prep} originates mainly from the band for ${\x}$, and the ``cone'' would be shrunk if ${\x}$ is further refined by other approaches, as shown in the right panel where a different range of $\x$ about $0.25\lesssim\x\lesssim0.35$ is adopted.
The ripples on the curve of $\widehat{P}$-$\widehat{\varepsilon}$ characterize the variation of $s^2$, as exemplified by the light-blue instance in which case there is an increasing of $s^2$, i.e., a possible peak occurring there.
The dotted black line represents configurations with $s^2>1$ (which violate the causality principle), while the black dashed line (marked by $P={\x}\varepsilon$) is the boundary given by $s^2/{\x}=1$,  which corresponds to the conformal limit since
\begin{equation}\label{ddf-gamma}
\gamma_{\rm{c}}\equiv\left.\frac{\d\ln P}{\d\ln\varepsilon}\right|_{\rm{center}}= s_{\rm{c}}^2/{\x}=1,
\end{equation}
see the definition for $\gamma$ given in Eq.\,(\ref{def_gamma}).
A few empirical EOSs (solid grey lines) are also shown for comparisons, see Ref.\,\cite{CL24-a} for detailed description on these EOSs.
A recent constraint on the NS EOS incorporating the pQCD effects\,\cite{Gorda2023} is shown by the dot-dashed cyan band ($\varepsilon_{\rm{c}}\approx1000\,\rm{MeV}/\rm{fm}^3$ and $P_{\rm{c}}\approx250\mbox{-}400\,\rm{MeV}\rm{fm}^3$ are taken).
Because the pQCD theory predicts an approximate conformal symmetry at very high densities ($\gtrsim 40\rho_{\rm{sat}})$,  the upper boundary of its constraining band is close to the conformal prediction $P={\x}\varepsilon$.

\begin{figure}[h!]
\centering
\includegraphics[height=6.2cm]{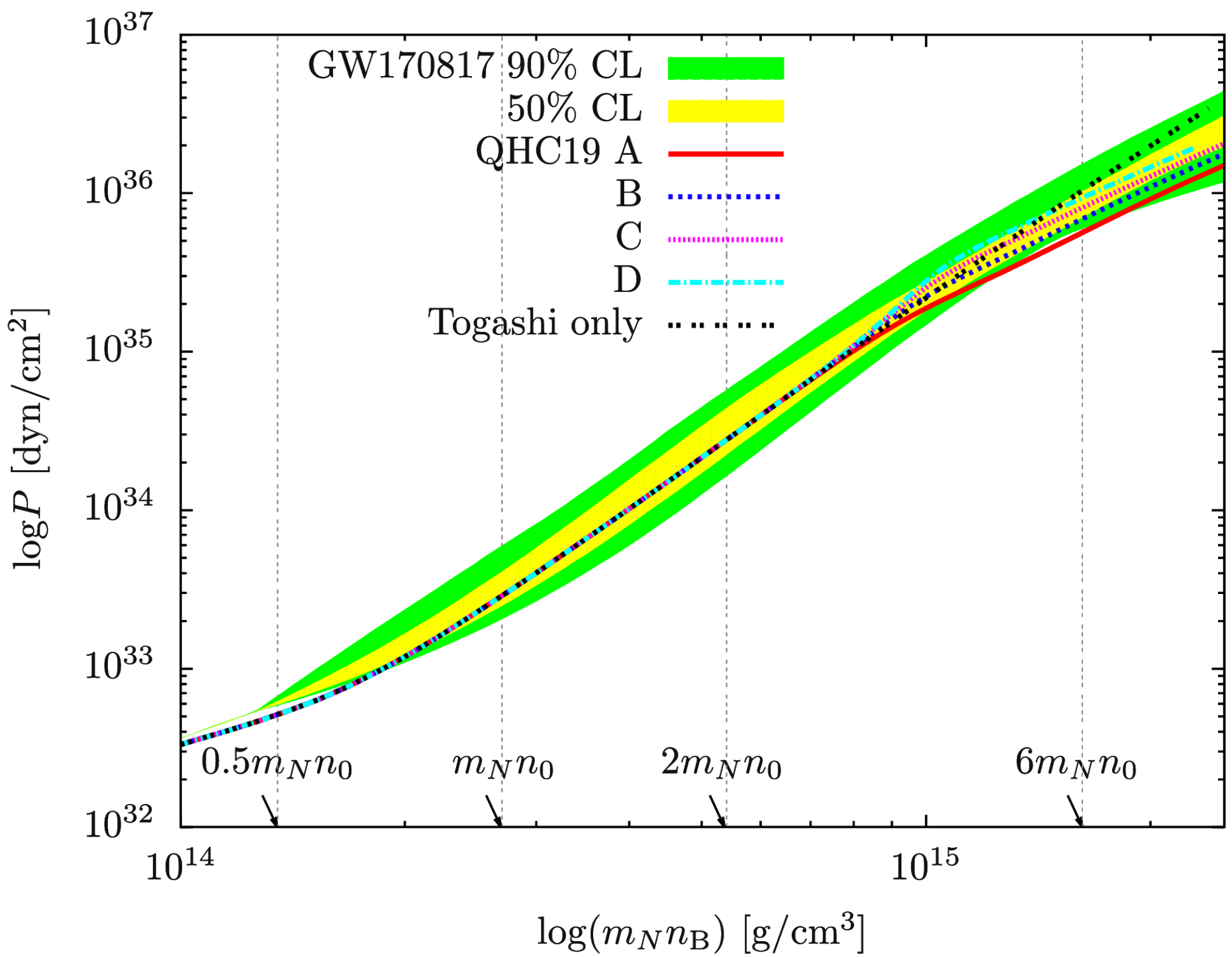}\quad
\includegraphics[height=6.2cm]{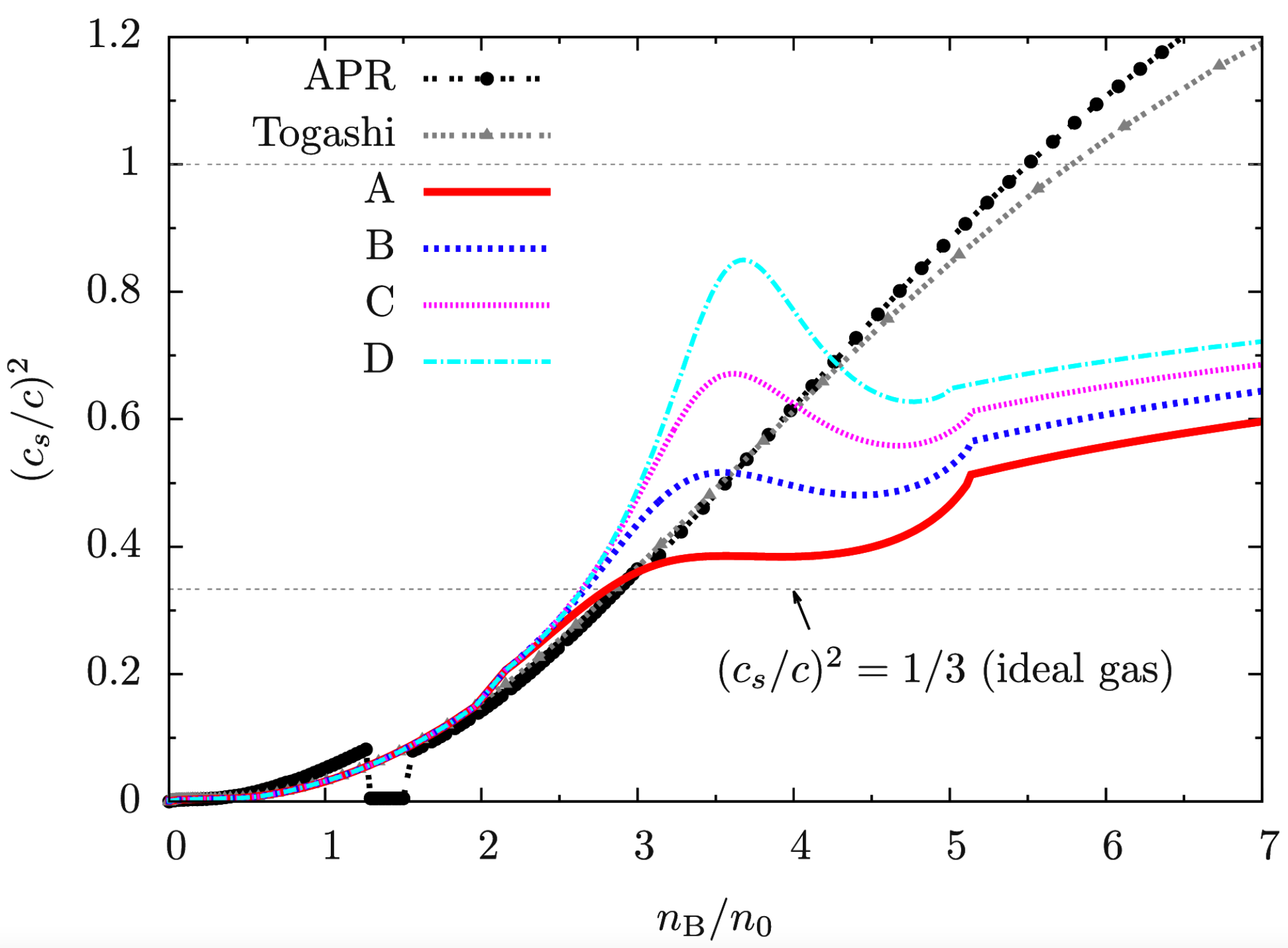}
\caption{(Color Online). Left panel: the crossover EOS interpolated smoothly from the hadron phase to quark matter phase with a transition density about $5\rho_{\rm{sat}}$. Right panel: the SSS as a function of density, notice that $s^2$ is not a constant at $\rho\gtrsim5\rho_{\rm{sat}}$.
Figures taken from Ref.\,\cite{Baym2019}.
}\label{fig_Baym19}
\end{figure}

{\color{xll}The core EOS shown by the plum band is directly inferred from the gravity (curved geometry of compact NSs) encapsulated in the TOV equations without any presumption about it, unlike in the traditional approach.}
Indeed, the EOSs from the two approaches match with each other as demonstrated by the hatched tan band in both panels of FIG.\,\ref{fig_s2_prep}.
Particularly, we have $0.925\lesssim\widehat{\varepsilon}\lesssim0.945$ (shown by two grey solid rectangles) for $\widehat{P}/{\x}\approx0.9$ by combining these two types of EOS.
{\color{xll}Moreover,  we have $s_{\rm{c}}^2\neq0$ according to Eq.\,(\ref{sc2-TOV}), therefore
a sharp phase transition (PT) near the center or equivalently a plateau on the $P$-$\varepsilon$ curve occurring there is basically excluded.}
However, a continuous crossover signaled by a smooth reduction of $s^2$ in NS cores (or equivalently a peaked behavior at some places near the center), see SECTION \ref{SEC_7}; and PTs occurring far from the centers are not excluded.
Moreover,  the polytropic index $\gamma_{\rm{c}}=s_{\rm{c}}^2/{\x}$ should not approach 1 due to the nonlinear feature of $s_{\rm{c}}^2$ shown in Eq.\,(\ref{sc2-TOV}). In fact, we have $\gamma_{\rm{c}}\geq4/3\approx1.33$, {\color{xll} 
implying that the matter in NS cores can hardly be conformal.} This finding is very different from the predictions in Refs.\,\cite{Ann23,Gorda2023}.
In addition, our prediction on the core EOS (plum band in FIG.\,\ref{fig_s2_prep}) is consistent with several empirical EOSs (grey lines) having a continuous crossover near the centers, see Ref.\,\cite{CL24-a} for details on the EOSs.
Interestingly, the EOSs of Ref.\,\cite{AFL2} and of Ref.\,\cite{Kap2021} fall out of the tan band (which are above the upper plum boundary in the left panel). The reason is that in these two models the quark matter EOS with approximate conformal symmetry (nearly constant $s^2$) is used at high densities (but still within the NS region), they would be closer to the conformal boundary in FIG.\,\ref{fig_s2_prep}.
On the other hand, in a recent study with an improved hadron-quark crossover EOS\,\cite{Baym2019} interpolated smoothly at about $5\rho_{\rm{sat}}$,  the predicted overall EOS is found to be consistent with our plum band but has certain deviation from the dashed-dotted cyan band, see the right panel of FIG.\,\ref{fig_s2_prep}.
In FIG.\,\ref{fig_Baym19}, we show the EOS of this crossover model as well as the SSS as a function of density. Notice that the $s^2$ in the quark matter phase is not a constant (which is the case in a conformal region).
{\color{xll}
Our discussions thus far indicate that the pQCD effects (with an approximate conformal symmetry) at extremely high densities may have little relevance for the dense NS matter EOS since it is fundamentally inconsistent with the TOV predictions on the EOS. Also, this finding does not  mean that there should be no appearance of quark matter in the cores of NSs.
}

Based on Eq.\,(\ref{zf-3}), we may obtain the SSS for finite $\mu$ straightforwardly as $s^2/{\x}=\d(\widehat{P}/{\x})/\d\mu$, e.g., by neglecting the finite-${\x}$ corrections (this is the Newtonian limit). We have ${\x}^{-1}\d s^2/\d\mu=\d^2(\widehat{P}/{\x})/\d\mu^2=32/15+8\mu/5>0$ for $-1\leq\mu\leq0$, i.e., the $s^2$ is a monotonically increasing function of $\widehat{\varepsilon}$, as expected.
We point out that one could directly work out the dependence of $s^2$ on $\phi=P/\varepsilon$.
Starting from $
\widehat{P}/{\x}\approx1+4\mu/3+16\mu^2/15+(4/3-4\mu/5)\mu{\x}$
 to order $\mu^2{\x}$ (see Eq.\,(\ref{zf-3}). For example for simplicity, we obtain 
\begin{align}
\phi\approx&\frac{{\x}}{1+\mu}\left[1+\frac{4\mu}{3}+\frac{16\mu^2}{15}+\left(
\frac{4}{3}
-\frac{4\mu}{5}\right)\mu{\x}\right]
\approx{\x}\left[1+\frac{\mu}{3}\left(4{\x}+1\right)+\frac{\mu^2}{15}\left(11-32{\x}\right)\right],
\end{align}
which gives inversely that 
\begin{equation}
\mu\approx \frac{3}{4{\x}+1}\left(\frac{\phi}{{\x}}-1\right)+
\frac{9}{5}\frac{32{\x}-11}{4{\x}+1}
\left(\frac{\phi}{{\x}}-1\right)^2.
\end{equation}
 Putting this $\mu=\mu(\phi)$ back into $s^2/{\x}\approx4/3+32\mu/15+4\mu^2/5+(4/3-8\mu/5-268\mu^2/45){\x}$ via Eq.\,(\ref{zf-3}) gives the $s^2$ profile near the NS center to order $(\phi/\phi_{\rm{c}}-1)^2$ as (with $\phi/\phi_{\rm{c}}\lesssim1$),
\begin{empheq}[box=\fbox]{align}
s^2/\phi_{\rm{c}}\approx&\frac{4}{3}+\frac{32}{5}\left(1-\frac{19}{4}{\x}\right)\left(\frac{\phi}{\phi_{\rm{c}}}-1\right)
-\frac{876}{25}\left(1-\frac{3439}{219}{\x}\right)\left(\frac{\phi}{\phi_{\rm{c}}}-1\right)^2,~~
\phi_{\rm{c}}\equiv \x.\label{DDDk}
\end{empheq}
We then find $[\phi_{\rm{c}}^{-1}\d s^2/\d \phi]_{\phi=\phi_{\rm{c}}}=(32/5)(1-19{\x}/4)$,  which may either be positive (Newtonian limit ${\x}\to0$) or negative,  depending on the magnitude of ${\x}$.
Moreover, since $s^2(\phi)$ as a function of $\phi$ of Eq.\,(\ref{DDDk}) is non-monotonic, implying $s^2$ and $\phi$ are two different attributes of NSs.
In fact, all the descriptions, namely the $\widehat{r}$-dependence, the $\mu$-dependence and the $\phi$-dependence, are equivalent on the peaked profile of $s^2$.  In this review, we may
adopt these representations interchangeably.
We discuss in more details on the peaked $s^2$ in SECTION \ref{SEC_7}.

\subsection{Central EOSs of superdense matter in canonical and massive NSs}\label{sub_1420}

In the above subsections of this section, we have drawn attention to the EOS and related issues for NSs at the TOV configuration. These analyses are mort relevant for massive NSs. Generally, for other NSs similar investigates are necessary. For this purpose, we extend the mass, radius and compactness scalings to stable NSs along the M-R curve and extract useful information on several NS properties.

In this and the following two subsections, we use a meta-model for generating randomly NS EOSs in a broad parameter space that can mimic diverse model predictions consistent will existing constraints from terrestrial experiments and astrophysical observations as well as general physics principles\,\cite{CL24-b,ZhangLi2021x,ZhangLiXu2018,ZhangLi2019ApJ,XieLi2020}. 
It is based on a so-called minimum model of NSs consisting of neutrons, protons, electrons and muons (npe$\mu$ matter) at $\beta$-equilibrium. Its most basic input is the EOS of isospin-asymmetric nucleonic matter in the form of energy per nucleon $E(\rho,\delta)=E_0(\rho)+E_{\rm{sym}}(\rho)\delta^2$, 
where the EOS of SNM $E_0(\rho)$ and the nuclear symmetry energy $E_{\rm{sym}}(\rho)$ are expanded around the saturation density $\rho_0$ according to Eq.\,(\ref{fE0}) and Eq.\,(\ref{fEsym}), respectively.

The total energy density is 
\begin{equation}
\varepsilon(\rho,\delta)=\left[E(\rho,\delta)+M_{\rm{N}}\right]\rho+\varepsilon_{\ell}(\rho,\delta),
\end{equation} 
where $M_{\rm{N}}\approx939\,\rm{MeV}$ and $\varepsilon_{\ell}(\rho,\delta)$ is the energy density of leptons from an ideal Fermi gas model\,\cite{TOV39-2}. The pressure $P(\rho,\delta)$ is $P(\rho,\delta)=\rho^2\d[\varepsilon(\rho,\delta)/\rho]/\d\rho$. The density profile of isospin asymmetry $\delta(\rho)$ is obtained by solving the $\beta$-equilibrium condition 
\begin{equation}
\mu_{\rm{n}}-\mu_{\rm{p}}=\mu_{\rm{e}}\approx\mu_\mu\approx4\delta E_{\rm{sym}}(\rho),
\end{equation}
and the charge neutrality requirement $\rho_{\rm{p}}=\rho_{\rm{e}}+\rho_\mu$, here the chemical potential $\mu_i$ for a particle $i$ is calculated from the energy density via
$\mu_i=\partial\varepsilon(\rho,\delta)/\partial\rho_i.$ With the $\delta(\rho)$ calculated consistently using the inputs given above, both the pressure $P$ and energy density $\varepsilon(\rho)=\varepsilon(\rho, \delta(\rho))$ become barotropic, i.e., depend on the density $\rho$ only. The EOS in the form of $P(\varepsilon)$ can then be used in solving the TOV equations. 
The core-crust transition density $\rho_{\rm{cc}}$ is determined self-consistently by the thermodynamic method\,\cite{Iida1997,XuJ}. In the inner crust with densities between $\rho_{\rm{cc}}$ and $\rho_{\rm{out}}\approx2.46\times10^{-4}\,\rm{fm}^{-3}$ corresponding to the neutron dripline we adopt the parametrized EOS $P=\alpha+\beta\varepsilon^{4/3}$\,\cite{Iida1997}; and for $\rho<\rho_{\rm{out}}$ we adopt the Baym--Pethick--Sutherland (BPS) and the Feynman--Metropolis--Teller (FMT) EOSs\,\cite{BPS71}.
In order to make a verification of the scalings studied in a broad EOS parameter space accommodating predictions of diverse classes of models, we select the saturation density $\rho_0$ to be $0.15\,\rm{fm}^{-3}\lesssim\rho_0\lesssim0.17\,\rm{fm}^{-3}$, the binding energy in the range of $-17\,\rm{MeV}\lesssim B\lesssim-15\,\rm{MeV}$,  the incompressibility for symmetric matter in $210\,\rm{MeV}\lesssim K_0\lesssim250\,\rm{MeV}$ and the skewness in $-400\,\rm{MeV}\lesssim J_0\lesssim0\,\rm{MeV}$. For the symmetry energy, we adopt $28\,\rm{MeV}\lesssim S\lesssim 36\,\rm{MeV}$ for its magnitude at $\rho_0$, $30\,\rm{MeV}\lesssim L\lesssim90\,\rm{MeV}$ for its slope, $-300\,\rm{MeV}\lesssim K_{\rm{sym}}\lesssim0\,\rm{MeV}$ for its curvature as well as $200\,\rm{MeV}\lesssim J_{\rm{sym}}\lesssim 1000\,\rm{MeV}$\,\cite{ZhangLi2021x,ZhangLiXu2018,ZhangLi2019ApJ,XieLi2020} for its skewness.
The EOS parameters generated randomly within the specified uncertainty ranges above are consistent with terrestrial experimental and contemporary astrophysical constraints.

\begin{figure}[h!]
\centering
\includegraphics[height=8.5cm]{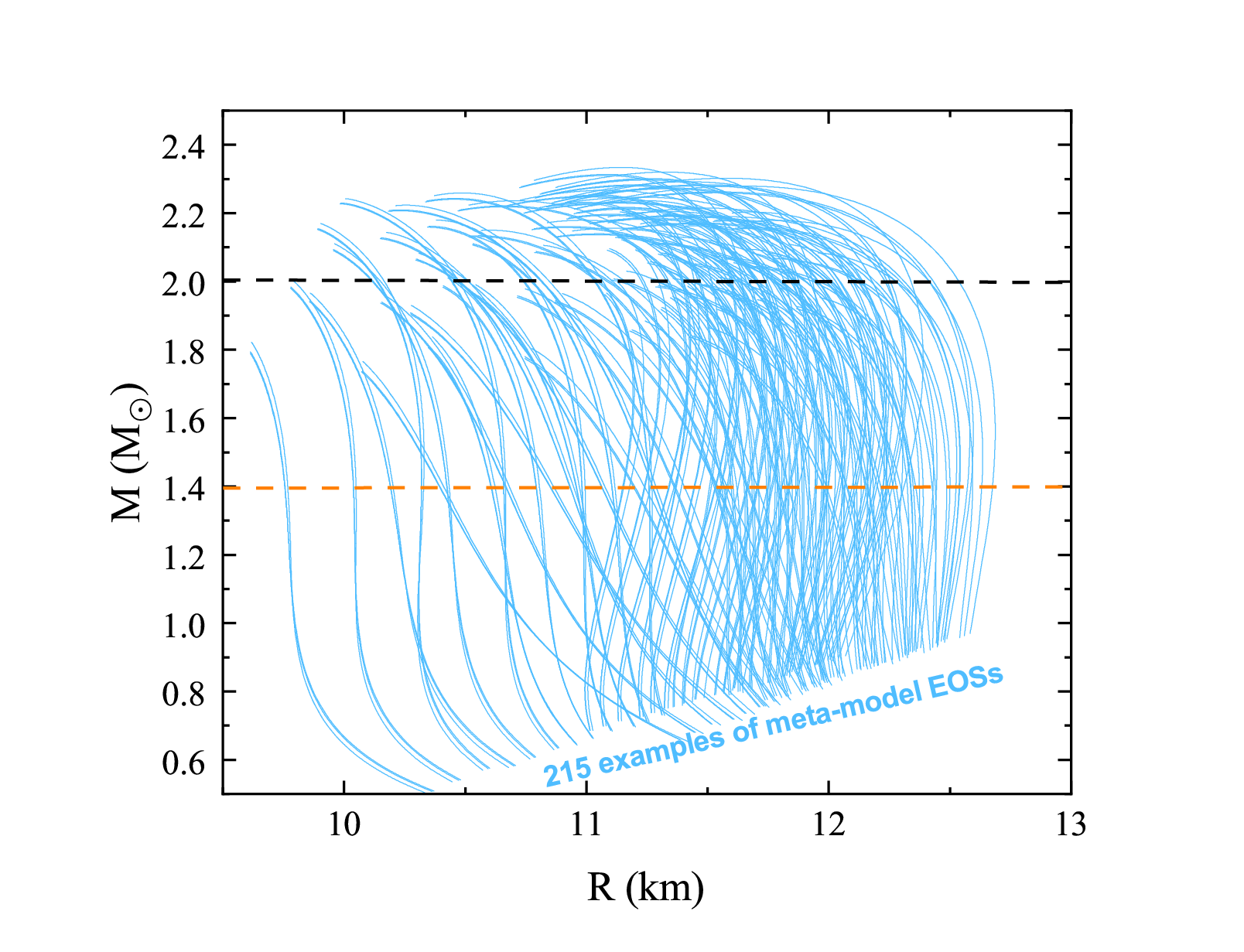}
\caption{(Color Online). Samples of the M-R curves generated by meta-model NS EOSs.
Figure taken from Ref.\,\cite{CL24-b}.}\label{fig_MR-ensem}
\end{figure}

FIG.\,\ref{fig_MR-ensem} gives 215 samples of the randomly generated M-R relation with its local derivative $\d M_{\rm{NS}}/\d R$ switching broadly between negative and positive values and also covering mass regions (``vertical shape of M-R'') where $\d M_{\rm{NS}}/\d R=\infty$ as well as cases where two NSs have the same radius. As discussed in detail in Ref.\,\cite{Li:2024imk}, similarly diverse predictions for the $M(R)$ curve have been made using different parametric meta-models, phenomenological models and/or microscopic theories. A changing sign of $\d M_{\rm{NS}}/\d R$ in a large range of mass or radius is often associated with the onset of new degrees of freedom or phases (e.g., strange/quark stars with hyperons or quark deconfinement), while a consistently positive one is normally for nucleonic or hadronic NSs, see, e.g., (1) the 40435 $M(R)$ curves from using a meta-model by parameterizing the high-density EOS with piecewise polytropes\,\cite{Ferreira:2024hxc}, (2) many samples of $M(R)$ curves from using hybrid models coupling various hadronic EOSs through a first-order phase transition to quark matter EOSs characterized by different speeds of sound in studying the possible formation of twin stars\,\cite{Han:2020adu}, and (3) many samples of $M(R)$ curves with EOSs encapsulating quarkyonic matter in NSs\,\cite{Zhao2020}. 
The $10^5$ EOSs we generated produce diverse $M(R)$ curves with features similar to those predicted by the EOSs mentioned above. While we can not guarantee that our meta-model EOSs cover every single NS EOS that has ever been proposed, we are confident that they are diverse enough to mimic most if not all existing NS EOSs in the literature.
\begin{figure}[h!]
\centering
\includegraphics[height=7.5cm]{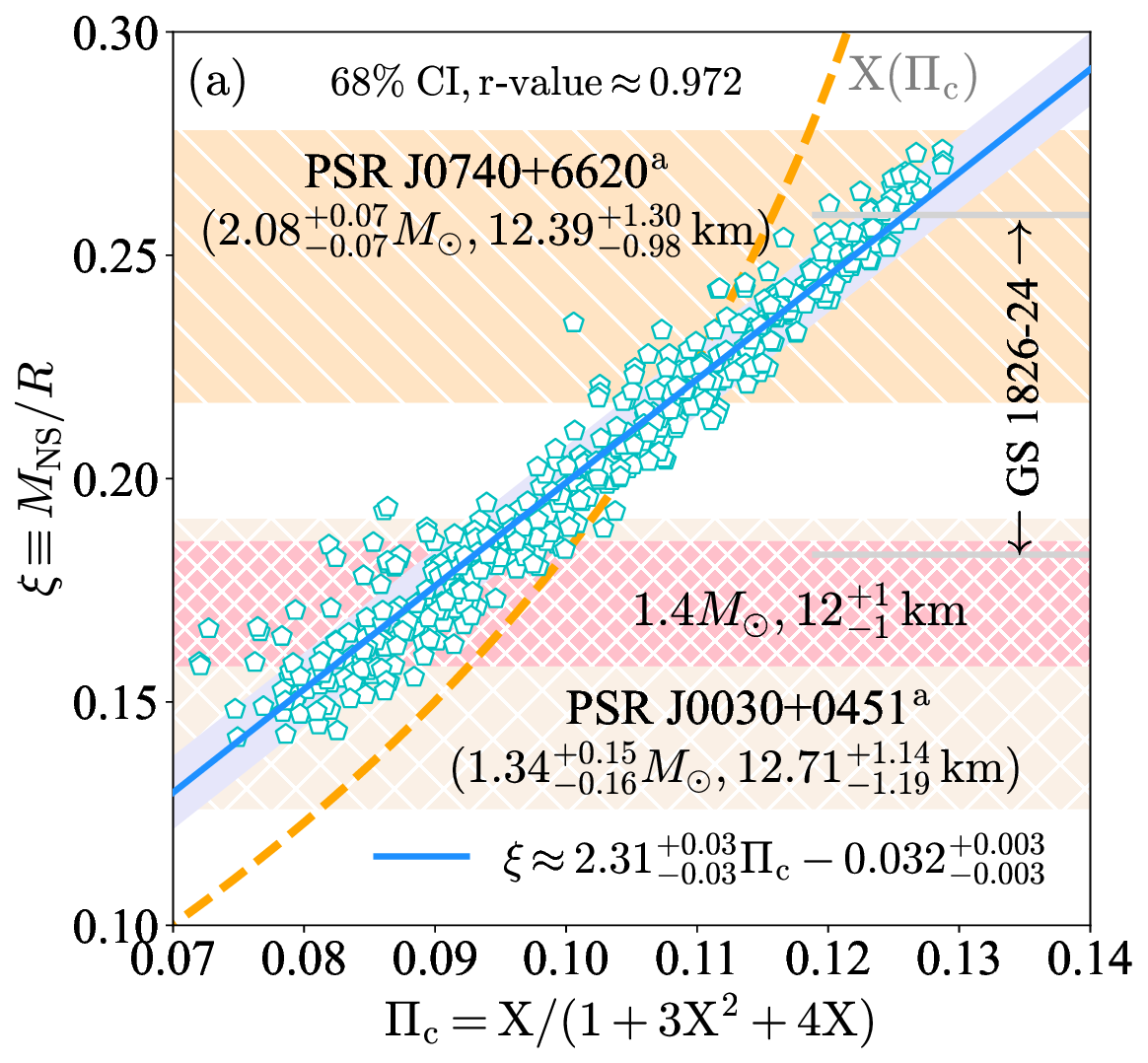}\qquad
\includegraphics[height=7.5cm]{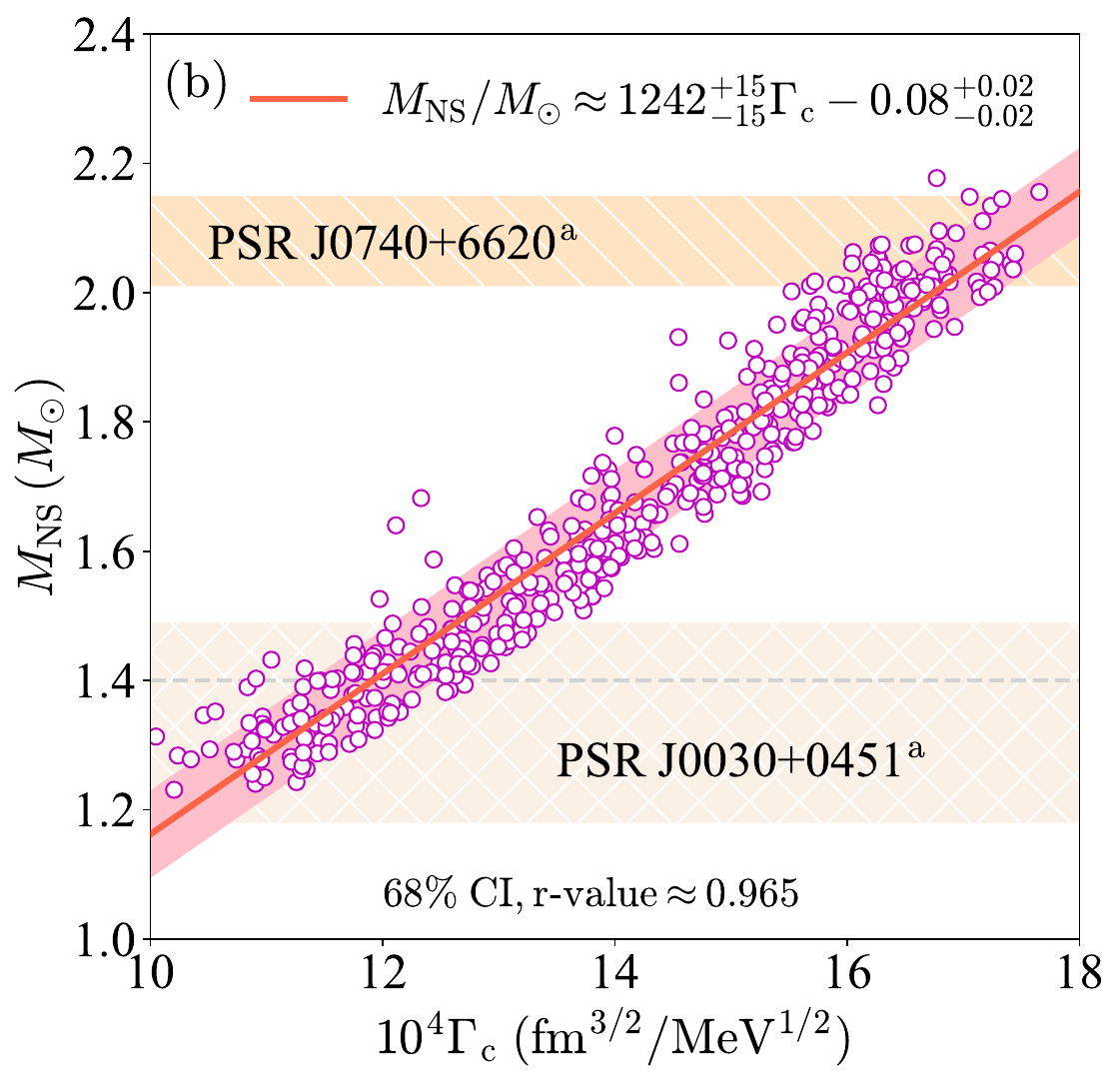}
\caption{(Color Online).  Left panel: the $\xi$-$\Pi_{\rm{c}}$ correlation using meta-model EOSs consistent with observational/experimental constraints.  The compactnesses for PSR J0740+6620\,\cite{Riley21}, PSR J0030+0451\,\cite{Riley19} and the NS in the X-ray burster GS 1826-24\,\cite{Zhou23} are shown. The function $\x(\Pi_{\rm{c}})$ is plotted by the dashed orange curve and the compactness for a typical canonical NS\,\cite{Brandes2023-a,Rich23} (with $R\approx12_{-1}^{+1}\,\rm{km}$) by the hatched pink band.  
Right panel: same as panel (a) but for the correlation  $M_{\rm{NS}}$-$\Gamma_{\rm{c}}$, the lavender and pink bands in panel (a) and (b) represents their 68\% confidence intervals (CIs).
Figures taken from Ref.\,\cite{CL24-b}.
}\label{fig_Phi_c}
\end{figure}

\begin{figure}[h!]
\centering
\includegraphics[height=7.5cm]{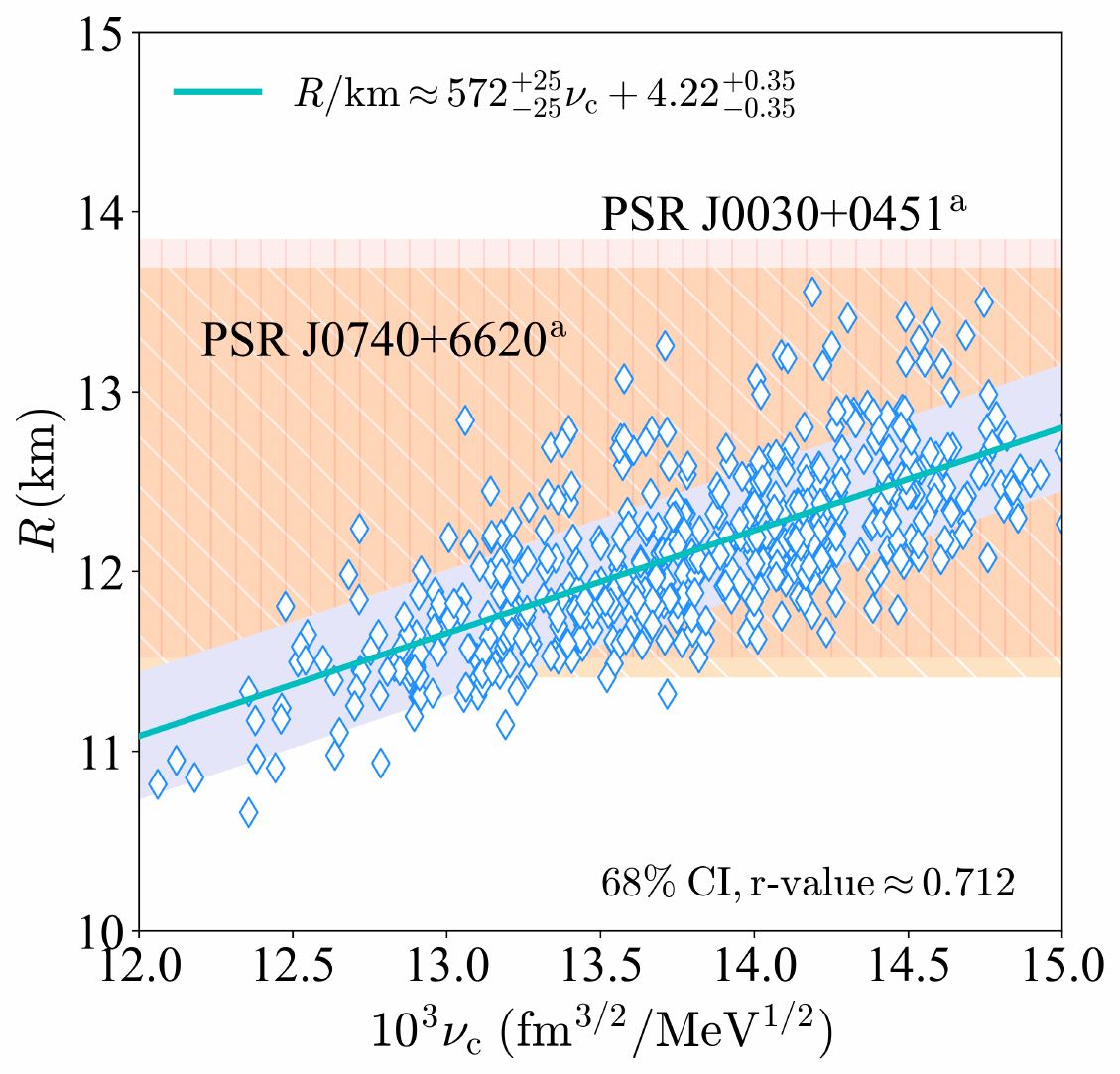}
\caption{(Color Online). The same as FIG.\,\ref{fig_Phi_c} but for the $R$-$\nu_{\rm{c}}$ correlation using the meta-model EOSs.}\label{fig_Rnc}
\end{figure}

To test the mass, radius and compact scalings of Eqs.\,(\ref{gk-mass}), (\ref{gk-radius}) and (\ref{gk-comp}), we randomly generate $10^5$ meta-model EOSs\,\cite{CL24-b}.
To explore the whole EOS parameter space currently allowed and present our results clearly, we select randomly one point on each mass-radius (M-R) curve from a given EOS (as shown in FIG.\,\ref{fig_MR-ensem}) within the mass range of $1.2M_{\odot}\lesssim M_{\rm{NS}}\lesssim2.2M_{\odot}$. The resulting scalings are shown in FIG.\,\ref{fig_Phi_c} and FIG.\,\ref{fig_Rnc}; here the panel (a) of FIG.\,\ref{fig_Phi_c} shows the compactness-$\Pi_{\rm{c}}$ scaling while the panel (b) is the mass-$\Gamma_{\rm{c}}$ scaling. In each panel, for clarity only 500 representative samples are shown while $10^5$ EOSs are used in calculating the scaling coefficients and their error bands.
The standard error (ste) and the coefficient of determination (the r-value) actually start converging quickly using about 300 samples.
In particular,  the ste for the compactness-$\Pi_{\rm{c}}$ (mass-$\Gamma_{\rm{c}}$, radius-$\nu_{\rm{c}}$) regression is about 0.03 (0.002$M_{\odot}$, 0.3\,km) and the r-value for the $\xi$-$\Pi_{\rm{c}}$ and $M_{\rm{NS}}$-$\Gamma_{\rm{c}}$ regressions is about 0.97 while that for the $R$-$\nu_{\rm{c}}$ is about 0.71\,\cite{CL24-b}.
The regression and its 68\% confidence interval (CI) are shown in the figures. Quantitatively, we have 
\begin{empheq}[box=\fbox]{align}
\xi\approx &A_\xi\Pi_{\rm{c}}+B_\xi\approx2.31_{-0.03}^{+0.03}\Pi_{\rm{c}}-0.032_{-0.003}^{+0.003},\label{gk-xi}\\
M_{\rm{NS}}/M_{\odot}\approx &A_{\rm{M}}+B_{\rm{M}}\approx1242_{-15}^{+15} \left(\frac{\Gamma_{\rm{c}}}{\rm{fm}^{3/2}/\rm{MeV}^{1/2}}\right)-0.08_{-0.02}^{+0.02},\label{gk-m}\\
R/\rm{km}\approx& A_{\rm{R}}\nu_{\rm{c}}+B_{\rm{R}}\approx 572_{-25}^{+25} \left(\frac{\nu_{\rm{c}}}{\rm{fm}^{3/2}/\rm{MeV}^{1/2}}\right)+4.22_{-0.35}^{+0.35},\label{gk-r}
\end{empheq}
here $\Gamma_{\rm{c}}$, $\nu_{\rm{c}}$ and $\Pi_{\rm{c}}$ (which is dimensionless) are defined in Eqs.\,(\ref{gk-mass}), (\ref{gk-radius}) and (\ref{gk-comp}), respectively.
We notice that dividing $R$ by $M_{\rm{NS}}$ in calculating the compactness $\xi$ largely diminishes the relatively large uncertainty in the radius scaling, in particular the prediction $\xi=2\Pi_{\rm{c}}$ of Eq.\,(\ref{gk-comp}) (indicated by the blue dash-dotted line) is quite close to the numerical result.
Moreover, the ste of $R$ about 0.3\,km is much smaller than that from the current NS observations.
Since $\x$ is limited to $\x\lesssim0.374$, see Eq.\,(\ref{Xupper}),  by causality applied in GR for strong-field gravity\,\cite{CLZ23-a}, we have from the scaling (\ref{gk-xi}) that,
\begin{equation}\label{xi_GR}
\boxed{
\xi\lesssim0.264_{-0.005}^{+0.005}\equiv \xi_{\rm{GR}},~~\mbox{for general stable NSs along M-R curve.}}
\end{equation} 
Obviously, this general limit for $\xi$ holding for all stellar models is consistent with (and smaller than) the one given by (\ref{upp-xi}) which is an upper limit for the NS compactness.
In the panel (a) of FIG.\,\ref{fig_Phi_c},  we show the compactnesses for PSR J0740+6620 and PSR J0030+0451 via NICER's simultaneous mass-radius observation, namely $M_{\rm{NS}}/M_{\odot}\approx2.08_{-0.07}^{+0.07}$ and $R/\rm{km}\approx12.39_{-0.98}^{+1.30}$ (at 95\% CI) for the former\,\cite{Riley21} and $M_{\rm{NS}}/M_{\odot}\approx1.34_{-0.16}^{+0.15}$ and $R/\rm{km}\approx12.71_{-1.19}^{+1.14}$ (at 68\% CI) for the latter\,\cite{Riley19}, both indicated by the superscript ``a''.
 Shown also are the compactness $\xi\approx0.183\sim0.259$ for the NS in GS 1826-24 directly from its surface gravitational red-shift measurement\,\cite{Zhou23} and the $\xi$ for a canonical NS with $R\approx12_{-1}^{+1}\,\rm{km}$\,\cite{Brandes2023-a,Rich23}. For a given $\xi$, one can directly obtain the $\Pi_{\rm{c}}$ from their scaling and the $\x$ via the function $\x(\Pi_{\rm{c}})$ (orange dashed curve) defined in Eq.\,(\ref{gk-comp}).
Similarly, the mass/radius bands for PSR J0740+6620\,\cite{Fon21} and PSR J0030+0451\,\cite{Riley19} are shown in the panel (b) of FIG.\,\ref{fig_Phi_c} and FIG.\,\ref{fig_Rnc}, respectively.

{\color{xll} It is necessary to point out here that as we include more terms in the expansions of $\heps$, $\hP$ and $\hM$, the relevant scalings should be modified accordingly.
For example, the NS mass scaling may read as 
\begin{equation}
    M_{\rm{NS}}\sim\frac{1}{\sqrt{\varepsilon_{\rm{c}}}}\left(\frac{\x}{1+3\x^2+4\x}\right)^{3/2}\cdot\left(1+\kappa_1\x+\kappa_2\x^2+\cdots\right),
\end{equation}
with the coefficients of high-order $\x$ terms, $\kappa_1$, $\kappa_2,\cdots$ determined by the dimensionless TOV equations themselves, or fitting predictions of realistic NS EOSs within the traditional approach. By investigating the scaled TOV equations alone, we found in Ref.\,\cite{CL24-c} that the next-to-leading order correction for the mass is $\kappa_1=18/25$ and the radius scaling still takes the present form of (\ref{gk-radius}).
Consequently, at the next-to-leading order the NS compactness scales as:
\begin{equation}
    \xi\sim\frac{\x}{1+3\x^2+4\x}\cdot\left(1+\frac{18}{25}\x\right).
\end{equation}
Since the first correction term $18\x/25$ is generally small especially since $\x\lesssim0.374$ (compared with 1) limited by causality, both the scalings (with or without the first correction) for $M_{\rm{NS}}$ (and $\xi$) work well for realistic NSs. Therefore, the NS properties (central pressure, central energy density and the maximum compactness) extracted/inferred are very similar and qualitatively equivalent.
See Ref.\,\cite{CL24-b} for some illustrations and quantitative comparisons of the results from the two scaling schemes with or without the first correction $18\x/25$. In this review, we use the scalings of (\ref{gk-mass}),  (\ref{gk-radius}) and (\ref{gk-comp}) and focus on the information thus extracted. Nevertheless, we note that the high-order corrections might have appreciable effects on some properties and/or correlations. They may thus need to be considered for some analyses especially as the precision of astrophysical observations further improve. Fortunately, the formalism we presented here allows people to systematically incorporate them when it is necessary for various physics purposes.  
}

\begin{figure}[h!]
\centering
\includegraphics[height=7.cm]{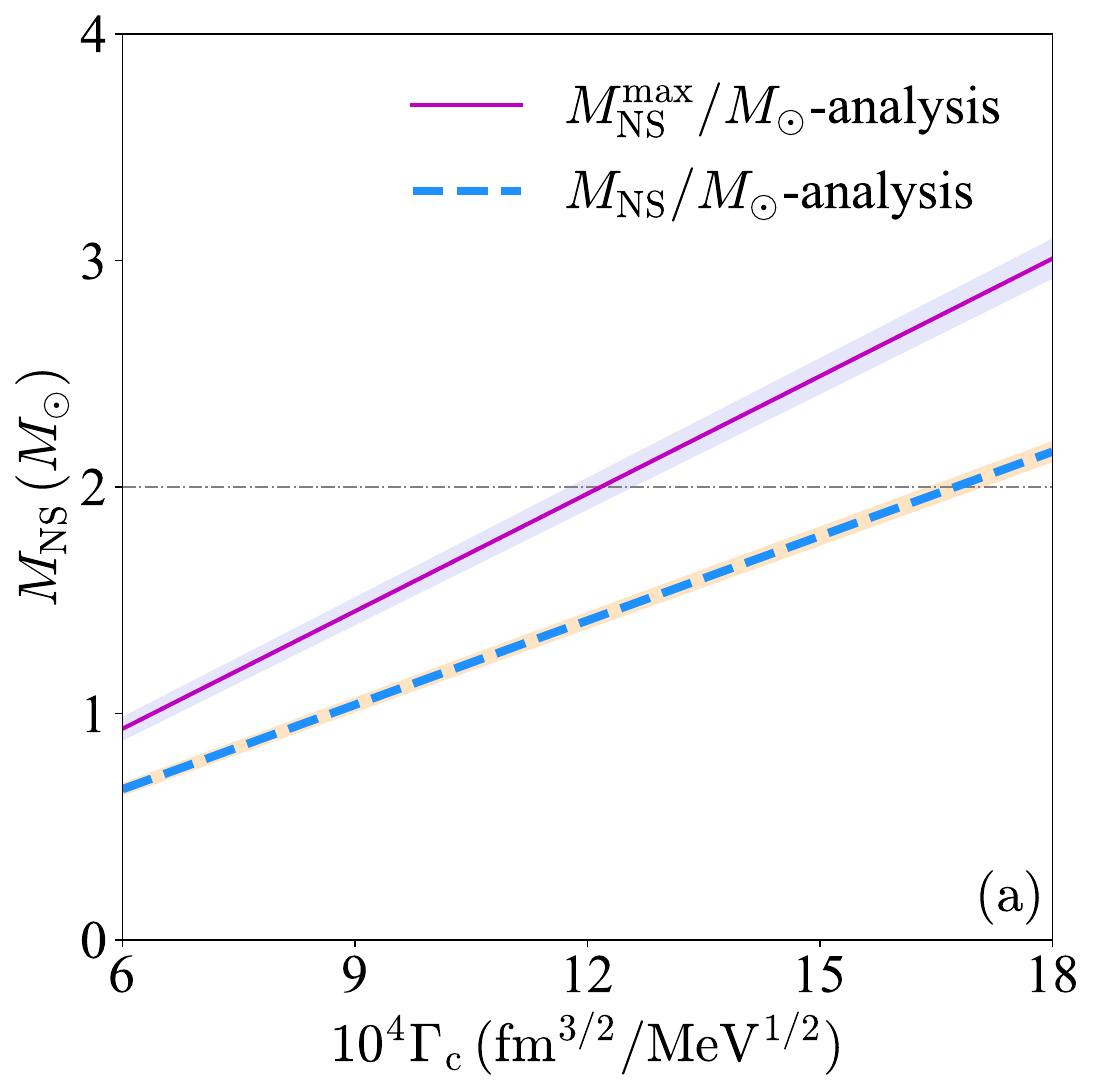}\qquad
\includegraphics[height=7.cm]{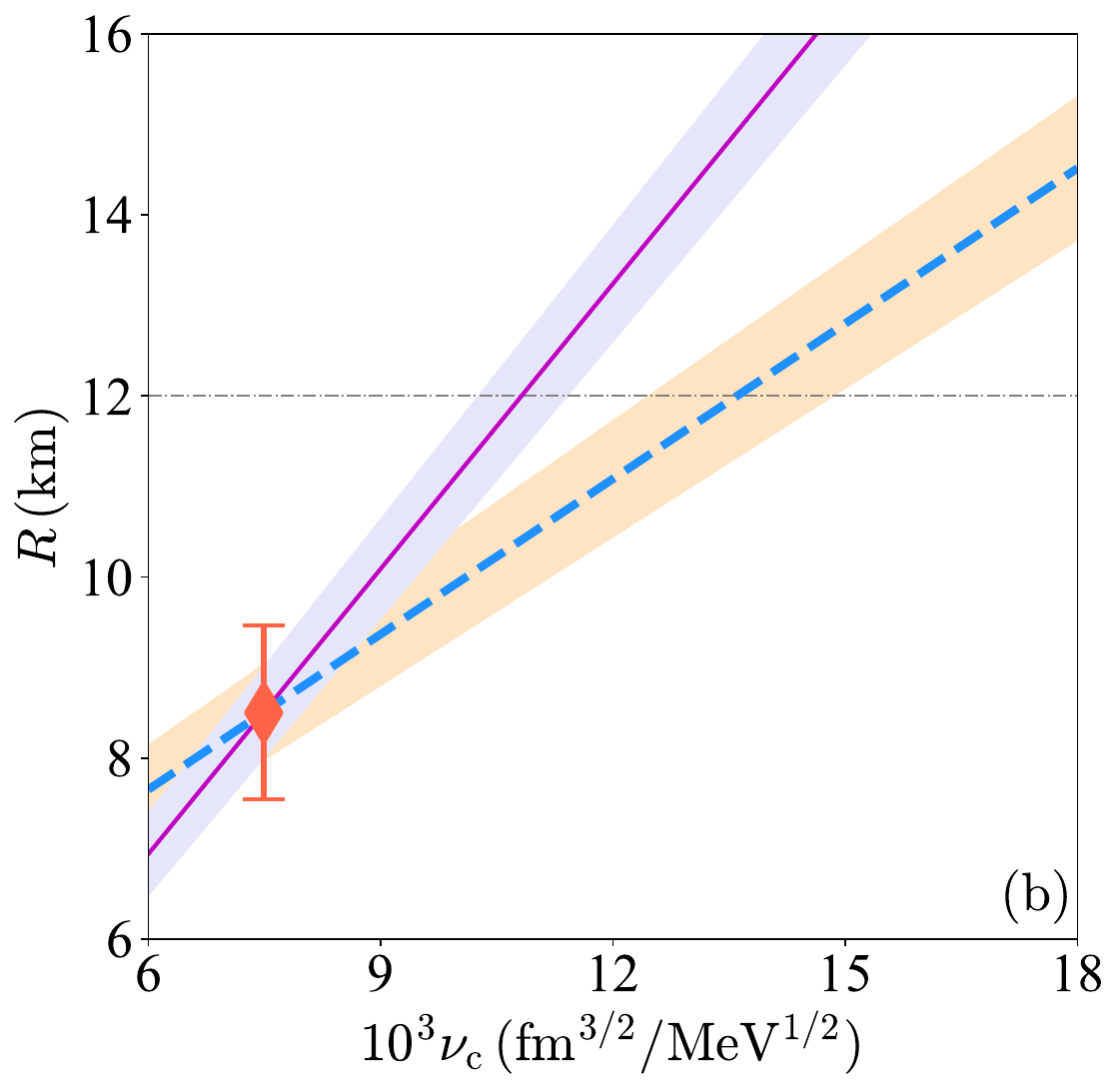}
\caption{(Color Online). Mass and radius scalings in the $M_{\rm{NS}}^{\max}/M_{\odot}$-analysis (using the NS masses and radii at the TOV configuration) and the $M_{\rm{NS}}/M_{\odot}$-analysis (considering the general stable NSs).}\label{fig_fMfR}
\end{figure}
Next, we compare the mass and radius scalings for two cases: (1) the one based on NSs at the TOV configuration (denoted as the $M_{\rm{NS}}^{\max}/M_{\odot}$-analysis) and (2) the other using the general stable NSs along the M-R curve (denoted as the $M_{\rm{NS}}/M_{\odot}$-analysis).
The mass-$\Gamma_{\rm{c}}$ and radius-$\nu_{\rm{c}}$ scalings for the two cases are shown in FIG.\,\ref{fig_fMfR}. As one expects, for a given NS mass depending on whether it is at the TOV configuration or not, the $\Gamma_{\rm{c}}$ factor extracted from the two analyses are different, as shown in the left panel. Specifically, the one from the $M_{\rm{NS}}^{\max}/M_{\odot}$-analysis is smaller than that from analyzing the $M_{\rm{NS}}/M_{\odot}$-scaling.
Considering the factor $\Pi_{\rm{c}}$ appearing in $\Gamma_{\rm{c}}=\varepsilon_{\rm{c}}^{-1/2}\Pi_{\rm{c}}^{3/2}$ is a slow-varying function of $\x$, this means the central energy density $\varepsilon_{\rm{c}}$ extracted from the $M_{\rm{NS}}^{\max}/M_{\odot}$-analysis is necessarily larger than that from the $M_{\rm{NS}}/M_{\odot}$-analysis.
Similar phenomenon holds for the radius scalings above (on the right side of) the intersection point with $R_{\rm{inter}}\approx R\approx R_{\max}\approx8.50\pm0.96\,\rm{km}$ (tomato diamond), as shown in the right panel of FIG.\,\ref{fig_fMfR}.
However, below the intersection point (on its left side), we have $\nu_{\rm{c}}^{[\rm{TOV}]}>\nu_{\rm{c}}^{[\rm{G}]}$ (with ``G'' abbreviating for generally stable NSs) for a given radius. This would be impossible for some critical radius $R_{\rm{crit}}<R_{\rm{inter}}$ except $\varepsilon_{\rm{c}}^{[\rm{TOV}]}<\varepsilon_{\rm{c}}^{[\rm{G}]}$. The latter condition is unphysical as the central energy density for the critically stable NS at TOV configuration should always be larger than that for a generally stable NS sharing the same radius. This then means that there exists a lower bound on NS radius (similar to the existence of an upper bound $M_{\rm{TOV}}$ for NS mass).
\begin{figure}[h!]
\centering
\includegraphics[height=7.5cm]{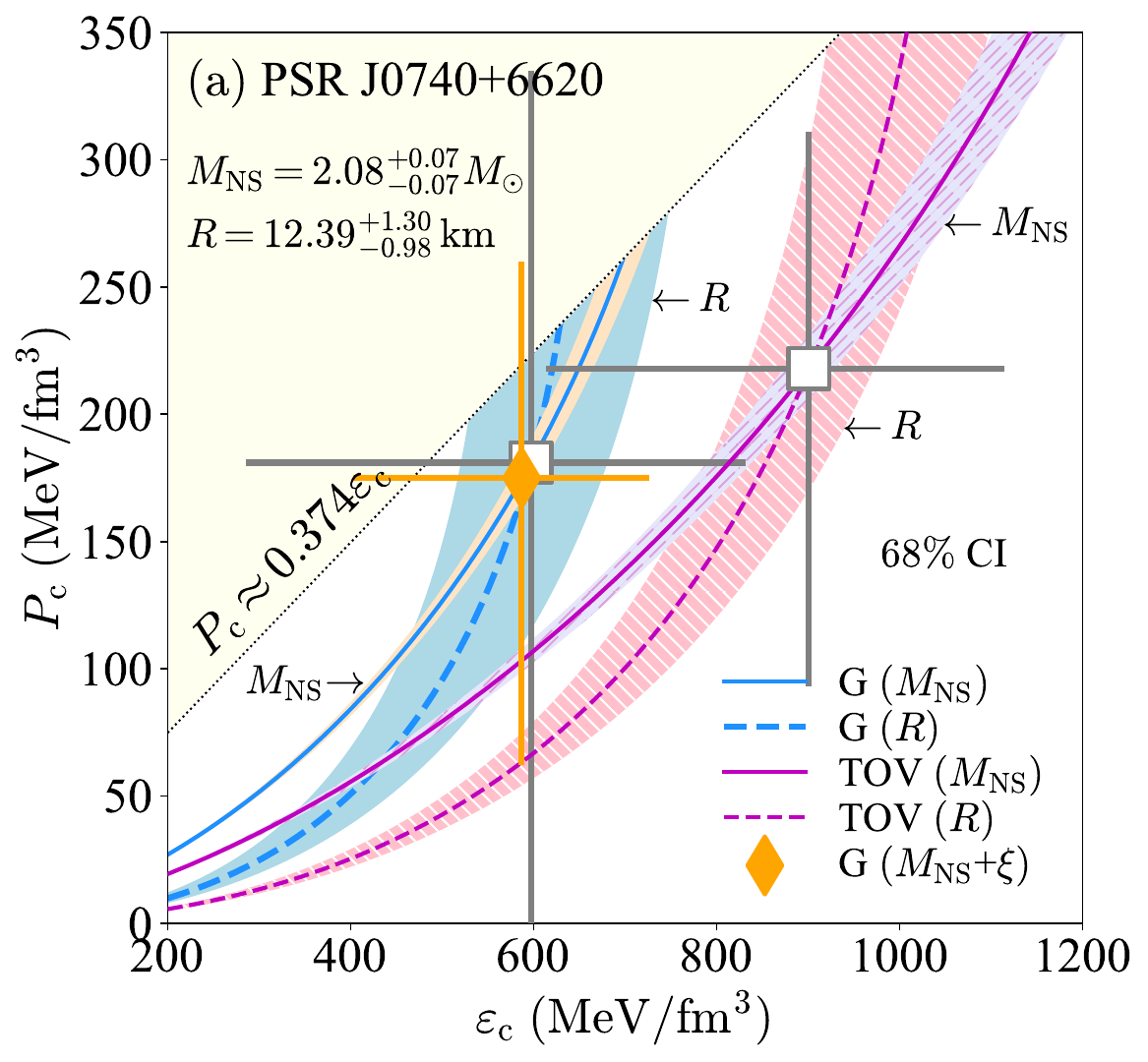}\qquad
\includegraphics[height=7.5cm]{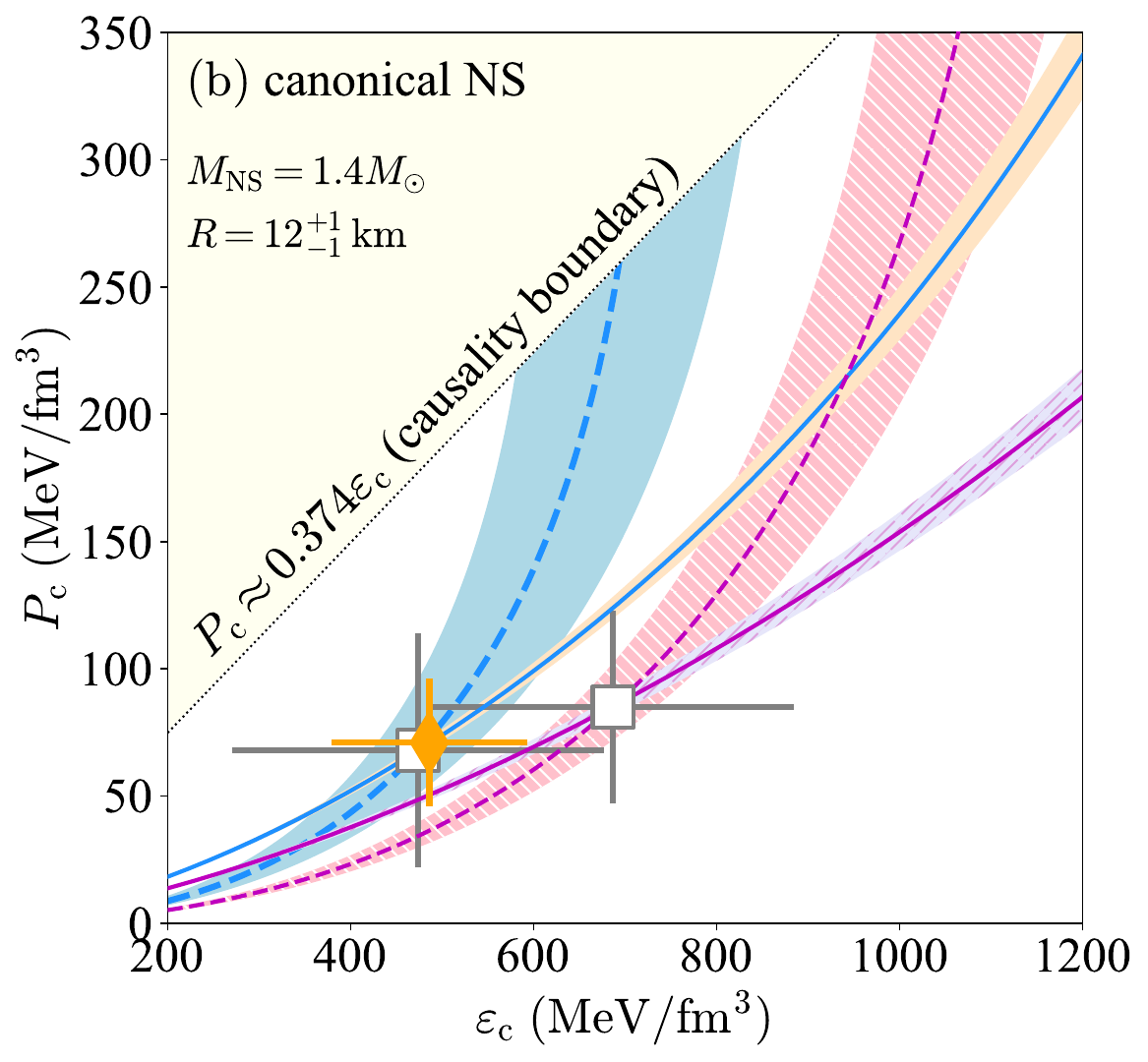}
\caption{(Color Online). Left panel: the central EOS for PSR J0740+6620\,\cite{Riley21} adopting the $M_{\rm{NS}}^{\max}/M_{\odot}$-analysis (indicated by ``TOV ($M_{\rm{NS}}$)'' and ``TOV ($R$)'') and the $M_{\rm{NS}}/M_{\odot}$-analysis (indicated by ``G ($M_{\rm{NS}}$)'' and ``G ($R$)''), respectively; here the grey shallow squares (with errorbars) are points $(\varepsilon_{\rm{c}},P_{\rm{c}})$ using these two analyses.
The orange solid diamond (with errorbars) is the inferred value for $\varepsilon_{\rm{c}}$ and $P_{\rm{c}}$ adopting the mass and compactness scalings (instead of the radius scaling) in the $M_{\rm{NS}}/M_{\odot}$-analysis.
Right panel: the same as the left panel but for a canonical NS.
}\label{fig_EOS_comp}
\end{figure}
\begin{table}[h!]
\renewcommand{\arraystretch}{1.5}
\centerline{\normalsize
\begin{tabular}{c|c|c|c|c|c} 
  \hline
  &&$\x$&$\varepsilon_{\rm{c}}$&$P_{\rm{c}}$&mass and radius\\\hline\hline
$M_{\rm{NS}}^{\max}/M_{\odot}$-analysis&PSR J0740+6620$^{\rm{a}}$&$0.24_{-0.07}^{+0.05}$&$901_{-287}^{+214}$&$218_{-125}^{+93}$&$2.08_{-0.07}^{+0.08}M_{\odot}$, $12.39_{-0.98}^{+1.30}\,\rm{km}$\\\hline
(mass+radius scalings) &PSR J0030+0451$^{\rm{a}}$&$0.11_{-0.02}^{+0.03}$&$550_{-178}^{+186}$&$58_{-31}^{+33}$&$1.34_{-0.16}^{+0.15}M_{\odot}$, $12.71_{-1.19}^{+1.14}\,\rm{km}$\\\hline
&PSR J0437-4715&$0.14_{-0.02}^{+0.02}$&$828_{-250}^{+166}$&$115_{-55}^{+37}$&$1.418_{-0.037}^{+0.037}M_{\odot}$, $11.36_{-0.63}^{+0.95}\,\rm{km}$\\\hline
&canonical NS&$0.12_{-0.02}^{+0.02}$&$687_{-197}^{+197}$&$85_{-38}^{+38}$&$1.4M_{\odot}$, $12_{-1}^{+1}\,\rm{km}$\\\hline\hline
$M_{\rm{NS}}/M_{\odot}$-analysis&PSR J0740+6620$^{\rm{a}}$&$0.30_{-0.18}^{+0.14}$&$597_{-312}^{+235}$&$181_{-204}^{+154}$&\\\hline
(mass+radius scalings)&PSR J0030+0451$^{\rm{a}}$&$0.12_{-0.03}^{+0.03}$&$350_{-160}^{+166}$&$40_{-29}^{+31}$&\\\hline
&PSR J0437-4715&$0.17_{-0.04}^{+0.04}$&$620_{-288}^{+192}$&$106_{-82}^{+54}$&\\\hline
&canonical NS&$0.14_{-0.04}^{+0.04}$&$474_{-204}^{+204}$&$68_{-46}^{+46}$&\\\hline\hline
$M_{\rm{NS}}/M_{\odot}$-analysis&PSR J0740+6620$^{\rm{a}}$&$0.30_{-0.10}^{+0.08}$&$588_{-184}^{+139}$&$175_{-113}^{+85}$&\\\hline
(mass+$\xi$ scalings)&PSR J0030+0451$^{\rm{a}}$&$0.13_{-0.03}^{+0.03}$&$411_{-178}^{+186}$&$52_{-32}^{+34}$&\\\hline
&PSR J0437-4715&$0.16_{-0.02}^{+0.02}$&$565_{-140}^{+94}$&$91_{-36}^{+24}$&\\\hline
&canonical NS&$0.15_{-0.02}^{+0.02}$&$487_{-108}^{+108}$&$71_{-25}^{+25}$&\\\hline\hline
    \end{tabular}}
        \caption{Central EOS for four NSs: PSR J0740+6620\,\cite{Riley21}, PSR J0030+0451\,\cite{Riley19}, PSR J0437-4715\,\cite{Choud24} and a canonical NS, here $\x$ is dimensionless and both $\varepsilon_{\rm{c}}$ and $P_{\rm{c}}$ are measured in $\rm{MeV}/\rm{fm}^3$.}\label{tab_cEOS}        
\end{table}

Once the mass or radius is known/observed for a NS, we can use the mass or radius scaling to determine the central EOS. This was already done using the $M_{\rm{NS}}^{\max}/M_{\odot}$-analysis for PSR J0740+6620 in Subsection \ref{sub_Densest}, see FIG.\,\ref{fig_eP-PSR740+6620}.
We can also do this similarly by adopting the $M_{\rm{NS}}/M_{\odot}$-analysis assuming all the NSs are stable along the M-R curve (without limiting them to be at the TOV configuration), the result for the central EOS for PSR J0740+6620 is shown in the left panel of FIG.\,\ref{fig_EOS_comp}, here the light-blue solid curve (associated with a tan band and indicated by ``G ($M_{\rm{NS}}$)'' and ``G ($R$)'') and the light-blue dashed curve (associated with a light-blue band and indicated by ``TOV ($M_{\rm{NS}}$)'' and ``TOV ($R$)'') are the predictions on the central EOS using the mass and radius scalings, respectively.
For comparisons, we also plot the central EOSs using the mass and radius scalings in the $M_{\rm{NS}}^{\max}/M_{\odot}$-analysis, by the solid magenta curve (associated with lavender hatched band) and the dashed magenta curve (associated with pink hatched band),  respectively.
In both analyses, i.e., the $M_{\rm{NS}}^{\max}/M_{\odot}$-analysis and the $M_{\rm{NS}}/M_{\odot}$-analysis, the uncertainties from the radius-scaling on the central EOS are much larger than those from the mass scaling, comparing it with the left panel of FIG.\,\ref{fig_MmaxS}.
Moreover, if both the mass and radius are known/observed, we can then determine both $\varepsilon_{\rm{c}}$ and $P_{\rm{c}}$ simultaneously. The results with error bars are shown by the shallow grey squares in FIG.\,\ref{fig_EOS_comp} and the numerical values are given in TAB.\,\ref{tab_cEOS}.
We see that the $P_{\rm{c}}$ and $\varepsilon_{\rm{c}}$ for PSR J0030+0451 (in the $M_{\rm{NS}}/M_{\odot}$-analysis) are consistent with those obtained in Ref.\,\cite{Brandes2023-a} using a Bayesian inference method.
Another feature is that the inferred values for $\varepsilon_{\rm{c}}$ and $P_{\rm{c}}$ in the $M_{\rm{NS}}/M_{\odot}$-analysis is smaller than those in the $M_{\rm{NS}}^{\max}/M_{\odot}$-analysis; as we discussed in the previous paragraphs regarding FIG.\,\ref{fig_fMfR}.
The right panel of FIG.\,\ref{fig_EOS_comp} shows the results similarly obtained for a canonical NS.
{\color{xll}We may point out that for light NSs such as a canonical one, it is more appropriate to infer the central EOSs using the $M_{\rm{NS}}/M_{\odot}$-analysis since they are generally far from the TOV configurations predicted by using various EOSs and are far below the maximum NS mass observed so far. On the hand, for massive NSs, e.g. PSR J0740+6620 or the second component of GW190814 with a mass of about 2.59$M_{\odot}$, an inference using the $M_{\rm{NS}}^{\max}/M_{\odot}$-analysis is preferred.}
In TAB.\,\ref{tab_cEOS}, the central EOSs for PSR J0030+0451\,\cite{Riley19} and PSR J0437-4715\,\cite{Choud24} are also given.

Since the radius scaling in the general $M_{\rm{NS}}/M_{\odot}$-analysis has quite a non-perfect regression feature (as indicated by FIG.\,\ref{fig_Rnc}), the prediction on $\varepsilon_{\rm{c}}$ and $P_{\rm{c}}$ has much larger uncertainties.
In order to improve the prediction on $\varepsilon_{\rm{c}}$ and $P_{\rm{c}}$, we may use the combination of the mass scaling and the compactness scaling, since both are very strong and model-independent (see FIG.\,\ref{fig_Phi_c}).
In particular, we can obtain $\x$ straightforwardly from $\xi=A_\xi \Pi_{\rm{c}}+B_\xi$ since now $\xi$ is available, after having $\x$, we then calculate the central energy density by
\begin{equation}\label{cda-1}
\varepsilon_{\rm{c}}=\left(\frac{\xi-B_\xi}{A_\xi}\right)^3\left(\frac{A_{\rm{M}}}{M_{\rm{NS}}/M_{\odot}-B_{\rm{M}}}\right)^2=\underbrace{\left(\frac{\x}{1+3\x^2+4\x}\right)^3}_{\mbox{upper bounded}}\left(\frac{A_{\rm{M}}}{M_{\rm{NS}}/M_{\odot}-B_{\rm{M}}}\right)^2,
\end{equation}
and consequently the central pressure $P_{\rm{c}}=\x\varepsilon_{\rm{c}}$.
The resulted values for $\varepsilon_{\rm{c}}$ and $P_{\rm{c}}$ are shown in FIG.\,\ref{fig_EOS_comp} by the orange diamonds and also in TAB.\,\ref{tab_cEOS} in the last four rows.
We can see that the prediction is very close to the value using the mass and radius scalings, but has a much smaller errorbar.

\subsection{Physics behind the slopes of NS mass-radius curves and their approximately ``vertical'' shapes}\label{sub_PsiVert}

\begin{figure}[h!]
\centering
\includegraphics[height=7.cm]{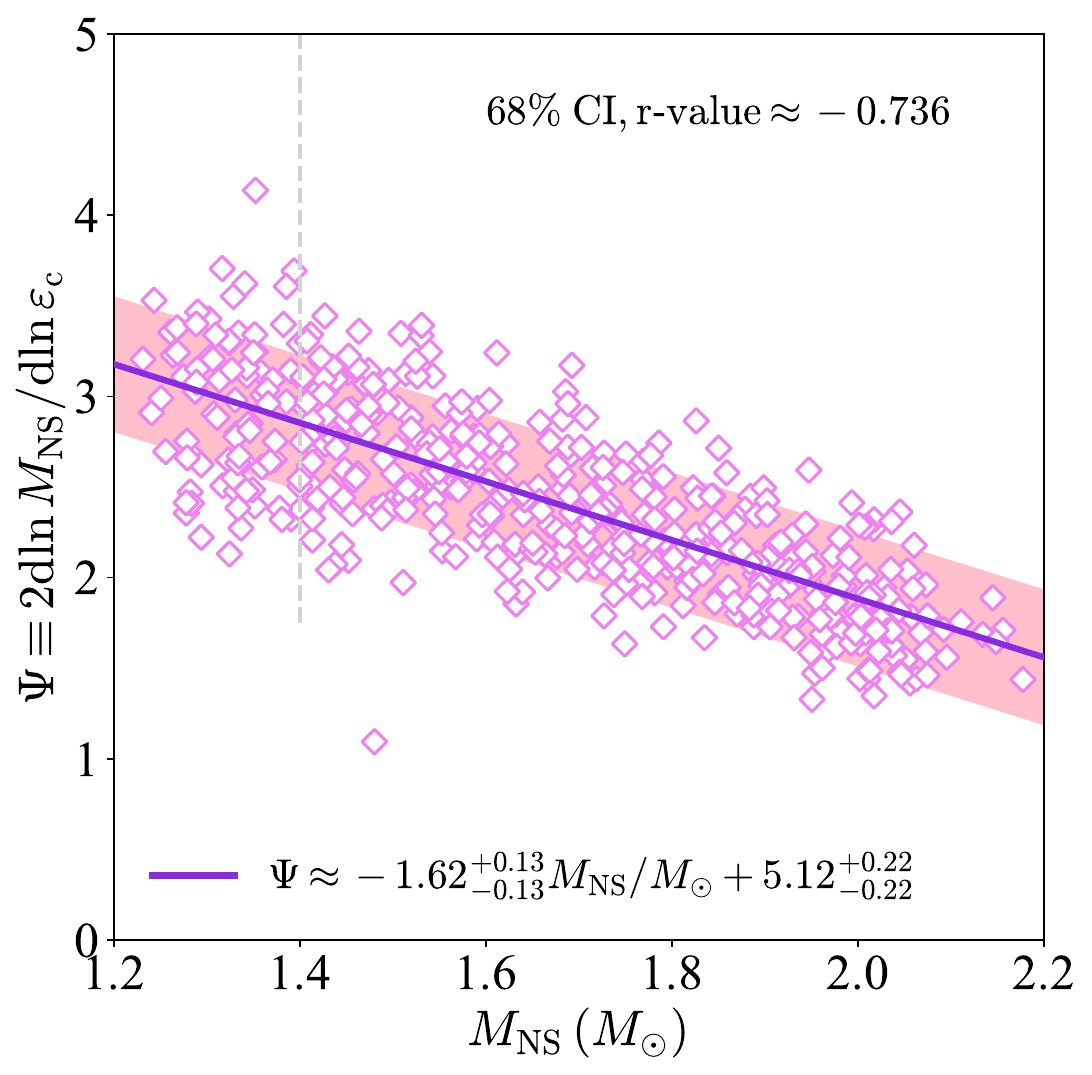}
\caption{(Color Online).  The same as FIG.\,\ref{fig_Rnc} but for the $\Psi$ dependence on NS mass $M_{\rm{NS}}$.
}\label{fig_k-fac}
\end{figure}

The factor $\Psi$ defined in Eq.\,(\ref{def-Psi}) is an important quantity characterizing the generally stable NSs with masses $M_{\rm{NS}}$.
Using the meta-model EOSs, one can similarly investigate quantitatively such dependence; the result is shown in FIG.\,\ref{fig_k-fac}.
Specifically, we have
\begin{equation}\label{gk-Psi}
\Psi\approx a_\Psi \left(\frac{M_{\rm{NS}}}{M_{\odot}}\right)+b_\Psi\approx-1.62_{-0.13}^{+0.13}\left(\frac{M_{\rm{NS}}}{M_{\odot}}\right)+5.12_{-0.22}^{+0.22}.
\end{equation}
The $M_{\rm{NS}}/M_{\odot}$-dependence of the $\Psi$-factor is expected to work better at low NS masses than high masses since for NSs with large $M_{\rm{NS}}/M_{\odot}$, they may depend on the (unknown value of) $M_{\rm{TOV}}$.
For example, for a model (theory) with $M_{\rm{TOV}}\approx2.3M_{\odot}$, the $\Psi$-factor for a NS with mass $\lesssim 2.3M_{\odot}$ is vanishingly small.
For a canonical NS, for example,  we have
$\Psi\approx2.85\pm0.29$.

\begin{figure}[h!]
\centering
\includegraphics[height=6.5cm]{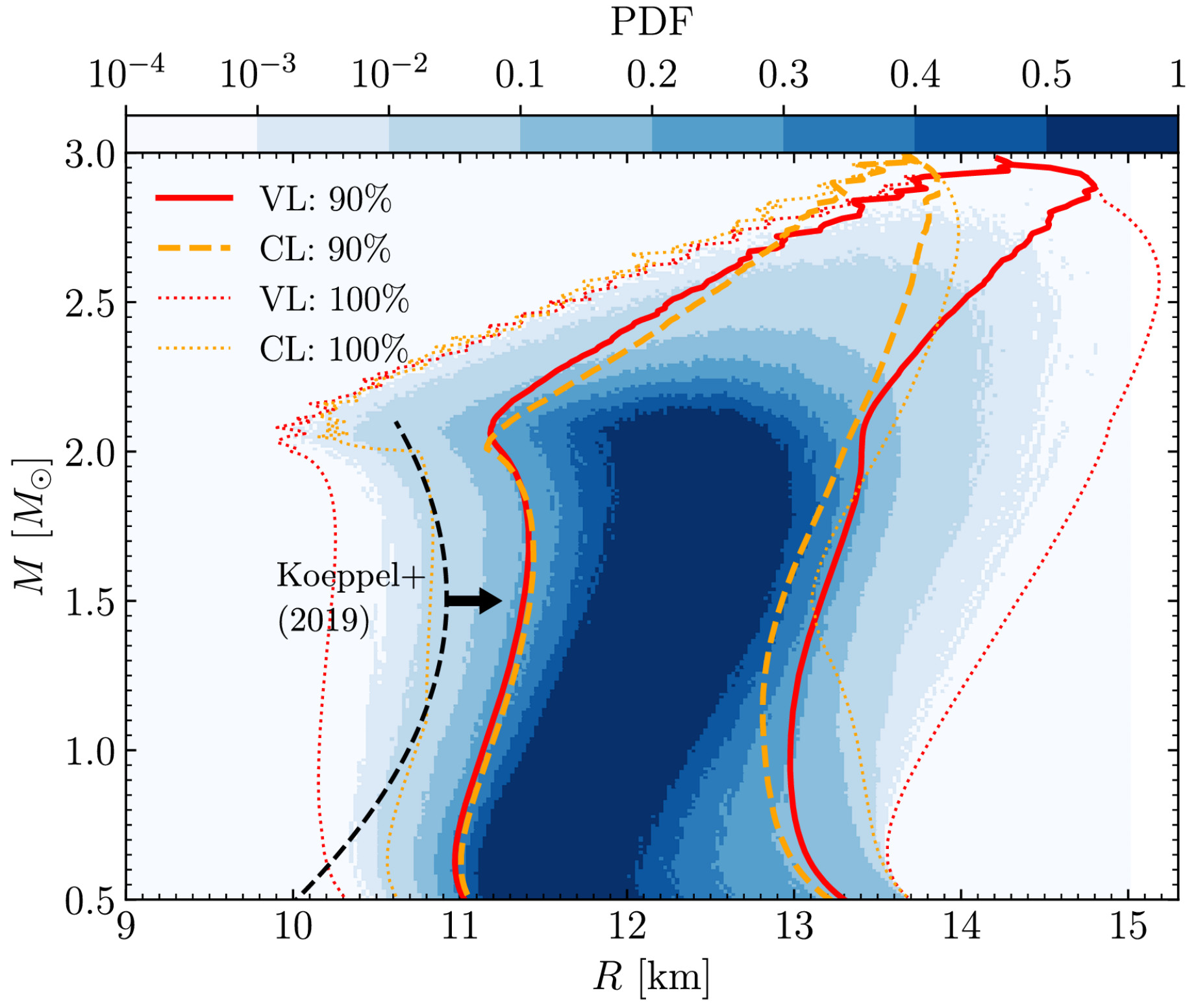}\qquad
\includegraphics[height=6.2cm]{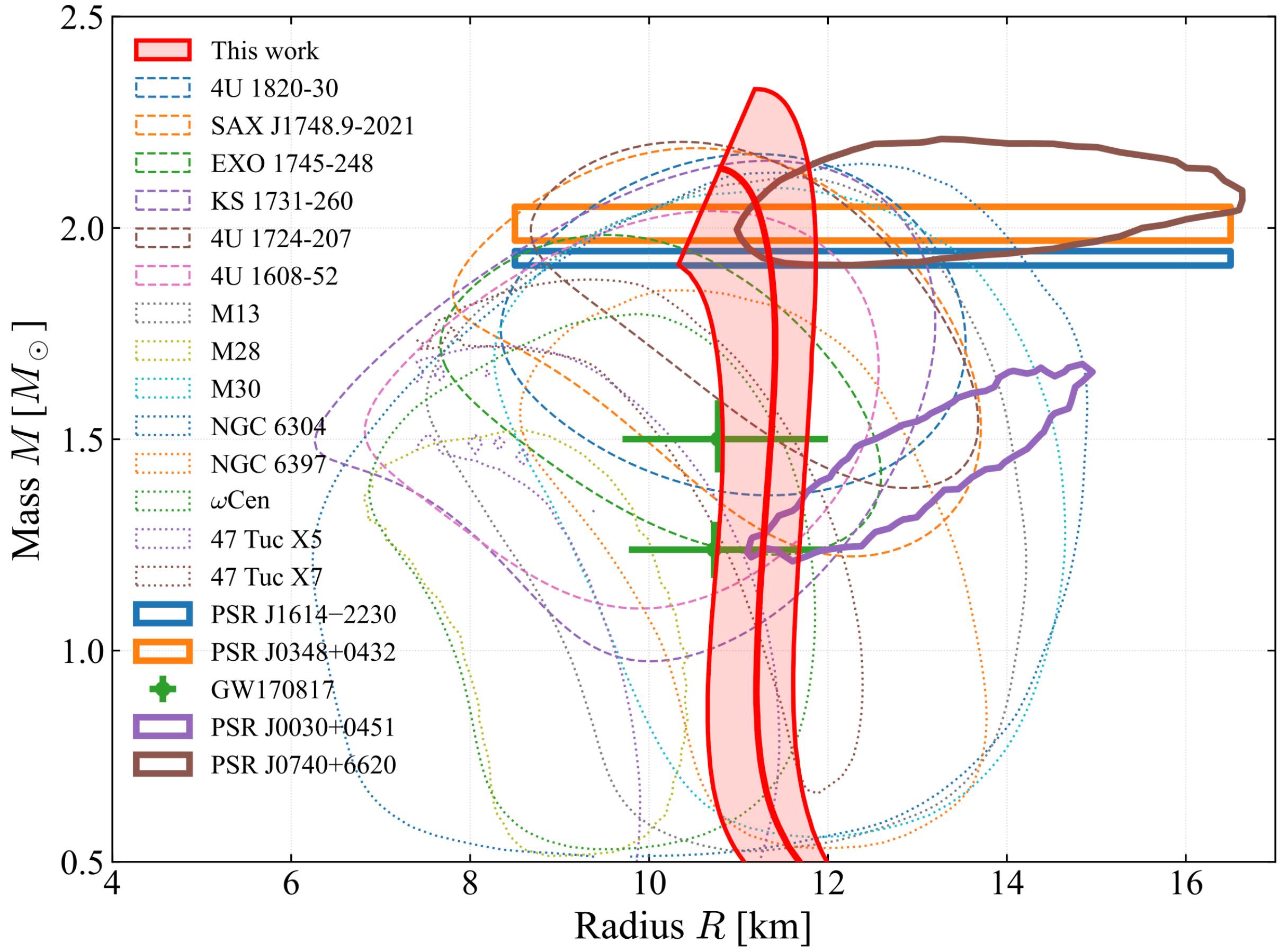}
\caption{(Color Online).  Inferred NS M-R relations using contemporary astrophysical constraints.
Figures taken from Ref.\,\cite{Jiang2023ApJ} (left panel) and from Ref.\,\cite{Fuji2024b} (right panel).
}\label{fig_MR-vert}
\end{figure}

The $\Psi$-$M_{\rm{NS}}$ scaling of FIG.\,\ref{fig_k-fac} implies that the $M_{\rm{NS}}$ increases with the central energy density $\varepsilon_{\rm{c}}$ according to an approximately power low (by treating $\Psi\approx\rm{const.}$):
\begin{equation}
\boxed{M_{\rm{NS}}/M_{\odot}\sim\varepsilon_{\rm{c}}^{\Psi/2}.}
\end{equation}
Therefore, mathematically the case with $\Psi>2$ implies a super-linear growth while that with $\Psi<2$ indicates a sub-linear growth.
The critical NS mass occurring at $\Psi_0=2$ is $M^0_{\rm{NS}}/M_{\odot}\approx1.93\pm0.21$.
For canonical NSs, the mass growth rate is about:
\begin{equation}
\mbox{at }M_{\rm{NS}}\approx1.4M_{\odot}:~~M_{\rm{NS}}/M_{\odot}\sim\varepsilon_{\rm{c}}^{1.43\pm0.14},
\end{equation}
while the growth rate of their radii is given by Eq.\,(\ref{fgk-dd}):
\begin{equation}
\mbox{at }
M_{\rm{NS}}\approx1.4M_{\odot}:~~R\sim\varepsilon_{\rm{c}}^{0.14\pm0.05}.
\end{equation}
According to Eq.\,(\ref{fgk-1}), $\d R/\d\varepsilon_{\rm{c}}=\d R/\d M_{\rm{NS}}\cdot\d M_{\rm{NS}}/\d\varepsilon_{\rm{c}}$, so $\d R/\d M_{\rm{NS}}=\d R/\d\varepsilon_{\rm{c}}\cdot(\d M_{\rm{NS}}/\d\varepsilon_{\rm{c}})^{-1}$, or
\begin{empheq}[box=\fbox]{align}\label{fgk-2}
\frac{\d R}{\d M_{\rm{NS}}}=\rm{const.}\times\left(1-\frac{2}{\Psi}\right)\cdot\varepsilon_{\rm{c}}^{-3^{-1}(1+\Psi)}
=
\rm{const.}\times\left(1-\frac{2}{\Psi}\right)\cdot \left(\frac{M_{\rm{NS}}}{M_{\odot}}\right)^{-(2/3)(1+\Psi^{-1})}
,
\end{empheq}
where the second relation follows from $M_{\rm{NS}}/M_{\odot}\sim\varepsilon_{\rm{c}}^{\Psi/2}$.
We then find for NSs with $\Psi\approx\Psi_0=2$ or equivalently $M^0_{\rm{NS}}/M_{\odot}\approx1.93\pm0.21$, the dependence of $R$ on $M_{\rm{NS}}$ is weak, i.e., $\d R/\d M_{\rm{NS}}\approx0$ (similarly $\d R/\d\varepsilon_{\rm{c}}\sim\varepsilon_{\rm{c}}^{-0.01\pm0.04}$ using Eq.\,(\ref{fgk-dd})), or $\d M_{\rm{NS}}/\d R\to+\infty$ for $M_{\rm{NS}}\to M_{\rm{NS}}^{0}$ from below and $\d M_{\rm{NS}}/\d R\to-\infty$ for $M_{\rm{NS}}\to M_{\rm{NS}}^0$ from above.
{\color{xll}This explains the empirical ``vertical'' shape seen in NS M-R sequences in both observations and many calculations, e.g., see the right panel\,\cite{Fuji2024b} of FIG.\,\ref{fig_MR-vert} or meta-model predictions shown in FIG.\ \ref{fig_MR-ensem}.
In addition, as $\Psi>2$ for light NSs, Eq.\,(\ref{fgk-2}) implies $\d R/\d M_{\rm{NS}}>0$; this means the ``vertical'' shape is slightly slanted from the lower-left corner to the upper-right corner,} see a typical inference on the M-R curve shown in the left panel of FIG.\,\ref{fig_MR-vert}.
In fact, the compilation of NS observational data shown in FIG.\,\ref{fig_NSMR-REV} also indicates such empirical fact. In short, our analyses above provide for the first time an EOS-model independent explanation for a well-known phenomenon in NS physics from analyzing the scaled TOV equations.

\begin{figure}[h!]
\centering
\includegraphics[width=12.cm]{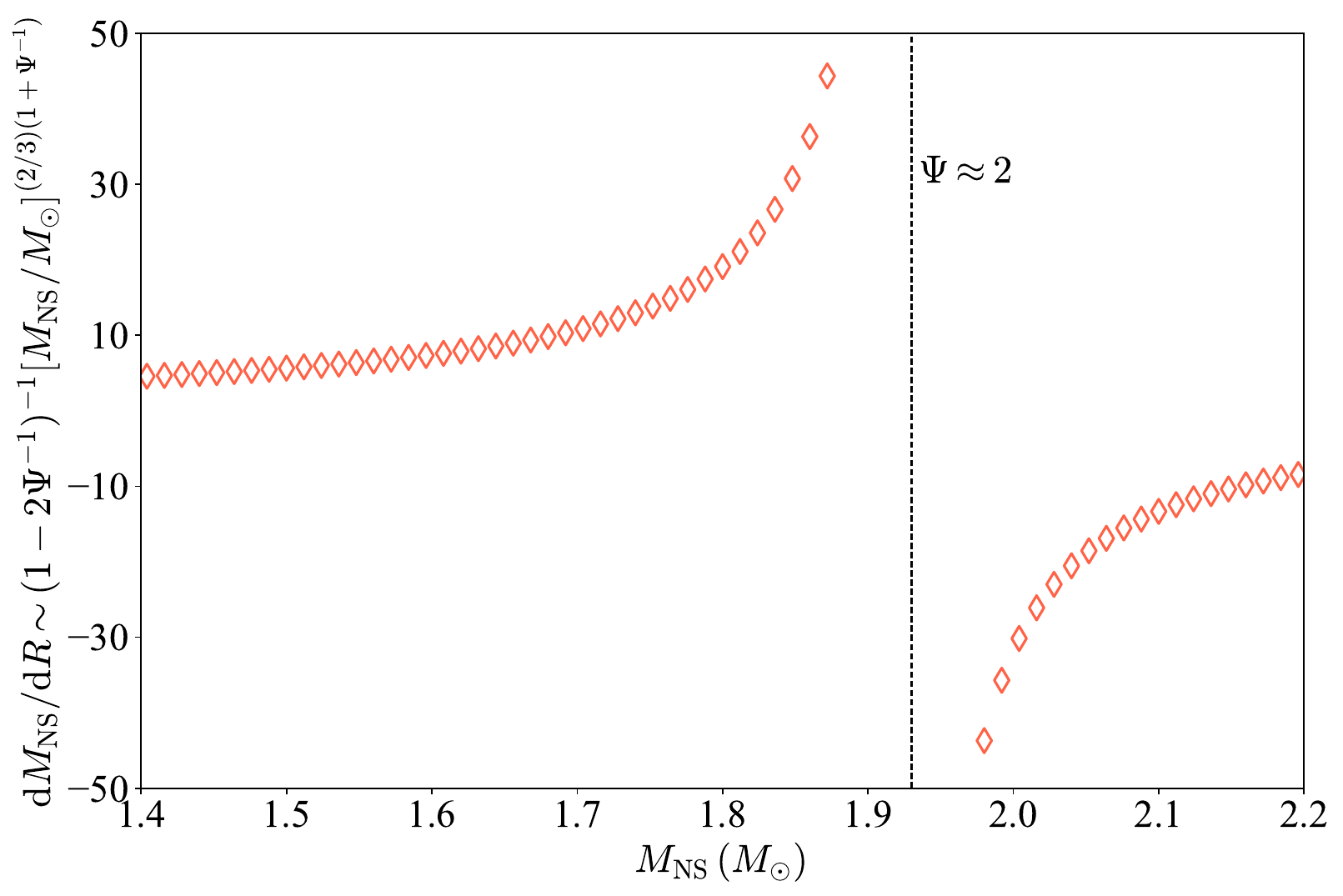}
\caption{(Color Online). Dependence of $(1-2\Psi^{-1})^{-1}(M_{\rm{NS}}/M_{\odot})^{(2/3)(1+\Psi^{-1})}\sim\d M_{\rm{NS}}/\d R$ on NS mass using Eq.\,(\ref{gk-Psi}).
}\label{fig_dMdR}
\end{figure}

Combined with Eq.\,(\ref{gk-Psi}), we can obtain the NS mass dependence factor $(1-2\Psi^{-1})^{-1}(M_{\rm{NS}}/M_{\odot})^{(2/3)(1+\Psi^{-1})}$ in its derivative $\d M_{\rm{NS}}/\d R$ with respect to $R$ according to Eq.\,(\ref{fgk-2}), see FIG.\,\ref{fig_dMdR}. Our analysis implies that the $\d M_{\rm{NS}}/\d R$ probably changes signs along the full M-R curve, see related discussions on this issue in Refs.\,\cite{Han:2020adu,Zhao2020,Ferr24,Li:2024imk}.
Similarly, using Eq.\,(\ref{fgk-dd}), we have:
\begin{equation}\label{fgk-8}
\boxed{
    {\d R}/{\d M_{\rm{NS}}}\sim\left(1-2\Psi^{-1}\right)
    \cdot R^{-2(1+\Psi^{-1})/(1-2\Psi^{-1})},}
\end{equation}
or,
\begin{equation}
    \boxed{{\d R}/{\d M_{\rm{NS}}}\approx3^{-1}\xi^{-1}\left(1-2\Psi^{-1}\right),}
\end{equation}
considering $\xi=M_{\rm{NS}}/R$.
Therefore, the $\d M_{\rm{NS}}/\d R\approx 3\xi(1-2\Psi^{-1})^{-1}$ may either be positive or negative although $\xi=M_{\rm{NS}}/R$ is always positive. This feature is similar to the relation between $\d P/\d\varepsilon=s^2$ (non-monotonic) and $P/\varepsilon=\phi$ (monotonic).
In particular, for canonical NSs with $R\approx12\pm1\,\rm{km}$, we have $\d M_{\rm{NS}}/\d R\approx1.73\pm0.44$ using $\Psi\approx2.85\pm0.29$ obtained previously.

The factor $\Psi$ is also very relevant for estimating the central SSS of generally stable NSs via Eq.\,(\ref{sc2-GG}), we give in Subsection \ref{sub_s2canon} the central SSS of a canonical NS and compare it with existing results in the literature.


\subsection{A counter-intuitive feature of NS mass scaling}\label{sub_E1_count}

\begin{figure}[h!]
\centering
\includegraphics[width=9.cm]{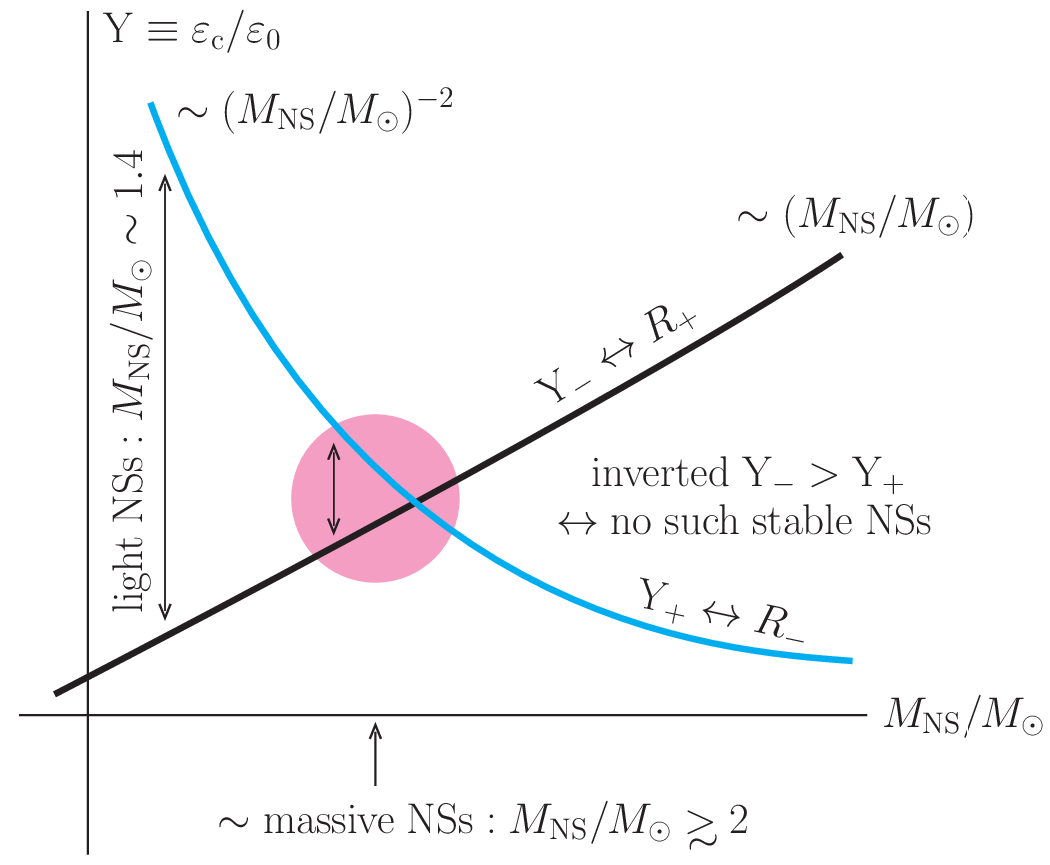}
\caption{(Color Online).  NS mass scaling implies that for a given mass $M_{\rm{NS}}$ there exists an upper limit $\y_+$ for the central energy density, roughly scaling as $\y_+\sim M_{\rm{NS}}^{-2}$, so massive NSs allow a lower upper limit for $\varepsilon_{\rm{c}}$.
On the other hand, the $\varepsilon_{\rm{c}}$ increases with $M_{\rm{NS}}$ due to the stability condition $\d M_{\rm{NS}}/\d\varepsilon_{\rm{c}}>0$, indicating a lower limit $\y_-$ exists for a given $M_{\rm{NS}}$.
Generally, one has $\y_-<\y_+$; and the difference $\y_+-\y_-$ decreases with $M_{\rm{NS}}$.
If the NS mass $M_{\rm{NS}}$ exceeds some critical value, then the relation between $\y_-$ and $\y_+$ would be inverted, implying the existence of a maximum NS mass.
}\label{fig_Ypm_sk}
\end{figure}

In this and the next subsections, we use our scalings as a tool to explore why and how a maximum NS mass emerges naturally in GR. The mass scaling of Eq.\,(\ref{gk-m}) has a very important (though counter-intuitive) feature: a heavier NS has a lower upper limit for the central energy density. It is a representation or reflection of the self-gravitating nature of compact stars.
Our scaling of (\ref{gk-m}) reveals such novel feature naturally.
According to Eq.\,(\ref{cda-1}) where $\Pi_{\rm{c}}=\x/(1+3\x^2+4\x)$ is upper bounded to about $0.002$ at $\x\approx0.374$, we have then
\begin{equation}\label{ci-1}
\boxed{
\y\equiv
\frac{\varepsilon_{\rm{c}}}{\varepsilon_0}\lesssim
\frac{21.71}{(M_{\rm{NS}}/M_{\odot} + 0.08)^2}\equiv \y_+\sim M_{\rm{NS}}^{-2},~~\mbox{under }\x\lesssim0.374\leftrightarrow\Delta_{\rm{c}}\gtrsim\Delta_{\rm{GR}}\approx-0.041,
}
\end{equation}
here $\varepsilon_0\approx150\,\rm{MeV}/\rm{fm}^3$ is the energy density at nuclear saturation density.
{\color{xll}For example, if one takes $\y\lesssim2.3$ for $M_{\rm{NS}}/M_{\odot}\approx3$, does it mean a $3M_{\odot}$ NS has a central density smaller than about $2.3\rho_0$? This looks unreasonable.
Therefore, on the contrary, the mass-$\Gamma_{\rm{c}}$ scaling indicates that there exists an upper bound for NS masses.}
The factor $\Pi_{\rm{c}}$ in Eq.\,(\ref{cda-1}) is crucial for upper bounding $\y$. If we neglect the GR correction term ``$3\x^2+4\x$'', then $\Pi_{\rm{c}}^{\rm{N},3}=\x^3\approx0.42$ (here ``N'' is for Newtonian limit and $\x$ is upper limited as $\x\leq3/4$). Consequently, $\Pi_{\rm{c}}^{\rm{N},3}/\Pi_{\rm{c}}^3\approx200$.
In fact, the inverse scaling between $\y$ and NS mass is not new, e.g., Ref.\,\cite{Lattimer10} constrained the upper limit of $\y$ according to the following relation\,\cite{Lattimer2012Ann}
\begin{equation}\label{Lat12Y}
    \y\lesssim51\left({M_{\rm{NS}}}/{M_{\odot}}\right)^{-2},
\end{equation}
see Eq.\,(\ref{LP11ult}).
In such case, assuming the existence of a $3M_{\odot}$ NS makes the $\y$ upper bounded as $\y\lesssim5.7$, which looks quite reasonable.
However, the $M_{\rm{TOV}}$ being smaller than $3M_{\odot}$ is the current community consensus.
On the other hand, (\ref{Lat12Y}) induces a weak restriction on $\y$ as $\y\lesssim9.6$ for $M_{\rm{NS}}^{\max}/M_{\odot}\approx2.3$. 
Our expression (\ref{ci-1}) gives a tight bound on $\y$ considering the upper limit of NS masses.
The upper limit on $\y$ is equivalent to a lower limit (denoted by $R_-$) on the radius $R$, since from our compactness scaling we have
\begin{equation}\label{ci-0}
\boxed{
R/\rm{km}=\frac{1.477M_{\rm{NS}}/M_{\odot}}{A_\xi\Pi_{\rm{c}}+B_\xi}\gtrsim5.58M_{\rm{NS}}/M_{\odot},}
\end{equation}
using the upper bounded nature of $\Pi_{\rm{c}}$.
Therefore we establish that
\begin{equation}\label{ll-1}
\y_+\leftrightarrow R_-,
\end{equation}
this is shown in FIG.\,\ref{fig_Ypm_sk}.

\begin{figure}[h!]
\centering
\includegraphics[width=8.cm]{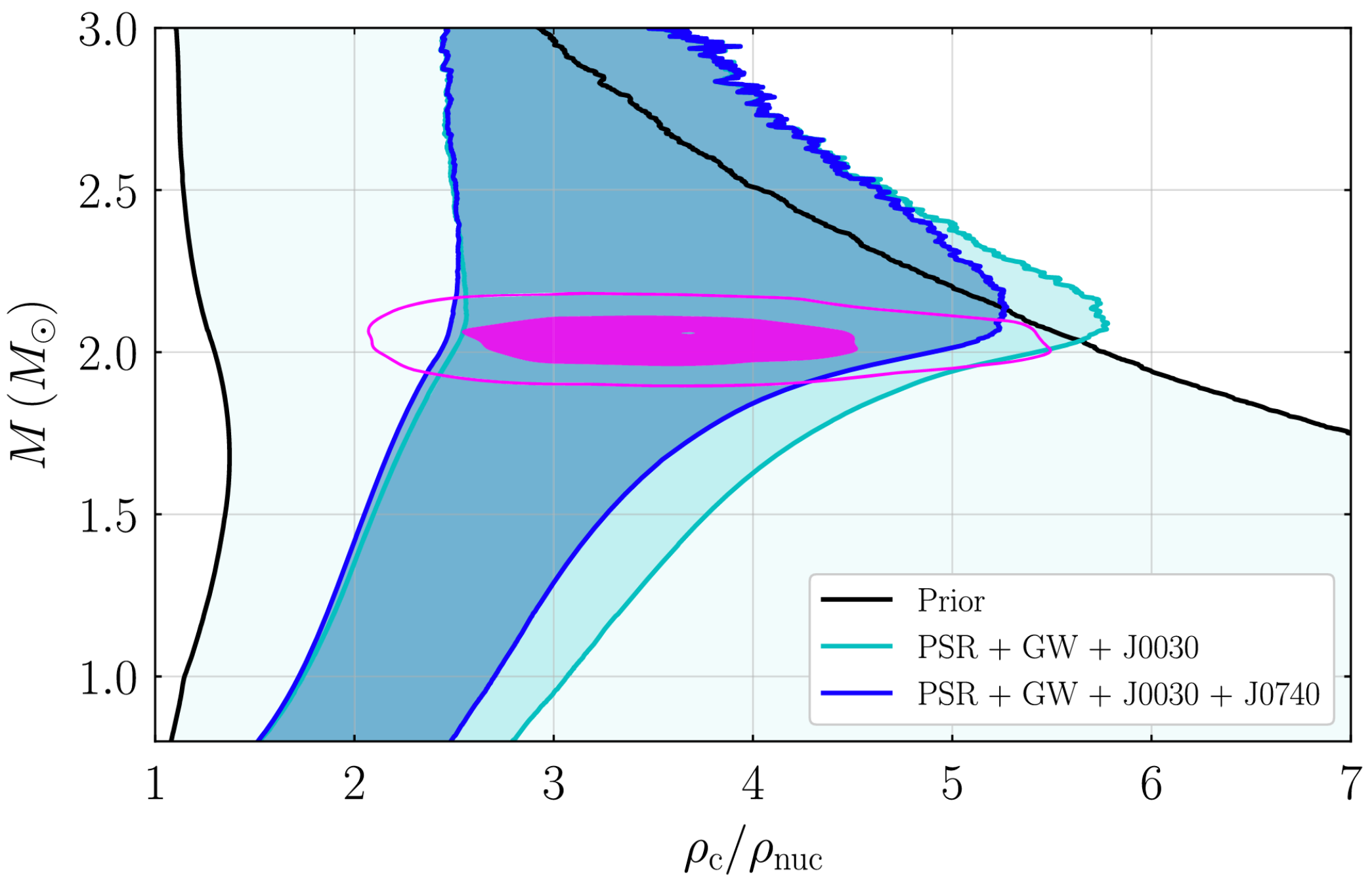}\quad
\includegraphics[width=8.2cm]{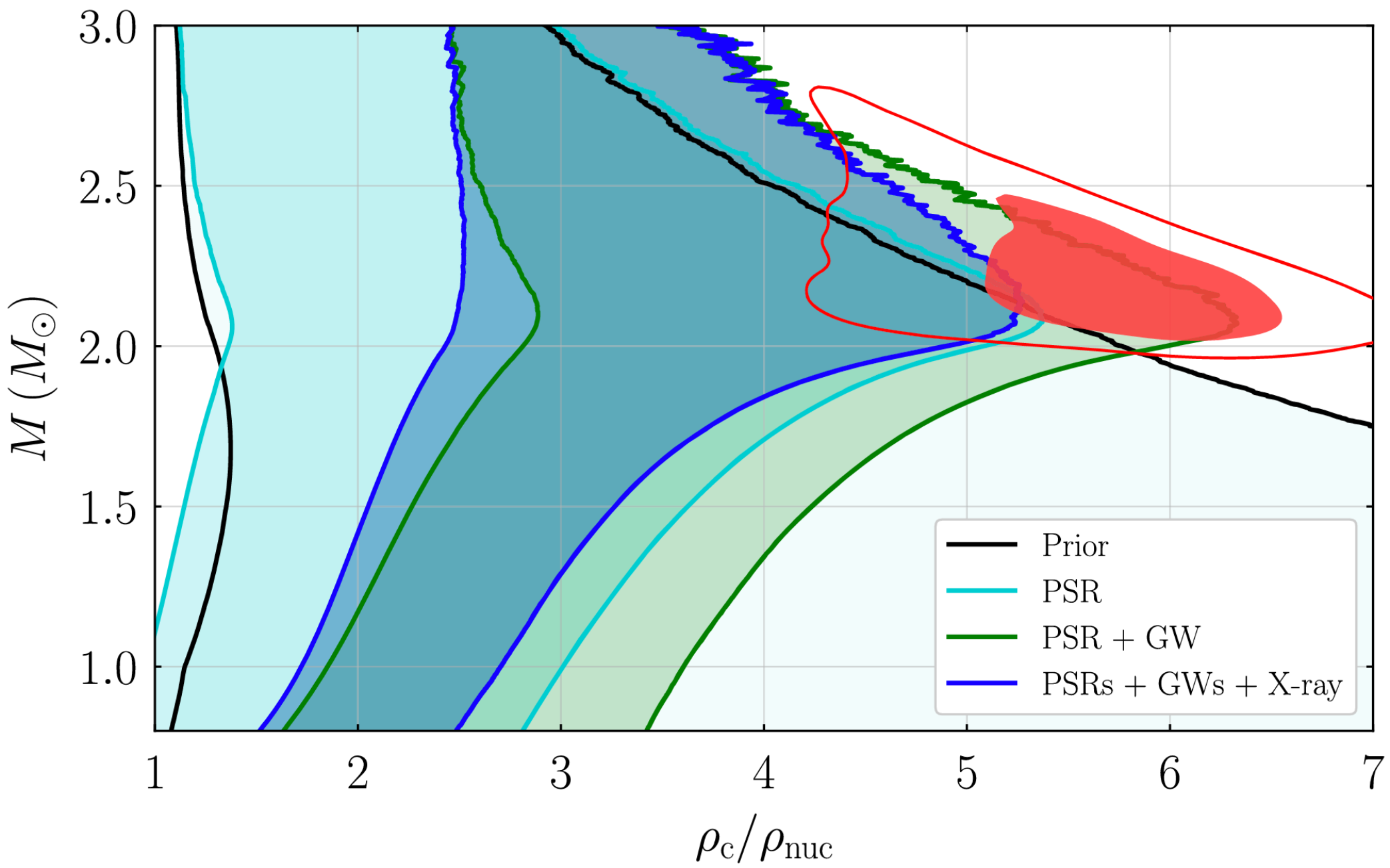}
\caption{(Color Online).  The central density as a function of NS mass, where the magenta (red)
contours in the left (right) panel show the 50\% and 90\% confidence level for the mass-central density posterior for PSR J0740+6620 (for the maximum-mass NS). Figure taken from Ref.\,\cite{Leg21}.
}\label{fig_Mrhoc-relta}
\end{figure}

On the other hand, for stable NSs, the derivative of NS mass with respect to the central (energy) density is positive\,\cite{Hartle78PR,Shapiro1983},
\begin{equation}\label{cond_stb}
\boxed{
\d M_{\rm{NS}}/\d\varepsilon_{\rm{c}}>0
\leftrightarrow\d\y/\d(M_{\rm{NS}}/M_{\odot})>0
,~~\mbox{for stable NSs along the M-R curve},}
\end{equation} see for example FIG.\,\ref{fig_Mrhoc-relta} for a recent inference on such relation using contemporary astrophysical observations.
This means there exists a lower limit $\y_-$ on the central energy density for a given NS mass, e.g., we have $\y_-\approx3.5$ from FIG.\,\ref{fig_Mrhoc-relta} for a $2M_{\odot}$ NS like PSR J0740+6620.
For a given $M_{\rm{NS}}$, our mass and compactness scalings enable us to write out explicitly that,
\begin{equation}\label{ci-2}
\boxed{
R/\rm{km}=\frac{\Sigma M_{\rm{NS}}/M_{\odot}}{\displaystyle A_\xi\left(\frac{M_{\rm{NS}}/M_{\odot}-B_{\rm{M}}}{A_{\rm{M}}}\sqrt{\y\varepsilon_0}\right)^{2/3}+B_\xi}
\approx\frac{1.477M_{\rm{NS}}/M_{\odot}}{0.106(M_{\rm{NS}}/M_{\odot}+0.08)^{2/3}\y^{1/3}-0.032},}
\end{equation}
here $\Sigma\equiv M_{\odot}/\rm{km}\approx1.477$.
Thus, a lower limit $\y_-$ corresponds to an upper limit $R_+$ for the radius, i.e.,
\begin{equation}\label{ll-2}
\y_-\leftrightarrow R_+,
\end{equation}
we also indicate this relation in FIG.\,\ref{fig_Ypm_sk}.

\begin{figure}[h!]
\centering
\includegraphics[width=11.cm]{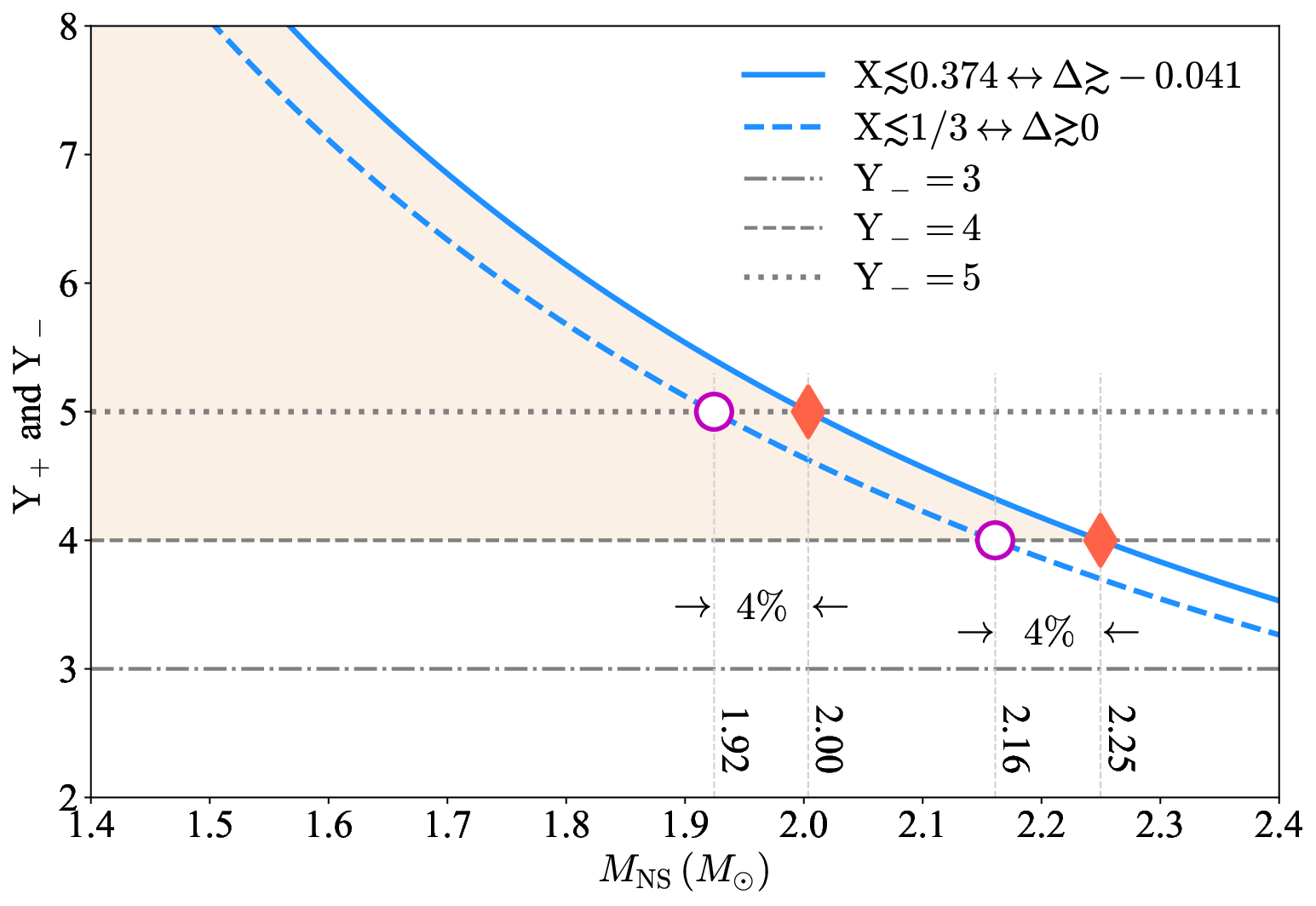}
\caption{(Color Online). An illustration on using $\y\lesssim\y_+$ of Eq.\,(\ref{ci-1}) to estimate some upper limit of $M_{\rm{NS}}$; here three $\y_-$'s are used (grey lines for $\y_-=3$, 4 and 5).
The effects of upper bound on $\x$ (either $\x\lesssim0.374$ or $\x\lesssim1/3$) are also shown.
The light-tan band shows the effective region for allowed NS masses in the case of $\y_-\approx4$ and $\x\lesssim0.374$.
}\label{fig_Ynp}
\end{figure}

{\color{xll} Considering all of the above in this subsection, we come to the following concluding points: (a) there should physically be an overlapped region for $\y$ as $\y_-\lesssim\y\lesssim \y+$,  and if $M_{\rm{NS}}/M_{\odot}$ exceeds some critical value no such overlapped region exists, this critical NS mass is an indication for the existence of $M_{\rm{TOV}}$; and (b) for NSs with masses smaller than the above critical value, relations (\ref{ll-1}) and (\ref{ll-2}) together set an effective region for the radius.}
For example, in FIG.\,\ref{fig_Ynp} we illustrate how to use $\y\lesssim\y_+$ of Eq.\,(\ref{ci-1}) to estimate some upper limits of $M_{\rm{NS}}$ adopting three $\y_-$'s (shown by grey lines); e.g., adopting $\y_-\approx4$ gives $M_{\rm{NS}}/M_{\odot}\lesssim2.25$ while using $\y_-\approx5$ we obtain $M_{\rm{NS}}/M_{\odot}\lesssim2.00$.
It is necessary to point out that these estimates are direct consequences of the self-gravitating nature of NSs considered in GR and the upper limit for $\x$ as $\x\lesssim0.374$ as well as the upper limit for $\Pi_{\rm{c}}$ correspondingly, as implied in Eq.\,(\ref{cda-1}).
If one requires $\x\lesssim1/3$ as predicted by pQCD at very high densities\,\cite{Kur10,Bjorken83}, then Eq.\,(\ref{ci-1}) should be replaced by $\y_+\approx20.09/(M_{\rm{NS}}/M_{\odot}+0.08)^2$.
In this case, $\y_-\approx4$ gives $M_{\rm{NS}}/M_{\odot}\lesssim2.16$, since a smaller upper bound on $\x$ implies the NS is less compact\,\cite{CL24-c}.
In the next subsection (Subsection \ref{sub_MTOV}), we may use all the mass, radius and compactness scalings explicitly to estimate the maximum mass of stable NSs.

Moreover, the above scheme for restricting the radius becomes more effective for massive NSs than for light ones, since for the latter the distance between $\y_-(\leftrightarrow R_+)$ and $\y_+(\leftrightarrow R_-)$ are quite large.
For example, for a canonical NS, $\y_+\approx9.9$ using relation (\ref{ci-1}) while the lower limit is around $\y_-\approx2.5$ (see FIG.\,\ref{fig_Mrhoc-relta}).
Therefore (\ref{ci-0}) and (\ref{ci-2}) together give us $R_-\approx7.8\,\rm{km}\lesssim R\lesssim R_+\approx13.4\,\rm{km}$, which is not quite useful.
On the other hand, the lower limit $\y_-$ for a $2.1M_{\odot}$ NS (such as PSR J0740+6620\,\cite{Fon21}) from FIG.\,\ref{fig_Mrhoc-relta} is around about $\y_-\approx3.7$ while the upper limit $\y_+$ from (\ref{ci-1}) is about $\y_+\approx4.6$, so in this case one can effectively restrict the $\y$ and similarly the radius $R$. In particular, we have $11.7\,\rm{km}\lesssim R\lesssim12.7\,\rm{km}$, which is quite close to the observed radius\,\cite{Miller21,Riley21,Ditt24,Salmi22,Salmi24}.
Similarly, considering PSR J2215+5135 with its mass about $2.15\,M_{\odot}$, we may obtain $12.0\,\rm{km}\lesssim R\lesssim 12.8\,\rm{km}$ by adopting a similar lower limit $\y_-\approx3.7$.
However, the estimate on the lower limit $\y_-$ in this scheme itself is non-trivial.

\begin{figure}[h!]
\centering
\includegraphics[width=11.cm]{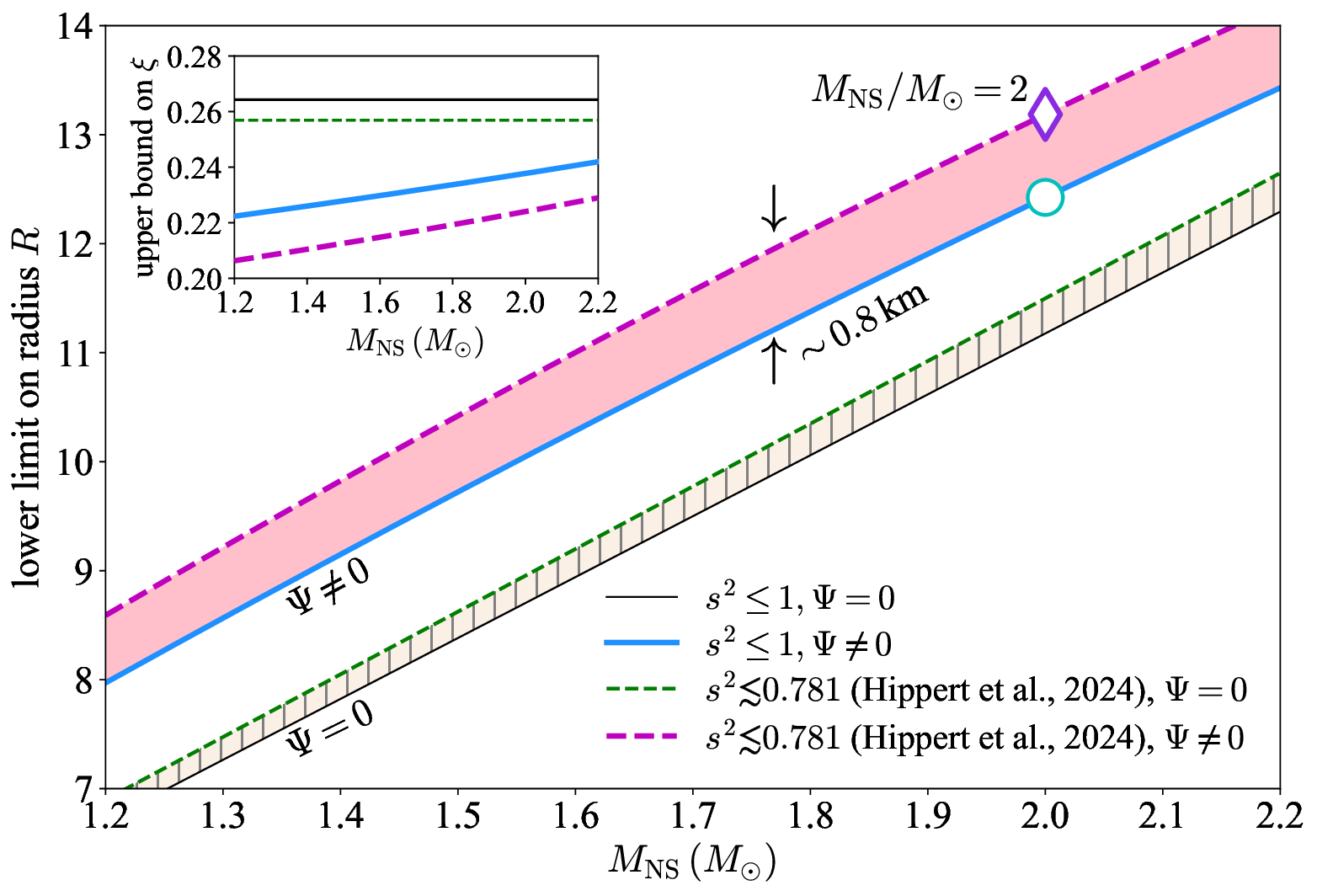}\\[0.25cm]
\includegraphics[width=8.5cm]{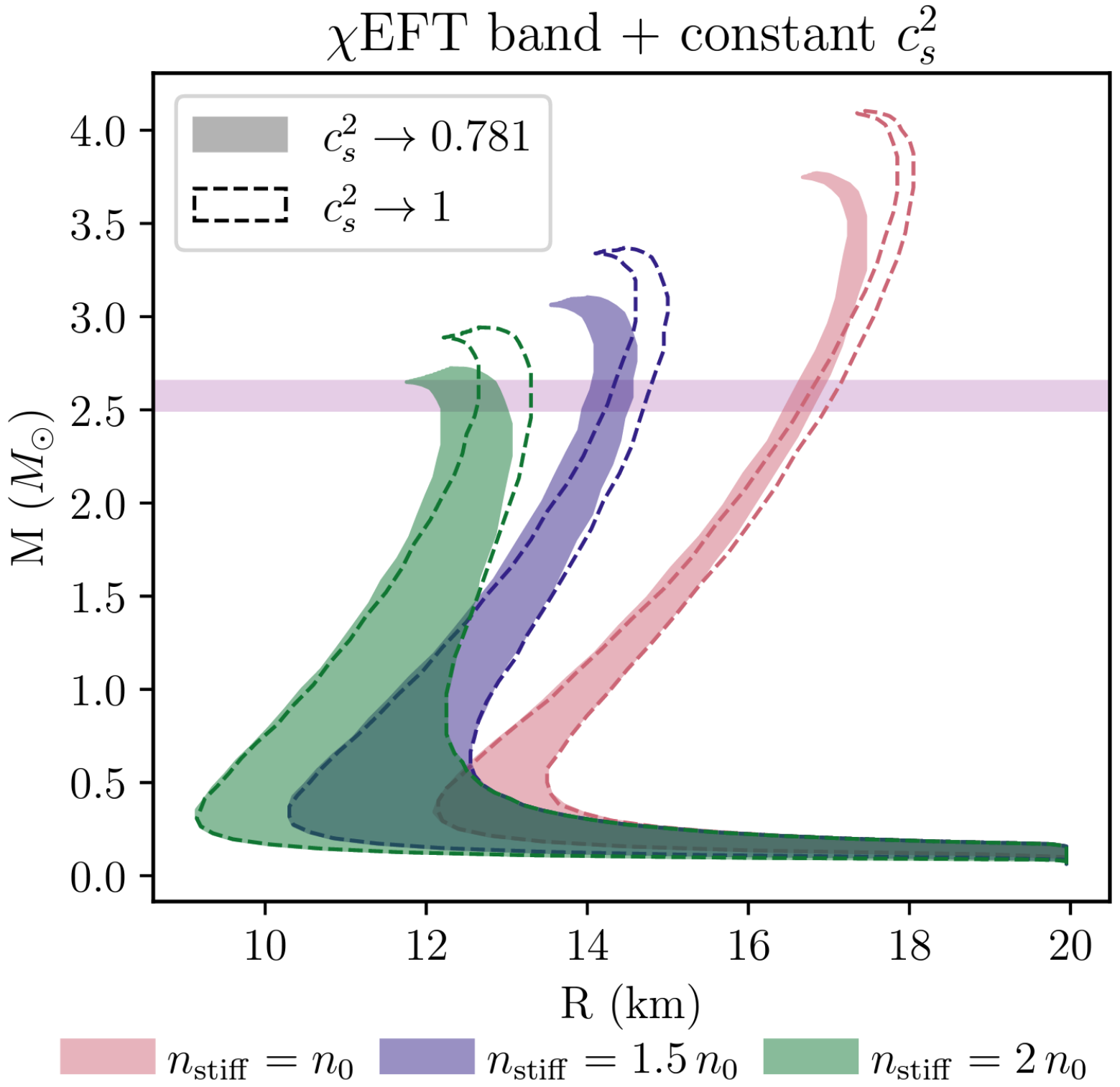}
\caption{(Color Online). Upper panel: influence of a finite $\Psi\neq0$ on the lower limit for $R$ (compare the two bands); in each band the solid line is for $s^2\leq1$ and the dashed line is for $s^2\lesssim0.781$\,\cite{Hipp24}; the inset plots the upper bound for $\xi$ as a function of $M_{\rm{NS}}/M_{\odot}$ (notice the solid black limit is given by Eq.\,(\ref{xi_GR})); two symbols on the upper band are the radii for $M_{\rm{NS}}/M_{\odot}\approx2$.
Lower panel: the impact of a lower upper limit for $s^2$ as $s^2\lesssim0.781$\,\cite{Hipp24} on the NS M-R curve. Figure taken from Ref.\,\cite{Hipp24}.
}\label{fig_Psi_on_R}
\end{figure}

In obtaining (\ref{ci-0}) the $\x\lesssim0.374$ (with $\Psi=0$) is used, i.e., $\xi\approx A_\xi\Pi_{\rm{c}}+B_\xi\lesssim0.264$ (of Eq.\,(\ref{xi_GR})) in the denominator is obtained at $\x\lesssim0.374$. A finite $\Psi$ of Eq.\,(\ref{def-Psi}) can be used to improve the upper limit for $\xi$ as a function of $M_{\rm{NS}}/M_{\odot}$ since now $\x$ is upper bounded to a lower value compared with $0.374$, and consequently the lower limit for $R$ for a given $M_{\rm{NS}}/M_{\odot}$.
The result is shown in the upper panel of FIG.\,\ref{fig_Psi_on_R}, here the two bands (pink and light-tan) are obtained with $\Psi=0$ and $\Psi\neq0$ (of Eq.\,(\ref{def-Psi})), respectively.
In each band, the solid (dashed) line is for $s^2\leq1$ ($s^2\lesssim0.781$); here the upper bound on $s^2$ as $s^2\lesssim0.781$ was given recently by considering nuclear matter transport constraints\,\cite{Hipp24}.
A smaller upper bound on $s^2$ corresponds to a softer EOS (and a smaller compactness as indicated in the inset), which may reduce the maximum NS mass (lower panel of FIG.\,\ref{fig_Psi_on_R}) and also generates a relatively larger $R$ for a given $M_{\rm{NS}}$.
For example, for a NS with $M_{\rm{NS}}/M_{\odot}\approx2$, the set with $\Psi\neq0$ and $s^2\lesssim0.781$ constrains $R\gtrsim13.2\,\rm{km}$ while that with $\Psi\neq0$ and $s^2\leq1$ gives $R\gtrsim12.4\,\rm{km}$, inducing an effect about 0.8\,km (see the two symbols on the upper band).
Future high-precision NSs radius measurements may help determine whether the $s^2$ should be upper bounded to a lower value compared with 1.
As a comparison, it is seen that the smaller upper bound on $s^2$ as $s^2\lesssim0.781$ almost has no impact on the radii of NSs with masses around $2M_{\odot}$\,\cite{Hipp24} as shown in the lower panel of FIG.\,\ref{fig_Psi_on_R}; this is different from our analysis.
Notice in their analysis\,\cite{Hipp24}, a constant SSS model near NS centers is adopted.
In addition, in our calculations with $\Psi=0$ (TOV configuration), the upper bound for $\xi$ is independent of $M_{\rm{NS}}/M_{\odot}$ (see the two thin lines in the inset).
For reference, we write out $\x$ as a function $s_{\rm{c}}^2$:
\begin{align}
    \x(s_{\rm{c}}^2,\Psi)\approx&\frac{3s_{\rm{c}}^2}{4+\Psi}
    \left[1-\frac{12(1+\Psi)}{(4+\Psi)^2}s_{\rm{c}}^2
    +\frac{18(1+\Psi)(4+13\Psi)}{(4+\Psi)^4}s_{\rm{c}}^4+\cdots
    \right]\notag\\
    \to &\frac{3s_{\rm{c}}^2}{4}\left(1-\frac{3}{4}s_{\rm{c}}^2+\frac{9}{32}s_{\rm{c}}^4+\cdots\right),
\end{align}
where the second relation follows from $\Psi=0$. 
A similar derivation for $\mathcal{M}(s_{\rm{c}}^2,\Psi)$ defined in Eq.\,(\ref{kl-1}) was given previously.
Similarly, the function $\Pi_{\rm{c}}$ of Eq.\,(\ref{gk-comp}) becomes
\begin{align}
    \Pi_{\rm{c}}(s_{\rm{c}}^2,\Psi)\approx&\frac{3s_{\rm{c}}^2}{4+\Psi}\left[1-\frac{12(5+2\Psi)}{(4+\Psi)^2}s_{\rm{c}}^2
    +\frac{9(344+298\Psi+71\Psi^2)}{(4+\Psi)^4}s_{\rm{c}}^4+\cdots\right]\notag\\
    \to&\frac{3s_{\rm{c}}^2}{4}\left(1 -\frac{15}{4}s_{\rm{c}}^2 + \frac{387}{32}s_{\rm{c}}^4+\cdots\right).
\end{align}
Reducing on the upper limit for $s^2$ (and therefore on $s_{\rm{c}}^2$) naturally leads to the reduction on $\x$ or $\Pi_{\rm{c}}$, so the compactness $\xi\sim \Pi_{\rm{c}}$ is also reduced; the NSs then become less compact (larger $R$ for a given $M_{\rm{NS}}$).

To this end, we point out that {\color{xll}in these analyses the radius scaling itself dose not come into play (either (\ref{ci-1}) or (\ref{ci-0})), although it contains extra information that the mass and compactness scalings do not have. It is encapsulated in the fitting coefficient $B_{\rm{R}}$,} its effects will be studied in the next subsection.

\subsection{Maximum masses of generally and critically stable NSs along M-R curves}\label{sub_MTOV}

In the previous subsections, we used together two of the three (mass, radius and the compactness) scalings to evaluate individually the $\varepsilon_{\rm{c}}$ and $P_{\rm{c}}$ or the compactness scaling alone to infer the $\x$.
At first glance, it seems that only two of the three scalings are independent. However, from the viewpoint of the numerical fitting of the scalings, using the three scalings as a whole could provide more information.
It could be understood as follows: {\color{xll}Since $B_{\rm{M}}\approx-0.08$ (see Eq.\,(\ref{gk-m})) is near zero, when we investigate the compactness scaling via $\xi=M_{\rm{NS}}/R$ the corresponding intercept $B_\xi\approx-0.032$ (see Eq.\,(\ref{gk-xi})) is also found to be near zero (as it should be). It means that effect of the two coefficients $A_{\rm{R}}$ and $B_{\rm{R}}$ (see Eq.\,(\ref{gk-r})) in the radius scaling is partially transferred into the coefficient $A_\xi$ in the compactness scaling, or, equivalently the effect of $B_{\rm{R}}$ is lost when considering the compactness scaling.}
Breaking the degeneracy in these combinations may provide us extra information on some relevant quantities.

How to understand that $M_{\rm{NS}}$ is upper bounded?
In this subsection, we use the full information contained in the mass, radius and compactness scalings, especially $A_{\rm{R}}$ and $B_{\rm{R}}$, to estimate the maximum NS mass allowed. As discussed briefly in the last subsection, NSs unavoidably have a maximum mass due to two competing factors: {\color{xll}the stability condition (\ref{cond_stb}) requires $\varepsilon_{\rm{c}}$ increase with $M_{\rm{NS}}$ while the self-gravitating property of NSs encapsulated in the mass scaling (see Eq.\,(\ref{ci-1})) requires (the upper bound for) $\varepsilon_{\rm{c}}$ be inversely proportional to $M_{\rm{NS}}$ roughly as $M_{\rm{NS}}^{-2}$, see the sketch given in FIG.\,\ref{fig_Ypm_sk}.
The counter-intuitive feature of the mass scaling can be traced back to the bound structure of $\Pi_{\rm{c}}^{3}$ in Eq.\,(\ref{cda-1}) which is an intrinsic property of GR.} The overlapped region satisfying both conditions on $\varepsilon_{\rm{c}}$ eventually shrinks, even to disappear at a critical NS mass (the maximum NS mass allowed) $\gtrsim2.2M_{\odot}$ (see FIG.\,\ref{fig_Mrhoc-relta}). 

\begin{figure}[h!]
\centering
\includegraphics[width=9.cm]{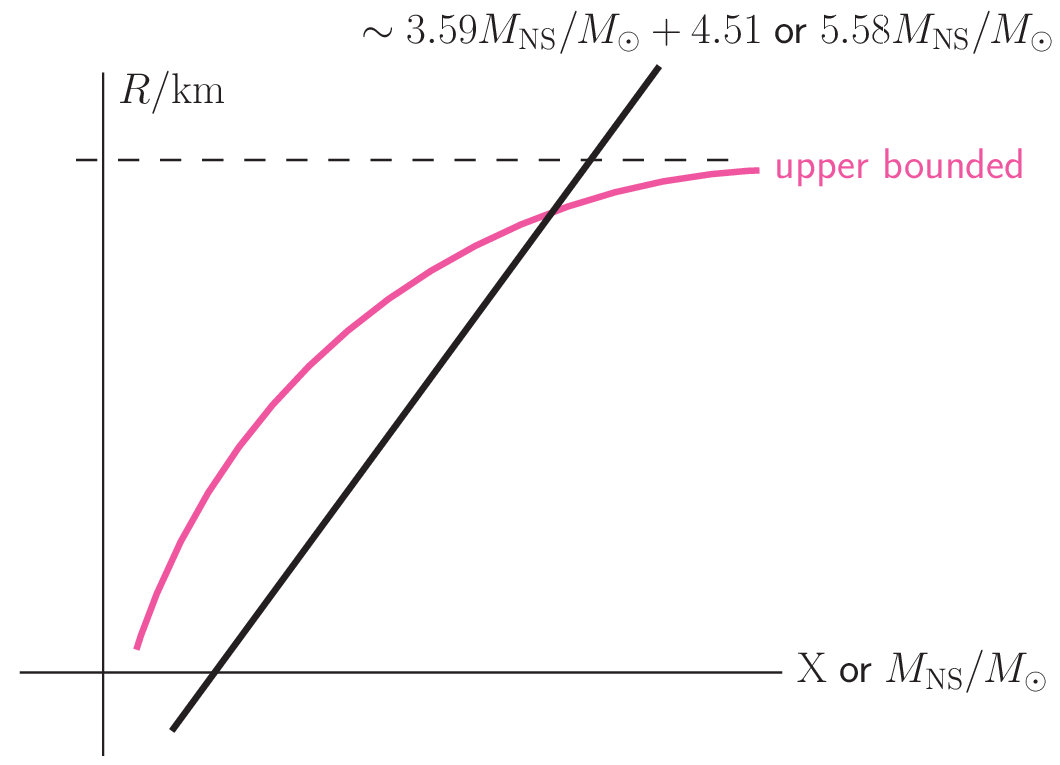}
\caption{(Color Online). The lower limit of NS radius $R$ linearly increases with $M_{\rm{NS}}/M_{\odot}$ according to Eq.\,(\ref{Rlin}), while the radius $R$ is upper bounded as shown by Eq.\,(\ref{rr2}); these two facts on $R$ imply that there exists a maximum NS mass allowed.
}\label{fig_R2sc}
\end{figure}

\begin{figure}[h!]
\centering
\includegraphics[height=7cm]{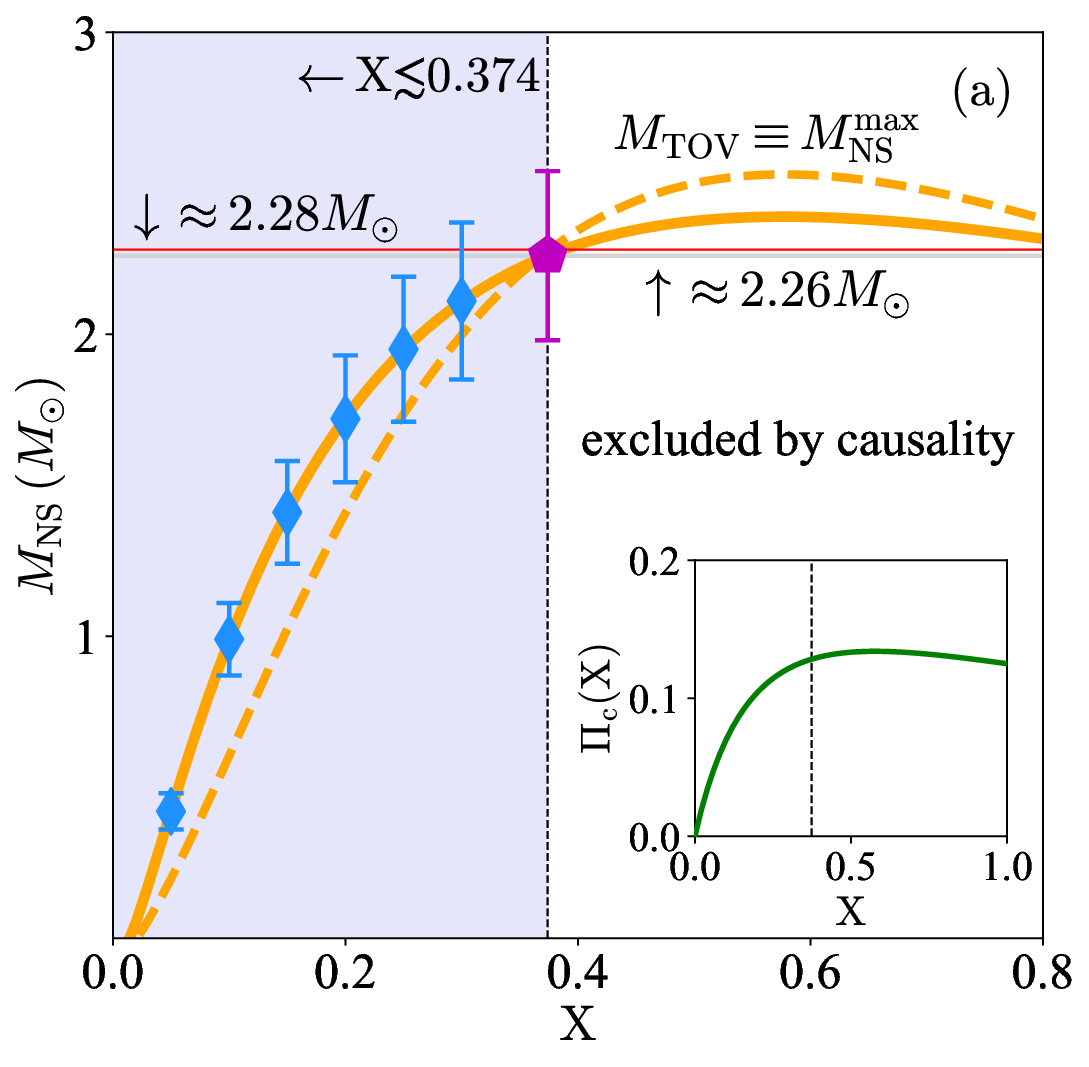}\qquad
\includegraphics[height=7.cm]{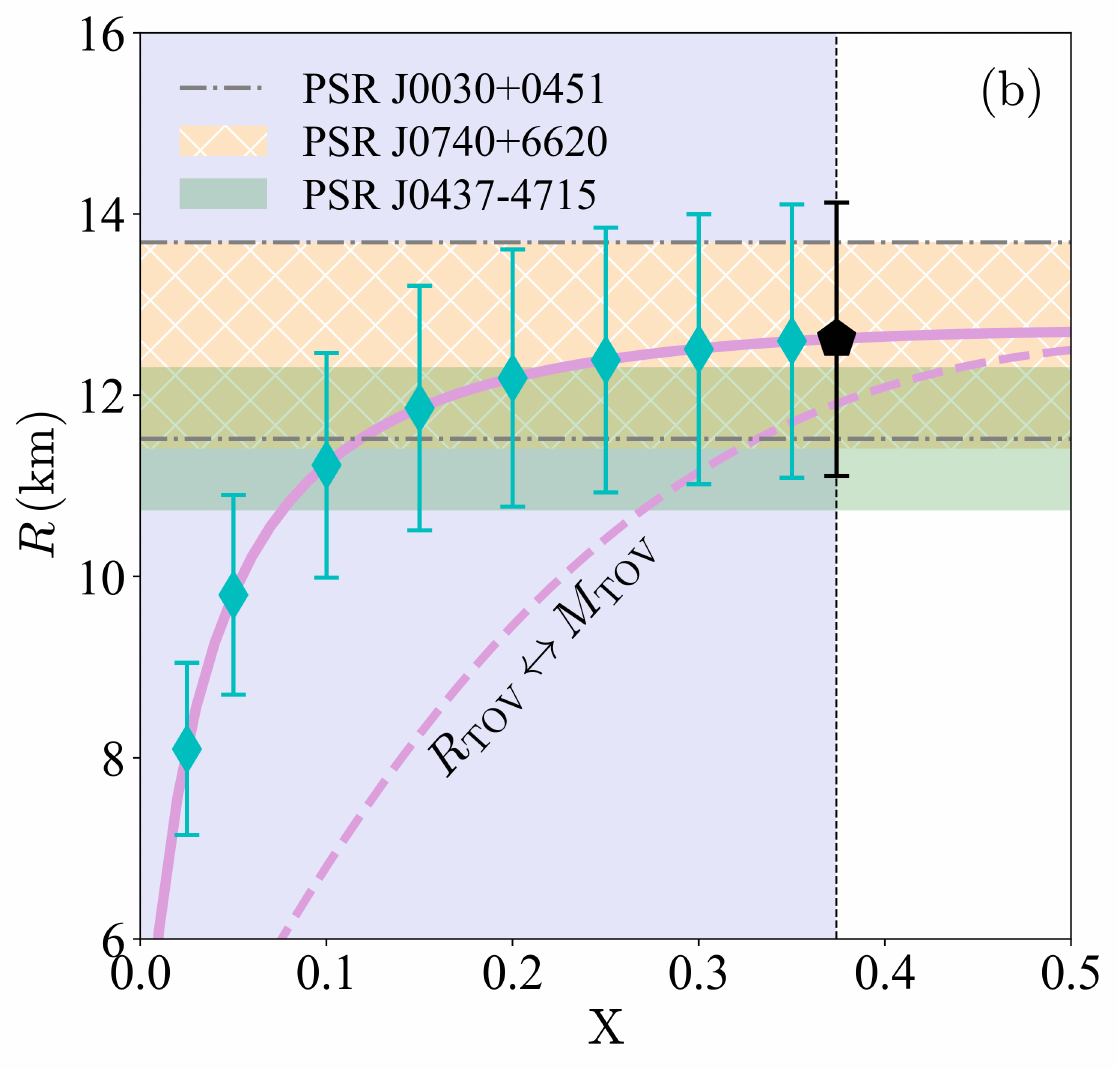}
\caption{(Color Online). Maximum masses for generally stable NSs and critically stable NSs (TOV configurations). The insert of the left panel plots the $\x$-dependence of $\Pi_{\rm{c}}$ from which we can clearly see that $\Pi_{\rm{c}}$ is upper bounded at $\x\approx0.374$.
}\label{fig_X-MR}
\end{figure}

How can we estimate an upper limit for $M_{\rm{NS}}$?
Firstly, we obtain the following condition by combining the mass and radius scalings of Eqs.\,(\ref{gk-m}) and (\ref{gk-r}),
\begin{equation}\label{Rlin}
\boxed{
R/\rm{km}\gtrsim3.59M_{\rm{NS}}/M_{\odot}+4.51,}
\end{equation}
this is similar to the Eq.\,(\ref{ci-0}) of $R/\rm{km}\gtrsim5.58M_{\rm{NS}}/M_{\odot}$ obtained via the compactness scaling of Eq.\,(\ref{gk-xi}) alone.
Therefore, the lower limit of $R$ is linearly proportional to NS mass $M_{\rm{NS}}/M_{\odot}$.
Next, combining all the three scalings of mass, radius and compactness enable us to write out,
\begin{equation}\label{rr1}
R/\rm{km}=\frac{(A_{\rm{M}}B_{\rm{R}}\Pi_{\rm{c}}-A_{\rm{R}}B_{\rm{M}})\Sigma }{(A_{\rm{M}}\Sigma -A_\xi A_{\rm{R}})\Pi_{\rm{c}}-A_{\rm{R}}B_\xi},
\end{equation}
here $\Sigma=M_{\odot}/\rm{km}$.
This expression for $R$ implies that it is upper bounded.
Specifically,  since $|A_{\rm{R}}B_{\rm{M}}|\ll |A_{\rm{M}}B_{\rm{R}}\Pi_{\rm{c}}|$ and $|A_{\rm{R}}B_\xi|\ll |(A_{\rm{M}}\Sigma-A_\xi A_{\rm{R}})\Pi_{\rm{c}}|$, we could approximate (\ref{rr1}) without loosing its main features:
\begin{empheq}[box=\fbox]{align}\label{rr2}
\frac{R\Sigma^{-1}}{\rm{km}}\approx&\underbrace{\frac{A_{\rm{M}}B_{\rm{R}}}{A_{\rm{M}}\Sigma -A_\xi A_{\rm{R}}}}_{\mbox{no $\Pi_{\rm{c}}$ factor} }\times\Bigg[1+\underbrace{
\overbrace{\left(\frac{A_{\rm{R}}B_\xi}{A_{\rm{M}}\Sigma -A_\xi A_{\rm{R}}}-\frac{A_{\rm{R}}B_{\rm{M}}}{A_{\rm{M}}B_{\rm{R}}}\right)}^{\mbox{negative:$-0.027\pm0.007$}}\frac{1}{\Pi_{\rm{c}}}}_{\mbox{upper bounded}}+\mathcal{O}\left(\frac{1}{\Pi_{\rm{c}}^2}\right)
\Bigg].
\end{empheq}
Since 
\begin{equation}
\left(\frac{A_{\rm{R}}B_\xi}{A_{\rm{M}}\Sigma -A_\xi A_{\rm{R}}}-\frac{A_{\rm{R}}B_{\rm{M}}}{A_{\rm{M}}B_{\rm{R}}}\right)<0,
\end{equation}
and $\Pi_{\rm{c}}^{-1}$ is lower bounded at $\x\approx0.374$,  the $R\Sigma^{-1}/\rm{km}$ is therefore upper bounded according to (\ref{rr2}).
Neglecting the correction in the square bracket of (\ref{rr2}), we have approximately the following upper bound
\begin{equation}\label{RRk}
\boxed{
    \frac{R\Sigma^{-1}}{\rm{km}}\approx\frac{A_{\rm{M}}B_{\rm{R}}}{A_{\rm{M}}\Sigma -A_\xi A_{\rm{R}}}.}
\end{equation}
\textcolor{xll}{These two facts\,---\,the lower limit of $R$ linearly increases with $M_{\rm{NS}}/M_{\odot}$ according to the Eq.\,(\ref{Rlin}) while $R$ is upper bounded by the Eq.\,(\ref{rr2})\,---\,together imply that $M_{\rm{NS}}/M_{\odot}$ is upper bounded, i.e., the $M_{\rm{NS}}$ could not increase without a limit,}
see FIG.\,\ref{fig_R2sc}.
The Eq.\,(\ref{rr2}) also explains the empirical finding that the radii of NSs with $M_{\rm{NS}}/M_{\odot}\approx1.8\pm0.4$ share similar values, i.e., the ``vertical'' shape on the M-R curve (see FIG.\,\ref{fig_MR-vert}, Eq.\,(\ref{fgk-2}) and the discussions around it earlier).
Similarly, we have
\begin{equation}
\boxed{
\left(\frac{M_{\rm{NS}}}{M_{\odot}}\right)=\frac{(A_\xi \Pi_{\rm{c}}+B_\xi)(A_{\rm{M}}B_{\rm{R}}\Pi_{\rm{c}}-A_{\rm{R}}B_{\rm{M}})}{( A_{\rm{M}}\Sigma -A_\xi A_{\rm{R}})\Pi_{\rm{c}}-A_{\rm{R}}B_\xi}
\approx\frac{A_\xi A_{\rm{M}}B_{\rm{R}}}{A_{\rm{M}}\Sigma-A_\xi A_{\rm{R}}}\Pi_{\rm{c}}\approx\left(\frac{R\Sigma^{-1}}{\rm{km}}\right)A_{\xi}\Pi_{\rm{c}},}
\label{mm1}
\end{equation}
here the second approximation follows under $B_{\rm{M}}\approx B_\xi\approx0$,  $|A_{\rm{R}}B_{\rm{M}}|\ll |A_{\rm{M}}B_{\rm{R}}\Pi_{\rm{c}}|$ as well as $|A_{\rm{R}}B_\xi|\ll |(A_{\rm{M}}\Sigma-A_\xi A_{\rm{R}})\Pi_{\rm{c}}|$.
We see from Eq.\,(\ref{RRk}) or Eq.\,(\ref{mm1}) {\color{xll}that the coefficients $A_{\rm{R}}$ and $B_{\rm{R}}$ (of Eq.\,(\ref{gk-r})) appear independently in the NS radius or mass expression. Therefore, the $B_{\rm{R}}$ is very relevant for estimating $R$ or the upper limit of $M_{\rm{NS}}$ (since the factor $\Pi_{\rm{c}}$ is upper bounded).}
Using the full expression of Eq.\,(\ref{mm1}) for NS mass and Eq.\,(\ref{rr1}) for the radius, we finally obtain the following two limits for NS masses and radii
\begin{equation}\label{EST-MR}
\boxed{
M_{\rm{NS}}/M_{\odot}\lesssim2.26\pm0.28,~~R/\rm{km}\lesssim12.62\pm1.51.}
\end{equation}
In the left panel of FIG.\,\ref{fig_X-MR}, we show the $\x$-dependence of $M_{\rm{NS}}/M_{\odot}$ of Eq.\,(\ref{mm1}) by the solid orange curve, here the solid magenta pentagon is the termination point at $\x\approx0.374$.
Similarly, we show in the right panel of FIG.\,\ref{fig_X-MR} the $\x$-dependence of the upper bound for $R$ of Eq.\,(\ref{rr1}) by the plum solid curve.

As an independent check of our method, we use the 104 EOSs (phenomenological/microscopic) of Ref.\,\cite{CLZ23-a} with $M_{\rm{NS}}^{\max}\equiv M_{\rm{TOV}}$ and $R_{\max}\equiv R_{\rm{TOV}}$ to determine/estimate the maximum mass configuration of NSs; the mass and radius scalings together with the regression coefficients are given in Eqs.\,(\ref{Mmax-G}) and (\ref{Rmax-n}), respectively.
Similarly, we can also do the compactness scaling using the same EOS samples, and the result is
\begin{equation}\label{xixi}
\xi_{\max}\equiv\xi_{\rm{TOV}}\approx A_\xi^{\max}\Pi_{\rm{c}}+B_\xi^{\max}
\approx2.59\Pi_{\rm{c}}-0.05.
\end{equation}
Then using the similar formulas as Eq.\,(\ref{mm1}) and Eq.\,(\ref{rr1}) for NSs at the TOV configuration, we have $
M_{\rm{TOV}}/M_{\odot}\approx2.28$ and $R_{\rm{TOV}}/\rm{km}\approx11.91$ under $\x\lesssim0.374$.
Since the number of EOSs used in Ref.\,\cite{CLZ23-a} is far less than $10^5$ used in our meta-modeling\,\cite{CL24-a},
the errorbars on the $M_{\rm{TOV}}$ and $R_{\rm{TOV}}$ are consequently large.
If we adopt the similar errorbar level as we estimate for the stable NS masses as in Eq.\,(\ref{EST-MR}), we may have
\begin{equation}\label{EST-MRTOV}
\boxed{
M_{\rm{TOV}}/M_{\odot}\approx 2.28\pm0.28,~~R_{\rm{TOV}}/\rm{km}\approx 11.91\pm1.51.}
\end{equation}
The regression (\ref{xixi}) implies that an upper limit about $0.283$ exists for $\xi_{\rm{TOV}}$, i.e., $\xi_{\rm{TOV}}\lesssim0.283$. The latter is highly consistent with the estimate (\ref{upp-xi}). This is not surprising since we used the same input information on the EOSs. For comparisons, added in the left panel of FIG.\,\ref{fig_X-MR} is the $\x$-dependence of $M_{\rm{TOV}}$ thus obtained by the dashed orange curve, and in the right panel the radius $R_{\rm{TOV}}$ by the dashed plum curve.
We see that the estimate $M_{\rm{TOV}}/M_{\odot}
\lesssim2.28$ (for the critically stable NSs) of (\ref{EST-MRTOV}) is consistent with $M_{\rm{NS}}/M_{\odot}\lesssim2.26$ (for general stable NSs) of (\ref{EST-MR}), as one expects since the latter should be smaller than the former.
In addition, it is interesting to notice that the $M_{\rm{TOV}}$ and $R_{\rm{TOV}}$ we estimated are quite consistent with the results from a very recent study\,\cite{Fan24} which predicted that $
M_{\rm{TOV}}/M_{\odot}\approx2.25_{-0.07}^{+0.08}$ and $R_{\rm{TOV}}/\rm{km}\approx11.90_{-0.60}^{+0.63}$ using two approaches: (1) modeling the mass distribution function of 136 NSs observed with a sharp cut-off at $2.28M_{\odot}$, and (2) Bayesian inference of the EOS from the currently available multi-messenger data of NSs considering constraints from the 
CEFT at low densities\,\cite{Essick2021} and the pQCD at very high densities. Incidentally, the NS mass cut-off they used is exactly the $M_{\rm{TOV}}$ we estimated above.

\begin{figure}[h!]
\centering
\includegraphics[width=16.cm]{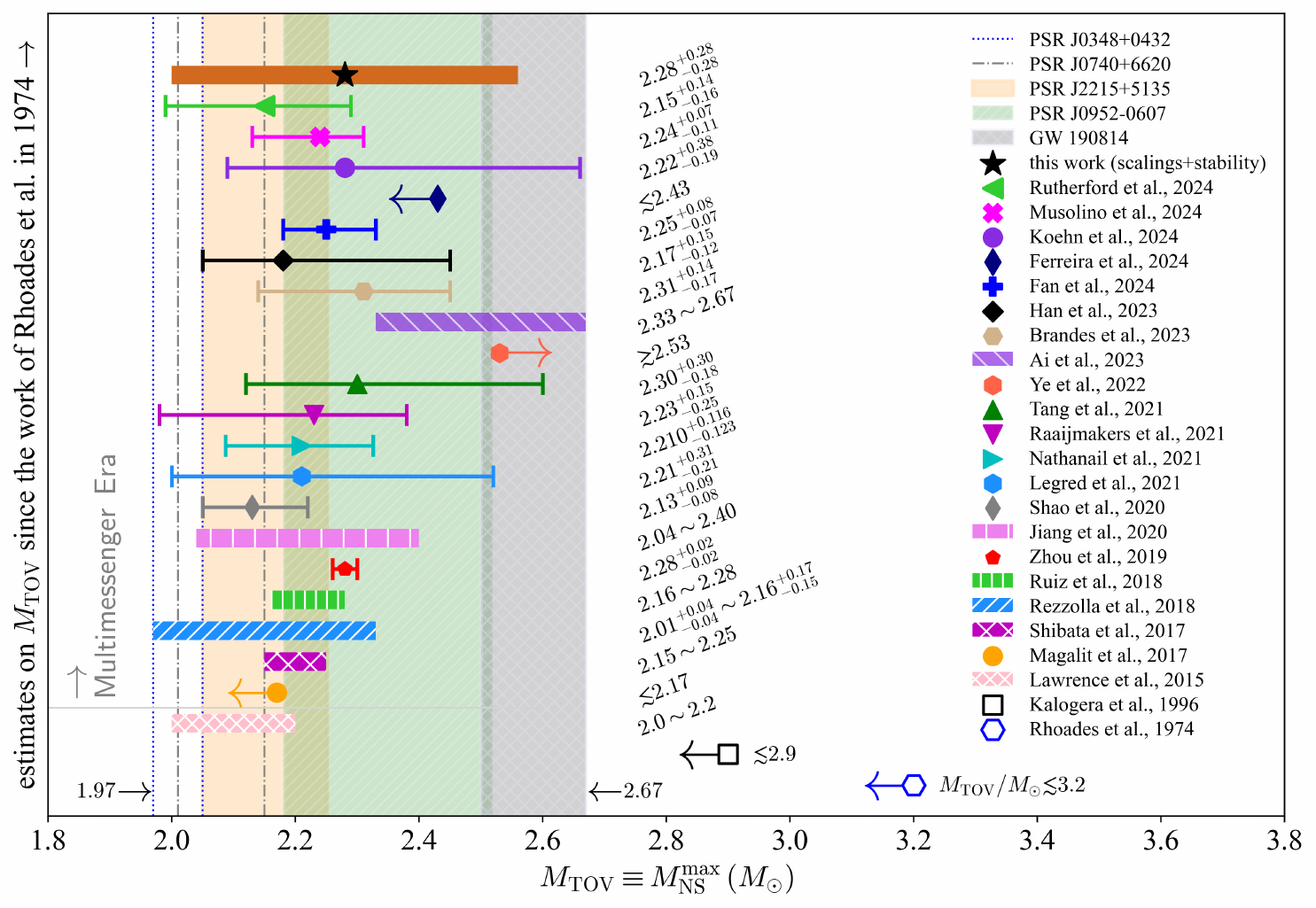}
\caption{(Color Online). A partial list of the estimated NS maximum-mass $M_{\rm{TOV}}$ of non-rotating NSs since 1974 after the seminal work of Rhoades and Ruffini\,\cite{Rho74}.
See the text for details on these constraints.
}\label{fig_MTOV-STAT}
\end{figure}

Estimating accurately the value of $M_{\rm{TOV}}$ has been an important and challenging task in NS physics since the seminal work of Rhoades and Ruffini in 1974\,\cite{Rho74} when they predicted that $M_{\rm{TOV}}\lesssim3.2M_{\odot}$. Since then a lot of works have been devoted to carrying out this task using various theories, models and data. Summarized in FIG.\,\ref{fig_MTOV-STAT} is a (partial) list of the available estimates of $M_{\rm{TOV}}$. In particular, besides Ref.\,\cite{Rho74},  Ref.\,\cite{Kalogera96} also obtained a secure upper bound $M_{\rm{TOV}}\lesssim2.9M_{\odot}$ based on the knowledge of nuclear EOS valid up to about $2\rho_0$. Similarly, Ref.\,\cite{Lawrence2015} gave $2.0\lesssim M_{\rm{TOV}}/M_{\odot}\lesssim2.2$ considering Gamma-ray bursts constraints.
Since the beginning of the multi-messenger era marked by the detection of gravitational waves from the NS-NS merger in GW170817\,\cite{Abbott2017,Abbott2018} in 2017 as well as the observations of massive NSs\,\cite{Ant13,Fon21,Sul24,Romani22}, more independent and novel investigations on $M_{\rm{TOV}}$ emerged. Naturally, the $M_{\rm{TOV}}$ is required to be at least as large as the mass of the currently observed most massive NS. Specifically, other estimates listed in FIG.\,\ref{fig_MTOV-STAT} are from the following studies: using the signals from GW170817\,\cite{Abbott2018}, the $M_{\rm{TOV}}$ was predicted to be about $2.01_{-0.04}^{+0.04}\lesssim M_{\rm{TOV}}/M_{\odot}\lesssim2.16_{-0.15}^{+0.17}$\,\cite{Rezzolla2018}; similarly Ref.\,\cite{Margalit2017} showed it should be $\lesssim2.17M_{\odot}$ by including the electromagnetic information; a consistent prediction of $2.16\lesssim M_{\rm{TOV}}/M_{\odot}\lesssim 2.28$ through general relativistic magnetohydrodynamics simulations was given in Ref.\,\cite{Ruiz2018}; Ref.\,\cite{Shibata2017} constrained $2.15\lesssim M_{\rm{TOV}}/M_{\odot}\lesssim2.25$ considering the absence of optical counterparts associated with relativistic ejecta; Ref.\,\cite{ZhouY2019} predicted a quite narrow $M_{\rm{TOV}}/M_{\odot}\approx2.28_{-0.02}^{+0.02}$ by constructing nuclear EOSs consistent with low-density microscopic calculations and heavy-ion collision data based on a non-relativistic energy density functional; 
Ref.\,\cite{Jiang2020ApJ} found $2.04\lesssim M_{\rm{TOV}}/M_{\odot}\lesssim2.4$ using the observational data of PSR J0030+0451 and of GW170817 while Ref.\,\cite{Shao2020} predicted $M_{\rm{TOV}}/M_{\odot}\approx2.13_{-0.08}^{+0.09}$ by including extra information from GRB 170817A/AT2017gfo; Ref.\,\cite{Leg21} similarly obtained $M_{\rm{TOV}}/M_{\odot}\approx2.21_{-0.21}^{+0.31}$ using a nonparametric EOS model constructed with astrophysical constraints; Ref.\,\cite{Nathanail2021} restricted the maximum mass to about $M_{\rm{TOV}}/M_{\odot}\approx2.210_{-0.123}^{+0.116}$ to be consistent with existing astrophysical/nuclear constraints although the GW190814 was studied in their inference; 
similarly Ref.\,\cite{Raaij2021} constrained $M_{\rm{TOV}}/M_{\odot}\approx2.23_{-0.25}^{+0.15}$ by analyzing the radio mass measurement of
PSR J0740+6620, the joint mass-radius estimate of PSR J0740
+6620, and combining GW170817, GW190425, AT2017gfo,
PSR J0740+6620 as well as PSR J0030+0451; Ref.\,\cite{Tang2021} adopted similar information as in Ref.\,\cite{Raaij2021} and found  $M_{\rm{TOV}}/M_{\odot}\approx2.30_{-0.18}^{+0.30}$ and Ref.\,\cite{Han2023} predicted $M_{\rm{TOV}}/M_{\odot}\approx2.17_{-0.12}^{+0.15}$; considering the GW190814 led Ref.\,\cite{Ye2022} to predict that $M_{\rm{TOV}}/M_{\odot}\gtrsim2.53$ and similarly Ref.\,\cite{Ai2023} found $2.33\lesssim M_{\rm{TOV}}/M_{\odot}\lesssim2.67$ at $1\sigma$ confidence level; Ref.\,\cite{Brandes2023-a} found $M_{\rm{TOV}}/M_{\odot}\approx2.31_{-0.17}^{+0.14}$ by incorporating the data of PSR J0952-0607; very recently Ref.\,\cite{Fan24} updated the constraint of Ref.\,\cite{Han2023} to be  $M_{\rm{TOV}}/M_{\odot}\approx2.25_{-0.07}^{+0.08}$ (with a precision about 3\%); 
in Ref.\,\cite{Ferr24} the authors constrained the maximum mass to be $\lesssim2.43M_{\odot}$ by restricting the slope $\d M_{\rm{NS}}/\d R$ to non-negative values in a certain mass range;
Ref.\,\cite{Mus24} found $M_{\rm{TOV}}/M_{\odot}\approx2.24_{-0.11}^{+0.07}$ by using a quasi-universal maximum-mass ratio between rotating and non-rotating stars;
Ref.\,\cite{Koehn2024} and Ref.\,\cite{Ruther2024} updated the constraints on $M_{\rm{TOV}}$ by including data of the newly announced NICER measurement of PSR J0437-4715\,\cite{Choud24} and found $M_{\rm{TOV}}/M_{\odot}\approx2.22_{-0.19}^{+0.38}$ and $M_{\rm{TOV}}/M_{\odot}\approx2.15_{-0.16}^{+0.14}$ (using a PP model), respectively;
Ref.\,\cite{Sneppen2024} restricted the NS EOS by using a method based on the abundance of helium to constrain the lifetime of NS merger remnants and found that $M_{\rm{TOV}}/M_{\odot}\lesssim2.3$.

Without any surprise, {\color{xll}all the constraints on $M_{\rm{TOV}}$ fall safely within the range $1.97M_{\odot}\lesssim M_{\rm{TOV}}\lesssim2.67M_{\odot}$, set by the lower limit of observed mass of PSR J0348+0432 (about $1.97M_{\odot}$) and the upper limit on the observed mass of the secondary component of GW190814 (about $2.67M_{\odot}$ at 68\% confidence level).}
Returning to our estimate of $M_{\rm{TOV}}$ in this work, its uncertainty ($\pm0.28$) is still quite large. In order to refine the prediction on $M_{\rm{TOV}}$ in our approach, the scalings may be improved by incorporating more terms in the expansions of $\hP$, $\heps$ and $\hM$ over $\hr$ (the current radius scaling is obtained by truncating the pressure as $\hP(R)\approx\x+b_2\hr^2=0$).
These improved/refined scalings should necessarily be verified by existing nuclear EOS models.

\setcounter{equation}{0}
\section{Strong-field Gravity in GR and Peaked Sound Speed Squared (SSS) in Massive NSs}\label{SEC_7}

In this section, we discuss in more details the possible peaked behavior of the SSS ($s^2$) in NSs, extending the brief review given in SECTION \ref{SEC_2}. Presently, this is quite a hot topic as the fundamental physics underlying the density or radius profile of SSS in NSs, especially whether it has a peak caused by what physics and where its peak is located, is still under debate. We first give in Subsection \ref{sub_s2ORDER} an order-of-magnitude analysis on the $s^2$ and the possible emergence of a peak in its density profile. Subsection \ref{sub_s2_bridge} is denoted to the role played by $s^2$ as a bridge linking the superdense NS matter and the strongly curved geometry in NSs. We then discuss analytically in Subsection \ref{sub_decomTA} a useful connection between the trace anomaly $\Delta$ and $s^2$.
The monotonic density profile of $s^2$ in Newtonian stars is studied in Subsection \ref{sub_s2_Newtonian}.
Subsection \ref{sub_VioCB} is devoted to study the violation of the conformal bound on $s^2$ for NSs at TOV configuration; the concept of average SSS will also be introduced in this subsection.
The central SSS for generally stable NSs, particularly that for canonical NSs, is investigated in Subsection \ref{sub_s2canon}.
Combining these two subsections, we conclude that the conformal bound for $s^2$ violates for both TOV NSs and generally stable ones. In Subsections \ref{sub_s2_1st} and \ref{sub_s2_2nd}, we give the first-order and the higher-order approximations of the $\mu$-expansion of $s^2$, respectively. The former subsection mainly deals with the issue of whether $s^2(\heps)$ is larger or smaller than its central value $s_{\rm{c}}^2$, while the latter one investigates where the possible peak in the radial profile of $s^2$ is located.
The peak in the derivative part of a $s^2$ decomposition is also given in Subsection \ref{sub_s2_2nd}.
Subsection \ref{sub_s2_peak} is devoted to a unified investigation on how the strong-field gravity may extrude a peak in the $s^2$ profile.
In Subsection \ref{sub_s2Esym}, we discuss the connection between the peaked $s^2$ in NSs and the high-density behavior of nuclear symmetry energy.
Finally, in Subsection \ref{sub_DenseQCD} we briefly discuss the emergence of a peaked $s^2$ profile in other dense systems, generalizing the results we have obtained for NSs.
As the trace anomaly $\Delta$ and SSS are closely connected, a few related issues to $s^2$ would be given in SECTION \ref{SEC_56} after we review the relevant status on $\Delta$.

\subsection{Qualitative features of SSS in stellar matter from order-of-magnitude analyses}\label{sub_s2ORDER}

Because $s^2$ is dimensionless (adopting units of $c=1$) and $P$ and $\varepsilon$ have the same dimensions, dimensional consideration implies that $s^2$ of Eq.\,(\ref{def-s2}) could be written generally as
\begin{equation}\label{s2_dimensionless}
\boxed{
s^2\equiv{\d P}/{\d\varepsilon}=\d\hP/\d\heps=\phi f(\phi),~~\phi=P/\varepsilon,}
\end{equation}
with $f(\phi)$ a dimensionless function depending on $\phi$ only.
A factor $\phi$ is separated out to the front in the above expression considering the fact that $s^2\to0$ if no matter exists $\phi\to0$.
The function $f$ can be expanded as a power series around $\phi\approx0$ (vacuum state) as 
\begin{equation}\label{s2_dim2}
f\approx f_0+f_1\phi+f_2\phi^2+\cdots,
\end{equation} with $f_0>0$ (required by the stability condition $s^2\geq0$).
Obviously, the Eq.\,(\ref{sc2-TOV}) has the form of (\ref{s2_dimensionless}).
Considering stars as white dwarfs (WDs), one has $P\lesssim 10^{22\mbox{-}23}\,\rm{dynes}/\rm{cm}^2\approx10^{-(11\mbox{-}10)}\,\rm{MeV}/\rm{fm}^3$ and $\varepsilon\lesssim10^{8\mbox{-}9}\,\rm{kg}/\rm{m}^3\sim10^{-6}\,\rm{MeV}/\rm{fm}^3$, thus $\phi\lesssim10^{-(5\mbox{-}4)}$. 
The typical radii of WDs are around $R\approx10^4\,\rm{km}$, so $\widehat{R}=R/Q\approx0.05$ and $\widehat{R}^2\approx3\times10^{-3}$.
In fact, the ratio $\phi$ of pressure over energy density could be even smaller for main-sequence stars like the sun. 
Specifically, the pressure and energy density in the solar core are about $10^{-16}\,\rm{MeV}/\rm{fm}^3$ and $10^{-10}\,\rm{MeV}/\rm{fm}^3$\,\cite{CL24-a}, respectively, and therefore $\phi\approx 10^{-6}$.
{\color{xll}These stars are Newtonian in the sense that GR effects are almost absent.}
Similarly, for NS matter around nuclear saturation density $\rho_0=\rho_{\rm{sat}}\approx0.16\,\rm{fm}^{-3}$, one estimates  the pressure as,
\begin{equation}
P(\rho_0)\approx P_0(\rho_0)+P_{\rm{sym}}(\rho_0)\delta^2
\approx3^{-1}L\rho_0\delta^2\lesssim3\,\rm{MeV}/\rm{fm}^3,
\end{equation}
using $L\approx60\,\rm{MeV}$ and the fact $\delta^2\lesssim1$.
Moreover, $P_0(\rho_0)$ is the pressure of SNM at $\rho_0$, which is zero by definition.
The energy density is similarly estimated as $
\varepsilon(\rho_0)\approx[E_0(\rho_0)+E_{\rm{sym}}(\rho_0)\delta^2+M_{\rm{N}}]\rho_0\approx150\,\rm{MeV}/\rm{fm}^3$, where $M_{\rm{N}}\approx939\,\rm{MeV}$ is the static mass of a nucleon, and $E_0(\rho_0)\approx-16\,\rm{MeV}$ and $E_{\rm{sym}}(\rho_0)\approx32\,\rm{MeV}$ are used; therefore $\phi\lesssim0.02$.
For reference, at $\rho\approx2\rho_0$, we can estimate $P(2\rho_0)\lesssim 20\,\rm{MeV}/\rm{fm}^3$ and $\varepsilon(2\rho_0)\approx300\,\rm{MeV}/\rm{fm}^3$, and thus $\phi\lesssim0.067$ by using Eq.\,(\ref{fE0}) and Eq.\,(\ref{fEsym})  and $S\approx32\,\rm{MeV}$, $ L\approx60\,\rm{MeV}$, $K_{\rm{sym}}\approx-100\,\rm{MeV},J_{\rm{sym}}\approx200\,\rm{MeV}$ and $K_0\approx230\,\rm{MeV}$ together with $J_0\approx-300\,\rm{MeV}$ to make the estimate.
In fact, the effect of $J_{\rm{sym}}$ for estimating $P/\varepsilon$ at $2\rho_0$ is $\lesssim2\%$.

\begin{figure}[h!]
\centering
\includegraphics[width=10.cm]{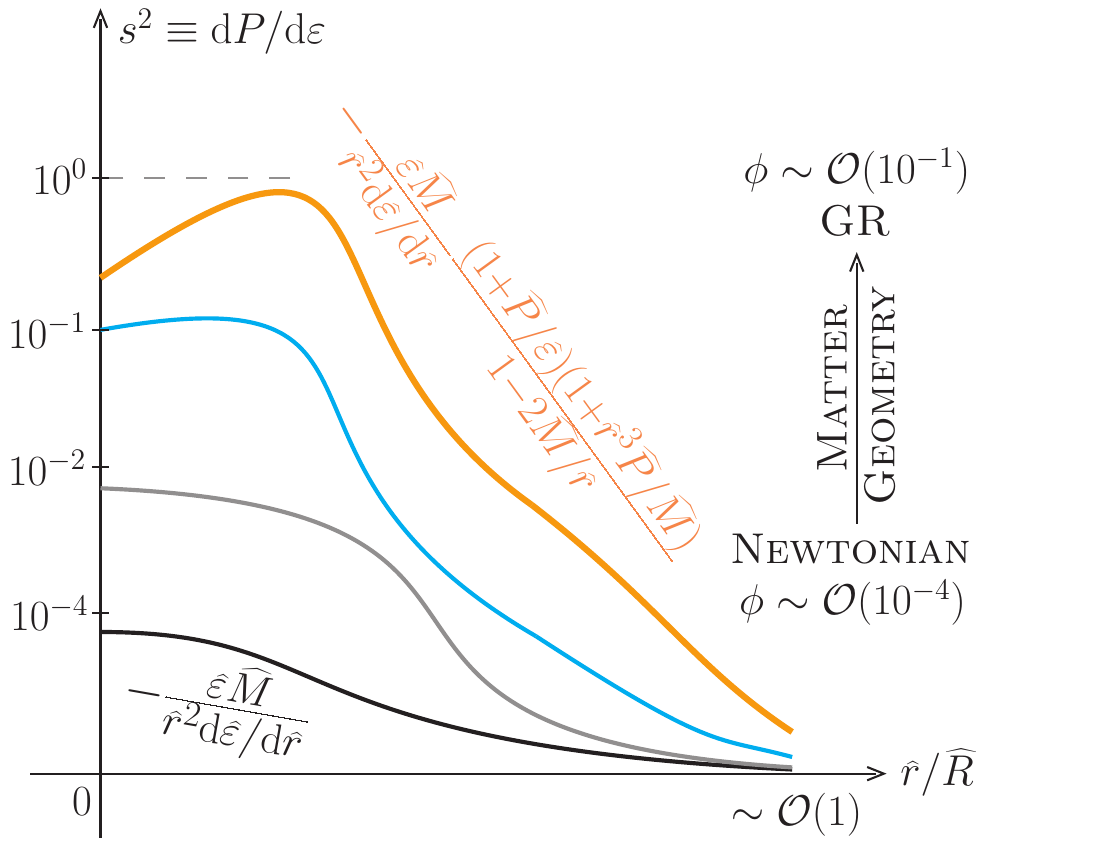}
\caption{(Color Online). A sketch of the evolution of radial profile of $s^2$ in NSs: at low densities characterized by small $\phi=P/\varepsilon$ (Newtonian limit), the $s^2$ monotonically decreases from the center to the surface (black line),  and as the $\phi$ increases approaching the GR case (orange line) a peak eventually emerges near the NS center.
The general expressions for $s^2$ are also captioned near the curves,  $c=1$ is adopted. Figure taken from Ref.\,\cite{CL24-a}.
}\label{fig_s2_GRNewt}
\end{figure}

For these systems one can safely neglect the terms $f_1\phi+f_2\phi^2+\cdots$ in the expansion of $f$, therefore $s^2\approx f_0\phi$ or equivalently $\hP=C_0\heps^{f_0}$ is obtained\,\cite{Shapiro1983}, here $0<C_0\ll\mathcal{O}(1)$ is an integration constant, indicating $s^2$ is probably an increasing function of $\heps$. The absence of a peak in $s^2$ in NSs around $\rho_{\rm{sat}}$ was confirmed by the CEFT\,\cite{Tews2018,Essick2021}, e.g., Ref.\,\cite{Essick2021} predicted that $s^2$ is monotonic around $\rho_{\rm{sat}}$ and $s^2_{\rm{sat}}\equiv s^2(\rho_{\rm{sat}})\approx0.03\sim\phi$ and similarly Ref.\,\cite{Tews2018} found that $0.03\lesssim s^2_{\rm{sat}}\lesssim0.05\sim\phi$ near $\rho_{\rm{sat}}$. 
These studies also indicate that 
\begin{equation}\label{f012}
f_0\approx s^2/\phi\gtrsim1\sim2.\end{equation}
In fact, we may argue $f_0\approx2$ by using the trace anomaly decomposition formula (see relation (\ref{s2_on_phi})).
On the other hand,  $\phi$ could be sizable $\gtrsim\mathcal{O}(0.1)$ in massive NSs especially in their cores, see TAB.\,\ref{tab_cEOS}, and possible peaks in $s^2$ may emerge.
In this sense, massive NSs like PSR J0740+6620\,\cite{Fon21,Riley21,Miller21,Salmi22}, PSR J2215+5135\,\cite{Sul24},  PSR J1614-2230\,\cite{Dem10} and PSR J0348+0432\,\cite{Ant13} provide us excellent opportunities to study the possible peaked structure in $s^2$ profiles.
See FIG.\,\ref{fig_s2_GRNewt} for a sketch of the $s^2$ radial profile evolved with $\phi$.

Consider the next-order correction $f_1\phi^2$ to the SSS, i.e., $s^2\approx f_0\phi+f_1\phi^2$, and perturb the solution $\hP=C_0\heps^{f_0}$ to $\hP\approx C_0\heps^{f_0}[1+\psi(\heps)]$ with $\psi(\heps)$ a small correction (since $\phi^2\ll1$ is small).
The correction $\psi(\heps)$ could be obtained:
\begin{equation}
\psi(\heps)\approx C_1+C_0\frac{f_1}{f_0-1}\heps^{f_0-1},
\end{equation}
the integration constant $C_1=0$ since $\psi(\heps)\to0$ if $\heps\to0$.
After that we obtain the $s^2$ containing the first-order correction in $\psi(\heps)$ via $s^2\approx f_0\phi$ and $\phi\approx C_0\heps^{f_0-1}[1+\psi(\heps)]$; and an extreme point $\heps_{\ast}$ exists if $f_1<0$:
\begin{equation}
\boxed{
\heps_{\ast}\approx\left(-\frac{2C_0f_1}{f_0-1}\right)^{-\frac{1}{f_0-1}},~~
s^2(\heps_{\ast})\approx C_0f_0\heps_{\ast}^{f_0-1}\left(1+\frac{C_0f_1}{f_0-1}\heps_{\ast}^{f_0-1}\right)=\frac{f_0(1-f_0)}{4f_1},~~f_0>1,}
\end{equation}
using $[{\d s^2}/{\d\heps}]_{\heps_{\ast}}=0$.
The condition $0\leq s^2(\heps_{\ast})\leq1$ implies $4^{-1}f_0(1-f_0)-1\lesssim f_1\lesssim 4^{-1}f_0(1-f_0)$.
Furthermore, it is a local maximum point since $[\d^2s^2/\d\widehat{\varepsilon}^2]_{\heps_{\ast}}$ is negative:
\begin{equation}
\left[\frac{\d^2 s^2}{\d\heps^2}\right]_{\heps_{\ast}}\approx C_0f_0\heps_{\ast}^{f_0-3}\left[f_0^2-3f_0+2+4C_0\left(f_0-\frac{3}{2}\right)f_1\heps_{\ast}^{f_0-1}\right]
=-C_0f_0\left(f_0-1\right)^2\left(-\frac{2C_0f_1}{f_0-1}\right)^{-\frac{f_0-3}{f_0-1}}<0.
\end{equation}
Using the range for $f_1$ limited just above, we have then
\begin{equation}
-\left(\frac{1}{2}\right)^{\frac{f_0-3}{1-f_0}}\left(C_0f_0\right)^{\frac{2}{f_0-1}}(1-f_0)^2
\lesssim
\left[\frac{\d^2 s^2}{\d\heps^2}\right]_{\heps_{\ast}}\lesssim-C_0f_0(1-f_0)^2\left(\frac{1}{2}\frac{1}{f_0-1}\right)^{\frac{f_0-3}{1-f_0}};
\end{equation}
and the second-order derivative would be negatively small ($\gtrsim-10^{-2}$), indicating the peak might be wide near $\heps_{\ast}$.
On the other hand, the $s^2$ is monotonically increasing function of $\heps$ if $f_1>0$.

\subsection{SSS as a bridge connecting superdense matter and strongly curved geometry in and around NSs}\label{sub_s2_bridge}

Using the dimensionless TOV equations of (\ref{TOV-ds}), the SSS $s^2$ can be written as
\begin{align}\label{st_s2}
\textsc{GR}:\;s^2
=\frac{\d\widehat{P}}{\d\widehat{\varepsilon}}
=-\frac{\widehat{\varepsilon}\widehat{M}}{\widehat{r}^2\d\widehat{\varepsilon}/\d\widehat{r}}\frac{(1+\widehat{P}/\widehat{\varepsilon})(1+\widehat{r}^3\widehat{P}/\widehat{M})}{1-2\widehat{M}/\widehat{r}}.
\end{align}
It shows that {\color{xll}not only dense NS matter {(characterized by $\widehat{P}$, $\widehat{\varepsilon}$ as well as their ratio $\widehat{P}/\widehat{\varepsilon}$)} but also the strongly curved geometry {(characterized by the factor $1-2\widehat{M}/\widehat{r}$)} could affect $s^2$.}
{In this sense, {\color{xll}the speed of sound acts as a bridge connecting properties of supra-dense matter with curved geometry of compact/massive NSs.}}
On the other hand, if one starts directly from the Newtonian evolution equation $\d\widehat{P}/\d\widehat{r}=-\widehat{\varepsilon}\widehat{M}/\widehat{r}^2$ without incorporating the SR and GR effects discussed at the beginning of SECTION \ref{SEC_2}\,\cite{Chan10-a}, we have
\begin{equation}\label{s2-N}
\textsc{Newtonian}:\;s^2
=\frac{\d\widehat{P}}{\d\widehat{\varepsilon}}
=-\frac{\widehat{\varepsilon}\widehat{M}}{\widehat{r}^2\d\widehat{\varepsilon}/\d\widehat{r}}.
\end{equation}
Because there is no factor of $\widehat{P}$ on the right side, i.e., the pressure (matter effect) could not affect the $s^2$ explicitly (though $\widehat{\varepsilon}$ contains implicitly the effects of $\widehat{P}$).
Moreover, since $1+\widehat{P}/\widehat{\varepsilon}>1$, $1+\widehat{r}^3\widehat{P}/\widehat{M}>1$ and $0<1-2\widehat{M}/\widehat{r}<1$, we have $(1+\widehat{P}/\widehat{\varepsilon})(1+\widehat{r}^3\widehat{P}/\widehat{M})/(1-2\widehat{M}/\widehat{r})>1$, i.e., $s^2$ in NSs with GR effects is enhanced compared with its Newtonian limit.
In addition, because $s^2\leq1$ due to the principle of causality, it could not always increase (even if matter/geometry effects are strong enough), implying that in certain circumstances $s^2$ may either increase or decrease with decreasing $\widehat{r}$ (or increasing $\widehat{\varepsilon}$) when going into the stellar core.

{After this general discussion on the evolutionary behavior of $s^2$, we now figure out under which circumstance a possible peak may emerge in $s^2$ density/radial profile.}
By inserting the perturbative expansions of $\widehat{P}$, $\widehat{\varepsilon}$ and $\widehat{M}$ into Eq.\,(\ref{st_s2}), one can obtain an expression for $s^2$ similar to Eq.\,(\ref{zf-3}) for the core EOS.
In particular, we have to order $\widehat{r}^2$,
\begin{equation}
\boxed{
 s^2\approx s_{\rm{c}}^2+l_2\widehat{r}^2,
~~
 l_2=\frac{2s_{\rm{c}}^2}{b_2}\left(b_4-s_{\rm{c}}^2a_4\right),}
\end{equation}
see Eq.\,(\ref{s2_r_exp}) for the $\hr$-expansion of $s^2$.
Because $s_{\rm{c}}^2>0$ (see Eq.\,(\ref{sc2-TOV})) and in order for $s^2$ to obtain a peak near $\widehat{r}=0$, it is necessary that $l_2>0$ and therefore $s^2(\hr)>s_{\rm{c}}^2$.
{Since $b_2=-6^{-1}(1+3{\x}^2+4{\x})<0$, see Eq.\,(\ref{ee-b2}),
the condition $l_2>0$ is equivalent to $b_4-s_{\rm{c}}^2a_4<0$,
or $a_4>b_4/s_{\rm{c}}^2>0$. By using the expression of $b_4$ given in Eq.\,(\ref{ee-b4}) which involves the coefficient $b_2$ of Eq.\,(\ref{ee-b2}), we then obtain an equivalent condition of $l_2>0$ as,
}
\begin{align}\label{def_a4ineq}
a_4>\frac{1}{12}\frac{1+3{\x}^2+4{\x}}{s_{\rm{c}}^2}\left({\x}
+\frac{4+9{\x}}{15s_{\rm{c}}^2}\right)\approx\frac{1}{80{\x}^2}\left(1
+\frac{17}{4}{\x}+\frac{9}{2}{\x}^2-\frac{13}{4}{\x}^3
-\frac{49}{2}{\x}^4+\cdots\right).
\end{align}
Besides,  $a_4$ should fulfill some extra general constraints: 
\begin{enumerate}[label=(\alph*)]
\item The decreasing of $\widehat{\varepsilon}\approx1+a_2\widehat{r}^2+a_4\widehat{r}^4$ with respective to $\widehat{r}$ (i.e., $\d\widehat{\varepsilon}/\d\widehat{r}<0$) requires necessarily $a_4<-a_2/2\widehat{R}^2$.
\item The increasing of $\widehat{M}$ with respect to $\widehat{r}$ (i.e., $\d\widehat{M}/\d\widehat{r}=\widehat{r}^2\widehat{\varepsilon}>0$) gives the criterion $a_4\widehat{R}^4>-1-a_2\widehat{R}^2$.
\end{enumerate}
Here $\widehat{R}\sim\mathcal{O}(1)$,  and $a_2=b_2/s_{\rm{c}}^2<0$ is fixed at a certain ${\x}$.
Meeting the above three criteria, namely inequality (\ref{def_a4ineq}), $a_4<-a_2/2\widehat{R}^2$ and $a_4\widehat{R}^4>-1-a_2\widehat{R}^2$ guarantees a peak in $s^2$ at some $\widehat{r}\neq0$ with certain ranges of ${\x}$, i.e.,
\begin{empheq}[box=\fbox]{align}\label{NS_cond}
&\mbox{Necessary and sufficient conditions for the appearance of a peak in $s^2$: }\notag\\
&\hspace*{3.cm}\frac{1}{12}\frac{1+3{\x}^2+4{\x}}{s_{\rm{c}}^2}\left({\x}
+\frac{4+9{\x}}{15s_{\rm{c}}^2}\right)\lesssim a_4\lesssim\frac{1}{\widehat{R}^4}
 \mbox{ with $\widehat{R}\sim\mathcal{O}(1)$}.
\end{empheq}
Moreover, the peaked profile of  $s^2(\hr)$ could easily be mapped onto that on $s^2(\widehat{\varepsilon})$ or $s^2(\rho)$.
In addition,  the necessity of a positive $a_4$ can be understood as follows.
Since a positive $b_4$ (of Eq.\,(\ref{ee-b4})) slows down the decrease of $\widehat{P}$ (due to $b_2<0$) when going out from the center; for the $s^2$ to be larger than $s_{\rm{c}}^2$, a positive $a_4$ is necessary to slow down the decrease of $\widehat{\varepsilon}$ (due to $a_2<0$) too since approximately 
\begin{equation}\label{s2app}
s^2\approx{\Delta \widehat{P}}/{\Delta\widehat{\varepsilon}},
\end{equation} obtained by two nearby points on the EOS curve.
In Subsections \ref{sub_s2_1st} and \ref{sub_s2_2nd},  we carry out detailed calculations of the $s^2$ profile to verify the qualitative analysis on the peaked behavior of $s^2$.
In particular, we find that $l_2<0$ holds (as in Eq.\,(\ref{for_s2Newt}) and Eq.\,(\ref{for_s2Newt-1})) in the Newtonian limit even if $a_4$ is positive, {i.e., no peak would emerge in $s^2$ profiles for Newtonian stars.} We thus conclude that it is the GR effect that extrudes a peak in the $s^2$ profile.

\begin{figure}[h!]
\centering
\includegraphics[height=7.cm]{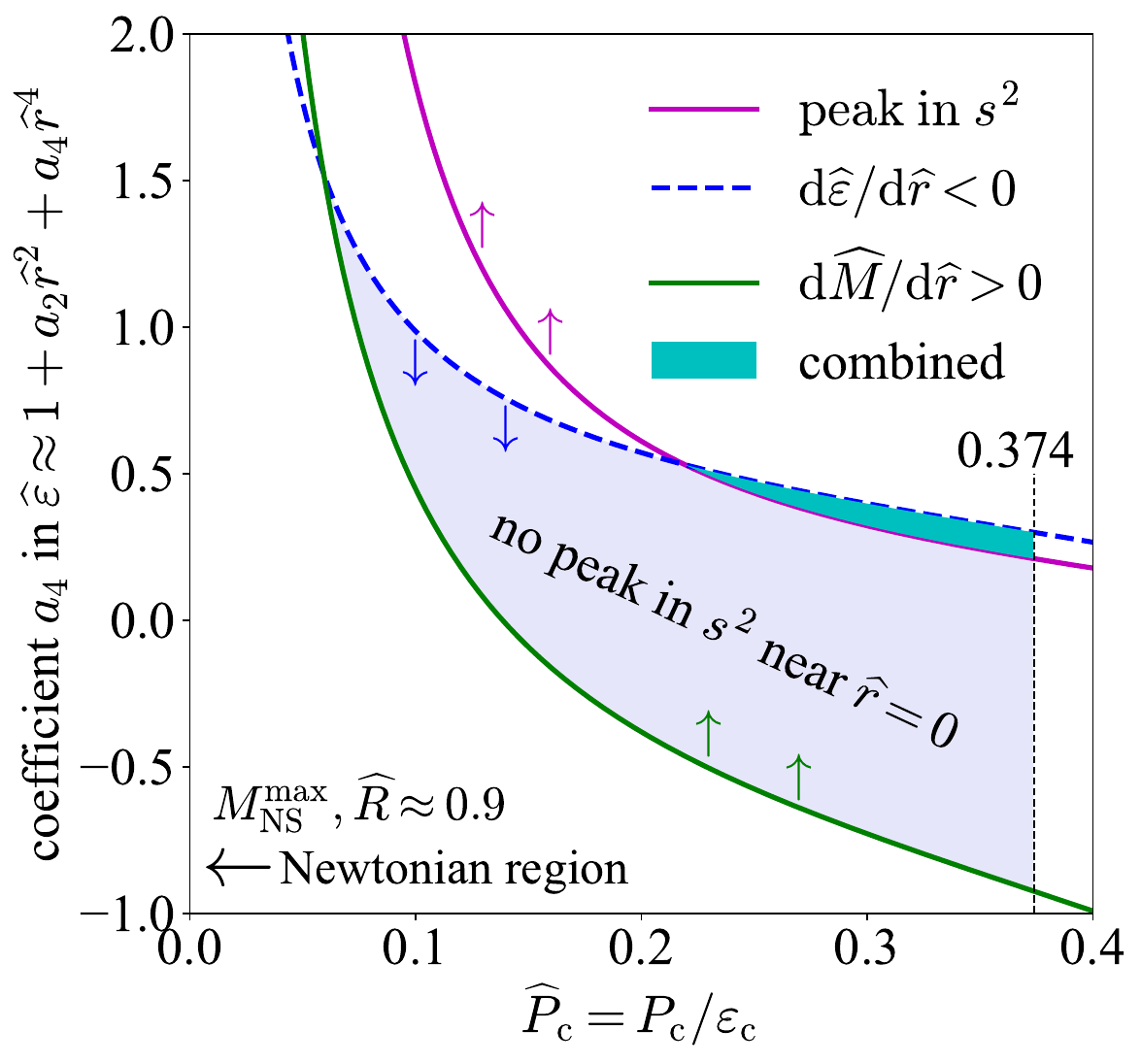}\qquad
\includegraphics[height=7.cm]{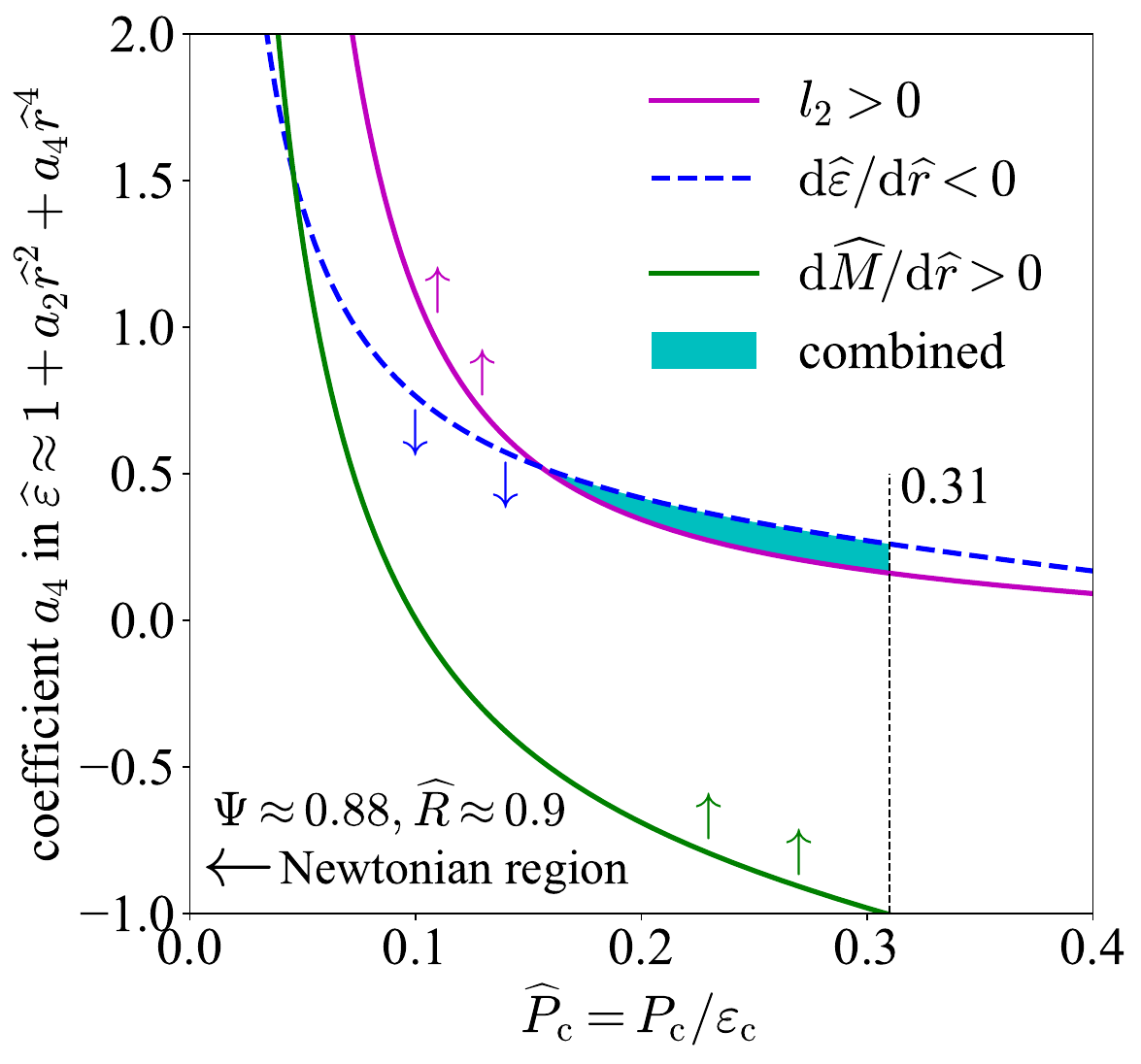}
\caption{(Color Online).  Left panel: combined region for $a_4$ (cyan band) for the onset of a peaked profile of $s^2$ {under the inequality (\ref{def_a4ineq}) and the criteria (a) and (b)},  here $\widehat{R}\approx0.9$ is adopted for illustrations.
The vertical dashed line marks ${\x}\approx 0.374$ by setting $s_{\rm{c}}^2$ of Eq.\,(\ref{sc2-TOV}) $\leq1$, $l_2$ appears in $s^2\approx s_{\rm{c}}^2+l_2\widehat{r}^2+\cdots$.
Right panel: same as the left panel but for NS configurations climbing the M-R curve (instead of being at the TOV configuration), here $\Psi\approx0.88$ is adopted. Figures taken from Ref.\,\cite{CL24-a}.
}\label{fig_a4peak}
\end{figure}

\begin{figure}[h!]
\centering
\includegraphics[height=6.5cm]{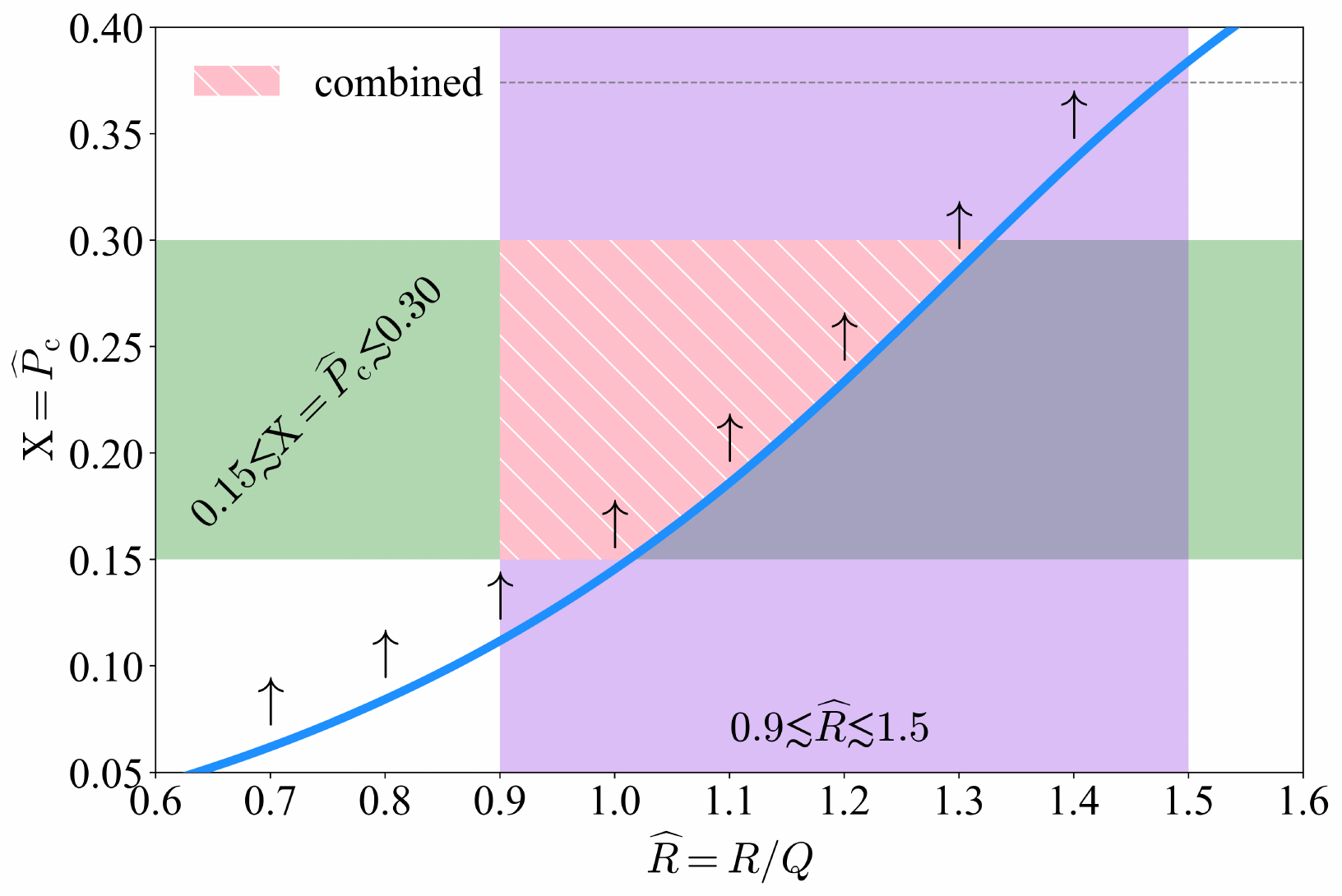}
\caption{(Color Online).  The dependence of the region for $\x$ which can generate a peaked $s^2$ on $\widehat{R}=R/Q$.
}\label{fig_a4_for_parameter}
\end{figure}

\begin{figure}[h!]
\centering
\includegraphics[width=11.cm]{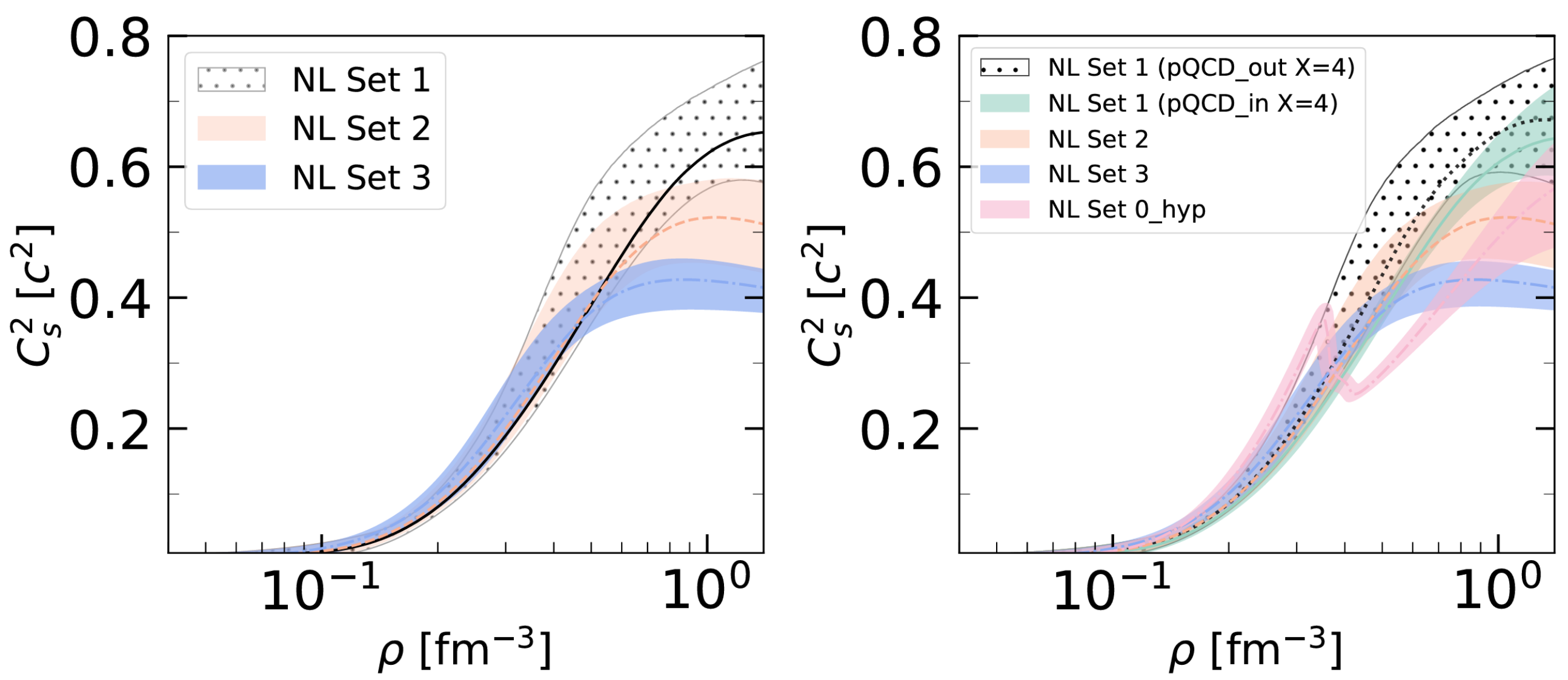}\\[0.5cm]
\includegraphics[width=14.cm]{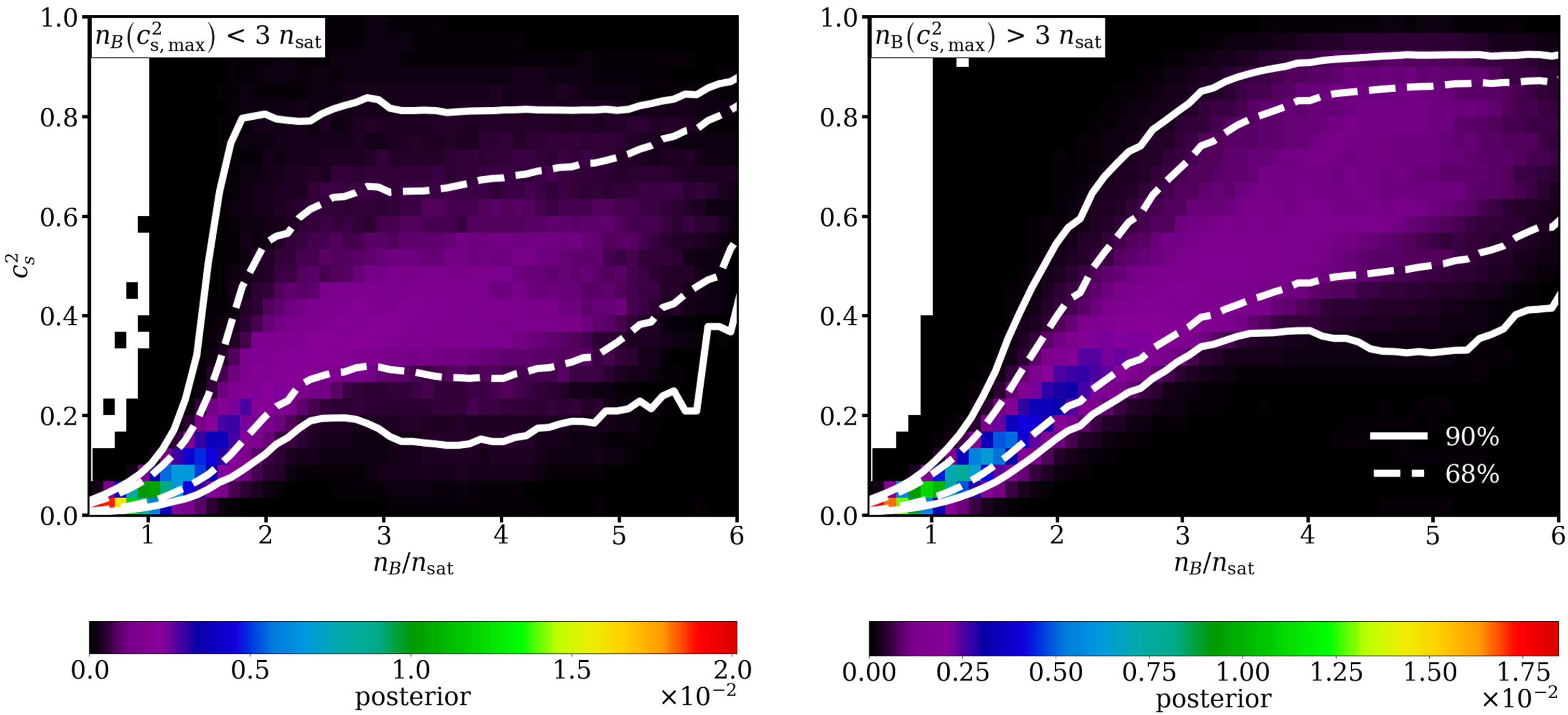}
\caption{(Color Online).  Upper panel: the $s^2$ as a function of baryon density using nonlinear relativistic mean field models without (left) or with pQCD constraints (right). Figure taken from Ref.\,\cite{Pro2024}.
Lower panel: posteriors for $s^2$ in cases when a global maximum is present below (left) and above (right) $3\rho_{\rm{sat}}=3n_{\rm{sat}}$. Figure taken from Ref.\,\cite{Mro23}.
}\label{fig_Pro23FIG}
\end{figure}

\begin{figure}[h!]
\centering
\includegraphics[width=15.cm]{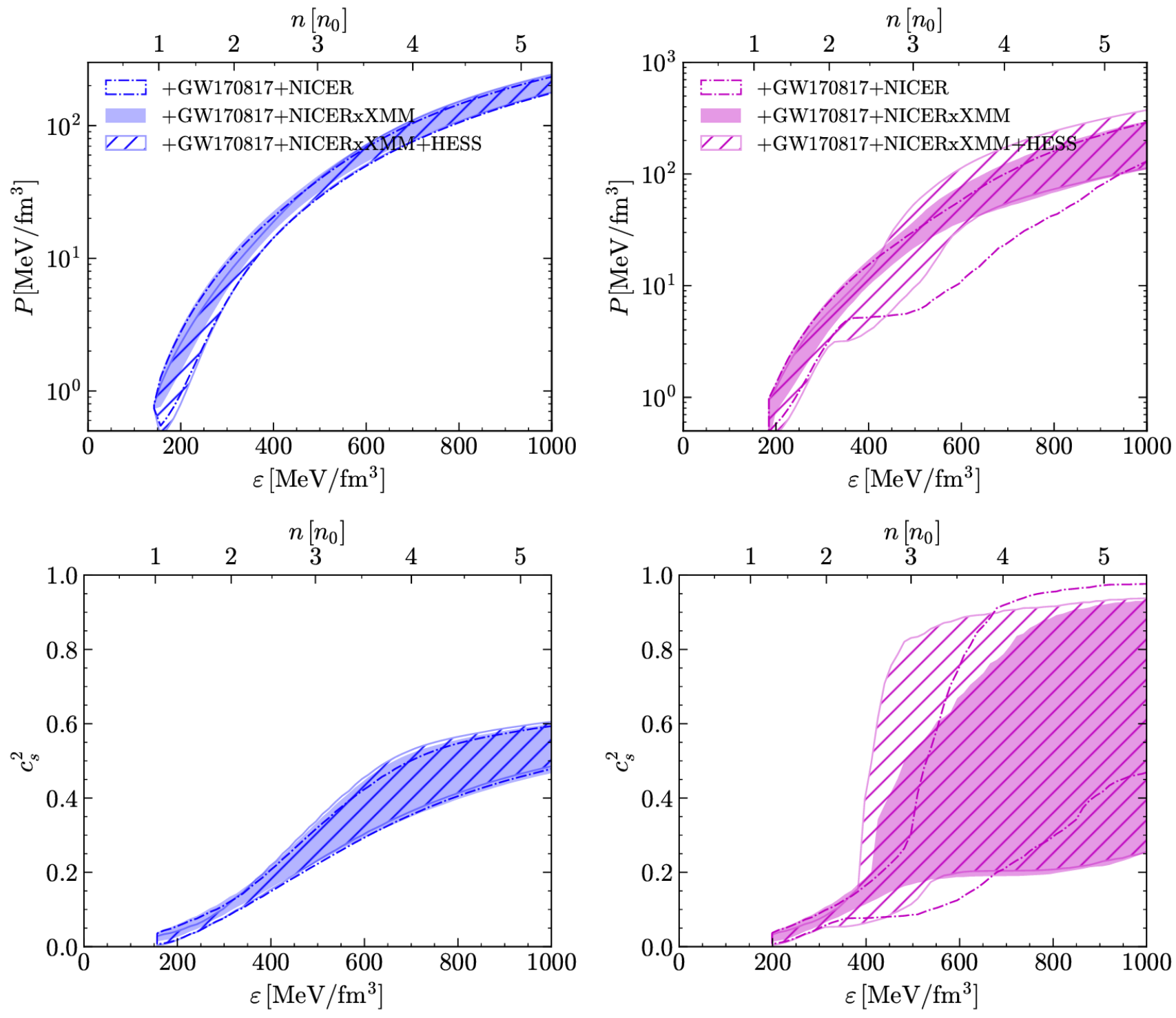}
\caption{(Color Online). The NS EOS and its SSS (left two panels) and those for hybrid stars (HSs, right two panels) considering differential astrophysical observations.
Figure taken from Ref.\,\cite{Miao24PRD}.
}\label{fig_Miao24s2}
\end{figure}

\begin{figure}[h!]
\centering
\includegraphics[width=9.cm]{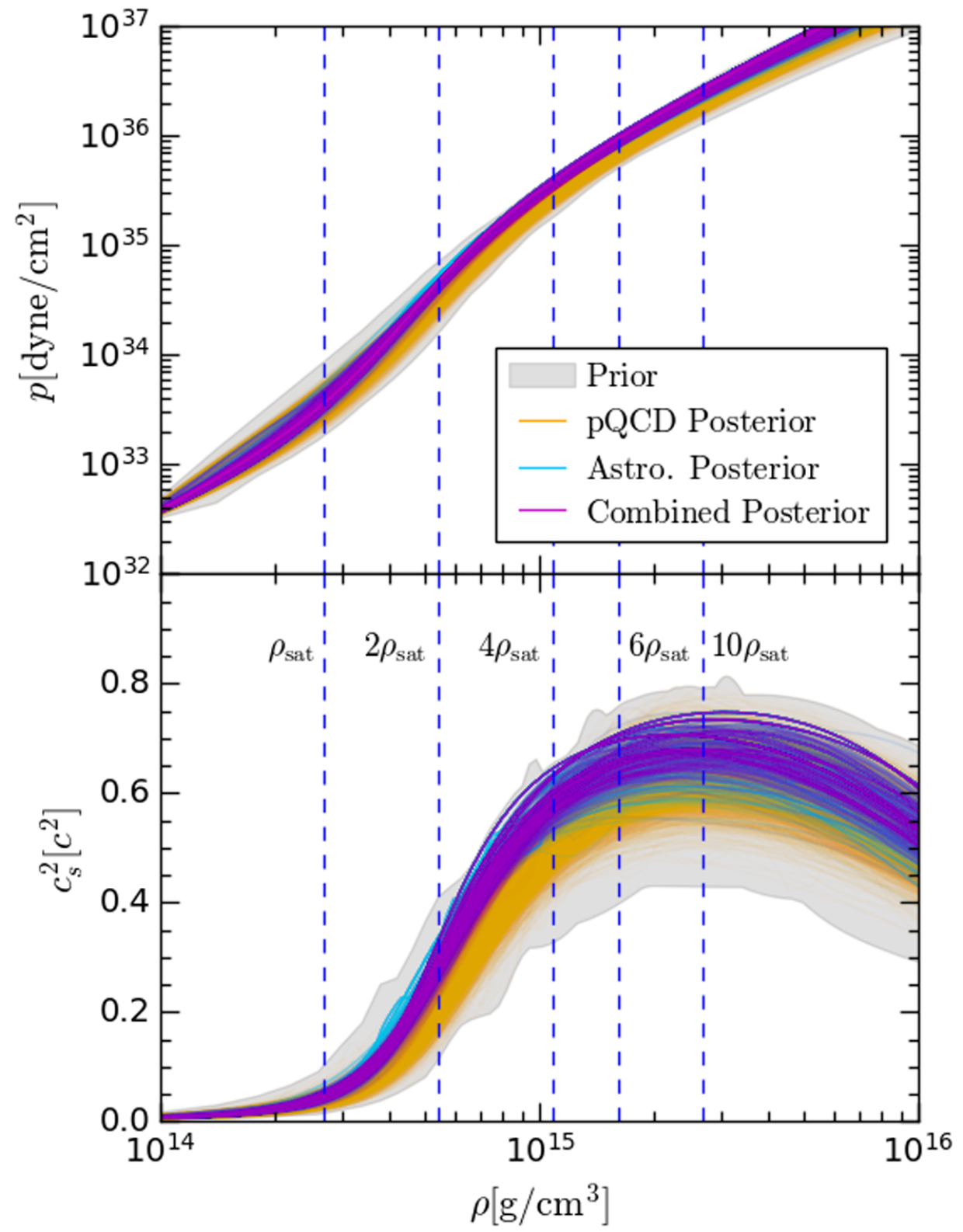}
\caption{(Color Online). NS EOS (upper panel) and the SSS $s^2$ (lower panel) constrained using a non-parametric inference utilizing NS astrophysical observations and pQCD effects. Figure taken from Ref.\,\cite{Cuceu2024}.
}\label{fig_Cuceu24s2}
\end{figure}

The resulted regions for $a_4$ are shown in the left panel of FIG.\,\ref{fig_a4peak}, here the colored arrows indicate whether the corresponding curve is a lower- or an upper-limit.
The criterion $\d\widehat{\varepsilon}/\d\widehat{r}<0$ {(criterion (a))} in the small-${\x}$ limit becomes $a_4\lesssim-16^{-1}\widehat{R}^{-2}{\x}^{-1}\sim\mathcal{O}({\x}^{-1})$ while the criterion $l_2>0$ {(inequality (\ref{def_a4ineq}))} gives $a_4\gtrsim80^{-1}{\x}^{-2}\sim\mathcal{O}({\x}^{-2})$, implying there would be no overlapped region for $a_4$ for such {small} ${\x}$, see the relation (\ref{NS_cond}).
In fact,  the overall condition $a_4\lesssim\widehat{R}^{-4}\approx1$ from criteria (a) and (b) certainly becomes incompatible with the magenta bound for ${\x}\lesssim0.1$ as it requires $a_4\gtrsim2$ for ${\x}\lesssim0.1$.
{This discrepancy becomes even larger as ${\x}$ decreases to even smaller values.}
On the other hand, the combined region for $a_4$ for the onset of a peak in $s^2$ profile may eventually emerge (shown as the cyan band) with increasing ${\x}$.  
The fact that the upper limit of (\ref{NS_cond}) should be greater than lower limit gives a critical lower $\x$ for a given $\widehat{R}$, and the result is shown in FIG.\,\ref{fig_a4_for_parameter}, here $0.15\lesssim\x\lesssim0.30$ is taken as a typical region for $\x$ in massive NSs and $0.9\lesssim\widehat{R}\lesssim1.5$ as the typical region for $\widehat{R}$.
A relatively smaller $\widehat{R}$ (corresponding to smaller $R$ or larger $Q$) is advantageous
for producing a larger region of $\x$ for generating a peaked $s^2$ profile; if $\widehat{R}$ is very large ($\approx1.5$) there would be no such $\x$.

Similarly, for normally stable NSs on the M-R curve, the expression for $b_2$ and the relation $a_2=b_2/s_{\rm{c}}^2$ do not change. However, the expression for $s_{\rm{c}}^2$ is modified from Eq.\,(\ref{sc2-TOV}) to Eq.\,(\ref{sc2-GG}). Taking $\varepsilon_{\rm{c}}\approx900\,\rm{MeV}/\rm{fm}^3$ and $\varepsilon_{\rm{c}}\approx400\,\rm{MeV}/\rm{fm}^3$ for a $2M_{\odot}$ NS and a canonical NS, respectively,  we may find an approximation $\Psi\approx0.88$ for a wide range of central energy densities.
The result is shown in the right panel of FIG.\,\ref{fig_a4peak} with $\Psi=0.88$. It is seen that the combined region for $a_4$ is similar to the one shown in the left panel of FIG.\,\ref{fig_a4peak} for $M_{\rm{NS}}^{\max}$.
In this case,  one has ${\x}\lesssim0.31$ by requiring $s_{\rm{c}}^2\leq1$.
In the following, we pay our attention to the $s^2$ profile for NSs at the TOV configuration.

To this end, we would like to point out that our analysis above is consistent with many model calculations of the $s^2$ profile. The shared qualitative conclusion is that not any dense matter EOS could generate a peaked $s^2$ profile.
This is because although $a_2$ is definitely negative, the coefficient $a_4$ could be either positive or negative (therefore no peaked structure),  i.e.,  the corresponding $s^2$ is a monotonic function (of energy density or radial distance).
For example, Ref.\,\cite{Pro2024} studied the $s^2$ using the nonlinear relativistic mean field models with or without considering the pQCD effects and found that $s^2$ may not generally have a peaked structure, as shown in the upper panel of FIG.\,\ref{fig_Pro23FIG}.
The result shown in FIG.\,\ref{fig_a4peak} is also consistent with a recent study showing that the peaked $s^2$ is not necessary\,\cite{Mro23}, even after considering several recent observational as well as theoretical constraints, see the lower panel of FIG.\,\ref{fig_Pro23FIG}, which shows that $s^2$ is probably monotonic within a wide range of baryon densities.
The monotonicity of $s^2$ was also found either for NSs or hybrid stars (NSs) using different astrophysical observations. For example, including the supernova remnant HESS J1731-347 data\,\cite{Doro22} only slightly increases the $s^2$ at moderate energy densities as shown in FIG.\,\ref{fig_Miao24s2} and does not induce a peaked behavior, though the uncertainties on $s^2$ are quite large.
Very recently in Ref.\,\cite{Cuceu2024}, the constraints on the NS EOS and SSS using a non-parametric inference utilizing astrophysical observations and pQCD effects found that the density where $s^2$ peaks is greater than $6\rho_{\rm{sat}}$, which may probably exceed the maximum NS densities, as shown in FIG.\,\ref{fig_Cuceu24s2}.
Including the astrophysical data was found to increase the pressure, the $s^2$ and its peak position, by comparing their yellow and violet bands in FIG.\,\ref{fig_Cuceu24s2}.

\subsection{Trace anomaly decomposition of SSS: general features}\label{sub_decomTA}

The trace anomaly decomposition introduced in Ref.\,\cite{Fuji22} is very useful/important to investigate the $s^2$ using the information on $\Delta$\,\cite{CL24-b}.
In this decomposition, we have\,\cite{Fuji22}
\begin{equation}\label{for1}
\boxed{
s^2=s^2_{\rm{deriv}}+s^2_{\rm{non-deriv}}=-\overline{\varepsilon}\frac{\d\Delta}{\d\overline{\varepsilon}}+\frac{1}{3}-\Delta,~~\overline{\varepsilon}\equiv\varepsilon/\varepsilon_0,}
\end{equation}
where the first term is the derivative part and the remaining terms together represent the non-derivative part; $\ep$ is the dimensionless energy density reduced with any selected energy density scale (in our demonstration we use $\varepsilon_0$).
We may understand the behavior (especially the peaked structure) of the $s^2$ by considering the feature of the derivative term $\Delta'\equiv\d\Delta/\d\overline{\varepsilon}$.

\begin{figure}[h!]
\centering
\includegraphics[width=13.5cm]{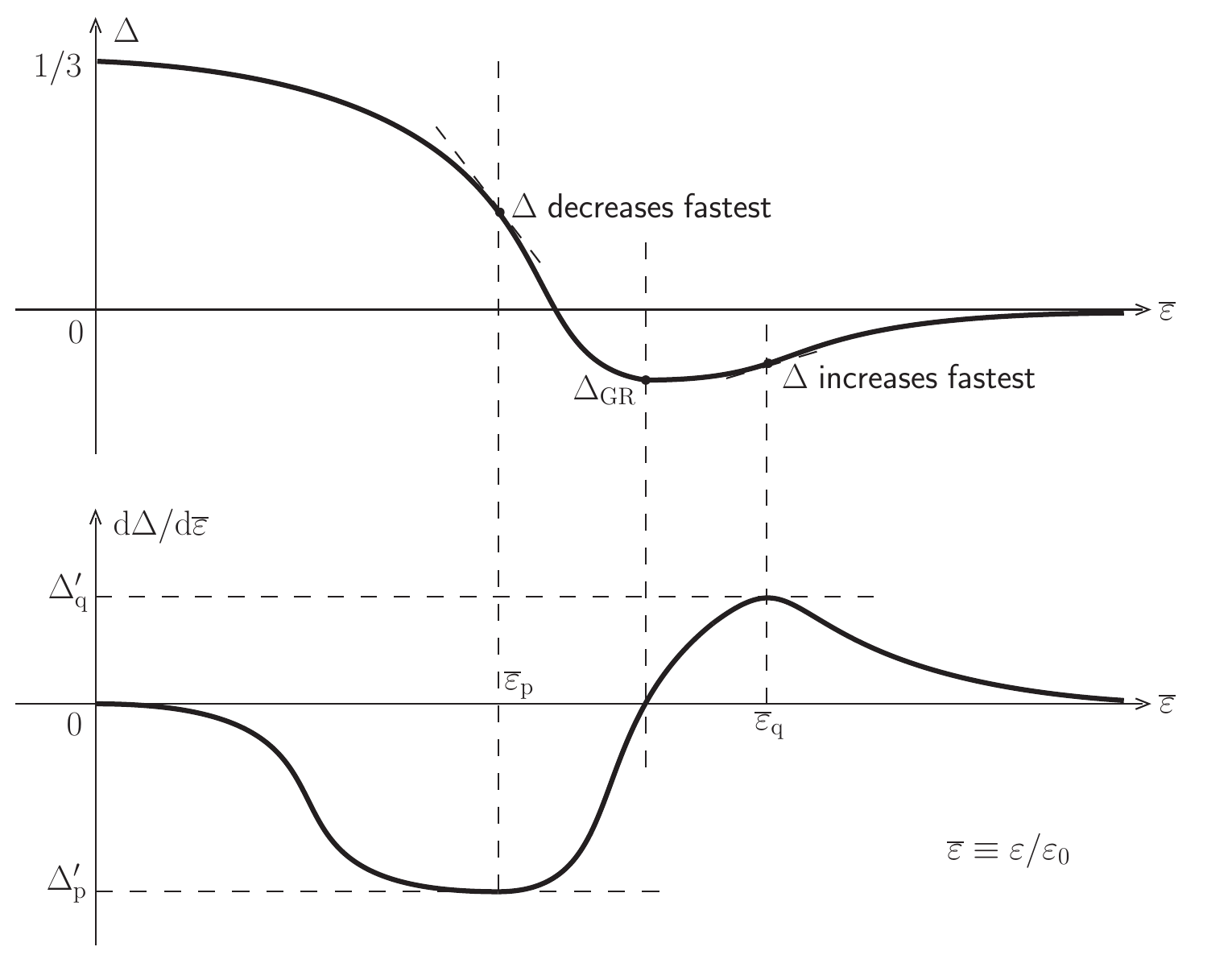}
\caption{Sketches of the trace anomaly $\Delta$ (upper panel) and its derivative $\d\Delta/\d\overline{\varepsilon}$ (lower panel) as functions of energy density.
This shape of $\Delta$ corresponds to the magenta curve in FIG.\,\ref{fig_Dp_sk}.
Figure taken from Ref.\,\cite{CL24-b}.
}\label{fig_s2phys}
\end{figure}

Generally, there is a point $\overline{\varepsilon}_{\rm{p}}$ where $\Delta$ decreases fastest; around this point-p we can expand its derivative $\Delta'$ as
\begin{equation}
\Delta'\approx\Delta_{\rm{p}}'+2^{-1}\Delta_{\rm{p}}'''\left(\overline{\varepsilon}-\ep_{\rm{p}}\right)^2,~~
\Delta_{\rm{p}}'<0,~~\Delta_{\rm{p}}'''>0,
\end{equation}
as the first-order derivative of $\Delta'$, namely $\Delta''$ at point-p is zero, see the lower panel of FIG.\,\ref{fig_s2phys}.
Since $\Delta_{\rm{p}}'''$ characterizes the curvature of $\Delta'$, its magnitude is small for a shallow-shaped $\Delta'$ but large for a sharp-shaped one. 
Correspondingly, the trace anomaly $\Delta$ itself around the point-p can be approximated by integration of $\Delta'$ over $\ep$, i.e., $
\Delta\approx\Delta_{\rm{p}}'(\overline{\varepsilon}-\ep_{\rm{p}})+6^{-1}\Delta_{\rm{p}}'''(\overline{\varepsilon}-\ep_{\rm{p}})^3+\Delta_{\rm{p}}$.
We could then obtain the $s^2$ using the formula (\ref{for1}).
Calculating its derivative $\d s^2/\d\ep$ gives
\begin{equation}\label{for2}
{\d s^2}/{\d\ep}=-2\Delta_{\rm{p}}'+\left(
3\ep_{\rm{p}}\ep-2\ep^2-\ep_{\rm{p}}^2
\right)\Delta_{\rm{p}}'''.
\end{equation}
We discuss it in the following two cases:
\begin{enumerate}[label=(\alph*)]
\item If the valley of the $\Delta'$ curve is shallow, i.e., $\Delta_{\rm{p}}'''$ is positively small, we can neglect the second term in (\ref{for2}), leading to $
{\d s^2}/{\d\ep}\approx-2\Delta_{\rm{p}}'>0$.
Thus, regardless of the shape of its derivative term $-\ep\d\Delta/\d\ep$ (e.g. it has a peak or equivalently a valley in $\d\Delta/\d\ep$), the $s^2$ is always a monotonically increasing function of $\ep$.

\item On the other hand, if $|\Delta_{\rm{p}}'|$ is much smaller than $\Delta_{\rm{p}}'''$,  i.e., the valley in $\Delta'$ is sharp, we can treat the first term in (\ref{for1}) as a perturbation and solve the equation $\d s^2/\d\ep=0$ for its extreme point $\ep_{\rm{p}}^{\ast}$. The result is
\begin{equation}\label{for3}
\boxed{
\ep_{\rm{p}}^{\ast}=\frac{3\ep_{\rm{p}}\Delta_{\rm{p}}'''+\sqrt{\ep_{\rm{p}}^2\Delta_{\rm{p}}'''^2-16\Delta_{\rm{p}}'\Delta_{\rm{p}}'''}}{4\Delta_{\rm{p}}'''}
\approx\ep_{\rm{p}}\left(1-\frac{2}{\ep_{\rm{p}}^2}\frac{\Delta_{\rm{p}}'}{\Delta_{\rm{p}}'''}\right)>\ep_{\rm{p}},~~\mbox{for small }\Delta_{\rm{p}}'.}
\end{equation}
Moreover, we have for the extreme value and second-order derivative of $s^2$ at $\ep_{\rm{p}}^\ast$ as
\begin{align}
s^2(\ep_{\rm{p}}^\ast)\approx&\frac{1}{3}-\Delta_{\rm{p}}-\ep_{\rm{p}}\Delta_{\rm{p}}',~~
\left.\frac{\d^2s^2}{\d\ep^2}\right|_{\ep_{\rm{p}}^\ast}\approx\Delta_{\rm{p}}'''\ep_{\rm{p}}\left(
\frac{8}{\ep_{\rm{p}}^2}
\frac{\Delta_{\rm{p}}'}{\Delta_{\rm{p}}'''}-1\right)<0.
\end{align}
The negativeness of the second-order derivative shows that it is a maximum point (peak) of $s^2$.
The correction in the bracket of (\ref{for3}) is positive since $\Delta_{\rm{p}}'$ and $\Delta_{\rm{p}}'''$ have the opposite signs, this means that {\color{xll}the peak in $s^2$ occurs on the right side of the peak in $-\ep\d\Delta/\d\ep$.}
We can also evaluate separately the decomposition terms $s_{\rm{deriv}}^2\equiv-\ep\d\Delta/\d\ep$ and $s_{\rm{non-deriv}}^2\equiv 3^{-1}-\Delta$ of $s^2$ at $\ep_{\rm{p}}^\ast$. The results are
\begin{equation}
s_{\rm{deriv}}^2(\ep_{\rm{p}}^\ast)\approx-\ep_{\rm{p}}\Delta_{\rm{p}}',~~s_{\rm{non-deriv}}^2(\ep_{\rm{p}}^\ast)\approx
\frac{1}{3}-\Delta_{\rm{p}},\mbox{ and therefore }
\frac{s^2(\ep_{\rm{p}}^\ast)}{s_{\rm{deriv}}^2(\ep_{\rm{p}}^\ast)}
=1-\frac{3^{-1}-\Delta_{\rm{p}}}{\ep_{\rm{p}}\Delta_{\rm{p}}'}>1.
\end{equation}
The latter relation means that {\color{xll}the height of the peak in $s^2$ is definitely larger than that in its derivative component $s_{\rm{deriv}}^2$.}
\end{enumerate}

The analysis for the point-q in FIG.\,\ref{fig_s2phys} where $\Delta'$ is a maximum is totally parallel. In particular, there exists a valley in $s^2$ described as:
\begin{equation}\label{valley-k}
\boxed{
\left.\frac{\d^2s^2}{\d\ep^2}\right|_{\ep_{\rm{q}}^\ast}\approx\Delta_{\rm{q}}'''\ep_{\rm{q}}\left(
\frac{8}{\ep_{\rm{q}}^2}
\frac{\Delta_{\rm{q}}'}{\Delta_{\rm{q}}'''}-1\right)>0,
\mbox{  at  }
\ep_{\rm{q}}^\ast\approx
\ep_{\rm{q}}\left(1-\frac{2}{\ep_{\rm{q}}^2}\frac{\Delta_{\rm{q}}'}{\Delta_{\rm{q}}'''}\right)>\ep_{\rm{q}},}
\end{equation}
since now $
\Delta_{\rm{q}}'>0$ and $\Delta_{\rm{q}}'''<0$.
This means that {\color{xll}the valley in $s^2$ still appears after that in its derivative part;} the new feature is that the valley in the derivative part is definitely negative, $
s_{\rm{deriv}}^2(\ep_{\rm{q}}^\ast)\approx-\ep_{\rm{q}}\Delta_{\rm{q}}'<0$.
Considering $\d\Delta/\d\ep\to0$ for large $\ep$,  the $s^2$ finally approaches its asymptotic value determined by pQCD theories\,\cite{Bjorken83,Kur10}.
The analyses above provide the physics basis for our ``educated guess'' in imagining the evolution of trace anomaly $\Delta$ and the corresponding $s^2$ to completely uncharted superdense region in FIG.\,\ref{fig_Dp_sk}.

By writing $\Delta(\ep)\approx 3^{-1}+a\ep+b\ep^2$ with $a<0$, we obtain $s^2\approx-2a\ep-3b\ep^2$ and $\phi=3^{-1}-\Delta(\ep)\approx-a\ep-b\ep^2$. Putting the $\ep(\phi)=-(2b)^{-1}\sqrt{a^2-4b\phi}$ with $b\leq a^2/4\phi$ from solving the latter relation into $s^2$ enables us to write the dependence of $s^2$ on $\phi$ (with $\phi\to0$ or the Newtonian limit) as:
\begin{equation}\label{s2_on_phi}
    s^2(\phi)\approx3\phi-\frac{a}{2b}\left(a+\sqrt{a^2-4b\phi}\right)
    \approx2\phi\left(1-\frac{b}{2a^2}\phi\right)+\mathcal{O}(\phi^3)
    ,~\rm{and}~~s^2/\phi\approx2\left(1-\frac{b}{2a^2}\phi\right).
\end{equation}
We then obtain $f_0=2$ (see relations (\ref{s2_dim2}) and (\ref{s2_dimensionless})), and magnitude of the correction $b\phi/2a^2\ll1/8$ since $b\ll a^2/4\phi$.
Whether $s^2(\phi)$ may develop a peak depends on the sign/magnitude of the coefficient $b$, e.g., we have $s^2(\phi_{\rm{pk}})\approx a^2/3b$ at $\phi_{\rm{pk}}\approx2a^2/9b$ in the case of $b>0$, so $\gamma_{\rm{pk}}=s^2(\phi_{\rm{pk}})/\phi_{\rm{pk}}\approx3/2$.
It is necessary to point out if the leading-order term in $\Delta(\overline{\varepsilon})-3^{-1}$ is not $a\overline{\varepsilon}$ but $a\overline{\varepsilon}^k$ with some constant $k$, then $f_0\approx1+k$.

\begin{figure}[h!]
\centering
\includegraphics[width=17.cm]{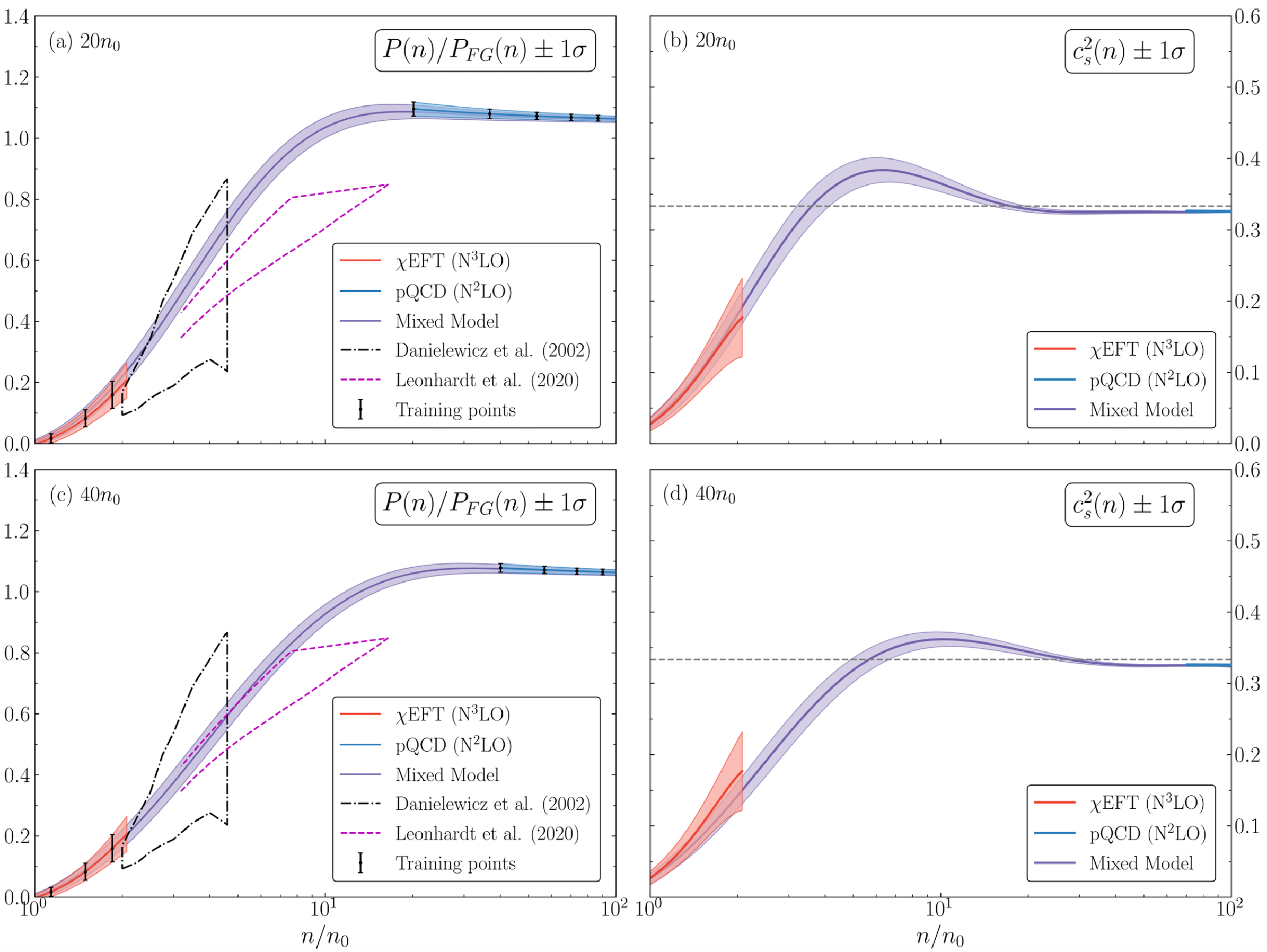}
\caption{(Color Online).  Bayesian model mixing for the reduced pressure (left two panels) and the SSS (right two panels) by  combining CEFT and pQCD results via a Gaussian process. Figure taken from Ref.\,\cite{Semp24}.
}\label{fig_Semp24FIG}
\end{figure}

Using what have learned from the above analysis about the trace anomaly and its energy density dependence, we can understand generally why a peak in $s^2$ may emerge.
{\color{xll}This is mostly because $\Delta\to1/3$ at the low-density limit and it reaches some limit $\Delta_{\rm{limit}}$ at large densities. The latter may be the pQCD value $\Delta_{\rm{pQCD}}= 0$ at extremely high densities or $\Delta_{\rm{GR}}\approx-0.041$ of Eq.\,(\ref{GRDelta}) accessible within realistic NSs.
If $\Delta_{\rm{limit}}\approx \Delta_{\rm{GR}}$, then a peak definitely emerges in $s^2$ at densities realizable in NSs. On the other hand, if $\Delta_{\rm{limit}}\approx\Delta_{\rm{pQCD}}=0$ and $\Delta$ is always positive then a peak may still emerge in $s^2$, but at some larger densities not accessible in realistic NSs.}
When does the pQCD constraint become relevant\,\cite{Zhou2024} for NSs? The answer may tell us whether/where the peaked structure in $s^2$ may occur.
For example, if the pQCD cutoff is set at a relatively lower density about $20\rho_0=20\rho_{\rm{sat}}$, then by combining the CEFT results at low-densities and the pQCD constraints starting at $20\rho_{\rm{sat}}$ Ref.\,\cite{Semp24} found that a peak emerges in $s^2$ at about $6.5\rho_{\rm{sat}}$ using a Bayesian model mixing method\,\cite{Qiu2024}. This emerging density is large but may still be accessible in some NSs, see the panel (b) in FIG.\,\ref{fig_Semp24FIG}. On the other hand, if the pQCD cutoff density is set to $40\rho_{\rm{sat}}$, then the corresponding peak position in $s^2$ is about $10.5\rho_{\rm{sat}}$ 
as shown in panel (d) of FIG.\,\ref{fig_Semp24FIG}.
This emerging density probably exceeds the maximum density realized in NSs according to many existing calculations, indicating that the $s^2$ in NSs is monotonic in this case. Similar conclusions also apply for the reduced pressure as indicated in panels (a) and (c) of FIG.\,\ref{fig_Semp24FIG}.

\begin{figure}[h!]
\centering
\includegraphics[width=13.cm]{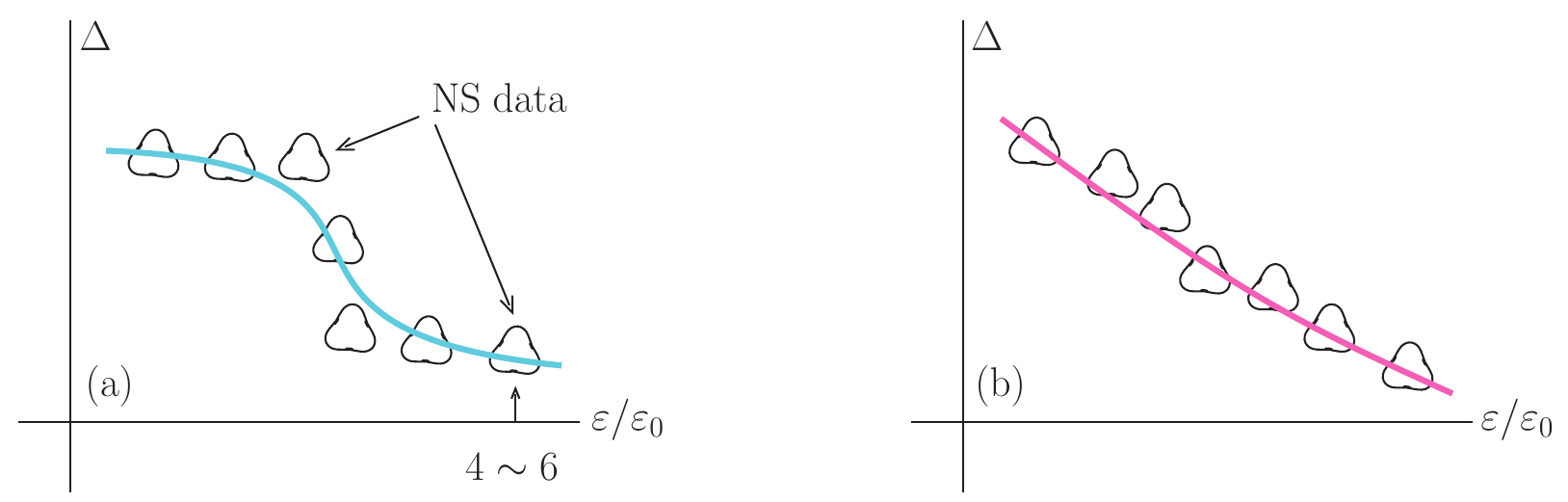}
\caption{(Color Online).  The trace anomaly $\Delta$ in the left panel (having a sharp reduction at some $\ep$ and becoming flat at both small and large $\ep$'s) may generate a peaked $s^2$;  while that in the right panel implies $\d\Delta/\d\overline{\varepsilon}\approx\rm{const.}<0$ and $s^2$ may thus increase monotonically with $\varepsilon/\varepsilon_0$. The solid curve in each panel represents the probable shape of $\Delta$.
Figure taken from Ref.\,\cite{CL24-b}.
}\label{fig_DEL_s2peak}
\end{figure}

\begin{table}[h!]
\renewcommand{\arraystretch}{1.5}
\centerline{\normalsize
\begin{tabular}{c|c|c|c|c|c|c} 
  \hline
$\Delta$ (by construction)&0.19&0.18&$0.17$&0.02&0.01&0.00\\\hline\hline
$\y=\varepsilon_{\rm{c}}/\varepsilon_0$ (by construction)&3.0&3.3&3.6&3.7&4.0&4.3\\\hline
$M_{\rm{NS}}/M_{\odot}$ (using scaling (\ref{gk-m}))&1.44&1.45&1.47&2.19&2.13&2.05\\\hline
$R/\rm{km}$ (using scaling (\ref{gk-xi}))&12.5&12.0&11.7&12.9&12.4&11.9\\\hline
\end{tabular}}
\caption{Two classes of NSs constructed by $\Delta$ and $\y$ using scalings of (\ref{gk-xi}), (\ref{gk-m}) and (\ref{gk-r}); the $\Delta$ for the canonical NSs is around 0.18 while that for massive NSs is about 0.01.
}\label{tab_artf_NS} 
\end{table}

\begin{figure}[h!]
\centering
\includegraphics[height=5.6cm]{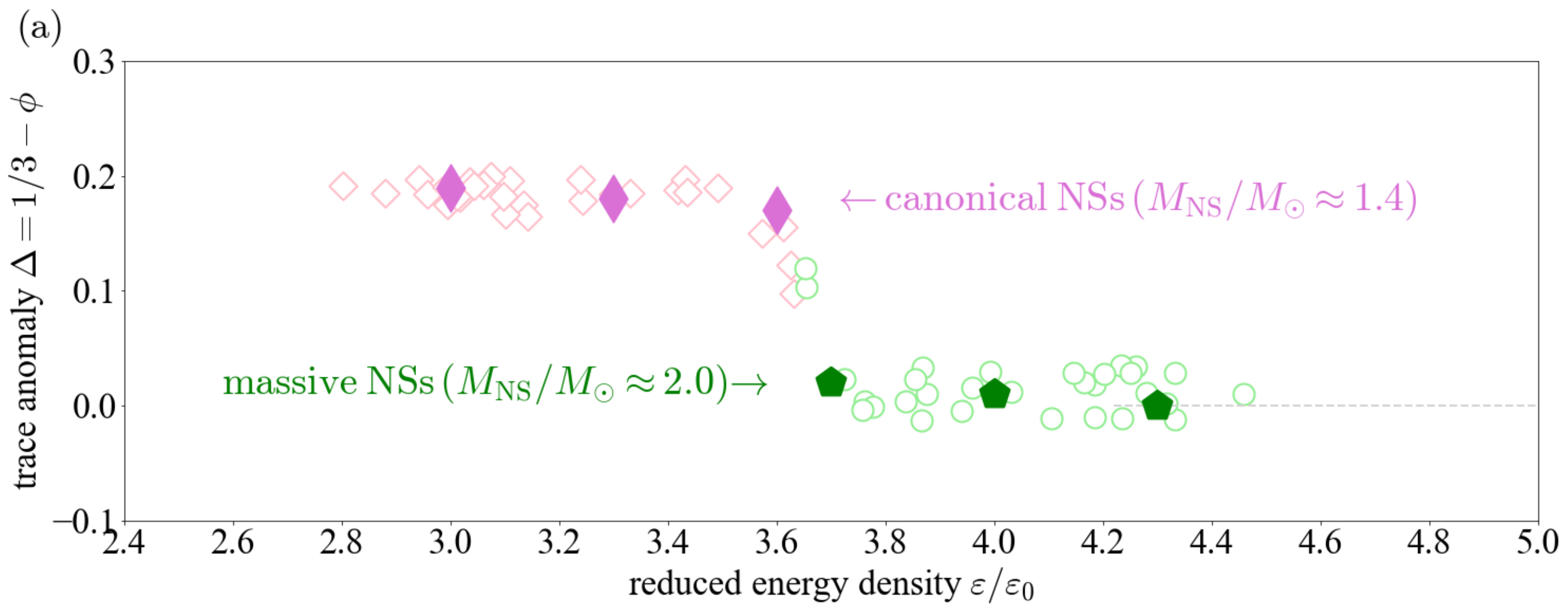}\\[0.5cm]
\includegraphics[height=8.cm]{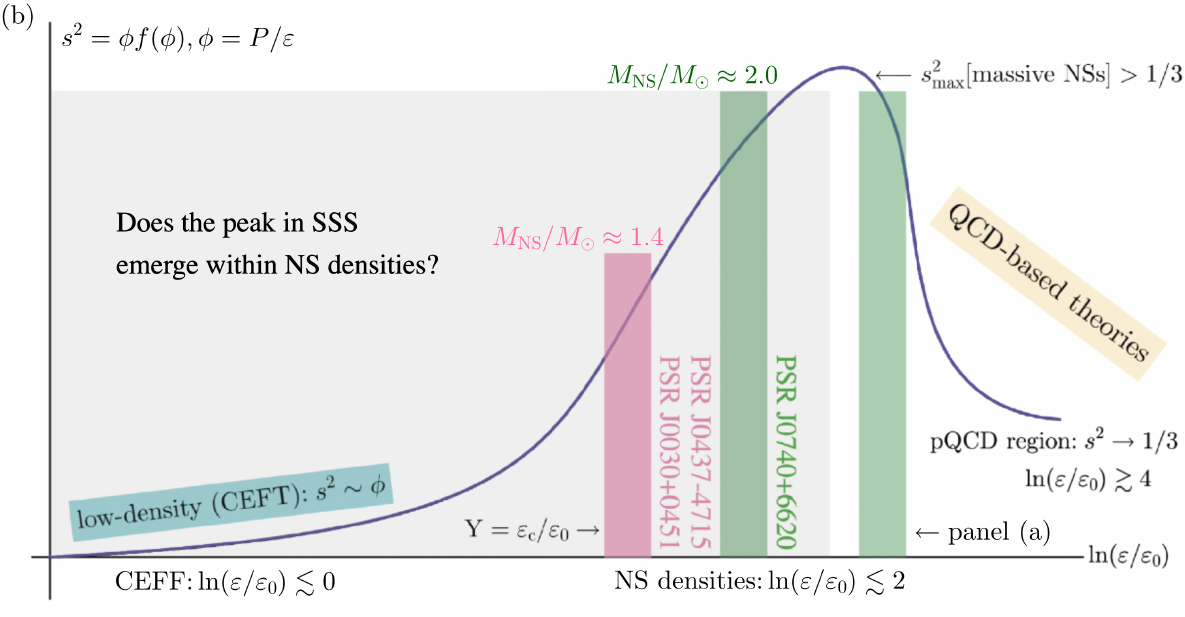}
\caption{(Color Online). Upper panel: two classes of six NS instances constructed in TAB.\,\ref{tab_artf_NS} where there is quick (clear) decreasing of $\Delta$ around $\overline{\varepsilon}\approx3.7$.
Lower panel: since the magnitude of $s^2$ in massive NSs is probably greater than the pQCD limit 1/3, there would be unavoidably a peak in $s^2$ as a function of energy density; whether such peak emerges within NS densities rely essentially on the observational data.
The right (left) green band for PSR J0740+6620 corresponds to the case where the peak emerges within (beyond) NS densities.
}\label{fig_artf_fffff}
\end{figure}

A more useful approach to infer convincing evidence on whether there exists a peak or not in $s^2$ is to rely on observational data (masses and radii) themselves if at all possible.
This idea is illustrated in FIG.\,\ref{fig_DEL_s2peak} using two typical shapes of $\Delta$ as a function of $\varepsilon/\varepsilon_0$.
In the left panel, we expect the $s^2$ to have a peak somewhere since there is a quick decreasing in $\Delta$ and two approximate plateaus at low and high energy densities, respectively. These features together will lead to a peak in $s^2$ as we just discussed above. While the $\Delta$ in the right panel indicates that $\d\Delta/\d\overline{\varepsilon}$ is approximately negative and constant, both the derivative part $-\overline{\varepsilon}\d\Delta/\d\overline{\varepsilon}$ and the non-derivative term $1/3-\Delta$ of $s^2$ then monotonically increase with $\varepsilon/\varepsilon_0$.
Which type of NS mass/radius observations may give such energy density dependence of $\Delta$? Scalings of (\ref{gk-xi}), (\ref{gk-m}) and (\ref{gk-r}) provide us the clue: If two classes of NSs, the canonical NSs with radii about 12\,km and massive NSs with $M_{\rm{NS}}/M_{\odot}\gtrsim2$ and similar radii, are accurately observed with large samples in the future, then we can study whether the $s^2$ as a function of energy density may have a peaked structure within NS densities.
We construct these two classes of NSs in TAB.\,\ref{tab_artf_NS}, the first three NSs (columns 2 to 4) have masses about 1.4$M_{\odot}$ and the last threes have $M_{\rm{NS}}/M_{\odot}\gtrsim2$; all the six NS instances have similar radii about $\gtrsim$12\,km.
We illustrate the corresponding $s^2$ in FIG.\,\ref{fig_artf_fffff}. If the $\Delta(\varepsilon)$ extracted from observational data takes the left pattern of FIG.\,\ref{fig_DEL_s2peak}, then there would be a peaked structure in $s^2$ within NS densities reached (e.g., the right green band for PSR J0740+6620). On the other hand, if the $\Delta(\varepsilon)$ takes the right pattern of FIG.\,\ref{fig_DEL_s2peak}, then within NS densities reachable the $s^2$ may monotonically increase (indicated by the gray background in the lower panel of FIG.\,\ref{fig_artf_fffff}). As discussed earlier, since $s^2\to1/3$ at very high densities and the $s^2$ in massive NSs is probably greater than 1/3, theoretically there would unavoidably be a peak in $s^2$ profile. However, this peak is outside the NS densities. We shall use this idea in Subsection \ref{sub_NSdatapeak} to explore the (possible) peaked structure in $s^2$ using available NS data, see FIG.\,\ref{fig_s2ab} and FIG.\,\ref{fig_fffff}. Unfortunately, the currently available NS data could not invariably generate a peaked $s^2$ profile\,\cite{CL24-b}.

\subsection{A reference: SSS in Newtonian stars as a monotonic function of radial distance from their centers}\label{sub_s2_Newtonian}

We now work out the specific expression for $s^2$ for Newtonian stars to be used as a reference.
Going away from the center by expanding the right side of Eq.\,(\ref{s2-N}) considering the perturbative expansions of $\heps$ of Eq.\,(\ref{ee-heps}) and $\widehat{M}$ of Eq.\,(\ref{ee-hM}),  we obtain the $s^2$ to order $\widehat{r}^4$ as\,\cite{CL24-b},
\begin{align}\label{for_s2Newt}
s^2 \approx& s_{\rm{c}}^2+l_2^{\rm{N}}\widehat{r}^2+l_4^{\rm{N}}\widehat{r}^4
\approx
s_{\rm{c}}^2+\left(12a_4s_{\rm{c}}^4-\frac{4}{15}\right)\widehat{r}^2+
\left(
144a_4^2s_{\rm{c}}^6+18a_6s_{\rm{c}}^4-\frac{62}{35}a_4s_{\rm{c}}^2+\frac{1}{60s_{\rm{c}}^2}\right)
\widehat{r}^4,
\end{align}
where a superscript ``N'' is added to remind us this is for the Newtonian case.
The Newtonian $s^2$ radial profile does not depend on ${\x}$ explicitly, although $s_{\rm{c}}^2$ may implicitly contain factors of ${\x}$.

\begin{figure}[h!]
\centering
\includegraphics[width=8.5cm]{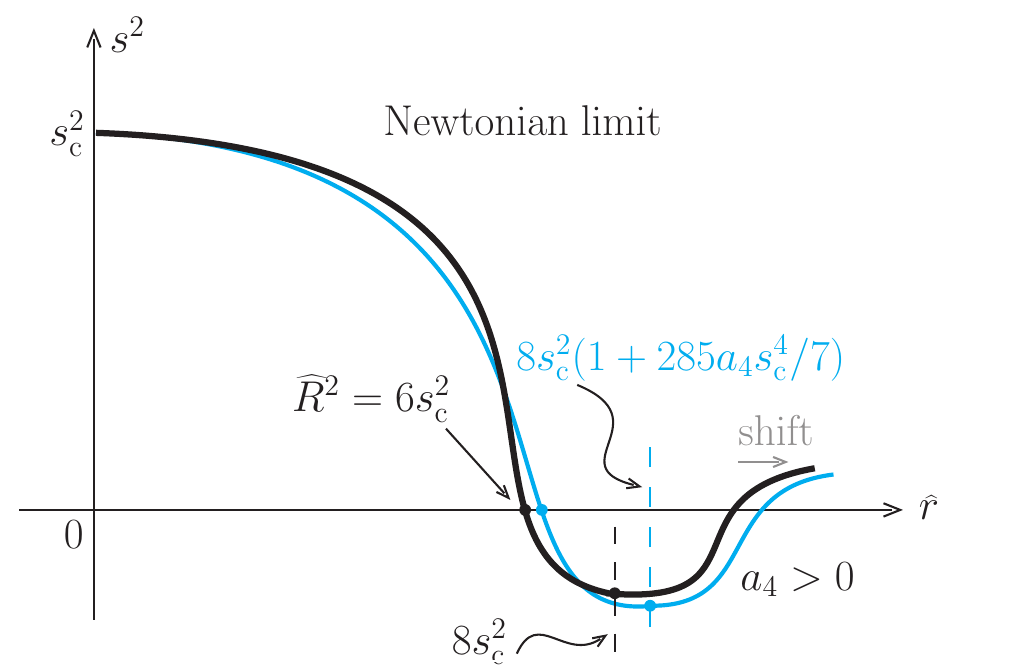}
\caption{(Color Online). Sketch of the Newtonian $s^2$ when expanding $s^2$ to order $\widehat{r}^4$ (black) as in Eq.\,(\ref{for_s2Newt-1}).
The correction due to a positive $a_4$ is also shown (light-blue) from Eq.\,(\ref{for_s2_a6}).
Figure taken from Ref.\,\cite{CL24-b}.
}\label{fig_s2New_r4}
\end{figure}

For Newtonian stars, the reduced radius $\widehat{R}$ is generally small; we have estimated in Subsection \ref{sub_s2ORDER} that for WDs $\widehat{R}\approx0.05$ and $\widehat{R}^2\approx3\times10^{-3}$ as well as $\phi=P/\varepsilon\approx10^{-4}\sim10^{-5}$.
Therefore it is conservative to use $\phi\lesssim\mathcal{O}(10^{-k})$ with $k\gtrsim4\mbox{-}5$ for Newtonian stars.
Moreover, considering $s_{\rm{c}}^2\approx4{\x}/3\sim {\x} \sim\mathcal{O}(10^{-k})$ with $k\gtrsim4\mbox{-}5$, we have 
\begin{equation}
a_2=b_2/s_{\rm{c}}^2\approx-1/6s_{\rm{c}}^2\lesssim\mathcal{O}\left(10^{k-1}\right),~~
12a_4s_{\rm{c}}^4\lesssim\mathcal{O}\left(10^{2-k}\right),
\end{equation}
the latter is obtained via $a_2\widehat{R}^2\gtrsim a_4\widehat{R}^4$ from the perturbative expansion of $\widehat{\varepsilon}$ and therefore $a_4\lesssim a_2/\widehat{R}^2\lesssim\mathcal{O}(10^{k+1})$.
Similarly, we have the order-of-magnitude scalings
\begin{equation}
144a_4^2s_{\rm{c}}^6\sim18a_6s_{\rm{c}}^4\lesssim\mathcal{O}\left(10^{5-k}\right),~~({62}/{35})a_4s_{\rm{c}}^2\lesssim\mathcal{O}\left(10^1\right),~~1/60s_{\rm{c}}^2\lesssim\mathcal{O}\left(10^{k-2}\right).
\end{equation}
Therefore, the $-4/15$ and $1/60s_{\rm{c}}^2$ terms denominate over others in Eq.\,(\ref{for_s2Newt}).
Consequently, 
\begin{equation}\label{for_s2Newt-1}
\boxed{
s^2\approx s_{\rm{c}}^2-({4}/{15})\widehat{r}^2+({60s_{\rm{c}}^2})^{-1}\widehat{r}^4.}
\end{equation}
Using this expression, we find $s^2$ takes its minimum as $s_{\min}^2=-{s_{\rm{c}}^2}/{15}$ at $\widehat{r}^2_{\min}=8s_{\rm{c}}^2$\,\cite{CL24-b}.
The vanishing of $s^2$ gives two special $\widehat{r}^2$ values, namely $6s_{\rm{c}}^2$ and $10s_{\rm{c}}^2$.
Since the stability condition requires $s^2\geq0$, we find $s^2$ is a decreasing function of $\widehat{r}\leq\widehat{R}$ with $\widehat{R}^2=6s_{\rm{c}}^2$, see FIG.\,\ref{fig_s2New_r4} for the sketch of $s^2$ in the Newtonian limit.
The decreasing feature of $s^2$ under Newtonian limit could also be seen from Eq.\,(\ref{zf-3}) since $s^2/{\x}\approx 4/3+32\mu/15$ for ${\x}\approx0$, which decreases with decreasing $\widehat{\varepsilon}=\mu+1$ (outward from the NS center).

The coefficient $l_4^{\rm{N}}\approx 1/60s_{\rm{c}}^2$ in Eq.\,(\ref{zf-3}) originated from $b_4$ of Eq.\,(\ref{ee-b4}) is fundamental for explaining the radial dependence of $s^2$ in the Newtonian limit.
This is because even when the next high-order term $l_6^{\rm{N}}\widehat{r}^6$ is included,  the conclusion will not change qualitatively.
Specifically, we have  $s^2\approx s_{\rm{c}}^2+l_2^{\rm{N}}\widehat{r}^2+l_4^{\rm{N}}\widehat{r}^4+l_6^{\rm{N}}\widehat{r}^6$:
\begin{empheq}[box=\fbox]{align}\label{for_s2_a6}
s^2\approx& s_{\rm{c}}^2+\left(12a_4s_{\rm{c}}^4-\frac{4}{15}\right)\widehat{r}^2
+
\left(
144a_4^2s_{\rm{c}}^6+18a_6s_{\rm{c}}^4-\frac{62}{35}a_4s_{\rm{c}}^2+\frac{1}{60s_{\rm{c}}^2}\right)
\widehat{r}^4\notag\\
&+\left[
1728a_4^3s_{\rm{c}}^8+432a_4a_6s_{\rm{c}}^6+\left(24a_8-\frac{744}{35}a_4^2\right)s_{\rm{c}}^4
-\frac{52}{15}a_6s_{\rm{c}}^2+\frac{1}{35}a_4
\right]\widehat{r}^6\notag\\
\approx&s_{\rm{c}}^2-({4}/{15})\widehat{r}^2+({60s_{\rm{c}}^2})^{-1}\widehat{r}^4
+(a_4/35)\widehat{r}^6,
\end{empheq}
which becomes exact as $s_{\rm{c}}^2\to0$.
{Here the term $a_4\widehat{r}^6/35$ denominates at order $\widehat{r}^6$ using the same order-of-magnitude estimates given above.}
Notice that there are no terms inversely proportional to $s_{\rm{c}}^2$ appear in the coefficient $l_6^{\rm{N}}$ (which is different from $l_4^{\rm{N}}$).
Using the full form of $s^2$ in Eq.\,(\ref{for_s2_a6}), we obtain:
\begin{align}
\widehat{r}_{\min}^2\approx&8s_{\rm{c}}^2\left(1+{285}a_4s_{\rm{c}}^4/7+1416a_6s_{\rm{c}}^6\right),\\
s_{\min}^2\approx&-\frac{s_{\rm{c}}^2}{15}\left(1+{288}a_4s_{\rm{c}}^4/7+9344a_6s_{\rm{c}}^6\right),\\
\widehat{R}^2\approx& 6s_{\rm{c}}^2\left(1+{144}a_4s_{\rm{c}}^4/7+1620a_6s_{\rm{c}}^6\right).
\end{align}
The corrections in the brackets are perturbations (compared with the leading ``1'') in the sense that $285a_4s_{\rm{c}}^4/7\lesssim\mathcal{O}(10^{3-k})\ll1$ and $1416a_6s_{\rm{c}}^6\lesssim\mathcal{O}(10^{7-2k})\ll1$ for $k\gtrsim4\mbox{-}5$, etc.
We show in FIG.\,\ref{fig_s2New_r4} the case $a_4>0$ by the light-blue line, from which we find that the overall shape (especially the monotonicity) does not change\,\cite{CL24-b}.

The feature of $s^2$ for Newtonian stars near $\widehat{\varepsilon}\approx1$ could be extracted straightforwardly from their structure equations.
Starting directly from Eq.\,(\ref{s2-N}), we obtain,
\begin{align}\label{prf}
\left(\frac{\d s^2}{\d\widehat{\varepsilon}}\right)_{\rm{N}}
=&-\frac{3\widehat{\varepsilon}}{\widehat{r}^3}\left(\frac{\d\widehat{\varepsilon}}{\d\widehat{r}}\right)^2\left(\frac{\widehat{r}^3\widehat{\varepsilon}}{3}-\widehat{M}\right)+
\frac{\widehat{\varepsilon}\widehat{M}}{\widehat{r}^2}
\left(\frac{\d\widehat{\varepsilon}}{\d\widehat{r}}\right)^{-3}
\left[\frac{\d^2\widehat{\varepsilon}}{\d\widehat{r}^2}
-\frac{1}{\widehat{r}}\frac{\d\widehat{\varepsilon}}{\d\widehat{r}}
\left(1+\frac{\widehat{r}}{\widehat{\varepsilon}}\frac{\d\widehat{\varepsilon}}{\d\widehat{r}}\right)
\right],
\end{align}
where $\widehat{M}=\int\d\widehat{r}\widehat{r}^2\widehat{\varepsilon}$.
Since $
[{\d}/{\d\widehat{r}}]({\widehat{r}^3\widehat{\varepsilon}}/{3}-\widehat{M})=3^{-1}\widehat{r}^3{\d\widehat{\varepsilon}}/{\d\widehat{r}}<0$ (notice $\d\widehat{M}/\d\widehat{r}=\widehat{r}^2\widehat{\varepsilon}$ and $\d\widehat{\varepsilon}/\d\widehat{r}<0$),
i.e., $\widehat{r}^3\widehat{\varepsilon}/3-\widehat{M}$ decreases with increasing of $\widehat{r}$ (starting from 0), therefore we have $\widehat{r}^3\widehat{\varepsilon}/3-\widehat{M}<[\widehat{r}^3\widehat{\varepsilon}/3-\widehat{M}]_{\widehat{r}=0}=0$. Moreover,  the factor in square bracket of Eq.\,(\ref{prf}) is negative, one then has $(\d s^2/\d\widehat{\varepsilon})_{\rm{N}}>0$ definitely (near $\widehat{r}=0$), i.e.,  $s^2$ is an increasing function of $\widehat{\varepsilon}$ near $\widehat{\varepsilon}\approx1$. This is equivalent to $l_2<0$ for Newtonian stars.

\subsection{Violation of the perturbative QCD conformal bound on SSS and the average SSS in NSs}\label{sub_VioCB}

\begin{figure}[h!]
\centering
\includegraphics[height=7.cm]{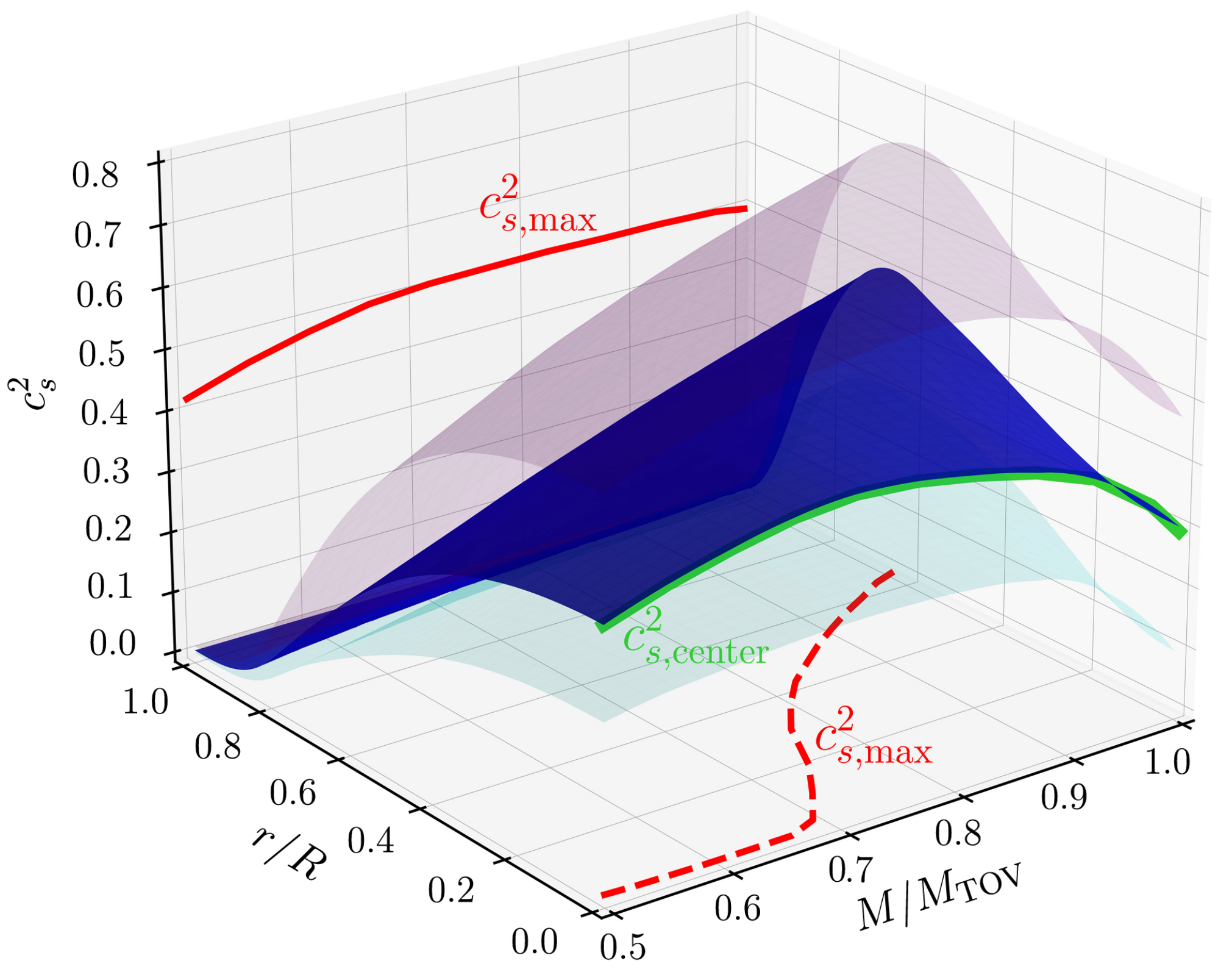}\quad
\includegraphics[height=6.2cm]{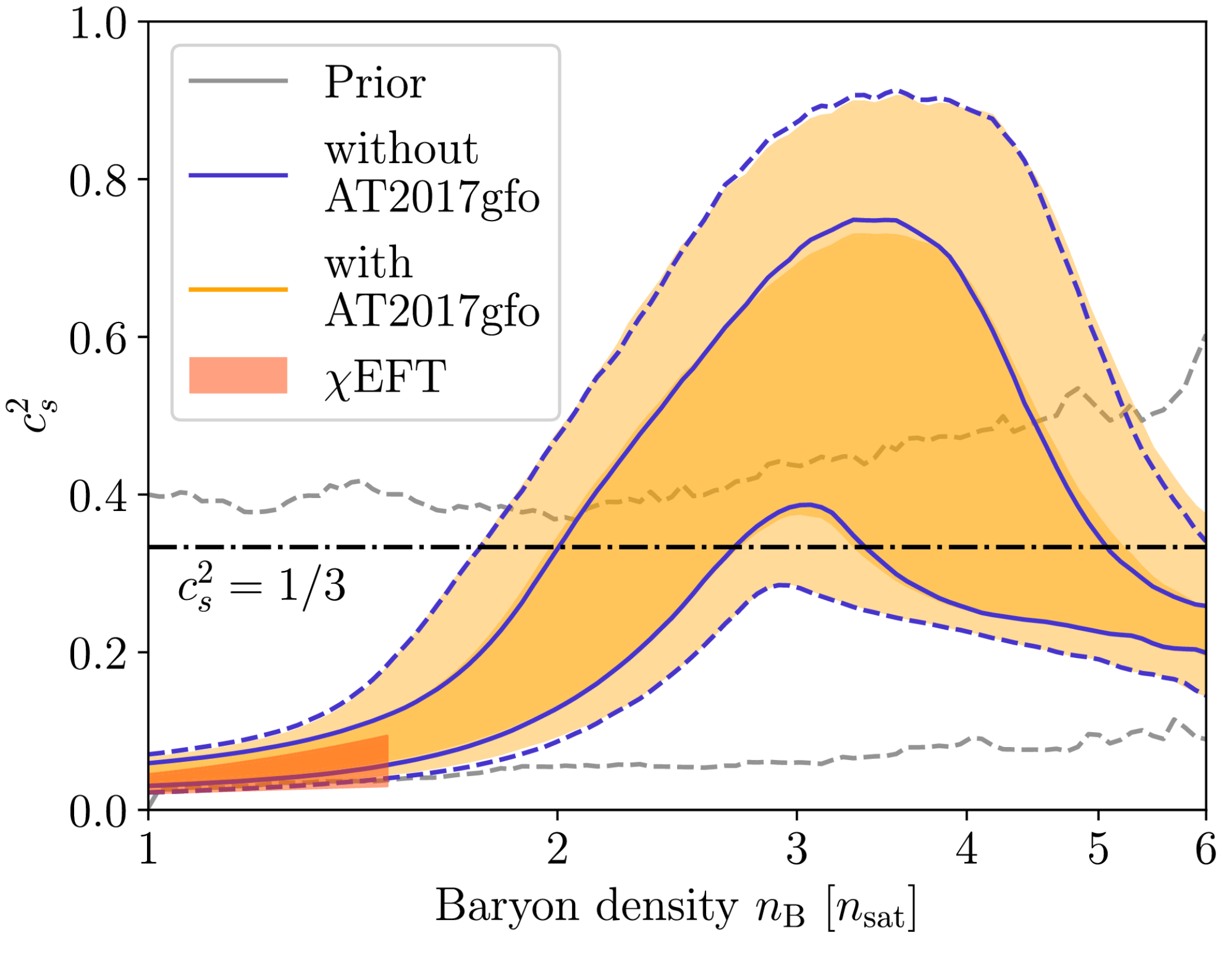}
\caption{(Color Online). Left panel: dependence of the SSS in NSs on NS mass and on the radial distance from the center. Figure taken from Ref.\,\cite{Ecker2022}. Right panel: posterior on the SSS as a function of baryon density with/without incorporating the constraints from AT2017gfo\,\cite{Abbott2017gfo}. Figure taken from Ref.\,\cite{Pang24}.
}\label{fig_s2-ff}
\end{figure}

In this subsection, we study the SSS in NS cores and the possible violation of the conformal bound on SSS, using our expression for the latter.
The central SSS of a NS at the maximum-mass configuration is given by Eq.\,(\ref{sc2-TOV}), where only the combination $\x$ is needed instead of the individual $P_{\rm{c}}$ and $\varepsilon_{\rm{c}}$.
For the $\x\approx0.24_{-0.07}^{+0.05}$ extracted for PSR J0740+6620\,\cite{CLZ23-a}, one obtains\,\cite{CLZ23-a},
\begin{equation}\label{sc2-6620}
\boxed{\mbox{PSR J0740+6620:}~~
s_{\rm{c}}^2\approx0.45_{-0.18}^{+0.14}.}
\end{equation}
This value is consistent with constraints on the SSS from other studies.
For example, we shown in the left panel of FIG.\,\ref{fig_s2-ff} the $s^2$ as a function of the radial distance from the center for general NSs\,\cite{Ecker2022},  using more than a million randomly generated EOSs satisfying all known theoretical and observational constraints.
In that work, the central SSS for a NS at the TOV configuration is found to be about $s_{\rm{c}}^2\approx0.3$\,\cite{Ecker2022}.
In addition, Ref.\,\cite{Ecker2022} also found that the maximum SSS $s_{\max}^2$ in NSs approaches some fixed value about $s_{\max}^2\approx0.5$. Considering that the maximum enhancement on the central SSS is less than about 10\%\,\cite{CLZ23-b} (see discussions given in the following subsections of this section) if we perturb $s^2$ to places near the center, we find that the maximum $s_{\max}^2$ based on (\ref{sc2-6620}) is similarly about 0.5.
In the right panel of FIG.\,\ref{fig_s2-ff}, a similar plot for the SSS as a function of baryon density is shown\,\cite{Pang24}. It is seen that at densities $\approx(5\pm1)\rho_0$ (which is similar as the central density in massive NSs), the $s^2$ is around 0.3, being consistent with both Ref.\,\cite{Ecker2022} and our constraint (\ref{sc2-6620}).

\begin{figure}[h!]
\centering
\includegraphics[width=7.5cm]{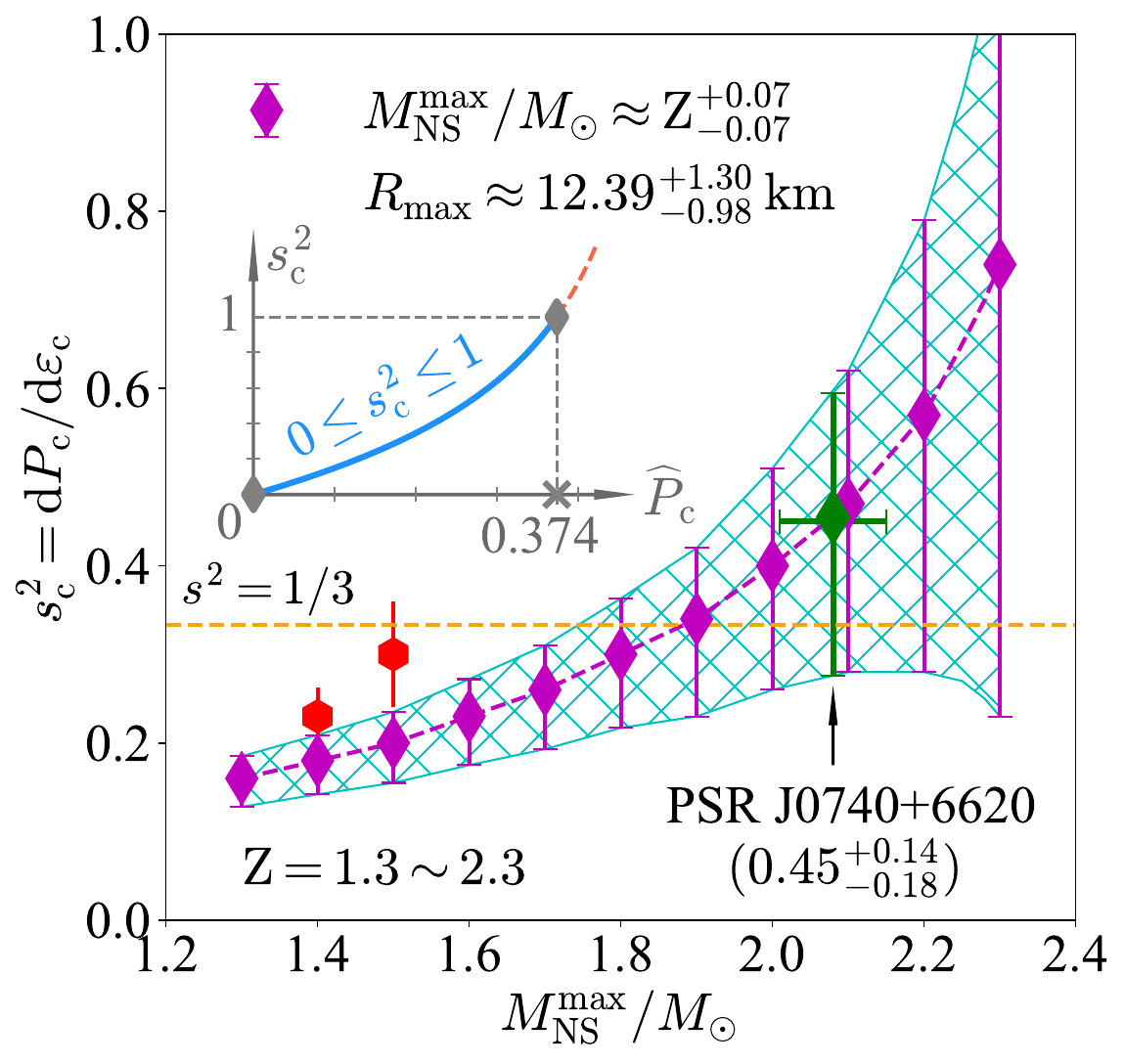}
\caption{(Color Online). Dependence of $s_{\rm{c}}^2$ on the maximum mass $M_{\rm{NS}}^{\max}/M_{\odot}$,  the conformal bound (CB) on SSS is indicated by the dashed orange line.
The inset plots the $s_{\rm{c}}^2$ as a function of ${\x}$, where the lightblue curve is allowed by conditions of stability and causality ($0\leq s_{\rm{c}}^2\leq1$).
Figure taken from Ref.\,\cite{CLZ23-b}.
}
\label{fig_sckk}
\end{figure}

\begin{figure}[h!]
\centering
\includegraphics[height=5.5cm]{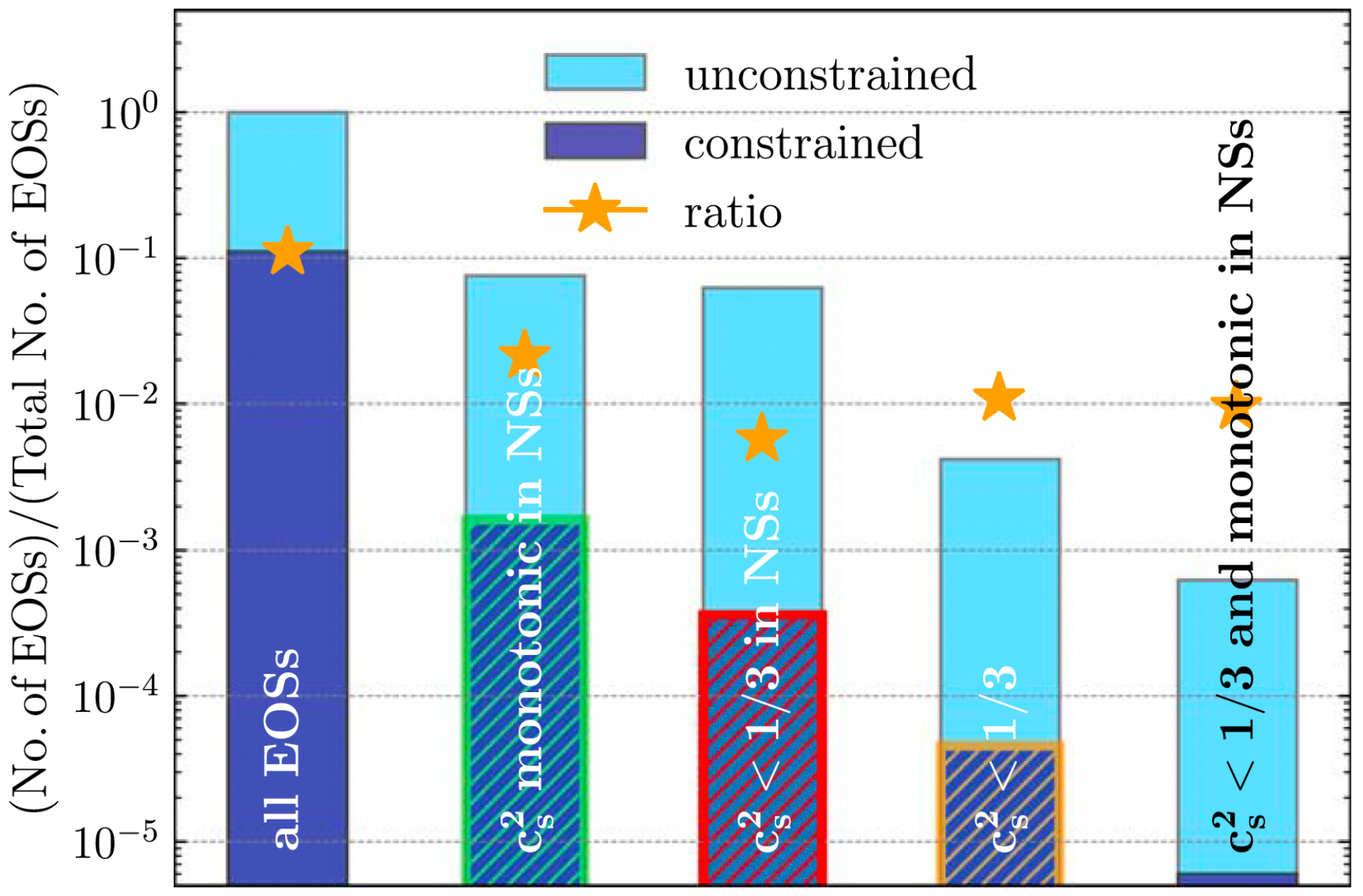}\\[0.5cm]
\includegraphics[height=7.4cm]{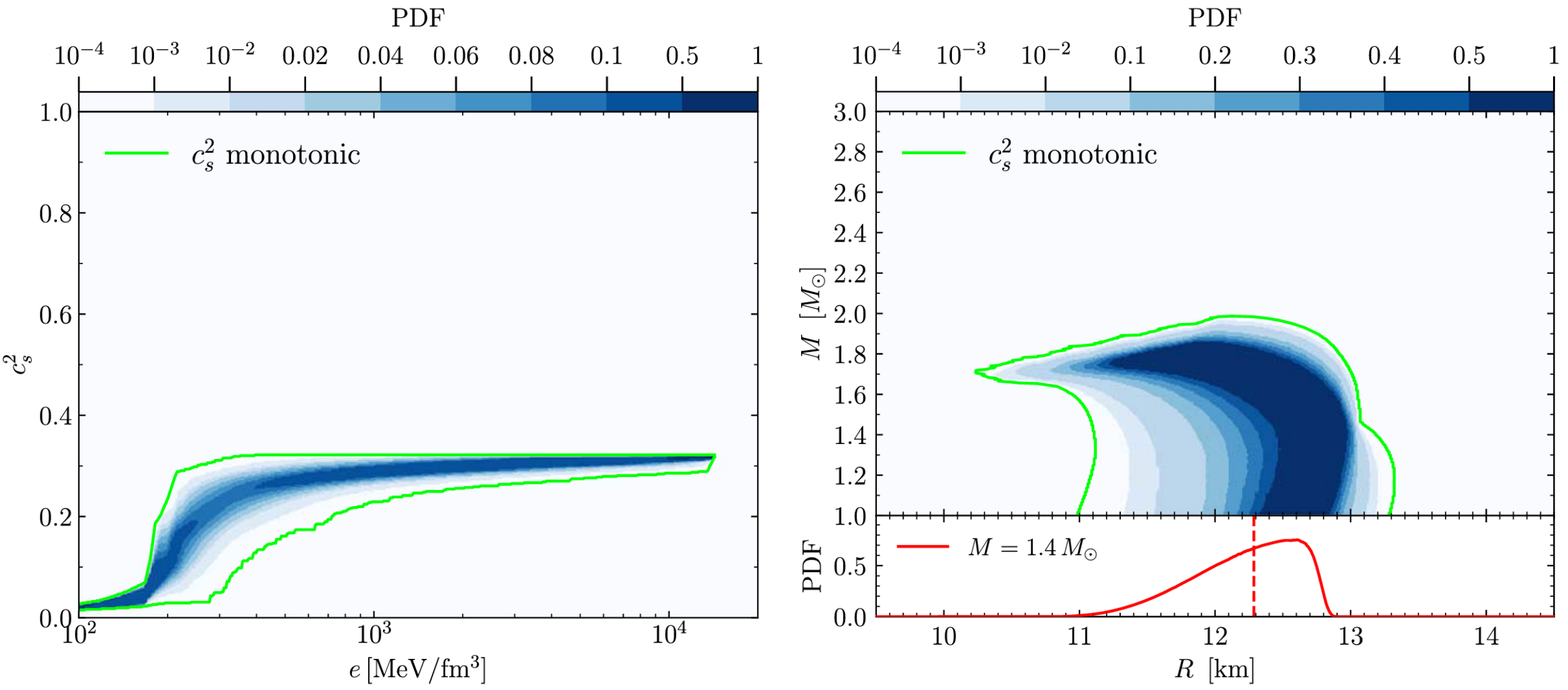}
\caption{(Color Online).  Upper panel: schematic representation of the various sets into which the total sample of EOSs can be decomposed, either when subject to the observational constraints (dark blue) or when not (light blue). 
Lower panel: PDFs for the SSS as a function of the energy density, and of the NS mass as a  function of the stellar radius; here the $s^2$ is monotonic and is limited in the range of $0\leq s^2\leq1/3$. Figures taken from Ref.\,\cite{Altiparmak2022}. }\label{fig_RN-s2-ALT}
\end{figure}

The central SSS for $\x\leq1/3$ is bounded as,
\begin{equation}
\boxed{
s_{\rm{c}}^2(\x\leq1/3)\leq7/9\approx0.778,~~\mbox{for NSs at the TOV configuration}.}
\end{equation}
However, it does not mean that an URFG has a SSS of $7/9$ since $\x=1/3$ (which holds only at the center) does not generally indicate $P=\varepsilon/3$ (which is a functional relationship for the whole range of $\varepsilon$).
The dependence of $s_{\rm{c}}^2$ on $\x$ could be straightforwardly transformed into its dependence on $M_{\rm{NS}}^{\max}/M_{\odot}$, under some assumptions.
Here, we assume that NSs with masses $\rm{Z}_{-0.07}^{+0.07}M_{\odot}$ where $\rm{Z}=1.3\mbox{$\sim$}2.3$ have similar radii as indicated by the NICER observations to extract the $\nu_{\rm{c}}$ from Eq.\,(\ref{Rmax-n}).
In particular,  NICER found that the radius of PSR J0740+6620 (mass $\approx2.08_{-0.07}^{+0.07}M_{\odot}$) is about $12.39_{-0.98}^{+1.30}\,\rm{km}$\,\cite{Riley21} while that of PSR J0030+0451 (mass $\approx1.34_{-0.16}^{+0.15}M_{\odot}$) is about $12.71_{-1.19}^{+1.14}\,\rm{km}$\,\cite{Riley19}.  They are very similar, indicating that the NS radius is less sensitive to the central EOS\,\cite{CLZ23-a}.
We therefore have $10^3\nu_{\rm{c}}\approx11.2\,\rm{fm}^{3/2}/\rm{MeV}^{1/2}$ by using $R_{\max}\approx12.39_{-0.98}^{+1.30}\,\rm{km}$\,\cite{Riley21}.
Perturbatively, we have for the SSS at NS center from expanding Eq.\,(\ref{sc2-TOV}):
\begin{align}
s_{\rm{c}}^2&\approx\frac{4}{3}{\x}\left(1+{\x}+\frac{3}{2}{\x}^2+3{\x}^3\right)+\mathcal{O}\left({\x}^5\right)
\approx\frac{4}{3}H\left(1+5H+\frac{57}{2}H^2+175H^3\right)+\mathcal{O}\left(H^5\right),\label{cdk-1}
\end{align}
where $H\approx0.052(M_{\rm{NS}}^{\max}/M_{\odot}+0.106)\ll1$.
Keeping only the leading-order term of (\ref{cdk-1}) gives the Newtonian prediction $s_{\rm{c}}^2=4{\x}/3$ from which one infers ${\x}\lesssim3/4$\,\cite{CL24-c}.
The GR contributions in the TOV equations have an effect about 100\% on the upper limit for ${\x}$.
Essentially,  the $s_{\rm{c}}^2$ increases as $M_{\rm{NS}}^{\max}$ increases as shown in FIG.\,\ref{fig_sckk}.
The results shown there provide us a straightforward way to infer the central SSS $s_{\rm{c}}^2$ once the radii/masses are known (measured/observed). It also implies that more compact NSs (characterized by $\xi$ or equivalently by $\x$ through Eq.\,(\ref{gk-comp})) have larger $s_{\rm{c}}^2$, e.g., NSs of masses $1.5M_{\odot}$ and $1.4M_{\odot}$ with radii 9.9$\sim$11.2\,km\,\cite{Ozel2016} and $11.0_{-0.6}^{+0.9}\,\rm{km}$\,\cite{Capano2020} have $s_{\rm{c}}^2\approx0.31_{-0.07}^{+0.07}$ and $s_{\rm{c}}^2\approx0.23_{-0.05}^{+0.03}$,  respectively, shown as red hexagons in FIG.\,\ref{fig_sckk}.
{\color{xll}Assuming a canonical NS at the TOV configuration is unreasonable based on our current knowledge from both astrophysical theories and observations, in order to estimate the central SSS of such NSs the factor $\Psi$ should be taken into account appropriately. We study this issue in Subsection \ref{sub_s2canon}, see Eq.\,(\ref{sc2-GG}) and Eq.\,(\ref{def-Psi}).
Consequently, although $\x$ is smaller for a canonical NS its $\Psi$ is greater than zero, and the $s_{\rm{c}}^2$ is larger than about $1/3$ estimated here; implying that the conformal bound for $s^2$ is also violated in canonical NSs.}

We also show the conformal bound (CB) on the SSS of $s^2=1/3$\,\cite{Hohler2009,Cherman2009} in FIG.\,\ref{fig_sckk} by the dashed orange line. It is seen that the CB for SSS tends to break down for $M_{\rm{NS}}^{\max}/M_{\odot}\gtrsim1.89$.
This mass threshold is very close to the most probable critical maximum-mass about $M_{\rm{NS}}^{\max}/M_{\odot}\approx1.87$ given in Ref.\,\cite{Bed15}, above which the CB is likely to break down, see FIG.\,\ref{fig_RN-s2-BS}.  Similarly, Ref.\,\cite{Altiparmak2022} predicted that the largest mass of NSs being consistent with the CB is $\lesssim1.99M_{\odot}$ indicated by the lower panel of FIG.\,\ref{fig_RN-s2-ALT}. In fact, the number of EOSs passing through all the astrophysical/microscopic physics constraints but with $s^2\leq1/3$ is very small as shown in the upper panel of FIG.\,\ref{fig_RN-s2-ALT}.
Furthermore, the CB $s_{\rm{c}}^2\leq1/3$ is {essentially broken} for ${\x}\gtrsim0.195$ from Eq.\,(\ref{sc2-TOV}). This finding is consistent with conclusions from related studies on the same issue, e.g., see Ref.\,\cite{McLerran2019,Tan2022-a,Tan2022-b,Altiparmak2022,Ecker2022,Ecker23,Alsing2018,Tews2018,Leon20,Miao2021}, indicating that Eq.\,(\ref{sc2-TOV}) grasps the main features of the SSS. However, we may emphasize that our method adopts no specific model for NS matter EOS and also make no assumptions about the composition of NSs (such as nucleons, hyperons and/or quarks), therefore providing a EOS-model independent way for investigating the $s^2$.
Actually,  we can deduce from Eq.\,(\ref{sc2-TOV}) that $s_{\rm{c}}^2\to1/3$ occurs earlier than ${\x}\to1/3$, explaining why the CB for SSS is highly likely {to break down} in NS cores\,\cite{CLZ23-a,CLZ23-b}.
In addition, since the NS EOSs are often softened considering the exotic components such as hyperons, our formula (\ref{sc2-TOV}) indicates that the CB in these NSs with non-nucleonic particles tends to be obeyed\,\cite{Stone21,Otto20,Mott21}.

Next, we emphasize that although TOV NSs with lower maximum-mass have $s_{\rm{c}}^2\lesssim1/3$, it does not mean that these NSs exhibit conformal symmetry (in their cores) since the latter only emerges at extremely high energies where the strong force coupling constant diminishes, leading to the ``asymptotic freedom'' of quarks and gluons. This symmetry is broken as the energy scale lowers. At the energy scales relevant to NS cores, the presence of massive particles and strong gravitational fields disrupt conformal symmetry. The latter being not exact even in a quark-gluon plasma is generally expected to break down as the system transitions to a hadronic phase including more massive particles. Therefore, the observation of $s_{\rm{c}}^2\lesssim1/3$ in low-mass NSs should not be interpreted as implying that the conformal symmetry is reached there\,\cite{CLZ23-b}.

\begin{figure}[h!]
\centering
\includegraphics[width=16.5cm]{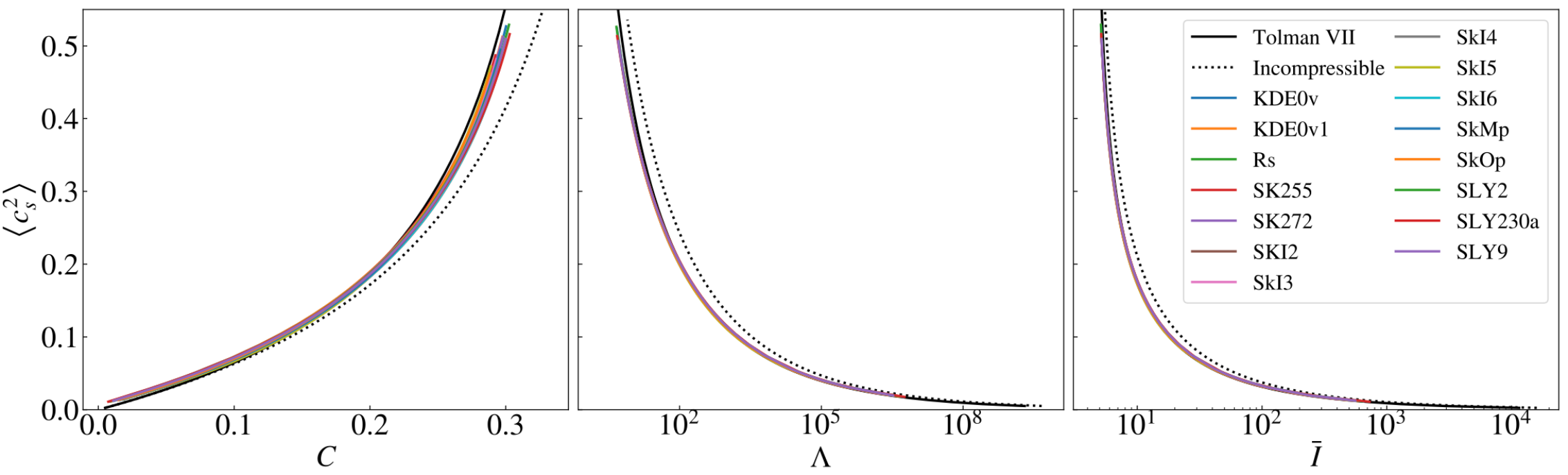}
\caption{(Color Online).  Relation between the $\x$ (also denoted by $\langle c_{\rm{s}}^2\rangle$) and the compactness $\xi$ (C), tidal deformability $\Lambda$, and dimensionless moment of inertia $\overline{I}$ using a set of realistic EOSs.
Figure taken from Ref.\,\cite{Saes2024}.
}\label{fig_ILC}
\end{figure}

\begin{figure}[h!]
\centering
\includegraphics[width=10.cm]{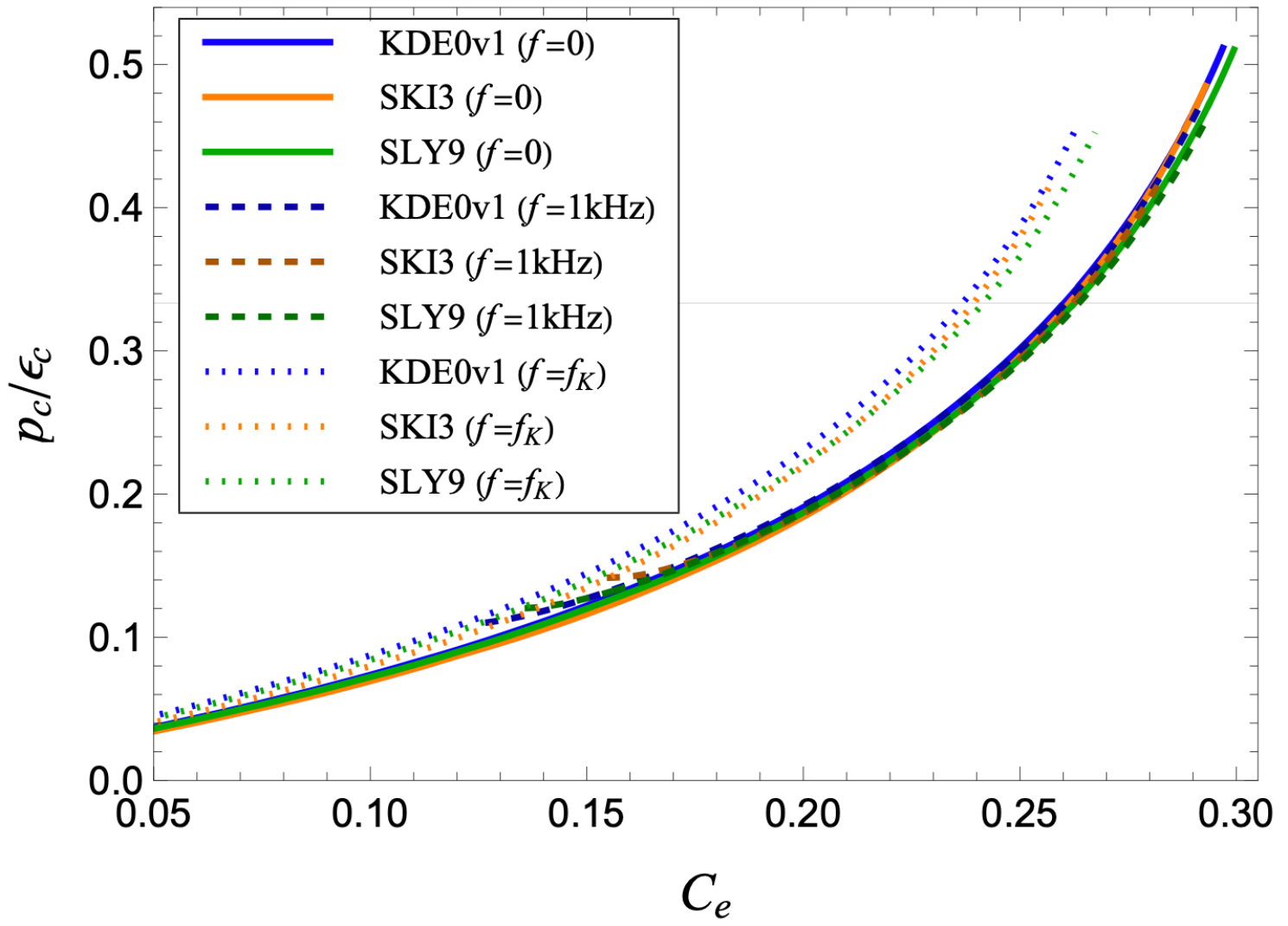}
\caption{(Color Online). Correlation between $\x$ (average SSS) and NS compactness using three empirical EOSs with/without rotations. Figure taken from Ref.\,\cite{Mendes2024}. 
}\label{fig_Mendes24X}
\end{figure}

We can work out the $\x$ expressed in terms of the compactness $\xi$ perturbatively.
Denoting the in-front coefficient of the scaling between $\xi$ and $\Pi_{\rm{c}}$ as $\tau$, i.e.,  $\xi\approx\tau\Pi_{\rm{c}}$, Eq.\,(\ref{gk-comp}) then gives 
\begin{equation}
\x
=-\frac{4\xi-\tau+\sqrt{4\xi^2-8\tau\xi+\tau^2}}{6\xi}
\approx\frac{\xi}{\tau}+4\left(\frac{\xi}{\tau}\right)^2
+19\left(\frac{\xi}{\tau}\right)^3+100\left(\frac{\xi}{\tau}\right)^4+\cdots.
\end{equation}
The value of $\tau$ is about $\tau\approx2$ by the scaling of Eq.\,(\ref{gk-comp}) and about $\tau\approx2.3$ by the numerical verification of the scaling between $\xi$ and $\x$ using meta-model EOSs, see the Subsection \ref{sub_1420}.
Therefore, the above relation could be approximated as $\x\approx\xi/2+\xi^2+19\xi^3/8+25\xi^4/4+\cdots$ from Eq.\,(\ref{gk-comp}) or $
\x\approx0.43\xi+0.76\xi^2+1.56\xi^3+3.56\xi^4+\cdots$ from the numerical verification;
these relations are also nearly universal.
Recently, Ref.\,\cite{Saes2024} studied the correlation between $\x$ (they called it the average SSS and denoted it as $\langle s^2\rangle$) and the compactness $\xi$, using both a few realistic EOSs and some simplified approximations like the incompressible fluid, the Tolman VII fluid and the Buchdahl fluid. Interestingly, for the latter three simplified cases, one can work out to order $\xi^2$ explicitly the following\,\cite{Saes2024}
\begin{empheq}[box=\fbox]{align}
\mbox{incompressible fluid:}&~~\x\approx\frac{\xi}{2}+\xi^2;\label{xixixi-1}\\
\mbox{Tolman VII fluid:}&~~\x\approx\frac{\xi}{2}+\frac{133\xi^2}{120};\label{xixixi-2}\\
\mbox{Buchdahl fluid:}&~~\x\approx\frac{\xi}{2}+\frac{5\xi^2}{4}.\label{xixixi-3}
\end{empheq}
The correlation between $\x\equiv\langle c_{\rm{s}}^2\rangle=\langle s^2\rangle$\,\cite{Saes2024} and the compactness $C\equiv\xi$ is shown in the left panel of FIG.\,\ref{fig_ILC} (with an uncertainty less than about 10\%). Also shown are the correlations between $\x$ and the tidal deformability $\Lambda$ (middle) and the dimensionless moment of inertial $\overline{I}$ (right). {\color{xll}In their work\,\cite{Saes2024}, the average speed of sound squared $\x=\langle c_{\rm{s}}^2\rangle$ has been interpreted as a measure of the mean stiffness of the EOS up to the central energy density of a given NS.}
A similar plot on the correlation between $\x$ and the NS compactness is shown in FIG.\,\ref{fig_Mendes24X}, using three empirical EOSs with/without considering rotations\,\cite{Mendes2024}.

Using our formula for the central SSS of Eq.\,(\ref{sc2-GG}), we have
\begin{equation}\label{EQ-1}
s_{\rm{c}}^2\approx\frac{4+\Psi}{3\tau}\xi+\frac{4}{3}\frac{5+2\Psi}{\tau^2}\xi^2+\frac{38+19\Psi}{\tau^3}\xi^3+\frac{100}{3}\frac{7+4\Psi}{\tau^4}\xi^4+\cdots.
\end{equation}
We have established that $\x$ is also a measure of compactness, i.e., $\x\sim\xi$, see Eq.\,(\ref{gk-comp}). We can reasonably say that the central stiffness $s_{\rm{c}}^2$ effectively characterizes the same physics as $\x$ (which is the average SSS as first shown in Ref.\,\cite{Saes2024}); and since $\x$ and $\xi$ are monotonically correlated, we have
at NS centers the correspondence 
\begin{equation}\label{EQ-CS}
\boxed{
\mbox{compactness }\xi\leftrightarrow \mbox{central pressure over energy density ratio $\x$ (average SSS) }\leftrightarrow\mbox{stiffness $s_{\rm{c}}^2$}.}
\end{equation}
This interpretation is consistent with that in Ref.\,\cite{Saes2024}.
However, it does not mean that the stiffness generally characterized by $s^2$ at places other than the origin ($\hr=0$) is also effectively equivalent to $\phi=P/\varepsilon$.
This is because the latter is probably a monotonically increasing function of $\hr$ when going toward the NS center, while the $s^2(\hr)$ may still have sizable probabilities to be larger than $s_{\rm{c}}^2$. Namely, $s^2(\hr)$ is not monotonic as a function of the radial distance $\widehat{r}$.
Similarly, although $\x$ or $s_{\rm{c}}^2$ (defined at $\hr=0$) characterizes the same physics as $\xi$, $\phi$ or $s^2$ (defined at finite $\hr$) generally can not be used to quantify the compactness.

\subsection{Central SSS in canonical NSs from NS compactness scaling}\label{sub_s2canon}

In this subsection, we estimate the central SSS for canonical NSs.
The relevant expression for $s_{\rm{c}}^2$ is Eq.\,(\ref{sc2-GG}),  and there are two approaches for determining $\x$: one from the mass and radius scalings and the other using the compactness scaling alone.
In the latter way, the process can be summarized as
\begin{equation}\label{can-sc2}
\xi\mbox{-scaling}\xrightarrow[\displaystyle\xi\approx A_\xi\Pi_{\rm{c}}+B_\xi]{\displaystyle M_{\rm{NS}}/R} \x\xrightarrow[\displaystyle s_{\rm{c}}^2=\x\left(1+\frac{1+\Psi}{3}\frac{1+3\x^2+4\x^2}{1-3\x^2}\right)]{\displaystyle M_{\rm{NS}}/M_{\odot}\to \Psi\mbox{ (FIG.\,\ref{fig_k-fac})}}s_{\rm{c}}^2.
\end{equation}
Obviously, this approach should give a prediction for $s_{\rm{c}}^2$ with small uncertainty, fundamentally because the compactness scaling is much strong than the radius scaling.

In order to use Eq.\,(\ref{can-sc2}) to calculate the central SSS in a normal NS, the knowledge of $\Psi$ especially its dependence on $M_{\rm{NS}}$ is desired, this is given in Subsection \ref{sub_PsiVert}.
Putting $\Psi\approx2.85\pm0.29$ for canonical NSs into the formula (\ref{can-sc2}) and considering $\x\approx0.15\pm0.02$ (see the last row of TAB.\,\ref{tab_cEOS}) using directly the $\xi$-scaling, we obtain
\begin{equation}\label{can-sc2-xi}
\boxed{
s_{\rm{c}}^2\approx0.47_{-0.09}^{+0.09},~~\mbox{for a canonical NS with }R\approx12_{-1}^{+1}\,\rm{km}.}
\end{equation}
On the other hand, if we use the mass and radius scalings to evaluate the $\x$ which is about $\x\approx0.14_{-0.03}^{+0.03}$ (see TAB.\,\ref{tab_cEOS}), the prediction on the central SSS is about $0.46\pm0.16$, which has a similar magnitude as (\ref{can-sc2-xi}) but induces a larger uncertainty. 

\begin{figure}[h!]
\centering
\includegraphics[width=8.5cm]{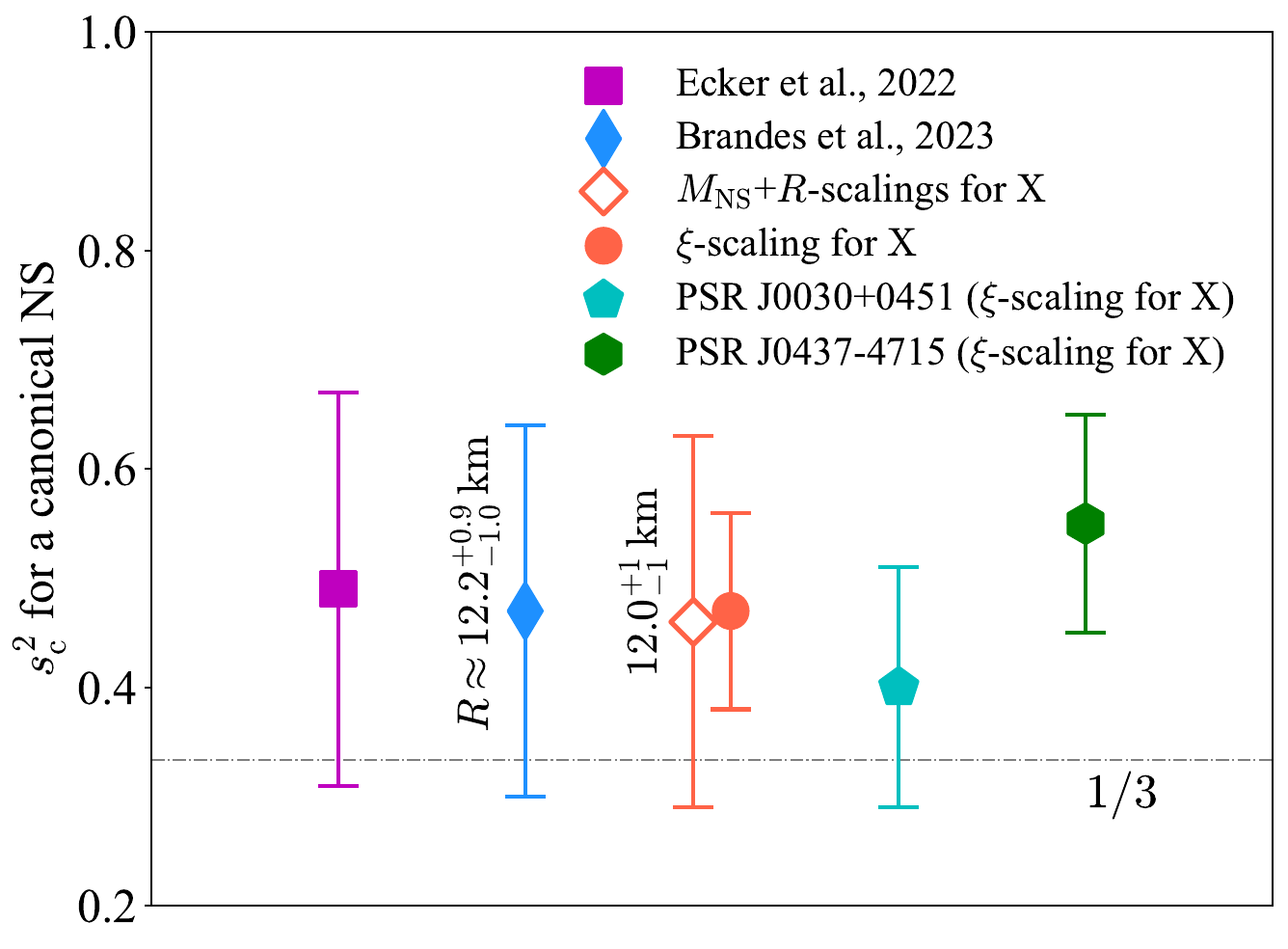}
\caption{(Color Online). Comparison on the $s_{\rm{c}}^2$ for a canonical NS.}\label{fig_sc2-can}
\end{figure}
Shown in FIG.\,\ref{fig_sc2-can} is a comparison of the central SSS obtained using our approach and some existing ones in the literature \,\cite{Ecker2022,Brandes2023-a}.
It is seen that our results are in excellently agreement with these investigations but have relatively smaller uncertainties.
Moreover, since PSR J0030+0451\,\cite{Riley19} and PSR J0437-4715\,\cite{Choud24} both have a mass near $1.4M_{\odot}$, they could be treated as canonical NSs. Using the observational NS masses and radii of these two pulsars (consequently one could obtain the compactness straightforwardly), we can evaluate the central SSS similarly; the results are given in TAB.\,\ref{tab_M14_sc2} and are plotted in FIG.\,\ref{fig_sc2-can}.
Since the observed radius of PSR J0437-4715 is much smaller than that of PSR J0030+0451, the inferred central SSS is correspondingly much larger, as the matter is deeply compressed in its core.
\begin{table}[h!]
\renewcommand{\arraystretch}{1.5}
\centerline{\normalsize
\begin{tabular}{c|c|c} 
  \hline
canonical NS&$s_{\rm{c}}^2$&radius\\\hline\hline
Ecker et al., 2022&$0.49\pm0.18$&\\\hline
Brandes et al., 2023&$0.47\pm0.17$&$12.2_{-1.0}^{+0.9}\,\rm{km}$\\\hline\hline
mass and radius scalings for $\x$&$0.46\pm0.16$&$12_{-1}^{+1}\,\rm{km}$\\\hline
$\xi$-scaling for $\x$ &$0.47\pm0.09$&\\\hline\hline
PSR J0030+0451 ($\xi$-scaling for $\x$)&$0.40\pm0.11$&$12.71_{-1.19}^{+1.14}\,\rm{km}$\\\hline
PSR J0437-4715 ($\xi$-scaling for $\x$)&$0.55\pm0.10$&$11.36_{-0.63}^{+0.95}\,\rm{km}$\\\hline
    \end{tabular}}
        \caption{SSS for a $1.4M_{\odot}$ NS at a 68\% confidence level. }\label{tab_M14_sc2}        
\end{table}

Our prediction for the central SSS of PSR J0437-4715 and a canonical NS even at the lower limit of the 68\% confidence level is greater than the conformal bound of $s^2=1/3$, implying that the latter is highly broken in a canonical NS. This conclusion is more reasonable than the one given in FIG.\,\ref{fig_sckk} since canonical NSs can hardly be treated as in the TOV configuration regarding obeservational facts.

\subsection{$1^{\rm{st}}$-order expansion of SSS in reduced energy density: is the central SSS greater than its surroundings?}\label{sub_s2_1st}

In this and the next subsections, we develop perturbative approximations for $s^2$ as a function of radial distance $\widehat{r}$ from the center (or of the reduced energy density $\mu=\widehat{\varepsilon}-1$).
This subsection is devoted to the issue whether $s_{\rm{c}}^2$ is the maximum value, i.e., whether $s^2(\widehat{\varepsilon})$ could be larger than $s_{\rm{c}}^2$ or not.
Then in Subsection \ref{sub_s2_2nd}, we estimate the location of the possible peak in $s^2(\widehat{\varepsilon})$.

Using the sum rules of (\ref{Peps-sumrule}), we can write the coefficient $a_4$ appearing in $\heps$ in terms of the coefficient $d_k$,
\begin{equation}\label{def-a4}
a_4={s_{\rm{c}}^{-2}}\left(b_4-a_2^2\sum_{k=1}^K2^{-1}{k(k-1)}d_k\right),
\end{equation}
Here $K$ is the effective truncation order of the expansions.
Putting the expression of $a_4$ of Eq.\,(\ref{def-a4}) into Eq.\,(\ref{s2_r_exp}) leads to the final expression for $s^2$ as,
\begin{empheq}[box=\fbox]{align}\label{oo-2}
s^2(\widehat{r})\approx s_{\rm{c}}^2\left[1+({2}/{b_2})\left(b_4-s_{\rm{c}}^2a_4\right)\widehat{r}^2\right]+\mathcal{O}\left(\widehat{r}^4\right)\approx
s_{\rm{c}}^2+2a_2D\widehat{r}^2
+\mathcal{O}(\hr^4),~~D=\sum_{k=1}^{K}2^{-1}{k(k-1)}d_k.
\end{empheq}
By using $\widehat{\varepsilon}(\widehat{r})\approx1+a_2\widehat{r}^2$,  we can transfer the dependence of $s^2$ on $\widehat{r}^2$ to that on $\widehat{\varepsilon}$:
\begin{equation}\label{s2eps}
s^2(\widehat{\varepsilon})\approx s_{\rm{c}}^2+2D\left(\widehat{\varepsilon}-1\right)=s_{\rm{c}}^2+2D\mu,
\end{equation}
where $\mu=\heps-1$, see Eq.\,(\ref{RE-small1}).
The approximation (\ref{s2eps}) is expected to be valid only near the NS centers; nevertheless, the sign of $D$ is sufficient for our purpose to deduce a relative relation between $s^2(\widehat{\varepsilon}$) and $s_{\rm{c}}^2$, i.e.,
\begin{equation}
\boxed{
D<0\leftrightarrow D\mu>0\leftrightarrow s_{\rm{c}}^2<s^2(\widehat{\varepsilon})
\leftrightarrow \mbox{``reduction of $s^2$\;toward\;NS\;centers''}.}
\end{equation}
In both Eq.\,(\ref{oo-2}) and Eq.\,(\ref{s2eps}) the zeroth-order term is the SSS at center,  therefore what we investigate is the possible crossover at zero temperature and densities realized in NSs with densities $\gtrsim5\rho_{\rm{sat}}$\,\cite{Brandes2023,Fuku2020,Fuji2023,ZhangLi2023a}, with a peaked behavior in $s^2$ (next subsection).
Besides the sum rules of (\ref{Peps-sumrule}),  the stability/causality condition for any $\widehat{\varepsilon}$, i.e., $0\leq s^2\leq 1$, put further constraints on the coefficients $\{d_k\}$:
\begin{equation}\label{oo-1}
\boxed{
0\leq d_1+2d_2\widehat{\varepsilon}+3d_3\widehat{\varepsilon}^2+\cdots\leq1,}
\end{equation}
which also guarantees the pressure $\widehat{P}$ never becomes negative.

For the TOV configuration $M_{\rm{NS}}^{\max}$ on the NS M-R curve, the inequality $\d^2M_{\rm{NS}}^{\max}/\d\varepsilon_{\rm{c}}^2<0$ gives a criterion (\ref{io-7}) with $s_{\rm{c}}^2$ given by Eq.\,(\ref{sc2-TOV}).
Inequality (\ref{io-7}) implies, e.g.,  for PSR J0740+6620 that $\d s_{\rm{c}}^2/\d{\x}\lesssim2.74$ using ${\x}\approx0.24$\,\cite{CLZ23-a}, which means if ${\x}$ increases by about 0.1, the increasing of $s_{\rm{c}}^2$ should be smaller than 0.274.
Moreover, we obtain using the relation $\d s^2/\d\widehat{P}=\d s^2/\d\widehat{\varepsilon}\cdot\d\widehat{\varepsilon}/\d\widehat{P}=\d^2\widehat{P}/\d\widehat{\varepsilon}^2\cdot\d\widehat{\varepsilon}/\d\widehat{P}$ an equivalent form of (\ref{io-7}),
\begin{equation}\label{io-5}
\boxed{
\left.\frac{\d^2\widehat{P}}{\d\widehat{\varepsilon}^2}\right|_{\widehat{\varepsilon}=\widehat{\varepsilon}_{\rm{c}}=1}=\sum_{k=1}^Kk(k-1)d_k=2D<\sigma_{\rm{c}}^2s_{\rm{c}}^2.}
\end{equation}
The coefficients $d_1$ and $d_2$ via conditions of (\ref{Peps-sumrule}) are given respectively by $
d_1=2{\x}-s_{\rm{c}}^2
+\sum_{k=3}^{K}(k-2)d_k$,  and
\begin{align}
d_2=-{\x}+s_{\rm{c}}^2
-\sum_{k=3}^{K}(k-1)d_k,\label{ref-d2}
\end{align}
In fact, $d_1$ should be zero (since $s^2=0$ for $\widehat{\varepsilon}=0$);
then we can solve for $d_3$ in terms of other $d_k$'s using the previous expression for $d_1$:
\begin{equation}\label{ref-d3}
d_3=-2{\x}+ s_{\rm{c}}^2-\sum_{k=4}^K(k-2)d_k,
\end{equation}
where we split the summation $\sum_{k=3}^K(k-2)d_k$ into $d_3+\sum_{k=4}^K(k-2)d_k$.
Therefore the coefficient $D$ becomes:
\begin{align}
\boxed{
D=\sum_{k=1}^K\frac{k(k-1)}{2}d_k
=d_2+3d_3+\sum_{k=4}^K\frac{k(k-1)}{2}d_k=2s_{\rm{c}}^2-3{\x}+\sum_{k=4}^K\frac{(k-2)(k-3)}{2}d_k.\label{ref-D}}
\end{align}
The first two terms in $D$ are deterministic while the last one has certain randomness (characterizing the uncertainties of the dense matter EOS);
for the special case of $K=3$ the coefficient $D$ becomes deterministic (see the following discussions).

The probability of $D<0$ could be estimated by combining the condition (\ref{oo-1}) of $0\leq s^2\leq1$,  inequality (\ref{io-5}) together with the general sum rules of (\ref{Peps-sumrule}), i.e.,
\begin{align}\label{io-6}
\boxed{\rm{prob}(D<0)
\approx\frac{\#[0\leq s^2\leq1\;\rm{and}\;2D<\sigma_{\rm{c}}^2s_{\rm{c}}^2\;\rm{and}\;D<0]}{\#[0\leq s^2\leq1\;\rm{and}\;2D<\sigma_{\rm{c}}^2s_{\rm{c}}^2]}, }
\end{align}
for the TOV configuration $M_{\rm{NS}}^{\max}$ of the NS M-R curve; here $\#$ counts for the number of samples of [$\dots$].
In the following, we uniformly sample the coefficients $d_k$ for $k\geq4$ within certain ranges fulfilling the requirements $0\leq s^2\leq 1$ and $2D<\sigma_{\rm{c}}^2s_{\rm{c}}^2$ and count the events of $D<0$ to estimate the probability of $D<0$.  The value of $d_2$ could be obtained correspondingly using Eq.\,(\ref{ref-d2}) and $d_3$ via Eq.\,(\ref{ref-d3}).
For a certain truncation order $K$,  different empirical ranges for the parameters $d_k$ are adopted to boost the sampling efficiency.

Considering the lower-order expansions with small $K$ is beneficial for a general analysis.
The simplest situation is $K=3$, i.e., $\widehat{P}\approx d_1\widehat{\varepsilon}+d_2\widehat{\varepsilon}^2+d_3\widehat{\varepsilon}^3\approx d_2\widehat{\varepsilon}^2+d_3\widehat{\varepsilon}^3$.
Correspondingly, we shall obtain $d_2=-s_{\rm{c}}^2+3{\x}$ and $d_3=s_{\rm{c}}^2-2{\x}$,
and the condition $D<0$ (see Eq.\,(\ref{ref-D})) is now equivalent to $
u_{\rm{c}}\equiv 2s_{\rm{c}}^2-3{\x}\approx-{{\x}}/{3}
<0$, with the approximation holding for small ${\x}$ as we have $s_{\rm{c}}^2\approx4{\x}/3$.
Considering Eq.\,(\ref{sc2-TOV}), we then obtain ${\x}\lesssim0.105$ in order to make $D<0$.
For ${\x}\gtrsim0.105$, there is no space for the occurrence of $D<0$, and the transition of the probability for $D<0$ from ${\x}\lesssim0.105$ to ${\x}\gtrsim0.105$ is sharp. This means that $s_{\rm{c}}^2$ is definitely larger than its surroundings for ${\x}\gtrsim0.105$.

Due to the importance of $D$ for our analysis, we explain/illustrate why the $D$ tends to be negative for small ${\x}$ using the next-order polynomial with $K=4$.
We now have two relations,  namely $d_2+d_3+d_4={\x}$ and $2d_2+3d_3+4d_4=s_{\rm{c}}^2$, from which we can solve for $d_2$ and $d_3$ to be expressed in terms of $d_4$.
The general condition $0\leq s^2\leq1$ now reads $0\leq2d_2\widehat{\varepsilon}+3d_3\widehat{\varepsilon}^2+4d_4\widehat{\varepsilon}^3\leq1$.
Putting the expressions of $d_2$ and $d_3$ into it gives the upper and lower limit for $d_4$, i.e., $d_4^{\rm{(l)}}\leq d_4\leq d_4^{\rm{(u)}}$, where
\begin{align}\label{def_d4ul}
d_4^{\rm{(l)}}(\heps)=&\frac{\widehat{\varepsilon}({\x}-s_{\rm{c}}^2)-{\x}+\widehat{\varepsilon}^{-1}}{4\widehat{\varepsilon}^2-3\widehat{\varepsilon}-1},~~
d_4^{\rm{(u)}}(\heps)=\frac{\widehat{\varepsilon}({\x}-s_{\rm{c}}^2)-{\x}}{4\widehat{\varepsilon}^2-3\widehat{\varepsilon}-1},
\end{align}
the dependence of the two quantities on $\heps$ is written explicitly.
Adding the deterministic term $u_{\rm{c}}=-3{\x}+2s_{\rm{c}}^2$, we obtain correspondingly $D^{\rm{(u/l)}}(\heps)=u_{\rm{c}}+d_4^{\rm{(u/l)}}(\heps)$ (see Eq.\,(\ref{ref-D})).

\begin{figure}[h!]
\centering
\includegraphics[width=4cm]{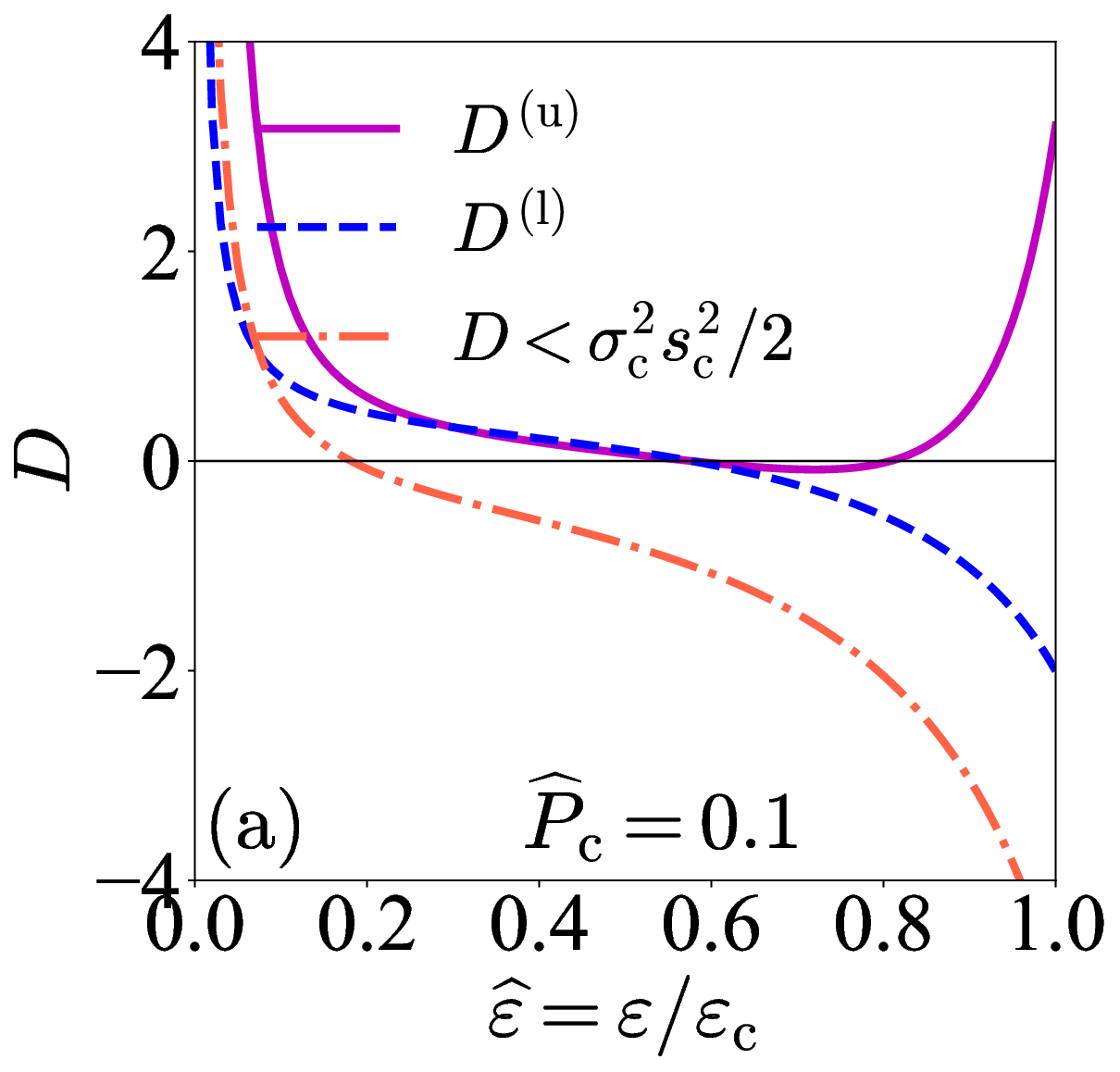}
\includegraphics[width=4cm]{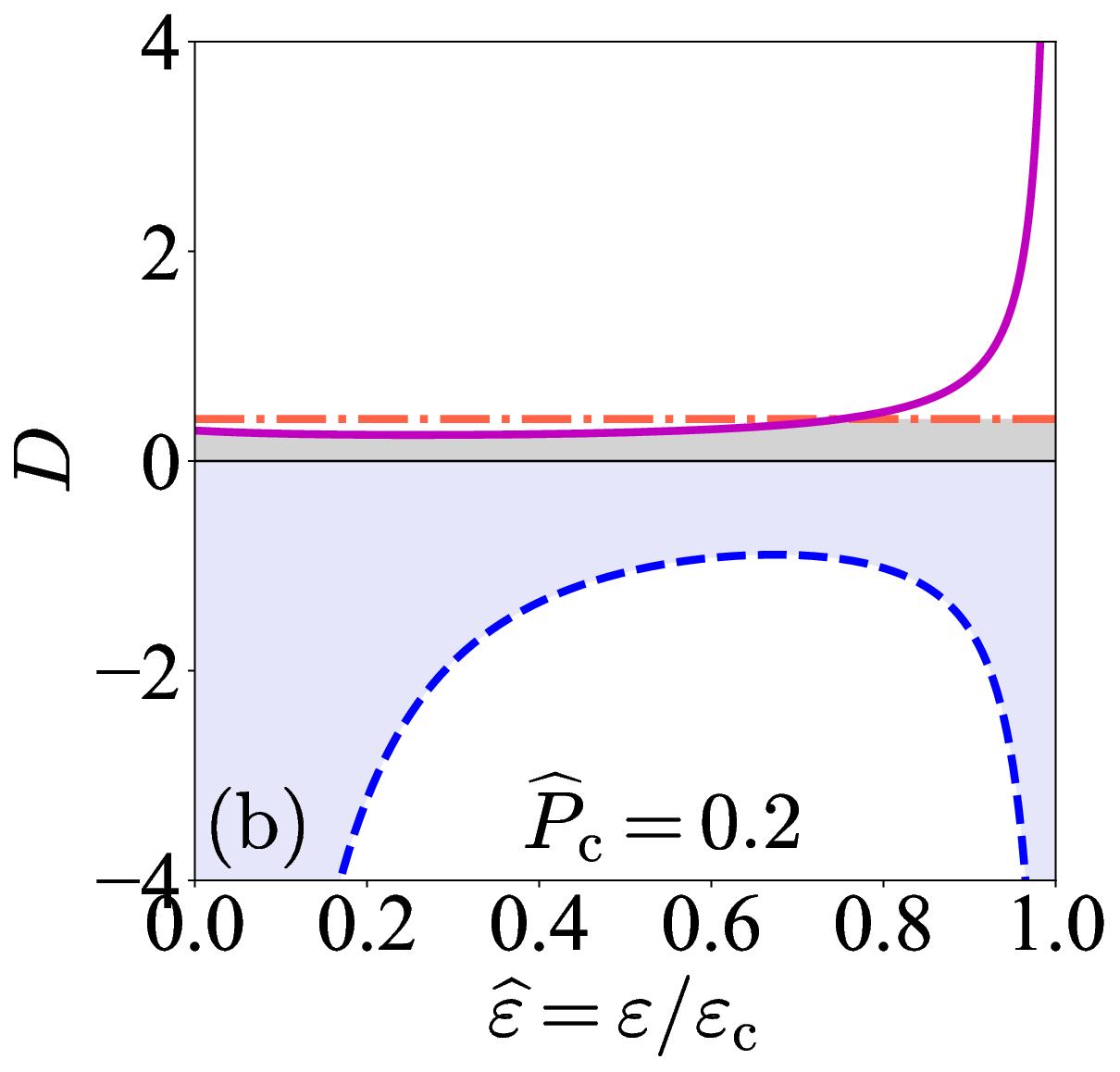}
\includegraphics[width=4cm]{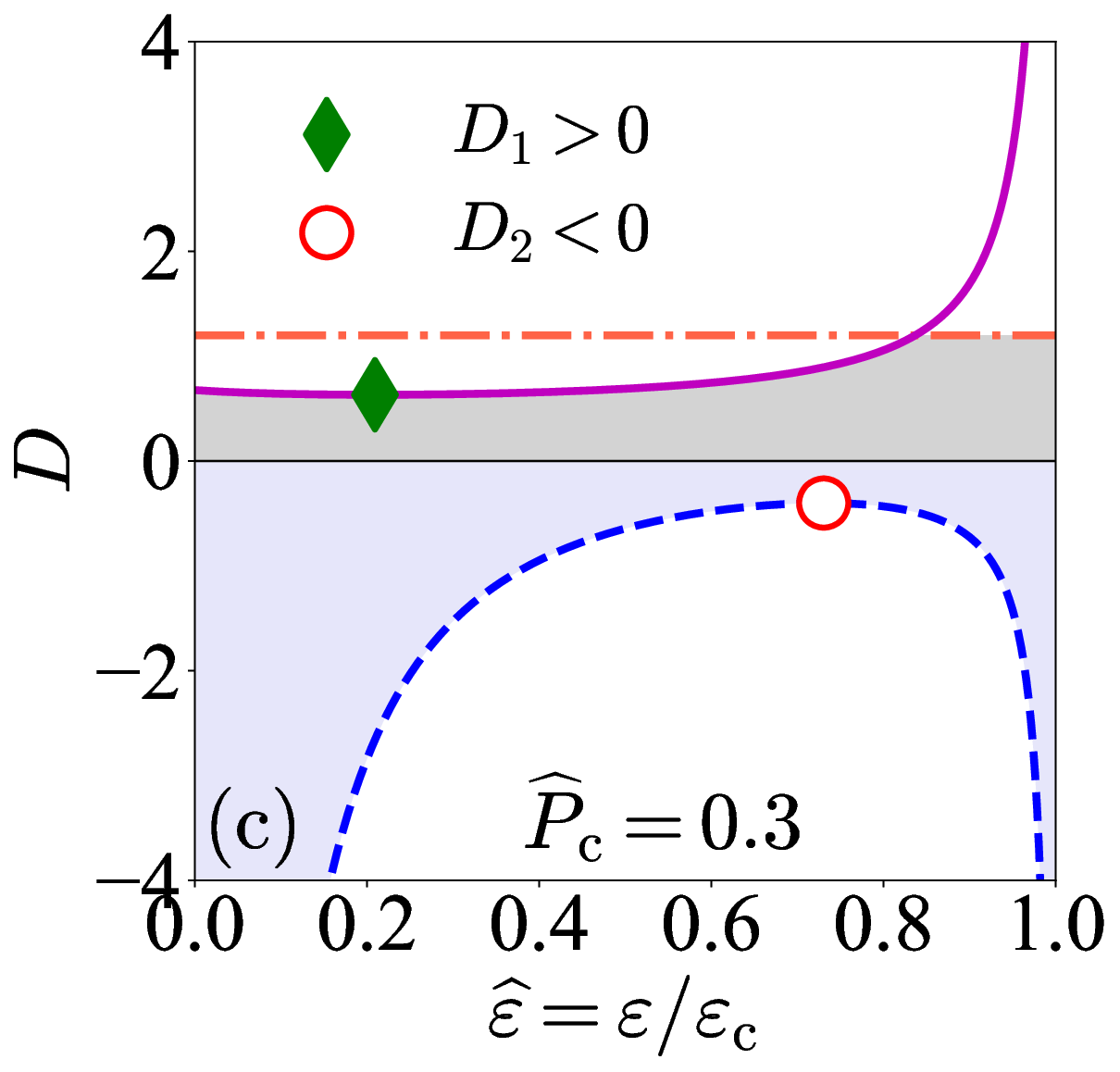}
\includegraphics[width=4cm]{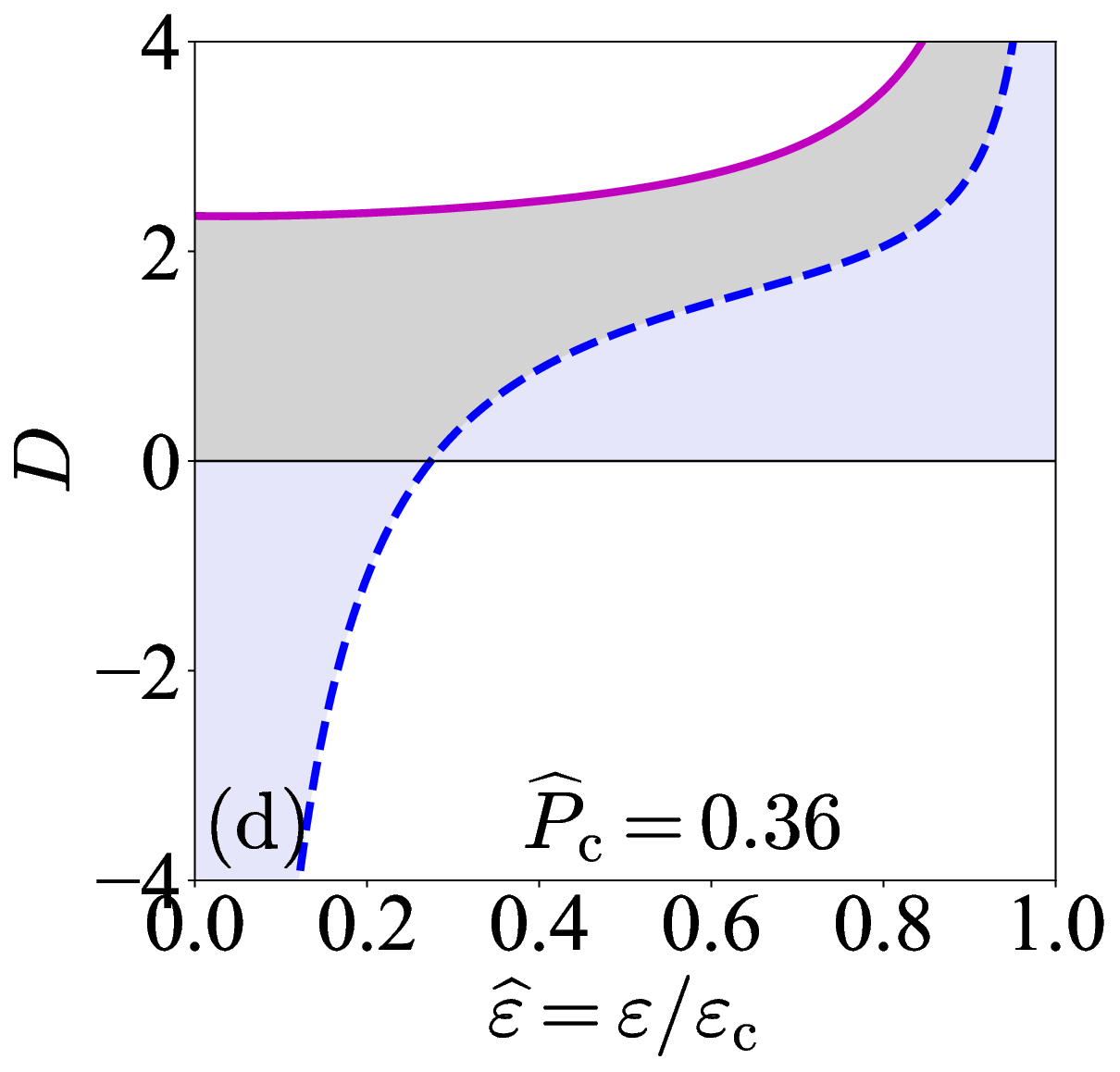}
\caption{(Color Online).  Explanations on why $D$ tends to be negative (positive) for small (large) ${\x}$, using $K=4$.See the text for details.
 Figures taken from Ref.\,\cite{CLZ23-b}.
}\label{fig_WhyD}
\end{figure}

\begin{figure}[h!]
\centering
\includegraphics[width=4.cm]{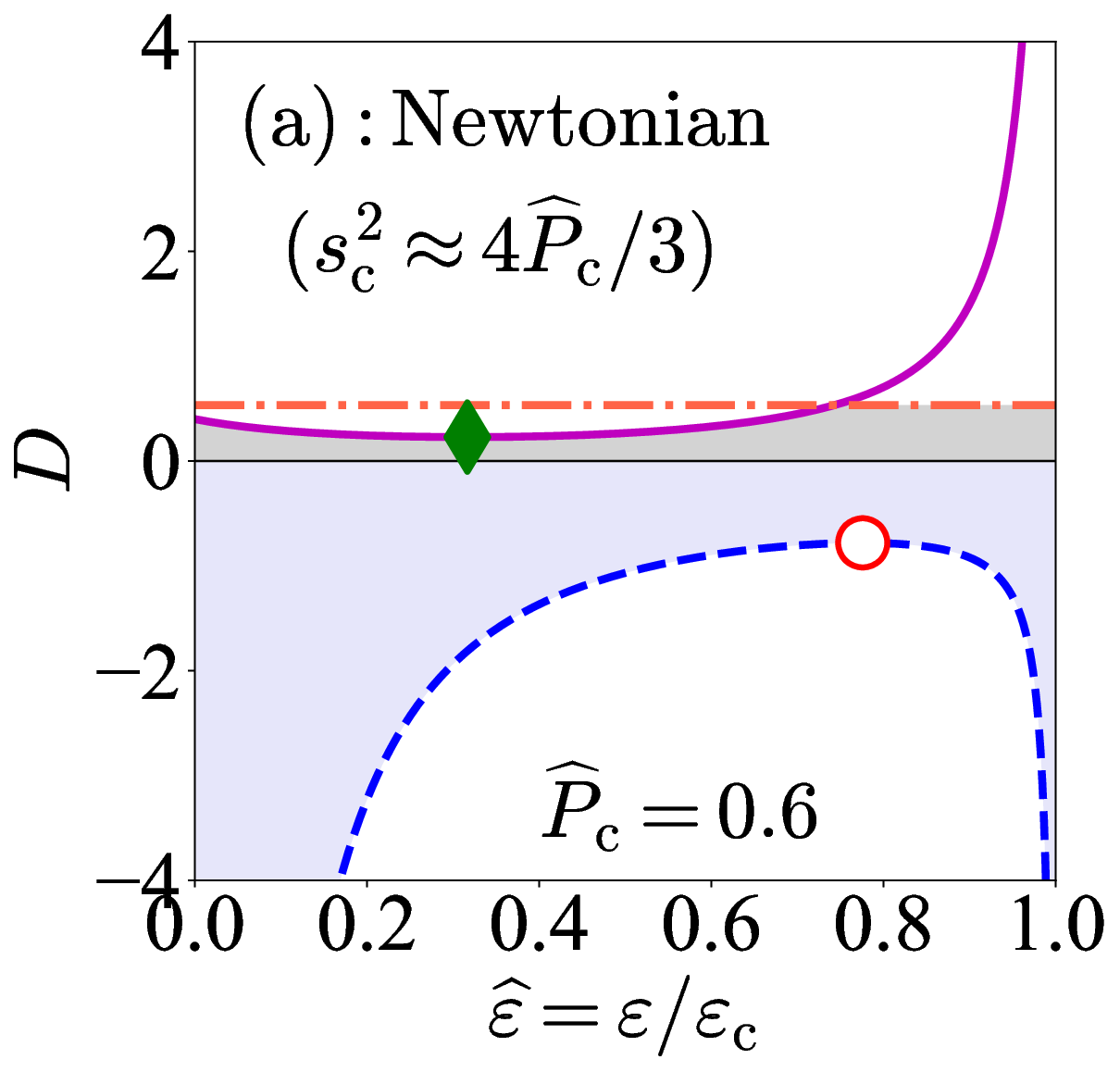}\quad
\includegraphics[width=4.cm]{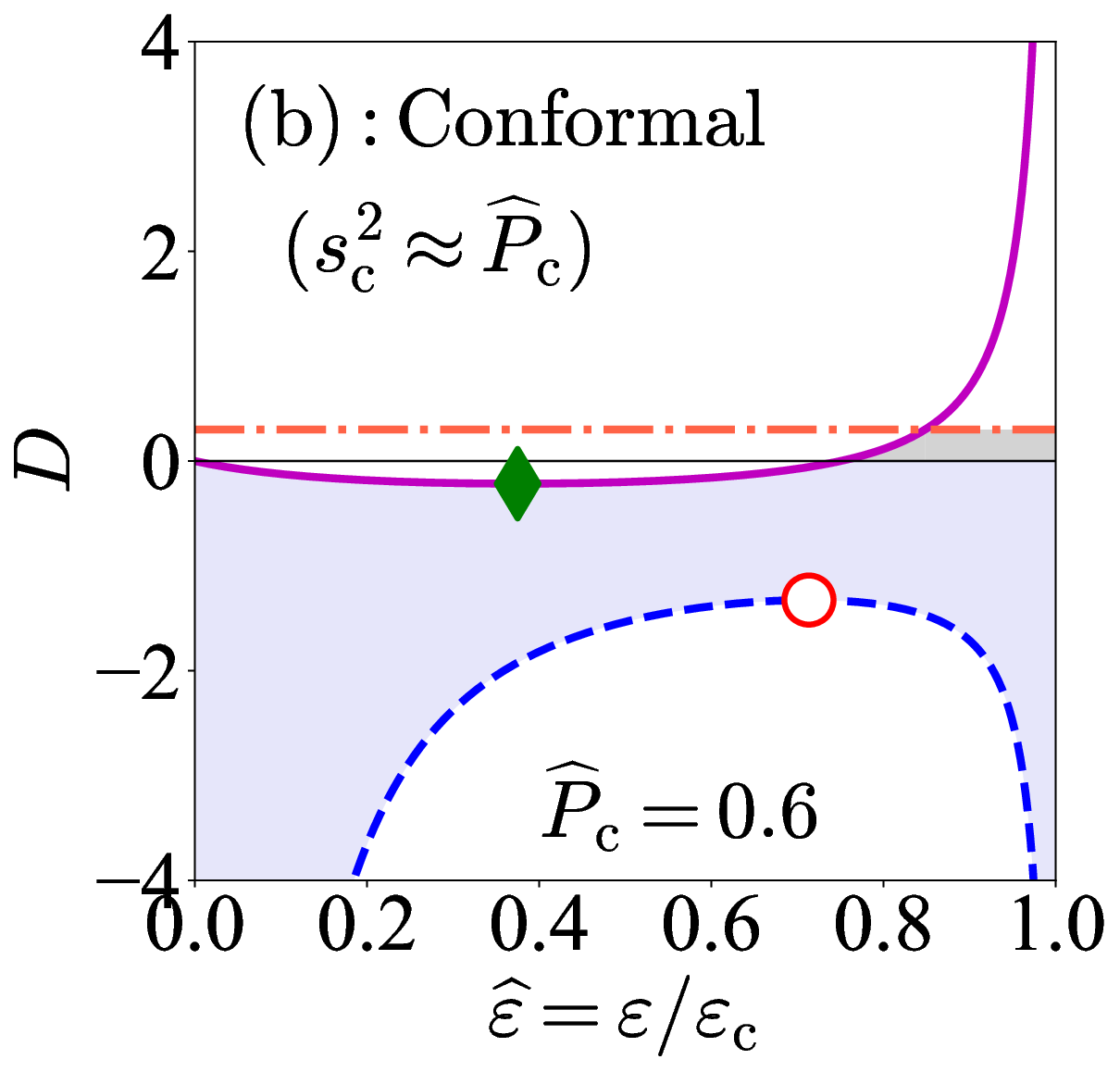}
\caption{(Color Online). The same as FIG.\,\ref{fig_WhyD} but adopting the Newtonian limit $s_{\rm{c}}^2\approx 4{\x}/3$ (left panel) and conformal limit $s_{\rm{c}}^2\approx{\x}$ (right panel). The principle of causality requires then ${\x}\lesssim3/4$ and ${\x}\lesssim1$ for these two situations, respectively.
Symbols and lines have the same meaning as in FIG.\,\ref{fig_WhyD}. Figures taken from Ref.\,\cite{CLZ23-b}.
}\label{fig_WhyDNL}
\end{figure}

\begin{figure}[h!]
\centering
\includegraphics[width=4.cm]{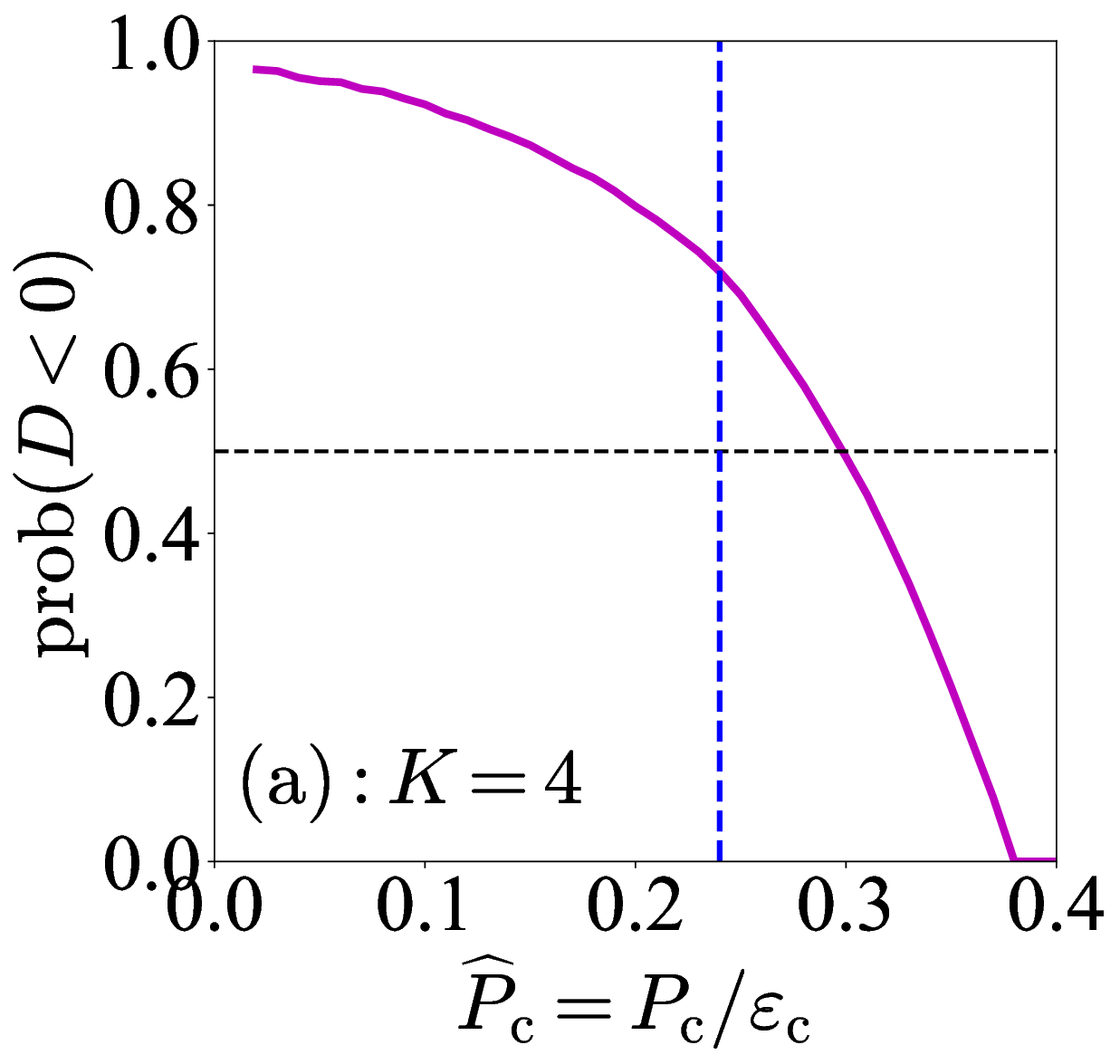}
\includegraphics[width=4.cm]{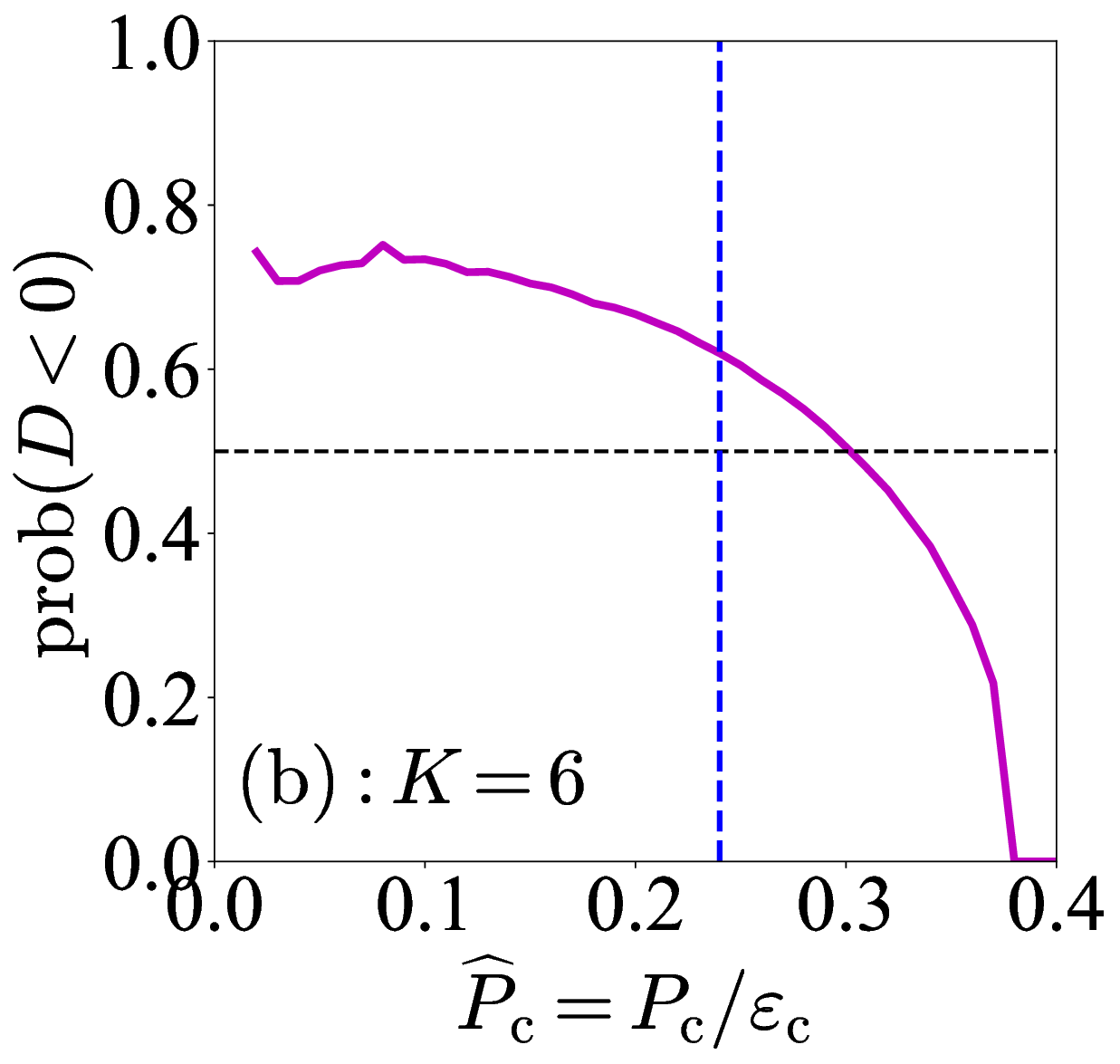}
\includegraphics[width=4.cm]{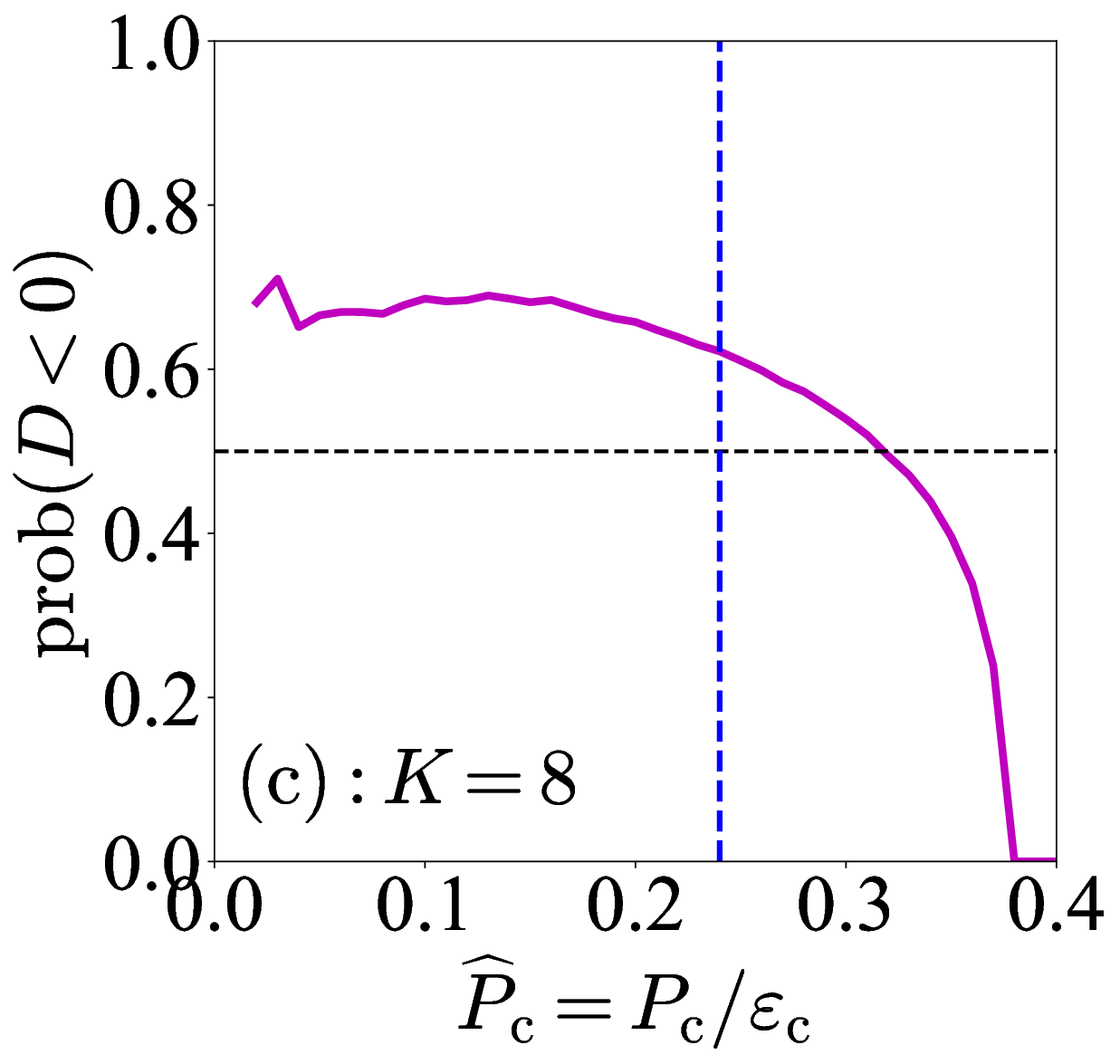}
\includegraphics[width=4.cm]{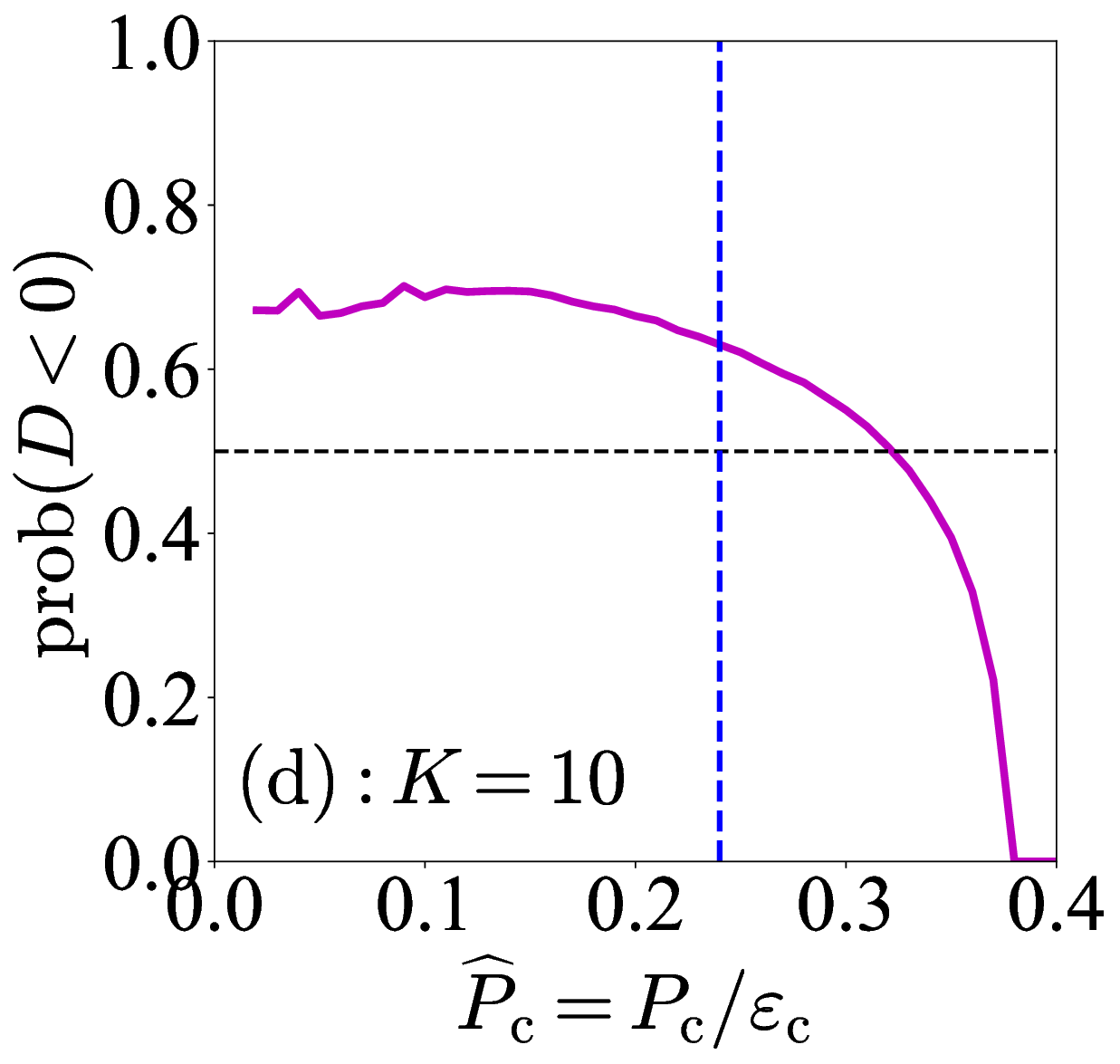}\\
\includegraphics[width=4cm]{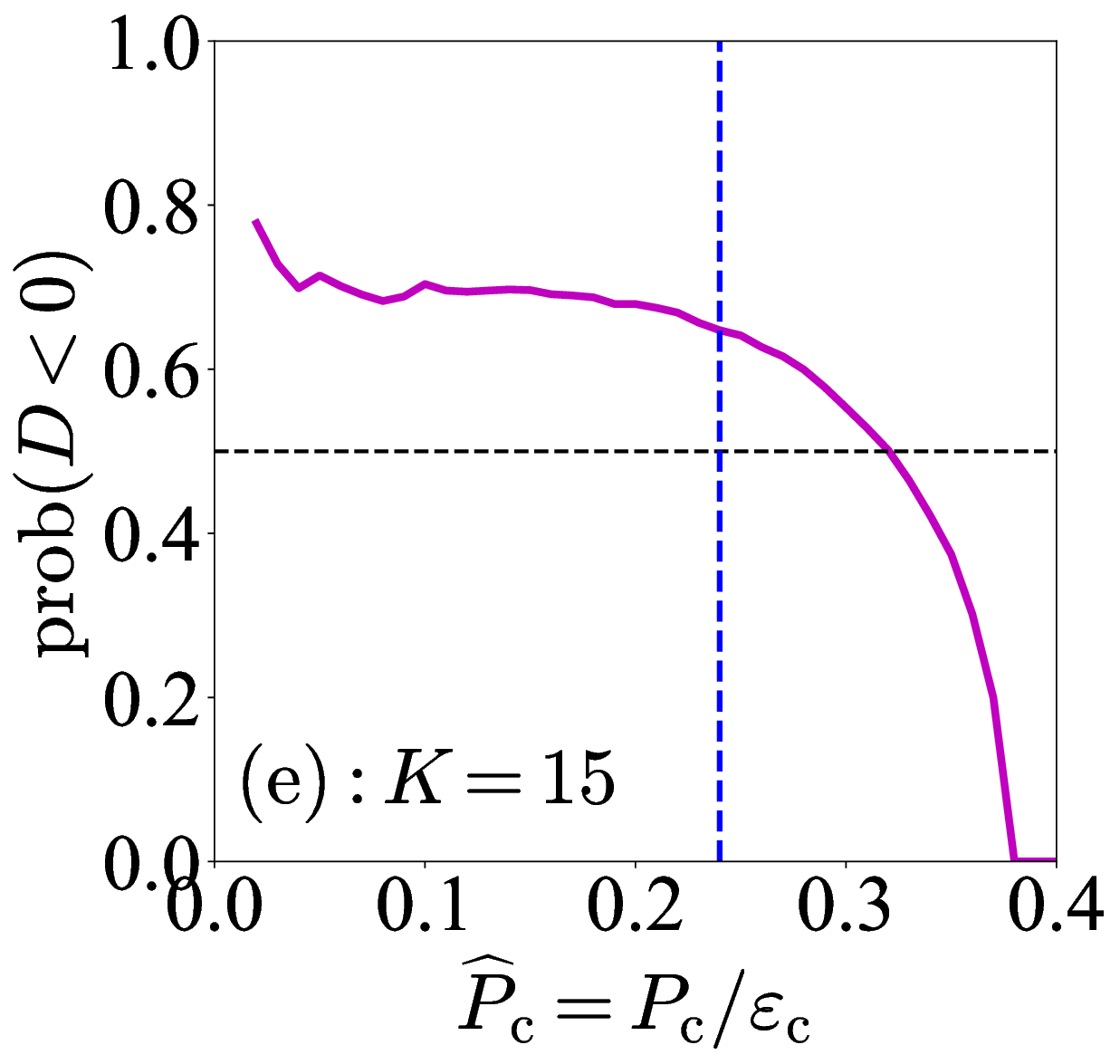}
\includegraphics[width=4.cm]{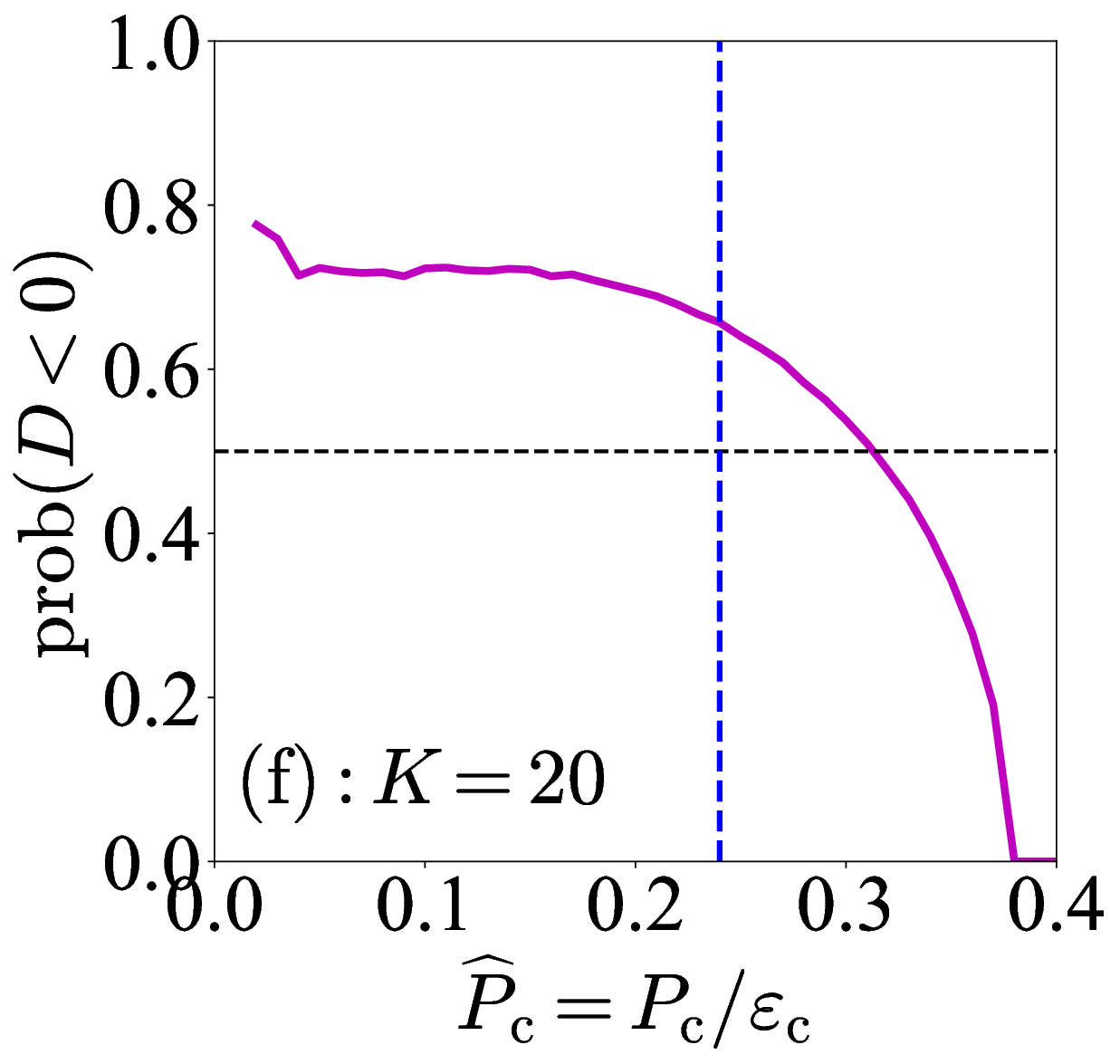}
\includegraphics[width=4.cm]{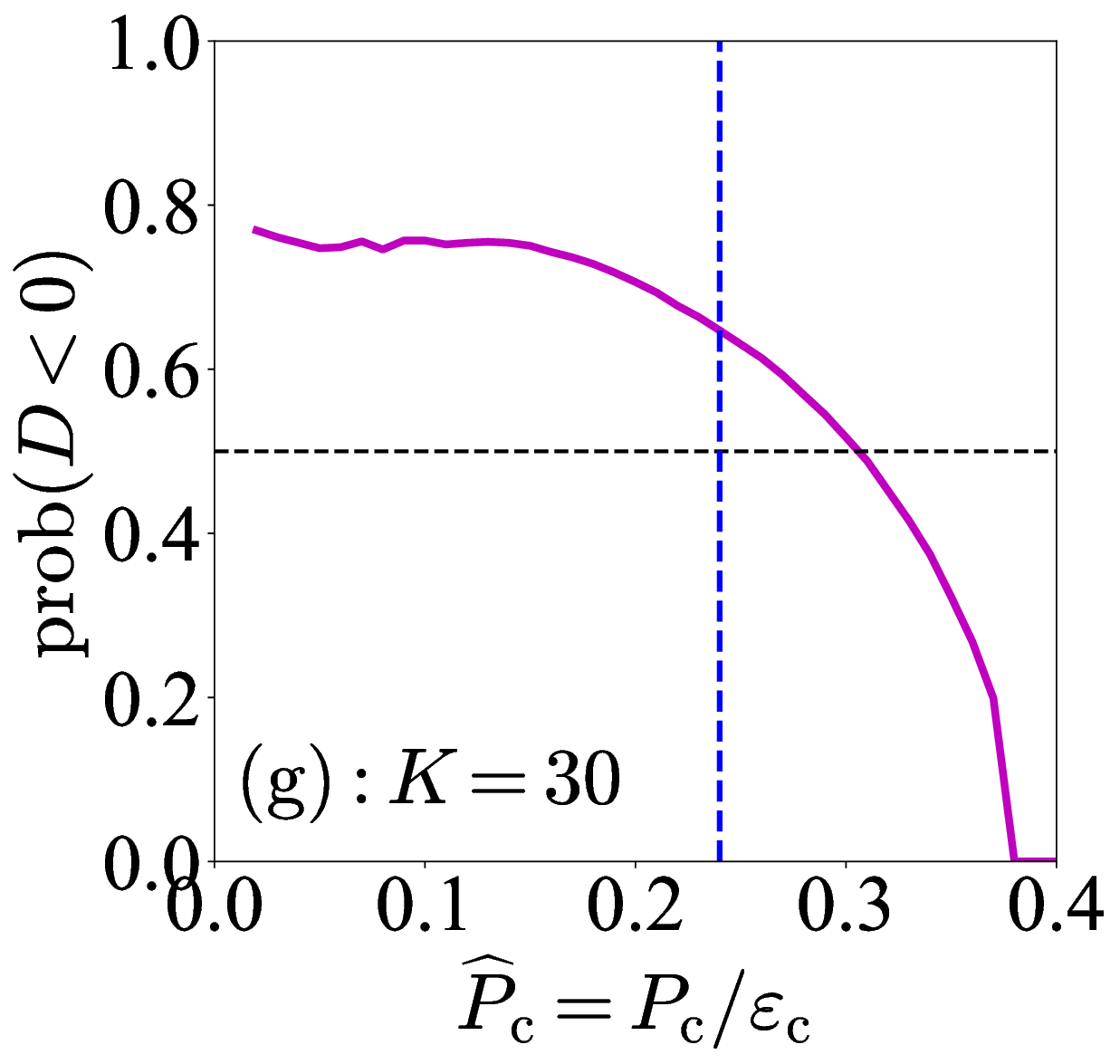}
\includegraphics[width=4.cm]{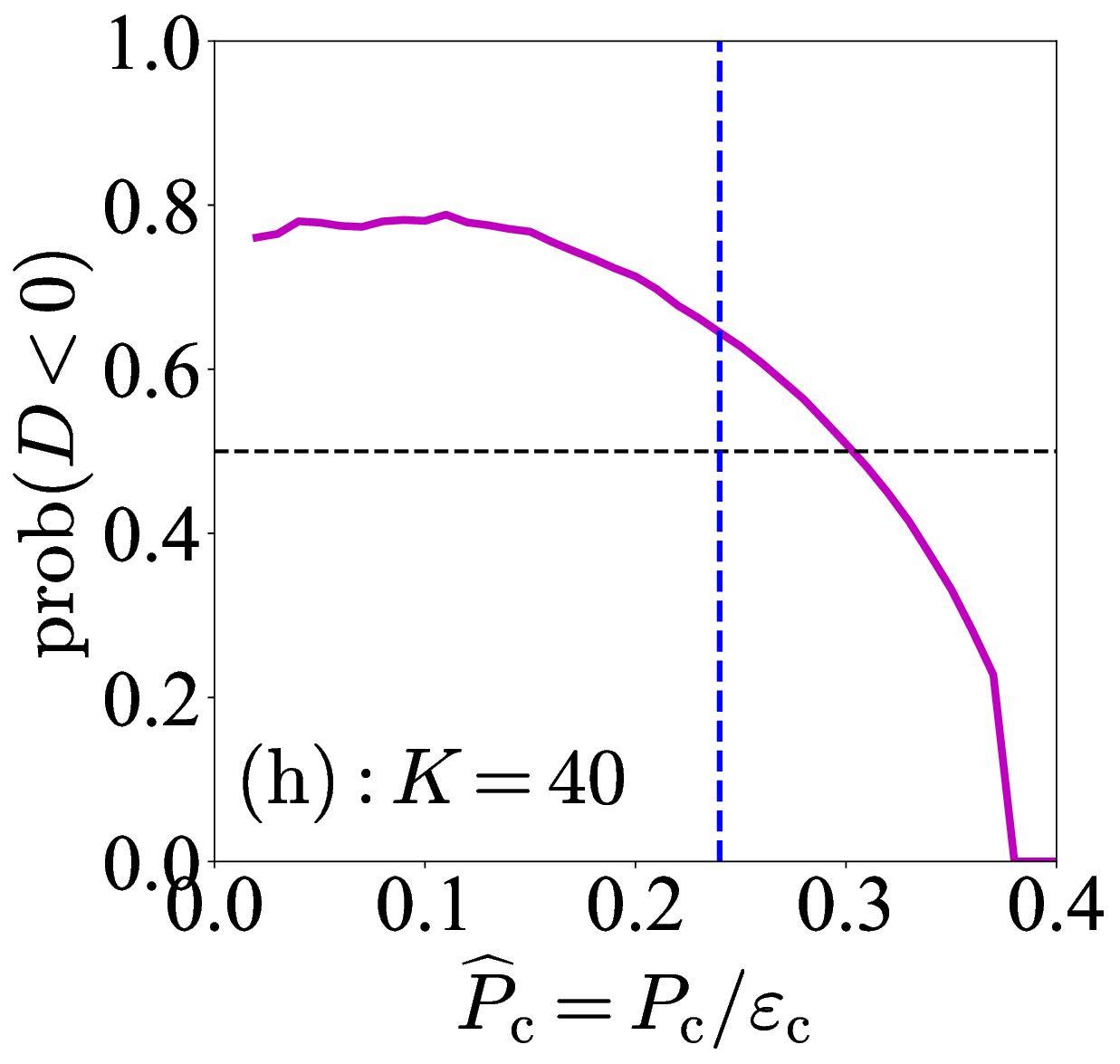}
\caption{(Color Online).  Probability of $D<0$ for different $K$; dashed vertical blue line in each panel is for ${\x}\approx0.24$ (PSR J0740+6620) and the black horizontal dashed line is plotted at $\rm{prob}(D<0)=50\%$. Figures taken from Ref.\,\cite{CLZ23-b}.
}\label{fig_NEG}
\end{figure}

The results are summarized in FIG.\,\ref{fig_WhyD}.
Here in each panel, the magenta (blue) line is for $D^{(\rm{u})}$ ($D^{\rm{(l)}}$), and the orange line marks the general boundary of $2D<\sigma_{\rm{c}}^2s_{\rm{c}}^2$ (see the estimate algorithm of (\ref{io-6})).
The positive and negative regions for $D$ are shown by the grey and lavender bands, respectively.
Moreover, $D\leq D^{(\rm{u})}$ indicates that $D$ should be smaller than the lowest point $D_1$ on the magenta line (solid green diamond in panel (c)); $D\geq D^{\rm{(l)}}$ implies that $D$ should be larger than the highest point $D_2$ on the blue line (shallow red circle in panel (c)).
When ${\x}$ is small (such as panels (a) and (b)), the magnitude of red circle is larger than that of the greed diamond, indicating that $D$ is more probable to be negative than be positive.
As ${\x}$ increases one eventually encounters $D_1\approx -D_2$ (as panel (c) shows), $D_2\approx0$ and $D_2\gtrsim 0$ (panel (d)).
The situation $D_1\approx -D_2$ roughly marks the value of ${\x}$ corresponding to equal probability of $D>0$ and $D<0$, this is about ${\x}\approx0.3$ (see FIG.\,\ref{fig_NEG} for simulation results).
When ${\x}$ increases even further, the probability of $D<0$ (of $D>0$) quickly decreases (increases); this is because the core matter is too stiff to be compressed further for large ${\x}$. 
Moreover, if one artificially adopts the Newton approximation for $s_{\rm{c}}^2$ as $s_{\rm{c}}^2\approx 4{\x}/3$ (panel (a) of FIG.\,\ref{fig_WhyDNL}) or the conformal limit $\gamma_{\rm{c}}=s_{\rm{c}}^2/{\x}\approx1\leftrightarrow s_{\rm{c}}^2\approx {\x}$ (panel (b) of FIG.\,\ref{fig_WhyDNL}), then even a large ${\x}\approx0.6$ may still tend to induce a negative $D$ (compared with FIG.\,\ref{fig_WhyD}).
In particular, the $D$ in the conformal limit is very likely negative, since both $D_1$ and $D_2$ are negative as indicated by the right panel of FIG.\,\ref{fig_WhyDNL}.
This means if one uses a (nearly) conformal EOS as the input, then the $s^2(\widehat{\varepsilon})$ is probably greater than $s_{\rm{c}}^2$; this strongly depends on the input assumption.
{\color{xll}In addition, if $s_{\rm{c}}^2$ is nearly a constant (independent of ${\x}$), then the condition $D<\sigma_{\rm{c}}^2s_{\rm{c}}^2/2$ alone tells that $D<0$ since $\sigma_{\rm{c}}^2=\d s_{\rm{c}}^2/\d{\x}=0$ (see the inequality (\ref{io-7})).
In this case if $s_{\rm{c}}^2=1$, then its nearby $s^2(\widehat{\varepsilon})$ keeps 1, otherwise $s^2(\widehat{\varepsilon})>s_{\rm{c}}^2$, i.e., the core is definitely softer than its surroundings.}
These results clearly show that the assumption on the input EOS may essentially affect the final inference on the EOS in NS cores, and the nonlinear dependence of $s_{\rm{c}}^2$ on ${\x}$ (of Eq.\,(\ref{sc2-TOV})) is fundamental to account for the eventual change on the sign of $D$, and therefore it may influence the possible continuous crossover occurring in cores of NSs.

\begin{figure}[h!]
\centering
\includegraphics[width=16.cm]{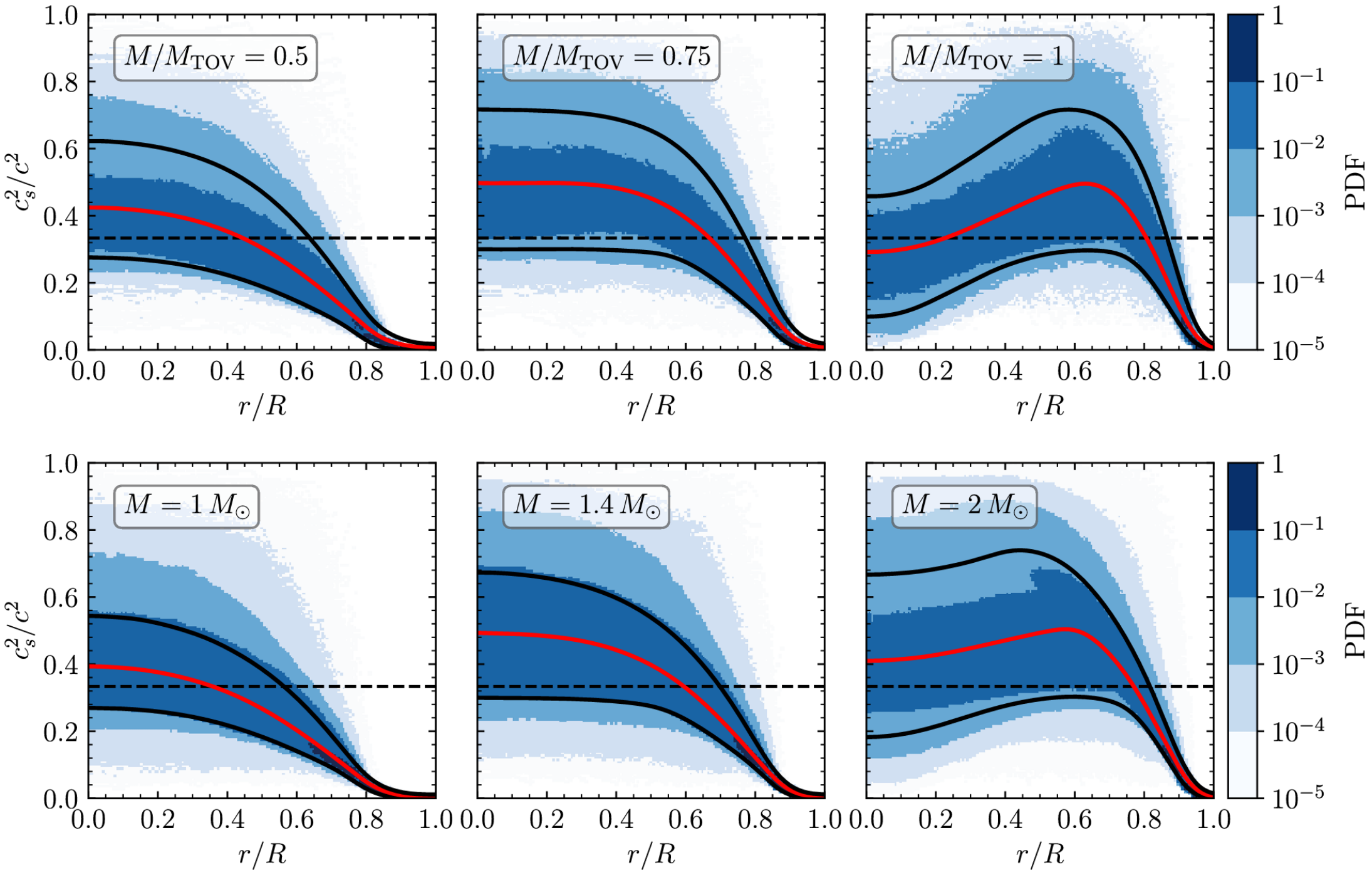}
\caption{(Color Online).  Radial dependence of $s^2$ with different NS masses. Figure taken from Ref.\,\cite{Ecker2022}.
}\label{fig_Ecker23s2r}
\end{figure}

The above examples with $K=3$ and $K=4$ involving the cubic and fourth-order polynomials of $\widehat{P}$ over $\widehat{\varepsilon}$ are two very special cases.  In order to analyze the 
probability of $D<0$ generally, we study more cases of varying $K$ as shown in FIG.\,\ref{fig_NEG} (each panel simulates $10^8$ uniform samples). 
It is seen that as the truncation order $K$ increases, the probability of $D<0$ eventually stabilizes (see panels (a) to (h) of FIG.\,\ref{fig_NEG} where $K$ varies from 4 to 40).
One finds from the curves that for ${\x}\lesssim0.3$ the probability of $D<0$ is larger than about 50\%, indicating that the continuous crossover probably occurs near the center.
For PSR J0740+6620 with its mass $\approx2.08\pm 0.07M_{\odot}$\,\cite{Fon21,Riley21,Miller21,Salmi22,Ditt24,Salmi24} being close to the theoretically predicted maximum NS mass (see FIG.\,\ref{fig_MTOV-STAT}) as well as ${\x}\approx0.24$ and $s_{\rm{c}}^2\approx0.45$\,\cite{CLZ23-a}, this probability is found to be larger than about 63\%. As ${\x}$ increases even further, the probability of $D<0$ decreases.
In the limiting case of $s_{\rm{c}}^2\to1$,  this probability is extremely small.

As a temporary summary, {\color{xll} a small $\x$ is advantageous for generating a negative $D$ and therefore a soft core. The physics reason for these behaviors could be traced back to the (deterministic) term $2s_{\rm{c}}^2-3{\x}\approx-{\x}/3$ in $D$ of Eq.\,(\ref{ref-D}), which is negative for small ${\x}$, though the summation term of Eq.\,(\ref{ref-D}) may be either positive or negative (EOS uncertainties).}
It  indicates that $\x$ is the relevant quantity for the onset of the continuous crossover ($s^2(\widehat{\varepsilon})>s_{\rm{c}}^2$) at zero temperature and high densities (in cores of cold NSs)\,\cite{Brandes2023,Fuku2020,Fuji2023,ZhangLi2023a}. It is likely to happen for large (therefore the ${\x}$ is relatively small) and massive NSs (so they are near the TOV configuration). 
At the Newtonian limit $s_{\rm{c}}\approx4{\x}/3$, since $u_{\rm{c}}=-3{\x}+2s_{\rm{c}}^2\approx-{\x}/3$ the coefficient $D$ tends more likely to be negative (see the left panel of FIG.\,\ref{fig_WhyDNL}). In this case,  we expect that the probability of $s^2(\widehat{\varepsilon})>s_{\rm{c}}^2$ is large.
However, as we demonstrated in Subsection \ref{sub_s2_Newtonian} the $s^2$ is probably monotonic for Newtonian stars, while the GR effects tend to reduce the probability $s^2(\widehat{\varepsilon})>s_{\rm{c}}^2$ for small ${\x}$.
The problem arises from the neglecting of $\Psi$ of Eq.\,(\ref{def-Psi}): For NSs not at TOV configuration (Newtonian stars) on the M-R curve, one has $\d^2M_{\rm{NS}}/\d\varepsilon_{\rm{c}}^2<0$ and therefore the criterion (\ref{io-6}) holds.
However, the deterministic term $u_{\rm{c}}=2s_{\rm{c}}^2-3{\x}$ in the coefficient $D$ of Eq.\,(\ref{ref-D}) should be modified to,
\begin{equation}
u_{\rm{c}}=2s_{\rm{c}}^2-3{\x}\approx-\frac{1-2\Psi}{3}{\x},~~\Psi>0,\end{equation}
by expanding Eq.\,(\ref{sc2-GG}) over ${\x}$ as $s_{\rm{c}}^2\approx(4+\Psi){\x}/3$ (Newtonian limit).
{\color{xll}A positive $\Psi$ tends to make the correction $D$ positive and therefore reduce the probability of $s_{\rm{c}}^2<s^2(\widehat{\varepsilon})$ (see Eq.\,(\ref{s2eps})).
For example, a canonical NS has $\Psi\approx2.85$ (see estimate given in Subsection \ref{sub_PsiVert}) and is less likely to have the crossover in the core than a massive NS (assuming they have similar radii).}
The above analysis is consistent with that Ref.\,\cite{Ecker2022} which predicted that the increasing of $s^2$ when going out from the NS center is reduced even to disappear as the NS mass decreases from $M_{\rm{NS}}^{\max}$ to about $1.4M_{\odot}$, as shown in FIG.\,\ref{fig_Ecker23s2r}, indicating a peaked $s^2$ only exists in massive NSs.
On the other hand, for massive NSs with similar radii of PSR J0740+6620 (thus the ${\x}$'s are also similar), the reduction of the probability for $D<0$ due to a (small) positive $\Psi$ is expected to be small as they are near the TOV configuration.
We investigate the problem more generally in Subsection \ref{sub_s2_peak} and {\color{xll} find that a sizable $\x$ neither being too small (near Newtonian limit) nor too large (near the causality limit $0.374$) is necessary and relevant to generating a peaked $s^2$ radial profile.}

\begin{figure}[h!]
\centering
\includegraphics[height=4.5cm]{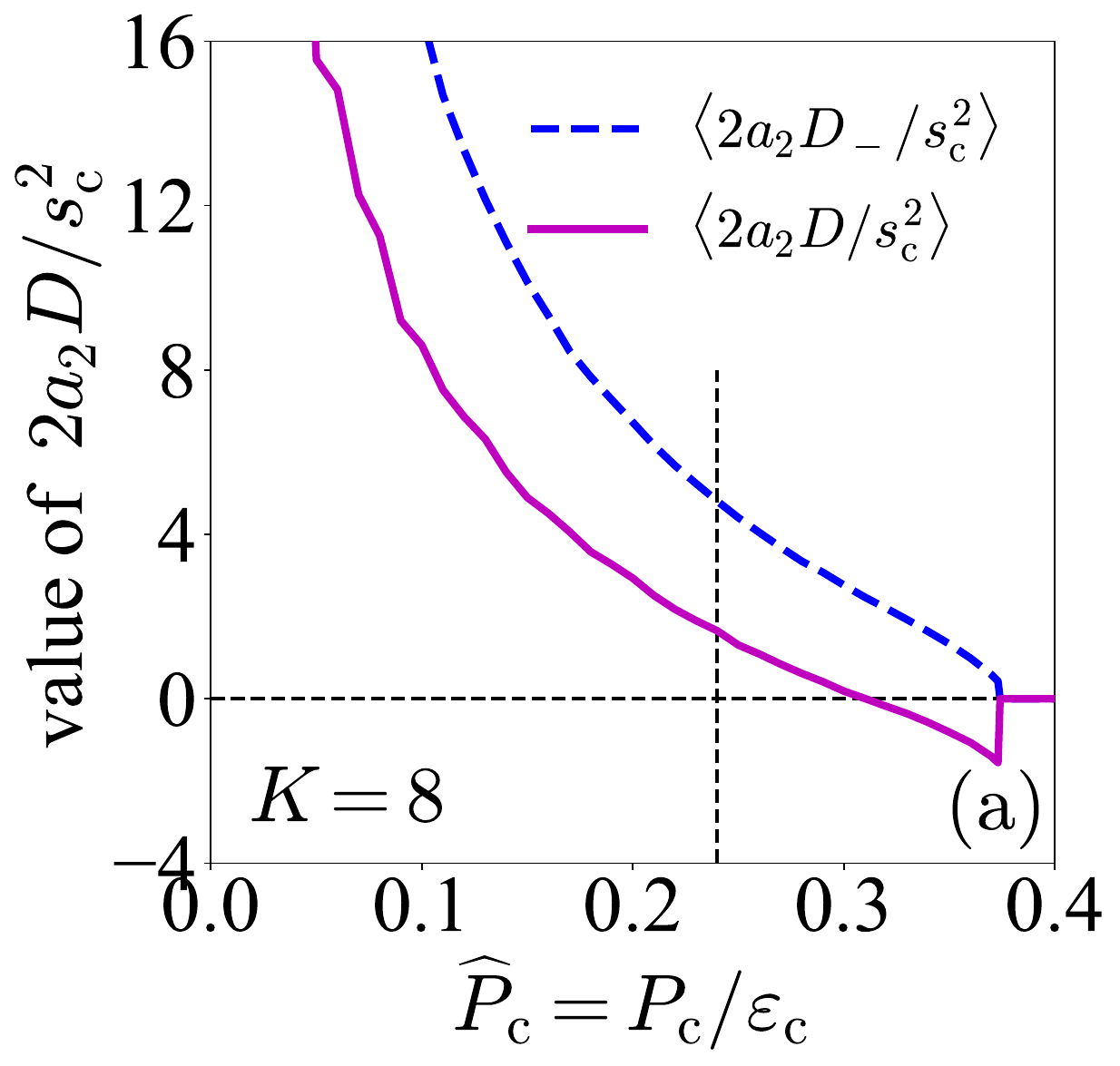}\qquad
\includegraphics[height=4.5cm]{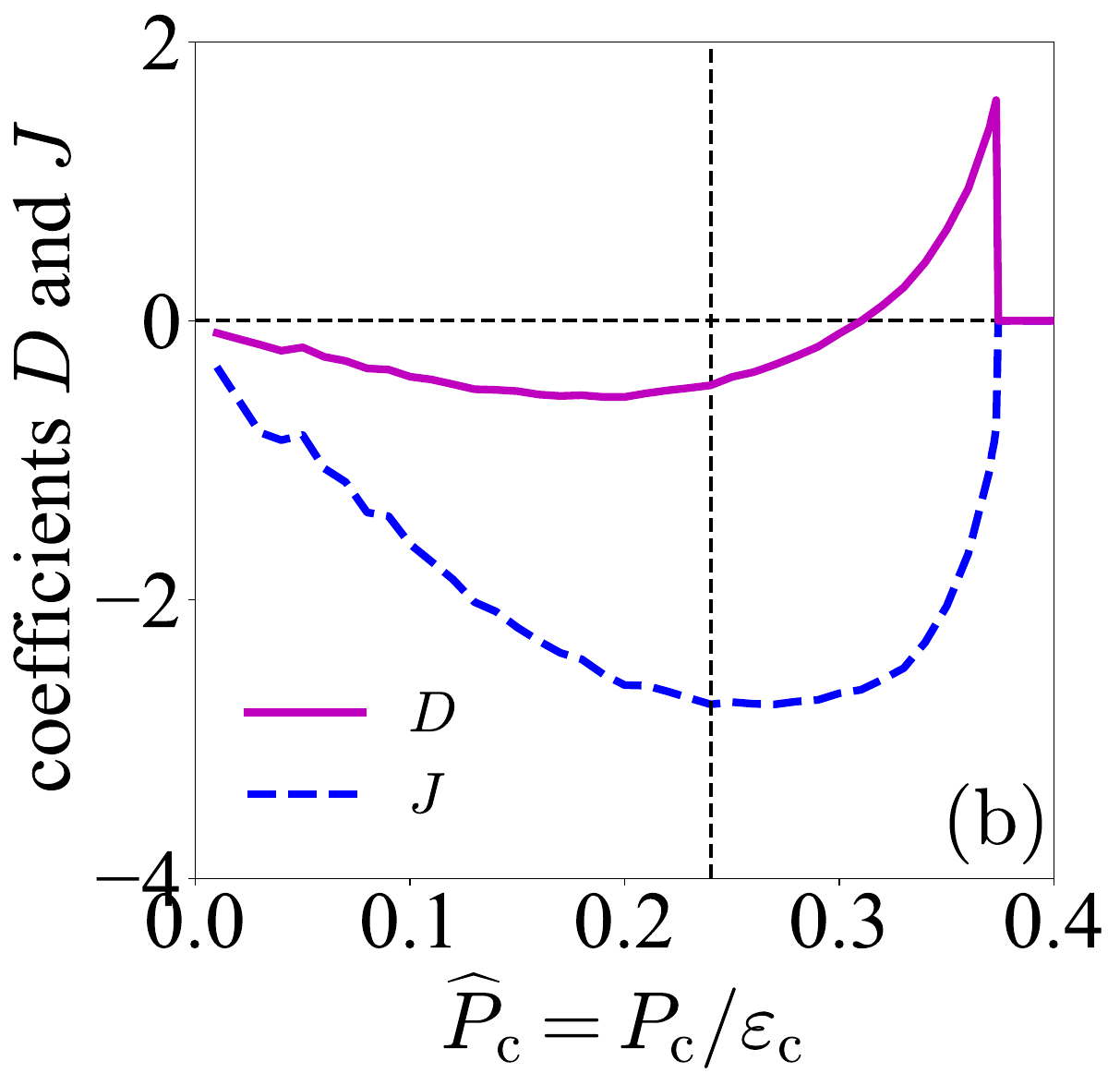}
\caption{(Color Online).  Averages of $2a_2D/s_{\rm{c}}^2$ and $2a_2D_-/s_{\rm{c}}^2$ as functions of ${\x}$ (panel (a)) and coefficients $D$ and $J$ (panel (b)) defined in Eq.\,(\ref{oo-2}) and Eq.\,(\ref{ref-J}), respectively, the vertical dashed line is at ${\x}=0.24$.
Figures taken from Ref.\,\cite{CLZ23-b}.
}\label{fig_AD}
\end{figure}

According to Eq.\ (\ref{oo-2}), the deviation of SSS from its central value is proportional to  $2a_2D\widehat{r}^2$. To evaluate the relative changing rate of SSS from its central value, 
we show in panel (a) of FIG.\,\ref{fig_AD} the $\langle 2a_2D/s_{\rm{c}}^2\rangle$ defined as
\begin{equation}
\left\langle {2a_2D}/{s_{\rm{c}}^2}\right\rangle=\left({2a_2}/{s_{\rm{c}}^2}\right)\sum_{k=\pm}\rm{prob}(D_k)D_k,
\end{equation}
here $D_k=D_{\pm}$ denotes the value of $D>0$ or $D<0$. For a comparison, the average value of $2a_2D_-/s_{\rm{c}}^2$ is also shown (blue dashed line); we then have $
\langle 2a_2D/s_{\rm{c}}^2\rangle\approx1.6$ for PSR J0740+6620\,\cite{Fon21,Riley21,Miller21,Salmi22,Ditt24,Salmi24} with ${\x}\approx0.24$, i.e.,  $s^2(\widehat{r})/s_{\rm{c}}^2\approx1+1.6\widehat{r}^2$ for small $\widehat{r}$.
Transforming it back using Eq.\,(\ref{s2eps}) gives us $s^2(\widehat{\varepsilon})/s_{\rm{c}}^2\approx1+2(D/s_{\rm{s}}^2)\mu\approx 1-2\mu= 3-2\widehat{\varepsilon}$, here $\widehat{\varepsilon}$ being very close to 1 is assumed and $D/s_{\rm{c}}^2\approx-1$ for PSR J0740+6620 is adopted.
Equivalently, we can rewrite the radial-dependence of the SSS by using Eq.\,(\ref{Rmax-n}) approximately as (by neglecting the intercept $0.64$),
\begin{equation}
{s^2}/{s_{\rm{c}}^2}\approx1+\frac{9.4{\x}\langle a_2D/s_{\rm{c}}^2\rangle}{1+3{\x}^2+4{\x}}\left(\frac{r}{R_{\max}}\right)^2,
\end{equation}
so numerically $ s^2/s_{\rm{c}}^2\approx1+0.85(r/R_{\max})^2$ for PSR J0740+6620.

\begin{figure}[h!]
\centering
\includegraphics[width=9.cm]{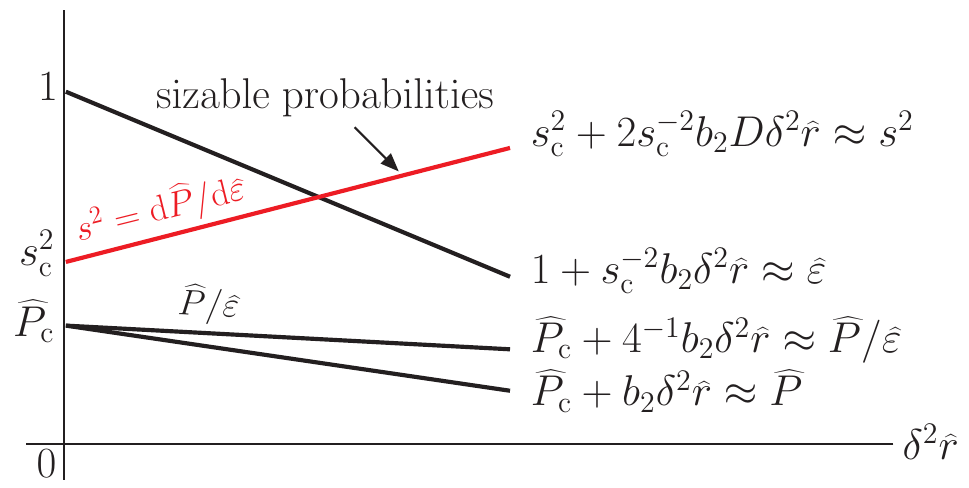}
\caption{(Color Online). Qualitative sketch of the reduction of $\widehat{P}$, $\widehat{\varepsilon}$,  $\widehat{P}/\widehat{\varepsilon}$ and $\widehat{\rho}$ (these quantities characterize the compactness/denseness of the matter) to order $\widehat{r}^2$ and the probable increasing of $s^2$ (characterizing the stiffness of the matter) near NS centers.
Figure taken from Ref.\,\cite{CLZ23-b}.
}\label{fig_re_pe}
\end{figure}

We summarize the radial dependence of the relevant quantities in FIG.\,\ref{fig_re_pe}. According to the perturbative expressions for the energy density $\varepsilon$ of Eq.\,(\ref{ee-heps}), the pressure $P$ of Eq.\,(\ref{ee-hP}), their ratio $P/\varepsilon$\,\cite{CL24-c} (see also equations of (\ref{all_dec1}) and (\ref{all_dec2})) and the baryon density $\rho$ (see Eq.\,(\ref{pk-4})), they are all decreasing functions of $\widehat{r}$ definitely.
While on the other hand, the SSS $s^2$ has sizable probabilities to be enhanced when going outward from NS centers, characterized by the coefficient $D$, see sketches shown in FIG.\,\ref{fig_re_pe}.
This demonstrates the fundamental difference between the stiffness (characterized effectively by $s^2$) and the denseness (characterized effectively by $\rho$, $P$, $\varepsilon$ and $\phi=P/\varepsilon$) or compactness (characterized by $\xi=M_{\rm{NS}}/R$). Although they are closely related, see FIG.\,\ref{fig_Compt} and relevant discussions around Eq.\,(\ref{EQ-CS}), the compactness $\x$ (the average SSS used in Ref.\,\cite{Saes2024}) is physically equivalent to the stiffness $s_{\rm{c}}^2$ only at NS centers, as $s_{\rm{c}}^2$ is monotonically varying with $\x$ (see Eq.\,(\ref{sc2-GG}) or Eq.\,(\ref{sc2-TOV})), so a denser NS ($\xi$ or $\x$ is larger) has a larger $s_{\rm{c}}^2$.
In another word regarding such attribute, the stiffness at places away from the center (characterized by $s^2$ instead of $s_{\rm{c}}^2$) is fundamentally different from the compactness. Moreover, the stiffness is basically different from the compactness; an illustrative example is that $s_{\rm{c}}^2$ (of Eq.\,(\ref{sc2-TOV})) approaches $1/3$ (stiffness) earlier than ${\x}\to1/3$ (compactness).
Similarly, while $\x$ is physically equivalent to $\xi$ when describing the compactness, the ratio $\phi$ is generally not equivalent to $\xi$ although they are closely related. In particular, Eq.\,(\ref{all_dec2}) gives $\phi\approx\x+ b_2\widehat{r}^2(1-\gamma_{\rm{c}}^{-1})$, we then have $\phi_{(1)}-\phi_{(2)}\approx \x_{(1)}-\x_{(2)}+\widehat{r}^2[b_2^{(1)}-b_2^{(2)}-(b_2^{(1)}/\gamma_{\rm{c}}^{(1)}-b_2^{(2)}/\gamma_{\rm{c}}^{(2)})]$ for two NSs labeled by (1) and (2), respectively, at $\widehat{r}$.
The relation between $\phi_{(1)}$ and $\phi_{(2)}$ is not definite even though that between $\x_{(1)}$ and $\x_{(2)}$ is;
e.g., a NS with larger $\x$ may have a smaller $\phi$ at a finite $\widehat{r}$, or vice versa.
Therefore, a larger $\xi$ corresponds to a larger $\x$, but it is not necessarily equivalent to a larger $\phi$ at a finite $\hr$.
For a given NS, $\phi$ monotonically changes with $\x$ at a fixed $\widehat{r}$ (at least near the center) since $b_2(1-\gamma_{\rm{c}}^{-1})>0$.

\subsection{$2^{\rm{nd}}$-order expansion of SSS in reduced energy density: where is its peak located?}\label{sub_s2_2nd}

The case of $D<0$ studied in the last subsection implies that there should exist a peak in $s^2$ somewhere along the path away from the center.  Since $\varepsilon$ is a decreasing function of $\widehat{r}$, the peak of $s^2$ is located at an energy density $\varepsilon_{\rm{pk}}<\varepsilon_{\rm{c}}$, e.g., Ref.\,\cite{Tak23} predicted that $\varepsilon_{\rm{pk}}\approx565\,\rm{MeV}/\rm{fm}^3$ or $\varepsilon_{\rm{pk}}/\varepsilon_0\approx3.8$ for massive NSs.
Take our over-simplified EOS with truncation order $K=3$ as an example,  the peak $\widehat{\varepsilon}_{\rm{pk}}$ of $s^2(\widehat{\varepsilon})$ can be obtained as $\widehat{\varepsilon}_{\rm{pk}}\approx5/6+{\x}+7{\x}^2/2+13{\x}^3\gtrsim5/6$ and $s^2(\widehat{\varepsilon}_{\rm{pk}})\approx25{\x}/18+5{\x}^2/9+17{\x}^3/6$ (with ${\x}$ being small $\lesssim0.105$) using $s^2\approx2(3{\x}-s_{\rm{c}}^2)\widehat{\varepsilon}+3(s_{\rm{c}}^2-2{\x})\widehat{\varepsilon}^3$.
Obviously,  the peak of $s^2$ could not be estimated through Eq.\,(\ref{s2eps}) since it is linear in $\widehat{\varepsilon}$ and the higher-order terms, e.g., $a_4$, $b_4$ and $a_6$, etc., are necessary for such estimate.

\begin{figure}[h!]
\centering
\includegraphics[width=4.cm]{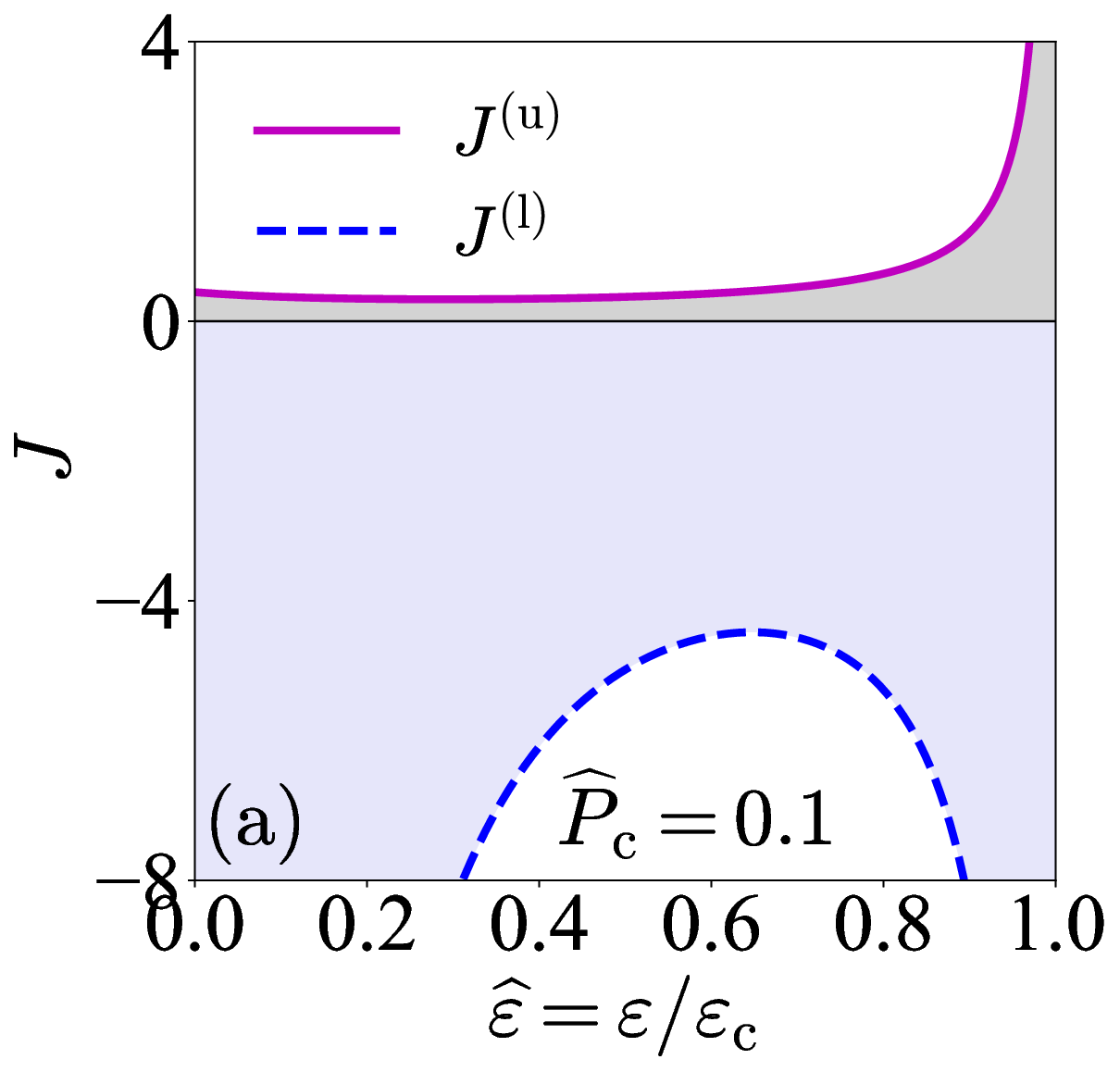}
\includegraphics[width=4.cm]{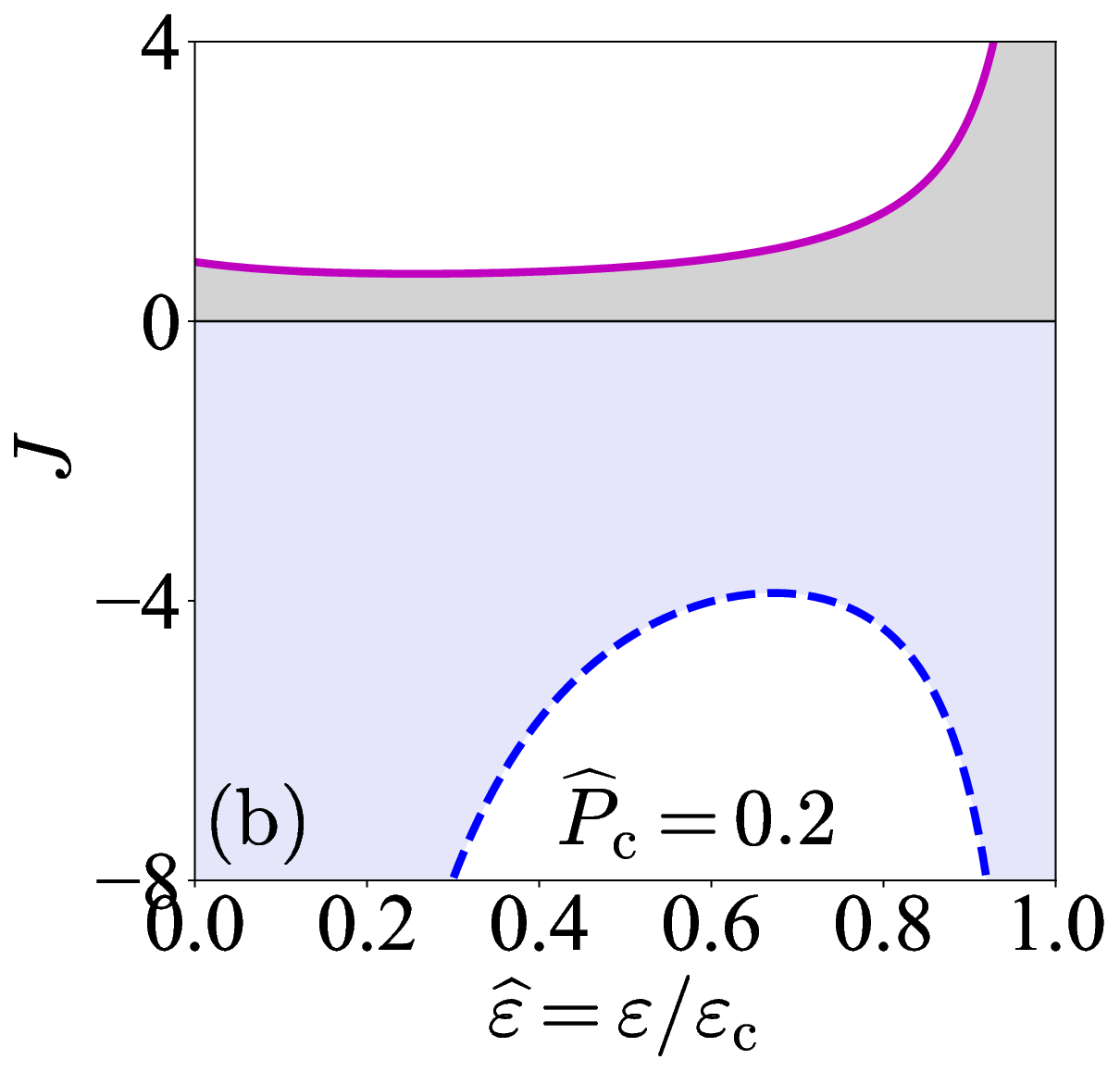}
\includegraphics[width=4.cm]{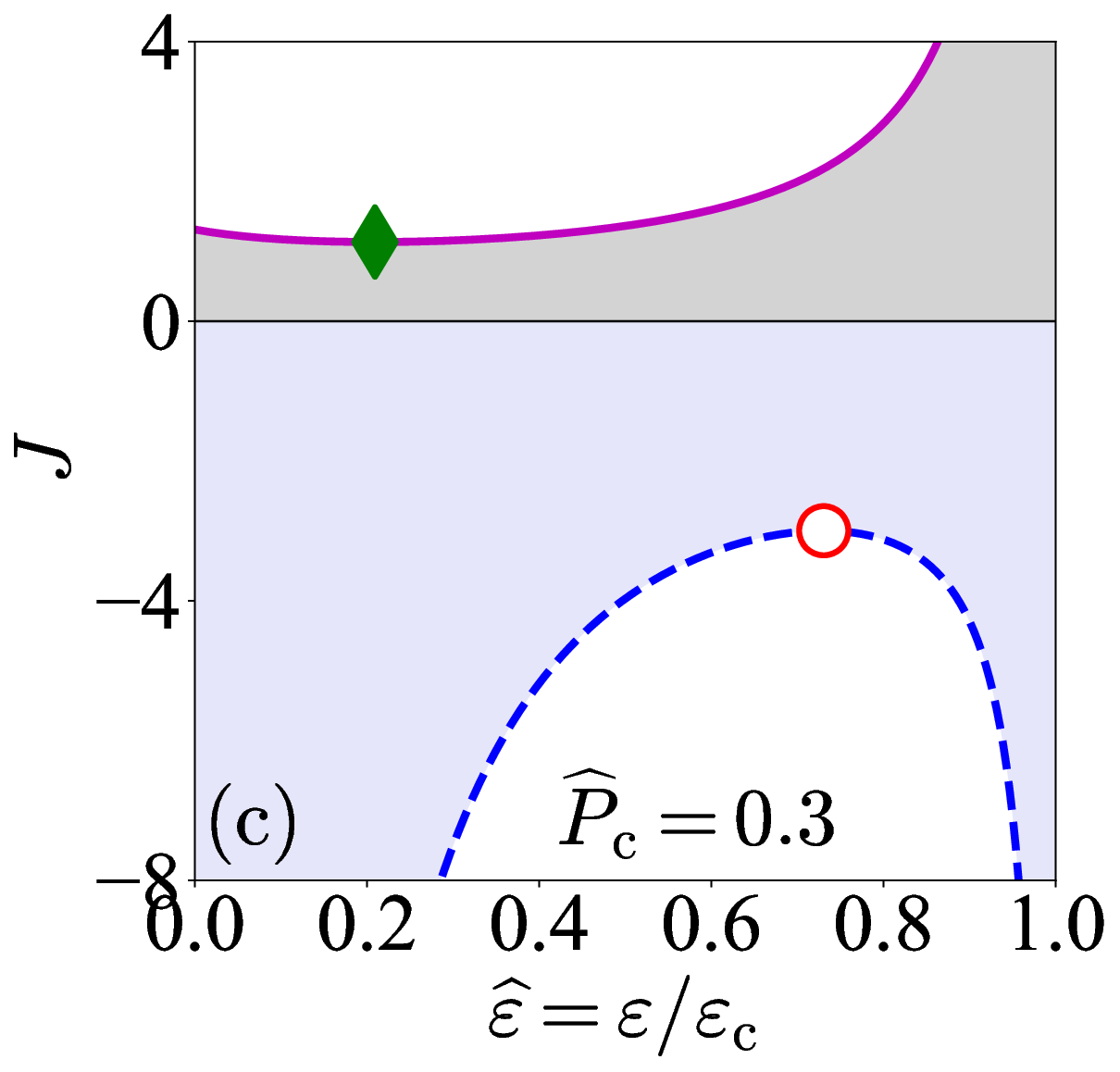}
\includegraphics[width=4.cm]{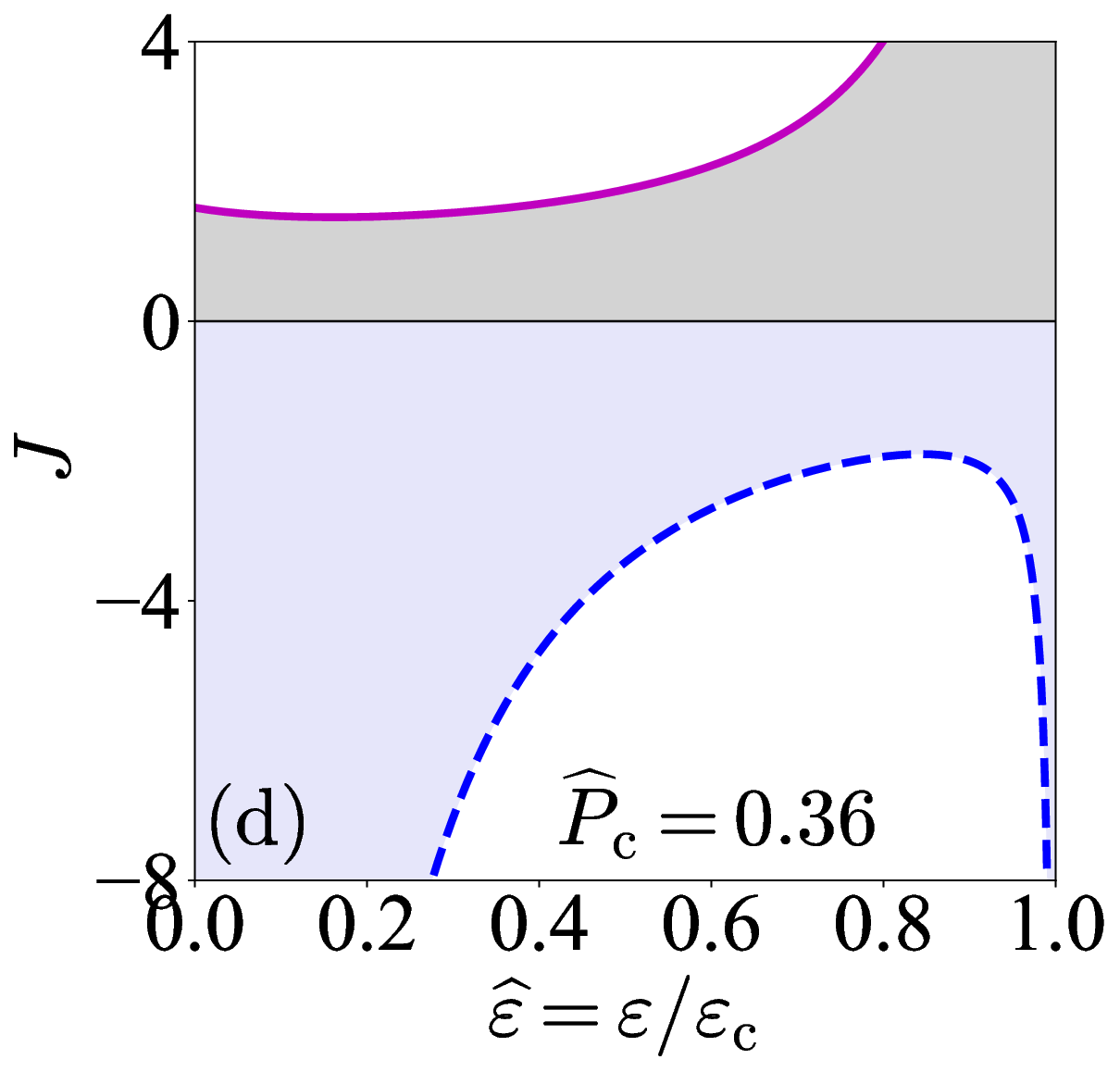}
\caption{(Color Online). The same as FIG.\,\ref{fig_WhyD} but for the coefficient $J$ of Eq.\,(\ref{ref-J}).
Figures taken from Ref.\,\cite{CLZ23-b}.
}\label{fig_WhyJ}
\end{figure}

We now use our perturbative expansions as a tool and include these higher-order contributions to estimate the peak of $s^2$ (near NS centers) if it exists.
When considering terms including order $b_6$-coefficient in the expansion of $\widehat{P}$ of Eq.\,(\ref{ee-hP}) and the $a_6$-term in the expansion of $\widehat{\varepsilon}$ of Eq.\,(\ref{ee-heps}), we derive the SSS from (\ref{s2_r_exp}) that,
\begin{align}\label{ef-1}
s^2/s_{\rm{c}}^2\approx&
1+\frac{2}{b_2}\left(b_4-s_{\rm{c}}^2a_4\right)\widehat{r}^2
+\underbrace{\frac{3}{b_2}\left[
\left(b_6-s_{\rm{c}}^2a_6\right)-\frac{4}{3}\frac{a_4}{a_2}\left(b_4-s_{\rm{c}}^2a_4\right)
\right]\widehat{r}^4}_{\mbox{term relevant for estimating the peak}}.
\end{align}
The coefficient $a_6$ could be similarly eliminated as $a_4$: expanding the reduced pressure $\widehat{P}=\sum_{k=1}^Kd_k\widehat{\varepsilon}^k$ with $\widehat{\varepsilon}=1+\sum_{k=1}^Ka_k\widehat{r}^k$ and comparing the coefficient of $\widehat{r}^6$ in the expansion $\widehat{P}={\x}+\sum_{k=1}^Kb_k\widehat{r}^k$.
The result is $
b_6-s_{\rm{c}}^2a_6=a_2^3J+2a_2a_4D$,
where $J$ is a quantity constructed from $d_k$'s defined in the expansion $\widehat{P}=\sum_{k=1}^Kd_k\widehat{\varepsilon}^k$,
\begin{equation}\label{ref-J}
\boxed{
J=\sum_{k=1}^K\frac{k(k-1)(k-2)}{6}d_k
=d_3+4d_4+10d_5+\cdots.}
\end{equation}
The coefficient $J$ has certain randomness like the coefficient $D$ (characterizing the uncertainties of the dense matter EOS).
We can similarly understand qualitatively the coefficient $J$ by using the polynomial $\hP$ of $\heps$ with order 4.
Under this assumption, we have $J=3d_4+(s_{\rm{c}}^2-{\x})/3$ where $d_4^{\rm{(l)}}\leq d_4\leq d_4^{\rm{(u)}}$ of (\ref{def_d4ul}), and therefore $J^{\rm{(l)}}\leq J\leq J^{\rm{(u)}}$.
As shown in FIG.\,\ref{fig_WhyJ}, we probably have $J<0$ though the allowed region of $J<0$ (of $J>0$) eventually shrinks (expands) as ${\x}$ increases.
We can rewrite the radial variation of $s^2$ using $J$ as,
\begin{equation}\label{ef-2}
s^2(\widehat{r})\approx s_{\rm{c}}^2+2a_2D\widehat{r}^2+\left(3a_2^2J+2a_4D\right)\widehat{r}^4.
\end{equation}
Taking the derivative $\d s^2/\d \widehat{r}$ and setting it to zero locates the peak of $s^2$ to
\begin{equation}\label{da-1}
\boxed{
\widehat{r}_{\rm{pk}}=\left(-\frac{a_2D}{3a_2^2J+2a_4D}\right)^{1/2}.}
\end{equation}
Here $a_2D>0$, so the expression under square root is positive if $J<-2a_4D/3a_2^2$ (with the latter being positive since $a_4D<0$).
A further quantitative estimate for $\widehat{r}_{\rm{pk}}$ depends strongly on the coefficients $a_4$, $D$ and $J$.

The $\hr$-dependence could be transferred to the $\mu$-dependence.
Inverting $\widehat{\varepsilon}\approx 1+a_2\widehat{r}^2+a_4\widehat{r}^4+a_6\widehat{r}^6$ gives us $\widehat{r}^2\approx(\mu/a_2)[1-(a_4/a_2^2)\mu+(2a_4^2/a_2^4-a_6/a_2^3)\mu^2]$, where $\mu\equiv\widehat{\varepsilon}-1<0$.
Putting this $\widehat{r}^2$ into the full expression for $s^2(\widehat{r})$ to order $\widehat{r}^4$ of Eq.\,(\ref{ef-2}) enables us to write the $s^2(\widehat{\varepsilon})=s^2(\mu)$ as,
\begin{align}\label{smu}
s^2(\mu)\approx& s_{\rm{c}}^2+2D\mu
+3J\mu^2
-\frac{2}{a_2^2}\left(
{3a_4}J+\frac{a_6}{a_2}D
\right)\mu^3,
\end{align} 
which generalizes (\ref{s2eps}).
Thus we have to order $\mathcal{O}(\mu^3)$,
\begin{align}\label{Pmu3}
\widehat{P}(\mu)\approx&{\x}
+s_{\rm{c}}^2\mu+\left.\frac{1}{2}\frac{\d s^2}{\d\mu}\right|_{\mu=0}\mu^2
+\left.\frac{1}{6}\frac{\d^2s^2}{\d\mu^2}\right|_{\mu=0}\mu^3
\approx{\x}+s_{\rm{c}}^2\mu+D\mu^2+J\mu^3.
\end{align}
In fact, Eq.\,(\ref{Pmu3}) could be obtained straightforwardly via the expansion $\widehat{P}=\sum_{k=1}^Kd_k\widehat{\varepsilon}^k$.
However, the coefficient $J$ (Eq.\,(\ref{ref-J})) now like the coefficient $D$ (Eq.\,(\ref{ref-D})) is not totally random, but is constrained through the general requirements $0\leq s^2\leq 1$ and $\sigma_{\rm{c}}^2s_{\rm{c}}^2>2D$ used in the algorithm of Eq.\,(\ref{io-6}), i.e., the randomness of $J$ and $D$ is effectively reduced by the physical constraints.
Therefore, we can estimate $J$ in a similar manner as for $D$.
Taking ${\x}\approx0.24$ (for PSR J0740+6620) we obtain an estimate $J\approx-2.7$ (see panel (b) of FIG.\,\ref{fig_AD}).
Combining $s_{\rm{c}}^2\approx0.45$ and $D\approx-0.45$ (panel (b) of FIG.\,\ref{fig_AD}) for ${\x}\approx0.24$, the $\widehat{P}(\widehat{\varepsilon})$ is found to be an increasing function of $\widehat{\varepsilon}$ for $\widehat{\varepsilon}\gtrsim0.7$. It basically excludes the possibility of having a plateau in $\widehat{P}$ at high densities near the center.
Including the last term in Eq.\,(\ref{smu}) may slightly shift $\widehat{\varepsilon}\gtrsim0.7$ to a higher value, here $a_6$ could be estimated via $b_6-s_{\rm{c}}^2a_6=a_2^3J+2a_2a_4D$ with the expression for coefficient $b_6$ given in Eq.\,(\ref{ee-b6}).
Moreover, the $D$-term and $J$-term contribute to $s^2(\mu)$ (and also $\widehat{P}(\mu)$) with different signs (since $\mu<0$) for ${\x}\lesssim0.3$ (panel (b) of FIG.\,\ref{fig_AD}), explaining the existence of a peak in $s^2(\mu)$.
One finds the peak of $s^2(\mu)$ from Eq.\,(\ref{smu}) (neglecting the last term) as,
\begin{equation}\label{s2pk}
\boxed{
\mu_{\rm{pk}}\approx-D/3J.}
\end{equation}  
We may find later that the possible peak in the derivative part of $s^2$ is shifted to a slightly lower value (see Eq.\,(\ref{tpk}) of the next subsection), being consistent with our general analysis given in Subsection \ref{sub_decomTA}.
Considering that $D/J$ is positive and smaller than 1 for $\x\lesssim0.3$, we may expand Eq.\,(\ref{da-1}) as:
\begin{equation}
    \widehat{r}_{\rm{pk}}\approx\sqrt{-\frac{1}{3a_2}\frac{D}{J}}\cdot\left(1-\frac{a_4}{3a_2^{3/2}}\frac{D}{J}\right),
\end{equation}
here $a_4/3a_2^{3/2}\lesssim\mathcal{O}(1)$.
Therefore $a_2\widehat{r}_{\rm{pk}}^2\approx-D/3J\approx\mu_{\rm{pk}}$, as expected.

\begin{figure}[h!]
\centering
\includegraphics[height=7.5cm]{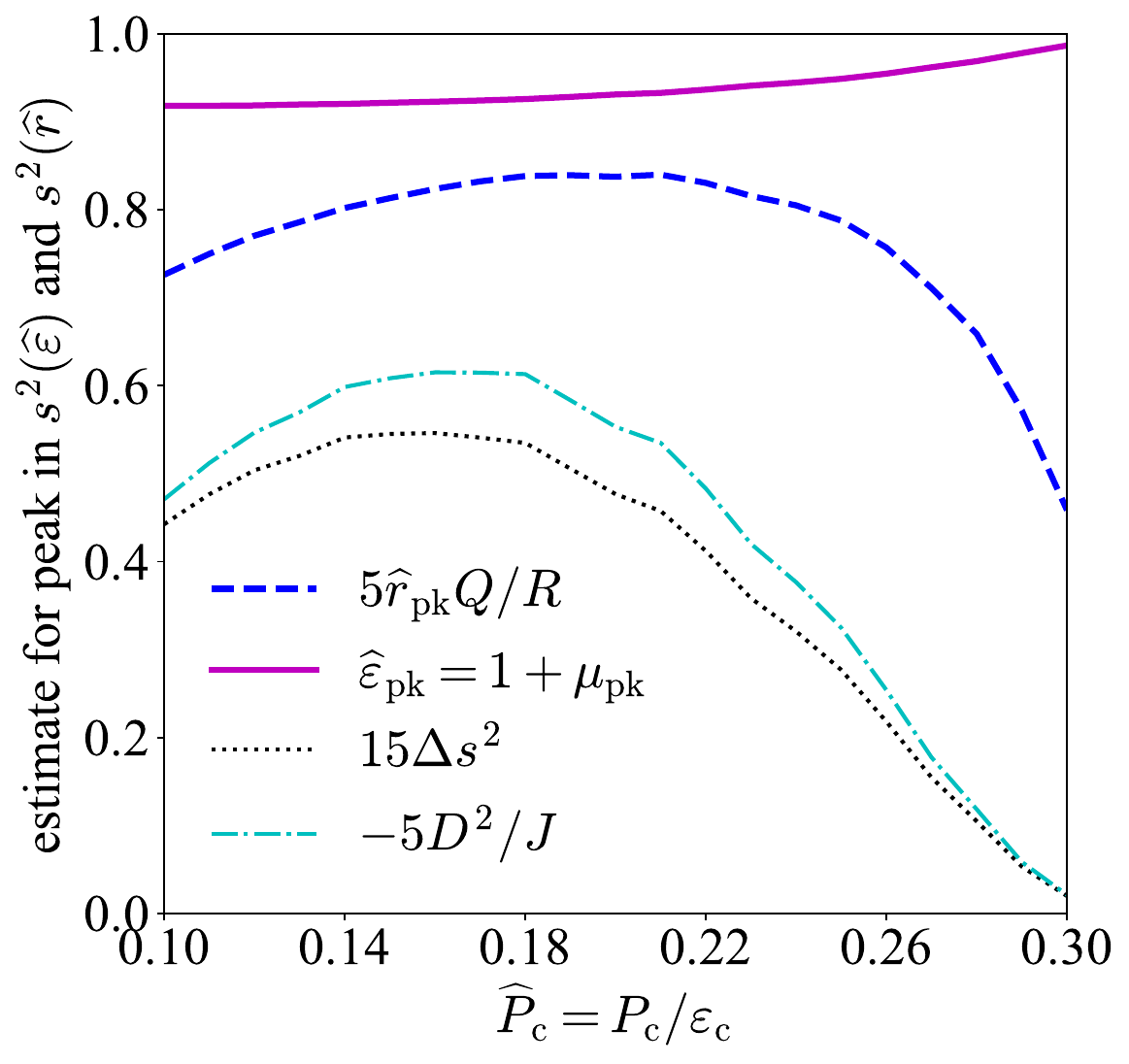}
\caption{(Color Online). Key quantities for estimating the strength and location of a peak in $s^2$ profile either for $s^2(\widehat{\varepsilon})$ or $s^2(\widehat{r})$,
the enhancement of $s^2$, i.e., $\Delta s^2=s^2(\widehat{r}_{\rm{pk}})-s_{\rm{c}}^2=-D^2/(3J+2a_4D/a_2^2)$ and its approximation $15\cdot(-D^2/3J)=-5D^2/J$ are shown; the coefficient $D$ becomes averagely negative when ${\x}\gtrsim0.3$ (see FIG.\,\ref{fig_AD}); the truncation order of the expansions is $K=8$.
Figure taken from Ref.\,\cite{CLZ23-b}.}
\label{fig_s2peak}
\end{figure}

\begin{figure}[h!]
\centering
\includegraphics[width=7.9cm]{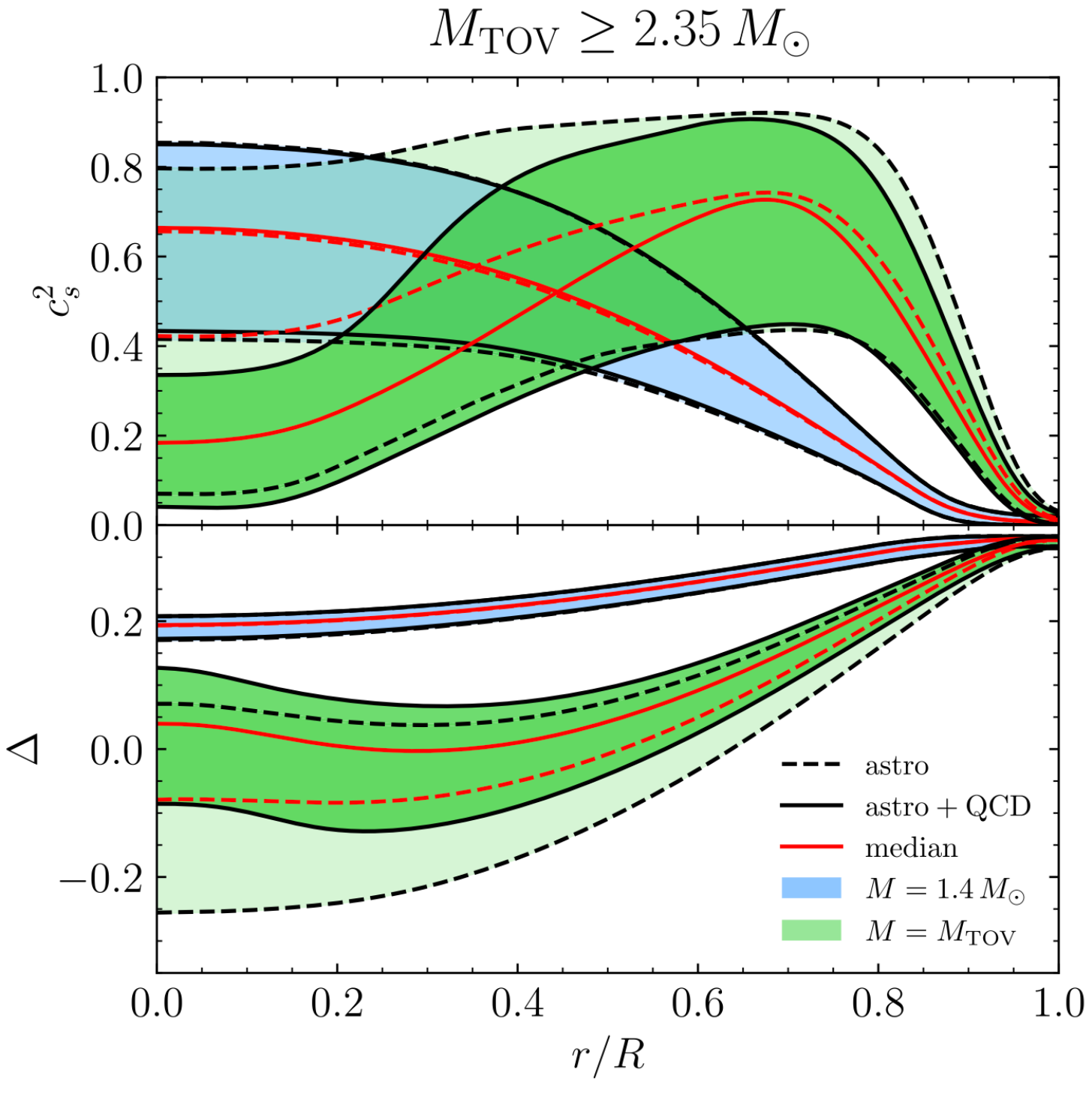}\qquad
\includegraphics[width=7.2cm]{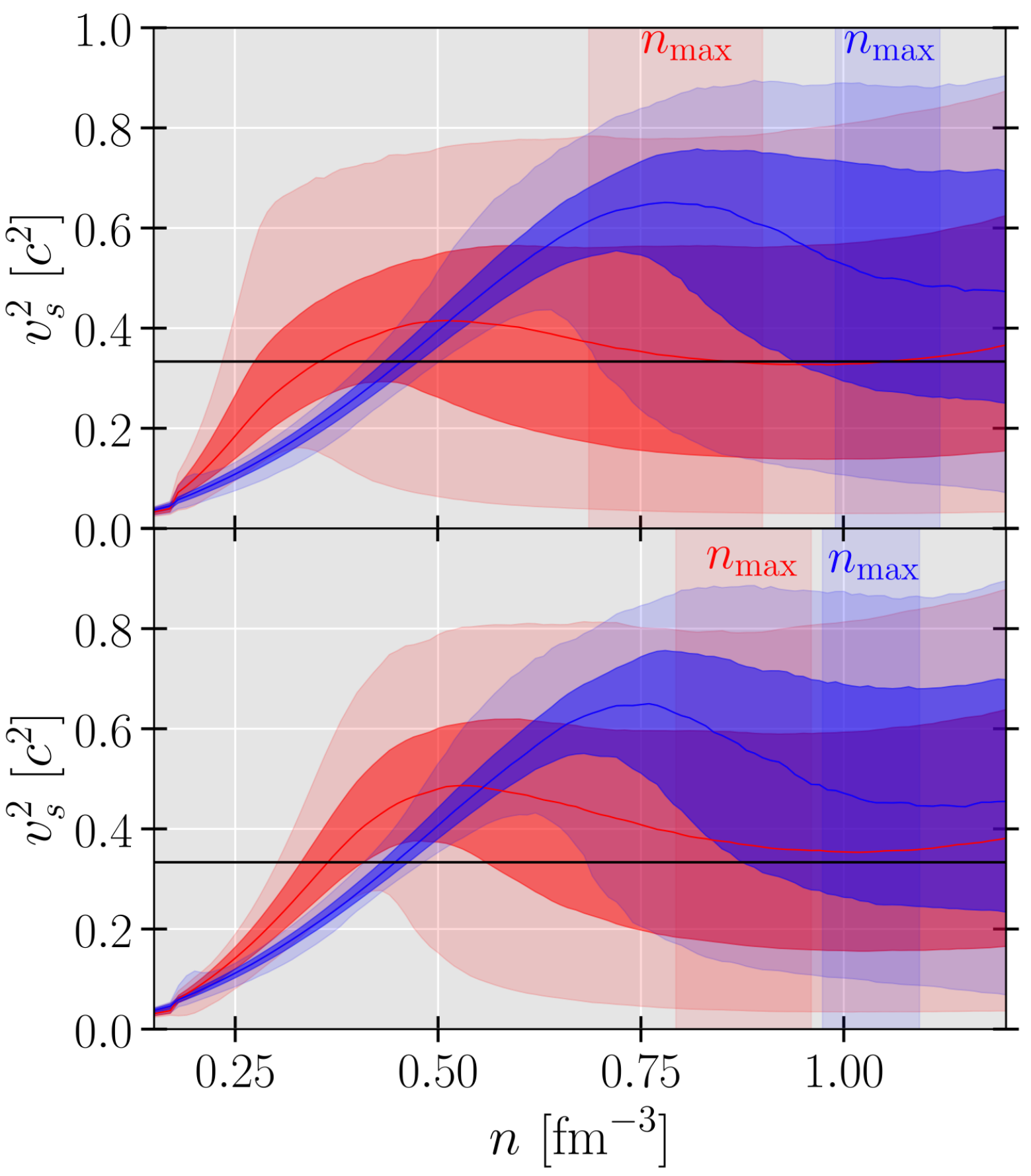}
\caption{(Color Online). Left panel: radial dependence of $s^2$ (lower plot) with the constraint $M_{\rm{TOV}}/M_{\odot}\gtrsim2.35$. Figure taken from Ref.\,\cite{Ecker23}. Right panel: density dependence of $s^2$ considering the
restriction $\d M_{\rm{NS}}/\d R<0$ for all NS mass range (blue) or $\d M_{\rm{NS}}/\d R\geq0$ for a certain mass range (red) on the M-R curve.
Figure taken from Ref.\,\cite{Ferr24}.}\label{fig_Ecker23Delta}
\end{figure}

\begin{figure}[h!]
\centering
\includegraphics[width=7.8cm]{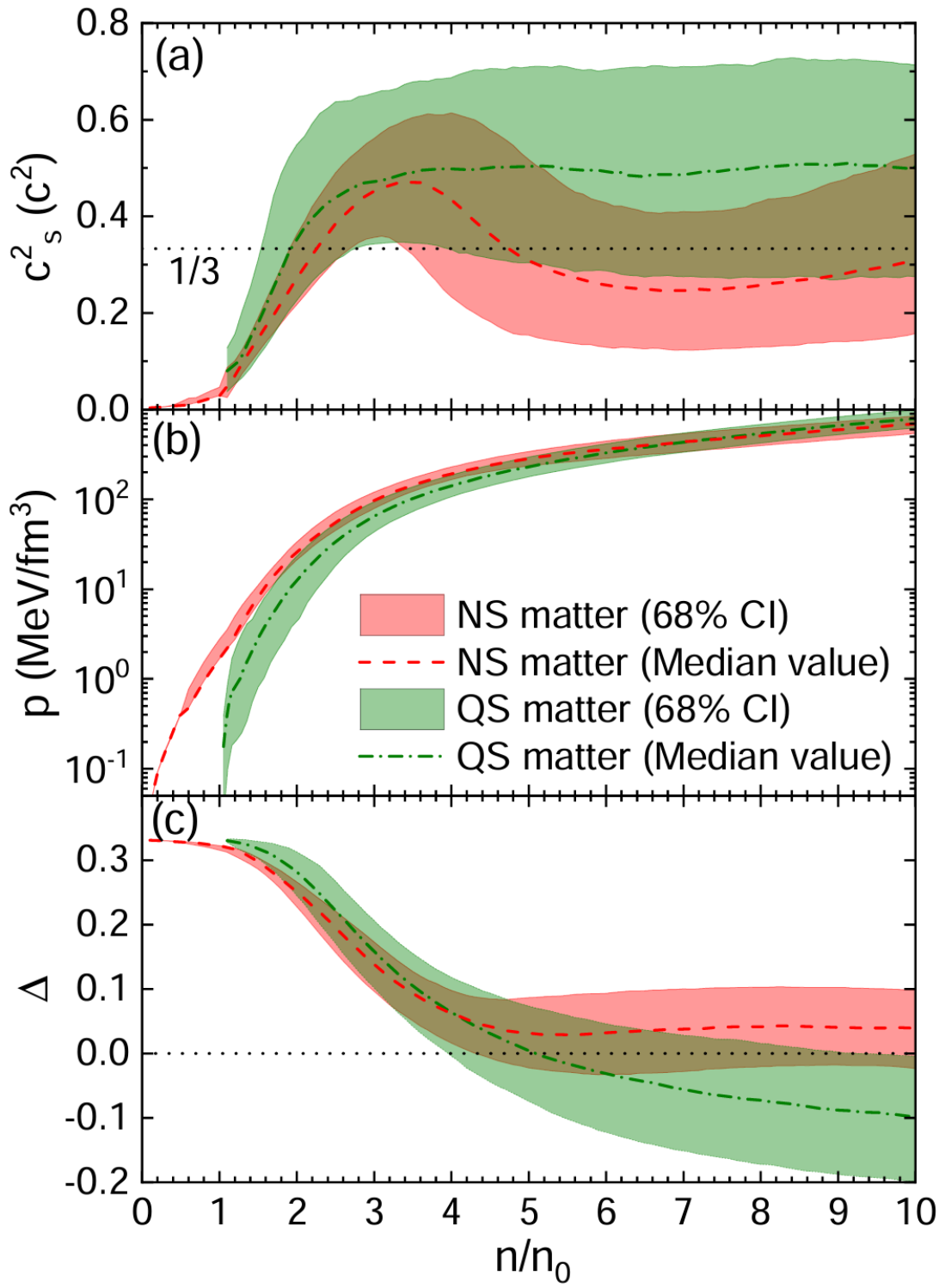}\qquad
\includegraphics[width=8.cm]{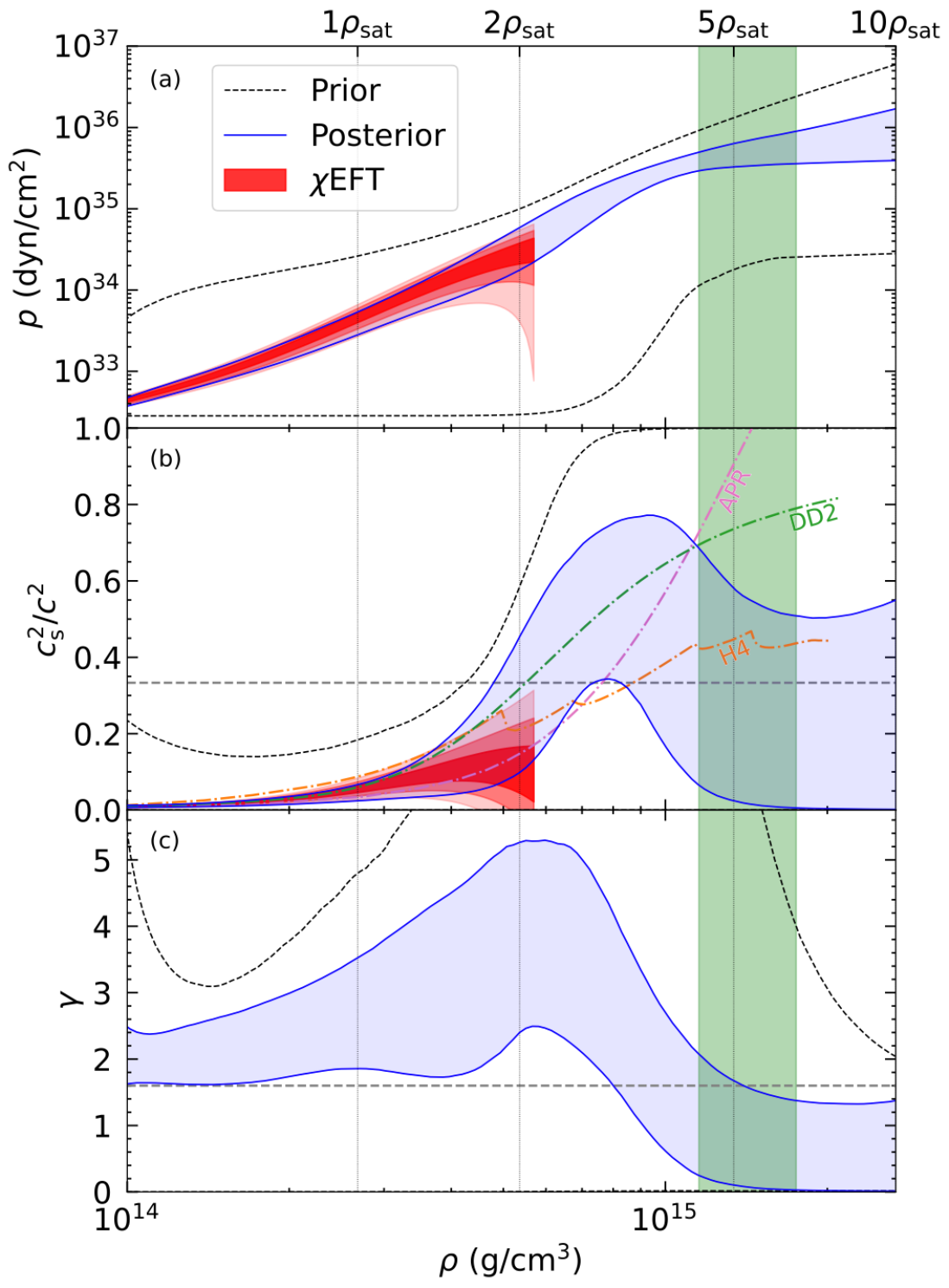}
\caption{(Color Online).  Left panel: $s^2$ as a function of baryon density (first line) for self-bound quark stars and NSs. Figure taken from Ref.\cite{Cao23}. Right panel: $s^2$ inferred by incorporating multi-messenger data of GW170817, PSR J0030+0451, PSR J0740+6620, and state-of-the-art theoretical progresses, including the information from CEFT and pQCD calculations. Figure taken from Ref.\,\cite{Han2023}.
}\label{fig_Cao23Delta}
\end{figure}

As an illustration, we take $\varepsilon_{\rm{c}}\approx901\,\rm{MeV}/\rm{fm}^3$ (which is the central energy density for PSR J0740+6620)\,\cite{CLZ23-a} and $R\approx 12.39\,\rm{km}$\,\cite{Riley21} while  let ${\x}$ be a free parameter to demonstrate the numerical results. The radius length is obtained as $Q=(4\pi G\varepsilon_{\rm{c}})^{-1/2}\approx8.7\,\rm{km}$; then $0\leq\widehat{r}Q/R\leq1$. Shown in FIG.\,\ref{fig_s2peak} are our estimates for the peak in $s^2$ for either $s^2(\widehat{\varepsilon})$ or $s^2(\widehat{r})$.
Specifically, $\widehat{\varepsilon}_{\rm{pk}}=1+\mu_{\rm{pk}}\approx1-D/3J$ using Eq.\,(\ref{s2pk}) and $\widehat{r}_{\rm{pk}}$ is given by Eq.\,(\ref{da-1}). We find the peak is very near the NS center, namely $\heps_{\rm{pk}}\gtrsim0.92$ and $r_{\rm{pk}}/R=\hr_{\rm{pk}}Q/R\lesssim0.15$.
The enhancement of $s^2$, i.e., $\Delta s^2=s^2(\widehat{r}_{\rm{pk}})-s_{\rm{c}}^2=-D^2/(3J+2a_4D/a_2^2)$ using Eq.\,(\ref{ef-2}) is also shown in FIG.\,\ref{fig_s2peak}, where $a_4\approx1$ is adopted for illustration.
Considering that $|2a_4D/a_2^2|\ll|3J|$ (see FIG.\,\ref{fig_AD}), we have approximately $\Delta s^2\approx-D^2/3J$ or directly using Eq.\,(\ref{s2pk}) as $\Delta s^2\approx 2D\mu_{\rm{pk}}+3J\mu_{\rm{pk}}^2\approx-D^2/3J$, see the cyan dash-dotted line for $15\cdot(-D^2/3J)=-5D^2/J$.
The difference between $\Delta s^2$ using Eq.\,(\ref{ef-2}) and Eq.\,(\ref{smu}) is that 
Eq.\,(\ref{ef-2}) is full to order $\widehat{r}^4$ while Eq.\,(\ref{smu}) is an approximation (even the $\mu^3$-term is included).
{\color{xll}It is seen that as the ${\x}$ increases, the $\widehat{\varepsilon}_{\rm{pk}}$ eventually approaches 1, implying that the peak $\widehat{\varepsilon}_{\rm{pk}}$ eventually moves to the center, and it may even disappear if the ${\x}$ increases further (causality limit).}
Actually for ${\x}\gtrsim0.3$ (see panel (b) of FIG.\,\ref{fig_AD}), the $D$ changes from being negative to positive and therefore both terms $2D\mu$ and $3J\mu^2$ in $s^2(\mu)$ become negative,
i.e., $s_{\rm{c}}^2$ is larger than its surroundings. 
This feature is also reflected in $\Delta s^2$ or $\widehat{r}_{\rm{pk}}Q/R$. For example, we have for ${\x}\to0.374$ that $s_{\rm{c}}^2\to1$ (Eq.\,(\ref{sc2-TOV})) and therefore $\Delta s^2\to0$ (no space under this limit for $s^2$ to be enhanced when going outward from the center).
In our illustration,  we find $(\Delta s^2)_{\max}\approx0.038$ occurring at about ${\x}\approx0.16$, therefore $(\Delta s^2)_{\max}/s_{\rm{c}}^2\approx14\%$ using $s_{\rm{c}}^2\approx0.26$.
Similarly, we have $\Delta s^2/s_{\rm{c}}^2\approx5\%$ for ${\x}\approx0.24$.
The enhancement of $s^2$ due to the peak is generally smaller than about 20\%; and it is consistent with predictions of Refs.\,\cite{Brandes2023-a,Cao23,Ferr24}.
These conclusions and numerical values would be verified in Subsection \ref{sub_s2_peak} using a slightly different but equivalently analysis.

\begin{figure}[h!]
\centering
\includegraphics[height=7.cm]{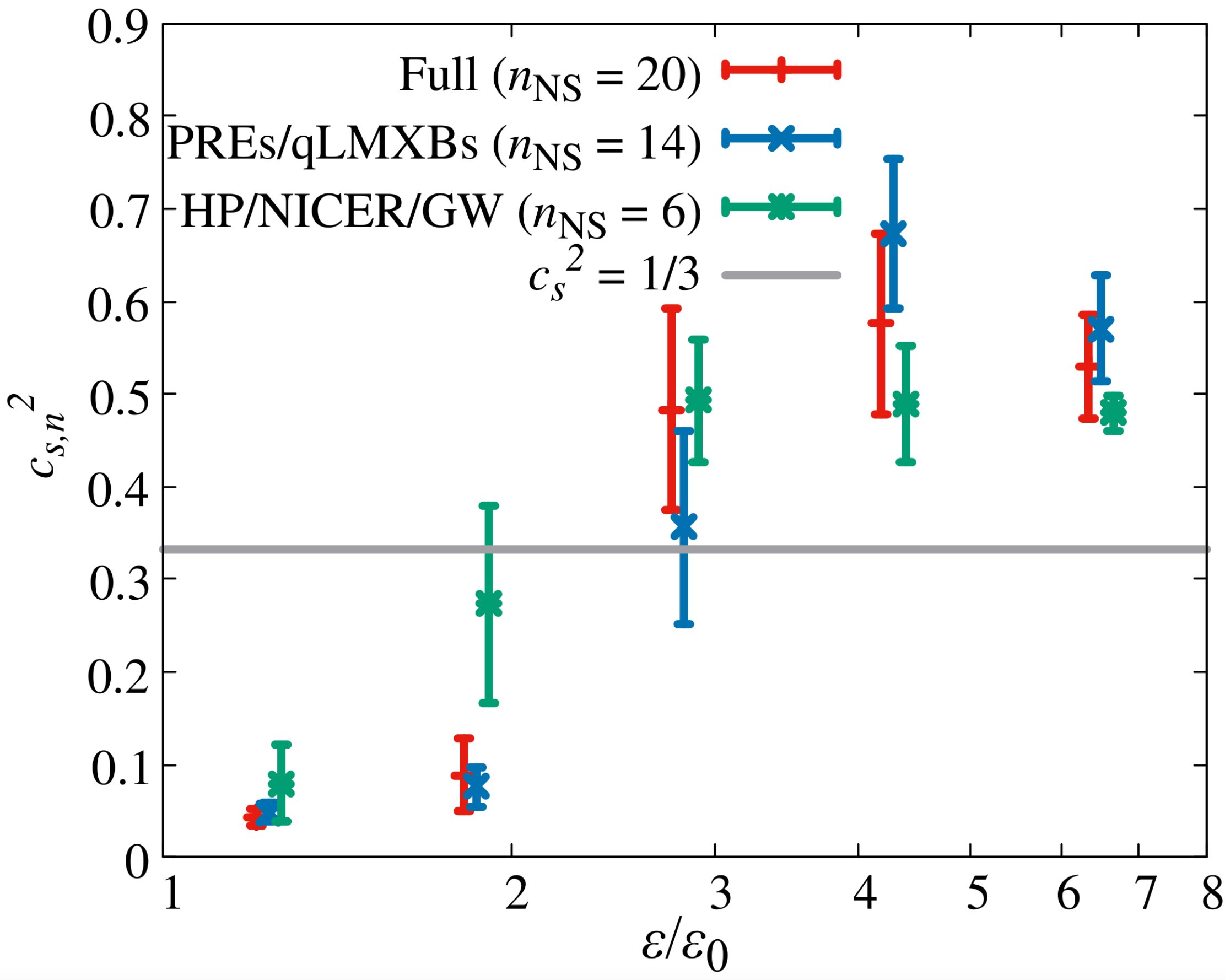}\\[0.5cm]
\includegraphics[height=10.cm]{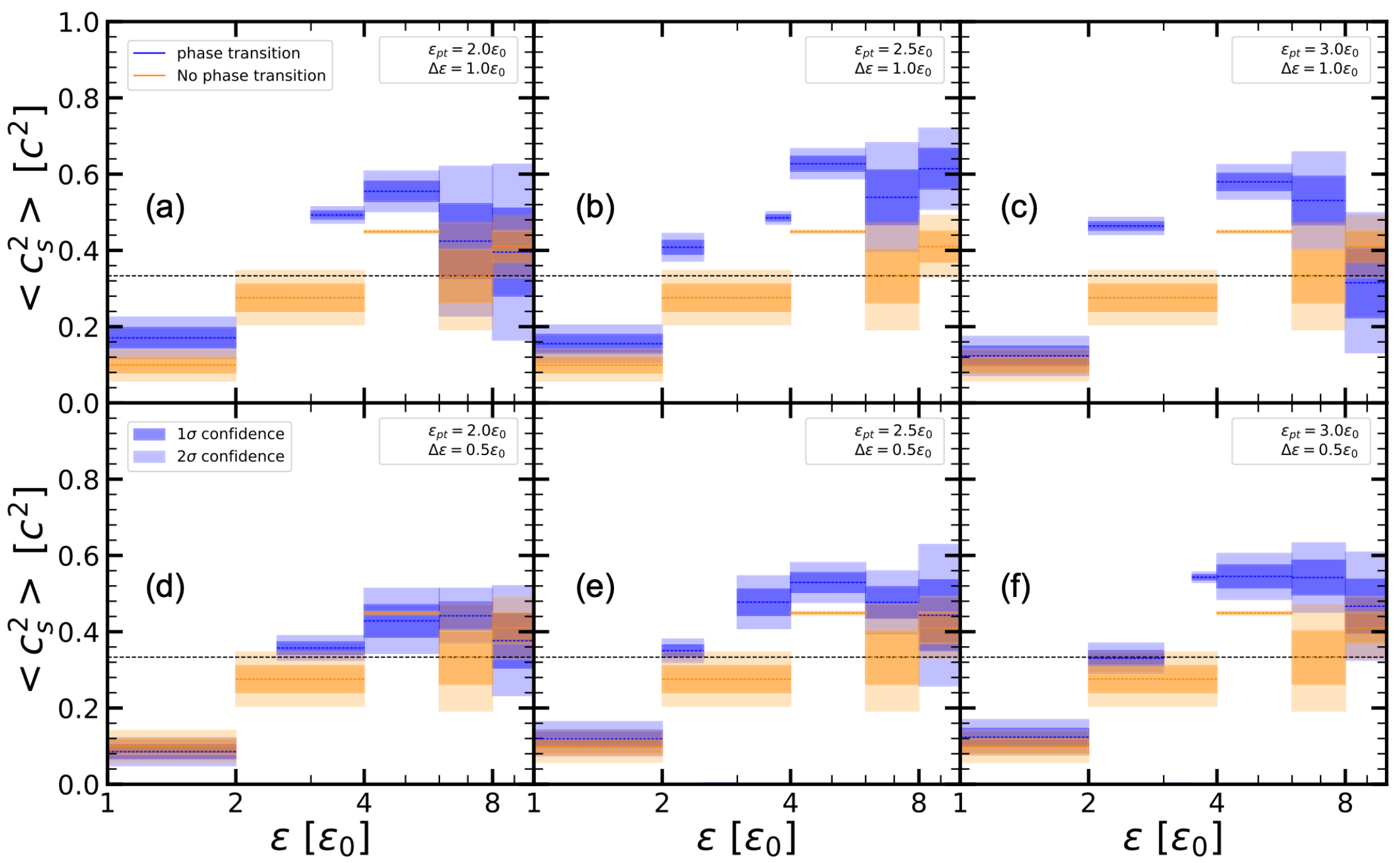}
\caption{(Color Online). Upper panel: the SSS in NSs using machine learning algorithms under different inference schemes.
Figure taken from Ref.\,\cite{Fuji2024b}.
Lower panel: a similar prediction on $s^2$ using deep neural network algorithms. Figure taken from Ref.\,\cite{HuJN24}.
}
\label{fig_Fuji24s2}
\end{figure}

We compare our estimates on the peak position and the ones existing in the literature: our $r_{\rm{pk}}/R\lesssim0.15$ is different from the one given by Ref.\,\cite{Ecker23}, which found that the peak of $s^2$ for TOV NSs under the constraint $M_{\rm{NS}}^{\max}\gtrsim2.35M_{\odot}$ is at about 2/3 of its radius, i.e., $r_{\rm{pk}}/R\approx2/3$, see the left panel of FIG.\,\ref{fig_Ecker23Delta} (see also FIG.\,\ref{fig_Ecker23s2r}).
Recently, Ref.\,\cite{Ferr24} found that the peak position in $s^2$ is at about (4-5)$\rho_{\rm{sat}}$ considering the constraint $\d M_{\rm{NS}}/\d R<0$ for all NS mass range (blue band in the right panel of FIG.\,\ref{fig_Ecker23Delta}), being much closer to NS centers than the previous predictions of the peak position at about $3\rho_{\rm{sat}}$\,\cite{Altiparmak2022,Ecker23}.
Similarly, Ref.\,\cite{Cao23} found $\rho_{\rm{pk}}/\rho_{\rm{c}}\approx0.98$ via $\rho_{\rm{c}}\approx0.56\,\rm{fm}^{-3}$ as well as $\rho_{\rm{pk}}\approx0.55\,\rm{fm}^{-3}$ for NSs with masses $\gtrsim2M_{\odot}$, thus $\widehat{\varepsilon}_{\rm{pk}}\lesssim0.98$ using the approximation (\ref{kkk}); the result is shown in the left panel of FIG.\,\ref{fig_Cao23Delta}.
By incorporating multi-messenger data of GW170817, PSR J0030+0451, PSR J0740+6620, and state-of-the-art theoretical progresses including the information from CEFT and pQCD calculations, the $s^2$ profile is predicted and the peak is found to be located at $\rho\lesssim5\rho_{\rm{sat}}$\,\cite{Han2023}, shown in  the right panel of FIG.\,\ref{fig_Cao23Delta}.
A Bayesian analysis\,\cite{Brandes2023-a} utilizing the CEFT and astrophysical constraints especially the PSR J052-0607 predicted the peak in $s^2$ is wide and very close to the center.
In the upper panel of FIG.\,\ref{fig_Fuji24s2},
the energy density dependence of $s^2$ using machine learning algorithms is shown adopting different inference schemes, one finds that $\widehat{\varepsilon}_{\rm{pk}}\gtrsim4$.
A similar investigation on $s^2$ using deep neural network algorithms also find the peak in $s^2$ (if existing) is at about (4-5)$\varepsilon_0$\,\cite{HuJN24}, see the lower panel of FIG.\,\ref{fig_Fuji24s2} (blue bands).
These comparisons suggest that accurately determining the peak position in $s^2$ will require further efforts, particularly using precise astrophysical data, such as observations of NS masses/radii, alongside advanced inference algorithms.

Our above qualitative and quantitative analyses/estimates on the peak of $s^2$ as well as the qualitative predictions for the $\widehat{P}(\widehat{\varepsilon})=\widehat{P}(\mu)$ are expected to be (nearly) model independent. This is because we rely only on (without using extra assumptions) the general conditions/requirements to make the prediction, these include the inequality $0\leq s^2\leq1$ of (\ref{oo-1}),  the inequality (\ref{io-5}), the sum rules of (\ref{Peps-sumrule}) together with formula (\ref{sc2-TOV}) and the bounds on the high-order coefficients such as $a_4$ (see Subsection \ref{sub_rhoc}) and $a_6$, etc.
It is necessary to emphasize that due to its perturbative nature, however, the above analysis on the $\widehat{\varepsilon}$-dependence of $s^2$ (or the signature of the continuous crossover) is only effective and qualitatively reasonable at finite distances $\widehat{r}$ being very close to NS centers ($\widehat{r}=0$).
Whether there exist sharp PTs at some large distances away from the centers (or at certain intermediate energy density $\varepsilon$) could not be excluded immediately. This issue should be in principle analyzed by including more higher-order terms in the expansions, and the results may have certain model dependence.
Furthermore, $\widehat{P}(\mu)<{\x}$ is definite for $\mu\approx0^-$ considering Eq.\,(\ref{Pmu3}) which is fundamentally different from $s^2(\mu)$ of Eq.\,(\ref{smu}),  since $D$ still has sizable probabilities to be positive and thus $s^2(\mu)<s_{\rm{c}}^2$, see the sketch given in FIG.\,\ref{fig_re_pe}.

We use the perturbative expansion of $P/\varepsilon$ and Eq.\,(\ref{smu}) to explain the existence of a peak in the derivative part $-t$ of the SSS\,\cite{Fuji22}, according to the trace anomaly decomposition.
Dividing Eq.\,(\ref{Pmu3}) by $\mu=\widehat{\varepsilon}-1$ gives the ratio $P/\varepsilon$ to order $\mu^2$ as $P/\varepsilon\approx{\x}-t_{\rm{c}}\mu+(D+t_{\rm{c}})\mu^2$ where $t_{\rm{c}}={\x}-s_{\rm{c}}^2$.
Consequently, the derivative part $-t$ of $s^2$ is obtained as to order $\mu^2$:
\begin{equation}-t(\widehat{\varepsilon})=-t(\mu)\approx-t_{\rm{c}}+\left(2D+t_{\rm{c}}\right)\mu+\left(3J-D-t_{\rm{c}}\right)\mu^2,
\end{equation} from which the peak could be found,
\begin{equation}\label{tpk}
\mu_{\rm{pk}}^{(-t)}=\widehat{\varepsilon}_{\rm{pk}}^{(-t)}-1
=\frac{1}{2}\frac{2D+t_{\rm{c}}}{D+t_{\rm{c}}-3J}.
\end{equation}
Taking for example ${\x}\approx0.16\mbox{$\sim$}0.30$ (being relevant for PSR J0740+6620), this peak is shown to be located at about $-\mu_{\rm{pk}}^{(-t)}\approx10\%\mbox{$\sim$}5\%$ (see discussions given after Eq.\,(\ref{Pmu3}) for the estimate on $D$ and $J$).

On the other hand, using the expansion $\phi=P/\varepsilon=s_{\rm{non-deriv}}^2$ just given here, we deduce that $t_{\rm{c}}/2(D+t_{\rm{c}})>0$, i.e., there would be no peak in the non-derivative part of $s^2$, or this part is an increasing function of $\varepsilon$\,\cite{Fuji22}.
The value of $-t$ at $\mu_{\rm{pk}}^{(-t)}$ is,
\begin{equation}
-t_{\rm{pk}}\equiv
-t(\mu_{\rm{pk}}^{(-t)})=\frac{4D^2+12Jt_{\rm{c}}-3t_{\rm{c}}^2}{4D-12J+4t_{\rm{c}}}.
\end{equation}
Then, we have $-t_{\rm{pk}}\approx0.15\mbox{$\sim$}0.35$ and the non-derivative part (of $s^2$), namely $\phi=P/\varepsilon\approx0.14\mbox{$\sim$}0.28$ for ${\x}\approx0.16\mbox{$\sim$}0.30$. Moreover, comparing $\mu_{\rm{pk}}^{(-t)}$ of Eq.\,(\ref{tpk}) and $\mu_{\rm{pk}}=-D/3J$ of $s^2(\mu)$ (see Eq.\,(\ref{s2pk})) leads us to 
\begin{equation}\label{tpk1}
\boxed{
\mu_{\rm{pk}}-\mu_{\rm{pk}}^{(-t)}=\frac{2}{9}\left(\frac{D}{J}\right)^2\cdot
\frac{1+t_{\rm{c}}/D+3Jt_{\rm{c}}/2D^2}{1-D/3J-t_{\rm{c}}/3J}
>0.}
\end{equation}
This means the location of the peak in $s^2$ (i.e., $\mu_{\rm{pk}}$) occurs at a higher energy density than that of the peak in $-t$ (i.e., $\mu_{\rm{pk}}^{(-t)}$),  see our general discussions given in Subsection \ref{sub_decomTA}.  Equivalently, the peak in $s^2$ is closer to the NS center than the peak in $-t$.
When higher order terms of $\mu$ are included in the expansions, the specific location of $\mu_{\rm{pk}}$ and $\mu_{\rm{pk}}^{(-t)}$ may vary, however the relation $\mu_{\rm{pk}}-\mu_{\rm{pk}}^{(-t)}>0$ would not be changed.

\subsection{Strong-field gravity extruding a peak in SSS profile: analytical and numerical demonstrations}\label{sub_s2_peak}

In a few subsections above, we studied the sign of $s^2(\heps)-s_{\rm{c}}^2$ and the possible peak in $s^2$ profile by simulating the coefficients $D$ and $J$. By doing so, several useful properties of the dense matter in NS cores have been revealed, e.g., how the coefficient $D$ affects the appearance of a peaked structure in $s^2$; how the coefficients $D$ and $J$, both characterizing the dense matter EOS, are related with each other, etc.
These aspects more or less involve the treatments of the dense matter EOS via the $D$ and $J$.
In order to investigate {\color{xll}the physical origin} of a peaked $s^2$ more apparently and with the least EOS-model dependence, we may start from the dimensionless TOV equations straightforwardly and make as little use as possible of the dense matter EOS. The Eq.\,(\ref{ef-1}) for SSS to order $\hr^4$ provides a useful starting point for this purpose. In the following, we summarize how we did this originally in Ref.\,\,\cite{CL24-b}.

Compared with the Newtonian case discussed in Subsection \ref{sub_s2_Newtonian}, the strong GR effects bring two modifications to $s^2$: (a) ${\x}$ can be sizable $\gtrsim\mathcal{O}(0.1)$ in NSs; and (b) the expression for $s^2$ is changed using the GR version (of Eq.\,(\ref{st_s2})).
Parallel to Eq.\,(\ref{prf}), we obtain $\d s^2/\d\widehat{\varepsilon}$ with GR as\,\cite{CL24-b},
\begin{empheq}[box=\fbox]{align}\label{PM}
\frac{\d s^2}{\d\widehat{\varepsilon}}
=&\frac{Y}{1-2\widehat{M}/\widehat{r}}\left\{
\left(\frac{\d s^2}{\d\widehat{\varepsilon}}\right)_{\rm{N}}
-\frac{\widehat{\varepsilon}\widehat{M}}{1-2\widehat{M}/\widehat{r}}\frac{2}{\widehat{r}^4}
\left(\frac{\d\widehat{r}}{\d\widehat{\varepsilon}}\right)^2\left(\widehat{r}^3\widehat{\varepsilon}-\widehat{M}\right)\right.\notag\\
&\hspace*{0.5cm}\left.-\frac{\widehat{\varepsilon}}{\widehat{M}}\frac{\d\widehat{r}}{\d\widehat{\varepsilon}}
\left(1+\frac{\widehat{r}^3\widehat{P}}{\widehat{M}}\right)^{-1}
\left[
\widehat{r}\widehat{M}s^2+\widehat{P}\frac{\d\widehat{r}}{\d\widehat{\varepsilon}}\left(3\widehat{M}-\widehat{r}^3\widehat{\varepsilon}\right)
\right]
-\frac{\widehat{M}}{\widehat{r}^2}\frac{\d\widehat{r}}{\d\widehat{\varepsilon}}
\left(1+\frac{\widehat{P}}{\widehat{\varepsilon}}\right)^{-1}\left(s^2-\frac{\widehat{P}}{\widehat{\varepsilon}}\right)
\right\},
\end{empheq}
where $Y=(1+\widehat{P}/\widehat{\varepsilon})(1+\widehat{r}^3\widehat{P}/\widehat{M})$ (not to be confused with $\y\equiv\varepsilon_{\rm{c}}/\varepsilon_0$ introduced earlier for the reduced central energy density) and $(\d s^2/\d\widehat{\varepsilon})_{\rm{N}}>0$ is given by Eq.\,(\ref{prf}). 
In Eq.\,(\ref{PM}),  besides $3\widehat{M}-\widehat{r}^3\widehat{\varepsilon}>0$ (proved after Eq.\,(\ref{prf})) and $s^2-\widehat{P}/\widehat{\varepsilon}=s^2-P/\varepsilon=s^2-\phi>0$ is the derivative part of $s^2$\,\cite{Fuji22}, we also have (for small $\widehat{r}$),
\begin{equation}
\widehat{r}^3\widehat{\varepsilon}-\widehat{M}
\approx\frac{2\widehat{r}^3}{3}\left(1+\frac{6}{5}a_2\widehat{r}^2\right)>0.
\end{equation}
This means each term in Eq.\,(\ref{PM}) contributes to $\d s^2/\d\widehat{\varepsilon}$ with definite sign.

The two terms  in Eq.\,(\ref{PM}) contributing negatively to the derivative $\d s^2/\d\widehat{\varepsilon}$ can be combined as,
\begin{align}\label{g-1}
-\frac{Y}{1-2\widehat{M}/\widehat{r}}\left(\frac{\d\widehat{r}}{\d\widehat{\varepsilon}}\right)^2&\left[\frac{\widehat{\varepsilon}\widehat{M}}{1-2\widehat{M}/\widehat{r}}\frac{2}{\widehat{r}^4}
\left(\widehat{r}^3\widehat{\varepsilon}-\widehat{M}\right)
+\frac{\widehat{P}\widehat{\varepsilon}}{\widehat{M}}\left(1+\frac{\widehat{r}^3\widehat{P}}{\widehat{M}}\right)^{-1}\left(3\widehat{M}-\widehat{r}^3\widehat{\varepsilon}\right)
\right]<0.
\end{align}
The first term in (\ref{g-1}) originated from ``$-2\widehat{M}/\widehat{r}$'' in Eq.\,(\ref{st_s2}) survives even when $\widehat{P}=0$ is considered (so it is a geometric correction);  while the second term is the geometry-matter coupling (from ``$\widehat{r}^3\widehat{P}/\widehat{M}$'' in Eq.\,(\ref{st_s2})), which disappears if $\widehat{P}\approx0$ is taken.
This demonstrates {\color{xll}that GR geometrical effects, the matter-geometry couplings can both effectively modify the $\widehat{\varepsilon}$-dependence (or equivalently the $\widehat{r}$-dependence) of $s^2$ compared with its Newtonian counterpart.
Specifically, they all tend to make the $s^2$ decrease with increasing $\widehat{\varepsilon}$ (i.e., $\d s^2/\d\widehat{\varepsilon}<0$).}
On the other hand,  $s^2>0$ itself contributes positively to the derivative $\d s^2/\d\widehat{\varepsilon}$ in Eq.\,(\ref{PM}),  namely
{\begin{align}\label{g-2}
&-\frac{Y}{1-2\widehat{M}/\widehat{r}}\left[\frac{\widehat{\varepsilon}}{\widehat{M}}\frac{\d\widehat{r}}{\d\widehat{\varepsilon}}
\left(1+\frac{\widehat{r}^3\widehat{P}}{\widehat{M}}\right)^{-1}
\widehat{r}\widehat{M}s^2
+\frac{\widehat{M}}{\widehat{r}^2}\frac{\d\widehat{r}}{\d\widehat{\varepsilon}}
\left(1+\frac{\widehat{P}}{\widehat{\varepsilon}}\right)^{-1}s^2\right]\notag\\
=&-\frac{\widehat{\varepsilon}\widehat{r} s^2}{1-2\widehat{M}/\widehat{r}}\left(\frac{\d\widehat{r}}{\d\widehat{\varepsilon}}\right)\left(1+\frac{2\widehat{P}}{\widehat{\varepsilon}}+\frac{\widehat{M}}{\widehat{\varepsilon}\widehat{r}^3}\right)>0.
\end{align}}This means $s^2$ itself tends to remove the peak, i.e., it tends to make $\d s^2/\d\widehat{\varepsilon}>0$ near $\widehat{r}=0$.
{Only the first term in the curry bracket of Eq.\,(\ref{PM}), namely $(\d s^2/\d\widehat{\varepsilon})_{\rm{N}}$ survives in the Newtonian limit, and all the other terms (proportional to $(\d\widehat{r}/\d\widehat{\varepsilon})^2$ and $\d\widehat{r}/\d\widehat{\varepsilon}$) disappear.
Therefore, the correction (\ref{g-2}) also disappears in $\d s^2/\d\widehat{\varepsilon}$ for Newtonian stars.}
The final sign of $\d s^2/\d\widehat{\varepsilon}$ is the result of a balance/competition between Eqs.\,(\ref{g-1}) and (\ref{g-2}). It may depend on the EOS model, and therefore not every dense matter EOS could induce a peaked $s^2$ profile, as expected.

A formula for $s^2\approx s_{\rm{c}}^2+l_2\widehat{r}^2+l_4\widehat{r}^4$ similar to Eq.\,(\ref{zf-3}) for the reduced pressure could be obtained for the GR case directly from (\ref{ef-1}) by expanding terms to linear order of ${\x}$ (the $s_{\rm{c}}^2$ is kept without being expanded over ${\x}$),
\begin{align}
s^2\approx&s_{\rm{c}}^2+\left[\left(12a_4s_{\rm{c}}^4-\frac{4}{15}\right)
-\left(48a_4s_{\rm{c}}^4+s_{\rm{c}}^2+\frac{3}{5}\right){\x}
\right]\widehat{r}^2\notag\\
&+
\left[\left(
144a_4^2s_{\rm{c}}^6+18a_6s_{\rm{c}}^4-\frac{62}{35}a_4s_{\rm{c}}^2+\frac{1}{60s_{\rm{c}}^2}
+\frac{1}{12}s_{\rm{c}}^2-\frac{1}{18}
\right)\right.\notag\\
&\hspace*{2cm}
+\left.\left(
\frac{1}{15s_{\rm{c}}^2}+\frac{1}{15}
-12a_4s_{\rm{c}}^4-72a_6s_{\rm{c}}^4
-1152a_4^2s_{\rm{c}}^6+\frac{116}{35}a_4s_{\rm{c}}^2
\right){\x}
\right]
\widehat{r}^4.\label{k-1}
\end{align}
Expanding $s_{\rm{c}}^2$ of Eq.\,(\ref{sc2-TOV}) over ${\x}$ as $s_{\rm{c}}^2\approx 4{\x}/3+4{\x}^2/3+\cdots$ further approximates (\ref{k-1}) as,
\begin{empheq}[box=\fbox]{align}
s^2\approx& \frac{4}{3}{\x}+\frac{4}{3}{\x}^2
+\left[-\frac{4}{15}-\frac{3}{5}{\x}
+\left(\frac{64a_4}{3}-\frac{4}{3}\right){\x}^2\right]\widehat{r}^2\notag\\
&+\left[
\frac{1}{80{\x}}-\frac{13}{720}+
\left(\frac{229}{1440}-\frac{248a_4}{105}\right){\x}
+\left(\frac{157}{360}+\frac{72a_4}{35}+32a_6\right){\x}^2
\right]\widehat{r}^4.
\label{for_1}
\end{empheq}

\begin{figure}[h!]
\centering
\includegraphics[height=6.5cm]{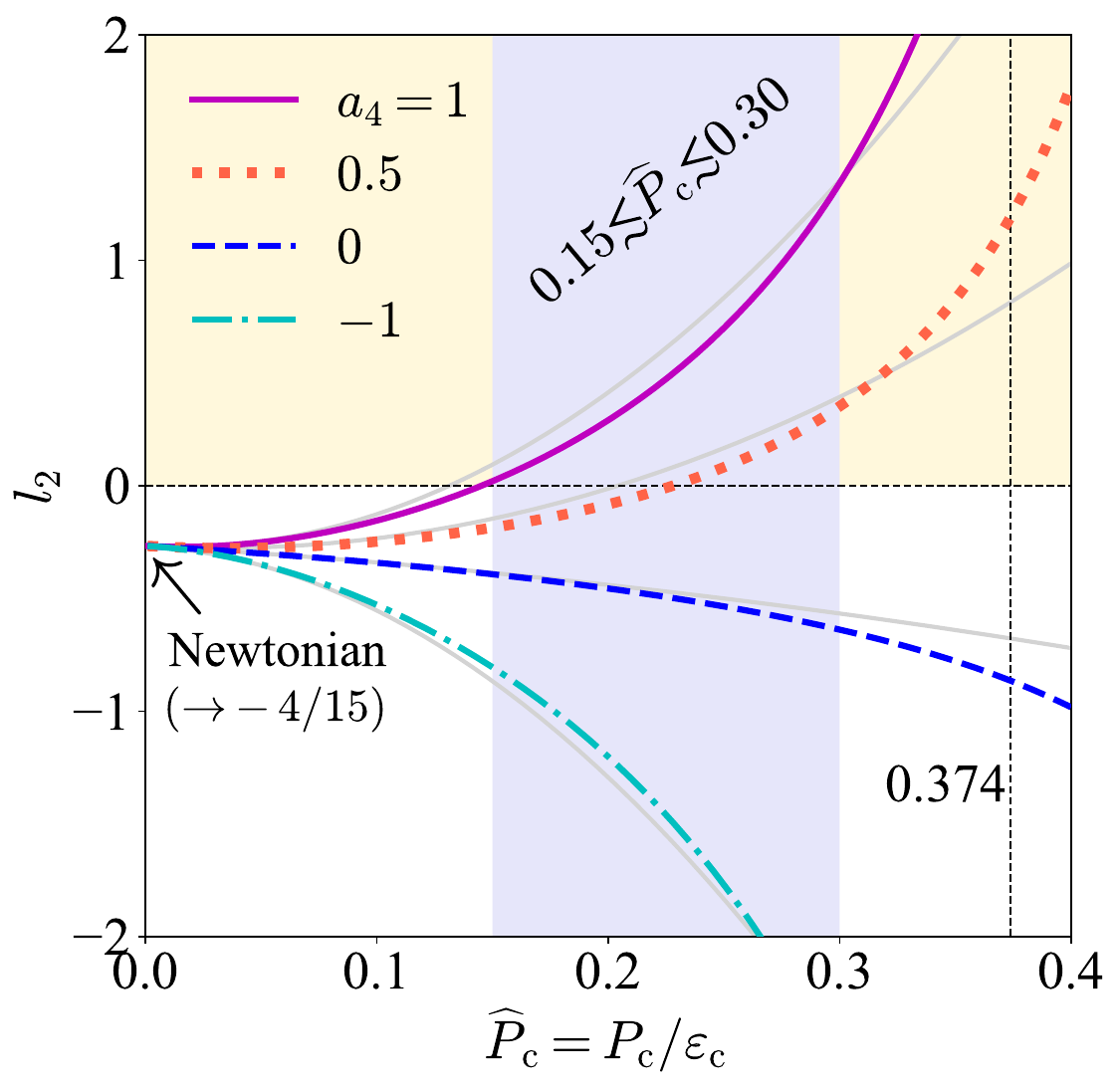}\qquad
\includegraphics[height=6.5cm]{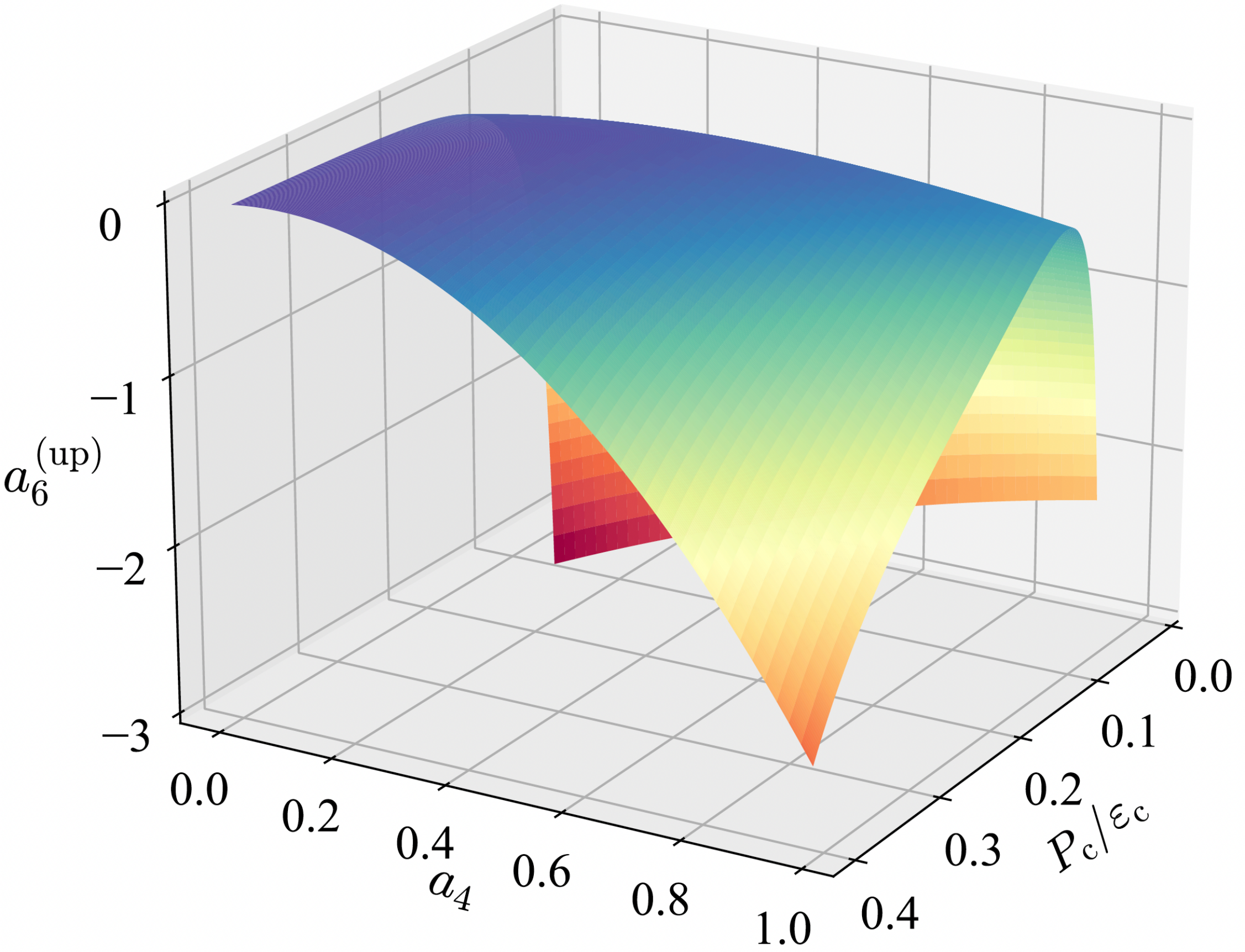}
\caption{(Color Online). Left panel: the coefficient $l_2$ as a function of ${\x}$ adopting different $a_4$ values, the grey line near each colored curve is for the approximated $l_2\approx-4/15-3{\x}/5+(64a_4/3-4/3){\x}^2$ to order ${\x}^2$.
The light-yellow background shows the necessary condition for a peaked SSS profile, i.e., $l_2>0$.
Right panel: the surface of the upper limit $a_6^{\rm{up}}$ as a function of $a_4$
and ${\x}$ in order to make the coefficient $l_4$ negative.
Figures taken from Ref.\,\cite{CL24-b}.
}\label{fig_s2_GR}
\end{figure}

\begin{figure}[h!]
\centering
\includegraphics[width=8cm]{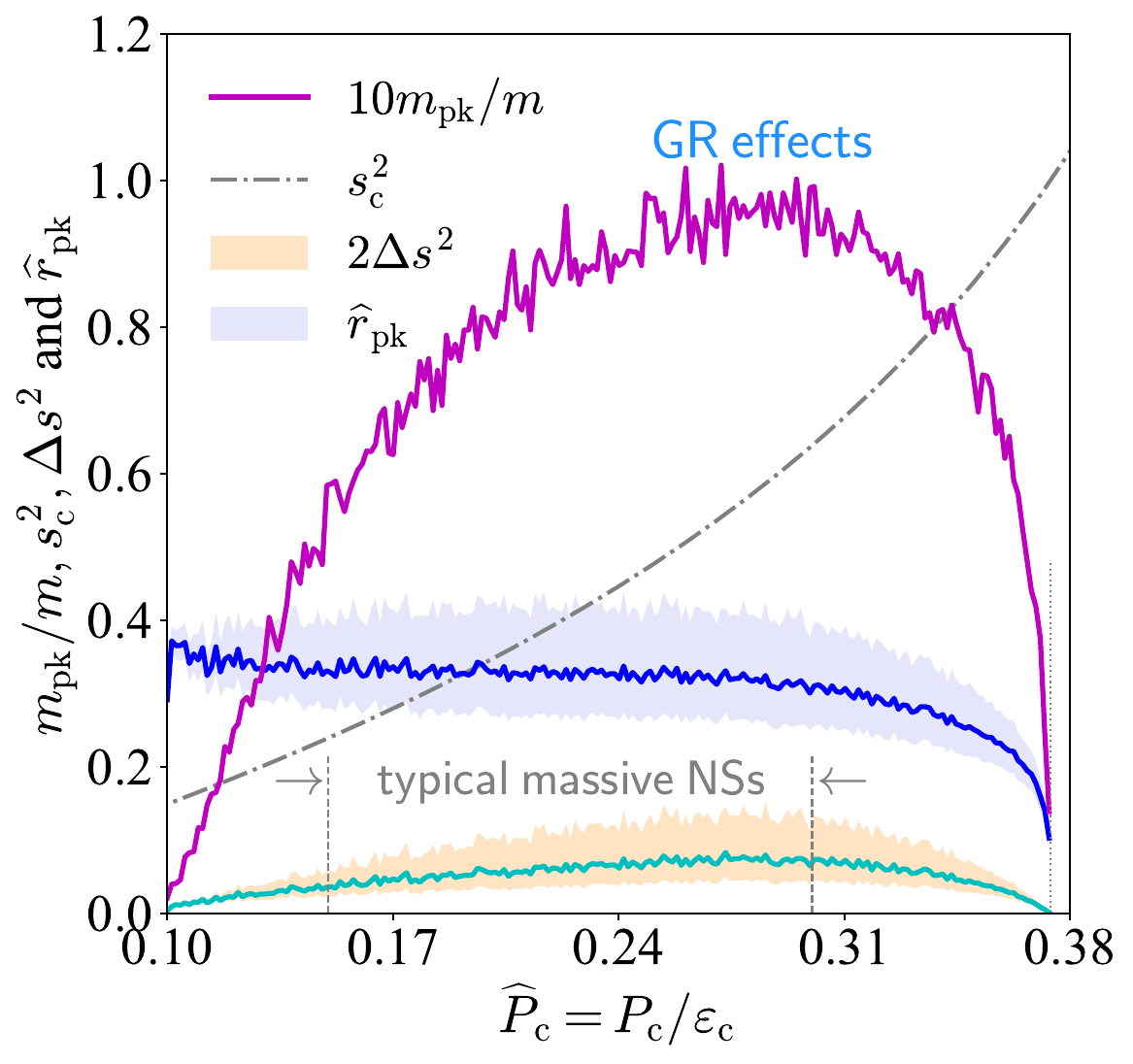}\qquad
\includegraphics[width=8cm]{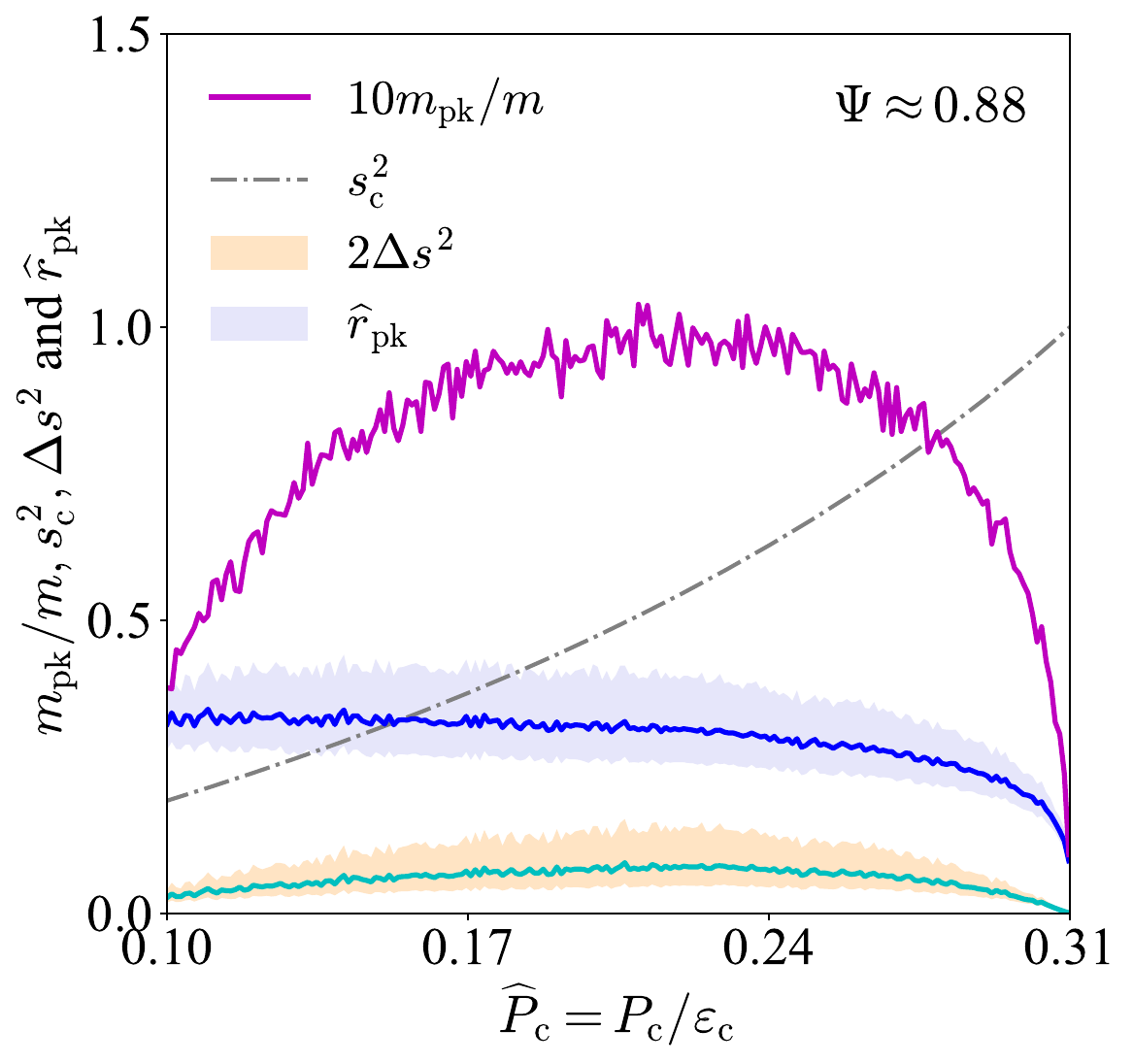}
\caption{(Color Online).  Left panel: probability of the occurrence of a peak in $s^2$ profile (magenta line),  the position of the peak $\widehat{r}_{\rm{pk}}=\sqrt{-l_2/2l_4}$ of $s^2$ (blue line), and (two times) the enhancement $\Delta s^2\equiv s_{\max}^2/s_{\rm{c}}^2-1=-l_2^2/4l_4s_{\rm{c}}^2$ on $s_{\rm{c}}^2$ (green line).
The background bands on $\widehat{r}_{\rm{pk}}$ (lavender) and $2\Delta s^2$ (orange) represent their 1$\sigma$ uncertainties.  The ${\x}$-dependence of $s_{\rm{c}}^2$ is also shown (grey dash-dotted line). Right panel: same as the left panel but with $\Psi=0.88$ for normally stable NSs on the M-R curve. Figures taken from Ref.\,\cite{CL24-a}.
}\label{fig_s2_peak_stat_prop}
\end{figure}

\begin{figure}[h!]
\centering
\includegraphics[width=7.5cm]{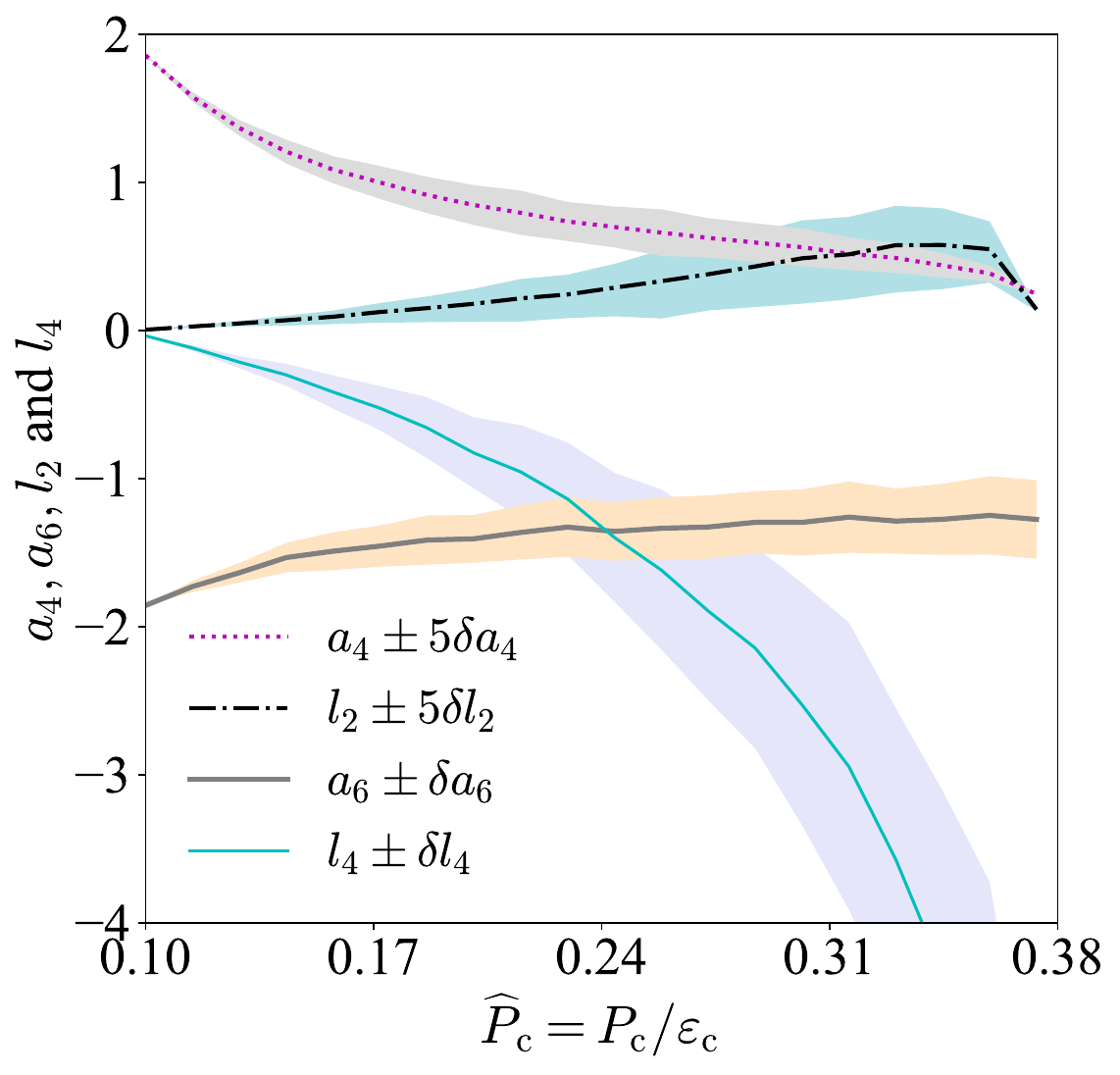}\qquad
\includegraphics[width=7.5cm]{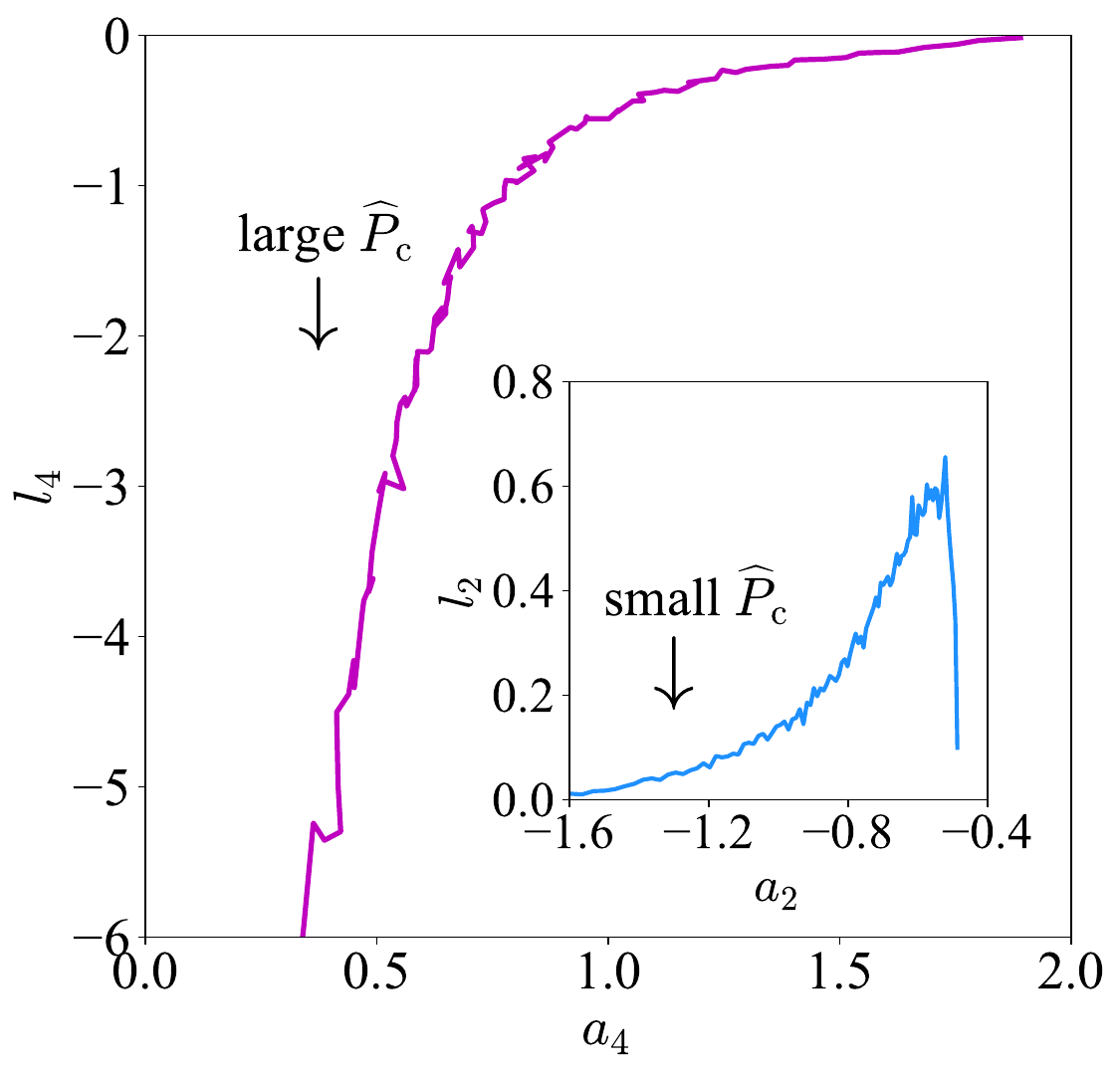}
\caption{(Color Online). Coefficients $a_4,a_6,l_2$ and $l_4$ as functions of ${\x}$.
The uncertainties on $a_4$ and $l_2$ are amplified by 5 times for a more clear visualization due to their relative smallness. 
Figure taken from Ref.\,\cite{CL24-a}.
Right panel: dependence of $l_4$ on $a_4$ and that of $l_2$ on $a_2$ (inset).
}\label{fig_s2_peak_stat_a4a6}
\end{figure}

\begin{figure}[h!]
\centering
\includegraphics[height=7.2cm]{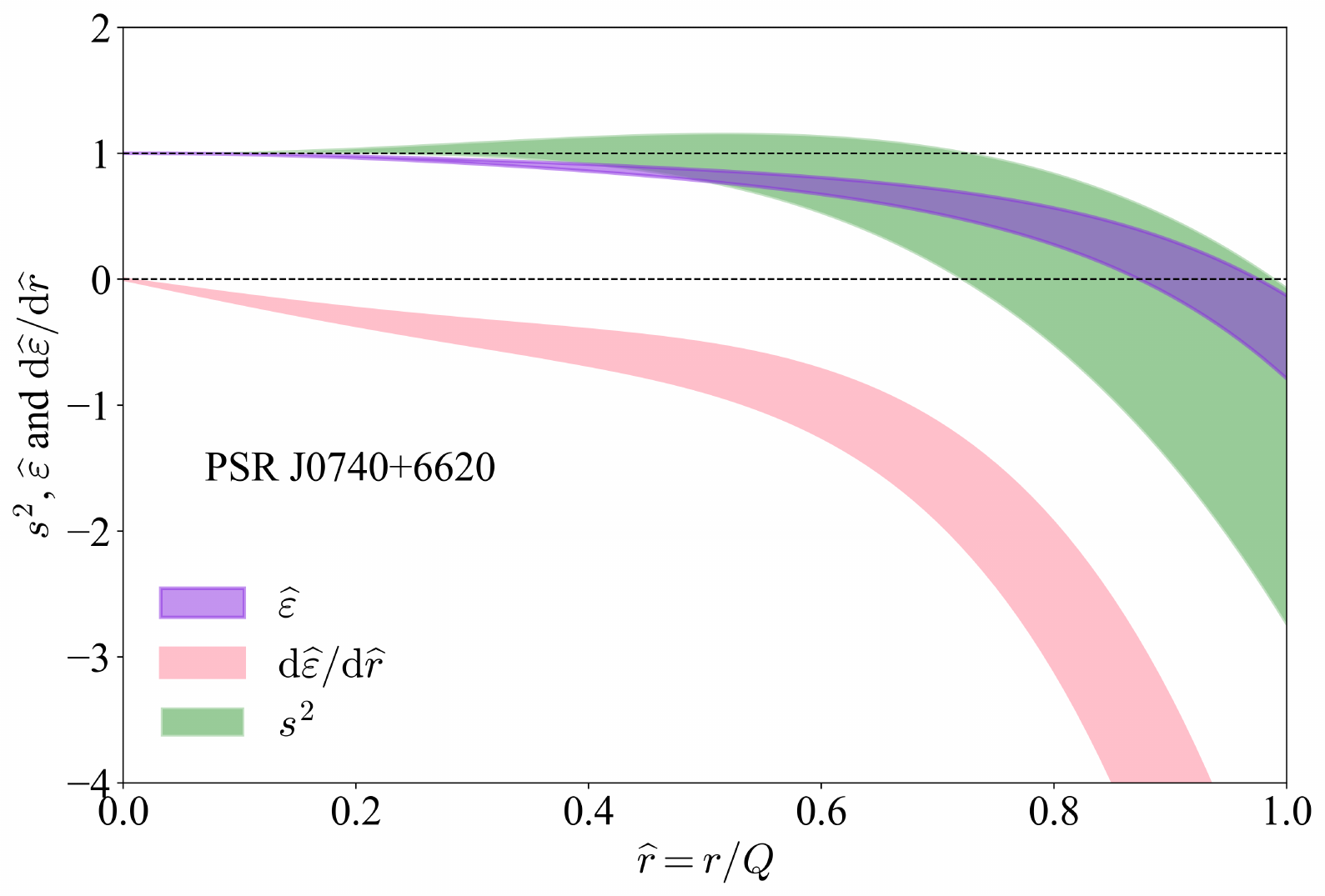}
\caption{(Color Online).  Radial dependence of $s^2$,  $\heps$ and $\d\heps/\d\hr$ for PSR J0740+6620 in the case that there exists a (wide) peak in $s^2$, using constrained $a_4$ and $a_6$ from FIG.\,\ref{fig_s2_peak_stat_a4a6}.
}\label{fig_eps-vs-r6602}
\end{figure}

\begin{figure}[h!]
\centering
\includegraphics[width=5.6cm]{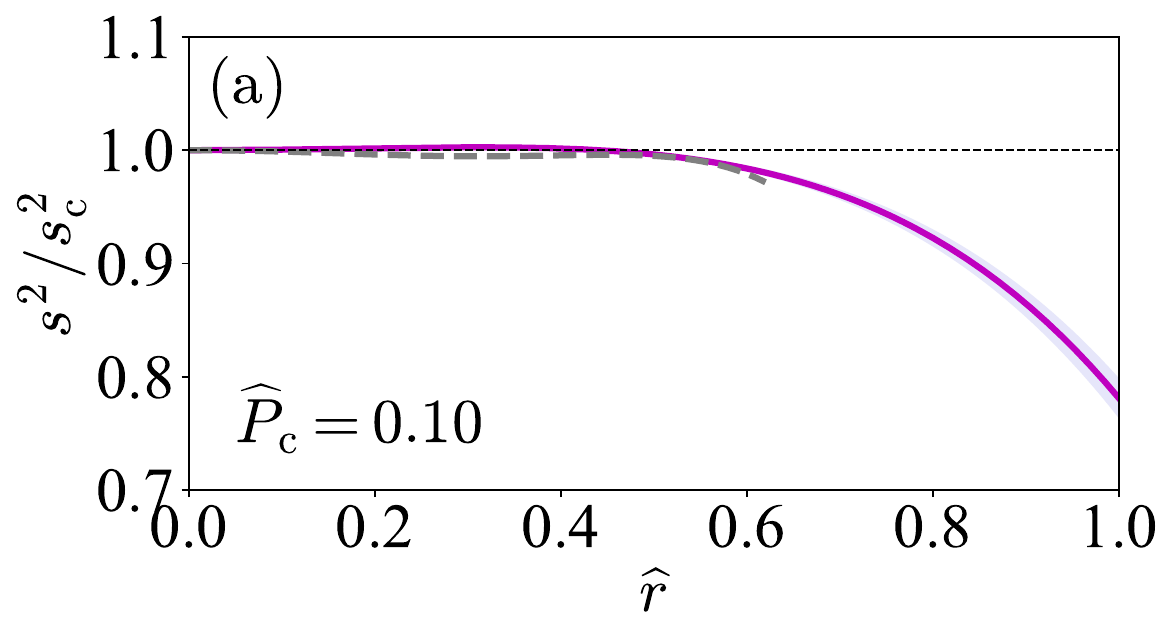}
\includegraphics[width=5.6cm]{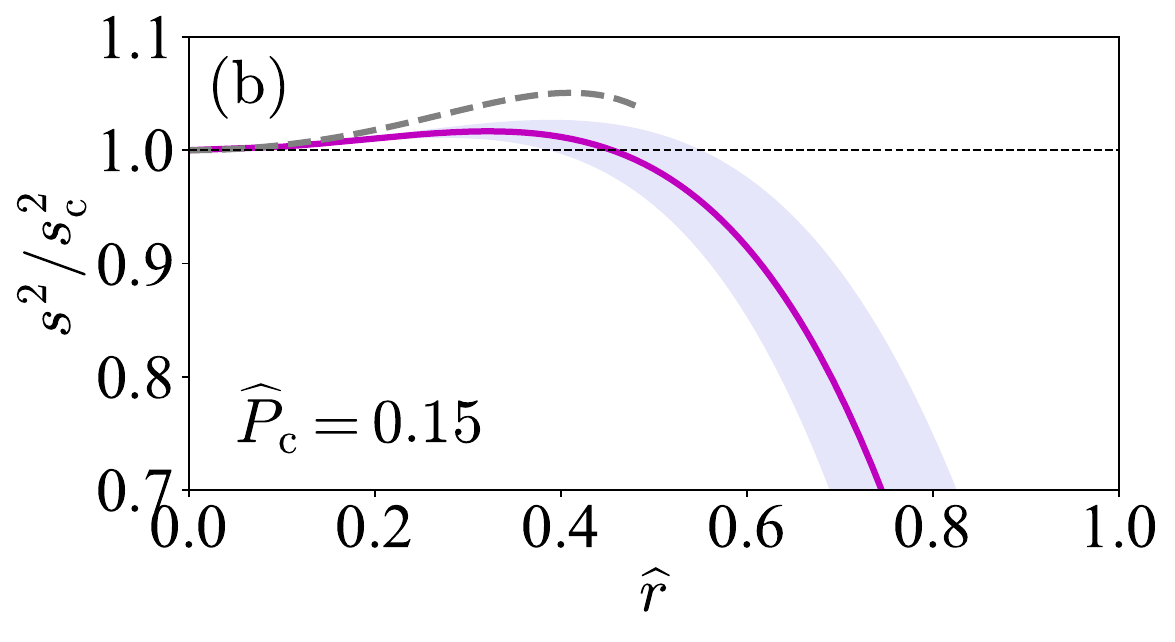}
\includegraphics[width=5.6cm]{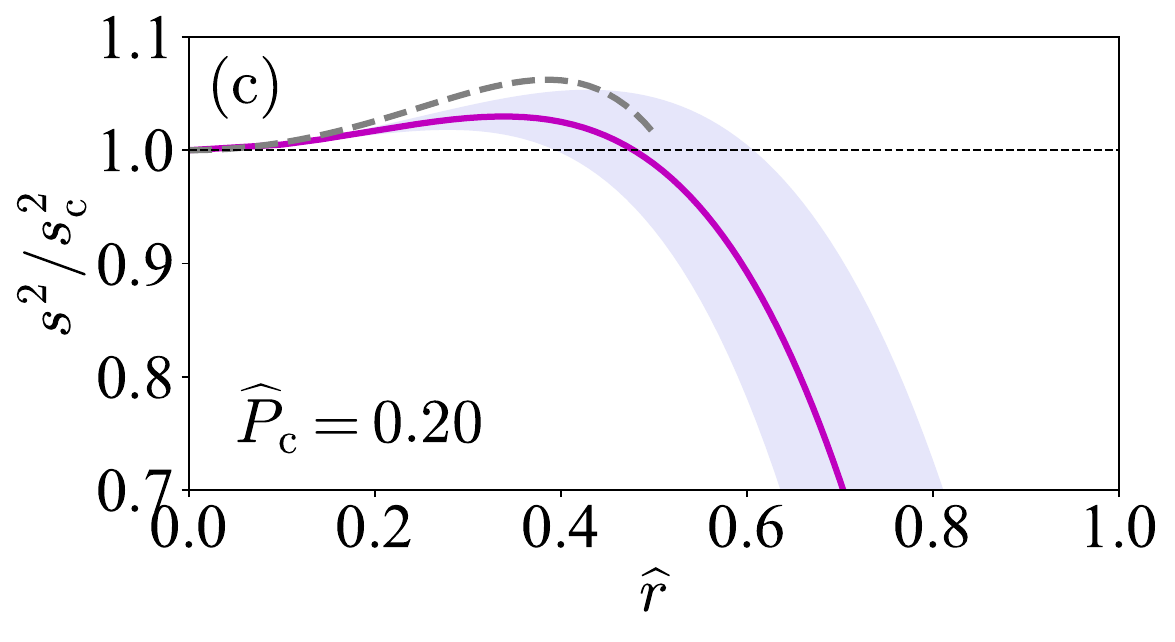}\\
\includegraphics[width=5.6cm]{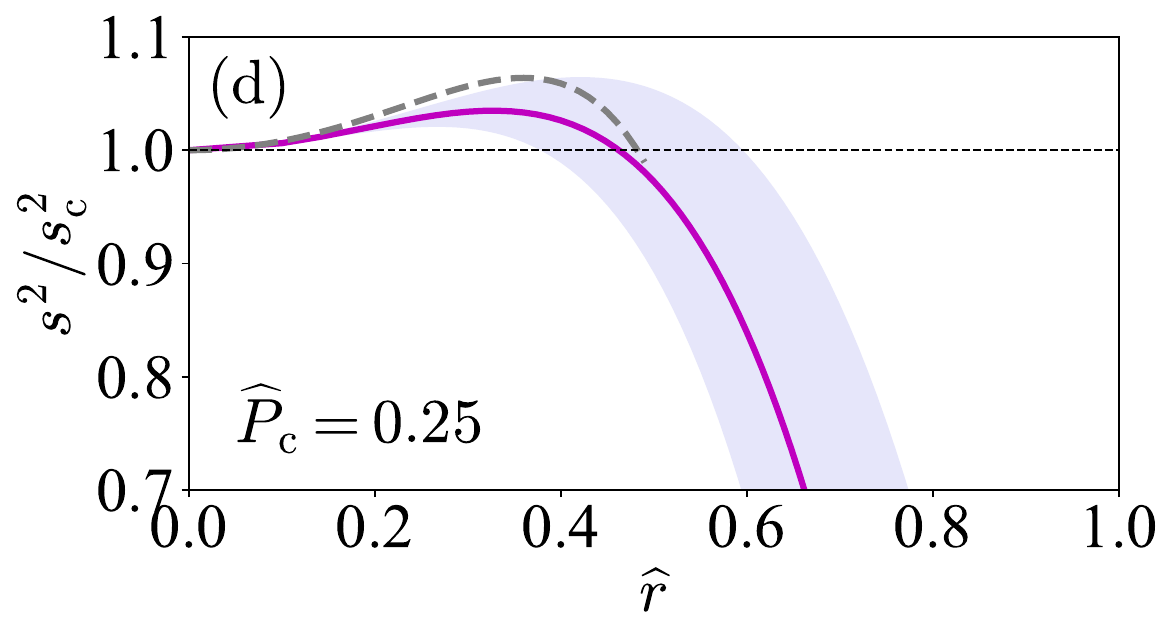}
\includegraphics[width=5.6cm]{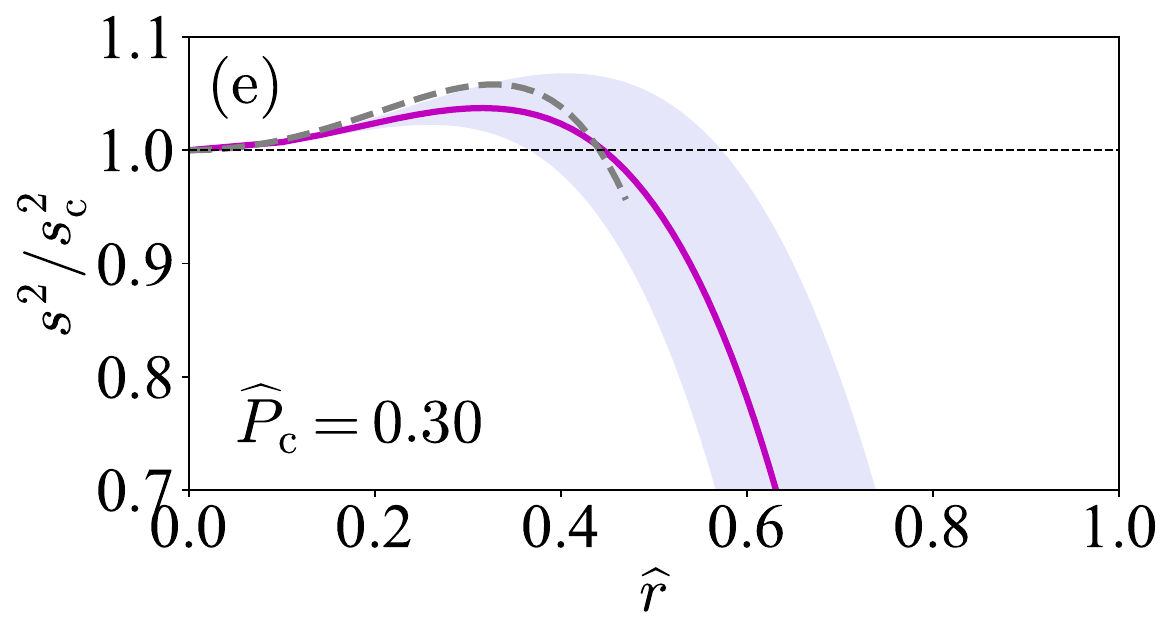}
\includegraphics[width=5.6cm]{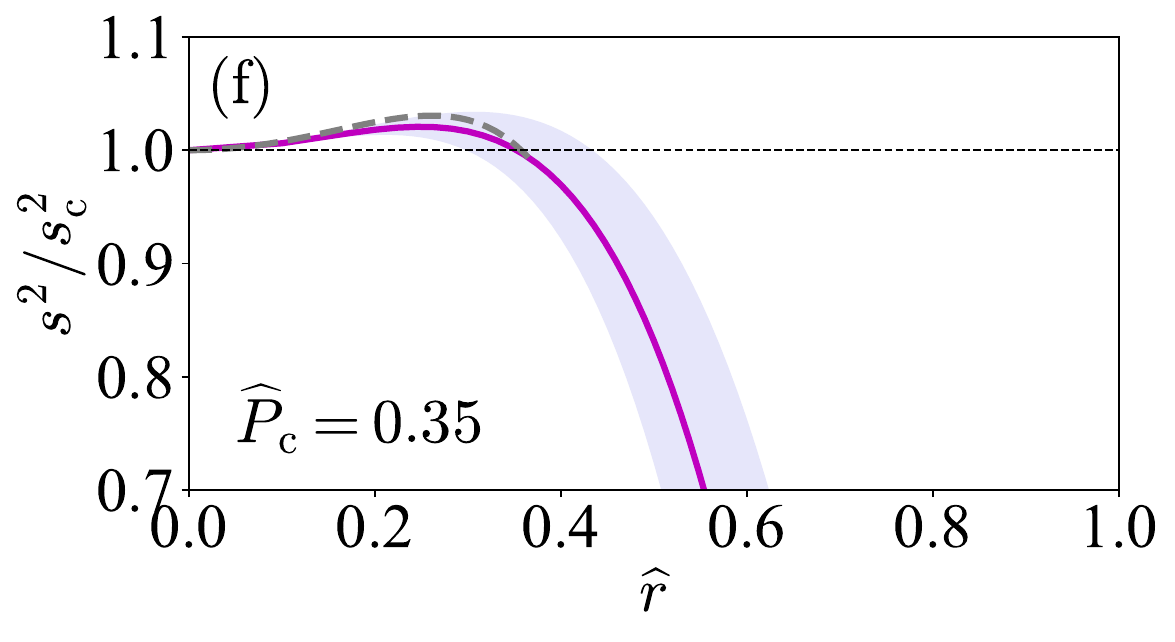}\\
\includegraphics[width=5.6cm]{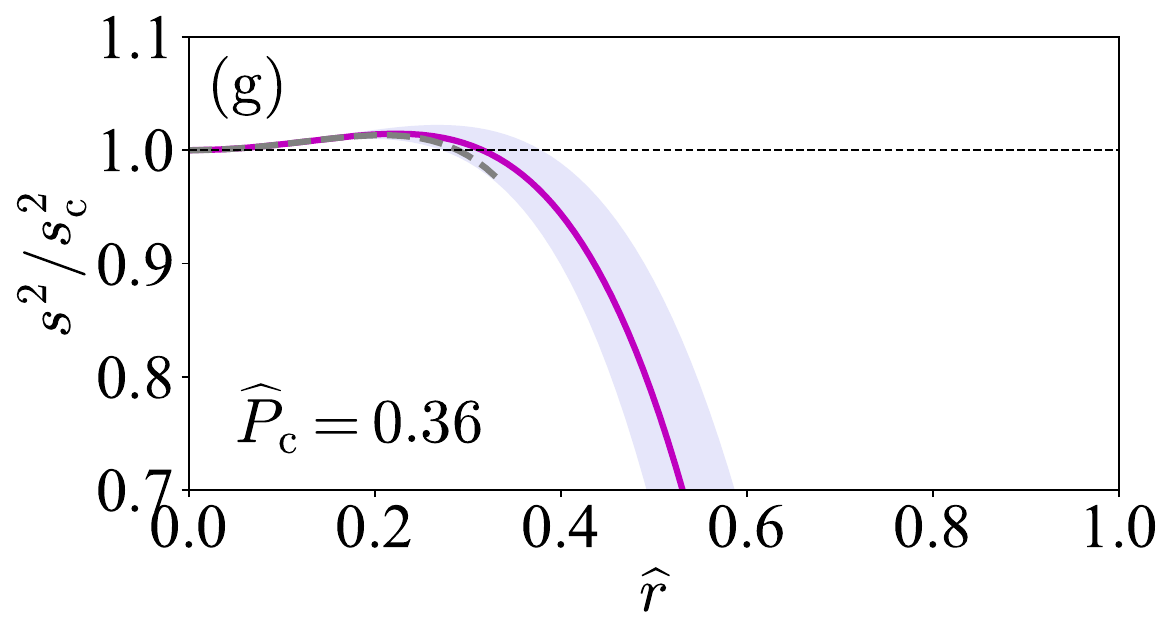}
\includegraphics[width=5.6cm]{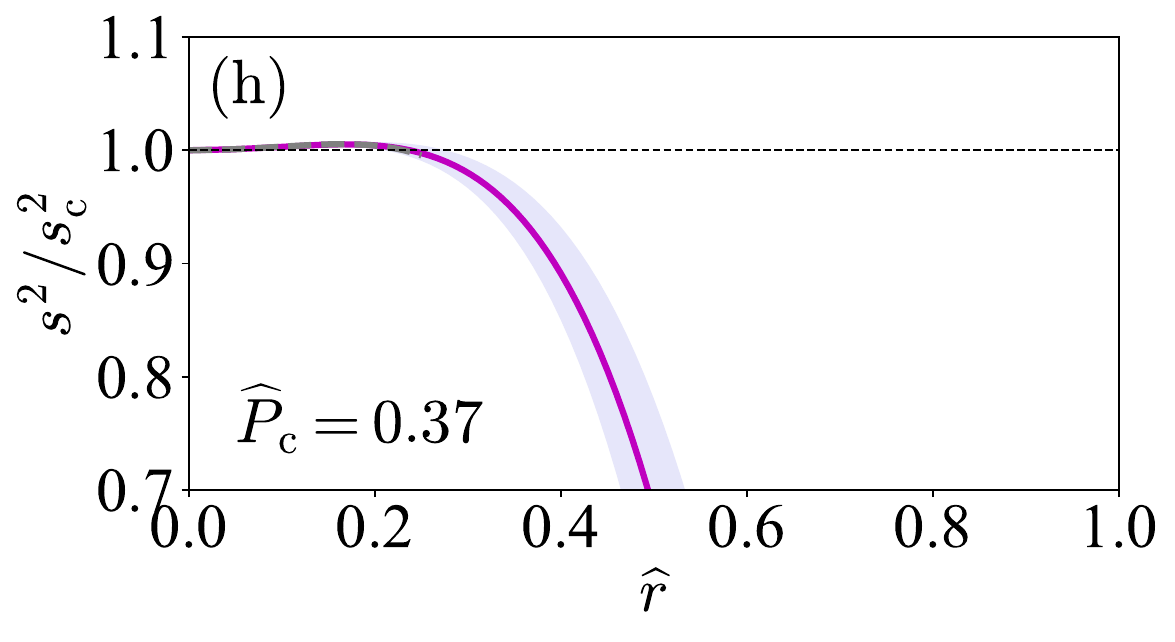}
\includegraphics[width=5.6cm]{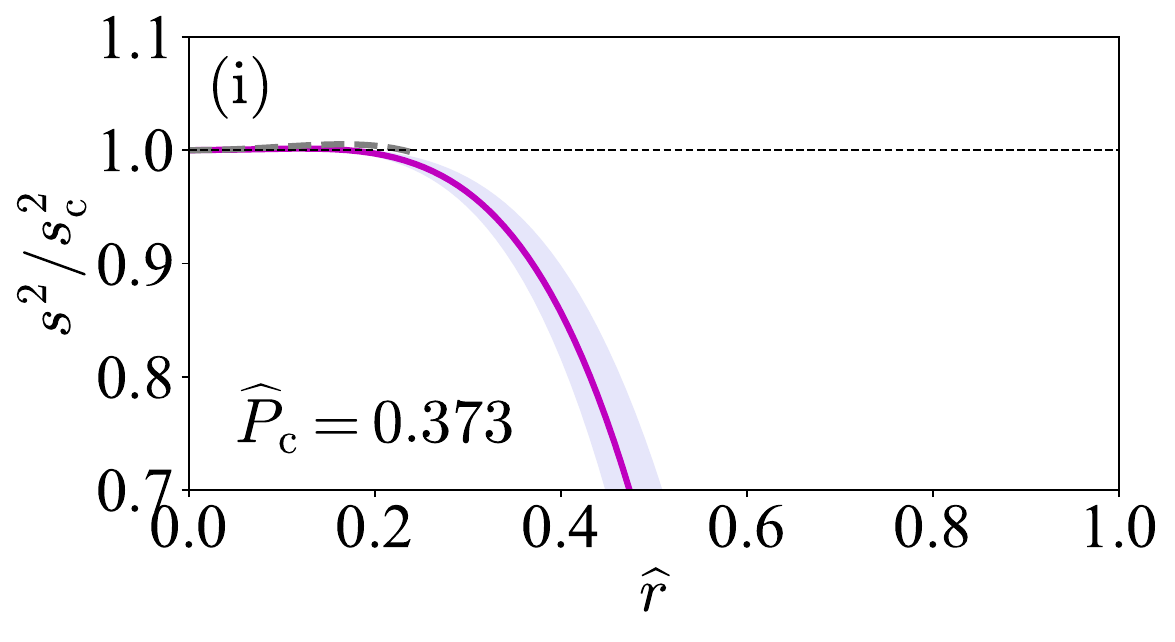}
\caption{(Color Online). $s^2/s_{\rm{c}}^2$ as a function of $\widehat{r}$ {with increasing values of ${\x}$ and with 1$\sigma$ uncertainties,  the grey dashed line is the one including contribution of $l_6\widehat{r}^6$, namely $s^2\approx s_{\rm{c}}^2+l_2\widehat{r}^2+l_4\widehat{r}^4+l_6\widehat{r}^6$.} Figures taken from Ref.\,\cite{CL24-a}.
}\label{fig_s2_peak_stat-af}
\end{figure}

We find that Eq.\,(\ref{k-1}) with GR is very different from Eq.\,(\ref{for_s2Newt}) for Newtonian stars even if ${\x}\approx0$ is taken (this is due to the geometrical correction ``$-2\widehat{M}/\widehat{r}$''), e.g., the terms ``$s_{\rm{c}}^2/12$'' and ``$-1/18$'' at order $\widehat{r}^4{\x}^0=\widehat{r}^4$ are new compared with Eq.\,(\ref{for_s2Newt}).
Similarly, new terms may emerge at other orders of ${\x}$.
Considering Eq.\,(\ref{for_1}), e.g., we find $a_4$ starts to operate at order ${\x}^2$ {(in the form of $64a_4{\x}^2/3$).}
Then for finite ${\x}\sim\mathcal{O}(0.1)$, the coefficient $l_2$ may take positive values.
For example, taking $a_4\approx1$ ($a_4\approx0.5$) and requiring $l_2>0$ leads to ${\x}\gtrsim0.15$ (${\x}\gtrsim0.23$), which are typical values of ${\x}$ in massive NSs.
Thus, a sizable ${\x}$ is necessary for inducing a peaked $s^2$ profile:

\begin{equation}
\boxed{
\mbox{a peaked $s^2$: }
l_2>0\rm{\;needs\;}a_4>0\rm{\;as\;well\;as\;}{\x}\gtrsim\mathcal{O}(0.1).
}
\end{equation}
However, $a_4>0$ in the Newtonian limit could still not generally make $l_2>0$ (see Subsection \ref{sub_s2_Newtonian} and especially Eq.\,(\ref{for_s2_a6})), we thus see clearly that ${\x}\gtrsim\mathcal{O}(0.1)$ is fundamental. The left panel of FIG.\,\ref{fig_s2_GR} gives an illustration of the ${\x}$-dependence of $l_2$ (adopting four different $a_4$'s), where $0.15\lesssim{\x}\lesssim0.30$ is marked as the range for ${\x}$ in typical massive NSs.
The grey line near each colored curve is for the approximated $l_2\approx-4/15-3{\x}/5+(64a_4/3-4/3){\x}^2$. Obviously, $l_2^{\rm{N}}=-4/15$ (see Eq.\,(\ref{for_s2Newt-1})).
Once the coefficient $l_2$ becomes positive, there would unavoidably be a peak in $s^2$ profile at some finite distance $\widehat{r}<\widehat{R}$ regardless of the higher-order coefficients (since near the surface the $s^2\to0$).
One can demonstrate that the coefficient $a_6$ needs to be negative if $l_4<0$ is required.
The resulted condition for $a_6$ could be obtained from Eq.\,(\ref{ef-1}), namely $l_4<0$,
or equivalently,
{\begin{equation}
a_6<\frac{b_6}{s_{\rm{c}}^2}+\frac{4}{3}\frac{a_4}{b_2}\left(s_{\rm{c}}^2a_4-b_4\right)\equiv a_6^{\rm{(up)}},
\end{equation}}where all the expressions (for $b_2,b_4$ and $b_6$ as well as $s_{\rm{c}}^2$) are available.
The $a_4$-dependence of $a_6^{\rm{(up)}}$ is shown in the right panel of FIG.\,\ref{fig_s2_GR}, from which
we find that $a_6<0$ and notice that both $a_4$ and $a_6$ are $\sim\mathcal{O}(1)$, e.g.,
for ${\x}\approx0.2$ and $a_4\approx1$, we then have $a_6\lesssim-1$.
If the coefficient $l_4$ is also positive, then an analysis of the even higher-order terms becomes necessary, e.g., the coefficient $l_6$, etc.
Nonetheless, the peak will emerge at some finite $\widehat{r}$.
In this review, we focus on $l_2>0$ and $l_4<0$, and point out the extension if the $l_6$-term is included when necessary.

\begin{figure}
\centering
\includegraphics[width=5.6cm]{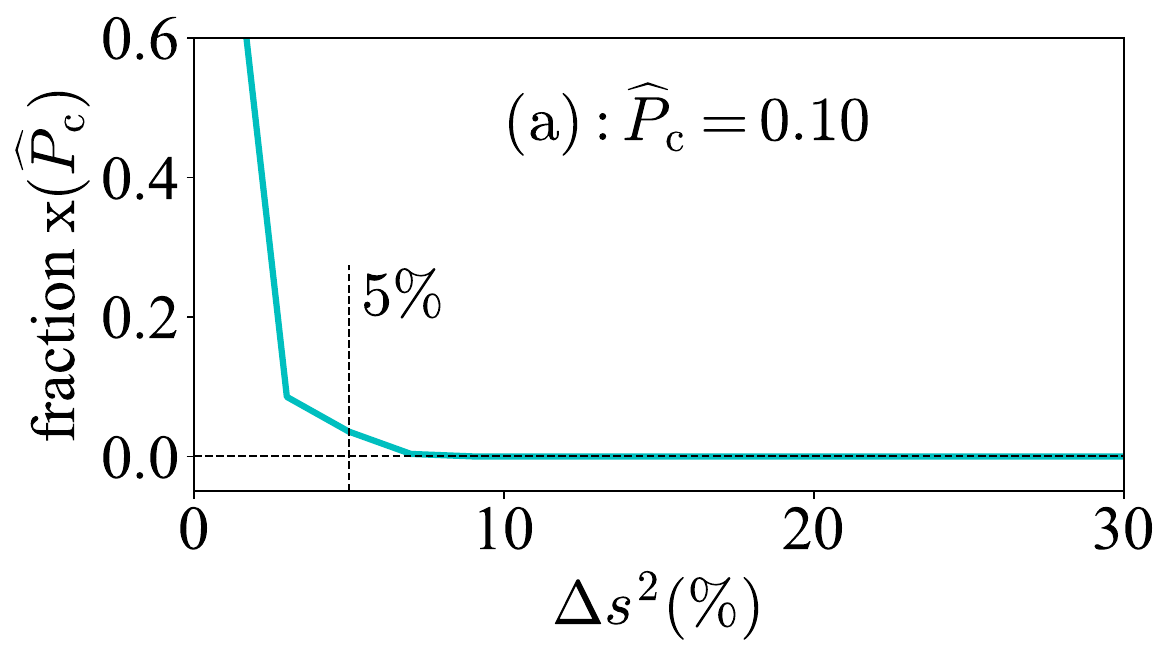}
\includegraphics[width=5.6cm]{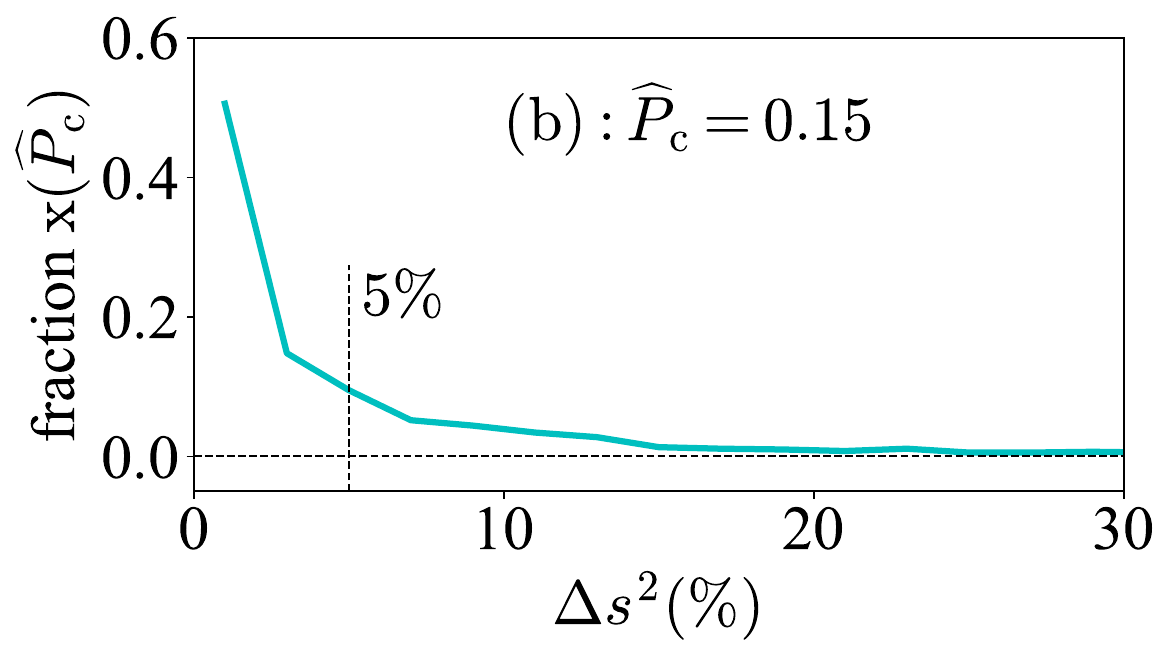}
\includegraphics[width=5.6cm]{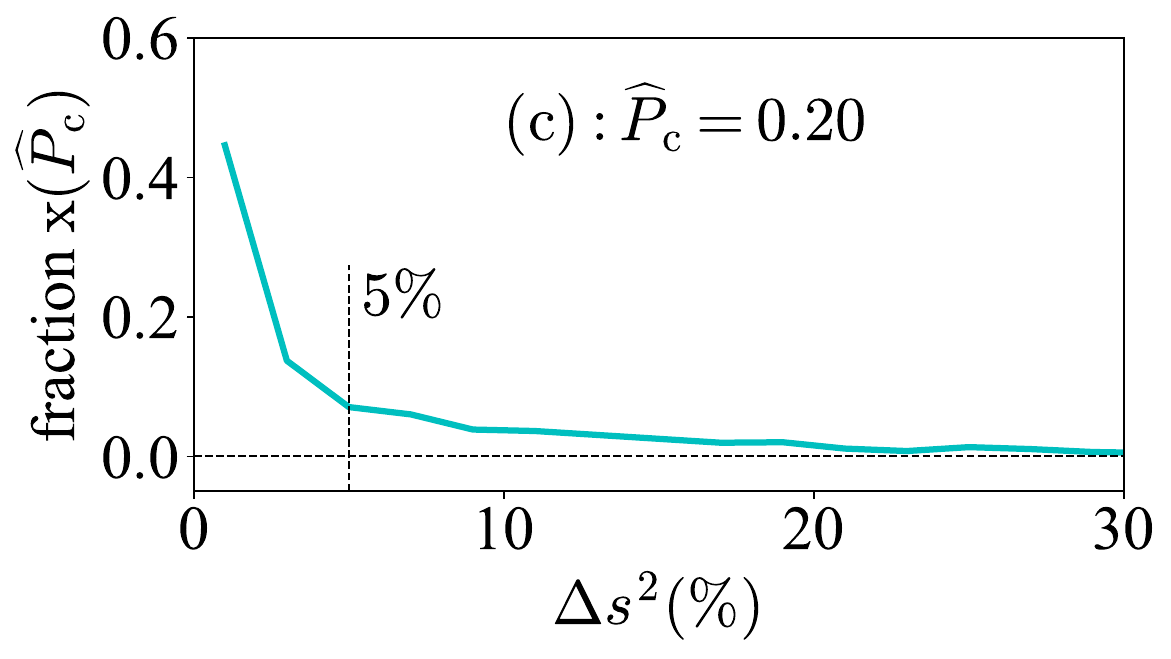}\\
\includegraphics[width=5.6cm]{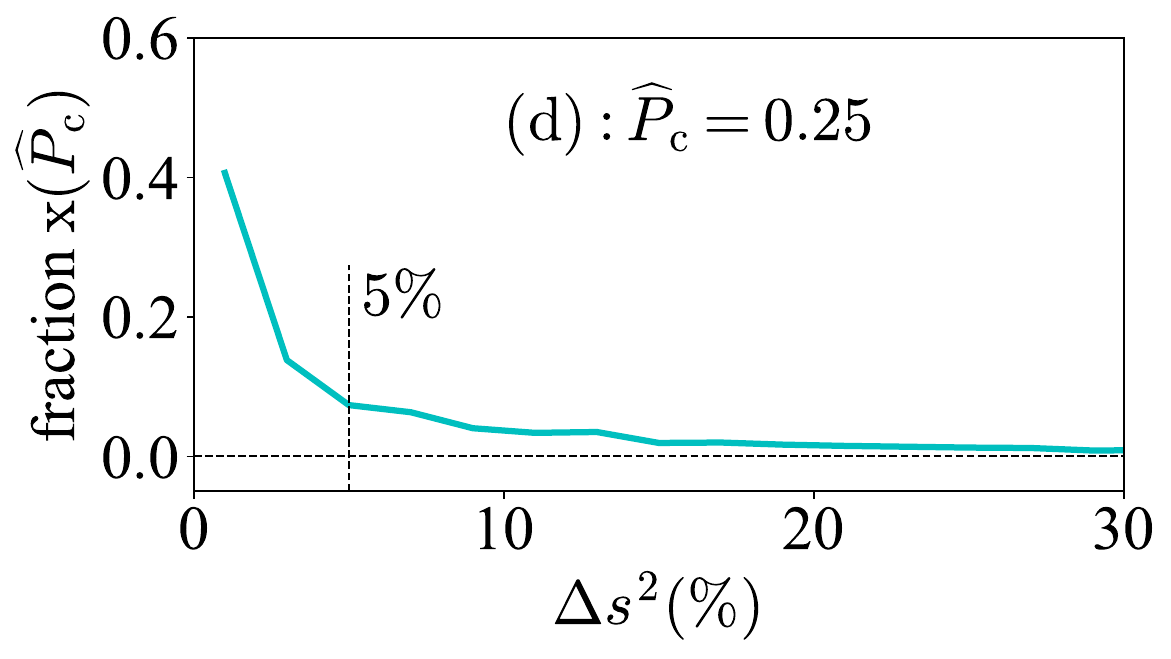}
\includegraphics[width=5.6cm]{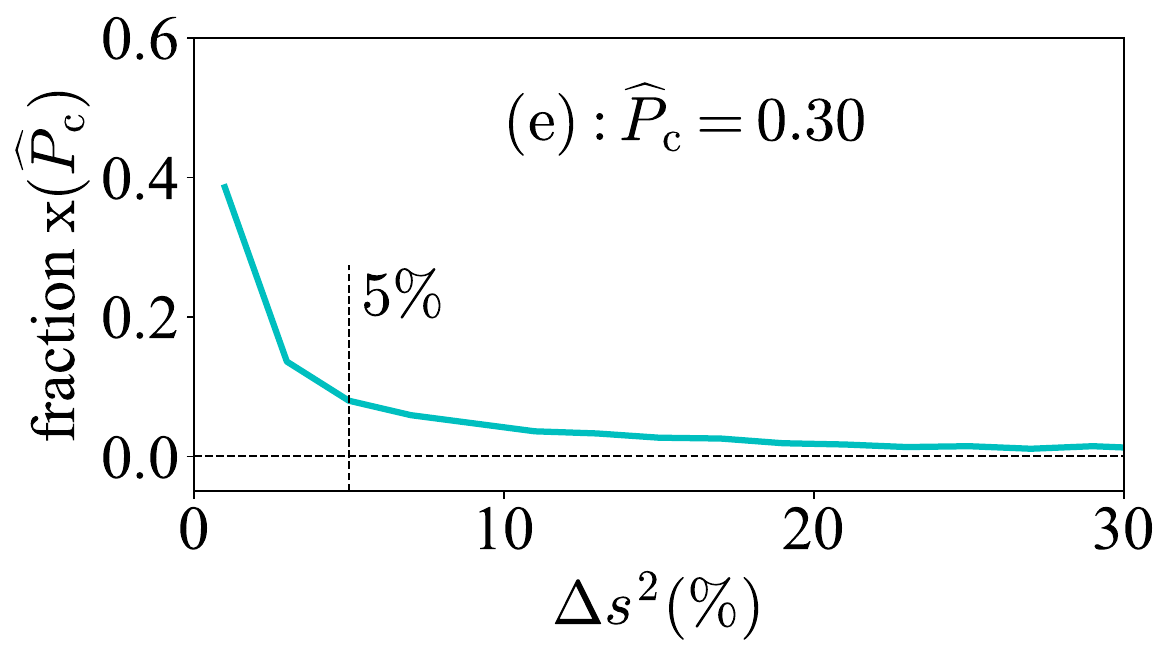}
\includegraphics[width=5.6cm]{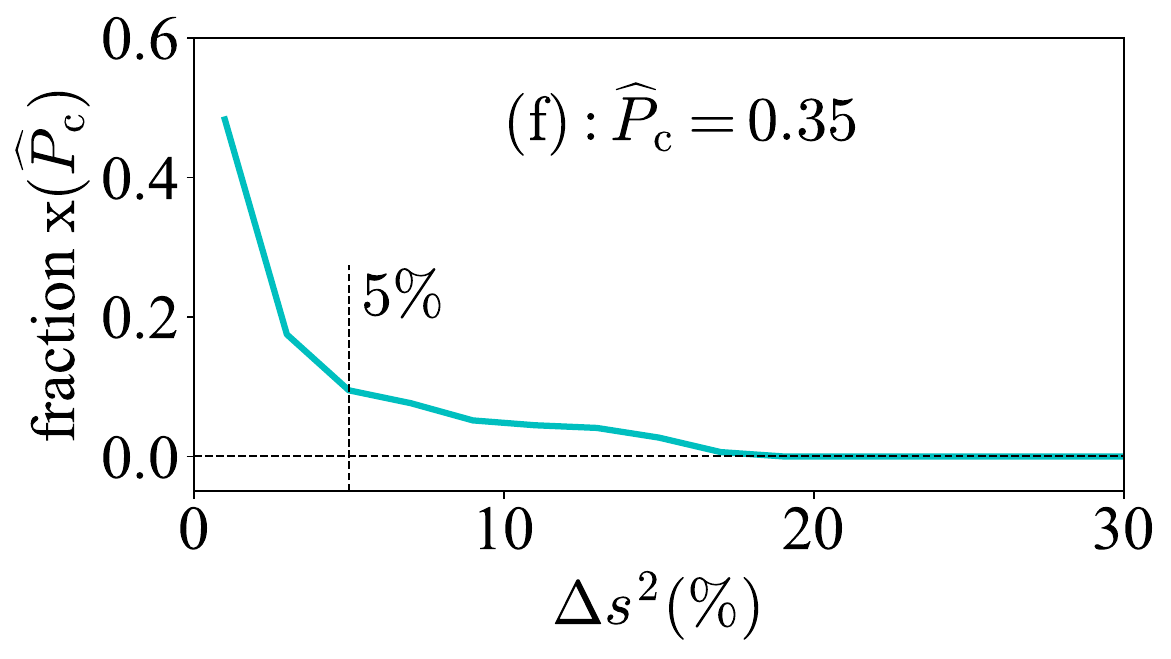}\\
\includegraphics[width=5.6cm]{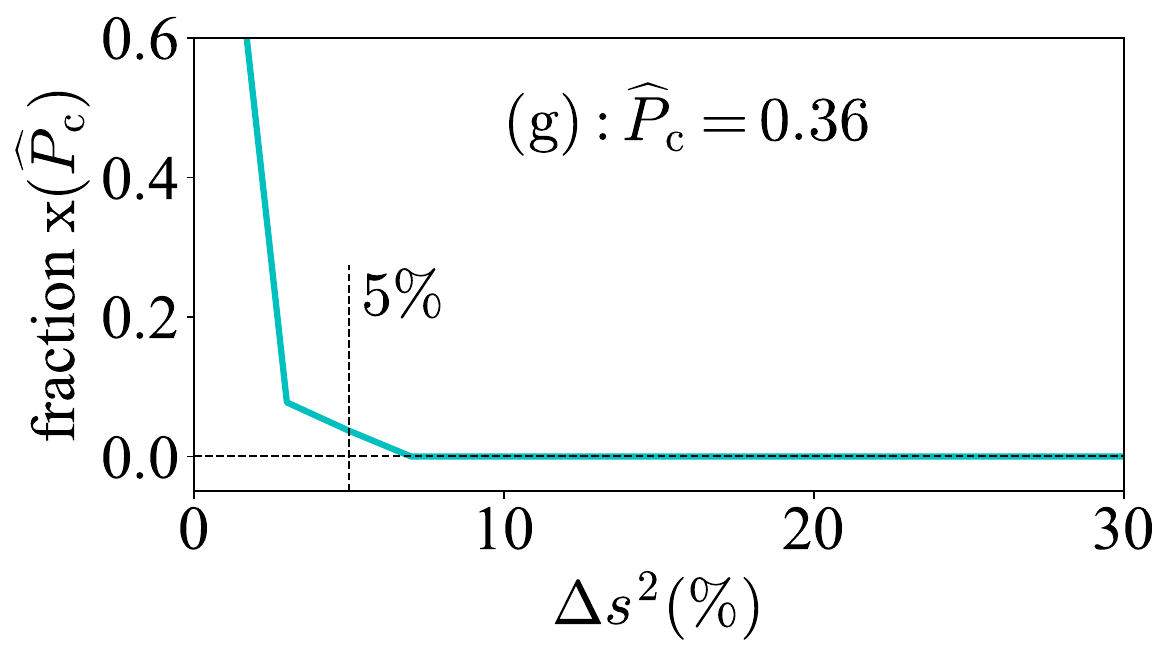}
\includegraphics[width=5.6cm]{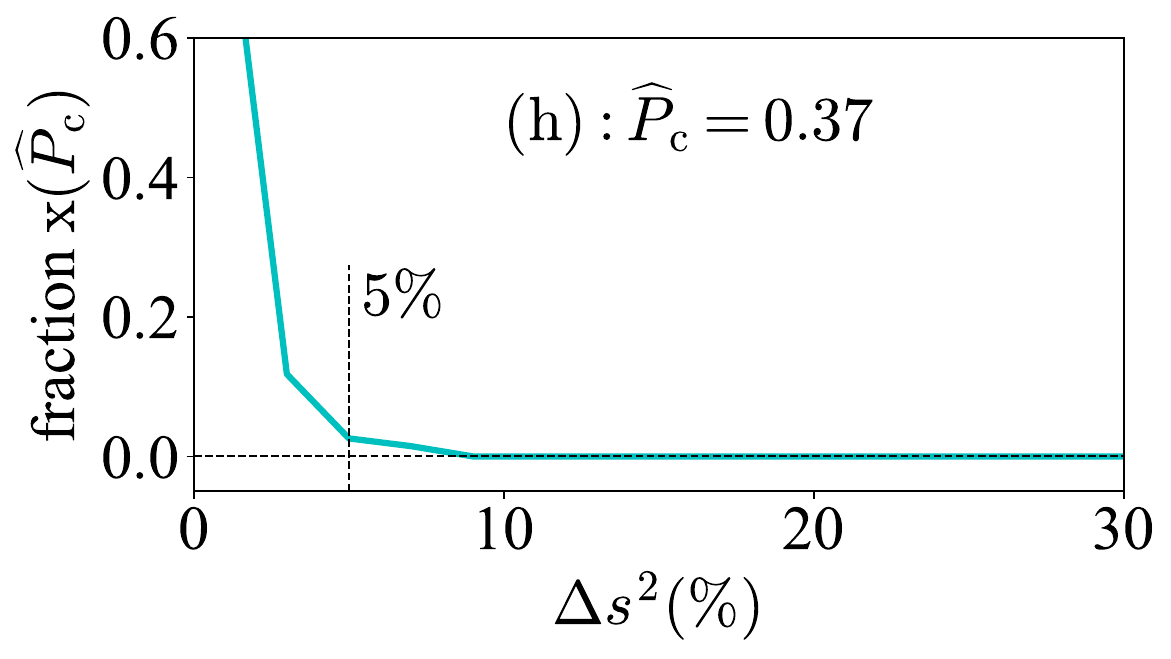}
\includegraphics[width=5.6cm]{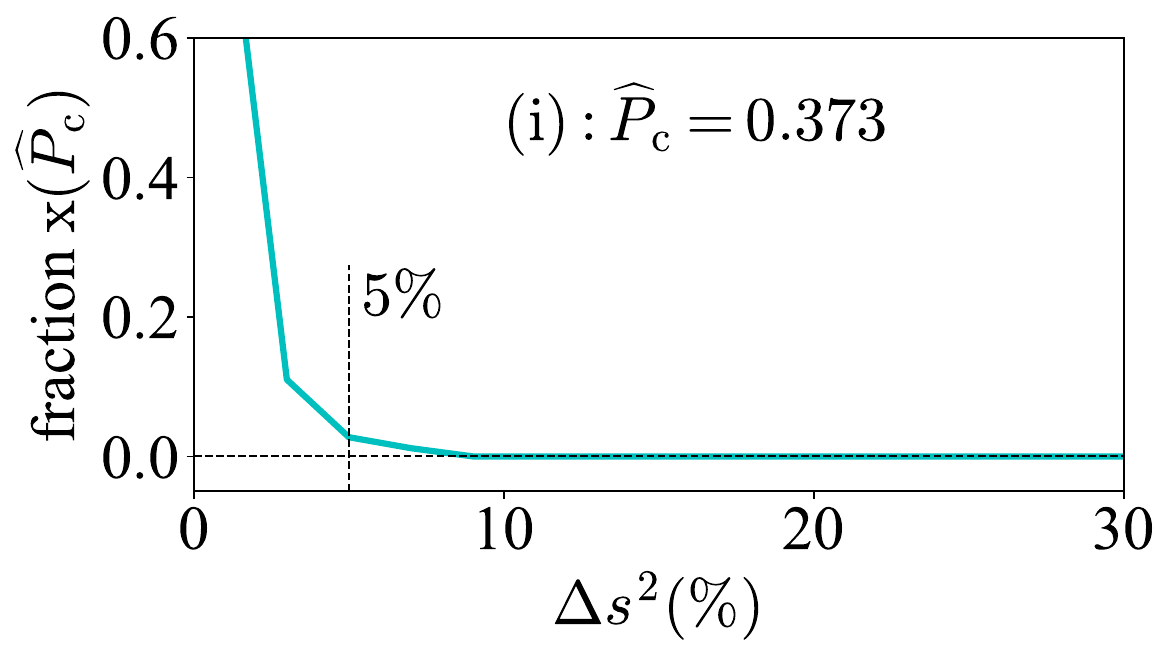}
\caption{(Color Online). Same as FIG.\,\ref{fig_s2_peak_stat-af} but for the fraction $\rm{x}({\x})$ for the enhancement $\Delta s^2$.
Figure taken from Ref.\,\cite{CL24-a}.
}\label{fig_s2_peak_stat-3af}
\end{figure}

Consequently,  we find the position of the peak using $s^2\approx s_{\rm{c}}^2+l_2\hr^2+l_4\hr^4$\,\cite{CL24-b}: 
\begin{equation}
\widehat{r}_{\rm{pk}}=\sqrt{-l_2/2l_4}.
\end{equation}
For each ${\x}$, we randomly sample $a_4\sim\rm{Unif}[0,2]$ and $a_6\sim\rm{Unif}[-2,2]$ by $m$ times, and count the number $m_{\rm{pk}}$ of samples where there is a peak in the $s^2$ profile while satisfying the causality condition $s_{\max}^2\leq1$, here\,\cite{CL24-b}
\begin{equation}\label{def-Ds2}
s_{\max}^2\equiv s^2(\widehat{r}_{\rm{pk}})=s_{\rm{c}}^2-l_2^2/4l_4.
\end{equation}
The probability is estimated as $m_{\rm{pk}}/m$, see the left panel of FIG.\,\ref{fig_s2_peak_stat_prop} for the simulated results.
For a relatively small ${\x}\approx0.1$, the probability is also small since the ${\x}$ is not large enough to make $l_2$ positive (even if $a_4$ is positive). The probability eventually increases with  increasing ${\x}$, but it becomes small again when the ${\x}$ is close to the causality limit about 0.374. This is because the matter is too stiff to be further compressed by gravity, i.e.,  there is no space for $s^2(\widehat{r})>s_{\rm{c}}^2$ since $s_{\rm{c}}^2$ itself is $\to1$\,\cite{CLZ23-b}.
Combining the above information with what we learned from FIG.\,\ref{fig_a4peak}, we find that the probability is relatively larger (compared with its surroundings) for $0.2\lesssim{\x}\lesssim0.35$, where the strong gravitational force effectively bends down the $s^2$.
Interestingly, ${\x}\approx0.24_{-0.07}^{+0.05}$\,\cite{CLZ23-a} for PSR J0740+6620 extracted directly from its observed mass and radius\,\cite{Fon21,Miller21,Riley21,Salmi22,Ditt24,Salmi24} is quite close to this region, indicating massive NSs with radii $\approx12\mbox{-}14\,\rm{km}$ are excellent objects to study the peaked $s^2$ profile.
The resulted $a_4$ and $a_6$ for ${\x}\approx0.24$ are found respectively to be about $a_4\approx0.71\pm0.03$ and $a_6\approx-1.37\pm0.19$, both are $\sim\mathcal{O}(1)$.
We show in FIG.\,\ref{fig_eps-vs-r6602} the radial dependence of $s^2$,  $\heps$ and $\d\heps/\d\hr$ for PSR J0740+6620 in the case that there exists a peak in $s^2$, using these constrained $a_4$ and $a_6$;
the peak in $s^2$ is weak and wide and the plots are effective at small $\hr\lesssim0.5$.
See the left panel of FIG.\,\ref{fig_s2_peak_stat_a4a6} for the general ${\x}$-dependence of the coefficients $a_4$, $a_6$, $l_2$ and $l_4$, where the coefficient $l_4$ (compared with $a_4,a_6$ and $l_2$) may take some large values (for large ${\x}$).
Furthermore, the coefficient $l_2$ starting from about zero at ${\x}\approx0.1$ implies that for even smaller ${\x}$ values there would be no peaked feature in the $s^2$ profile.
The right panel of FIG.\,\ref{fig_s2_peak_stat_a4a6} shows the dependence of $l_4$ on $a_4$ and that of $l_2$ on $a_2$ (inset), from which we find that $l_2$ positively correlates with $a_2$ and $l_4$ positively correlates with $a_4$.
Moreover, while $a_2$, $a_4$,  $l_2$ are $\sim\mathcal{O}(1)$ the coefficient $l_4$ may take some large values (especially for large ${\x}$).

\begin{figure}[h!]
\centering
\includegraphics[width=5.6cm]{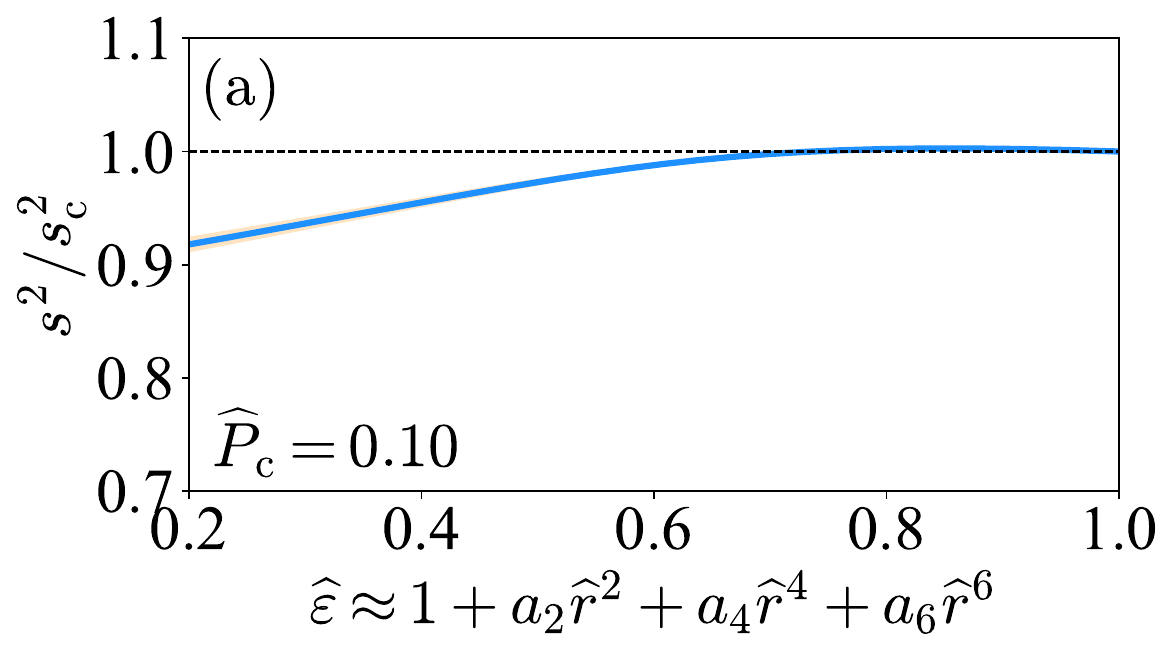}
\includegraphics[width=5.6cm]{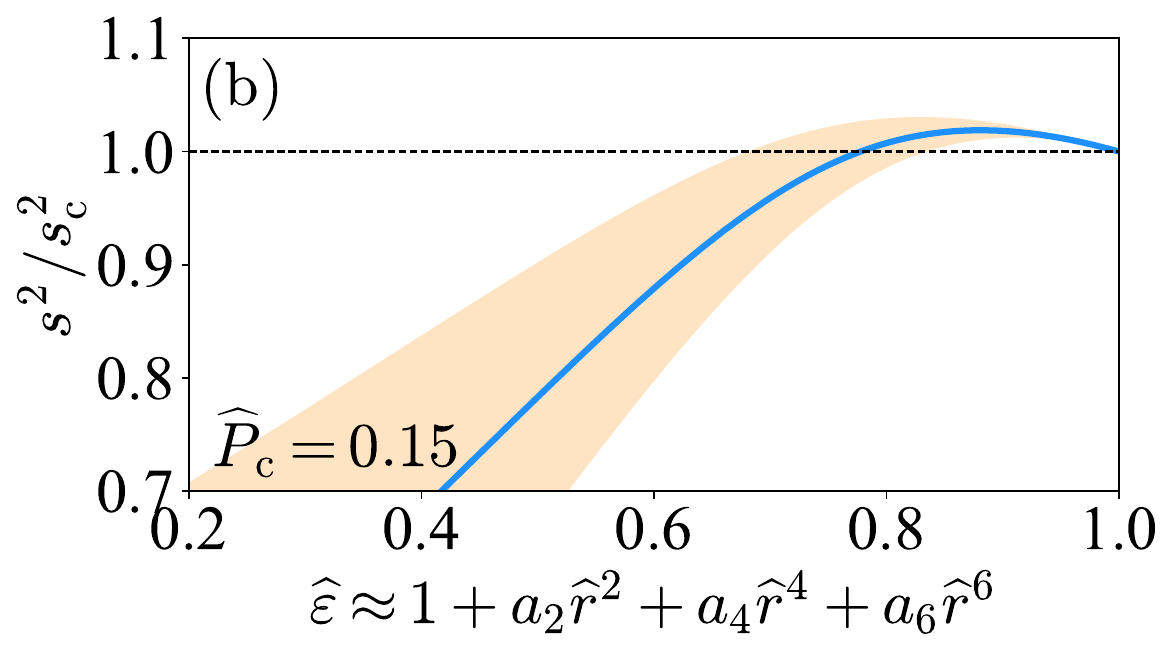}
\includegraphics[width=5.6cm]{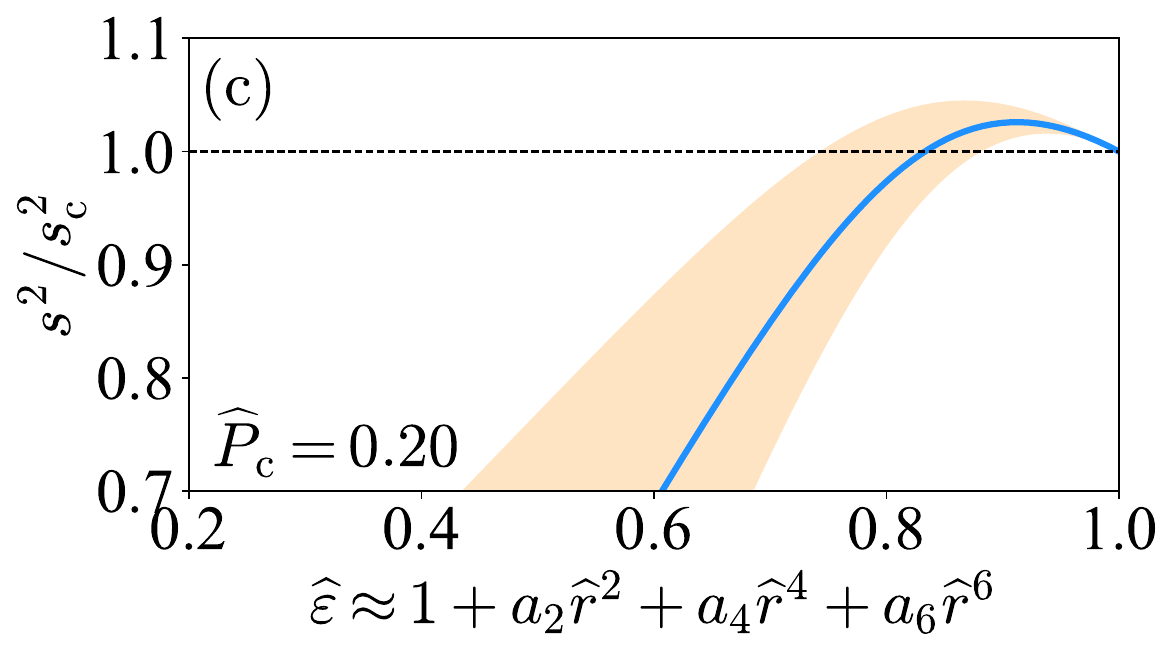}\\
\includegraphics[width=5.6cm]{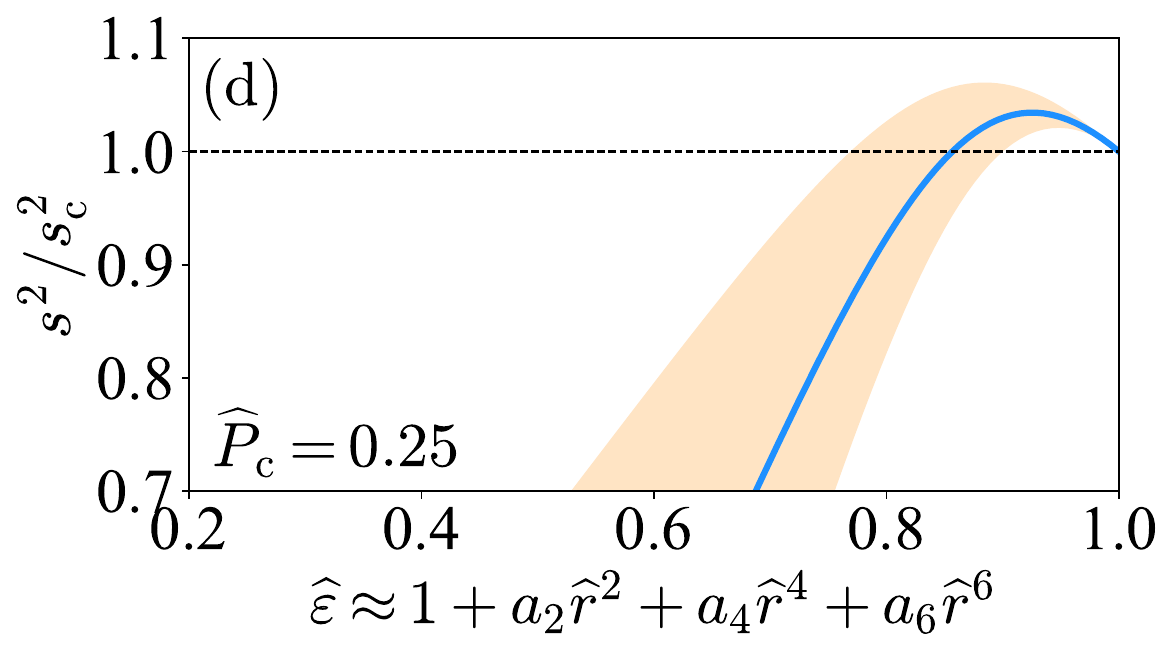}
\includegraphics[width=5.6cm]{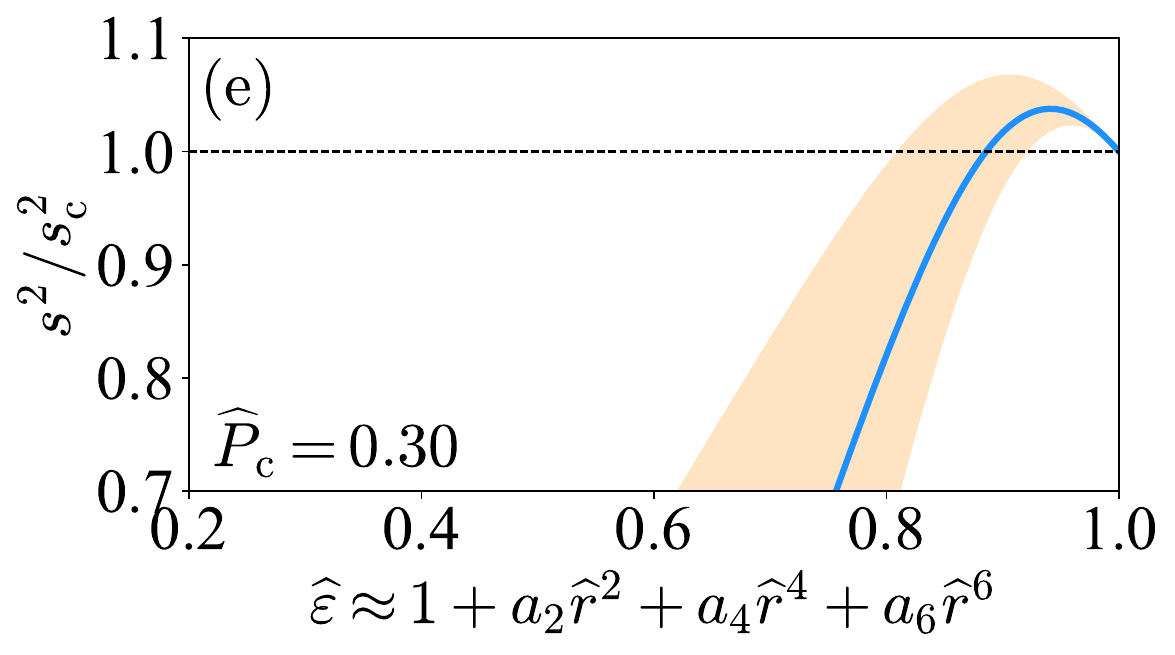}
\includegraphics[width=5.6cm]{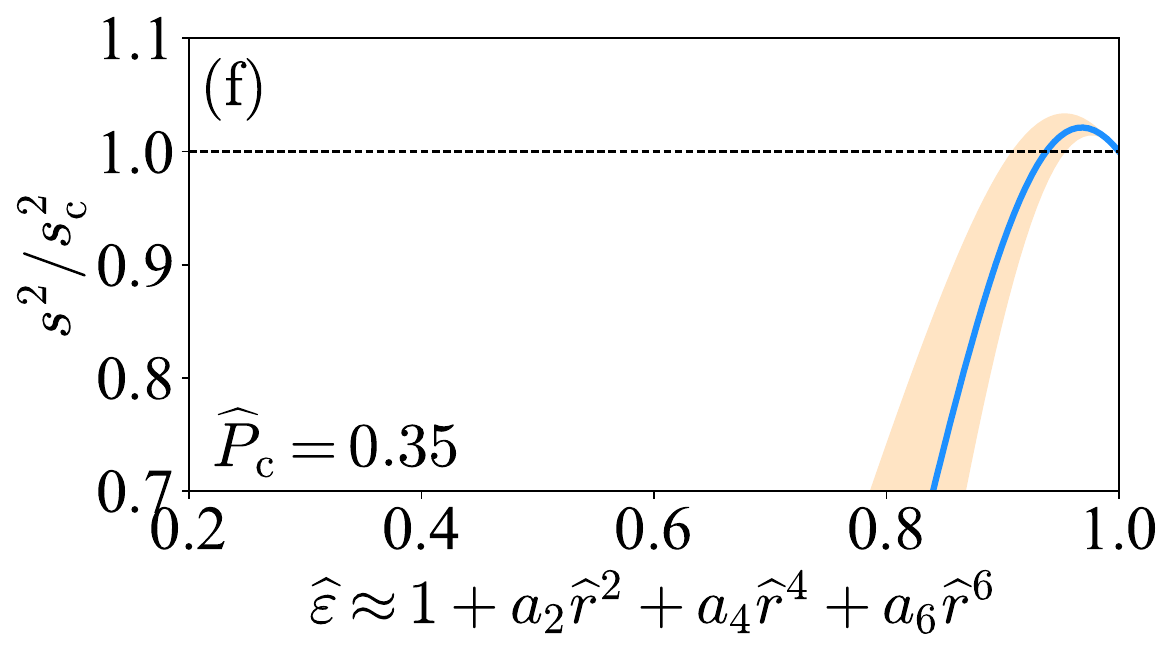}\\
\includegraphics[width=5.6cm]{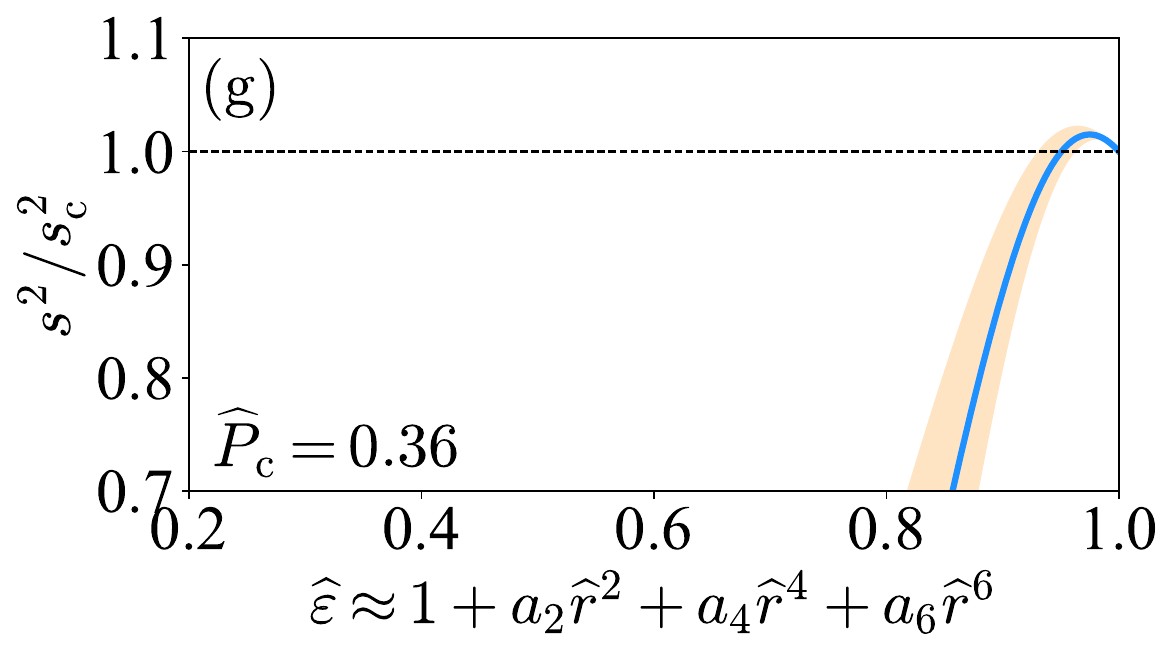}
\includegraphics[width=5.6cm]{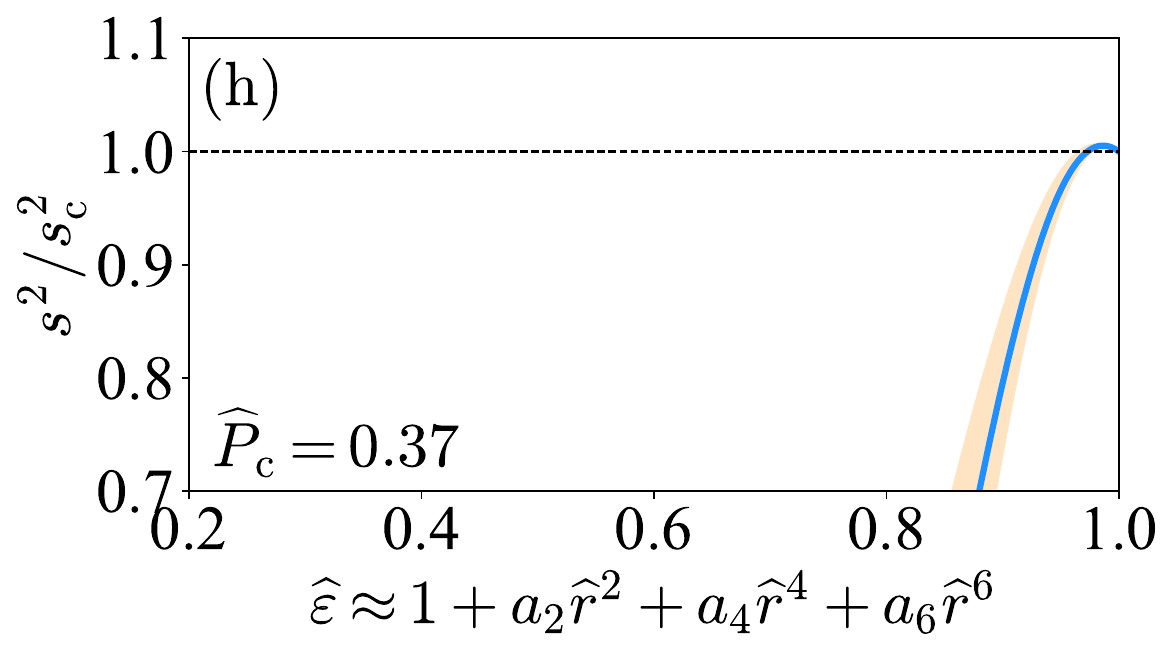}
\includegraphics[width=5.6cm]{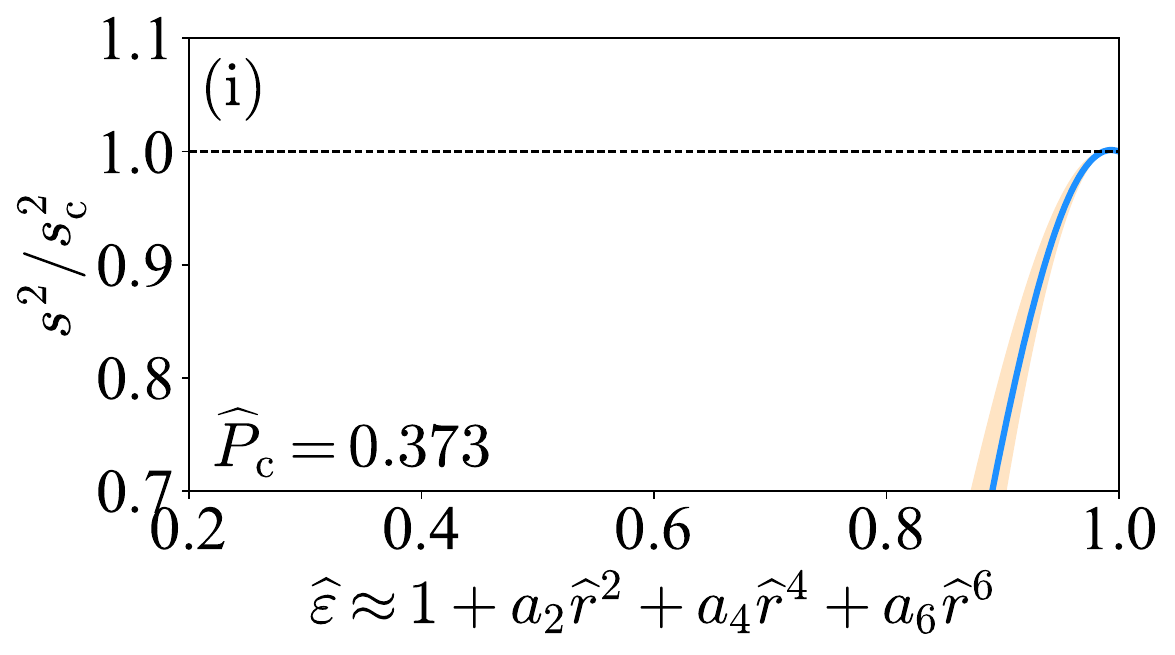}
\caption{(Color Online). Same as FIG.\,\ref{fig_s2_peak_stat-af} but for the $s^2/s_{\rm{c}}^2$ as a function of $\widehat{\varepsilon}$ (near NS centers).  Figures taken from Ref.\,\cite{CL24-a}.
}\label{fig_s2_peak_stat-1af}
\end{figure}

\begin{figure}[h!]
\centering
\includegraphics[width=5.6cm]{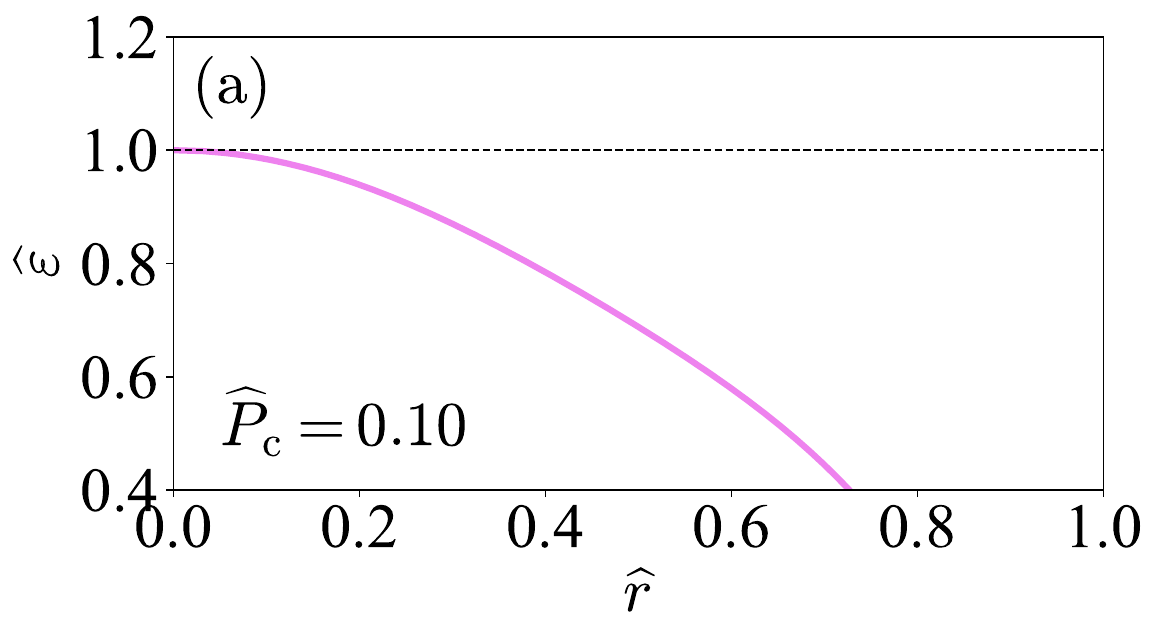}
\includegraphics[width=5.6cm]{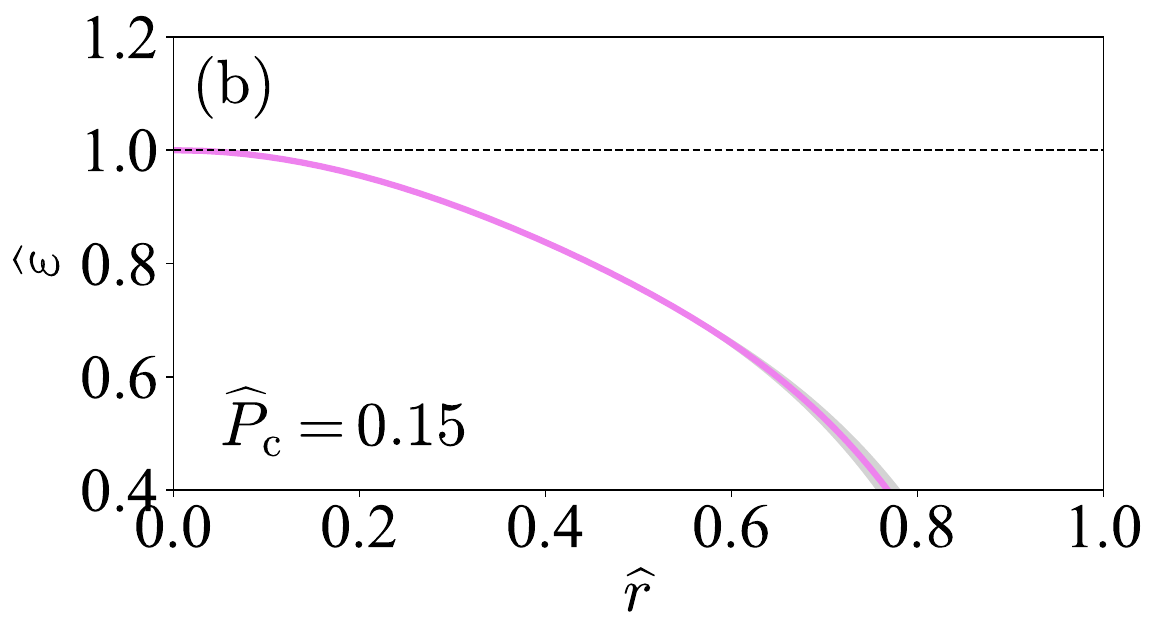}
\includegraphics[width=5.6cm]{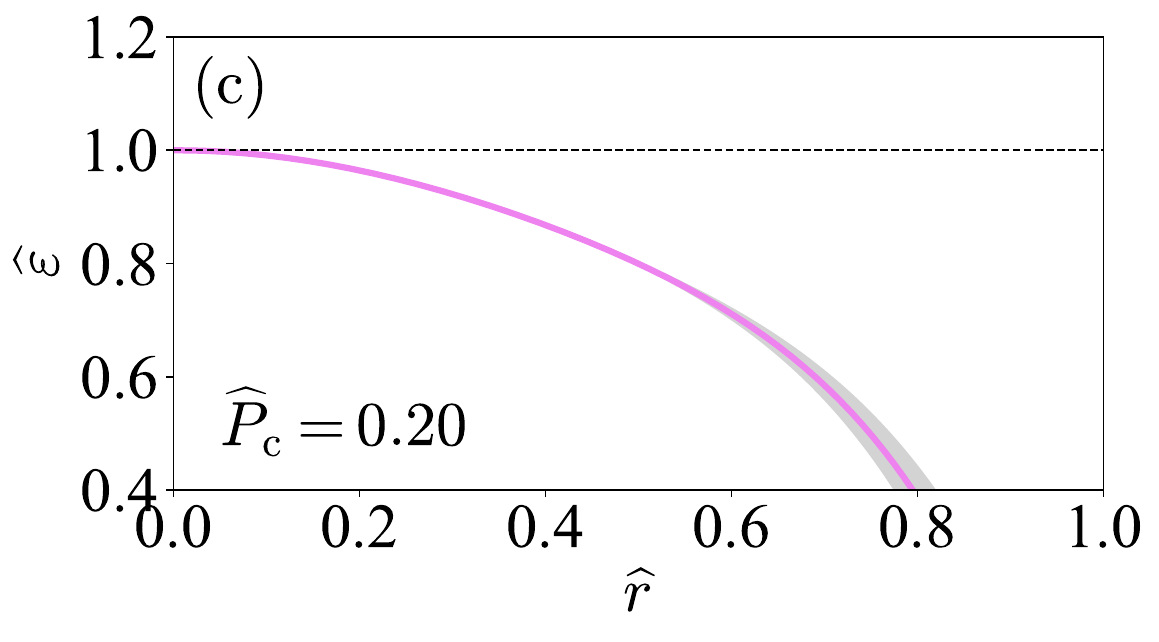}\\
\includegraphics[width=5.6cm]{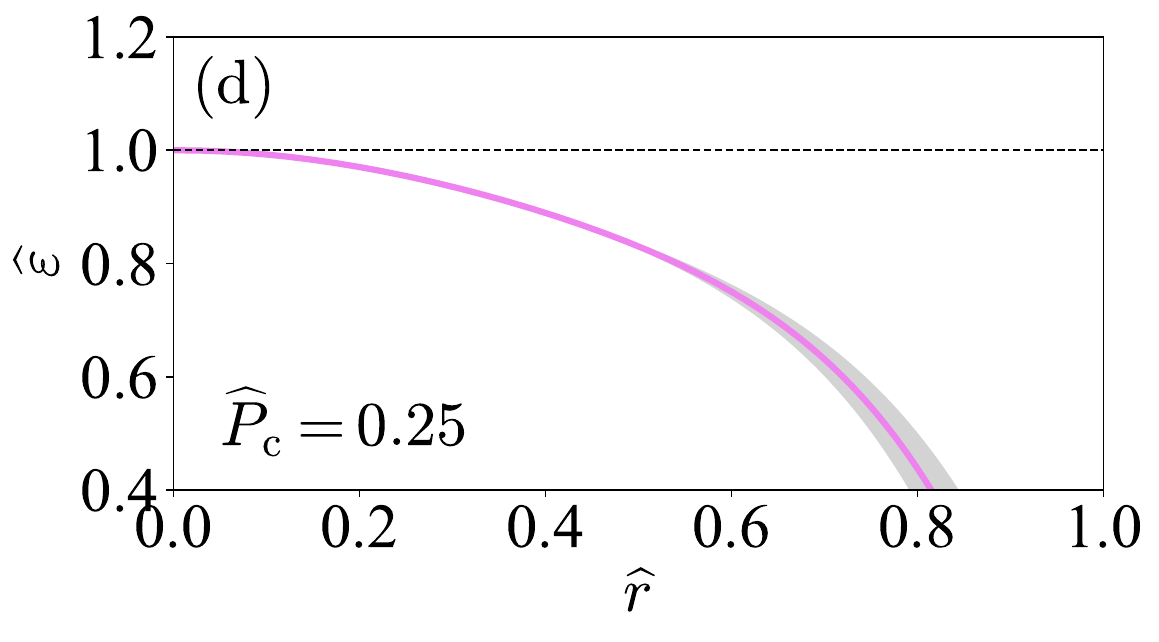}
\includegraphics[width=5.6cm]{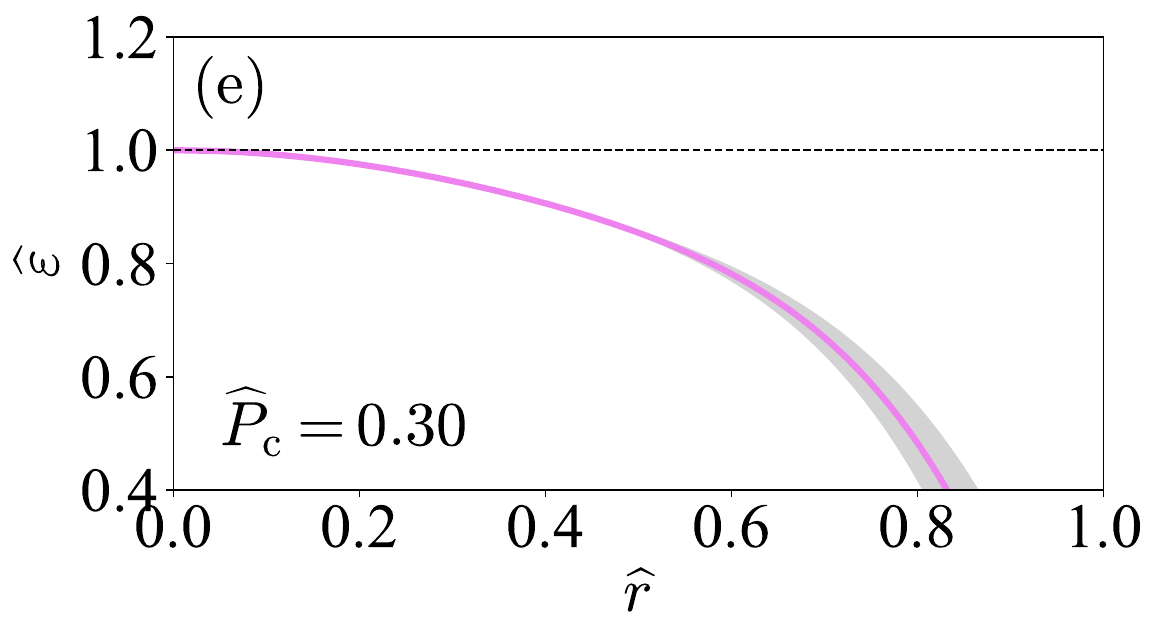}
\includegraphics[width=5.6cm]{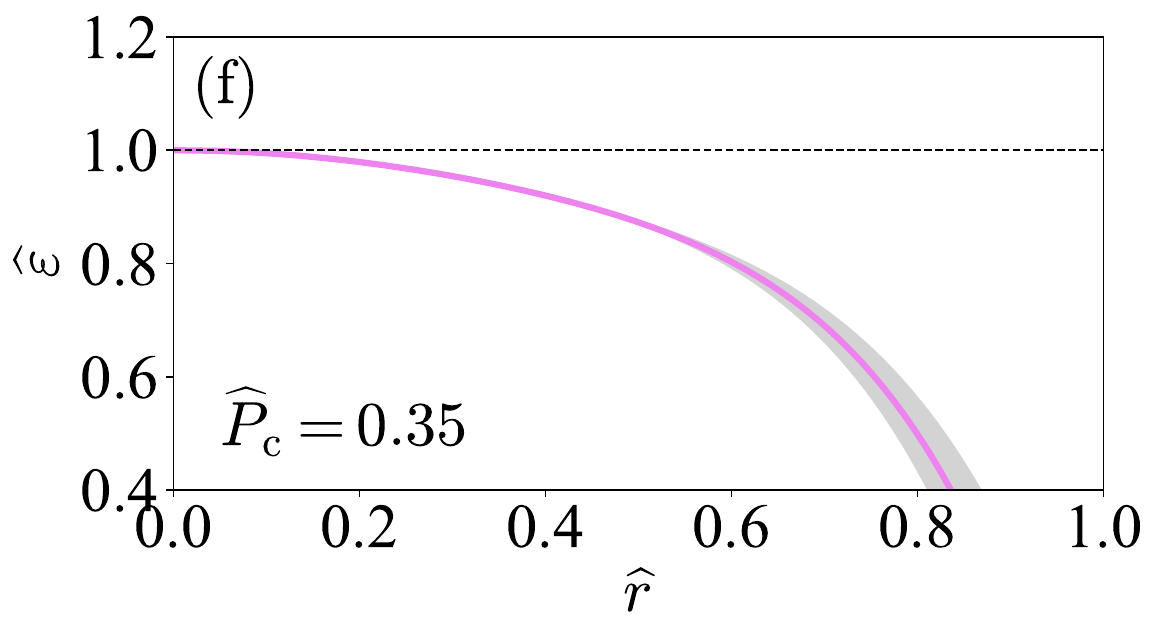}\\
\includegraphics[width=5.6cm]{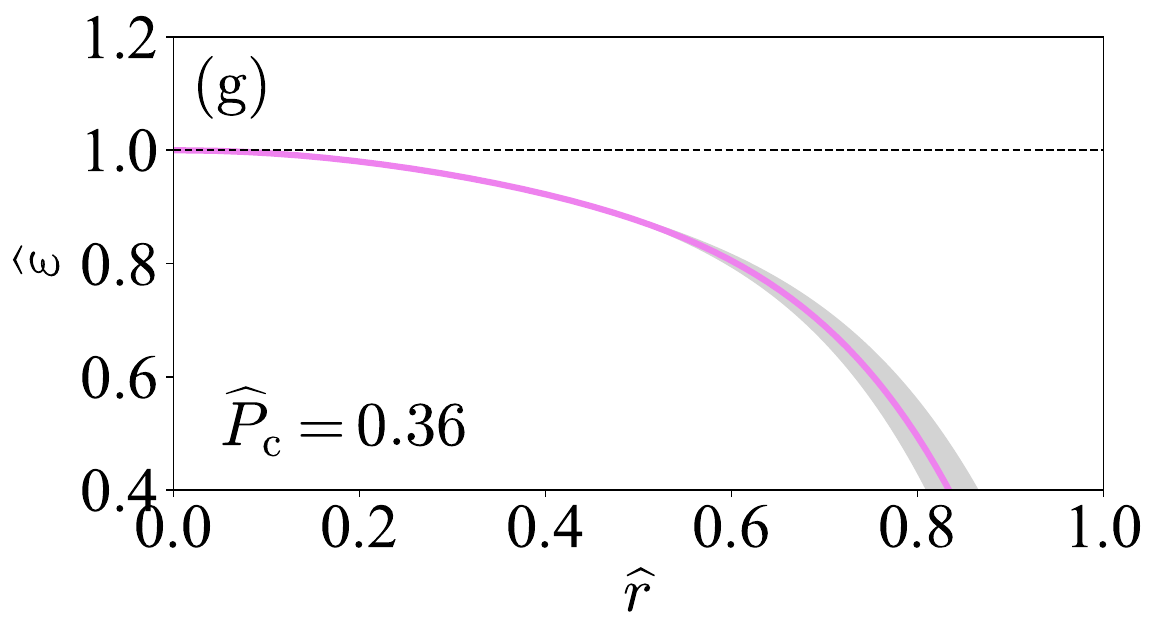}
\includegraphics[width=5.6cm]{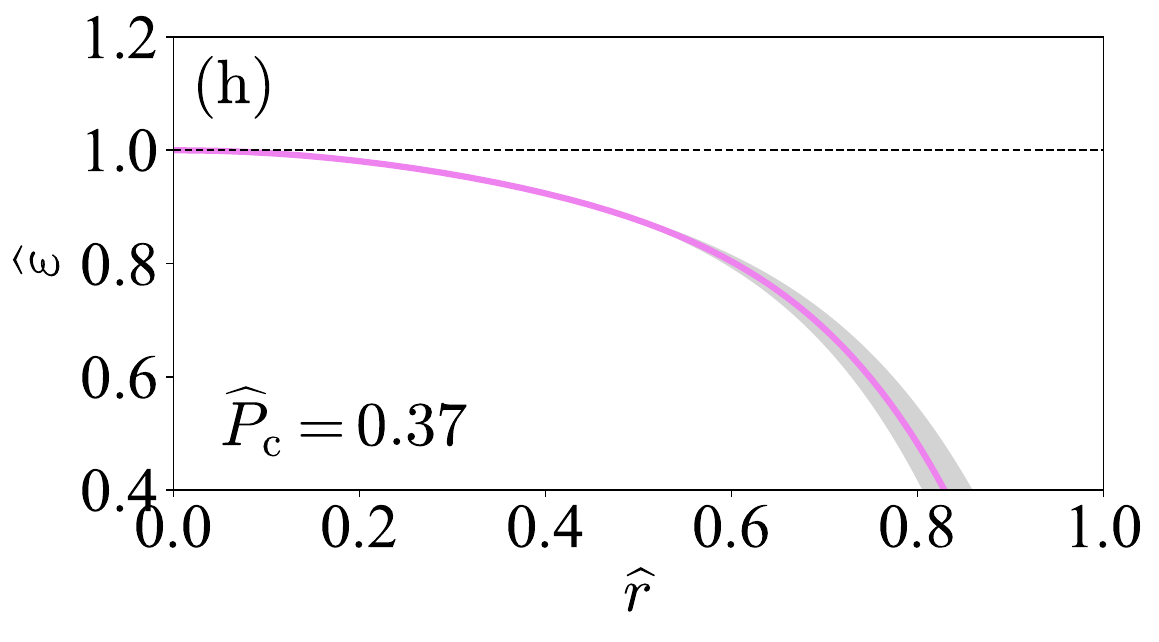}
\includegraphics[width=5.6cm]{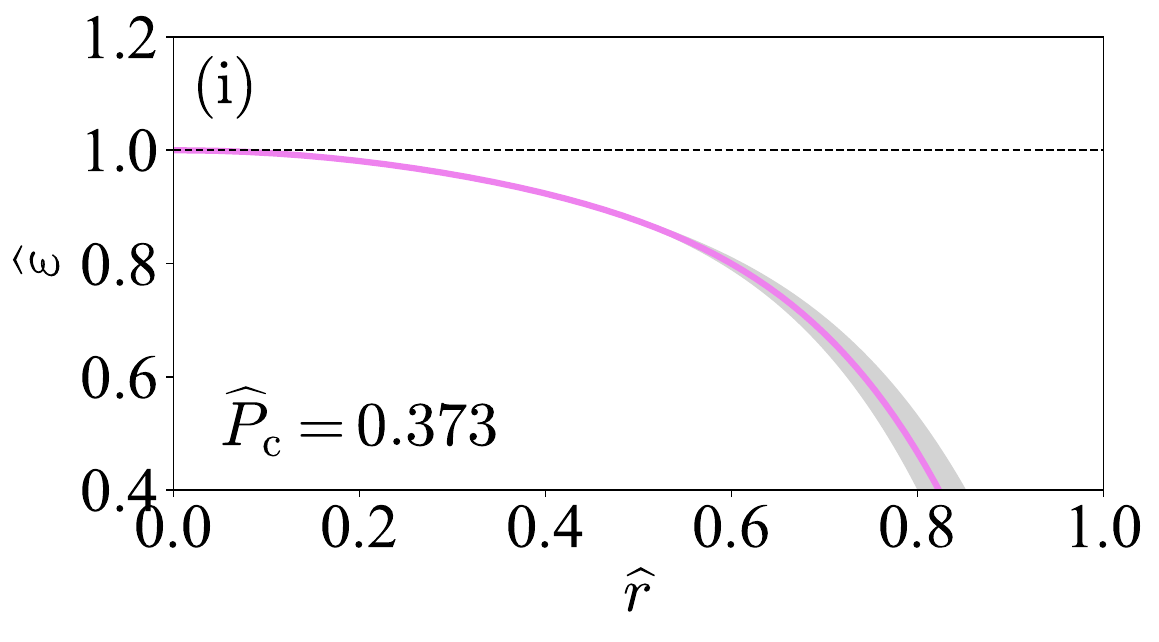}
\caption{(Color Online). Same as FIG.\,\ref{fig_s2_peak_stat-af} but for $\widehat{\varepsilon}\approx 1+a_2\widehat{r}^2+a_4\widehat{r}^4+a_6\widehat{r}^6$ as a function of $\widehat{r}$. Figures taken from Ref.\,\cite{CL24-a}.
}\label{fig_s2_peak_stat-2af}
\end{figure}

Also shown in the left panel of FIG.\,\ref{fig_s2_peak_stat_prop} is the peak position $\widehat{r}_{\rm{pk}}$ which is located at about $0.35\lesssim\widehat{r}_{\rm{pk}}\lesssim0.4$ for a wide range of ${\x}$.
Since the reduced radius $\widehat{R}=R/Q\approx1$, this means the peak locates at about $0.35\sim0.4$ times the radius from the center.
Moreover,  we show in the left panel of FIG.\,\ref{fig_s2_peak_stat_prop} the (average) enhancement of $s_{\rm{c}}^2$, namely (see Eq.\,(\ref{def-Ds2}))\,\cite{CL24-b},
 \begin{equation}
 \boxed{
\Delta s^2\equiv s_{\max}^2/s_{\rm{c}}^2-1=-l_2^2/4l_4s_{\rm{c}}^2,}
\end{equation}
which is generally small about $\lesssim5\%$;
and this is consistent with FIG.\,\ref{fig_s2peak}.
FIG.\,\ref{fig_s2_peak_stat-af} shows the radial profiles of $s^2$ adopting nine different ${\x}$ values specified in the plots.
The $s^2$ is probably a monotonically decreasing function of $\widehat{r}$ for small ${\x}\lesssim0.1$ (panel (a) of FIG.\,\ref{fig_s2_peak_stat-af}), the peak eventually emerges/develops as ${\x}$ increases to about ${\x}\approx0.30$ (panels (b) to (e)) and then tends to disappear for even large ${\x}\gtrsim0.35$ (panels (f)-(i)).
We may also find that at the two limiting sides of ${\x}$ (i.e., small ${\x}\lesssim\mathcal{O}(0.1)$ and ${\x}\gtrsim0.35$) the fraction $\rm{x}({\x})$ of large enhancement $\Delta s^2(\gtrsim15\%)$ is extremely small, and most of the fraction is accumulated at $\Delta s^2\lesssim5\%$, as shown in FIG.\,\ref{fig_s2_peak_stat-3af}.

If the contribution $l_6\widehat{r}^6$ (which involves the coefficient $a_8$) is included as $s^2\approx s_{\rm{c}}^2+l_2\widehat{r}^2+l_4\widehat{r}^4+l_6\widehat{r}^6$,
then the equation determining the peak position $\widehat{r}_{\rm{pk}}$ becomes ($l_2>0$),
\begin{equation}\label{kk-1}
\boxed{
l_2+2l_4\widehat{r}^2+3l_6\widehat{r}^4=0.}
\end{equation}
The expression for $l_6$ could be obtained straightforwardly like that for $l_4$ (see Eq.\,(\ref{ef-1})), 
\begin{equation}
    l_6=\frac{4}{b_2}\left[
    \left(b_8-s_{\rm{c}}^2a_8\right)-\frac{3}{2}\frac{a_4}{a_2}\left(b_6-s_{\rm{c}}^2a_6\right)-\left(\frac{3}{2}\frac{a_6}{a_2}-2\left(\frac{a_4}{a_2}\right)^2\right)\left(b_4-s_{\rm{c}}^2a_4\right)
    \right],
\end{equation}
where the expression for $b_8$ is
\begin{align}\label{ee-b8}
    b_8=&-\frac{1}{648}\left(1+3\x^2+4\x\right)\left(1-3\x-\frac{27}{2}\x^3\right)-\left(\frac{19}{1620}+\frac{\x}{54}+\frac{\x^2}{90}+\frac{7\x^3}{120}\right)a_2\notag\\
    &-\left(\frac{4}{225}+\frac{\x}{150}\right)a_2^2-\left(\frac{11}{756}-\frac{\x}{252}-\frac{\x^2}{12}\right)a_4-\frac{3a_2a_4}{70}-\left(\frac{1}{18}+\frac{5\x}{36}\right)a_6,
\end{align}
see relevant expressions of (\ref{ee-b2}), (\ref{ee-b4}) and (\ref{ee-b6}) for coefficients $b_2$, $b_4$ and $b_6$.
The magnitude of $b_8$ is even smaller than that of $b_2$, $b_4$ or $b_6$ (see FIG.\,\ref{fig_b2b4b6}).
Using Eq.\,(\ref{kk-1}), we immediately obtain the peak position $\widehat{r}_{\rm{pk}}$ as $\widehat{r}_{\rm{pk}}^2=[-l_4+(l_4^2-3l_2l_6)^{1/2}]/3l_6$ if $l_4>0$ and $\widehat{r}_{\rm{pk}}^2=[-l_4-(l_4^2-3l_2l_6)^{1/2}]/3l_6$ for $l_4<0$ (both satisfy the conditions $l_6\leq l_4^2/3l_2$ and $\widehat{r}_{\rm{pk}}^2>0$).  
By sampling $a_8\sim\rm{Unif}[-2,2]$, one can similarly analyze the peaked behavior of $s^2$, and the results are shown by the dashed grey lines in FIG.\,\ref{fig_s2_peak_stat-af}.
It is seen that the correction $l_6\widehat{r}^6$ slightly increases the enhancement $\Delta s^2$ while has almost no effect on the location of the peak $\widehat{r}_{\rm{pk}}$.
Moreover, using $\widehat{\varepsilon}\approx 1+a_2\widehat{r}^2+a_4\widehat{r}^4+a_6\widehat{r}^6$, one finds that $85\%\lesssim\widehat{\varepsilon}_{\rm{pk}}\lesssim95\%$  for $0.15\lesssim{\x}\lesssim0.35$.
Furthermore, we show in FIG.\,\ref{fig_s2_peak_stat-1af} the $s^2/s_{\rm{c}}^2$ as a function of $\widehat{\varepsilon}$.
The peak location $\widehat{\varepsilon}_{\rm{pk}}$ in energy density is consistent with our previous studies, i.e., $\widehat{\rho}\equiv
\rho/\rho_{\rm{c}}\approx
\widehat{\varepsilon}-\mu(1+4\mu/3){\x}(1-{\x})
$, see Eq.\,(\ref{kkk}), from which we inferred that $\rho_{\rm{pk}}/\rho_{\rm{c}}\approx\widehat{\varepsilon}_{\rm{pk}}\lesssim95\%$\,\cite{CLZ23-b}; and this is also consistent with findings in a few recent studies by others\,\cite{Mro23,Cao23,Ferr24} and consistent with findings of FIG.\,\ref{fig_s2peak}.
Finally, we show in FIG.\,\ref{fig_s2_peak_stat-2af} the $\widehat{r}$-dependence of $\widehat{\varepsilon}(\hr)\approx 1+a_2\widehat{r}^2+a_4\widehat{r}^4+a_6\widehat{r}^6$, where the uncertainty is small for $\widehat{r}\approx0$ and it is consistent with FIG.\,\ref{fig_s2_prep_r}, see also the results of FIG.\,\ref{fig_eps-vs-r6602}.

\begin{figure}[h!]
\centering
\includegraphics[height=5.6cm]{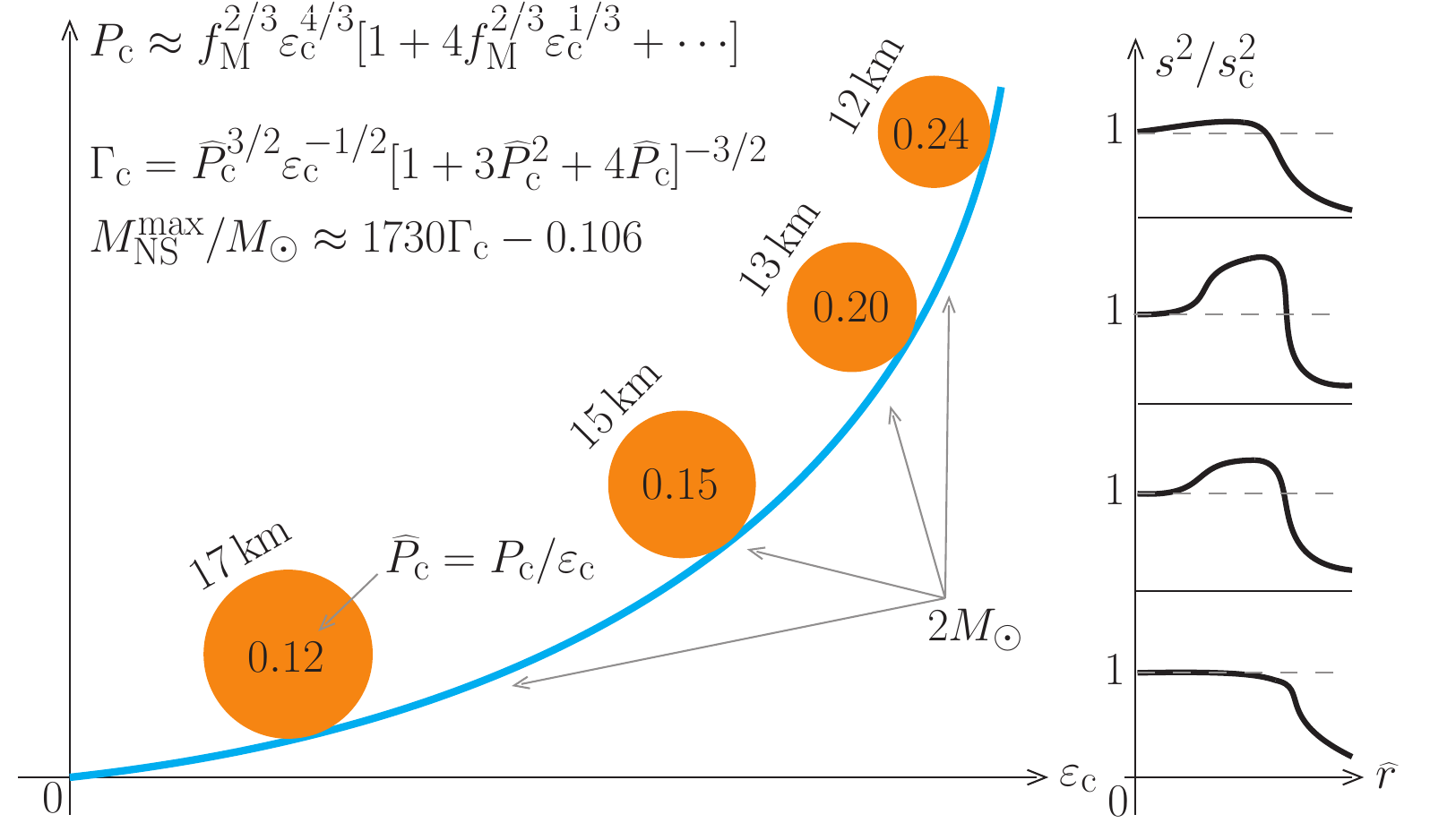}\qquad
\includegraphics[height=5.6cm]{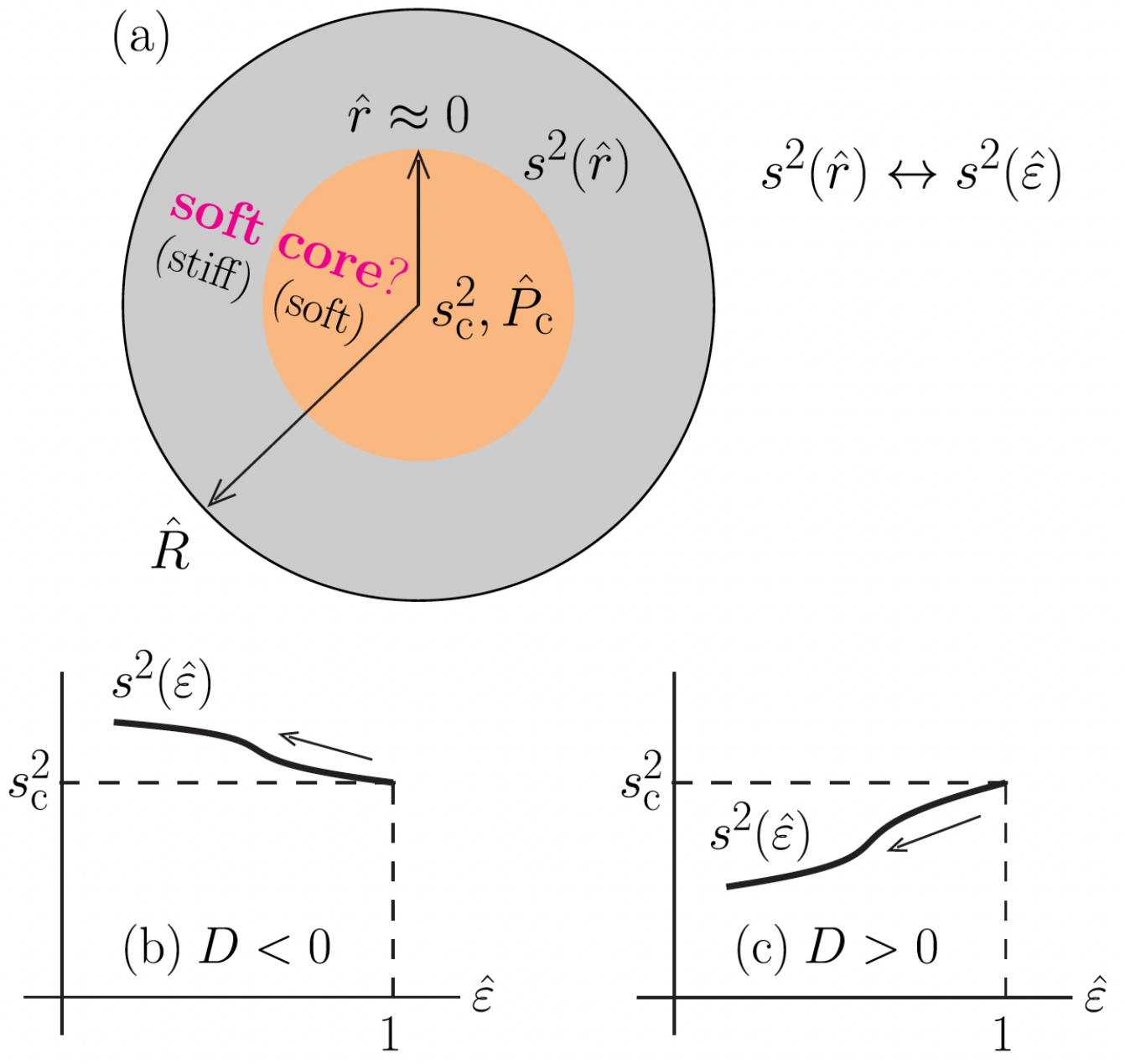}
\caption{(Color Online).  Left panel: central EOS of a $M_{\rm{NS}}^{\max}=2M_{\odot}$ NS and the sketch of variation of the peaked behavior of $s^2$ near $\widehat{r}=0$, here $f_{\rm{M}}$ is defined in Eq.\,(\ref{P1-pert}). Figure taken from Ref.\,\cite{CL24-b}.
Right panel: sketch of the SSS near NS centers (sub-panel (a)). 
The sign of coefficient $D$ (see Eq.\,(\ref{ref-D})) characterizes the possibility of $s_{\rm{c}}^2<s^2(\widehat{\varepsilon})$ (indication of a crossover/soft core, sub-panel (b)) or  $s^2_{\rm{c}}>s^2(\widehat{\varepsilon})$ (the center EOS is stiffer than its surroundings, sub-panel (c)). 
Figure taken from Ref.\,\cite{CLZ23-b}.
}\label{fig_EOS_sc2}
\end{figure}

For normal stable NSs on the M-R curve,  we could analyze the probability for the $s^2$ profile to have a peak near $\widehat{r}=0$ as done for the $M_{\rm{NS}}^{\max}$ configuration; the result is shown in the right panel of FIG.\,\ref{fig_s2_peak_stat_prop}.
We find that the probability is relatively large for $0.17\lesssim{\x}\lesssim0.25$ (compared with its low-${\x}$ surroundings), while on the other hand the location of the peak $\widehat{r}_{\rm{pk}}$ (lavender band) and the enhancement $\Delta s^2$ (tan band) are very similar to those of the left panel of FIG.\,\ref{fig_s2_peak_stat_prop}.
Combining the results of the right panels of FIG.\,\ref{fig_a4peak} and FIG.\,\ref{fig_s2_peak_stat_prop} and 
considering ${\x}\lesssim0.15$\,\cite{Brandes2023-a} for a canonical NS (see TAB.\,\ref{sstab}),  we see that the probability of occurrence of a peak in $s^2$ for such NSs is probably very low.
These features are consistent with the findings of Ref.\,\cite{Ecker2022} and FIG.\,\ref{fig_Ecker23s2r}, which predicted that the $s^2$ in canonical NSs is a monotonically decreasing function of $\widehat{r}$.

\begin{figure}[h!]
\centering
\includegraphics[width=16.cm]{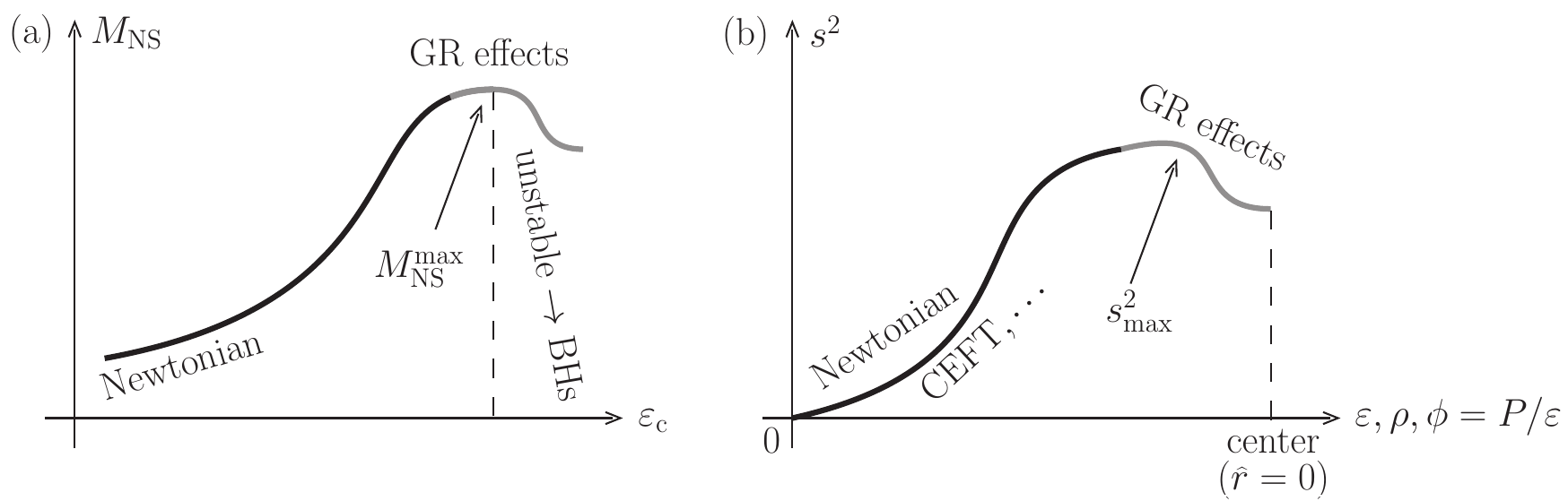}
\caption{(Color Online). Sketch of the $\varepsilon_{\rm{c}}$-dependence of NS mass $M_{\rm{NS}}$ (left panel) and the energy density $\varepsilon$-,  the density $\rho$- or the $\phi$-dependence of $s^2$ (right panel). Figures taken from Ref.\,\cite{CL24-b}.
}\label{fig_s2peak_sk}
\end{figure}

Now we have a coherent picture for generating a potential peak in $s^2$.
Consider a NS with mass $M_{\rm{NS}}^{\max}=2M_{\odot}$ fixed but its radius $R$ could vary.
The central EOS ${P}_{\rm{c}}$-$\varepsilon_{\rm{c}}$ is determined by the correlation (\ref{P1-pert}), see the light-blue line shown in the left panel of FIG.\,\ref{fig_EOS_sc2}.
{\color{xll}The ${\x}$ eventually increases with decreasing radius $R$  (e.g., from 17\,km to 12\,km) since the NS becomes more compact, accompanied with the emergence, enhancement and disappearing of the peak in $s^2$.}
The resulted ${\x}$'s are also shown in FIG.\,\ref{fig_EOS_sc2}, e.g., for $R\approx13\,\rm{km}$ we have ${\x}\approx0.20$, using the correlation (\ref{Rmax-n}).
So there may exist a peak in $s^2$ near $\widehat{r}=0$ when $R$ is relatively small (e.g., 12\,km) and the peak becomes enhanced when $R$ increases a little further (e.g., to 13\,km or 15\,km). 
If $R$ is very small, e.g., $R\lesssim 11\,\rm{km}$, then the $\x$ approaches the causality limit 0.374, in this case the peak disappears.
In essence, the profile of $s^2$ is analogous to the M-R curve where the GR leads to a maximum-mass point beyond which the NSs are unstable against collapsing into BHs, sketched in the left panel of FIG.\,\ref{fig_s2peak_sk}.
{\color{xll}The strong-gravitational force changes $\phi=P/\varepsilon$ from being small $\lesssim10^{-5}$ in Newtonian star to being sizable $\gtrsim\mathcal{O}(0.1)$ and extrudes a peak in the $s^2$ profile,} as sketched in the right panel of FIG.\,\ref{fig_s2peak_sk} (and numerically demonstrated in FIG.\,\ref{fig_s2_peak_stat-af}).
If $R$ is extra-ordinarily large (e.g., 17\,km and correspondingly ${\x}\approx0.12$ from Eqs.\,(\ref{Mmax-G}) and (\ref{Rmax-n})), the peak in $s^2$ may reduce and even disappear as illustrated in the left panel of FIG.\,\ref{fig_EOS_sc2}. 
\begin{figure}[h!]
\centering
\includegraphics[width=16.cm]{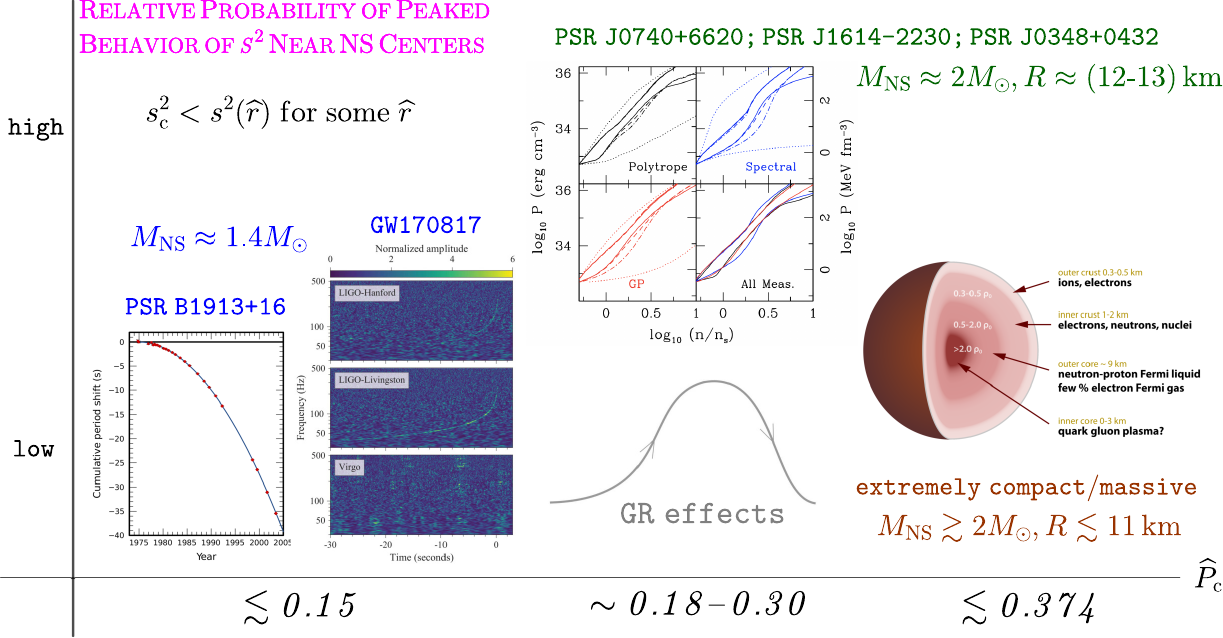}
\caption{(Color Online). An illustration of the relative probability of seeing a peaked behavior of $s^2$ profile near NS centers.
There tends to be no peaks in $s^2$ for light NSs like the canonical ones involved in GW170817 (left); the peak eventually emerges for massive NSs with radii about 12-13\,km such as PSR J0740+6620 (middle) and then disappears for more massive and very compact NSs with $R\lesssim11\,\rm{km}$ (right).
}\label{fig_cart_pcsk}
\end{figure}

The above discussions highlight the key role played by the central pressure ${\x}$ in determining the relative probabilities of seeing a peaked structure of $s^2$ in NSs with different masses and/or radii, as sketched in FIG.\,\ref{fig_cart_pcsk}. There is little chance to see a peak in the density/radius profile of $s^2$ for light NSs like the canonical ones involved in GW170817 (left) because of the low ${\x}$ reached. The peak eventually emerges for massive NSs with radii about 12-13\,km such as PSR J0740+6620 (middle) and then disappears for more massive and very compact NSs with $R\lesssim11\,\rm{km}$ (right).
Our main conclusion here is also consistent with the finding of a recent analysis\,\cite{Ecker2022,Ecker23} using as input in solving the TOV equations in the traditional approach the EOS from
nuclear theory and pQCD and imposing observational
constraints. In fact, their predictions imply that ${\x}$ is relevant for determining the structure of $s^2$. Nevertheless, 
we emphasize that not every EOS can induce a peaked $s^2$ profile, just similar to the fact that not every EOS could induce a maximum mass for NSs below the causality limit, e.g.,  the URFG predicts a linear M-R relation $M_{\rm{NS}}=3R/14$\,\cite{Lightman1975}.

Compact massive NSs (such as PSR J0740+6620 with its ${\x}\approx0.24$) provide excellent opportunities for studying the density or radius profile of $s^2$. {\color{xll}Our studies also predict that more massive and very compact NSs (with radii $\lesssim11\,\rm{km}$) may not have a peaked $s^2$ as we illustrated in FIG.\,\ref{fig_cart_pcsk}. Therefore, pulsars like PSR J0740+6620 are presently the best objects for exploring the possible peaked structure of $s^2$ density or radius profile.}
Looking forward, we expect the masses and radii of massive NSs as well as properties of post-merger remnants of NSs from future high-frequency GW observations as well as high-precision radius measurements using both X-rays and GWs will be very useful for inferring features of the $s^2$ profile especially near $\widehat{r}=0$.

\begin{figure}[ht!]
\centering
\includegraphics[height=8.cm]{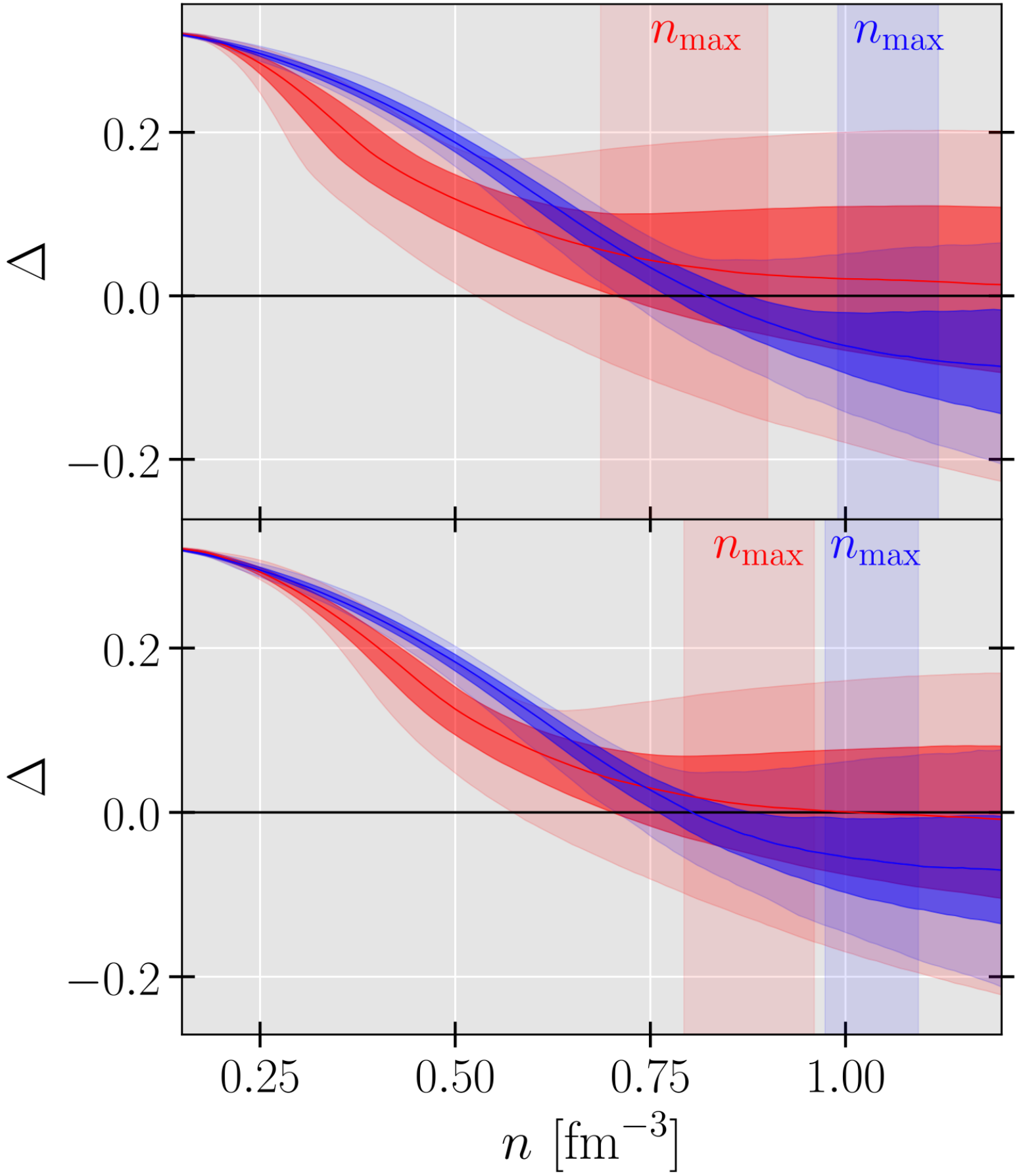}\hspace{1.cm}
\includegraphics[height=8.cm]{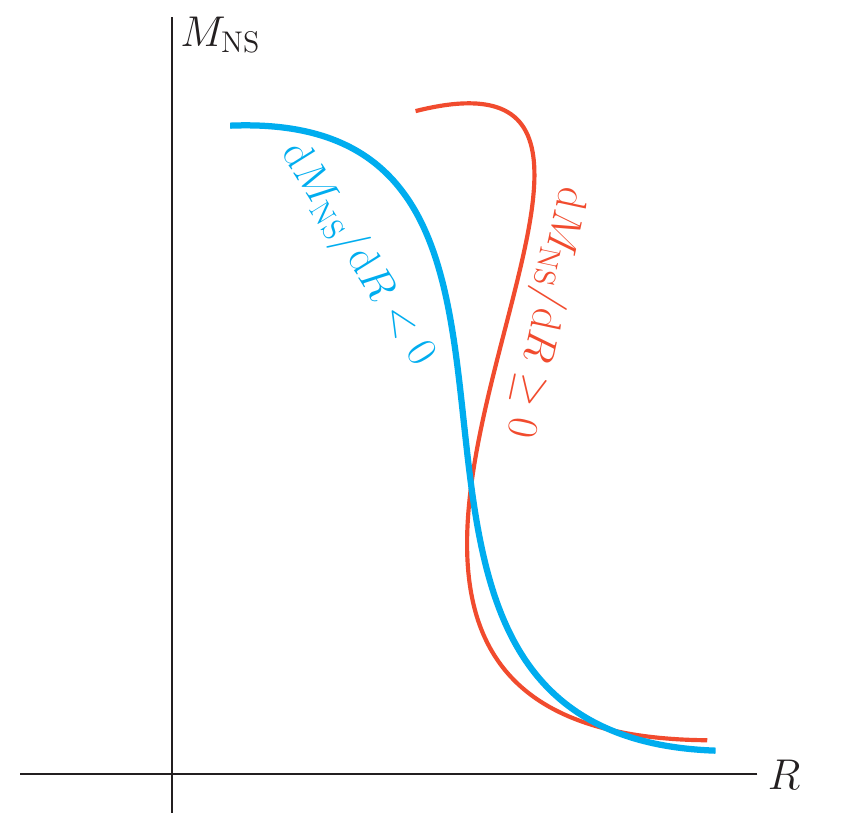}
\caption{(Color Online). Left panel: density dependence of $\Delta$ inferred under the constraint $\d M_{\rm{NS}}/\d R<0$ for all NS masses (blue) or $\d M_{\rm{NS}}/\d R\geq0$ for a certain mass range (red); inference in the bottom figure with astrophysical constraints. Figure taken from Ref.\,\cite{Ferr24}.
Right panel: two types of M-R curves classified by using the derivative $\d M_{\rm{NS}}/\d R$ for NS masses between about $1\,M_{\odot}$ and $M_{\rm{TOV}}$ to help understand the behavior of the trace anomaly in the left panel as well as the maximum baryon densities shown in the right panel of FIG.\,\ref{fig_Ecker23Delta}.
Figure taken from Ref.\,\cite{CL24-c}.
}\label{fig_MRdd}
\end{figure}

Bearing in mind that the strong gravity in GR plays an important/relevant role in extruding a peaked $s^2$ profile, our analysis above based on solving the dimensionless TOV equations without using any input nuclear EOS does not rely on the pQCD constraint at all. In this regard, besides the reviews given in SECTION \ref{SEC_2} (summarized in FIG.\,\ref{fig_RN-s2-summary}) and other places in this section, it is interesting to note that very recently in Ref.\,\cite{Ferr24} the authors classified their 40435 $M(R)$ curves into three groups with the $\d M_{\rm{NS}}/\d R$ negative, infinity and positive for NS masses between about 1.2$M_{\odot}$ and $M_{\rm{TOV}}$. Their EOSs are generated in a way similar to our meta-model with the high-density part of their EOSs parameterized with piecewise polytropes. They then explored how the global and/or local $\d M_{\rm{NS}}/\d R$ slopes at 1.4$M_{\odot}$ and two heavier mass points can inform us about $M_{{\rm{TOV}}}$ and the corresponding radius $R_{{\rm{TOV}}}$. In turn, observational constraints on the latter are expected to constrain the sign of $\d M_{\rm{NS}}/\d R$ globally or locally and thus the underlying dense matter EOS. In particular, they analyzed the SSS as a function of density with/without considering the constraint $\d M_{\rm{NS}}/\d R<0$ at certain mass range on the NS M-R curves. The latter condition may effectively affect both the position and strength of the peak in $s^2$, as shown in the right panel of FIG.\,\ref{fig_Ecker23Delta} where the blue band is for $\d M_{\rm{NS}}/\d R<0$ while the red band for $\d M_{\rm{NS}}/\d R\gtrsim0$. 
The $\d M_{\rm{NS}}/\d R < 0$ case has larger central densities probably because the matter is compressed more efficiently by gravity, see the right panel of FIG.\,\ref{fig_MRdd} for a sketch to illustrate why the set with $\d M_{\rm{NS}}/\d R<0$ may induce a larger central baryon density. A larger maximum baryon density (corresponding to a more compact NS) essentially generates a deeper $\Delta$, see the left panel of FIG.\,\ref{fig_MRdd}. Since the investigation of Ref.\,\cite{Ferr24} does not rely on specific EOS models, it thus provides an interesting viewpoint on $s^2$ directly from the general NS M-R curves.

\subsection{Effects of high-density nuclear symmetry energy on the density profile of SSS in NSs}\label{sub_s2Esym}

In the study of Ref.\,\cite{ZhangLi2023a}, it was found that by changing the high-density behavior of nuclear symmetry energy within the minimum model of NSs consisting of nucleons, electrons and muons at $\beta$-equilibrium the density profile of $s^2$ naturally shows a peak. 

\begin{figure}[h!]
\centering
\includegraphics[width=7.5cm]{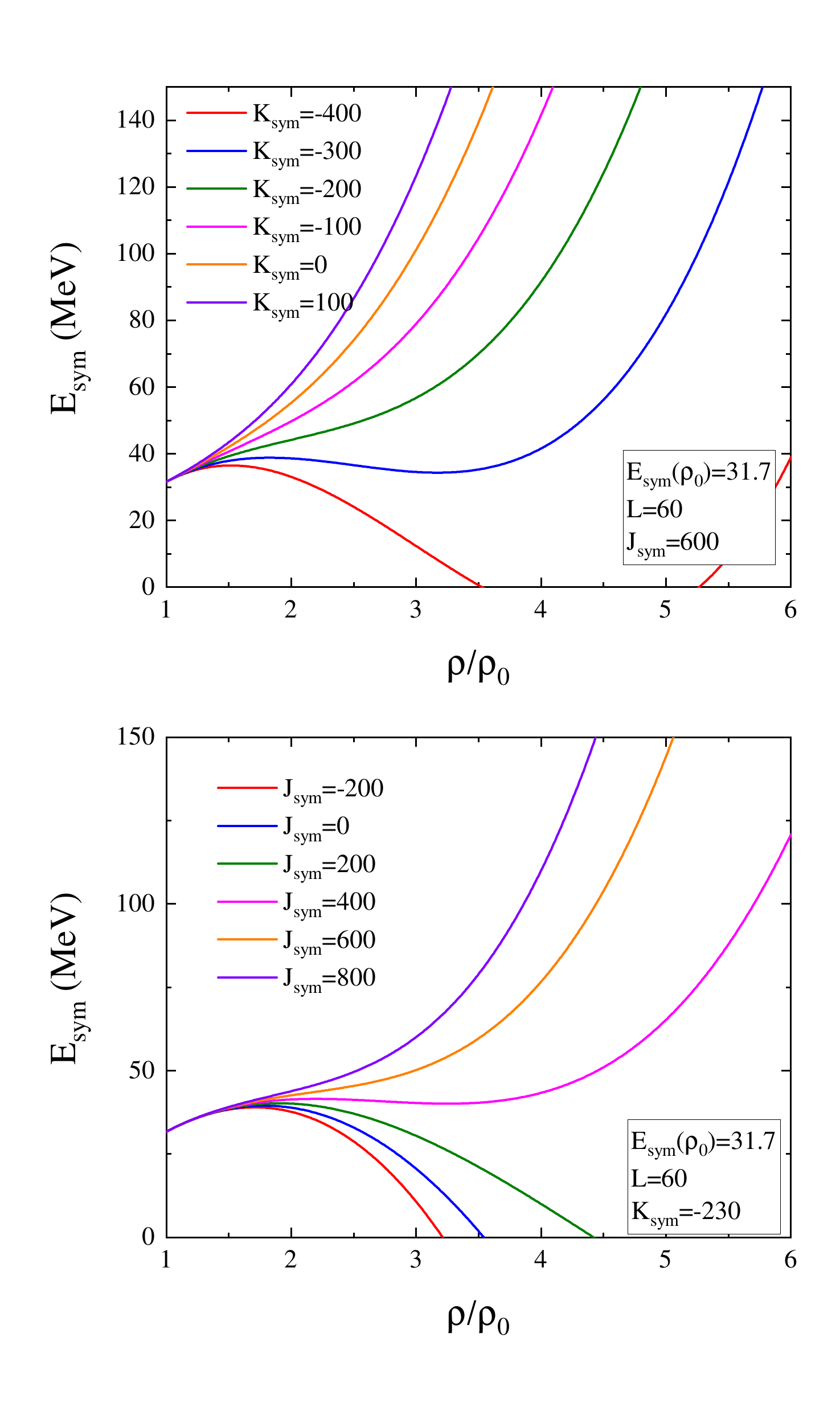}\qquad\quad
\vstretch{1.1}{\includegraphics[width=7.5cm]{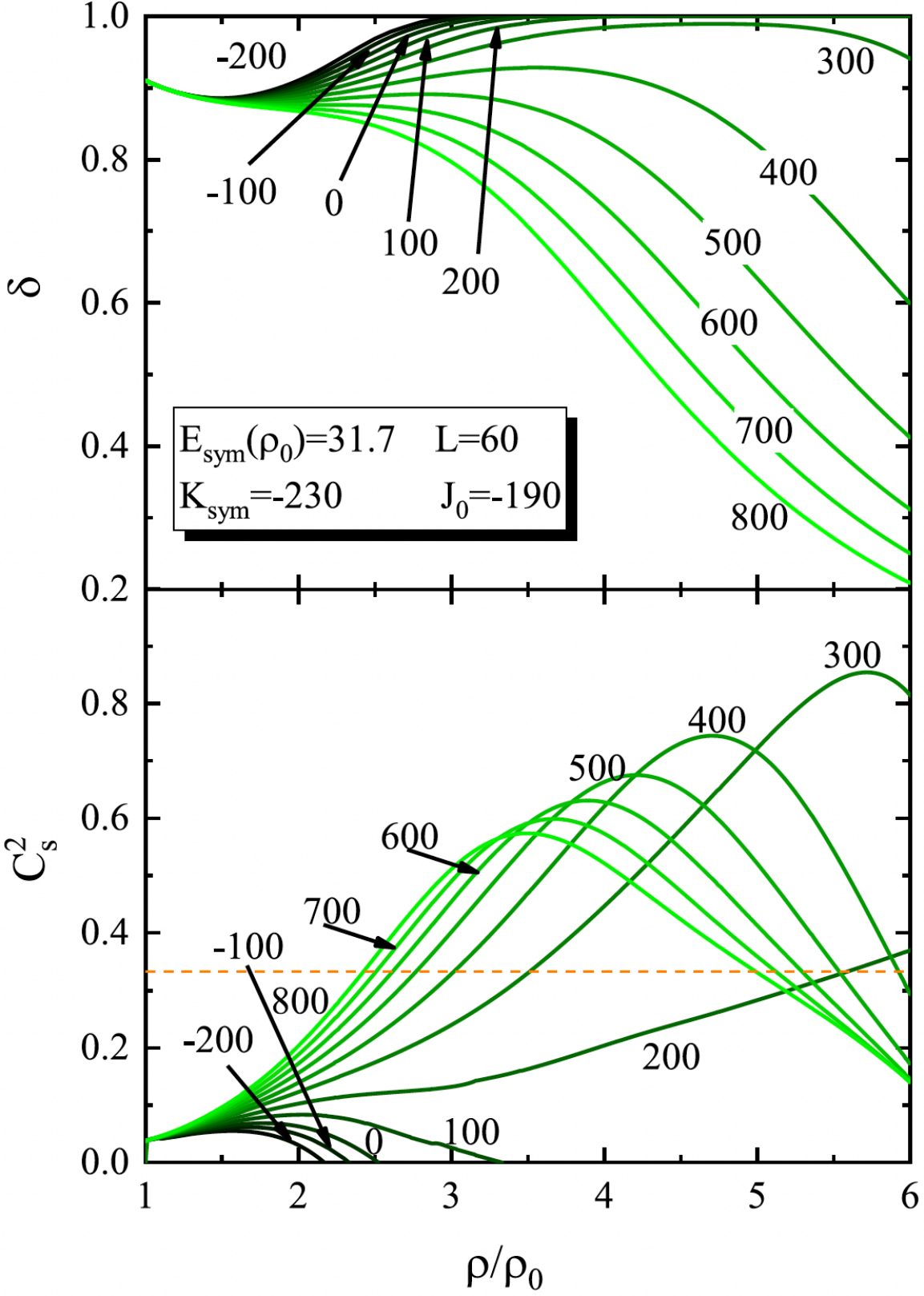}}
\caption{(Color Online). Left panel: High-density behaviors of nuclear symmetry energy by changing its curvature $K_{\rm{sym}}$ (upper left) and skewness $J_{\rm{sym}}$ (lower left) within their current uncertain ranges while fixing all the other parameters at their currently known most probable empirical values. Right panel: effects of varying the skewness
$J_{\rm{sym}}$ between $-200$ and $+800$\,MeV 
of nuclear symmetry energy on (1) (upper right): the density profile of the isospin asymmetry $\delta$ in NSs at $\beta$-equilibrium and (2) (lower right) the appearance of a peak in the density profile of $s^2$ in NSs. Figures taken from Ref.\,\cite{ZhangLi2023a}.
}\label{fig-ZL}
\end{figure}

As illustrated in the left panels of FIG.\ \ref{fig-ZL}, the high-density behavior of nuclear symmetry energy can be varied by modifying its curvature $K_{\rm{sym}}$ and/or skewness 
$J_{\rm{sym}}$ (see Eq.\,(\ref{fEsym})) within their current uncertain ranges while fixing all the others at their currently known most probably empirical values. Due to the $\beta$-equilibrium condition and the $E_{\rm{sym}}(\rho)\delta^2$ term in the energy per nucleon in neutron-rich nucleonic matter, it is well known that there exists the so-called isospin fractionation phenomenon, namely wherever $E_{\rm{sym}}(\rho)$ is low, the isospin asymmetry $\delta$ there would be high or vice versa to minimize the total energy and keep the isospin equilibrium. More specifically, 
considering two nearby regions with local densities $\rho_1$ and $\rho_2$, respectively, the chemical equilibrium condition requires $E_{\rm sym}(\rho_1)\delta(\rho_1)=E_{\rm sym}(\rho_2)\delta(\rho_2)$. Thus, for a given $E_{\rm{sym}}(\rho)$ function, the local isospin asymmetries at different densities will adjust themselves according to the relative symmetry energies in those regions. This is well illustrated in the upper right panel of 
of FIG.\,\ref{fig-ZL} by varying the $J_{\rm{sym}}$ between $-200$ and +800\,MeV. 
It is seen that the super-stiff $E_{\rm{sym}}(\rho)$ with $J_{\rm{sym}}$=800\,MeV leads to a vanishing $\delta$ (SNM) at high densities. While the super-soft $E_{\rm{sym}}(\rho)$ with negative $J_{\rm{sym}}$ leads to $\delta=1$ (pure neutron matter) at high densities. Most importantly, in both cases the product of $E_{\rm{sym}}(\rho)$ and $\delta$ at high densities reach zero as required by the $\beta$-equilibrium and charge neutrality conditions. The density profile of the isospin asymmetry $\delta$ in NSs can essentially influence the appearance of the peak in $s^2$, as shown in the lower-right panel of FIG.\,\ref{fig-ZL}. It is clearly seen that regardless of the high-density behavior of $E_{\rm{sym}}(\rho)$  (super-soft or super-stiff), the equilibrium speed of sound $C^2_s(\rho)$ in NSs at $\beta$-equilibrium always show a peak with its position depending on the high-density behavior of nuclear symmetry energy characterized by mostly $K_{\rm{sym}}$ and $J_{\rm{sym}}$.

Thus, within the simplest NS model consisting of nucleons, electrons and muons only without introducing the degree of freedom of quarks could account for a peaked structure in the density profile of $s^2$.
We can understand this phenomenon by working out the expression for $a_4$ in expanding the reduced energy density $\heps$\,\cite{CL24-a}.
By introducing the reduced densities $\widehat{\rho}=\rho/\rho_{\rm{c}}$ and $\widehat{\rho}_0=\rho_0/\rho_{\rm{c}}=\widehat{\rho}_{\rm{sat}}$ with $\rho_{\rm{c}}$ being the central density in NSs, we have $
\widehat{\rho}\approx
1+\beta_1\widehat{r}^2+(\beta_2 a_4-\beta_3)\widehat{r}^4
$, which defines the coefficients $\beta_1$, $\beta_2$ and $\beta_3$ from Eq.\,(\ref{pk-4}).
Based on these relations, we can express the energy density $\varepsilon=[E(\rho)+M_{\rm{N}}]\rho$, or its reduced form as $
\widehat{\varepsilon}
={\varepsilon}/{\varepsilon_{\rm{c}}}
=[E(\rho)+M_{\rm{N}}]({\rho_{\rm{c}}}/{\varepsilon_{\rm{c}}})\widehat{\rho}$.
Putting the $\widehat{\rho}$ into $\chi=(\rho-\rho_0)/3\rho_0$ and further putting $\chi$ back into $E_0(\rho)$ as well as $E_{\rm{sym}}(\rho)$ gives the $\widehat{\varepsilon}$ as a function of $\widehat{r}^2$\,\cite{CL24-a}:
\begin{align}\label{deff-2}
\widehat{\varepsilon}\approx& \left({\rho_{\rm{c}}}/{\varepsilon_{\rm{c}}}\right)
\left[M_{\rm{N}}+E_0(\rho_0)+2^{-1}K_0\chi^2+6^{-1}J_0\chi^3
+\left(S+L\chi+2^{-1}K_{\rm{sym}}\chi^2+6^{-1}J_{\rm{sym}}\chi^3\right)\delta^2\right].
\end{align}
For our purpose here, we need the expression of the $\widehat{r}^4$-term in Eq.\,(\ref{deff-2}) which itself involves $a_4$ due to the expression for $\widehat{\rho}$.
Equaling it with $a_4\widehat{r}^4$ gives a self-consistent equation for determining $a_4$ (both sides of which involve $a_4$).

\begin{figure}
\centering
\includegraphics[height=4.cm]{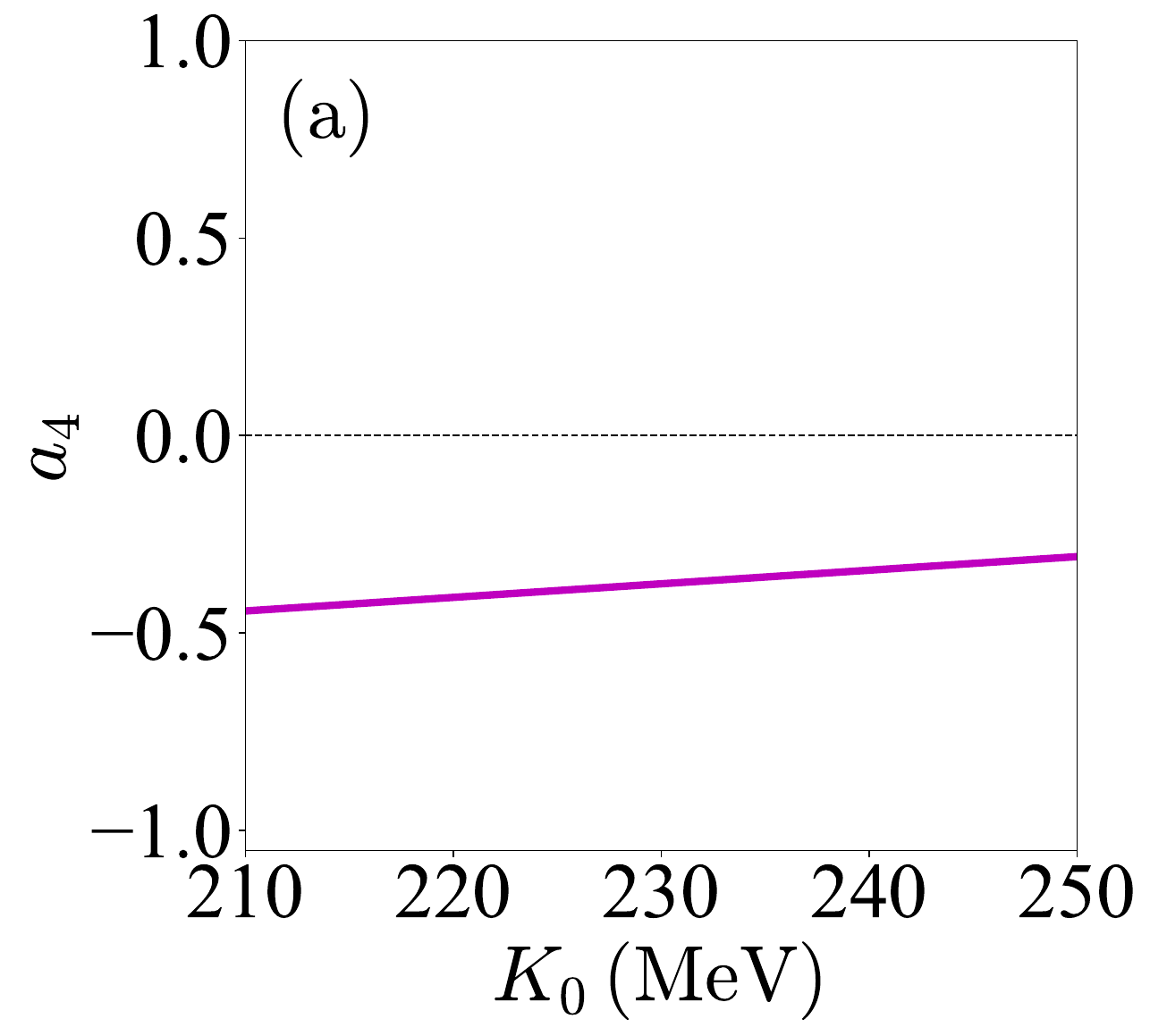}
\includegraphics[height=4.cm]{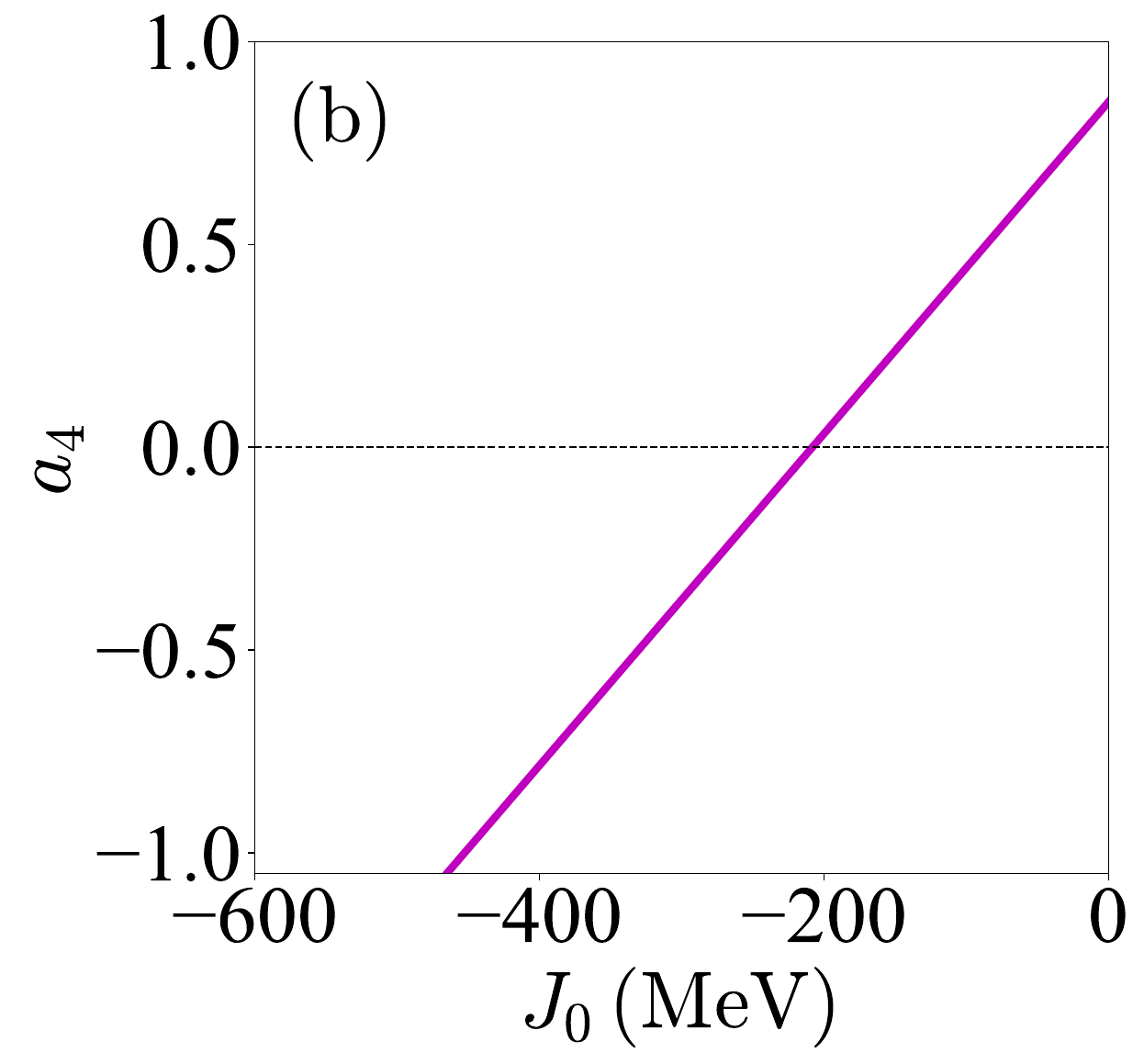}
\includegraphics[height=4.cm]{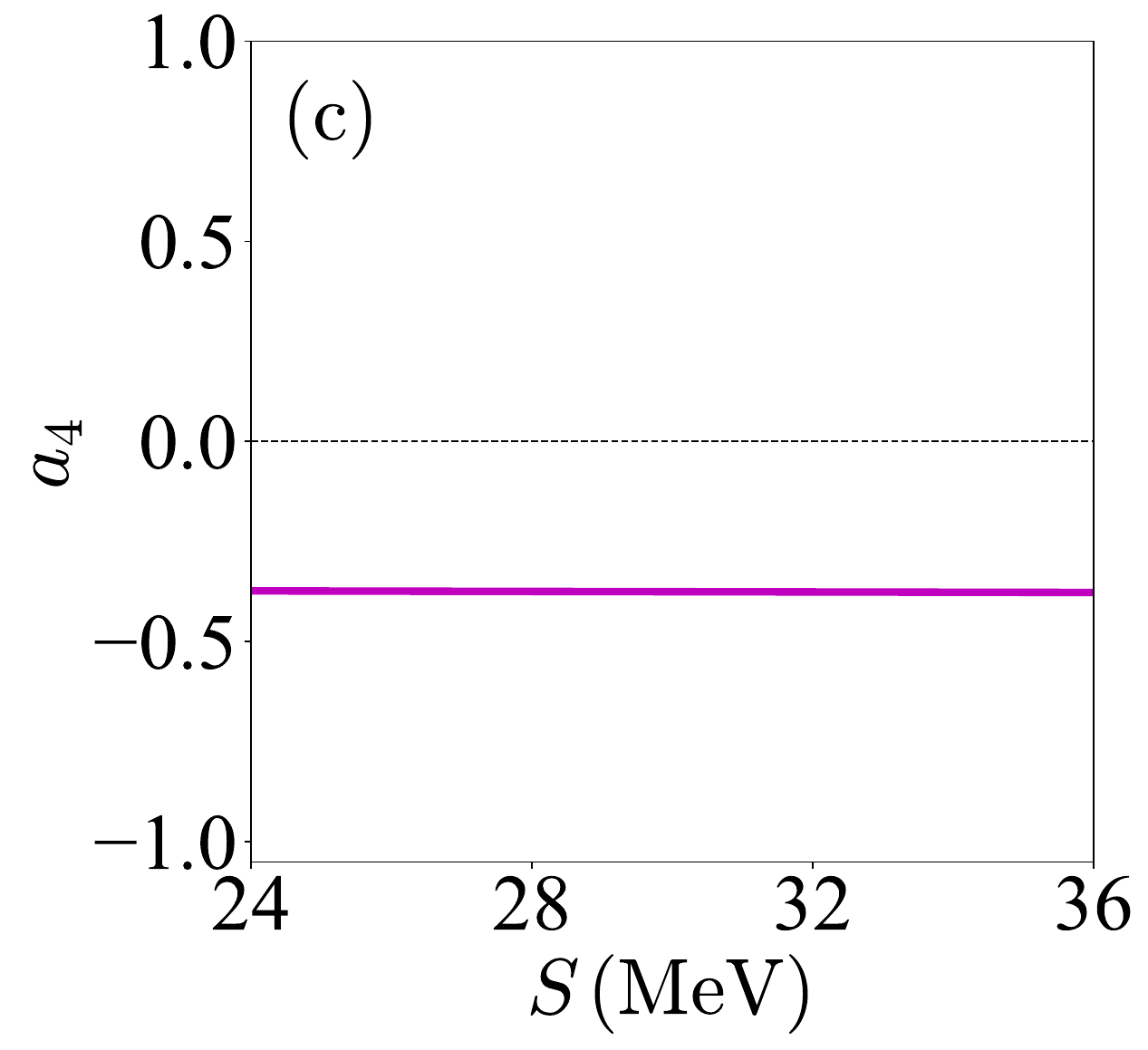}\\
\includegraphics[height=4.cm]{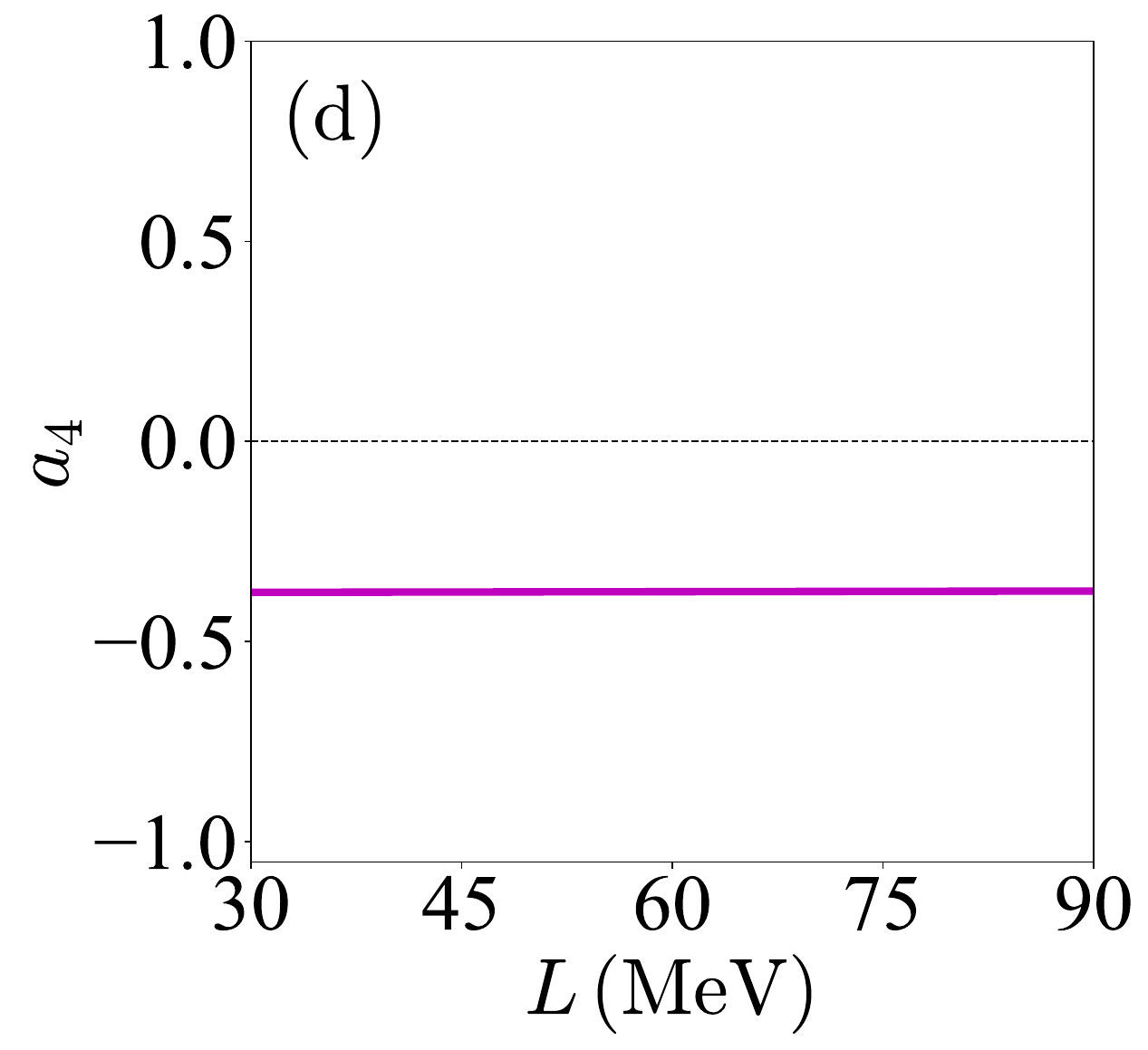}
\hspace{0.0cm}
\includegraphics[height=4.cm]{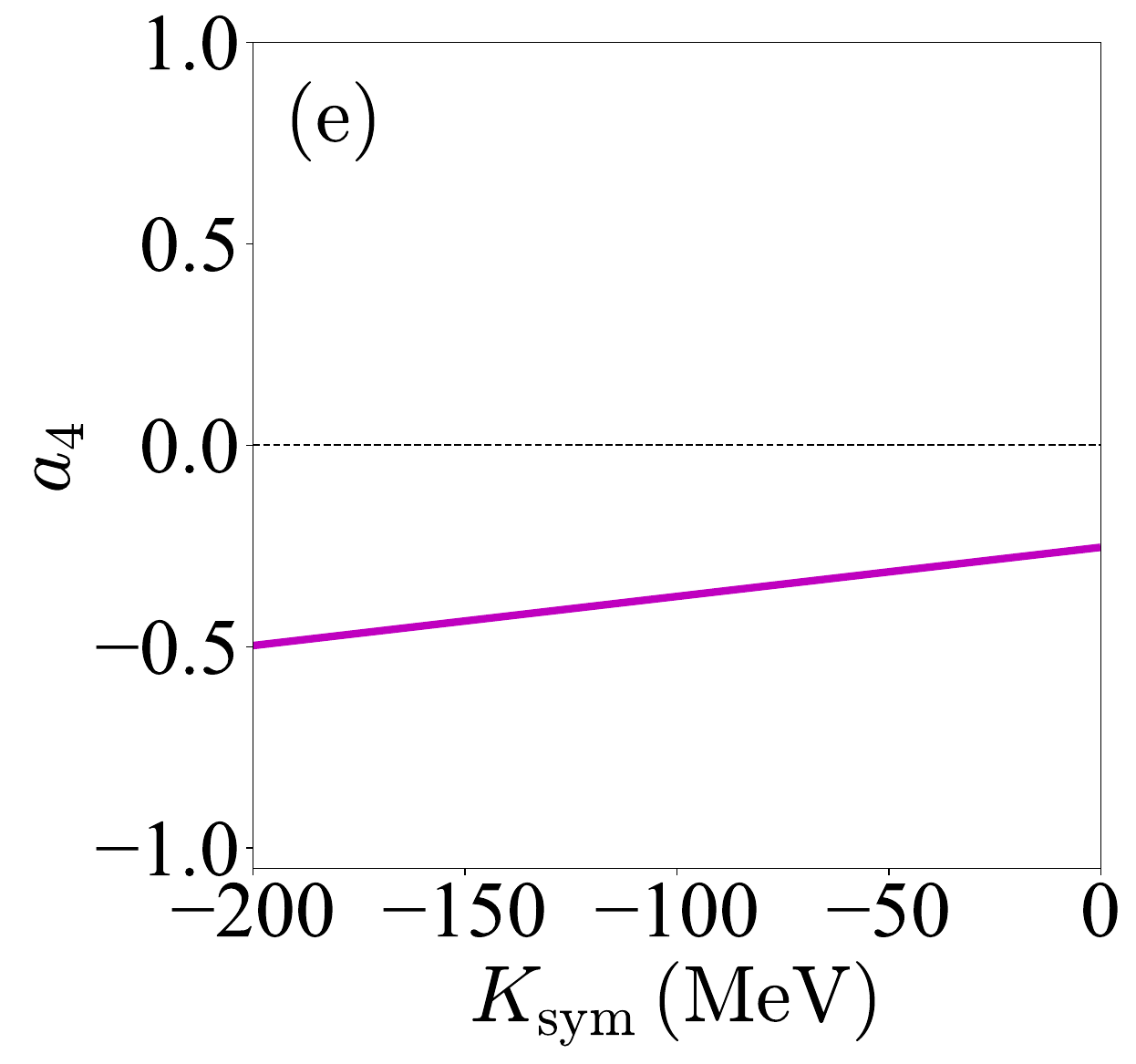}
\hspace{0.0cm}
\includegraphics[height=4.cm]{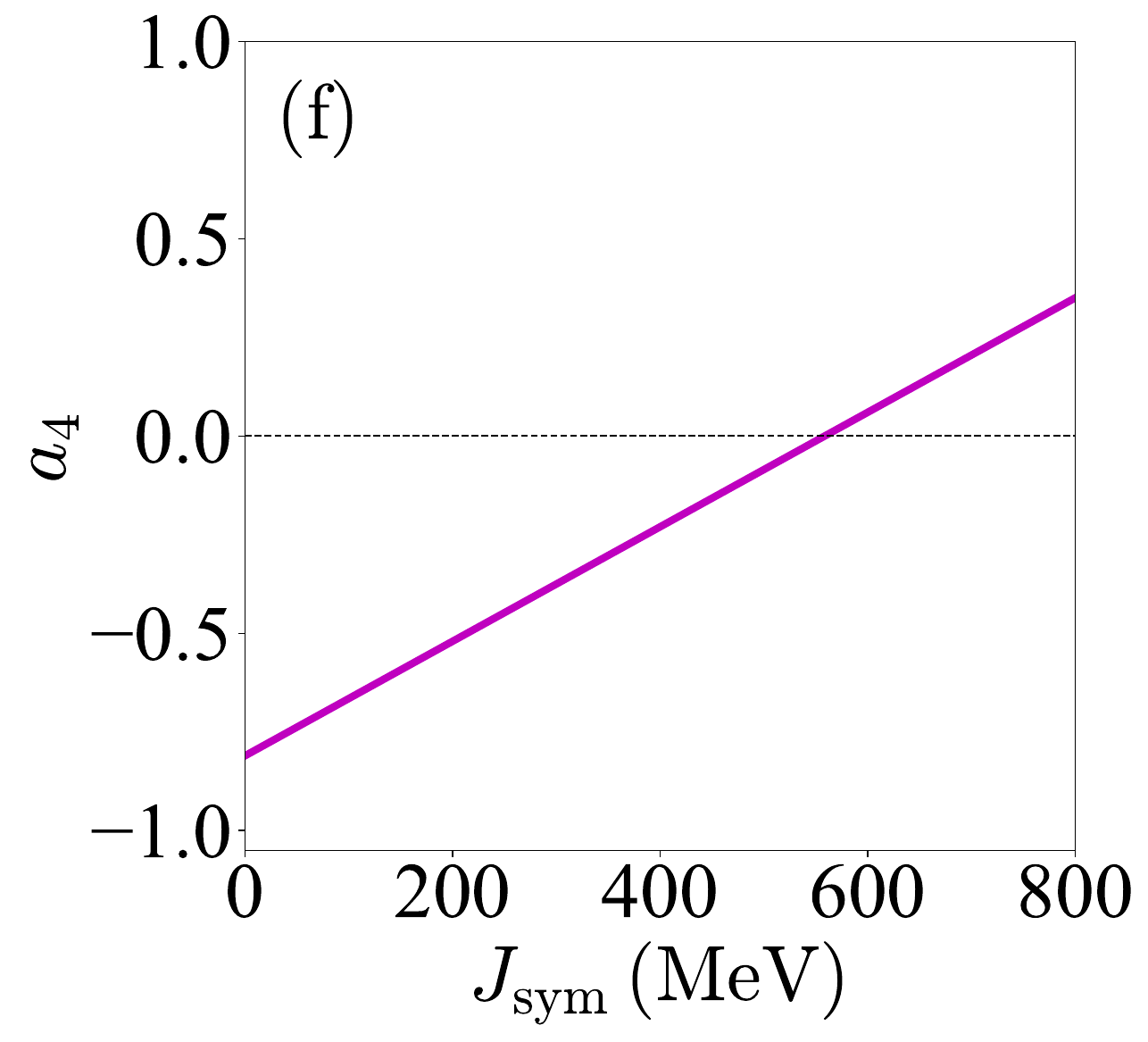}
\caption{(Color Online).  Dependence of $a_4$ on the empirical EOS parameters $K_0$, $J_0$, $S$, $L$, $K_{\rm{sym}}$ and $J_{\rm{sym}}$ according to Eq.\,(\ref{fff}),  see text for the details on the choice/determination of the parameters. Figures taken from Ref.\,\cite{CL24-a}.
}\label{fig_a4_ll}
\end{figure}

The expression for $a_4$ is quite complicated and since the magnitude of $\overline{M}_{\rm{N}}\equiv M_{\rm{N}}+E_0(\rho_0)\approx923\,\rm{MeV}$ (using $M_{\rm{N}}\approx939\,\rm{MeV}$ and $E_0(\rho_0)\approx-16\,\rm{MeV}$) is generally larger than $K_0\approx230\,\rm{MeV}$, $J_0\approx-300\,\rm{MeV}$, $S\approx30\,\rm{MeV}$, $L\approx 60\,\rm{MeV}$, $K_{\rm{sym}}\approx-100\,\rm{MeV}$ and $J_{\rm{sym}}\approx300\,\rm{MeV}$\,\cite{LCXZ2021}, one can introduce small quantities $K_0/\overline{M}_{\rm{N}}$, $J_0/\overline{M}_{\rm{N}}$, $S/\overline{M}_{\rm{N}}$, $L/\overline{M}_{\rm{N}}$, $K_{\rm{sym}}/\overline{M}_{\rm{N}}$ and $J_{\rm{sym}}/\overline{M}_{\rm{N}}$ to expand $a_4$ to dig out the main features of its dependence on the empirical EOS characteristics\,\cite{CL24-a}:
\begin{align}
a_4\approx&-\frac{\beta_3}{54(\beta_3\beta_2-1)^2\widehat{\rho}_0^3}
\left\{+
2
\left(\left(\beta_3\beta_2-1\right)\beta_1^2+\frac{2\beta_2\beta_3}{3}\right)
\left[\left(\frac{J_0}{\overline{M}_{\rm{N}}}\right)+
\left(\frac{J_{\rm{sym}}}{\overline{M}_{\rm{N}}}\right)\delta^2
\right]\right.\notag\\
&\hspace*{0.cm}-3\left(\left(\beta_3\beta_2-1\right)\beta_1^2+\beta_2\beta_3\right)
\left[\left(\frac{J_0}{\overline{M}_{\rm{N}}}\right)-3\left(\frac{K_0}{\overline{M}_{\rm{N}}}\right)
+\left(\left(\frac{J_{\rm{sym}}}{\overline{M}_{\rm{N}}}\right)-3\left(\frac{K_{\rm{sym}}}{\overline{M}_{\rm{N}}}\right)\right)\delta^2\right]\widehat{\rho}_0\notag\\
&\hspace*{0.cm}
+\left(\left(\beta_3\beta_2-1\right)\beta_1^2+2\beta_2\beta_3\right)
\left[\left(\frac{J_0}{\overline{M}_{\rm{N}}}\right)-6\left(\frac{K_0}{\overline{M}_{\rm{N}}}\right)
+\left(
\left(\frac{J_{\rm{sym}}}{\overline{M}_{\rm{N}}}\right)-6\left(\frac{K_{\rm{sym}}}{\overline{M}_{\rm{N}}}\right)
+18\left(\frac{L}{\overline{M}_{\rm{N}}}\right)
\right)\delta^2\right]\widehat{\rho}_0^2\notag\\
&\hspace*{0.cm}
-54\beta_2\beta_3\left[\left(\beta_3\beta_2-1\right)+\frac{1}{162}\left(\frac{J_0}{\overline{M}_{\rm{N}}}\right)
-\frac{1}{18}\left(\frac{K_0}{\overline{M}_{\rm{N}}}\right)\right.\notag\\
&\hspace{3.cm}
\left.\left.
+\left(
\frac{1}{162}\left(\frac{J_{\rm{sym}}}{\overline{M}_{\rm{N}}}\right)-\frac{1}{18}\left(\frac{K_{\rm{sym}}}{\overline{M}_{\rm{N}}}\right)+\frac{1}{3}\left(\frac{L}{\overline{M}_{\rm{N}}}\right)-\left(\frac{S}{\overline{M}_{\rm{N}}}\right)
\right)\delta^2\right]\widehat{\rho}_0^3\right\},\label{fff}
\end{align}
here $\beta_3=\overline{M}_{\rm{N}}\rho_{\rm{c}}/\varepsilon_{\rm{c}}$.
For a given set of nuclear EOS parameters $(K_0$, $J_0$, $S$, $L$, $K_{\rm{sym}}$, $J_{\rm{sym}}$) and a central energy density $\varepsilon_{\rm{c}}$ (or a central density $\rho_{\rm{c}}$), 
the NS mass as well as its radius could be obtained.

A main feature of Eq.\,(\ref{fff}) is that $J_0$ and $J_{\rm{sym}}$ are the leading-order contributions to $a_4$, since they appear at $\widehat{\rho}_0^0$ order in the curly brackets. Other characteristics such as $K_0$, $L$,  $K_{\rm{sym}}$, etc., appear at higher-orders of $\widehat{\rho}_0$. As an example, taking  $\rho_{\rm{c}}\approx5\rho_0\approx0.8\,\rm{fm}^{-3}$ (equivalently $\widehat{\rho}_0=\rho_0/\rho_{\rm{c}}\approx1/5$) and $K_0\approx230\,\rm{MeV}$, $J_0\approx-300\,\rm{MeV}$, $S\approx30\,\rm{MeV}$, $L\approx 60\,\rm{MeV}$, $K_{\rm{sym}}\approx-100\,\rm{MeV}$ together with $J_{\rm{sym}}\approx300\,\rm{MeV}$\,\cite{LCXZ2021}, then $\delta\approx0.6$ is obtained at $\rho_{\rm{c}}$, the central energy density $\varepsilon_{\rm{c}}$ and the central pressure $P_{\rm{c}}$ are obtained as $\varepsilon_{\rm{c}}\approx 863\,\rm{MeV}/\rm{fm}^3$ and $P_{\rm{c}}\approx 182\,\rm{MeV}/\rm{fm}^3$, respectively. Consequently, $\beta_3=\overline{M}_{\rm{N}}\rho_{\rm{c}}/\varepsilon_{\rm{c}}\approx0.86$,  ${\x}\approx0.21$ and $\beta_1\approx-0.74$, $\beta_2\approx0.86$ and $\beta_3\approx0.10$.
In FIG.\,\ref{fig_a4_ll}, we show the dependence of $a_4$ on the nuclear EOS parameters $K_0$, $J_0$, $S$, $L$, $K_{\rm{sym}}$ and $J_{\rm{sym}}$ according to Eq.\,(\ref{fff}), within their currently known uncertainty ranges.
Besides $a_4\sim\mathcal{O}(1)$, we find $a_4$ can take either positive or negative values, depending on the empirical parameters characterizing the high-density EOS.
For example, $a_4$ can be positive with the increasing skewness parameter $J_0$ of SNM or the skewness parameter $J_{\rm{sym}}$ of the symmetry energy, e.g., $J_{\rm{sym}}\gtrsim560\,\rm{MeV}$ is needed for $a_4\gtrsim0$ (while fixing other parameters). It is seen that a more positive $J_{\rm{sym}}$ or $J_0$ makes $a_4$ more positive, tending to generate the peak in $s^2$ at a smaller density\,\cite{ZhangLi2023a}.

\begin{figure}[h!]
\centering
\includegraphics[height=5.cm]{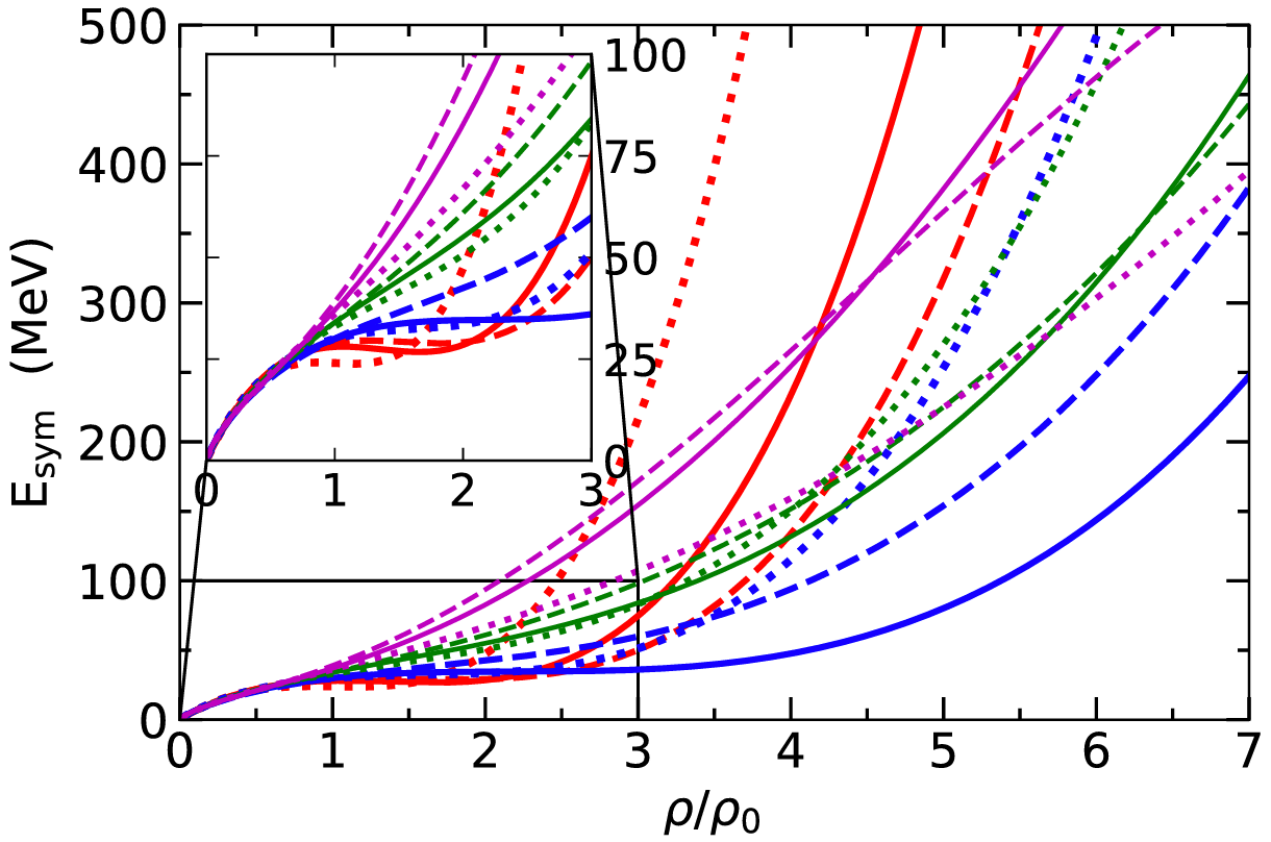}\qquad
\includegraphics[height=5.cm]{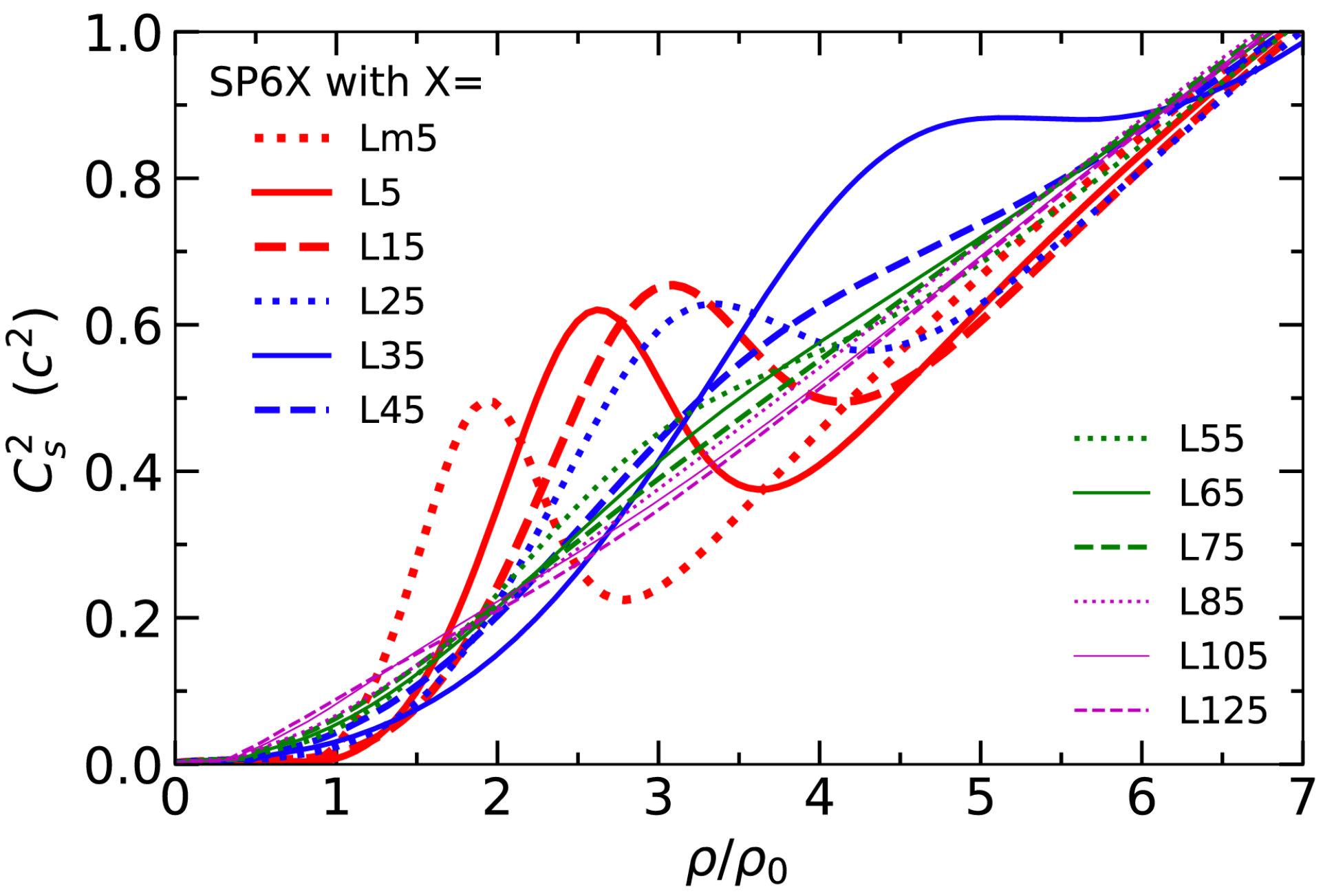}
\caption{(Color Online). Similar to FIG.\,\ref{fig-ZL} but demonstrating effects of the slope parameter $L$ on SSS. Figure taken from Ref.\,\cite{SPei24}.
}\label{fig_SP24FIG}
\end{figure}
In a very recent study, Ref.\,\cite{SPei24} explored the SSS as a function of number density by varying the slope $L$ of nuclear symmetry energy (left panel of FIG.\,\ref{fig_SP24FIG}). They found that for $L\lesssim35\,\rm{MeV}$, a peak emerges in $s^2$ at densities accessible in some normal NSs, as indicated in the right panel of FIG.\,\ref{fig_SP24FIG}.
However, in their work\,\cite{SPei24} not only $L$ varies with different interaction sets but also $K_{\rm{sym}}$ and $J_{\rm{sym}}$ are different. Thus, it is not quite clear which are really the deterministic quantities for the peaked $s^2$. Using our analysis of FIG.\,\ref{fig_a4_ll}, the effects of $S\equiv E_{\rm{sym}}(\rho_0)$ and $L$ on $a_4$ are small. Therefore, these low-order parameters may not be the important quantities for generating a peaked $s^2$ density profile. Nevertheless, we notice that the $J_{\rm{sym}}$'s in their interaction sets with smaller $L$'s are relatively larger than those corresponding to larger $L$'s\,\cite{SPei24}, indicating that  $J_{\rm{sym}}$ may be the hidden agent causing the peak in the density profile of $s^2$.

\begin{figure}[h!]
\centering
\includegraphics[width=7.5cm]{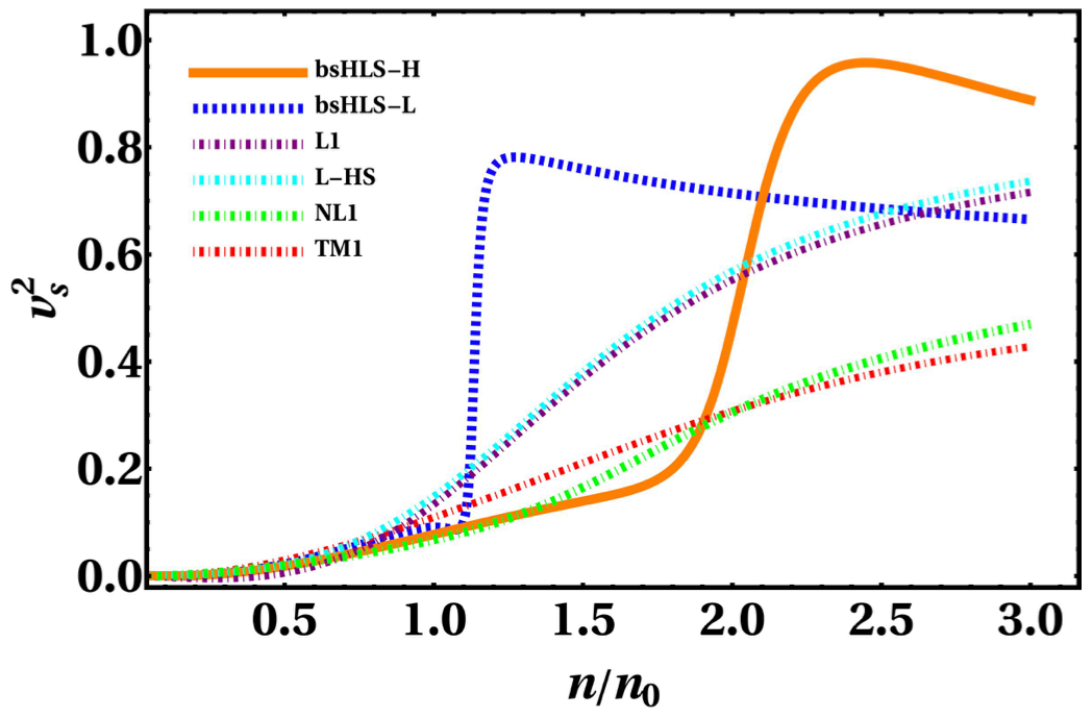}
\caption{(Color Online). The SSS obtained by using an effective theory implemented with a conformal compensator (nonlinear realization of scale symmetry) regarded as the source of the lightest scalar meson. Figure taken from Ref.\,\cite{YLMa24}.
}\label{fig_YLMa24s2}
\end{figure}
Finally, it is also interesting to note that a peaked $s^2$ without retrospect to phase transition or crossover from hadron to exotic configuration was recently found by using an effective theory implemented with a conformal compensator (nonlinear realization of scale symmetry) regarded as the source of the lightest scalar meson\,\cite{YLMa24}, see FIG.\,\ref{fig_YLMa24s2}.
In their work, the peaked position was found to be located at about $(1\mbox{-}2.5)\rho_0$, and the result may be related to the issue on how the hidden scale symmetry manifests in nuclear medium from the unitarity limit in dilute matter to dilaton limit in compact star matter.

\subsection{SSS in supradense matter from lattice QCD or other QCD-based effective theories}\label{sub_DenseQCD}

As we discussed in the last few subsections, the strong-field gravity in GR realized in NSs plays an essential role in generating a peaked structure in $s^2$ at densities accessible in NSs.
On the other hand as we mentioned earlier, properties of supra-dense matter have also been studied extensively from QCD-based theories. Our knowledge on superdense matter confined by strong-field gravity in NSs from the two directions have to match. It is thus useful to review briefly the SSS $s^2$ obtained for QCD matter simulated on a lattice or from other QCD-based effective theories. Because of our limited knowledge on this topic, we have no intention to review comprehensively all the relevant works along this line. We refer the interested readers to Refs.\,\cite{Brandt23} and references therein for more detailed discussions.
In the following, we give a few examples that have been very useful for us to explore and understand the density profiles of $s^2$ in NSs.
When necessary and possible, we make connections between the predictions of these QCD-based theories with what our perturbative analyses of the scaled TOV equation have taught us in the previous sections.

\begin{figure}[h!]
\centering
\includegraphics[height=5.cm]{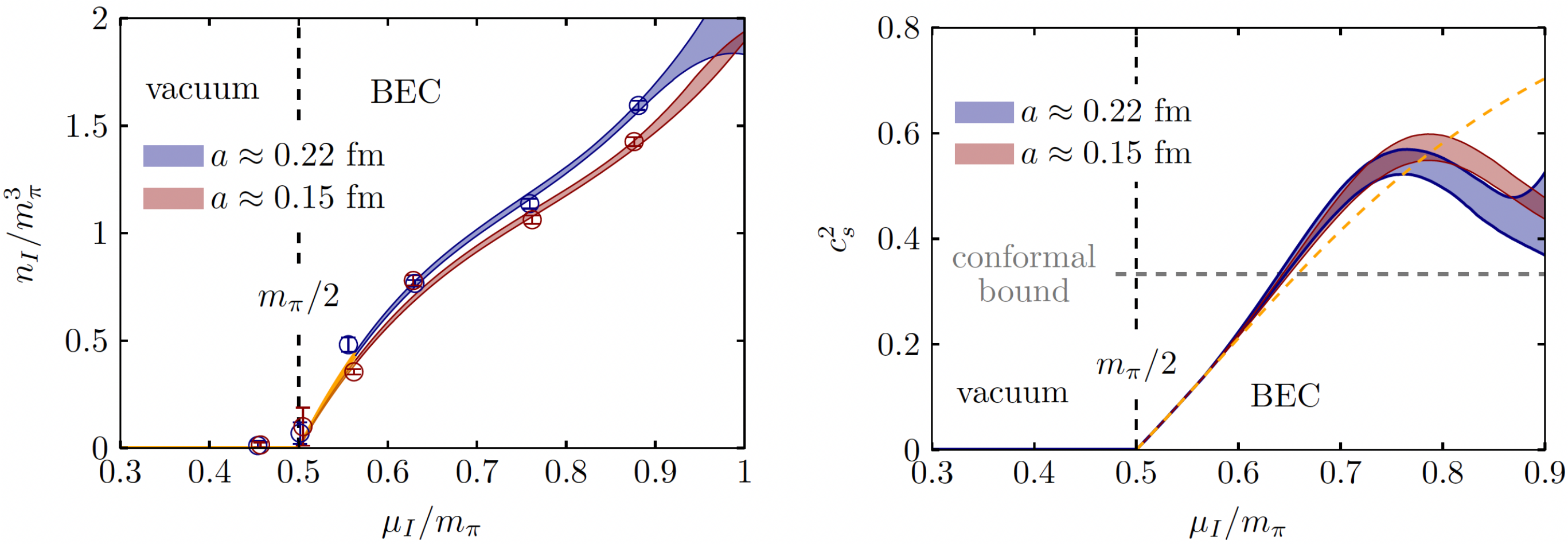}
\caption{(Color Online).  Left panel: relation between the isospin density $n_{\rm{I}}$ (reduced by $m_{\pi}^3$) and the isospin chemical potential $\mu_{\rm{I}}$ (reduced by $m_{\pi}$) for a $N_{\rm{f}}=(2+1)$-flavor QCD matter using the space-time lattice of sizes $(24^3\times32)\,\rm{fm}^4$ (with lattice spacing about $a\approx0.22\,\rm{fm}$) and $(32^3\times48)\,\rm{fm}^4$ ($a\approx 0.15\,\rm{fm}$), respectively.
Right panel: same as that in the left panel but for the $s^2$ as a function of $\mu_{\rm{I}}/m_\pi$.  
A physical mass about 140\,MeV for pions is used.
Figures taken from Ref.\,\cite{Brandt23}.
}\label{fig_Brandt23BD-1}
\end{figure}

\begin{figure}[h!]
\centering
\includegraphics[height=5.cm]{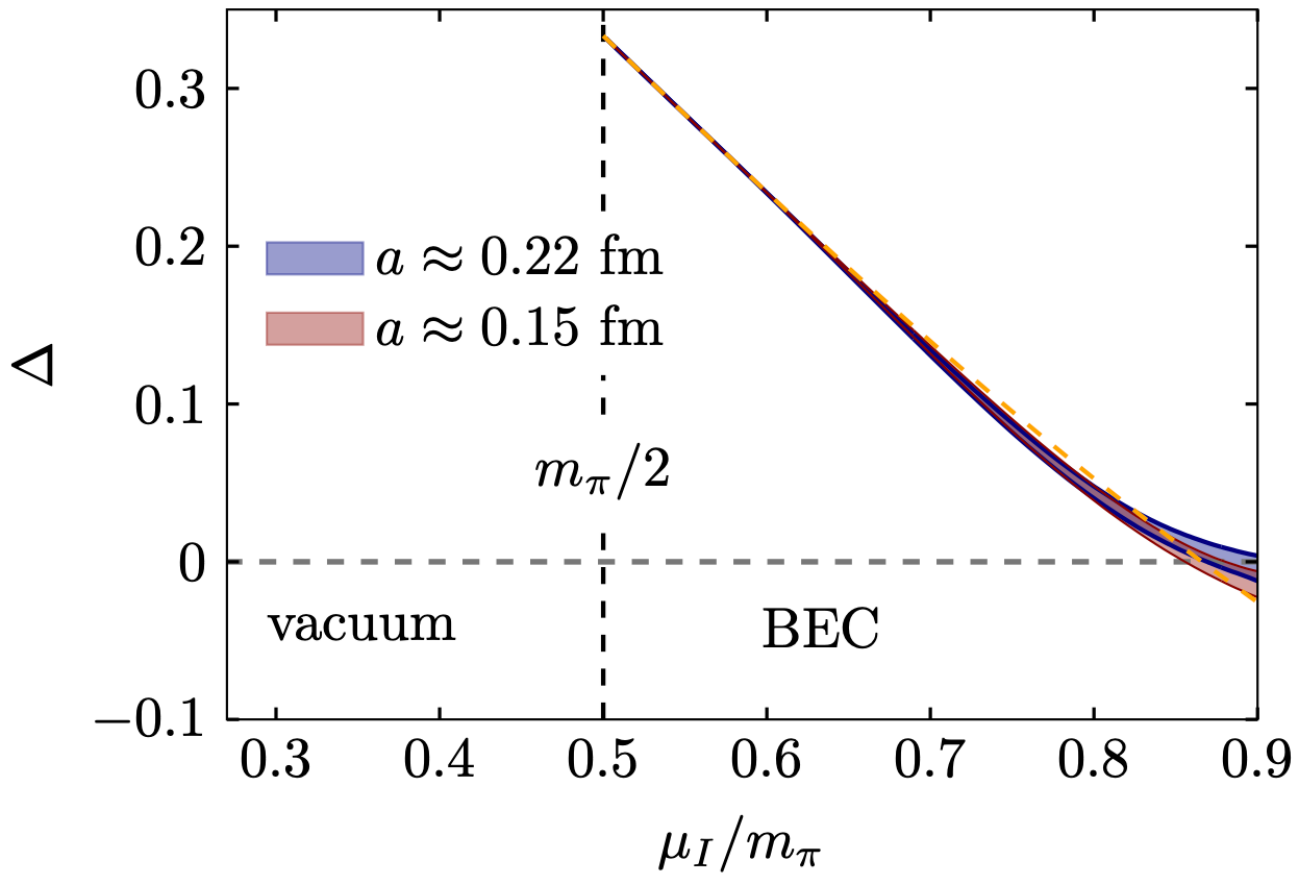}\quad
\includegraphics[height=5.1cm]{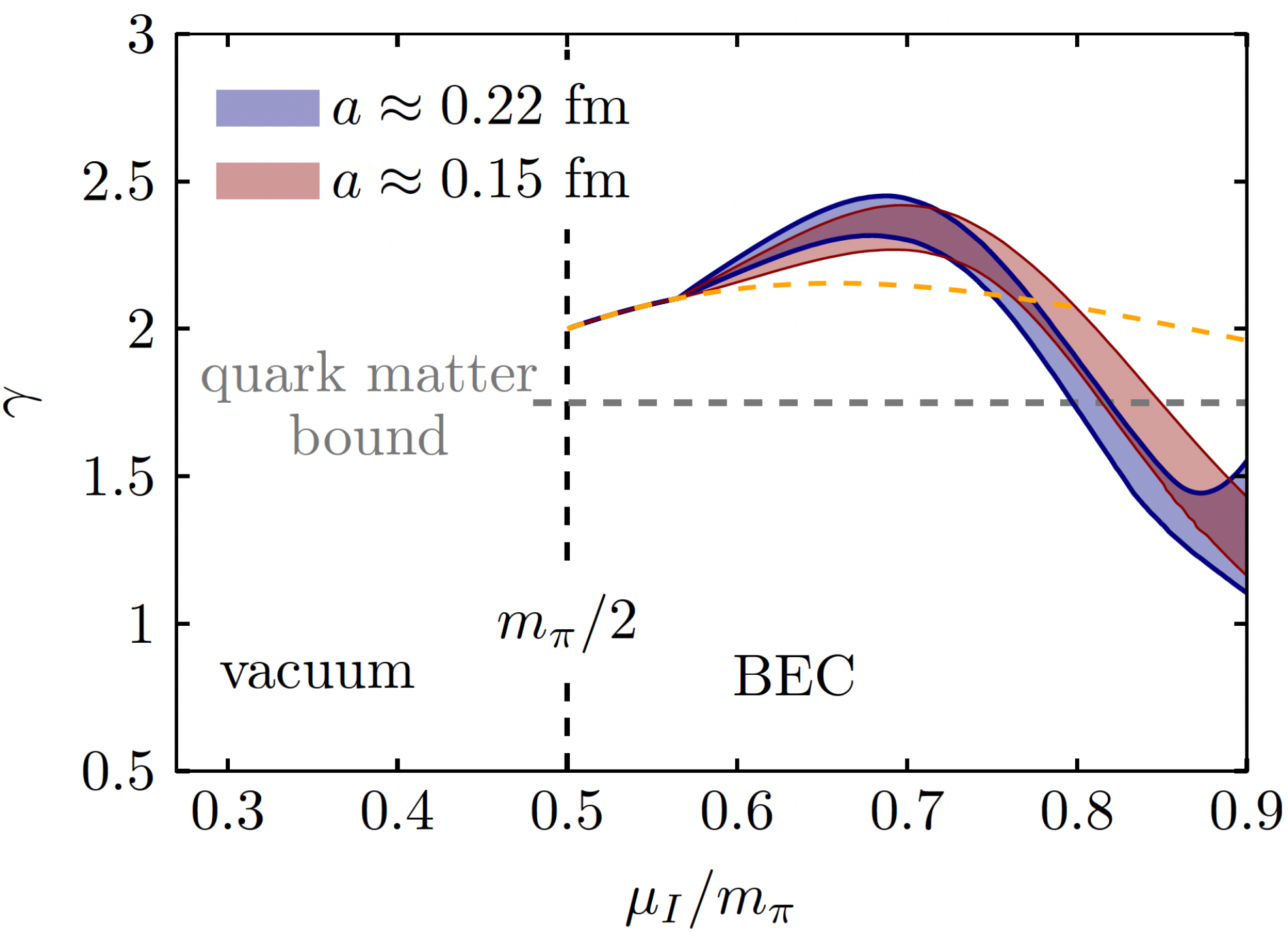}
\caption{(Color Online).  Same as FIG.\,\ref{fig_Brandt23BD-1} but for the dimensionless trace anomaly $\Delta$ (left panel) and the polytropic index $\gamma=s^2/\phi$ with $\phi=P/\varepsilon$.
Figures taken from Ref.\,\cite{Brandt23}
}\label{fig_Brandt23FIG}
\end{figure}

\begin{figure}[h!]
\centering
\hspace{0.2cm}\includegraphics[height=6.5cm]{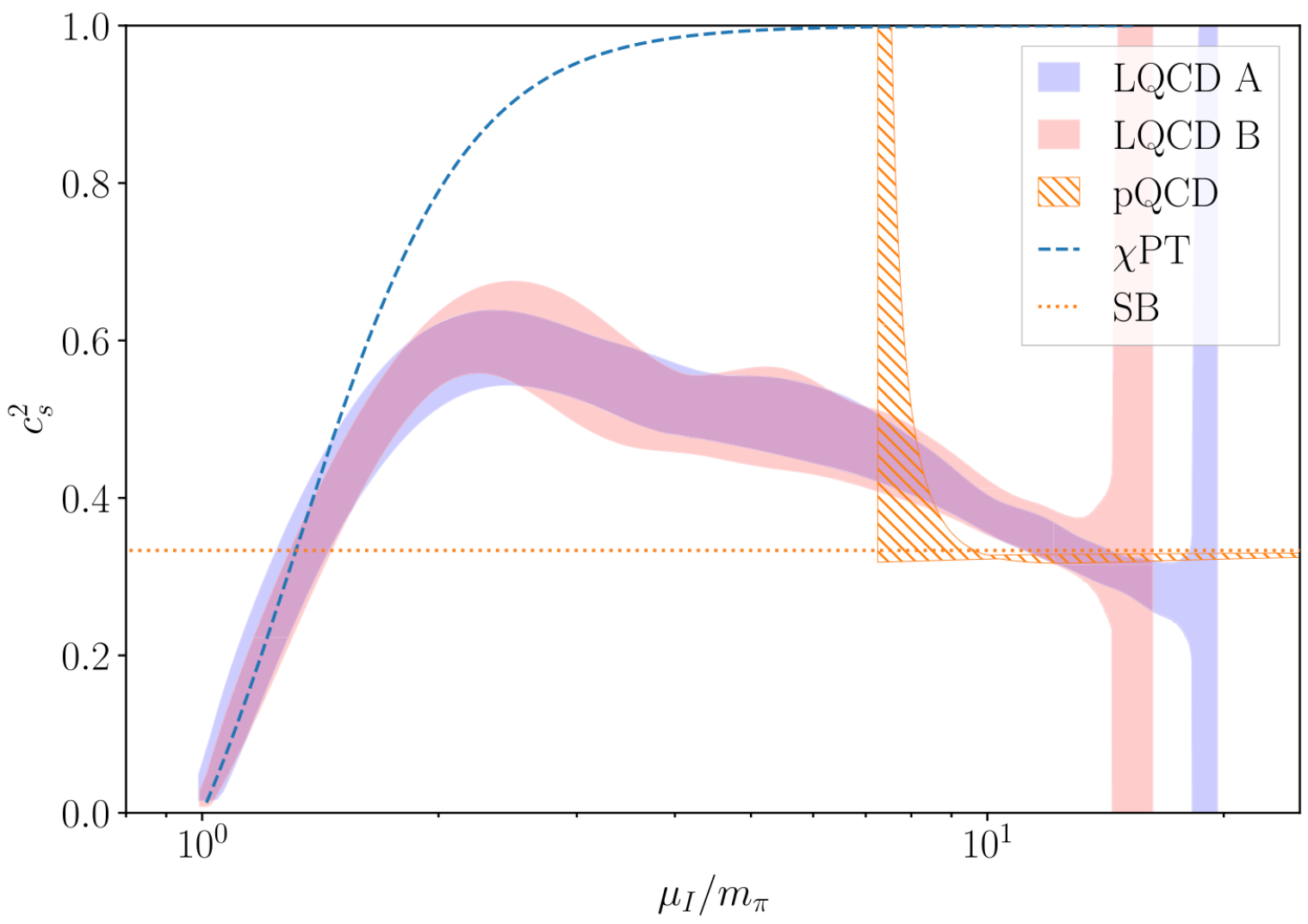}\\
\includegraphics[height=6.5cm]{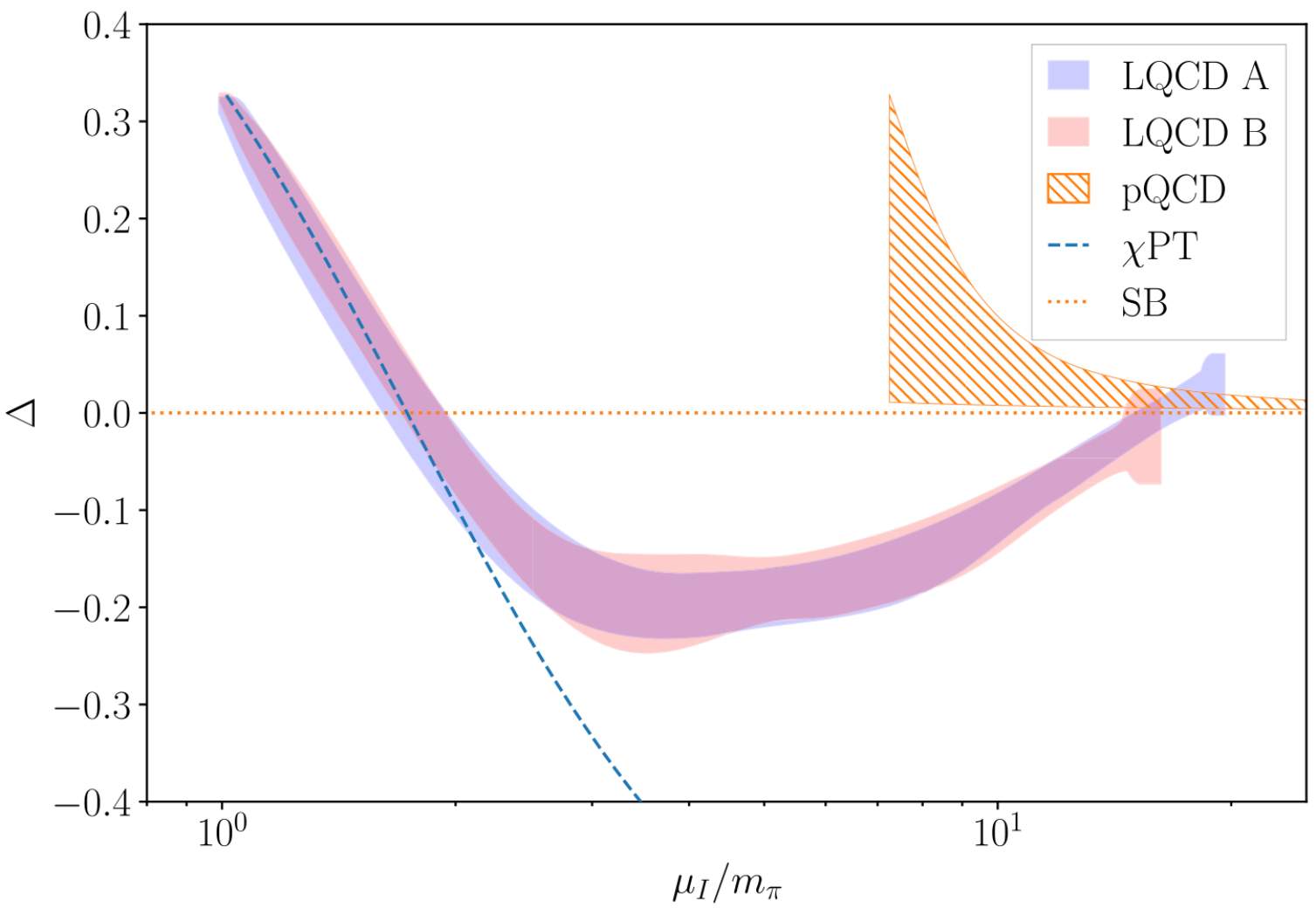}
\caption{(Color Online).  Speed of sound and the dimensionless trace anomaly obtained from Ref.\,\cite{Abb23}, where two lattices $(14.4^3\times8.8)\,\rm{fm}^4$ and $(15.8^3\times11.6)\,\rm{fm}^4$ are adopted; the pion mass is set as 170\,MeV.
Figures taken from Ref.\,\cite{Abb23}.
}\label{fig_Abb23FIG}
\end{figure}

Firstly, in Ref.\,\cite{Brandt23}, the lattice QCD simulations were performed by employing $N_{\rm{f}} = 2 + 1 $ flavors of dynamical staggered quarks at physical masses and using different lattice spacings.
Their results are shown in FIG.\,\ref{fig_Brandt23BD-1}. 
The left panel of FIG.\,\ref{fig_Brandt23BD-1} shows the relation between isospin density $n_{\rm{I}}$ (in unit of $m_{\pi}^3$) and the isospin chemical potential $\mu_{\rm{I}}$ (in unit of $m_{\pi}$), here the isospin density $n_{\rm{I}}$ is obtained via $n_{\rm{I}}=(T/V)\partial\ln\mathcal{Z}/\partial\mu_{\rm{I}}$ with $\mathcal{Z}$ the partition function in the grand ensemble; the full $\mu_{\rm{I}}$-dependence of the EOS could then be obtained.
Shown in the right panel of FIG.\,\ref{fig_Brandt23BD-1} is the isospin chemical potential dependence of the SSS $s^2$; it increases at small $\mu_{\rm{I}}\approx0.5m_{\pi}$ eventually to exceed the conformal bound 1/3 and develops a peak at $\mu_{\rm{I}}\approx0.77m_\pi\approx 108\,\rm{MeV}$ using $m_{\pi}\approx140\,\rm{MeV}$.
The peak value of $s^2$ is about $s_{\max}^2\approx0.57$ for $a\approx0.15\,\rm{fm}$.
The dimensionless trace anomaly $\Delta=1/3-\phi=1/3-P/\varepsilon$ and the polytropic index $\gamma=s^2/\phi$ are shown similarly in FIG.\,\ref{fig_Brandt23FIG}. The $\Delta$ at the peak position of $s^2$ is about 0.07 and is likely to be negative for even larger $\mu_{\rm{I}}$.
{\color{xll}We notice that $\Delta\approx0.07$ implies that the ratio of pressure to energy density $\phi\approx0.26$ is large, while the corresponding energy density of the system is about $\varepsilon_{\rm{pk}}\approx 30\,\rm{MeV}/\rm{fm}^3$ that is far smaller than those in NS cores and is even smaller than $\varepsilon_0\approx150\,\rm{MeV}/\rm{fm}^3$.
This means although the energy density is small, the large ratio $\phi$ implies that the matter is highly compressed (due to the lattice spacing set). In this sense, the emergence of a peaked $s^2$ is consistent with our analysis, i.e., the ``artificially strong gravity'' between quarks on the lattice effectively extrudes the peak in $s^2$ density profile.}
Similarly, the $\gamma$ parameter reaches its maximum about 2.4 at $\mu_{\rm{I}}\approx0.7m_{\pi}$ and approaches eventually to the conformal limit 1 at even larger isospin chemical potentials.

\begin{figure}[h!]
\centering
\includegraphics[height=7.cm]{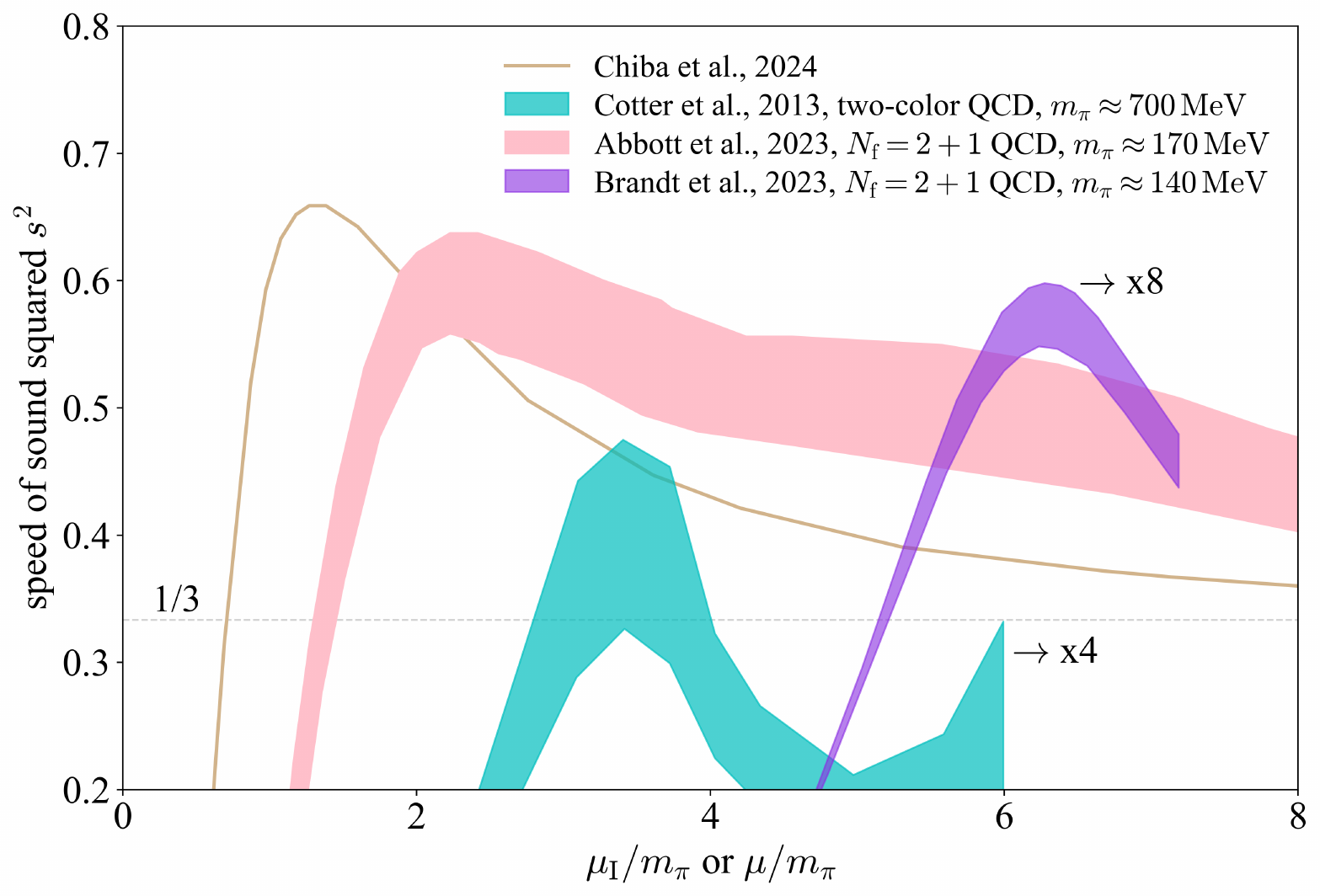}
\caption{(Color Online).  The chemical potential dependence of $s^2$ obtained from Refs.\,\cite{Cotter13,Abb23,Brandt23}, here the results of Ref.\,\cite{Brandt23} and of Ref.\,\cite{Cotter13} are stretched along the horizontal axis to be comparable with that of Ref.\,\cite{Abb23}.
}\label{fig_s2-QCD}
\end{figure}

In a recent study, Ref.\,\cite{Abb23} adopted two lattices of sizes $14.4^3\times8.8\,\rm{fm}^4$ and $15.8^3\times11.6\,\rm{fm}^4$ to simulate the QCD matter with a pion mass about 170\,MeV.
An exceeding of the $s^2$ over 1/3 was also verified as shown in the upper panel of FIG.\,\ref{fig_Abb23FIG}. In particular, $s^2$ exceeds 1/3 for $1.5\lesssim \mu_{\rm{I}}/m_\pi\lesssim14$ and takes the maximum value about 0.6 at $\mu_{\rm{I}}/m_\pi\approx2$. This result is consistent with that in Ref.\,\cite{Brandt23} but extends to a wider range of isospin densities (or isospin chemical potentials).
The isospin chemical potential was implemented using the grand canonical partition function in Ref.\,\cite{Brandt23} and so the systematic uncertainties are different from those of Ref.\,\cite{Abb23}.
Studies of Ref.\,\cite{Abb23} found that the asymptotic agreement wit the pQCD constraints requires very large values of the isospin chemical potential about $\mu_{\rm{I}}\gtrsim2\,\rm{GeV}\approx12m_{\pi}$.
A lower bound $\gtrsim-0.2$ on the trace anomaly was observed as shown in the lower panel of FIG.\,\ref{fig_Abb23FIG}. This finding is consistent with the result of our analysis based on the TOV equations that $\Delta$ is likely to be negative in cores of massive NSs.

Summarized in FIG.\,\ref{fig_s2-QCD} are the predictions on the SSS ($s^2$) from Refs.\,\cite{Abb23,Brandt23}, and Ref.\,\cite{Cotter13} using a two-color QCD matter (cyan band). We find from the cyan band that a peak emerges in $s^2$ at about $\mu/m_\pi\approx0.8$ with its value also exceeding the conformal bound, here $m_\pi\approx700\,\rm{MeV}$ was adopted and the plot is taken from Ref.\,\cite{Hipp24} using the data of Ref.\,\cite{Cotter13}.
Moreover, a prediction on $s^2$ (tan solid line) in an isospin-dependent QCD is given by using a quark-meson model\,\cite{Chiba24}. This model is renormalizable and eliminates high density artifacts in models with the ultraviolet cutoffs, here $m_\pi\approx140\,\rm{MeV}$ is used.

\begin{figure}[h!]
\centering
\includegraphics[height=6.cm]{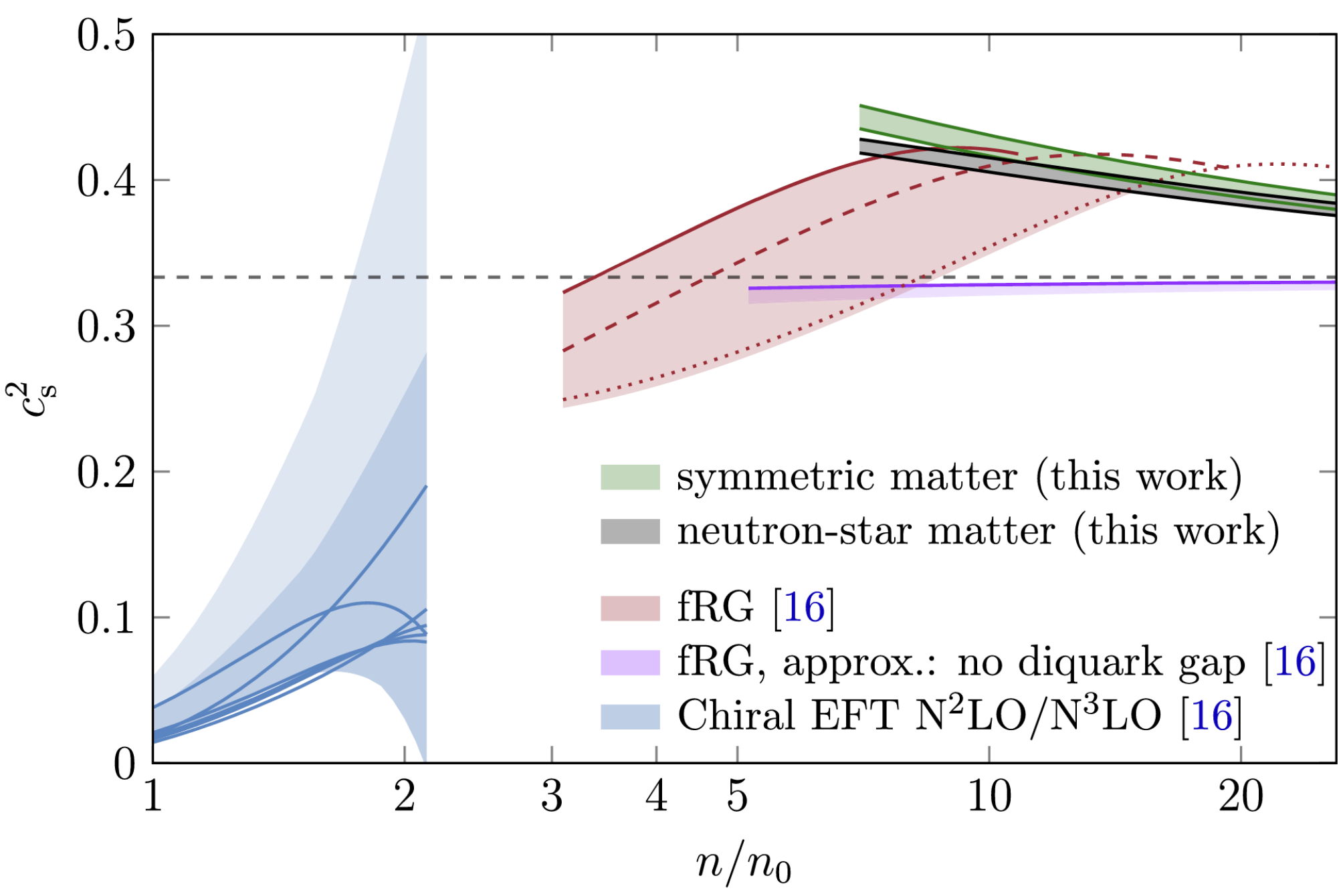}
\caption{(Color Online).  Indication of the emergence of a peak in $s^2$ at densities $\lesssim10\rho_0$ by combining the prediction from an effective field theory constrained by QCD renormalization-group (RG) flows (blue band) and that by a functional-RG (fRG) analysis (pink band)\,\cite{Leon20} together with the low-density CEFT.
Figures taken from Ref.\,\cite{Braun22}.
}\label{fig_FRGs2}
\end{figure}

\begin{figure}[h!]
\centering
\includegraphics[height=5.cm]{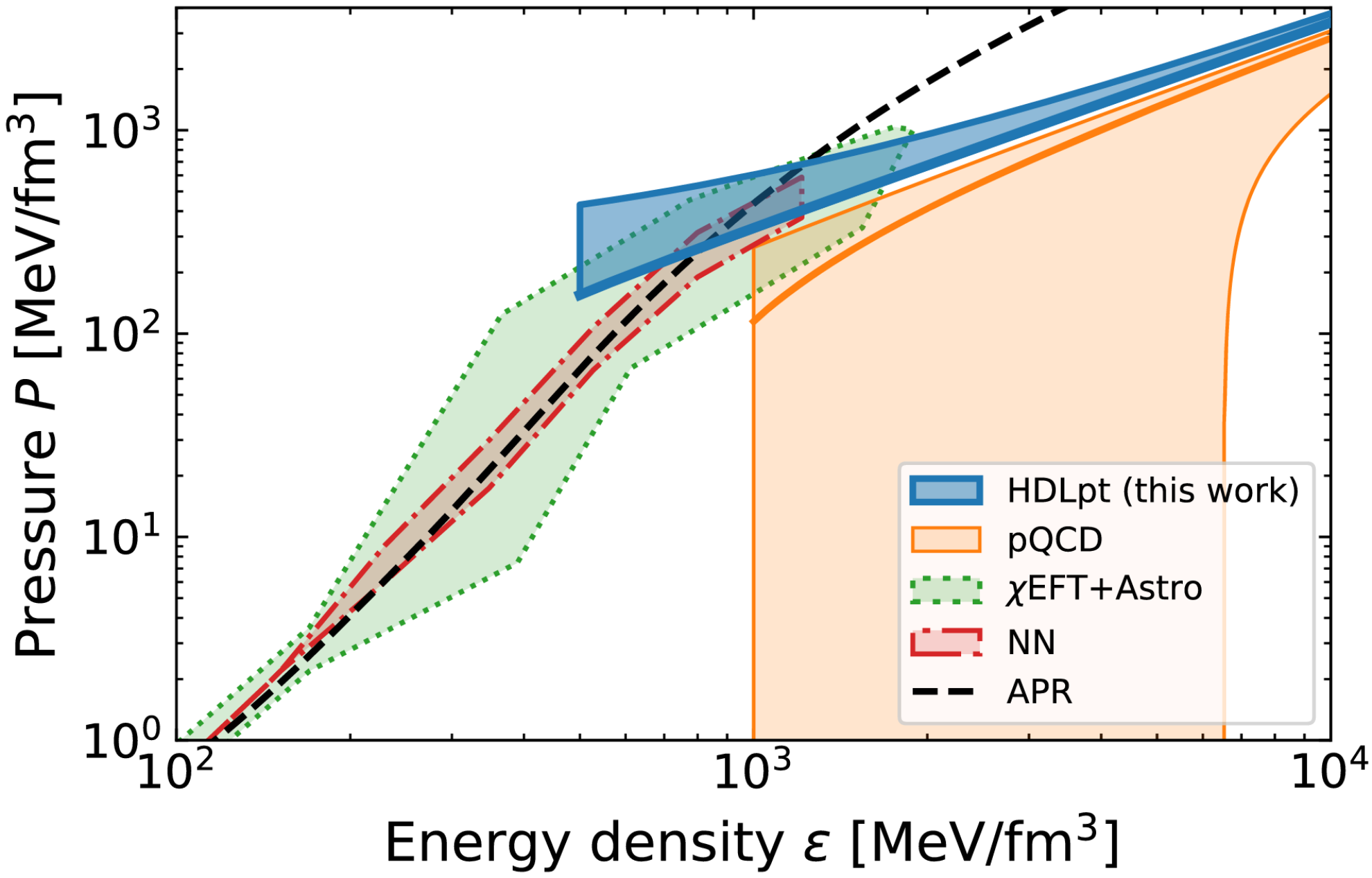}\qquad
\includegraphics[height=5.cm]{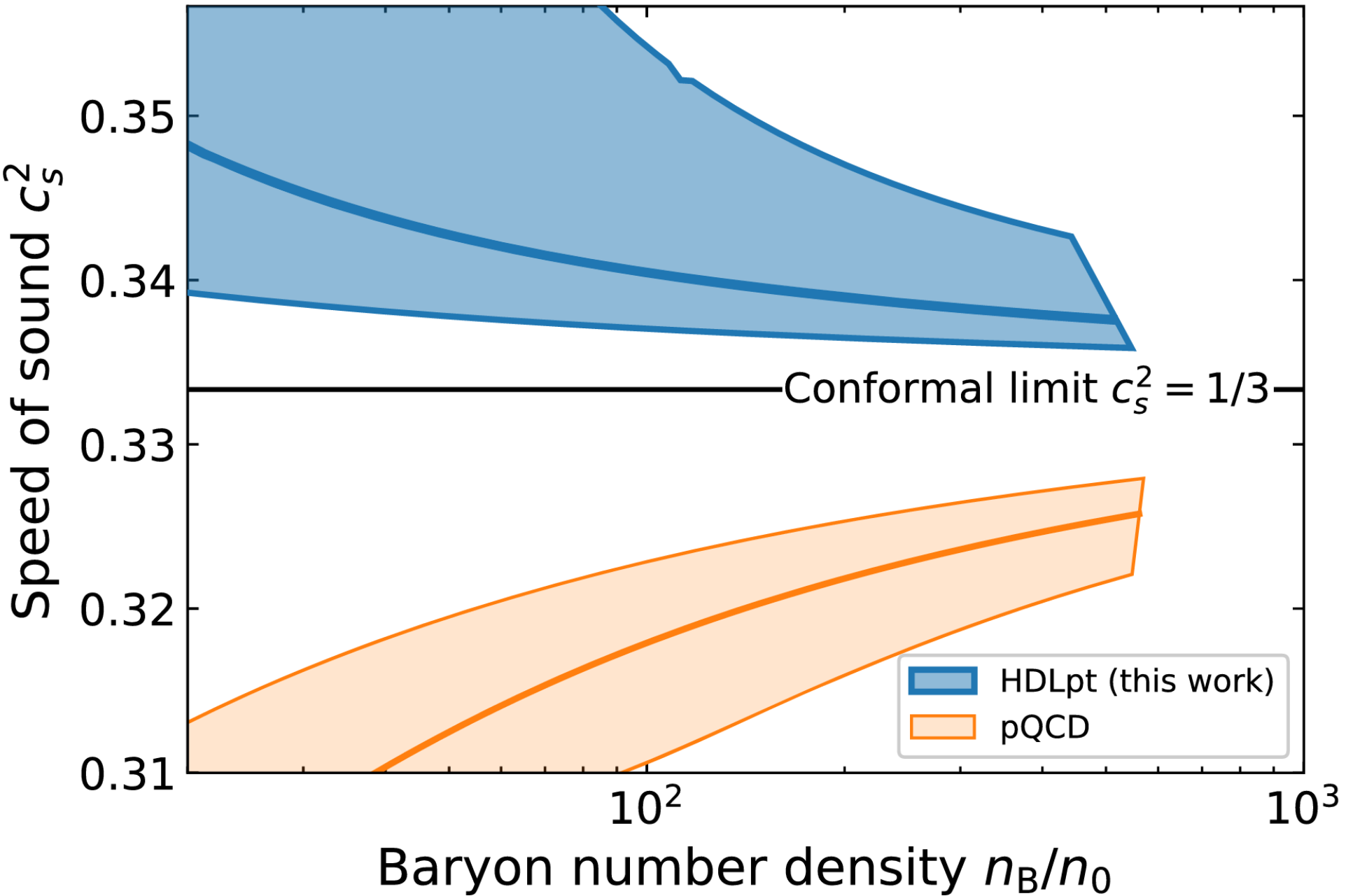}
\caption{(Color Online). The EOS and $s^2$ of cold and dense QCD matter obtained in a resummed perturbation theory based on hard thermal/dense loops. Figures taken from Ref.\,\cite{Fuji22HDL}.
}\label{fig_HDLs2}
\end{figure}

\begin{figure}[h!]
\centering
\includegraphics[height=5.cm]{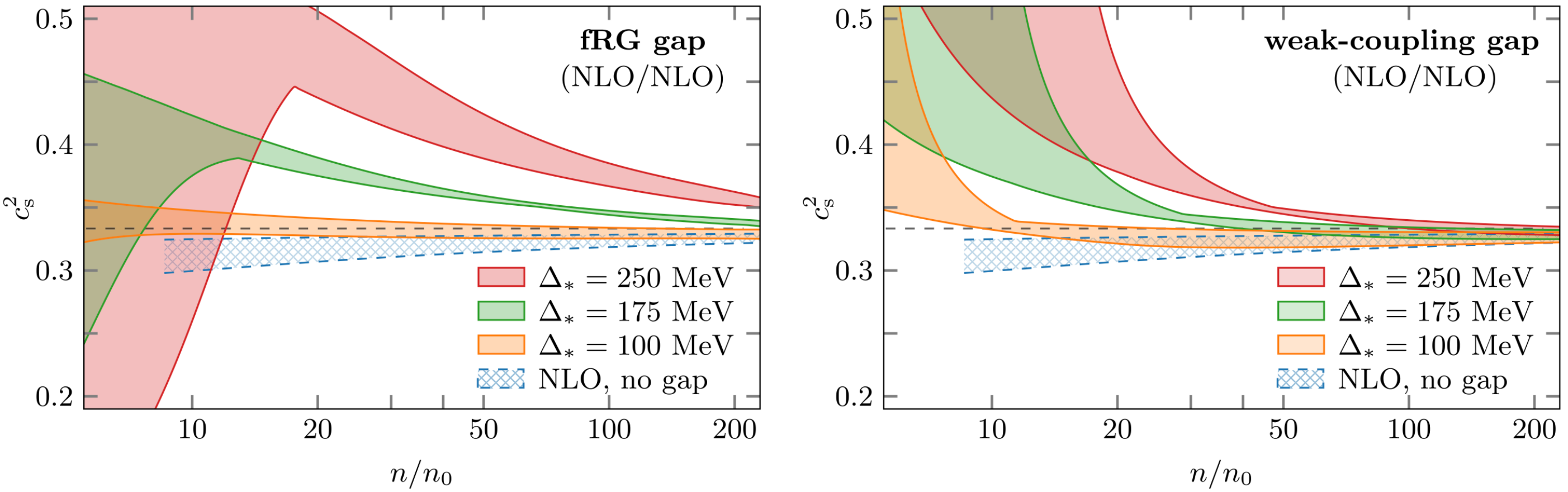}
\caption{(Color Online). The SSS obtained in a two-flavor color-superconducting quark matter at next-to-leading order, using different gap parameters. 
Figure taken from Ref.\,\cite{GGB24}.
}\label{fig_GGB24}
\end{figure}

\begin{figure}[h!]
\centering
\hspace*{0.4cm}
\includegraphics[height=5.5cm]{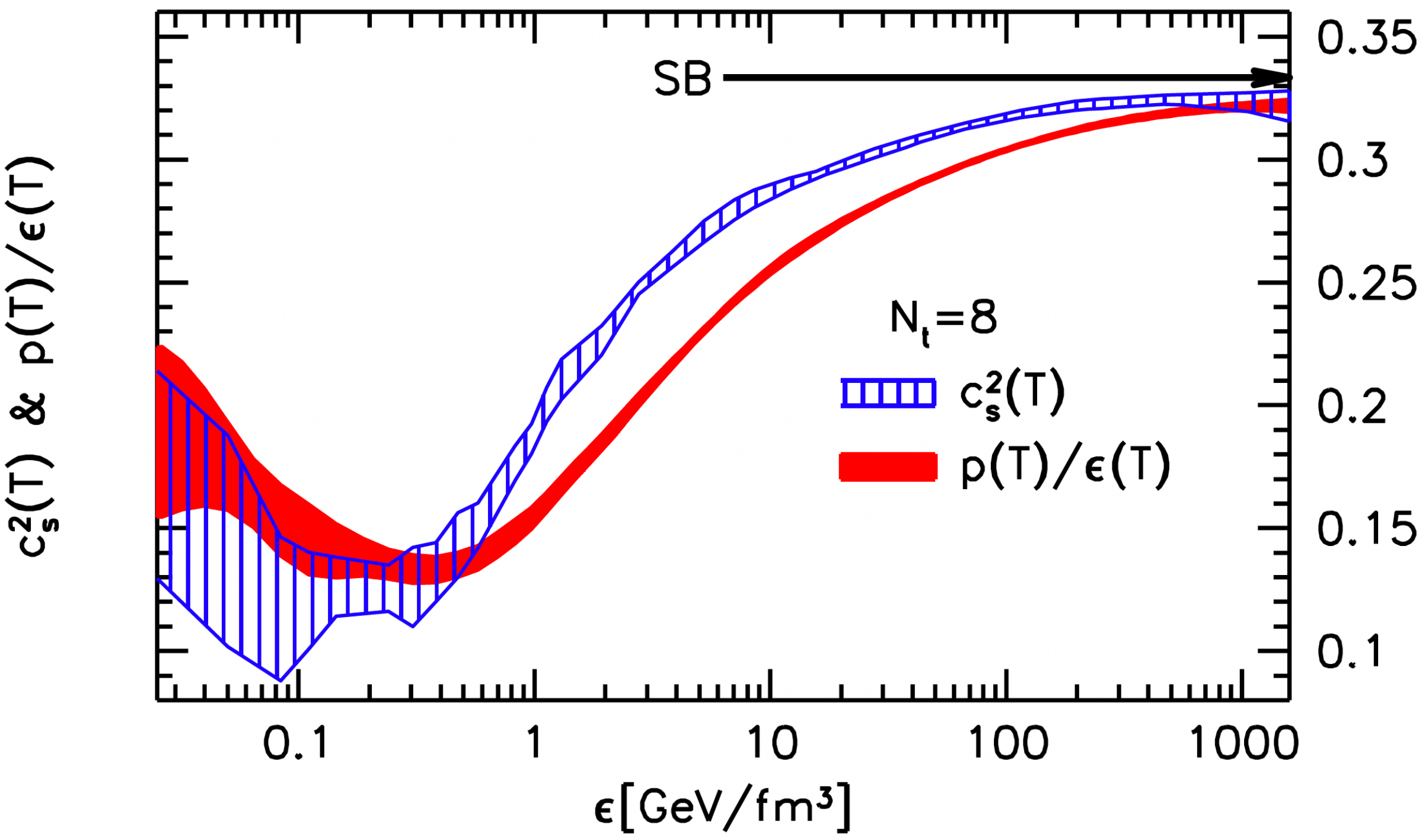}\\[0.5cm]
\includegraphics[height=5.5cm]{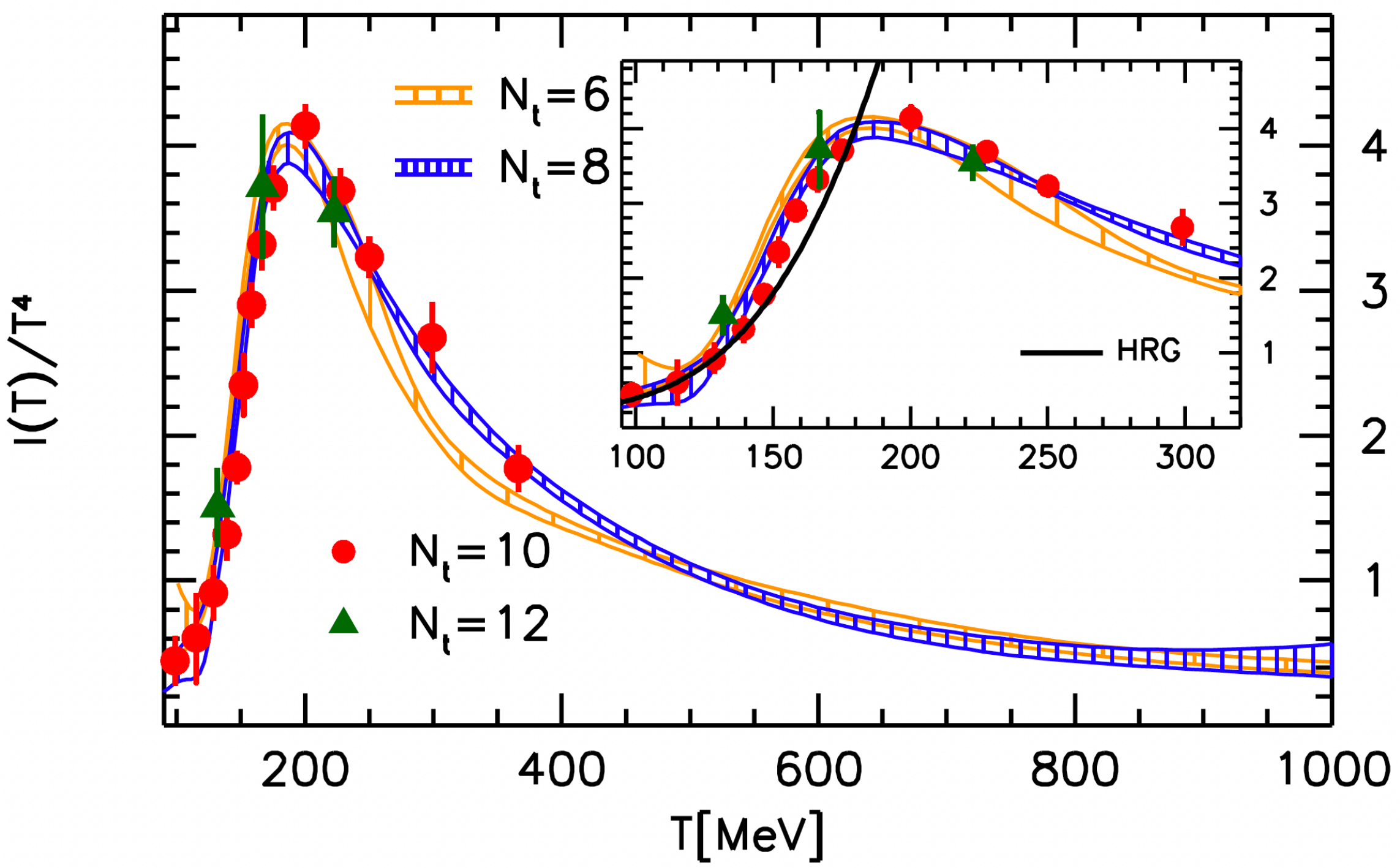}
\caption{(Color Online).  Speed of sound as well as $\phi=P/\varepsilon$ for QCD matter at finite temperature with $N_{\rm{f}}=2+1$,  here $I=\varepsilon-3P$ is the dimensional trace anomaly.
Figures taken from Ref.\,\cite{Bors10JHEP}.
}\label{fig_Bors10s2FIG}
\end{figure}

\begin{figure}[h!]
\centering
\includegraphics[height=6.cm]{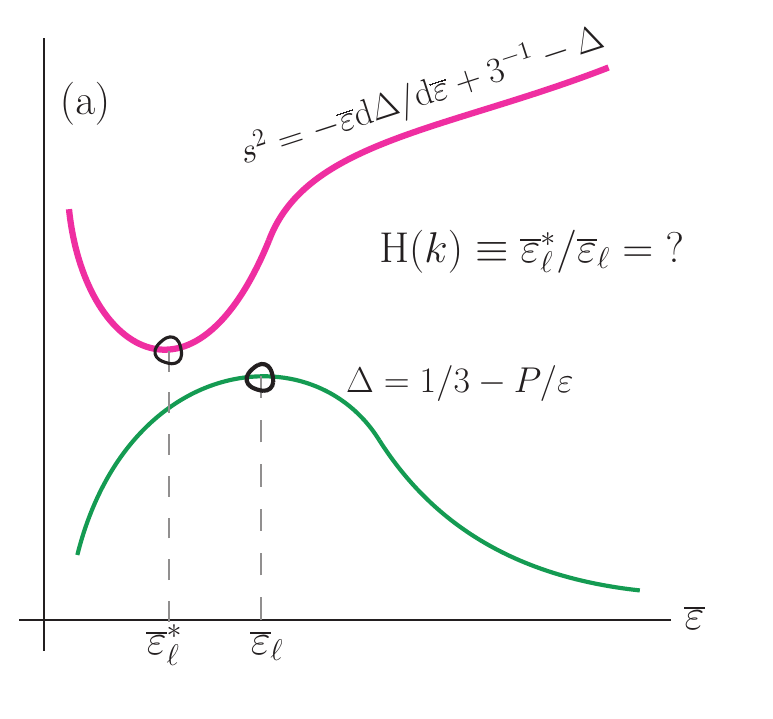}\quad
\includegraphics[height=6.cm]{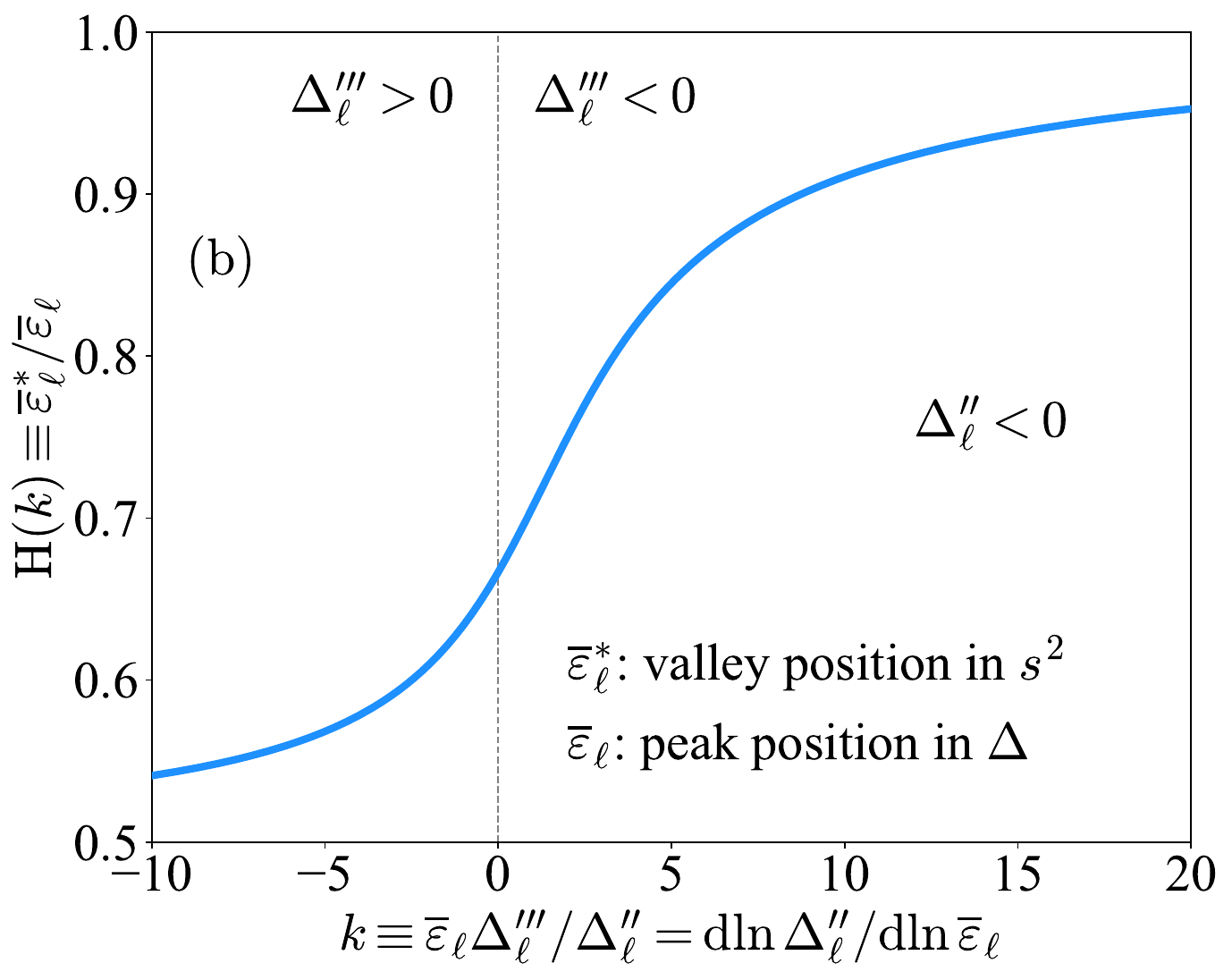}
\caption{(Color Online).  Left pane: sketch of the relation between $\ep_{\ell}^{\ast}$ (valley position in $s^2$) and $\ep_{\ell}$ (peak position in $\Delta$).
Right panel: function $\rm{H}(k)$ defined in Eq.\,(\ref{ratk}) by considering the trace anomaly $\Delta(\ep)\approx\Delta_{\ell}+2^{-1}\Delta_{\ell}''(\ep-\ep_{\ell})^2+6^{-1}\Delta_{\ell}'''(\ep-\ep_{\ell})^3$ where $\ep_{\ell}$ is the peak position in $\Delta$ so $\Delta_{\ell}''$ is negative.
}\label{fig_func_Wk}
\end{figure}

\begin{figure}[h!]
\centering
\includegraphics[height=5.5cm]{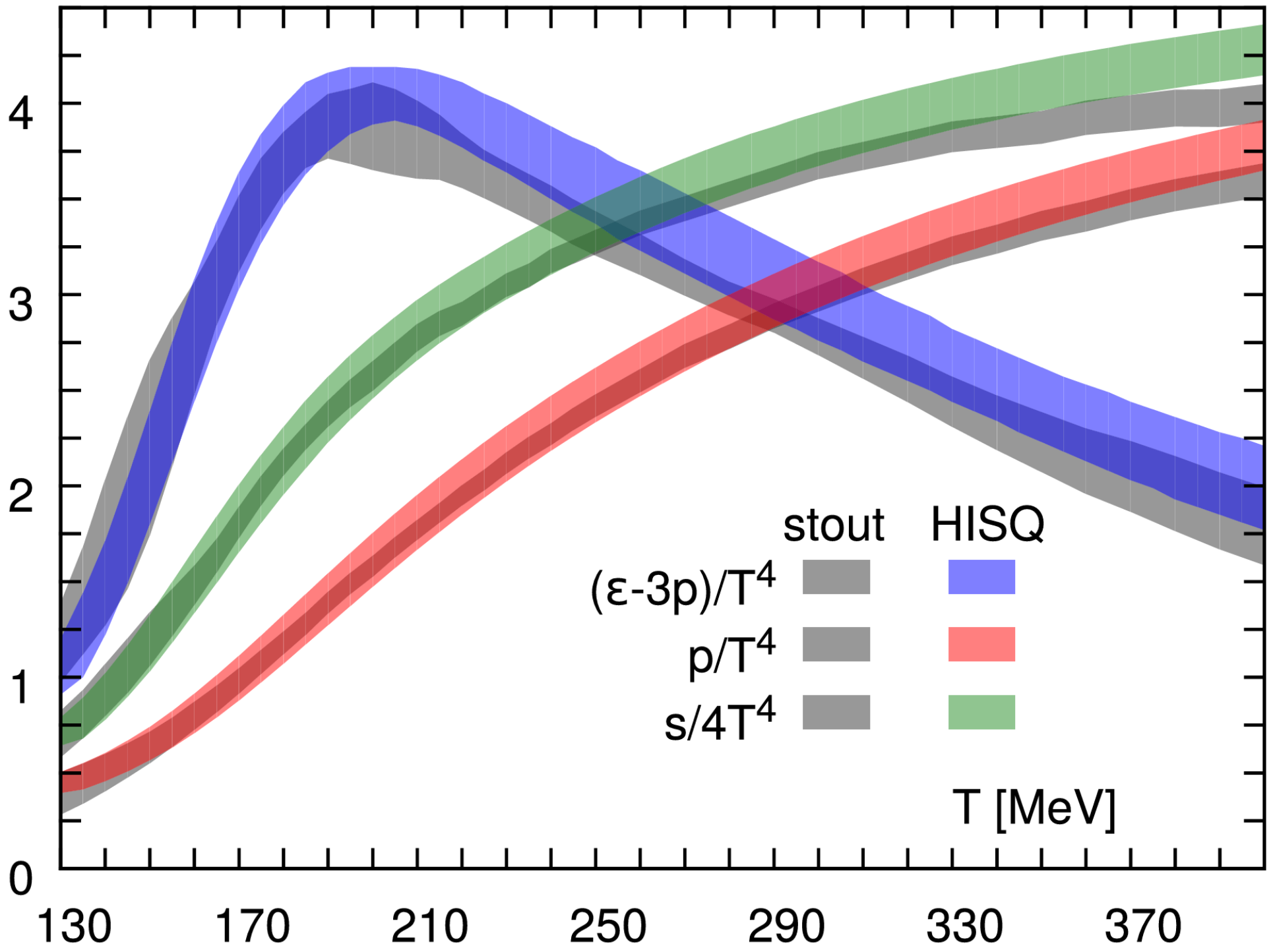}\qquad
\includegraphics[height=5.5cm]{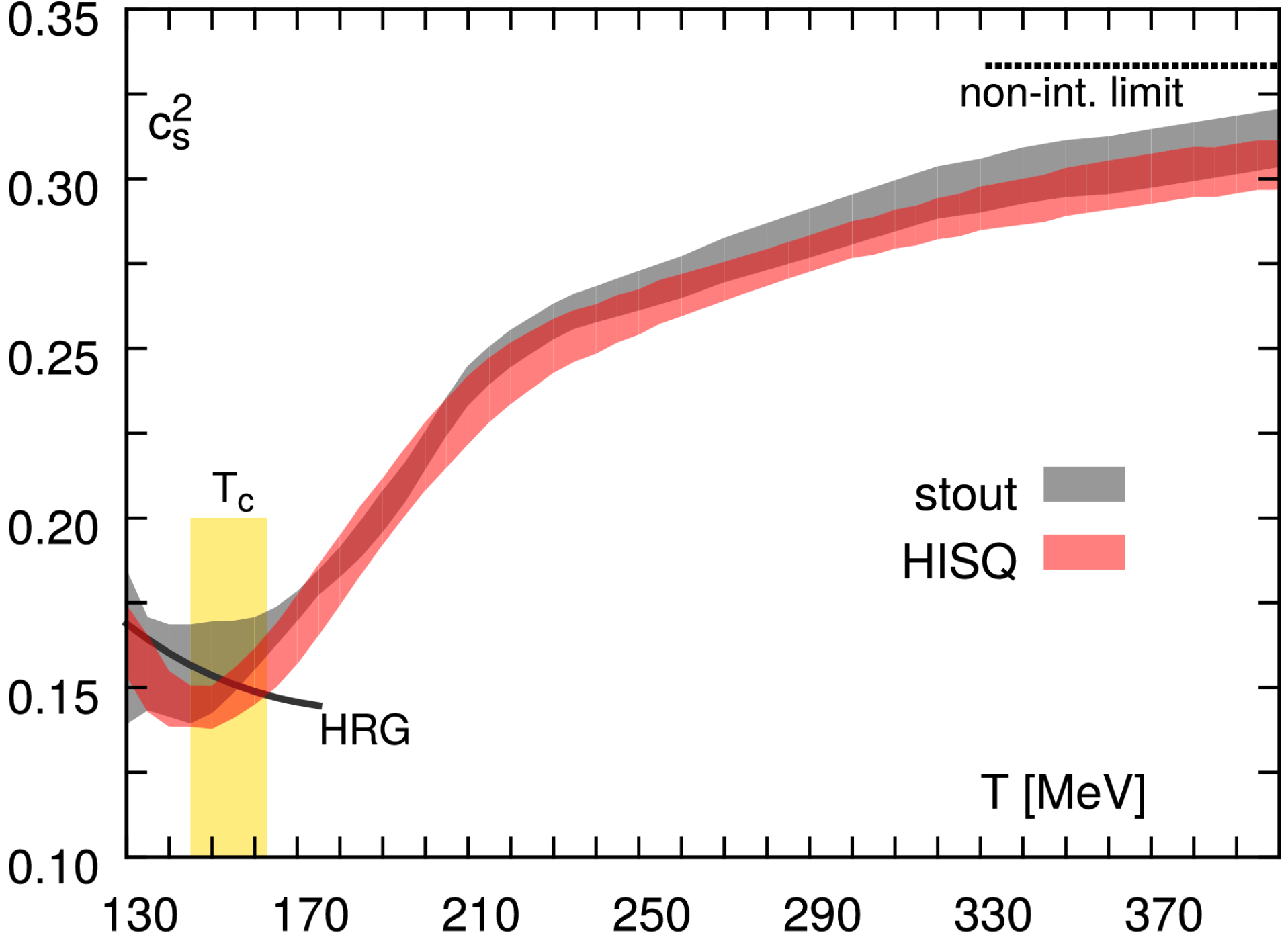}
\caption{(Color Online).  Similar as FIG.\,\ref{fig_Bors10s2FIG} using a highly improved staggered quark action. Figures taken from Ref.\,\cite{Bazavov14}.
}\label{fig_BZFIG}
\end{figure}

In Ref.\,\cite{Braun22}, a peak occurring at about $\rho/\rho_0\lesssim10$ in $s^2$ is implied by combining the prediction on the SSS profile from a functional renormalization group (fRG) analysis\,\cite{Leon20} combined with the Fierz-complete setting, the low-density CEFT and an effective field theory constructed by the QCD RG flows\,\cite{Braun22}. The predicted $s^2$ is shown in FIG.\,\ref{fig_FRGs2}.
Notice that such large densities $\approx10\rho_0$ is probably inaccessible even in very massive NSs. Also notice that in this investigation the pQCD constraints (being effective at densities $\approx40\rho_0$) are not included.
{\color{xll}The conclusions of Ref.\,\cite{Braun22} imply that QCD effects (either perturbatively at high densities or effectively by RG flows) may not be the origin of a peaked structure in $s^2$ in NSs, partially supporting our finding that the strong-field gravity in GR is responsible for it.}
Moreover, it also implies that the $s^2$ may approach the conformal bound from above (which is different from the pQCD prediction), this was verified recently by a resummed perturbation theory using hard thermal/dense loops\,\cite{Fuji22HDL}, as shown in FIG.\,\ref{fig_HDLs2}.
Other ingredients may also affect the pattern how $s^2$ approaches the conformal limit. For example, different gap parameters in a two-flavor color-superconducting quark matter (calculated at next-to-leading order) may effectively affect such pattern\,\cite{GGB24}, namely $s^2$ increases with the gap parameter $\Delta_{\ast}$ at a given density, as shown in FIG.\,\ref{fig_GGB24}, see also Refs.\,\cite{Fuji24PRDgap,Fukushima2024XXX}.

It is also interesting to note that the speed of sound and trace anomaly have been studied for QCD matter at finite temperature.
For example, the $s^2$, $\phi=P/\varepsilon$ and the dimensional trace anomaly $I=\varepsilon-3P$ are calculated on a lattice using $N_{\rm{f}}=2 + 1$ staggered flavors and one-link stout improvement\,\cite{Bors10JHEP}; the results are shown in FIG.\,\ref{fig_Bors10s2FIG}.
From the lower panel one can see that there exists a peak in $I/T^4$; this peak may correspondingly induce a peak in $\Delta$ as a function of energy density $\varepsilon$.

Using a similar method of Subsection \ref{sub_decomTA} (see FIG.\,\ref{fig_s2phys}), we can analyze and understand the influence of the above results (shown in FIG.\,\ref{fig_Bors10s2FIG}) on the possible valley structure of $s^2$. Denote the peak in $\Delta$ as $\Delta_{{\ell}}$, we have ($\Delta_{\ell}'\equiv \Delta'(\ep_{\ell})=0$)
\begin{equation}
\Delta(\ep)\approx\Delta_{{\ell}}+\frac{1}{2}\Delta_{{\ell}}''\left(\ep-\ep_{{\ell}}\right)^2,~~
\Delta_{{\ell}}\equiv \Delta(\ep_{\ell})>0,~~\Delta_{{\ell}}''\equiv \Delta''(\ep_\ell)<0,
\end{equation} since ``$\ell$'' is the local maximum point on $\Delta$, see the left panel of FIG.\,\ref{fig_func_Wk}.
Using the decomposition formula (\ref{for1}), we than write out the SSS as $s^2\approx-\Delta_{{\ell}}''\ep(\ep-\ep_{{\ell}})+3^{-1}-\Delta_{{\ell}}-2^{-1}\Delta_{{\ell}}''(\ep-\ep_{{\ell}})^2$; taking derivative of this $s^2$ with respective to $\ep$ and making it be zero gives the position of the local minimum:
\begin{equation}\label{dkfg}
\ep_{\ell}^{\ast}\approx{2\ep_{{\ell}}}/{3}\leftrightarrow\ep_{\ell}^{\ast}/\ep_{\ell}\approx2/3.
\end{equation}{\color{xll}This means that the valley in $s^2$ occurs on the left side of that in $\Delta$.} Moreover, the second-order derivative at $\ep_{\ell}^{\ast}$ is $[\d^2s^2/\d\ep^2]_{\ep_{\ell}^{\ast}}\approx-3\Delta_{{\ell}}''>0$, i.e., it is really a local minimum point. 

Ref.\,\cite{Bazavov14} found the valley (crossover region) in $s^2$ is at about $T\approx 154\,\rm{MeV}$ (right panel of FIG.\,\ref{fig_BZFIG}), corresponding to $\ep_{\ell}^{\ast}\approx2$; while that in $I$ is at about $T\approx190\,\rm{MeV}$ (lower panel of FIG.\,\ref{fig_Bors10s2FIG} or the left panel of FIG.\,\ref{fig_BZFIG}), corresponding to $\ep_{\ell}\approx9$.
Therefore $\ep_{\ell}^{\ast}/\ep_{\ell}\approx2/9$, which has sizable deviation from the estimate (\ref{dkfg}), indicating we need to add next-order terms to the trace anomaly $\Delta(\ep)$.
Using similar steps, we found that 
\begin{equation}
\boxed{
\rm{H}(k)\equiv\left(\frac{\ep_{\ell}^{\ast}}{\ep_{\ell}}\right)=\frac{3}{4}\left(1-\frac{1}{k}\right)+\frac{\sqrt{k^2-2k+9}}{4k},\label{ratk}}
\end{equation}
by including $6^{-1}\Delta_{\ell}'''(\ep-\ep_{\ell})^3$ in $\Delta(\ep)$;
the $k\equiv {\ep_{\ell}\Delta_{\ell}'''}/{\Delta_{\ell}''}$ in function $\rm{H}(k)$ could takes all real values. From the above relation, we obtain inversely that,
\begin{equation}
\boxed{
k\equiv \frac{\ep_{\ell}\Delta_{\ell}'''}{\Delta_{\ell}''}=\frac{\d\ln\Delta_{\ell}''}{\d\ln\ep_{\ell}}
=\left[2-3\left(\frac{\ep_{\ell}^{\ast}}{\ep_{\ell}}\right)\right]
\left[1-3\left(\frac{\ep_{\ell}^{\ast}}{\ep_{\ell}}\right)+2\left(\frac{\ep_{\ell}^{\ast}}{\ep_{\ell}}\right)^2\right]^{-1},}\end{equation}
here $\d\ln\Delta_{\ell}''/\d\ln\ep_{\ell}\equiv[\d\ln\Delta_{\ell}''(\ep)/\d\ln\ep]_{\ep=\ep_{\ell}}$.
The second-order derivative of $s^2$ at $\ep_{\ell}^{\ast}$ is
\begin{equation}
\boxed{
\left.\frac{\d^2s^2}{\d\ep^2}\right|_{\ep=\ep_{\ell}^{\ast}}
=-\Delta_{\ell}''\sqrt{k^2-2k+9}>0,~~\mbox{since}~~\Delta_{\ell}''<0,}
\end{equation}
i.e., it is a local minimum point in $s^2$.
For $k=0$, $[\d^2s^2/\d\ep^2]_{\ep_{\ell}^{\ast}}$ reduces to $-3\Delta_{\ell}''>0$ as given above.

The ratio $\ep_{\ell}^{\ast}/\ep_{\ell}$ of (\ref{ratk}) monotonically increases with $k$ and takes the minimum value of 1/2 when $k\to-\infty$, see the right panel of FIG.\,\ref{fig_func_Wk}. Therefore, in order to make $\ep_{\ell}^{\ast}/\ep_{\ell}$ be close to about $2/9$, a large negative $k$ is advantage. This means $k<0$ or $\Delta_{\ell}'''>0$ (a positive skewness of $\Delta$ at point ``$\ell$'') and in the mean while $k^{-1}$ is small. Consequently, we have
\begin{equation}
\ep_{\ell}^{\ast}/\ep_{\ell}\approx2^{-1}\left(1-k^{-1}-2k^{-2}-2k^{-3}+2k^{-4}+10k^{-5}\right);~~\mbox{for large and negative }k.
\end{equation}
We also give the expansions for $\ep_{\ell}^{\ast}/\ep_{\ell}$ under other two limits as:
\begin{align}
\ep_{\ell}^{\ast}/\ep_{\ell}
\approx&1 -
\frac{1}{k}+\frac{1}{k^2}+\frac{1}{k^3}-\frac{1}{k^4}-\frac{5}{k^5},~~\mbox{for large and positive }k;
\\
\ep_{\ell}^{\ast}/\ep_{\ell}\approx&
\frac{2}{3}\left(
1 +\frac{1}{18}k+\frac{1}{162}k^2-\frac{1}{1458}k^3-\frac{5}{13122}k^4-\frac{1}{39366}k^5\right),~~\mbox{for small }k\approx0.
\end{align}
Generally,  we have from (\ref{ratk}) that,
\begin{equation}\label{epast}
\boxed{
\frac{\ep_{\ell}^{\ast}\leftrightarrow\mbox{valley position in SSS $s^2$}}{\ep_{\ell}\leftrightarrow\mbox{peak position in trace anomaly $\Delta$}}:~~
\left\{
\begin{array}{ll}
1/2\leq \ep_{\ell}^{\ast}/\ep_{\ell}\leq2/3&~~~~\Delta_{\ell}'''\geq0;\\
2/3\leq\ep_{\ell}^{\ast}/\ep_{\ell}\leq 1 &~~~~\Delta_{\ell}'''\leq0,
\end{array}
\right.}
\end{equation}
see the right panel of FIG.\,\ref{fig_func_Wk} for the $k$-dependence of the function $\rm{H}(k)$.

The $s^2$ at $\ep_{\ell}^{\ast}$ is
\begin{equation}
\boxed{
s^2(\ep_{\ell}^{\ast})=\frac{1}{3}-\Delta_{\ell}-\frac{1}{96}\frac{\Delta_{\ell}''\ep_{\ell}^2}{k^2}
\left(3+k-\sqrt{k^2-2k+9}\right)^2\left(6-k+\sqrt{k^2-2k+9}\right),~~k= \frac{\ep_{\ell}\Delta_{\ell}'''}{\Delta_{\ell}''}.}
\end{equation}
Therefore for large and negative $k$, it reduces to
\begin{equation}
s^2(\ep_{\ell}^{\ast})\approx\frac{1}{3}-\Delta_{\ell}+\ep_{\ell}^2\Delta_{\ell}''\left(\frac{k}{12}-\frac{1}{8}\right)
=\frac{1}{3}-\Delta_{\ell}+\frac{\ep_{\ell}^3\Delta_{\ell}'''}{12}\left(1-\frac{3}{2k}\right)\approx
\frac{1}{3}-\Delta_{\ell}+\frac{\ep_{\ell}^3\Delta_{\ell}'''}{12},~~\mbox{large negative }k,
\end{equation}
so $s^2(\ep_{\ell}^{\ast})\geq0$ implies $\Delta_{\ell}\lesssim 3^{-1}+12^{-1}\ep_{\ell}^3\Delta_{\ell}'''$.
While for large and positive $k$, 
\begin{equation}
s^2(\ep_{\ell}^{\ast})\approx\frac{1}{3}-\Delta_{\ell}-\frac{1}{96}\frac{\Delta_{\ell}''\ep_{\ell}^2}{k^2}\left(80-\frac{96}{k}\right)\approx\frac{1}{3}-\Delta_{\ell}-\frac{5\Delta_{\ell}''}{6}\left(\frac{\Delta_{\ell}''}{\Delta_{\ell}'''}\right)^2,~~\mbox{large positive }k,
\end{equation}
so the non-negativeness of $s^2$ implies $\Delta_{\ell}\lesssim3^{-1}-5\Delta_{\ell}^{\prime\prime,3}/6\Delta_{\ell}^{\prime\prime\prime,2}$.
On other hand, for small $k$, we have
\begin{equation}
s^2(\ep_{\ell}^{\ast})\approx\frac{1}{3}-\Delta_{\ell}-\left(\frac{1}{6}-\frac{5k}{81}\right)\ep_{\ell}^2\Delta_{\ell}''\approx \frac{1}{3}-\Delta_{\ell}-\frac{\ep_{\ell}^2\Delta_{\ell}''}{6},~~k\approx0,
\end{equation}
and now  $s^2(\ep_{\ell}^{\ast})\geq0$ gives $\Delta_{\ell}\lesssim3^{-1}-6^{-1}\ep_{\ell}^2\Delta_{\ell}''$.

Similarly, if we include the fourth-order correction of $\Delta_{\ell}''''$ (from $24^{-1}\Delta_{\ell}''''(\ep-\ep_{\ell})^4$) and treat $\Delta_{\ell}''''$ as small, the result is given by,
\begin{equation}
\boxed{
\ep_{\ell}^{\ast}/
\ep_{\ell}\approx\rm{H}(k)\left(1+\frac{\ep_{\ell}^2\Delta_{\ell}''''}{6\Delta_{\ell}''}\frac{(2-5\rm{H}(k))(1-\rm{H}(k))^2}{\rm{H}(k)\sqrt{k^2-2k+9}}\right),}
\end{equation}
where the function $\rm{H}(k)$ is defined in (\ref{ratk}).
The overall factor involving $k$ in the bracket is negative,  so a negative $\Delta_{\ell}''''$ (since $\Delta_{\ell}''<0$) may help reduce the value of $\ep_{\ell}^{\ast}/\ep_{\ell}$. In fact, considering large negative $k$ as discussed in the above, we may obtain,
\begin{align}
\ep_{\ell}^{\ast}/
\ep_{\ell}\approx&\rm{H}(k)\left[1
+\frac{\ep_{\ell}^2\Delta_{\ell}''''}{24\Delta_{\ell}''}\frac{1}{k}\left(1-\frac{1}{k}-\frac{20}{k^2}+\cdots\right)\right]\notag\\
\approx&\frac{1}{2}\left[1-\frac{1}{k}\left(1-\frac{\ep_{\ell}^2\Delta_{\ell}''''}{24\Delta_{\ell}''}\right)
-\frac{2}{k^2}\left(1+\frac{\ep_{\ell}^2\Delta_{\ell}''''}{24\Delta_{\ell}''}\right)
-\frac{2}{k^3}\left(1+\frac{7\ep_{\ell}^2\Delta_{\ell}''''}{16\Delta_{\ell}''}\right)+\cdots\right]\notag\\
\approx&\frac{1}{2}\left[1-\frac{1}{k}-\frac{2}{k^2}-\frac{2}{k^3}+\frac{\ep_{\ell}^2\Delta_{\ell}''''}{24\Delta_{\ell}''}\frac{1}{k}\left(1-\frac{2}{k}-\frac{21}{k^2}\right)+\cdots\right]\notag\\
=&\frac{1}{2}\left[1-\frac{1}{k}-\frac{2}{k^2}-\frac{2}{k^3}+\frac{1}{24}\left(
\frac{\d\ln\Delta_{\ell}'''}{\d\ln\ep_{\ell}}\right)\left(1-\frac{2}{k}-\frac{21}{k^2}\right)+\cdots\right].
\end{align}
Similarly, for large and positive $k$, we have
\begin{align}
\ep_{\ell}^{\ast}/\ep_{\ell}
\approx&1-\frac{1}{k}+\frac{1}{k^2}+\frac{1}{k^3}
-\frac{\ep_{\ell}^2\Delta_{\ell}''''}{2\Delta_{\ell}''}\frac{1}{k^3}\left(1-\frac{8}{3k}-\frac{8}{3k^2}\right)
+\cdots\notag\\
\approx&1-\frac{1}{k}+\frac{1}{k^2}+\frac{1}{k^3}
-\frac{1}{2}\left(\frac{\d\ln\Delta_{\ell}'''}{\d\ln\ep_{\ell}}\right)
\left(\frac{\d\ln\Delta_{\ell}''}{\d\ln\ep_{\ell}}\right)^{-2}\left(1-\frac{8}{3k}-\frac{8}{3k^2}\right)+\cdots,
\end{align}
while for small $k$,
\begin{equation}
\ep_{\ell}^{\ast}/\ep_{\ell}
\approx\frac{2}{3}\left[1+\frac{1}{18}k+\frac{1}{162}k^2-\frac{1}{1458}k^3
-\frac{1}{81}\left(\frac{\d\ln\Delta_{\ell}'''}{\d\ln\ep_{\ell}}\right)
\left(\frac{\d\ln\Delta_{\ell}''}{\d\ln\ep_{\ell}}\right)\left(1+\frac{1}{36}k-\frac{2}{27}k^2-\frac{4}{243}k^3\right)\right].
\end{equation}

\begin{figure}[h!]
\centering
\includegraphics[width=11.5cm]{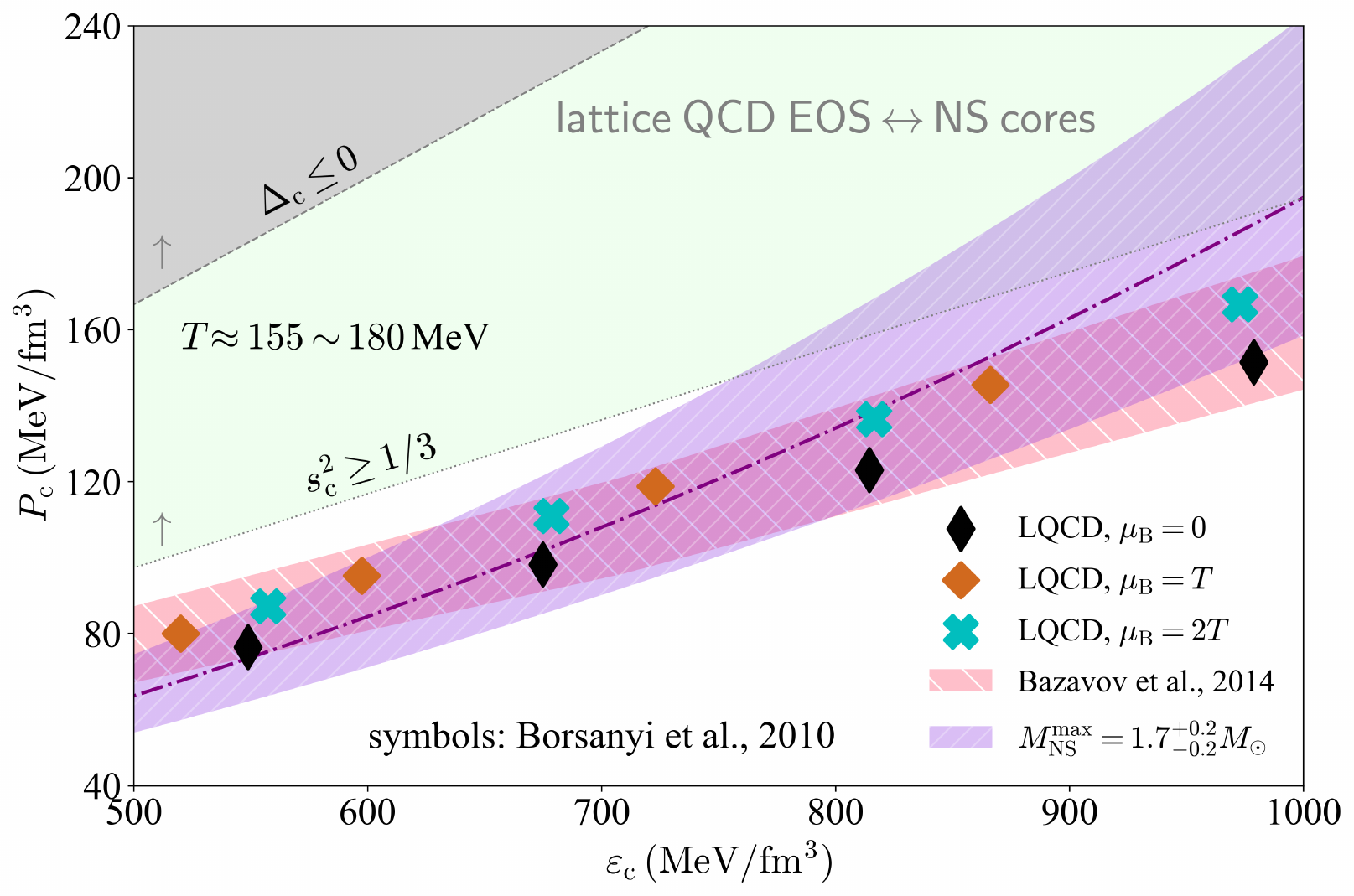}
\caption{(Color Online).  The EOS for finite-temperature QCD matter obtained in Ref.\,\cite{Bazavov14} (hatched pink band) and that from Ref.\,\cite{Bors21PRL} at three different baryon chemical potential $\mu_{\rm{B}}$ (shown by solid symbols).
The hatch plum band is the central EOS of NSs at TOV configuration with $M_{\rm{NS}}^{\max}\approx1.7_{-0.2}^{+0.2}M_{\odot}$.
}\label{fig_NS-QCD-EOS}
\end{figure}

\begin{figure}[h!]
\centering
\includegraphics[width=11.5cm]{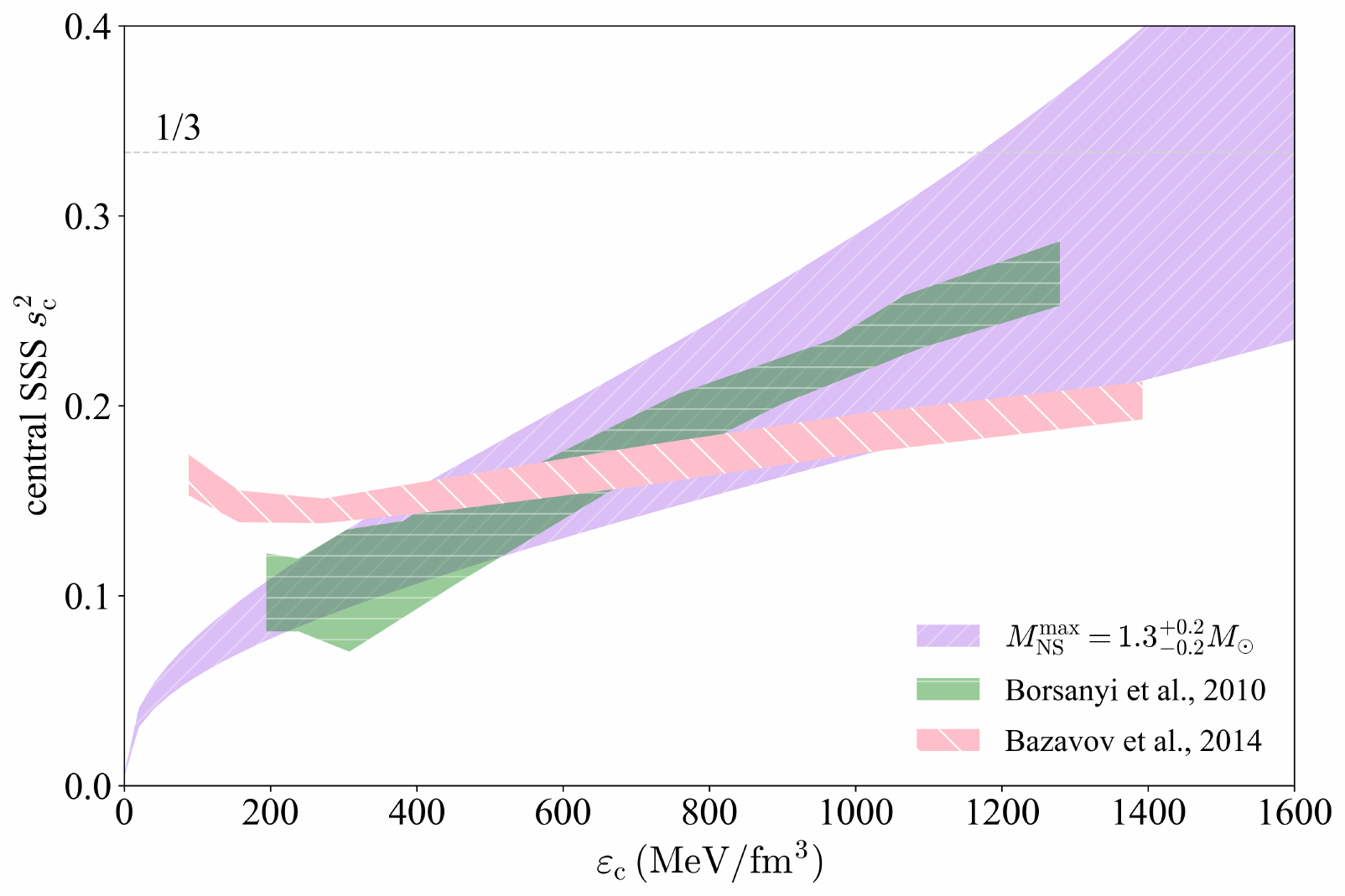}
\caption{(Color Online).  Comparison between the $s^2$ (as a function of energy density) obtained from the finite-temperature lattice QCD simulations\,\cite{Bors10JHEP,Bazavov14} and the SSS at NS centers (varying with the central energy density); a small NS mass about $1.3_{-0.2}^{+0.2}M_{\odot}$ is necessary to make the comparison be harmonic.
}\label{fig_s2_comparison_LQCD}
\end{figure}
\begin{figure}[h!]
\centering
\includegraphics[height=8.cm]{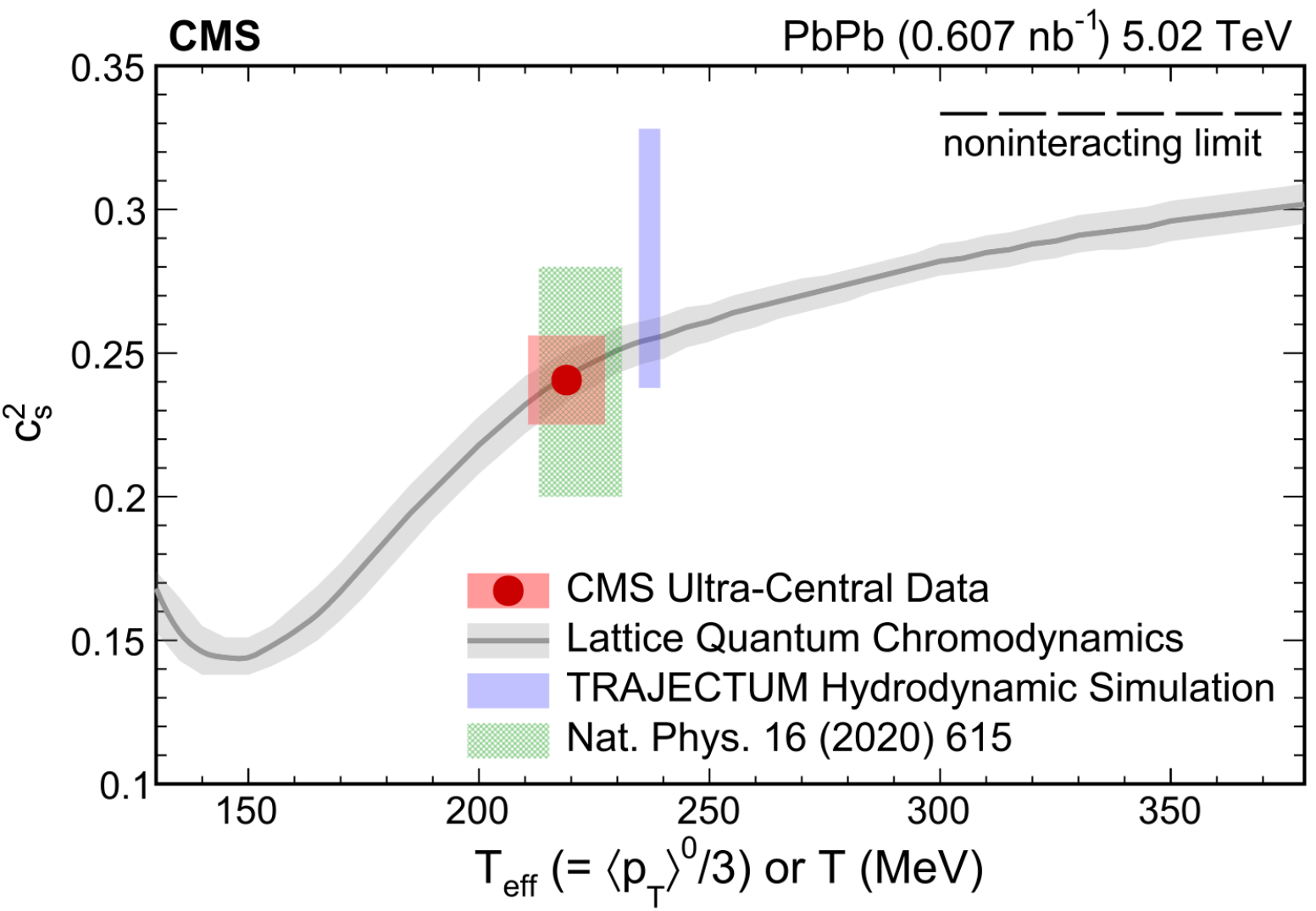}
\caption{(Color Online).  The speed of sound squared as a function of the effective
temperature $T_{\rm{eff}}$ with the CMS data point obtained from ultra-central PbPb collision data at $\sqrt{s_{\rm{NN}}} \approx 5.02\,\rm{TeV}$. Figure taken from Ref.\,\cite{CMS24}.
}\label{fig_CMS24s2}
\end{figure}
We show in FIG.\,\ref{fig_NS-QCD-EOS} the EOS for QCD matter at finite temperature obtained from Ref.\,\cite{Bazavov14} (hatched pink band and FIG.\,\ref{fig_BZFIG}) and that from Ref.\,\cite{Bors21PRL} at three different baryon chemical potential $\mu_{\rm{B}}$ (shown by solid symbols);
the central EOS of NSs at the TOV configuration with $M_{\rm{NS}}^{\max}\approx1.7_{-0.2}^{+0.2}M_{\odot}$ is also shown by the hatched plum band.
{\color{xll}The comparison in FIG.\,\ref{fig_NS-QCD-EOS} provides us an alternative on understanding why the conformal bound on $s^2=1/3$ is not violated in such lattice QCD simulations. This is because the analogous NS EOS (though not perfectly consistent with the lattice-simulated EOS) is not stiff enough to reach the boundary of $s_{\rm{c}}^2=1/3$ (indicated by the light-green band). It also implies that the dimensionless trace anomaly $\Delta$ is positive in the simulations as a $1.7M_{\odot}$ NS at the TOV configuration could even hardly reach the boundary of $\Delta_{\rm{c}}\leq0$ (shown by the light-grey band).}
We can similarly understand the realization of $s^2<1/3$ in these finite-temperature lattice QCD simulations\,\cite{Bors10JHEP,Bors21PRL,Bazavov14} by directly comparing the SSS (as a function of energy density) obtained with the one predicted for NSs; the result is shown in FIG.\,\ref{fig_s2_comparison_LQCD}. It is seen that {\color{xll}a relatively small NS mass about $1.3_{-0.2}^{+0.2}M_{\odot}$ at the TOV configuration is necessary/relevant to make the SSS's obtained from the two sides be harmonic with each other. This means that the EOS of QCD matter at finite temperature from these lattice simulations is similar as that in the core of a light NS; whose $\varepsilon_{\rm{c}}$ can hardly exceed $1\,\rm{GeV}/\rm{fm}^3$, thus $s^2<1/3$.}

Finally, it is necessary to mention that measuring the speed of sound in hot and dense matter created during ultra-relativistic heavy-ion collision has been an ongoing endeavor. Some interesting results have been obtained from several experiments. For example, using the CMS/LHC data obtained from ultra-central lead-lead collisions at $\sqrt{s_{\rm{NN}}} \approx 5.02\,\rm{TeV}$; the SSS in the hot and dense quark-gluon matter is found to be about $0.241\pm0.018$ with an effective medium temperature about $T_{\rm{eff}}\approx219\pm8\,\rm{MeV}$\,\cite{CMS24}, see FIG.\,\ref{fig_CMS24s2} for the temperature dependence of the $s^2$. 
These results are quite consistent with an earlier measurement which revealed that $s^2\approx0.24\pm0.04$ at $T_{\rm{eff}}\approx222\pm9\,\rm{MeV}$\,\cite{Gardim2020}.
We shall not discuss more on this issue as our primary motivation here is to note an important alternative approach (besides studying NSs) for understanding why both the conformal bound $s^2<1/3$ and positiveness of trace anomaly $\Delta>0$ are fulfilled in the lattice QCD simulations\,\cite{Bors10JHEP,Bors21PRL,Bazavov14}.

\setcounter{equation}{0}
\section{Gravitational Bound on Trace Anomaly and Related Issues}\label{SEC_56}

In this section, we study the gravitational bound on the trace anomaly $\Delta$ and some related issues.
In Subsection \ref{sub_Xgen} we briefly discuss the upper (lower) bound on the ratio $\phi=P/\varepsilon$ (thus the corresponding dimensionless trace anomaly $\Delta\equiv1/3-\phi$) in NS cores. This topic was recently reviewed by us in Ref.\,\cite{CL24-c}. For completeness, here we recall the main points. Subsection \ref{sub_conformal} is devoted to the question whether the core matter in NSs could be conformal based on some empirical characteristics. Then in Subsection \ref{sub_xiDelta}, we show that the compactness scaling straightforwardly probe the trace anomaly $\Delta$ in a model-independent manner. 
As a direct consequence, we then study in Subsection \ref{sub_NSdatapeak} whether the currently available NS data (mass and radius) could invariably generate a peaked SSS profile
using the trace anomaly decomposition.
Since $\x$ is closely related to the NS compactness $\xi$ (as shown in Eq.\,(\ref{gk-comp})), the upper bound on $\x$ may naturally induce a relevant bound on $\xi$ and consequently impact the causality boundary.
A causality boundary for stable NSs could basically be described as
\begin{equation}\label{cauB-des}
\boxed{
\mbox{causality boundary: }
R/\rm{km}\gtrsim\mbox{some minimum value for a given }M_{\rm{NS}}. }
\end{equation}
The simplest (though unrealistic) boundary is the Schwarzshild relation $R_{\rm{Sch}}=2M$ derived from the BH event.
Consequently,  it leads to the boundary $R_{\rm{Sch}}/\rm{km}\gtrsim2.95M_{\rm{NS}}/M_{\odot}$.
However, such boundary has little relevance for our research purpose here since BHs are not NSs.
More efficient and tight causality boundaries exist, based on simple stellar models or using empirical considerations.
We briefly discuss the existing causality boundaries for NSs in Subsection \ref{sub_cau_BF}, and then give an upper bound for the compactness in Subsection \ref{sub_UpperCompt} using our novel scalings, see Eq.\,(\ref{upp-xi}).
Then in Subsection \ref{sub_10km}, we give a new causality boundary using our mass and radius scalings obtained in SECTION \ref{SEC_4} for NSs at the TOV configuration. We study in Subsection \ref{sub_ultimate} the ultimate energy density and pressure allowed in realistic NSs.
Finally, we discuss the implications of the conjecture of a positive $\Delta$ on the peaked structure of $s^2$ in Subsection \ref{sub_conjecture}.

\subsection{Lower bound on trace anomaly due to strong-field gravity in GR}\label{sub_Xgen}

We have shown earlier with Eq.(\ref{Xupper}) and Eq.\,(\ref{sc2-TOV}) that the $\x=P_{\rm{c}}/\varepsilon_{\rm{c}}$ is upper bounded as $\x\lesssim0.374$. This result is obtained under the specific condition that it gives the upper limit for $\phi=P/\varepsilon$ at the centers of NSs at the TOV configuration.
By generalizing this special situation, we can show that\,\cite{CLZ23-b,CL24-c}
\begin{equation}\label{Xupper-GEN}
\boxed{
\mbox{near/at the centers of generally stable NSs along the M-R curve:}~~
\phi=P/\varepsilon=\widehat{P}/\widehat{\varepsilon}\leq \x\lesssim0.374.}
\end{equation}
An upper bound on $\phi$ sizably deviating from 1 implies that the EOS of core NS matter is significantly nonlinear. As we discussed earlier, this is because $s^2\leq1$ due to causality is equivalent to $\phi\equiv P/\varepsilon\leq1$ only for a linear EOS of the form $P=\zeta\varepsilon$ with $\zeta$ being some constant.
It is well known that the strong-field gravity in GR is fundamentally nonlinear, the EOS of NS matter especially in its core is thus also nonlinear. Therefore, the causality condition $s^2\leq1$ is expected to be appreciably different from $\phi\leq1$, and it may also effectively render the upper bound for $\phi$ to be smaller than 1.

The EOS of nuclear matter may be strongly nonlinear depending on both the internal interactions and the external environment/constraint of the system. This means that $\phi\leq1$ is necessary but not sufficient to ensure supra-dense matter in all NSs always stay casual. For example, the EOS of noninteracting degenerate Fermions (e.g., electrons) can be written in the polytropic form $P=K\varepsilon^{\beta}$\,\citep{Shapiro1983} where $\beta=5/3$ for non-relativistic and $\beta=4/3$ for extremely relativistic electrons; consequently $\phi\leq\beta^{-1}<1$.
Similarly, a long time ago Zel'dovich considered the EOS of an isolated ultra-dense system of baryons interacting through a vector field\,\citep{Zeldovich61}. In this case, $P=\varepsilon\sim\rho^2$, here $\rho$ is the baryon number density; therefore $P/\varepsilon\leq1$.
The EOS of dense nuclear matter where nucleon interactions are dominated by the $\omega$-meson exchange in the Walecka model\,\citep{Walecka1974} is an example of this type, namely\,\cite{CaiLi2022PRC-RK}
\begin{equation}
    P^{(\omega)}\approx\varepsilon^{(\omega)}\approx\frac{1}{2}g_{\omega}^2\omega^2\approx\frac{1}{2}\left(\frac{g_{\omega}}{m_{\omega}}\right)^2\rho^2,
\end{equation}
here $m_{\omega}$ is the mass of $\omega$-meson and $g_{\omega}$ is the $\omega$-nucleon coupling constant.
More generally, however, going beyond the vector field, the baryon density dependence of either $P(\rho)$ or $\varepsilon(\rho)$ could be very complicated and nontrivial; and the resulting EOS $P(\varepsilon)$ could also be significantly nonlinear.

The upper bound for $\phi$ is a fundamental quantity encapsulating essentially the strong-field properties of gravity in GR. Its accurate determination can help improve our understanding about the nature of gravity\,\cite{Hoyle:2003dw}.
An upper bound on $\phi$ substantially different from 1 then vividly characterizes how GR affects the superdense matter existing in NSs.
Indeed, there are fundamental physics issues regarding both the strong-field gravity and supra-dense matter as well as their couplings. For example, what is gravity? This has been one of the most fundamental questions unanswered yet. Massive NSs provide a natural laboratory to test possible answers to this question. It is well known that a gravity-matter duality exists in theories describing NS properties, see, e.g., Refs.\,\cite{Psaltis:2008bb,Shao:2019gjj} for reviews. The duality reflects a deep connection between microscopic physics of elementary particles in supra-dense matter and the nature of super-powerful gravity in and around massive NSs.
Developing a unified theory for the matter-gravity system in compact objects has been a long standing goal of cosmology. They both have to be fully understood to finally unravel mysteries associated with the most compact visible objects in the Universe.
Interestingly, some NS observational properties have been shown to have the potential to help break the gravity-matter duality, see, e.g., Refs.\,\cite{DeDeo:2003ju,Wen:2009av,Lin:2013rea,He:2014yqa,Yang:2019rxn}. 

\begin{figure}[h!]
\centering
\includegraphics[width=10.5cm]{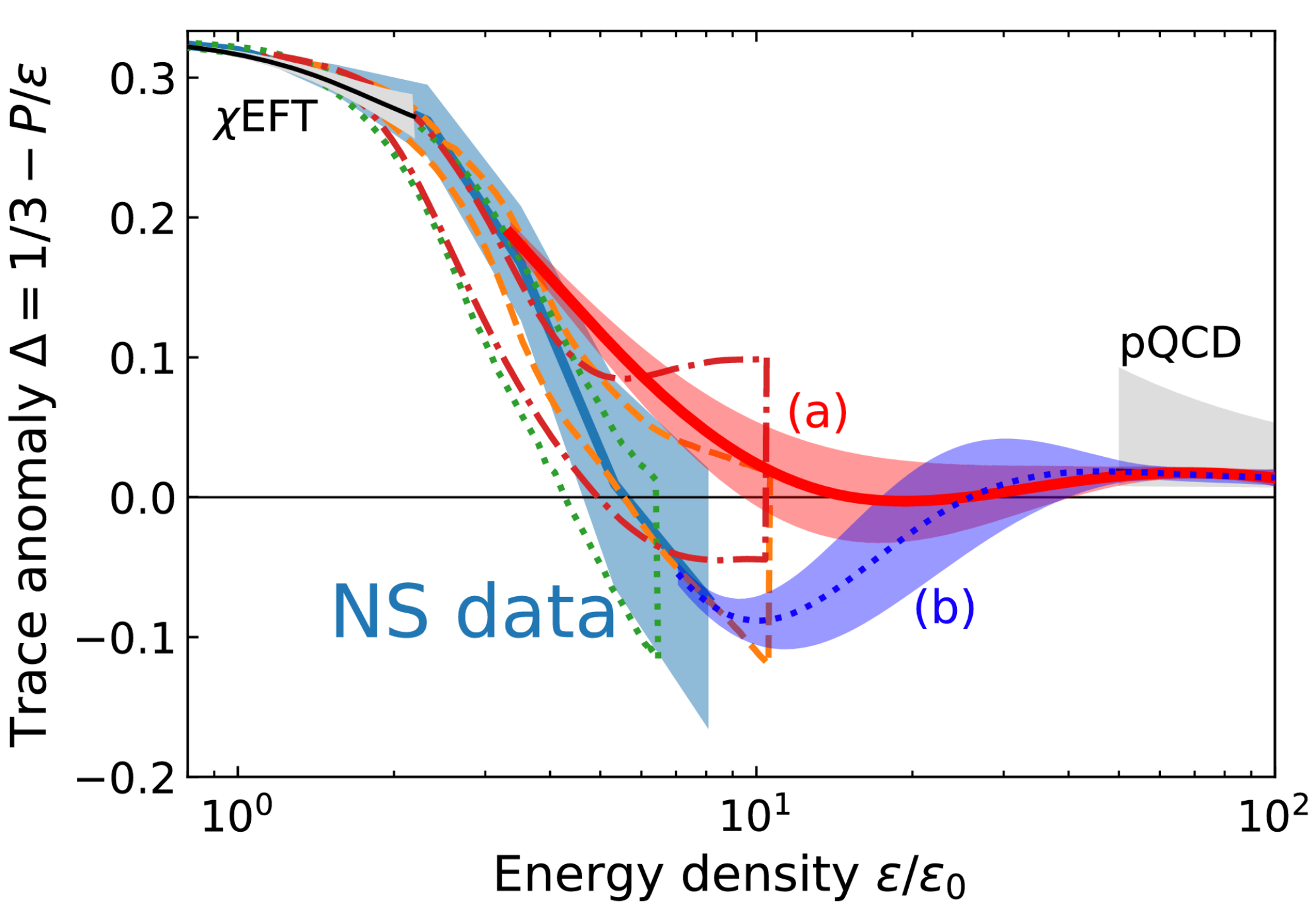}
\caption{(Color Online).  Trace anomaly $\Delta$ as a function of energy density $\varepsilon/\varepsilon_0$, here the $\Delta$ in NSs tends to be negative although the pQCD prediction on it approaches zero, $\varepsilon_0\approx150\,\rm{MeV}/\rm{fm}^3$ is the energy density at nuclear saturation density.
Figure taken from Ref.\,\cite{Fuji22}.
}\label{fig_Fuji22Delta}
\end{figure}

Back to the upper bound (\ref{Xupper-GEN}), we want to emphasize that its validity is limited to small $\widehat{r}$ due to the perturbative nature of expanding $\widehat{P}(\widehat{r})$ and $\widehat{\varepsilon}(\widehat{r})$. Whether $\phi=P/\varepsilon$ could exceed such upper limit at even larger distances away from NS centers depends on the joint analysis of $s^2$ and $P/\varepsilon$,  e.g., by including more higher-order contributions of the expansions\,\cite{CLZ23-b}. The upper bound $P/\varepsilon\lesssim0.374$ (at least near the NS centers) is an intrinsic property of the TOV equations at the leading order of their expansions. At the next-to-leading order, 
the upper bound of $P/\varepsilon$ was found to increase by about 2\% to $0.381$\,\cite{CL24-b}. The corresponding formalism, however, becomes too complicated for our analytical analyses to remain transparent and tractable at this time. In this review, we thus stay within the leading-order expansions. The upper bound of $P/\varepsilon$
embodies the strong-field aspects of gravity in GR, especially the strong self-gravitating nature\,\cite{CL24-c}.
In this sense, there is no guarantee {\it a prior} that this bound is consistent with all microscopic nuclear EOSs (either relativistic or non-relativistic). This is mainly because the latter were conventionally constructed without considering the limit set by the strong-field gravity. The robustness of such gravitational upper bound for $\phi=P/\varepsilon$ can be tested only by observable astrophysical quantities/processes under strong-field gravity, 
such as NS M-R data, NS-NS mergers and/or NS-BH mergers\,\citep{BS2010,Shibata2015,Baiotti2017,Kyutoku2021}. As mentioned earlier, in the NS matter-gravity inseparable system, it is the total action that determines the matter state and NS structure. Thus, to our best knowledge\,\cite{CL24-c}, there is no physics requirement that the EOS of superdense matter created in vacuum from high-energy heavy-ion collisions or other terrestrial laboratory experiments where effects of gravity can be neglected to be the same as that in NSs as nuclear matter in the two situations are in very different environments. 
Furthermore, we point out that (\ref{Xupper-GEN}) is general from analyzing perturbatively analytical solutions of the scaled TOV equations without using any specific nuclear EOS. 

Because the TOV equations are results of a hydrodynamical equilibrium of NS matter in the environment of a strong-field gravity from extremizing the total action of the matter-gravity system, features revealed from analyzing the gravitational aspects of the TOV equations must be matched by the nuclear EOS. This requirement can then put strong constraints on the latter. In particular, the upper bound for $\phi$ as $\phi\lesssim0.374$ of Eq.\,(\ref{Xupper-GEN}) may enable us to limit the density dependence of nuclear EOS relevant for NS modelings, see, e.g. Ref.\,\cite{CL24-c} for an illustrative example. Of course, logically for the same reasons mentioned above, the EOS extracted from NSs can only be compared to that from terrestrial nuclear experiments at the negligible-gravity limit. Thus, in our opinion, the nuclear EOS is environment-dependent and maybe nothing is wrong if the nuclear EOSs from studying massive NSs are very different from those extracted from high-energy heavy-ion collisions experiments on earth. Moreover, the difference may be invaluable for understanding the nature of both strong-field gravity and supra-dense matter.

\begin{figure}[ht!]
\centering
\includegraphics[width=14.cm]{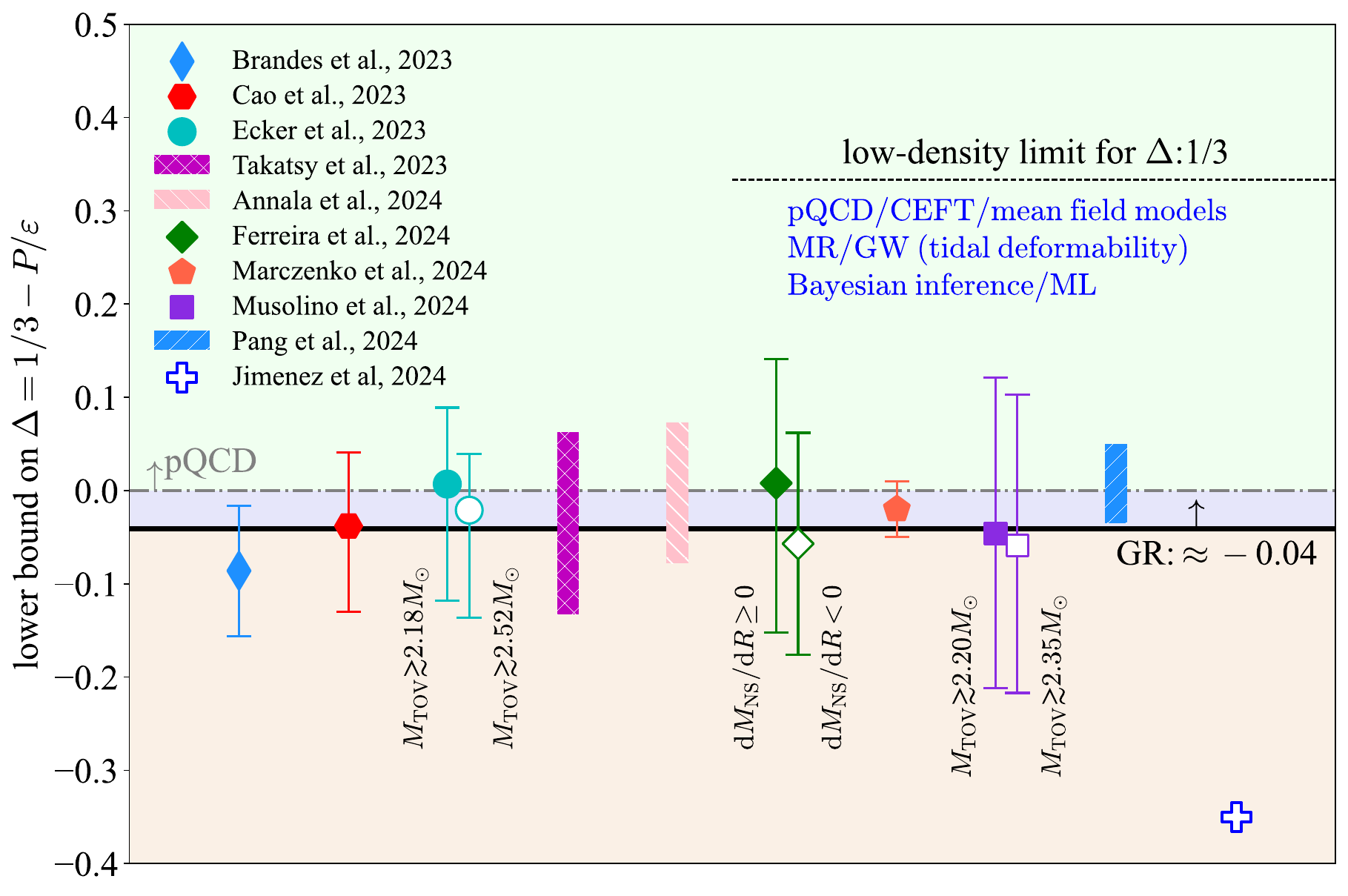}
\caption{(Color Online). Summary of current constraints on the lower bound of trace anomaly $\Delta$ in NSs from different analyses with respect to the pQCD (dot-dashed line) and GR (black solid line) predictions. See Ref.\,\cite{CL24-c} for details. Figure taken from Ref.\,\cite{CL24-c}.
}\label{fig_TAcomp}
\end{figure}

The upper bound on $\phi$ of Eq.\,(\ref{Xupper-GEN}) induces a lower bound on the dimensionless trace anomaly\,\cite{CL24-c}
\begin{equation}\label{GRDelta}
\boxed{
\Delta\geq \Delta_{\rm{GR}}\approx-0.04.}
\end{equation}
It is very interesting to notice that such GR bound on $\Delta$ is very close to the one predicted by perturbative QCD (pQCD) at extremely high densities owning to the realization of approximate conformal symmetry of quark matter\,\citep{Bjorken83,Fuji22}, as shown in FIG.\,\ref{fig_Fuji22Delta} using certain NS modelings. A possible negative $\Delta$ in NSs was first pointed out in Ref.\,\cite{Fuji22}, since then several studies on this issue have been made. We have summarized the constraints on $\Delta$ recently in Ref.\,\,\cite{CL24-c} and the result is shown again in FIG.\,\ref{fig_TAcomp}\,\cite{CL24-c}. We refer the interested readers to Ref.\,\cite{CL24-c} for more details. Finally, we emphasize that the trace anomaly is very useful as its energy density decomposition gives the SSS. We have discussed this issue in Subsection \ref{sub_decomTA} and will investigate in Subsection \ref{sub_NSdatapeak} whether the current available NS data (mass and radius) can invariably generate a peaked $s^2$ density or radius profile.

\subsection{Can the supradense matter at NS centers be (nearly) conformal?}\label{sub_conformal}

As we discussed in previous sections that a linear EOS in the form of $\hP=\zeta\heps$ with $\zeta=\rm{const.}$ is inconsistent with the TOV equations.
In this subsection, we explore further the conformal properties of the NS matter in their cores. A basic quantity to characterize the conformality is the polytropic index defined in Eq.\,(\ref{def_gamma}).
The condition of conformality in NSs is expressed as
\begin{equation}\label{conf_cond_1}
\boxed{
\Delta\to1/3 \mbox{~~and~~} s^2\to1/3,~~\mbox{or equivalently }\gamma\to1.}
\end{equation}

For example,  Ref.\,\cite{Ann23} claimed that for a $2M_{\odot}$ NS the polytropic index $\gamma$ approximately equals to 2 while that for a NS at the TOV configuration is very close to 1. The latter means that NSs at the TOV configuration contain conformal dense matter in their cores.
However, the $\gamma$ parameter in NS centers, actually, can not be unity since $\Delta_{\rm{c}}$ (or equivalently $\x$) and $s_{\rm{c}}^2$ could not approach zero (or 1/3) and 1/3 simultaneously, as shown by our Eq.\,(\ref{sc2-TOV}).
Physically, due to the nonlinear nature of the central EOS in NSs, it is likely that $\gamma\neq1$ (actually $\gamma=1$ if and only if $P\propto\varepsilon$).
In particular,  it is straightforward to find that $4/3\leq\gamma_{\rm{c}}\lesssim2.67$ using our formula (\ref{sc2-TOV}). For example, considering ${\x}\approx0.24_{-0.07}^{+0.05}$ for PSR J0740+6620\,\cite{CLZ23-a} leads to $\gamma_{\rm{c}}\approx1.86_{-0.21}^{+0.22}$.
Moreover, for $\Delta_{\rm{c}}=0$ the $\gamma_{\rm{c}}$ index is $7/3\approx2.33$.
These values are summarized in TAB.\,\ref{sstab} for comparisons.

\begin{table}[h!]
\renewcommand{\arraystretch}{1.5}
\centerline{\normalsize
\begin{tabular}{c|c|c|c|c|c} 
  \hline
Quantity&Range&$\x\approx0.18$&$0.24$&$1/3$&$0.374$\\\hline\hline
       $\Delta_{\rm{c}}=1/3-{\x}$ &$-0.041\lesssim\Delta_{\rm{c}}\leq1/3$&$0.15$&$0.09$&$0$&$-0.041$\\\hline
       $s_{\rm{c}}^2=\d{P}_{\rm{c}}/\d\varepsilon_{\rm{c}}$&$0\leq s_{\rm{c}}^2\leq1$&$0.31$&$0.45$&$7/9$&$1$\\\hline
       $\gamma_{\rm{c}}=s_{\rm{c}}^2/{\x}$&$4/3\leq\gamma_{\rm{c}}\lesssim2.67$&$1.68$&1.86&$7/3$&$2.67$\\\hline
            $-t_{\rm{c}}=s_{\rm{c}}^2-{\x}$&$0.63\lesssim t_{\rm{c}}\leq0$&$0.13$&$0.21$&$4/9$&$0.63$\\\hline
       $\Theta_{\rm{c}}=(\Delta_{\rm{c}}^2+t_{\rm{c}}^2)^{1/2}$&$0.19\lesssim\Theta_{\rm{c}}\lesssim0.63$&$0.19$&0.22&$4/9$&$0.63$\\\hline
       \end{tabular}}
        \caption{Ranges of quantities relevant for the trace anomaly, the values for them at four reference ${\x}$ are shown (last four columns), $t_{\rm{c}}=\d\Delta_{\rm{c}}/\d\ln\varepsilon_{\rm{c}}={\x}-s_{\rm{c}}^2$ is the logarithmic
derivative of $\Delta_{\rm{c}}$. Table taken from Ref.\,\cite{CLZ23-b} with slight modifications.}\label{sstab} 
\end{table}

\begin{figure}[h!]
\centering
\includegraphics[width=11.cm]{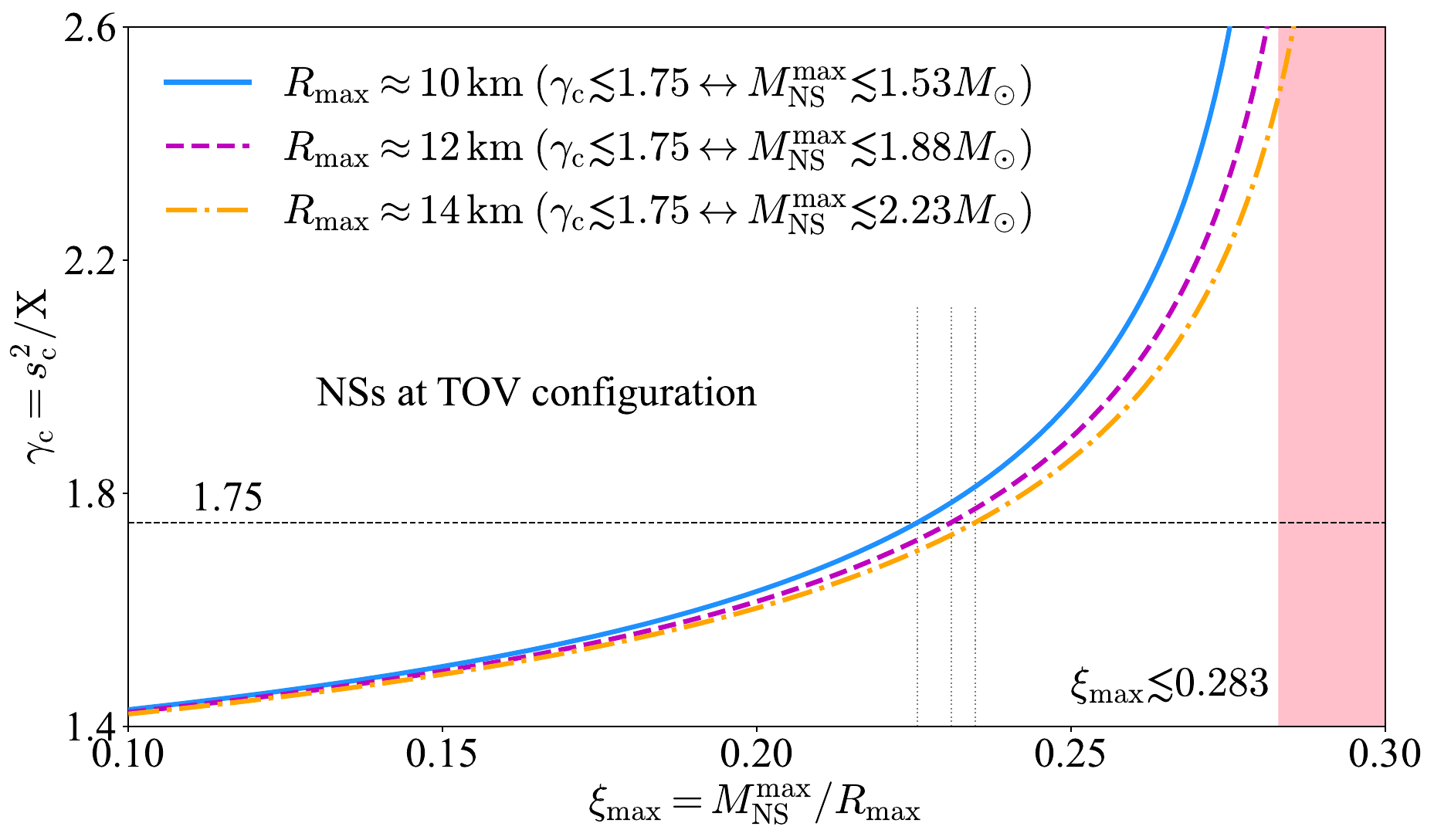}
\caption{(Color Online). Relation between $\gamma_{\rm{c}}$ and $\xi_{\max}$ connected by Eq.\,(\ref{gamma-xi}); the three vertical gray lines mark the critical compactness for $\gamma_{\rm{c}}\approx1.75$, giving the corresponding maximum NS mass (under a given radius).
}\label{fig_gamma-xi-rela}
\end{figure}

We can also establish the relation between $\gamma_{\rm{c}}$ and the NS compactness $\xi_{\max}$ for NSs at TOV configuration, the result is
\begin{equation}\label{gamma-xi}
\boxed{
\gamma_{\rm{c}}
=\frac{2}{3}\frac{(6\Lambda-1)\sqrt{1-8\Lambda+4\Lambda^2}+1-10\Lambda+6\Lambda^2}{(4\Lambda-1)\sqrt{1-8\Lambda+4\Lambda^2}+1-8\Lambda+4\Lambda^2}
\approx
\frac{4}{3}\left(1+\Lambda+\frac{11}{2}\Lambda^2+34\Lambda^3\right)+\cdots,}
\end{equation}
where,
\begin{equation}
\Lambda\approx\frac{105}{173}
\frac{\xi_{\max}/\Sigma+0.106/(R_{\max}/\rm{km})}{1-0.64/(R_{\max}/\rm{km})}\lesssim\frac{105}{173}\frac{\xi_{\max}}{\Sigma},
\end{equation}
with $\Sigma=M_{\odot}/\rm{km}\approx1.477$.
The relation (\ref{gamma-xi}) connects the polytropic index directly with the NS compactness.
For example, we have $\gamma_{\rm{c}}\approx1.88$ for $\xi_{\max}\approx0.25$ and $R_{\max}\approx12.5\,\rm{km}$; and similarly in order that $\gamma_{\rm{c}}\lesssim1.75$ one needs $\xi_{\max}\lesssim0.231$ ($\xi_{\max}\lesssim0.233$) or equivalently $M_{\rm{NS}}^{\max}\lesssim1.88M_{\odot}$ ($M_{\rm{NS}}^{\max}\lesssim2.05M_{\odot}$) using $R_{\max}\approx12\,\rm{km}$ ($R_{\max}\approx13\,\rm{km}$).
See FIG.\,\ref{fig_gamma-xi-rela} for the relation between $\gamma_{\rm{c}}$ and $\xi_{\max}$.
Including the first-order correction in (\ref{gamma-xi}), we have $\gamma_{\rm{c}}\approx4/3+0.55\xi_{\max}$, which is close to $4/3+89M_{\rm{NS}}/105R$ obtained under the assumption that both $\varepsilon$ and $\rho$ are constants in the interior of the stars\,\cite{Shapiro1983-aa}. Notice the $\gamma$ defined in Ref.\,\cite{Shapiro1983-aa} is $(s^2/\phi)(1+\phi)$ so it is slightly larger than $s^2/\phi$.
For NSs not at the TOV configuration, we shall use Eq.\,(\ref{sc2-GG}), and then
\begin{equation}
\gamma_{\rm{c}}=1+\frac{1+\Psi}{3}\frac{1+3\x^2+4\x}{1-3\x^2}
\geq\frac{4+\Psi}{3}\geq4/3.\end{equation}
{\color{xll}This analysis is model-independent, reflecting the intrinsic property encapsulated in the TOV equations. Therefore, the TOV equations alone imply that the $\gamma$ could not be 1 in NS centers.}
In particular, we have $\gamma_{\rm{c}}\approx1+(1+\Psi)/3$ for small $\x$ (Newtonian limit). It is consistent with the relation between the polytropic index $n$ (appearing in $P\sim\varepsilon^{1+1/n}$) and $\Psi$ we have established in the relation (\ref{n-Psi}).

\begin{figure}[h!]
\centering
\includegraphics[height=6.cm]{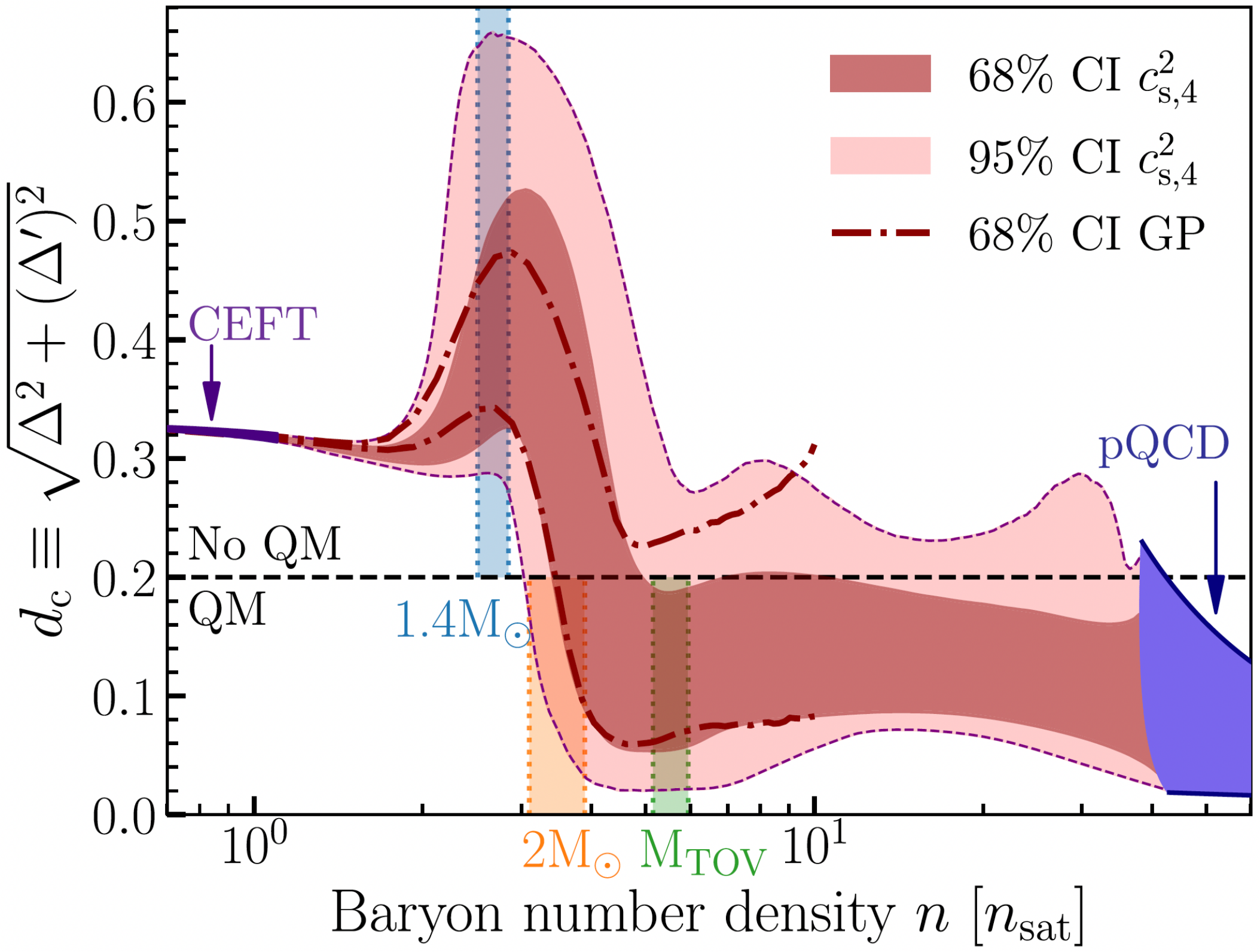}\qquad
\includegraphics[height=6.4cm]{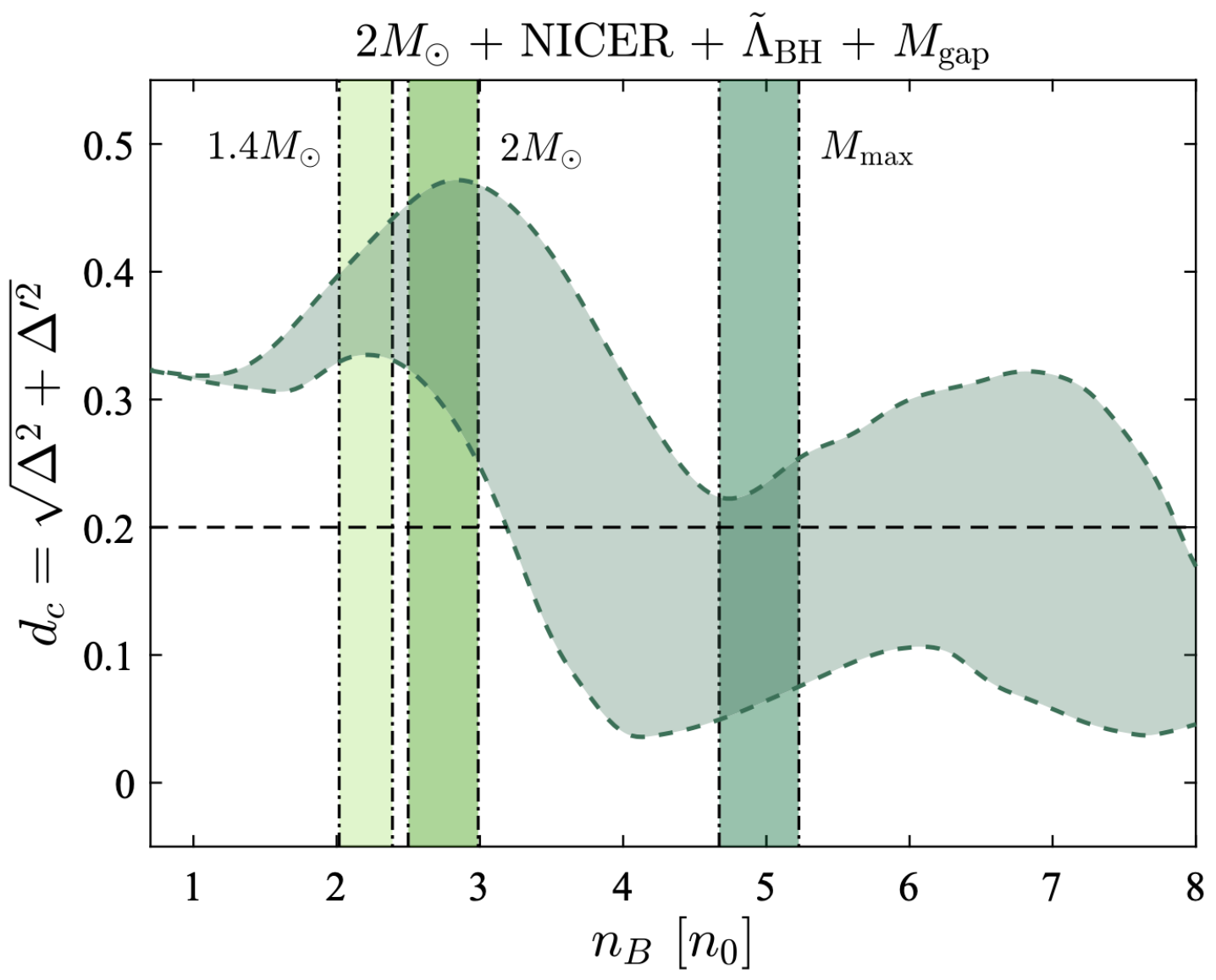}
\caption{(Color Online).  The measure $\Theta\equiv d_{\rm{c}}$ as a function of baryon density, here $\Delta'=s^2(1/\gamma-1)=\phi-s^2$. Figures taken from Ref.\,\cite{Ann23} and Ref.\,\cite{Tak23}, respectively.
}\label{fig_Ann24d}
\end{figure}

\begin{figure}[h!]
\centering
\includegraphics[height=7.cm]{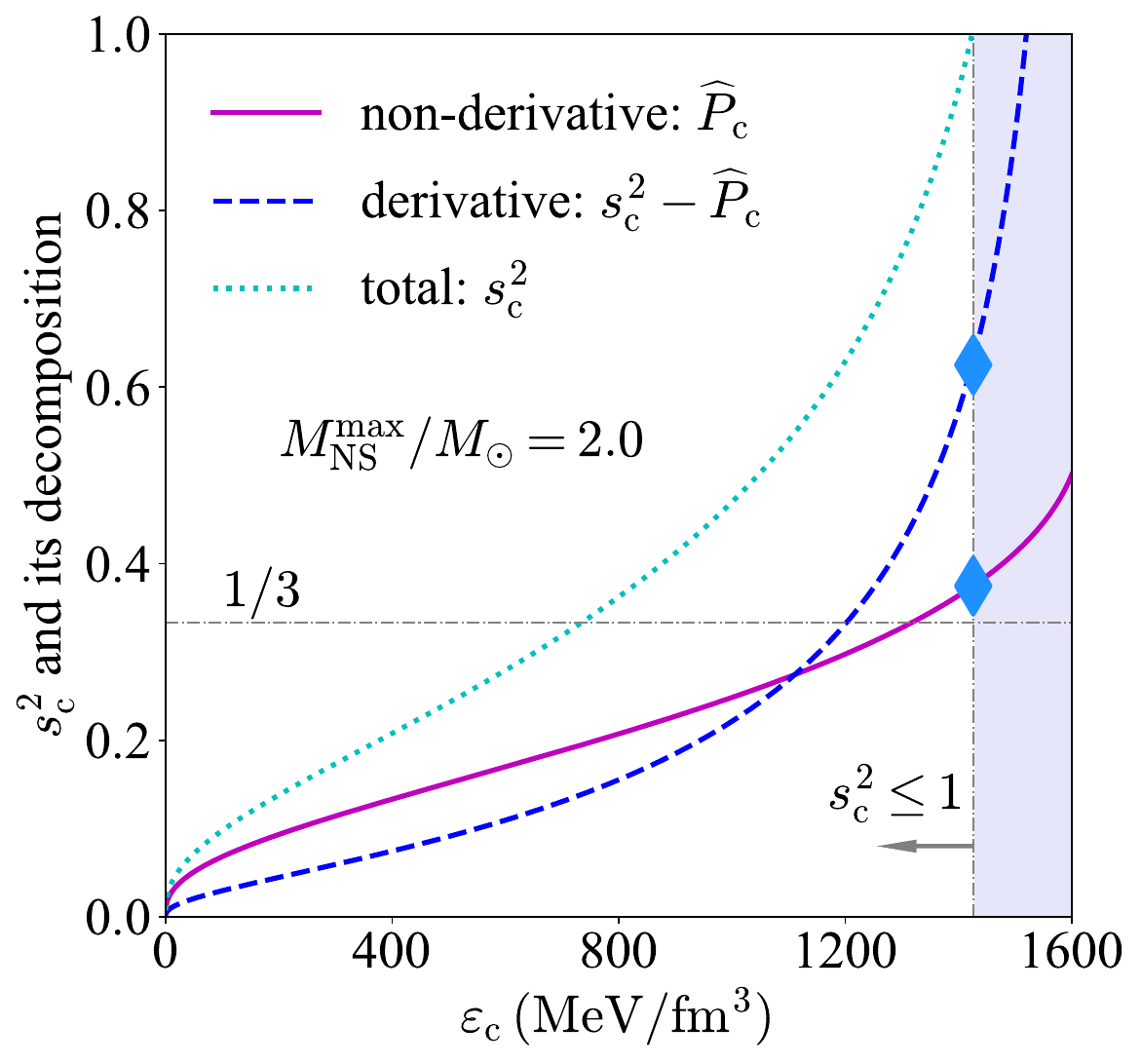}\qquad
\includegraphics[height=7.cm]{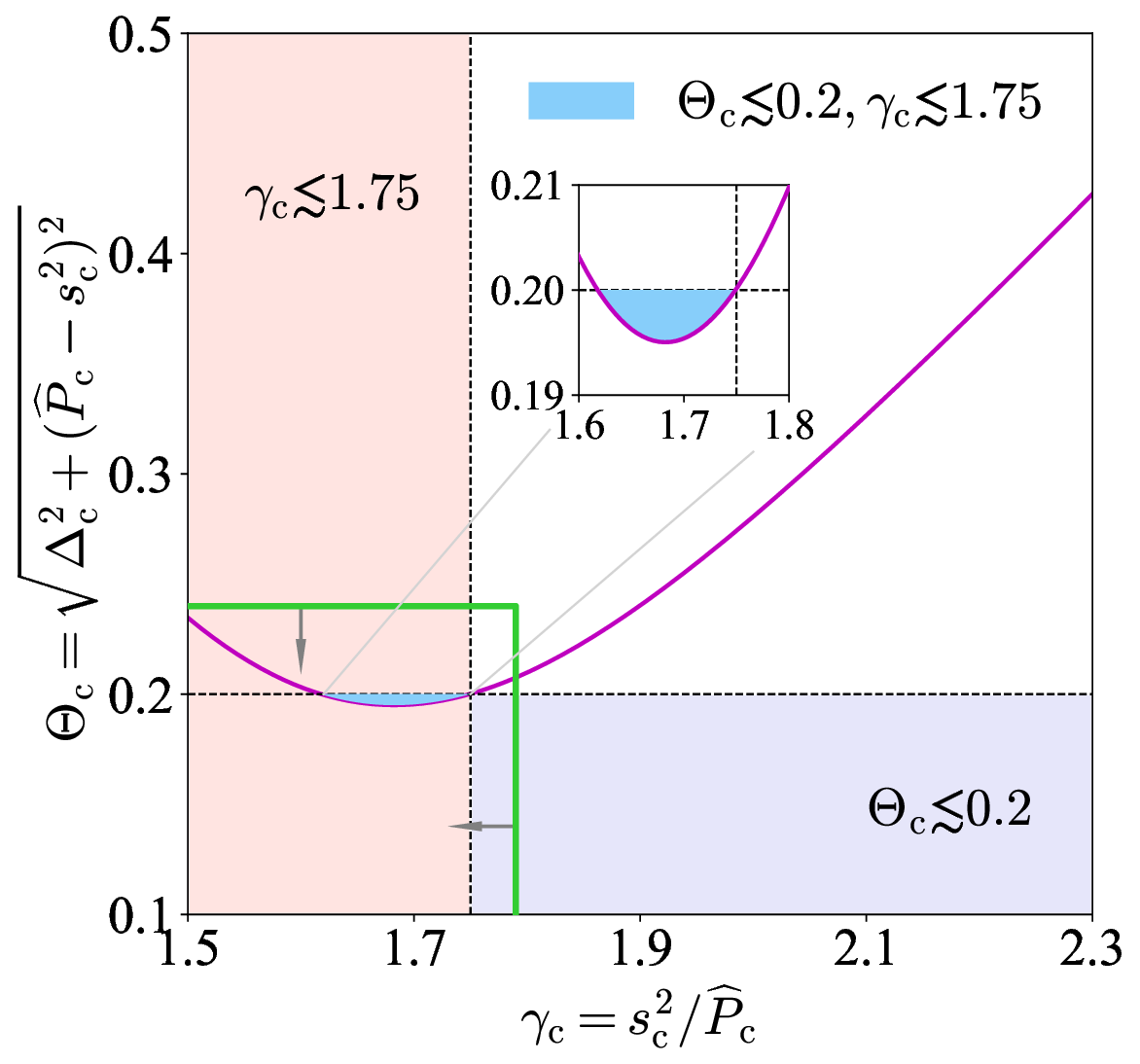}
\caption{(Color Online).  Left panel: decomposition of $s_{\rm{c}}^2$ into its non-derivative part ($\x$) and derivative part ($s_{\rm{c}}^2-{\x}$) for $M_{\rm{NS}}^{\max}/M_{\odot}=2$.
Solid diamonds characterize the boundary of $s_{\rm{c}}^2\leq1$ (left of the lavender band).
Right panel: dependence of $\Theta_{\rm{c}}$ on the polytropic index $\gamma_{\rm{c}}$ for the maximum-mass configuration $M_{\rm{NS}}^{\max}$, where in the lightblue region two criteria $\Theta_{\rm{c}}\lesssim0.2$ and $\gamma_{\rm{c}}\lesssim1.75$ hold simultaneously.
The constraints from Ref.\,\cite{Ann23} on $\gamma$ and $\Theta$ are shown by the green solid lines associated with arrows. The inset amplifies the lightblue region. Figures taken from Ref.\,\cite{CLZ23-b}.
}\label{fig_tc}
\end{figure}

In order to relax the conformality condition of (\ref{conf_cond_1}),  Ref.\,\cite{Ann23} furthermore defined a quantity 
\begin{equation}\label{def_Theta_1}
\boxed{
\Theta\equiv \sqrt{\Delta^2+(P/\varepsilon-s^2)^2}=\sqrt{\left(1/3-\phi\right)^2+\left(\phi-s^2\right)^2},}
\end{equation} to measure the conformality and found it should be $\lesssim0.65$ for all NSs. The {\color{xll}empirical} criterion $\Theta\lesssim0.2$ (besides $\gamma\lesssim1.75$ in conventional analysis) was then adopted\,\cite{Ann23} to identify the nearly-conformality at a given density, see the left panel of FIG.\,\ref{fig_Ann24d}.
The right panel of FIG.\,\ref{fig_Ann24d} shows a similar prediction for the $\Theta_{\rm{c}}$ as a function of baryon density using NICER's measurements and the secondary component of GW190814\,\cite{Abbott2020}. The predicted value for $\Theta_{\rm{c}}$ is slightly smaller than 0.2 at the TOV configuration.
Based on our formula (\ref{sc2-TOV}), we find that $0.19\lesssim\Theta_{\rm{c}}\lesssim0.63$ with the minimum (maximum) value obtained at ${\x}\approx0.18$ (${\x}\approx0.374$) and $\Theta_{\rm{c}}\approx0.22_{-0.03}^{+0.09}$ for ${\x}\approx0.24_{-0.07}^{+0.05}$ (PSR J0740+6620).
We summarize in TAB.\,\ref{sstab} these values at four reference ${\x}$ (0.18, 0.24, $1/3$ and 0.374), where $t_{\rm{c}}\equiv\d\Delta_{\rm{c}}/\d\ln\varepsilon_{\rm{c}}={\x}-s_{\rm{c}}^2$ as the logarithmic
derivative of $\Delta_{\rm{c}}$\,\cite{Ann23} (with respect to $\varepsilon_{\rm{c}}$) is also given. We have $t_{\rm{c}}\to0$ if $\gamma_{\rm{c}}\to1$ (the so-called conformal limit of matter), therefore $-t_{\rm{c}}$ characterizes the deviation from the conformal limit.  In Ref.\,\cite{Fuji22}, $-t=s^2-P/\varepsilon$ and $1/3-\Delta=P/\varepsilon$ are decomposed as the derivative and non-derivative parts of $s^2$, respectively, see more detailed discussions in Subsection \ref{sub_decomTA}.
Equivalently,  one has $
\Theta=[{\Delta^2+t^2}]^{1/2}$.
See FIG.\,\ref{fig_tc} for an example where $M_{\rm{NS}}^{\max}/M_{\odot}=2$ is adopted for the illustration (using Eq.\,(\ref{Mmax-G}) to solve for ${\x}$ and $s_{\rm{c}}^2$).
According to Eq.\,(\ref{sc2-TOV}), we find both the non-derivative and derivative parts of $s_{\rm{c}}^2$ are increasing functions of $\varepsilon_{\rm{c}}$, and take their maximum values at ${\x}\approx0.374$.
The solid diamonds in the left panel of FIG.\,\ref{fig_tc} characterize the boundary ${\x}\lesssim0.374$,  i.e., the curves above $\varepsilon_{\rm{c}}\gtrsim1.42\,\rm{GeV}/\rm{fm}^3$ (for $M_{\rm{NS}}^{\max}/M_{\odot}=2$) violate the causality condition $s_{\rm{c}}^2\leq1$, see Subsection \ref{sub_ultimate} and FIG.\,\ref{fig_ULTeps}.
For other values of $M_{\rm{NS}}^{\max}$, the shapes of the curves in FIG.\,\ref{fig_tc} are similar.

In TAB.\,\ref{sstab}, the column with ${\x}\approx0.18$ is also shown as the measure $\Theta_{\rm{c}}$ takes its minimum at this ${\x}$.
Interestingly, if one requires $\Theta_{\rm{c}}\approx0.2$, then two solutions ${\x}\approx0.21$ and ${\x}\approx0.16$ should be obtained,  the corresponding $\gamma_{\rm{c}}$ is found to be about 1.75 and 1.62, respectively.
We have approximately
\begin{equation}
\Theta_{\rm{c}}\approx1/3-\eta+2\eta^2/3+13\eta^3/6+41\eta^4/8+\cdots,\end{equation} where $0.5\gtrsim\eta\equiv1-4/3\gamma_{\rm{c}}\geq0$ acts as a small-expansion quantity (since $2.67\gtrsim\gamma_{\rm{c}}\geq4/3$).
We notice that ${\x}\approx0.189$\,\cite{CLZ23-a} and correspondingly $\Theta_{\rm{c}}\approx0.195$ and $\gamma_{\rm{c}}\approx1.70$ are obtained if the radius about $13.7\,\rm{km}$ was adopted for PSR J0740+6620\,\cite{Miller21}.
{\color{xll}These numerical values indicate that empirically the near-conformality may be possible in the core of very massive NSs.
However,  the likelihood for realizing conformality near the centers is small if the criteria $\Theta_{\rm{c}}\lesssim0.2$ and $\gamma_{\rm{c}}\lesssim1.75$\,\cite{Ann23} were adopted simultaneously,}  as shown in the right panel of FIG.\,\ref{fig_tc}, where the constraints $\gamma\lesssim1.79$ and $\Theta\lesssim0.24$ from Ref.\,\cite{Ann23} are shown (green solid lines associated with arrows) for comparison.
In a very recent study, Ref.\,\cite{Marquez2024} found that $\Theta_{\rm{c}}$ is greater than 0.2 for a hadronic matter using an extended NJL model.
Furthermore, the right panel of FIG.\,\ref{fig_tc} supports the conclusion that the criterion $\Theta_{\rm{c}}\lesssim0.2$ is more restrictive than $\gamma_{\rm{c}}\lesssim1.75$\,\cite{Ann23}.

The inference on $\Theta_{\rm{c}}$ is largely affected by the pQCD effects. For example, in Ref.\,\cite{Mus24} the constraint for $\Theta_{\rm{c}}$ changes from 0.138 to 0.161 (with an enhancement about 17\%) if the pQCD effects are removed.
Moreover, considering a canonical NS with $R\approx12_{-1}^{+1}\,\rm{km}$,  we have in Subsection \ref{sub_s2canon} that $s_{\rm{c}}^2\approx 0.47\pm0.09$, $\x\approx0.15\pm0.02$, $\Psi\approx2.85\pm0.29$, therefore 
\begin{equation}
\boxed{
\Theta_{\rm{c}}\approx0.38\pm0.05,~~\mbox{for a canonical NS with }
R\approx12_{-1}^{+1}\,\rm{km},}
\end{equation}
where $(3^{-1}-\x)^2$ contributes less than $(s_{\rm{c}}^2-\x)^2$ to $\Theta_{\rm{c}}$. 
We can also obtain the polytropic index about $\gamma_{\rm{c}}\approx3.26\pm0.23$.
The prediction on $\Theta_{\rm{c}}$ for a canonical NS is consistent with those from Ref.\,\cite{Ann23} and Ref.\,\cite{Tak23}, so the dense matter in cores of canonical NSs could not be conformal according to the empirical criterion $\Theta_{\rm{c}}\lesssim0.2$\,\cite{Ann23}.
If we consider the canonical pulsar PSR J0437-4715\,\cite{Choud24} with a small radius $R\approx11.36_{-0.63}^{+0.95}\,\rm{km}$, the $\Theta_{\rm{c}}\approx0.21\pm0.01$ could be obtained.
{\color{xll}For NSs at the causality limit (without considering their masses and radii individually), i.e., $s_{\rm{c}}^2\to1$ and $\x\to0.374$, TAB.\,\ref{sstab} gives $\Theta_{\rm{c}}\approx0.63$ (much larger than the empirical criterion). Such NSs are the densest since they could not be compressed even further. In this sense, the densest matter in NSs could hardly be conformal, this may be fundamentally different from the compressed QCD matter at very high baryon densities.}

Using the notations of Subsection \ref{sub_s2_2nd}, we investigate how the polytropic index and the empirical conformality measure $\Theta$ of Eq.\,(\ref{def_Theta_1}) behave at finite $\hr$\,\cite{CLZ23-b}, extending the above discussions.
In particular, we can work out using the $\hr$-dependence of the relevant quantities that:
\begin{align}
\gamma/\gamma_{\rm{c}}\approx&1
+\frac{b_2}{s_{\rm{c}}^2}\left(1+\frac{2D}{s_{\rm{c}}^2}-\frac{s_{\rm{c}}^2}{{\x}}\right)\widehat{r}^2
\approx1-\frac{3D}{16{\x}^2}\widehat{r}^2
,\label{gr-1}\\
\Theta/\Theta_{\rm{c}}\approx&1
+\frac{b_2}{s_{\rm{c}}^2}\frac{3t_{\rm{c}}(1+3s_{\rm{c}}^2-6{\x}-6D)}{1+9s_{\rm{c}}^2-6{\x}(1+3s_{\rm{c}}^2)+18{\x}^2}\widehat{r}^2
\approx1+\frac{1-6D}{8}\widehat{r}^2
,\label{Tr-1}
\end{align}
where the coefficients in front of $\widehat{r}^2$ in $\gamma/\gamma_{\rm{c}}$ and $\Theta/\Theta_{\rm{c}}$ are both positive, the second approximation for each quantity keeps only the leading-order term in ${\x}$.
Numerically, we then have $\Theta/\Theta_{\rm{c}}\approx1+3.8\widehat{r}^2$ and $\gamma/\gamma_{\rm{c}}\approx1+2.3\widehat{r}^2$ for ${\x}\approx0.24$ using $D\approx-0.45$ (panel (b) of FIG.\,\ref{fig_AD}).
For small ${\x}\to0$,  Eq.\,(\ref{Tr-1}) gives approximately $\Theta/\Theta_{\rm{c}}\approx1+\widehat{r}^2/8$.
Similarly, the $\mu$-dependence of $\gamma$ and $\Theta$ could be obtained,
\begin{align}
\gamma/\gamma_{\rm{c}}\approx&1+\left(\frac{t_{\rm{c}}}{{\x}}+\frac{2D}{s_{\rm{c}}^2}\right)\mu
\approx1+\frac{3D}{2{\x}}\mu,\label{gm-1}\\
\Theta/\Theta_{\rm{c}}\approx&1+
\frac{3t_{\rm{c}}+9s_{\rm{c}}^2(t_{\rm{c}}+2D)-18{\x}(t_{\rm{c}}+D)}{1+9s_{\rm{c}}^2-6{\x}(1+3s_{\rm{c}}^2)+18{\x}^2}\mu
\approx1+(6D-1){\x}\mu.
\label{Tm-1}
\end{align}
The coefficients in front of $\mu$ in Eqs.\,(\ref{gm-1}) and (\ref{Tm-1}) are both negative, e.g.,  we have $\gamma/\gamma_{\rm{c}}\approx1-2.9\mu$ and $\Theta/\Theta_{\rm{c}}\approx1-4.8\mu$ for ${\x}\approx0.24$. 
Therefore, Eq.\,(\ref{gr-1})-Eq.\,(\ref{Tm-1}) together generalize the conclusion of the right panel of FIG.\,\ref{fig_tc}: {\color{xll}If the dense matter at centers of massive NSs was not conformal,  it is likely its nearby surroundings are also not conformal\,\cite{CLZ23-b}.}

Combining the leading-order terms of Eq.\,(\ref{gm-1}) and Eq.\,(\ref{Tm-1}) or those of Eq.\,(\ref{gr-1}) and Eq.\,(\ref{Tr-1}) leads us to the correlation between the $\gamma$ and $\Theta$ parameters as,
\begin{equation}\label{TGG}
\boxed{
\frac{\Delta\Theta}{\Delta\gamma}\approx\left(4-\frac{2D}{3}\right){\x}^2\approx\left(4-\frac{2D}{3}\right)\x^2,}
\end{equation}
which holds near NS centers. It tells us that the two positive quantities $\Delta\Theta\equiv\Theta/\Theta_{\rm{c}}-1$ and $\Delta\gamma\equiv\gamma/\gamma_{\rm{c}}-1$ are positively correlated since ($4-2D/3>0$ according to panel (b) of FIG.\,\ref{fig_AD}).

\subsection{Unraveling the central trace anomaly from compactness scaling
nearly independent of EOS models}\label{sub_xiDelta}

The strong linear correlation between $\xi$ and $\Pi_{\rm{c}}$ indicated by Eq.\,(\ref{gk-comp}) and numerically verified in Eq.\,(\ref{gk-xi}) using the meta-model EOSs enables us to read off the $\x$ straightforwardly from the $\xi$ obtained either using NS mass-radius observation or the measurement on red-shift $z$ (see Eq.\,(\ref{RF}) for the formula). Therefore, the central trace anomaly $\Delta_{\rm{c}}=1/3-\x$ can be obtained straightforwardly from the observational data. Since the ratio $\phi=P/\varepsilon$ increases towards NS centers and reaches its maximum value $\x$ there, the trace anomaly takes its minimum $\Delta_{\rm{c}}$ there.
A lower bound for trace anomaly of Eq.\,(\ref{GRDelta}) from the GR limit $\x\lesssim0.374$\,\cite{CLZ23-a} is basically equivalently to the limit (\ref{xi_GR}) for the compactness. All the currently known NSs are safely consistent with inequality (\ref{xi_GR});
any future observations violating (\ref{xi_GR}) may imply a deeper trace anomaly than the $\Delta_{\rm{GR}}$ realized in NSs.

\begin{figure}[h!]
\centering
\includegraphics[height=7.5cm]{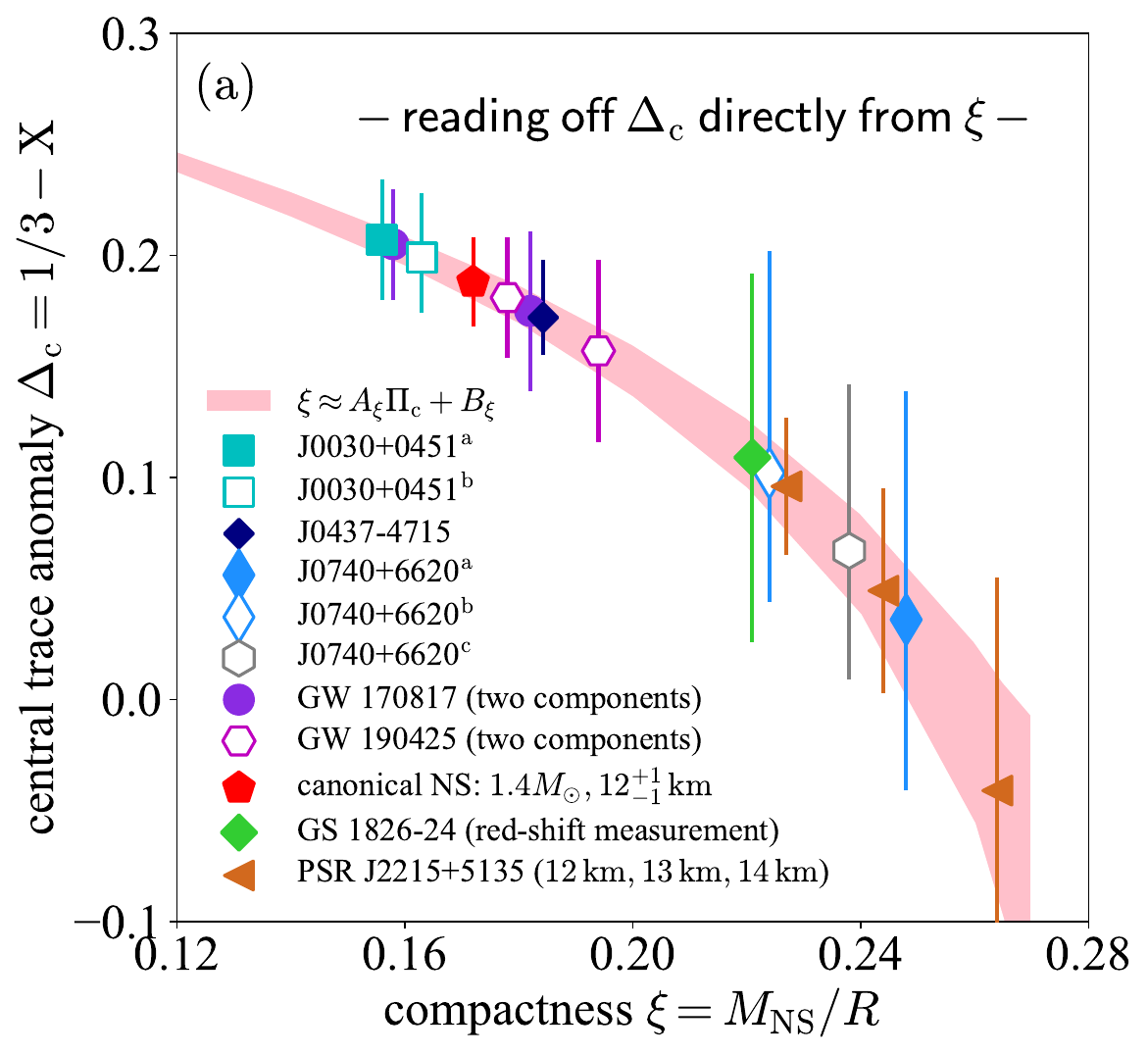}\qquad
\includegraphics[height=7.5cm]{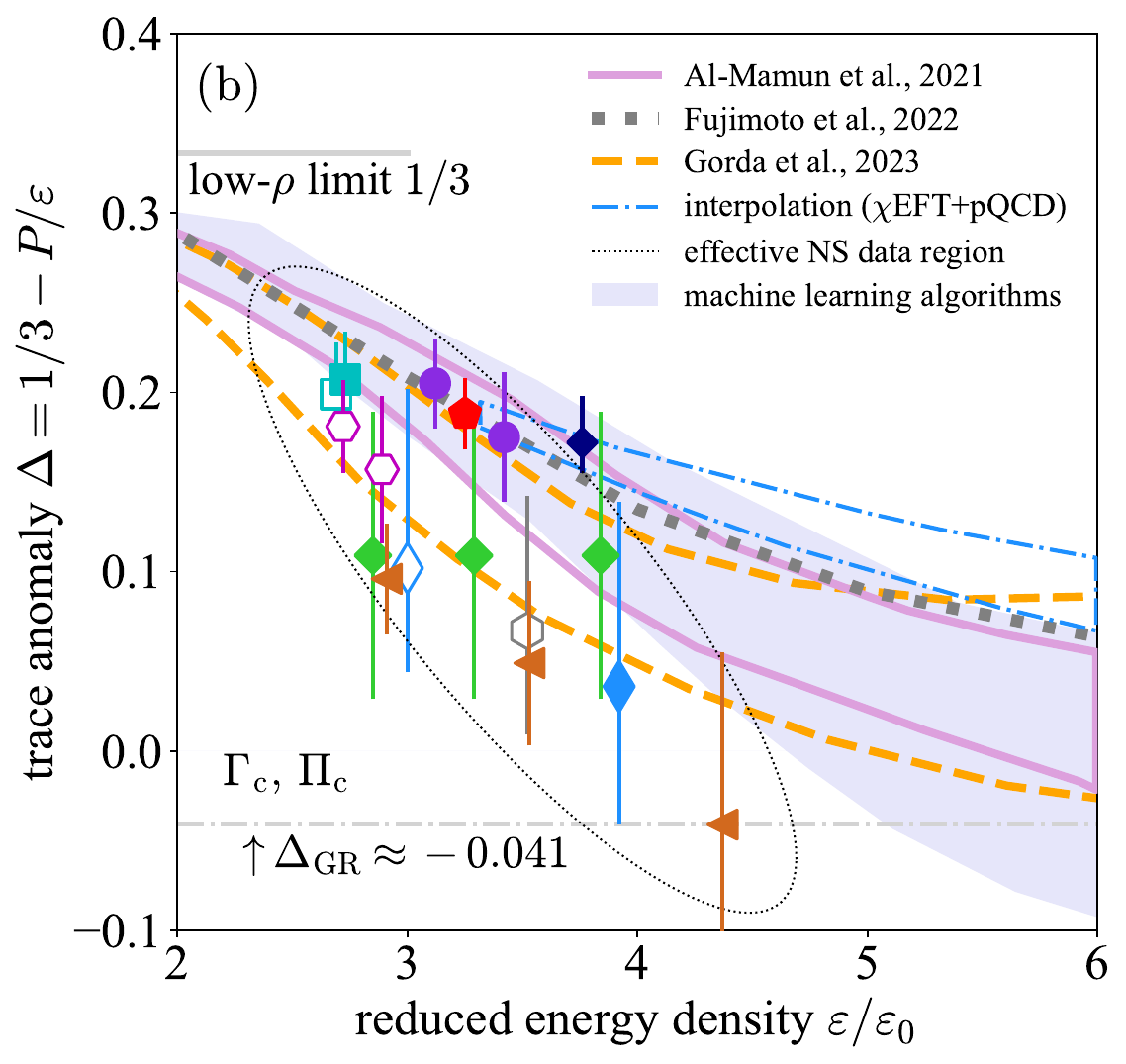}
\caption{(Color Online). Left panel: central trace anomaly $\Delta_{\rm{c}}$ as a function of compactness $\xi$ from inverting the compactness scaling (pink band) in comparison with observational data indicated. Right panel: energy density dependence of the trace anomaly where the trace anomalies from a few empirical NS EOSs via different input data/inference algorithms are also given. The GR bound of (\ref{GRDelta}) is plotted by the grey dash-dotted line, see  text for more details. Figures taken from Ref.\,\cite{CL24-b}.
}\label{fig_Delta}
\end{figure}
\renewcommand{\arraystretch}{1.5}
\begin{table}[h!]
\centerline{\normalsize
\begin{tabular}{c|c|c|c} 
  \hline
NS&$\x\equiv P_{\rm{c}}/\varepsilon_{\rm{c}}$&$\y\equiv \varepsilon_{\rm{c}}/\varepsilon_0$&~~Ref.~~\\\hline\hline
PSR J0030+0451$^{\rm{a}}$&$0.126_{-0.024}^{+0.026}$&$2.73_{-1.21}^{+1.25}$&\,\cite{Riley19}
\\\hline\hline
PSR J0030+0451$^{\rm{a}}$&$0.134_{-0.029}^{+0.025}$&$2.69_{-1.12}^{+1.07}$&\,\cite{Miller19}\\\hline
PSR J0437+4715&$0.161_{-0.026}^{+0.017}$&$3.76_{-1.05}^{+0.92}$&\,\cite{Choud24}\\\hline
PSR J0740+6620$^{\rm{a}}$&$0.297_{-0.103}^{+0.077}$&$3.92_{-1.19}^{+0.90}$&\,\cite{Riley21}\\\hline
PSR J0740+6620$^{\rm{b}}$ &$0.231_{-0.102}^{+0.058}$&$3.00_{-1.61}^{+0.93}$&\,\cite{Miller21}\\\hline
PSR J0740+6620$^{\rm{c}}$&$0.267_{-0.075}^{+0.057}$&$3.52_{-1.08}^{+0.84}$&\,\cite{Salmi22}\\\hline\hline
GW 170817 ($M_{\rm{NS}}^{(1)}$)&$0.159_{-0.036}^{+0.036}$&$3.42_{-1.37}^{+1.37}$&\,\cite{Abbott2018}\\\hline
GW 170817 ($M_{\rm{NS}}^{(2)}$)&$0.128_{-0.025}^{+0.025}$&$3.12_{-1.21}^{+1.21}$&\,\cite{Abbott2018}\\\hline\hline
GW 190425 ($M_{\rm{NS}}^{(1)}$)&$0.176_{-0.055}^{+0.055}$&$2.89_{-1.44}^{+1.44}$&\,\cite{Abbott2020-a}\\\hline
GW 190425 ($M_{\rm{NS}}^{(2)}$)&$0.152_{-0.037}^{+0.037}$&$2.72_{-1.15}^{+1.15}$&\,\cite{Abbott2020-a}\\\hline\hline
canonical NS&$0.146_{-0.020}^{+0.020}$&$3.25_{-0.79}^{+0.79}$&\,\cite{Rich23}\\\hline\hline
GS 1826-24 ($12\,\rm{km}$)&$0.224_{-0.082}^{+0.082}$&$3.84_{-2.16}^{+2.16}$&\,\cite{Zhou23}\\\hline
GS 1826-24 ($13\,\rm{km}$)&$0.224_{-0.082}^{+0.082}$&$3.29_{-1.82}^{+1.82}$&\,\cite{Zhou23}\\\hline
GS 1826-24 ($14\,\rm{km}$)&$0.224_{-0.082}^{+0.082}$&$2.85_{-1.61}^{+1.61}$&\,\cite{Zhou23}\\\hline\hline
PSR J2215+5135 ($12\,\rm{km}$)&$0.374_{-0.080}^{+0.080}$&$4.37_{-0.71}^{+0.71}$&\,\cite{Sul24}\\\hline
PSR J2215+5135 ($13\,\rm{km}$)&$0.283_{-0.039}^{+0.039}$&$3.53_{-0.57}^{+0.57}$&\,\cite{Sul24}\\\hline
PSR J2215+5135 ($14\,\rm{km}$)&$0.237_{-0.026}^{+0.026}$&$2.91_{-0.47}^{+0.47}$&\,\cite{Sul24}\\\hline\hline
    \end{tabular}}
        \caption{$\x=P_{\rm{c}}/\varepsilon_{\rm{c}}$ and $\y\equiv\varepsilon_{\rm{c}}/\varepsilon_0$ for the 17 NS instances,  the central trace anomaly is $\Delta_{\rm{c}}=1/3-\x$.
        }\label{tab_X}        
\end{table}

In panel (a) of FIG.\,\ref{fig_Delta}, we show the $\xi$-dependence of $\Delta_{\rm{c}}$ by inverting $\xi\approx A_\xi\Pi_{\rm{c}}(\x)+B_\xi$ (pink band).
Here 15 NS instances are shown,  these include NSs in GW170817\,\cite{Abbott2017,Abbott2018} and GW190425\,\cite{Abbott2020-a},  two alternative inferences of the radius using somewhat different approaches for PSR J0030+0451\,\cite{Riley19,Miller19} by superscript ``a,b'' and three for PSR J0740+6620\,\cite{Riley21,Miller21,Salmi22} by ``a,b,c''; the newly announced PSR J0437-4715\,\cite{Choud24} whose mass $1.418_{-0.037}^{+0.037}$ is quite accurate so the errorbar on its central trace anomaly is relatively small;
the NS in X-ray burster GS 1826-24\,\cite{Zhou23},  a canonical NS with radius $R\approx12_{-1}^{+1}\,\rm{km}$\,\cite{Brandes2023-a,Rich23}; and three central trace anomalies for PSR J2215+5135\,\cite{Sul24}. 
See FIG.\,\ref{fig_NSMR-REV} for the related observational data.
Though there is currently no observational constraint on the radius of PSR J2215+5135, we can use the $\xi_{\rm{GR}}$ or $\Delta_{\rm{GR}}$ to limit its radius to $R\gtrsim12\,\rm{km}$. Here three typical radii,  namely 12\,km, 13\,km and 14\,km are adopted for an illustration.
The error bar of $\Delta_{\rm{c}}$ is mainly due to the uncertainty of $\xi$ itself as the correlation $\xi$-$\Pi_{\rm{c}}$ is strong and model independent.
For instance,  the error bar of $\Delta_{\rm{c}}$ for a canonical NS with $R\approx12_{-1}^{+1}\,\rm{km}$ is apparently smaller than that for the two NSs in GW 170817\,\cite{Abbott2018} as the latter have larger mass uncertainties although they share similar radii (compare the red solid pentagon and dark-violet solid circles).
Besides the $\xi$-$\Pi_{\rm{c}}$ scaling, the mass-$\Gamma_{\rm{c}}$ scaling of Eq.\,(\ref{gk-mass}) and the specific form of Eq.\,(\ref{gk-m}) further gives individually the values of $P_{\rm{c}}$ and $\varepsilon_{\rm{c}}$ if both the $M_{\rm{NS}}$ and $R$ (or one of them together with the compactness) are observationally known. 
In order to obtain $\x$ and $\varepsilon_{\rm{c}}$ for the NS in GS 1826-24 (only its compactness is known), we adopt three typical radii (12\,km, 13\,km and 14\,km), the same as that for PSR J2215+5135, see the supplementary material of Ref.\,\cite{CL24-b}.
These 17 NS instances are also displayed in panel (b) of FIG.\,\ref{fig_Delta}. They are enclosed roughly by the dotted black ellipse (treated as the effective region of NS data). 
The values for $\x$ and $\y$ for these NS instances are given in TAB.\,\ref{tab_X}.

For comparisons, also shown in the right panel of FIG.\,\ref{fig_Delta} are the trace anomalies obtained/constrained by a few contemporary state-of-the-art NS EOS modelings 
using different input data and/or inference algorithms. These include the NS EOS inference\,\cite{Gorda2023} incorporating the pQCD impact (dashed orange band), a Bayesian inference of NS EOS\,\cite{Mam2021} combining the electromagnetic and gravitational-wave signals (plum solid band), the interpolation\,\cite{Fuji22} between low-density chiral effective field theories\,\cite{Essick2021} (CEFT) and high-density pQCD constraints\,\cite{Bjorken83} (between the dash-dotted light-blue lines),  a minimal parametrization\,\cite{Fuji22} between $\Delta$ and $\varepsilon/\varepsilon_0$ (grey dotted line) accounting for NS data and the NS EOS\,\cite{Fuji22} inferred via machine learning algorithms (lavender band).
Our results of ($\y\equiv\varepsilon_{\rm{c}}/\varepsilon_0$, $\Delta_{\rm{c}}=1/3-\x)$ for the 17 NS instances put stringent constraints on the theoretical NS EOSs. 
As shown in panel (b) of FIG.\,\ref{fig_Delta}, apparently the $\Delta$'s from some NS EOS modelings have sizable tensions with the limits set by the observational data based on our scaling analyses, especially for massive NSs. It implies that some ingredients in NS EOS inferences may need to be revised. For example, the NS EOS model incorporating pQCD effect\,\cite{Gorda2023} can well explain the $(\y,\Delta_{\rm{c}})$'s of PSR J0030+0451, GW 190425 and GS 1826-24 (with the radii $R\approx12\,\rm{km}$ and 13\,km,  the rightest two green diamonds). However, it could hardly account for the results for PSR J0740+6620 for all three radii\,\cite{Riley21,Miller21,Salmi22},  the two NSs in GW 170817\,\cite{Abbott2018} as well as those for PSR J2215+5135\,\cite{Sul24} (for all three $R$'s). 
In particular, it has apparent deviation from the data of the newly announced PSR J0437-4715\,\cite{Choud24}. Because the latter has a relatively smaller radius, its $\y\approx3.76$ is larger. Similarly, the NS EOS modeling with both the electromagnetic and gravitational-wave signals included\,\cite{Mam2021} can effectively explain the data of GW 170817, the canonical NS and PSR J0437-4715. However, it has certain tensions with our results based on observations for PSR J0030+451 and PSR J0740+6620, GW 190425 and the redback spider pulsar PSR J2215+5135. On the other hand, the interpolation\,\cite{Fuji22} between low-density CEFT\,\cite{Essick2021} and high-density pQCD\,\cite{Bjorken83} predicts a quite large $\Delta$ compared with what we extracted from PSR J0740+6620, the NS in GS 1826-24 and PSR J2215+5135 observations; however, it is consistent with $\Delta_{\rm{c}}$ of PSR J0437-4715.
More interestingly, although GW 190425 executes weaker limits on NS radii\,\cite{Abbott2020-a},  it effectively puts useful constraints on the trace anomaly $\Delta$. 

Compared with the predictions of existing NS EOS modelings, our analysis shows that the actual $\Delta$ is deeper.
Physically, this is because the $\y=\varepsilon_{\rm{c}}/\varepsilon_0$ allowed is actually smaller for heavier NSs, fundamentally due to the $M_{\rm{NS}}$-$\Gamma_{\rm{c}}$ correlation (which we discussed in more details in Subsection \ref{sub_1420} around FIG.\,\ref{fig_Ypm_sk}), see the three chocolate triangles for PSR 2215+5135\,\cite{Sul24}, here $\y\lesssim4.37$ (from the mass scaling) and $R\gtrsim12\,\rm{km}$ (via $\Delta\gtrsim\Delta_{\rm{GR}}$ of (\ref{GRDelta}) or (\ref{xi_GR})) are obtained.
Similarly, if we use the even heavier black widow pulsar PSR J0952-0607\,\cite{Romani22} with $2.35M_{\odot}$ as the input data, its radius $R$ should be $\gtrsim13.1\,\rm{km}$ via $\Delta\gtrsim\Delta_{\rm{GR}}$; and therefore $\y\lesssim3.68$.
{\color{xll}These examples clearly illustrate that including data from massive NSs in the analysis necessarily pulls down the $\Delta$-curve, even if the radius constraints are weak.}
Moreover, assuming different radii for GS 1826-24 horizontally shifts the green diamond. Nevertheless, the overall shape of $\Delta(\varepsilon/\varepsilon_0)$ remains almost unchanged due to the strong constraints set by PSR J0030+0451 and PSR J0740+6620.
The extracted $\x$ and therefore $\Delta_{\rm{c}}$ from either mass-radius observations or the red-shift measurement are very useful in analyzing the SSS in NSs as we shall demonstrate in Subsection \ref{sub_NSdatapeak}.

\begin{figure}[h!]
\centering
\includegraphics[width=11.cm]{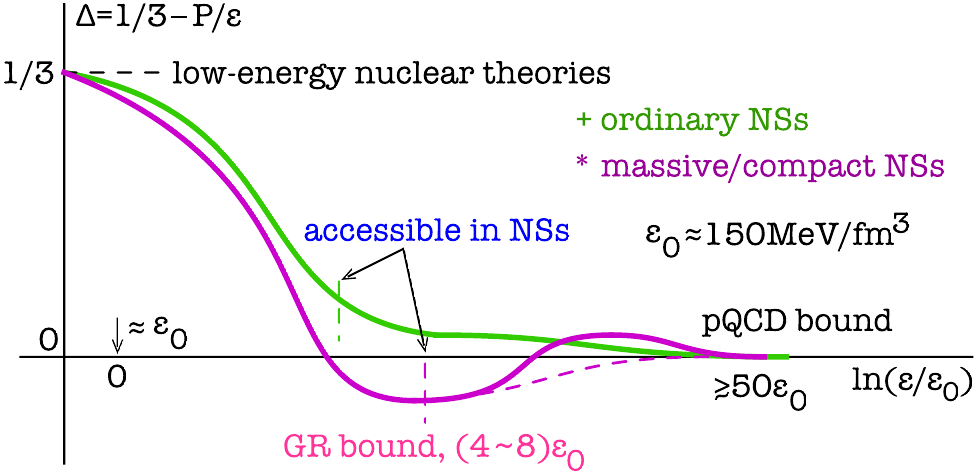}
\caption{(Color Online). Sketch of the patterns for $\Delta=1/3-P/\varepsilon$ in NSs. The $\Delta$ is well constrained around the fiducial density $\varepsilon_0\approx150\,\rm{MeV}/\rm{fm}^3$ by low-energy nuclear theories and is predicted to vanish due to conformality of the matter at $\varepsilon\gtrsim50\varepsilon_0$ (or equivalently $\rho\gtrsim40\rho_0$) using pQCD theories.
Massive and compact NSs provide a unique opportunity to probe the negativeness of $\Delta$ in certain energy density regions, where the GR bound on $\Delta$ with $\varepsilon$ being around ($4\mbox{$\sim$}8)\varepsilon_0$ is expected to be more relevant than the pQCD bound ($\varepsilon\gtrsim50\varepsilon_0$) in these NSs. Figure taken from Ref.\,\cite{CLZ23-b}.
}\label{fig_Dp_sk}
\end{figure}

Another interesting feature of the right panel of FIG.\,\ref{fig_Delta} is that $\Delta$ is probably positive considering all NS data analyzed. This feature may have some important consequences. Sketched in FIG.\,\ref{fig_Dp_sk}\,\cite{CLZ23-b} are two possible evolution curves of the trace anomaly $\Delta$. Their behaviors at intermediate-high energy densities regions are mostly based on educated guess with the wish of reaching eventually the pQCD prediction of $\Delta=0$ at extremely large energy densities $\varepsilon\gtrsim50\varepsilon_0\approx7.5\,\rm{GeV}/\rm{fm}^3$\,\cite{Fuji22,Kur10} or equivalently $\rho\gtrsim40\rho_0$.
Unfortunately, the latter is far larger than the energy or baryon density reachable in NSs based on our current knowledge. Thus, the pQCD prediction is possibly relevant but not fundamental for explaining the observed $\phi=P/\varepsilon\gtrsim1/3$ in massive NSs (either using microscopic or phenomenological models).
On the other hand, we have demonstrated that a GR bound on ${\x}=P_{\rm{c}}/\varepsilon_{\rm{c}}$ and $P/\varepsilon$ (near NS centers) naturally emerges, leading to the lower limit $\Delta_{\rm{GR}}\approx -0.04$, when dissecting perturbatively the TOV equations without using any specific EOS model\,\cite{CLZ23-a}. Thus, it is logical to say that {\color{xll}the GR bound on $\Delta$ with the $\varepsilon$ being roughly around (4$\sim$8$)\varepsilon_0$ is likely more relevant/fundamental than the pQCD prediction at  $\varepsilon\gtrsim50\varepsilon_0$ for nuclear EOS in NSs, although the latter may influence the extraction of $\Delta$ in some models\,\cite{Mus24}. It is also necessary to point out that understanding whether these two specific bounds on $\Delta$ ($\Delta\gtrsim\Delta_{\rm{GR}}\approx -0.04$ or $\Delta_{\rm{pQCD}}=0$ being very close to each other) are inner-related may deepen our understanding on the connection between GR and the microscopic theories of elementary particles.}
Our above discussions and findings are probably not out of bounds as ultimately properties of compact objects are determined by the Hamilton's principle using the total action of the whole system including gravity, matter (including nuclear matter, dark matter and energy) and their couplings\,\cite{11questions}. Some of these issues were discussed in Subsection \ref{sub_DenseQCD}.

As a negative $\Delta$ is unlikely to be observed in ordinary NSs (e.g., NSs with masses $\sim 1.7M_{\odot}$ or canonical NSs) as indicated by 
FIG.\,\ref{fig_Delta}, the real evolution of $\Delta$ is probably more like the green curve in FIG.\,\ref{fig_Dp_sk} at intermediate energy densities. An (unconventional) exception may come from light but very compact NSs, e.g., a $1.7M_{\odot}$ NS with a radius about 9.3\,km is expected to have its $\Delta_{\rm{c}}\approx-0.02$, since $\varepsilon_{\rm{c}}\approx1.86\,\rm{GeV}/\rm{fm}^3$ together with $P_{\rm{c}}\approx654\,\rm{MeV}/\rm{fm}^3$ should be obtained via the mass and radius scalings of (\ref{Mmax-G}) and (\ref{Rmax-n}) and so $\x\approx0.351$.
On the other hand, massive and compact NSs (masses $\gtrsim2M_{\odot}$) have the most relevance to observe a negative $\Delta$ (as indicated by the magenta curve in FIG.\,\ref{fig_Dp_sk}), but how the negative $\Delta$ evolves to the pQCD bound is up for imagination about properties of extraordinarily superdense matter\,\cite{CLZ23-b}.
Interestingly, both the green and magenta curves for the imagined $\Delta$ evolutions are closely connected with the density-dependence of the SSS, see Subsection \ref{sub_decomTA} for a detailed discussion.

\subsection{Can the currently available NS observational data invariably generate a peaked SSS density profile?}\label{sub_NSdatapeak}

In Subsection \ref{sub_xiDelta} and FIG.\,\ref{fig_Delta} we established the strong correlation between NS compactness $\xi$ and the factor $\Pi_{\rm{c}}=\x/(1+3\x^2+4\x)$; therefore either mass-radius observation or red-shift measurement can directly probe the $\x$ via this $\xi$-$\x$ scaling without using any nuclear EOS model.
Consequently, the (central) trace anomaly could be evaluated largely model-independently\,\cite{CL24-b}. Using the formula (\ref{for1}), the SSS profile and its decomposition can then be studied mostly based on the astrophysical observations alone.

\begin{figure}[h!]
\centering
\includegraphics[width=8cm]{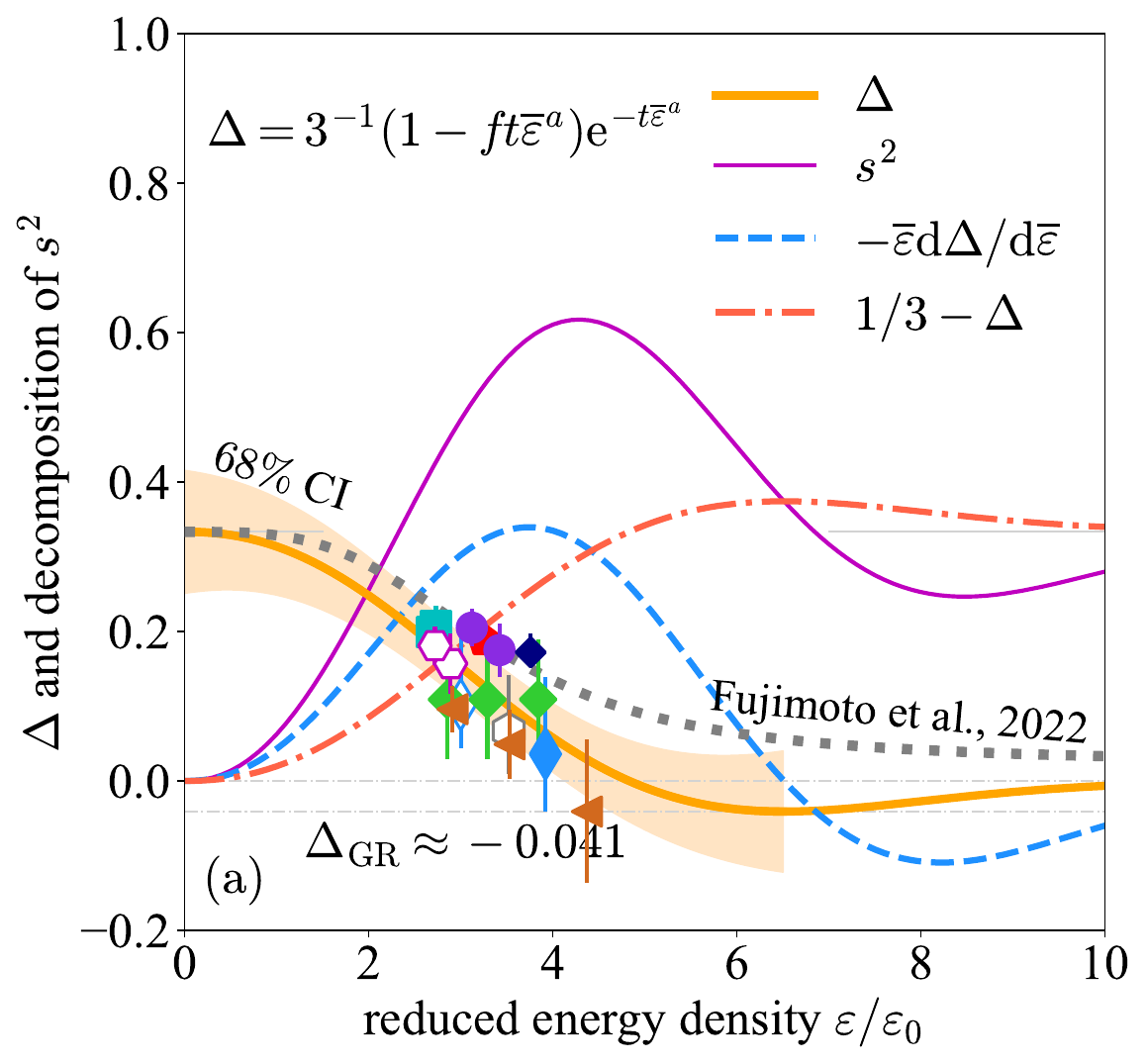}\quad
\includegraphics[width=8cm]{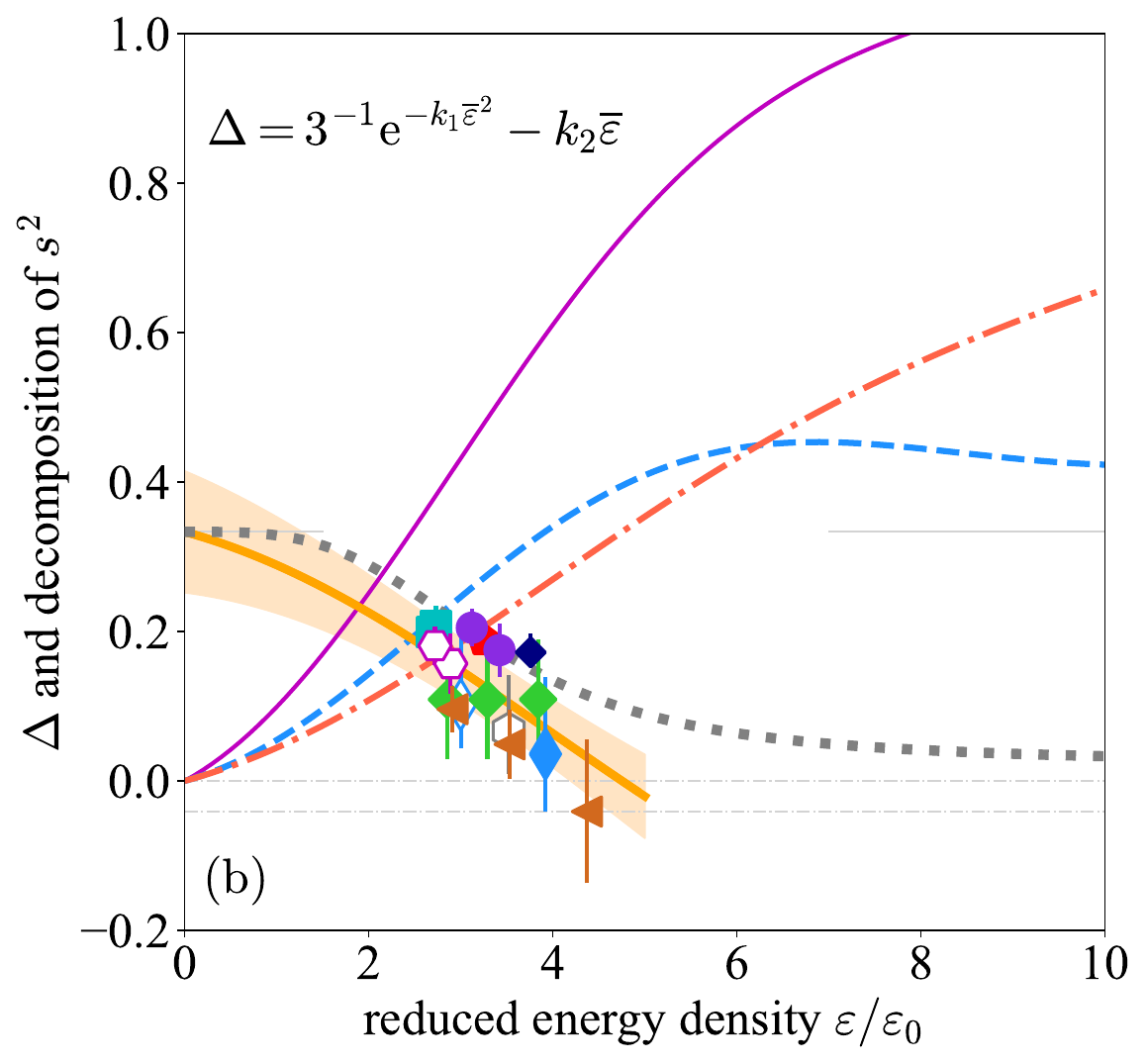}
\caption{(Color Online). The trace anomaly $\Delta$, $s^2$ and its decomposition via Eq.\,(\ref{for1}) using two effective parametrizations for $\Delta$ in panel (a) and (b), respectively.
The 17 NS instances of panel (b) of FIG.\,\ref{fig_Delta} are plotted with the tan band for its 68\% CI regression in each panel, here $\overline{\varepsilon}=\varepsilon/\varepsilon_0$.
The grey dotted line represents the minimal parametrization of Ref.\,\cite{Fuji22}.
Figures taken from Ref.\,\cite{CL24-b}.
}\label{fig_s2ab}
\end{figure}

In panel (a) of FIG.\,\ref{fig_s2ab}, we show a parametrization of trace anomaly $\Delta$:
\begin{equation}\label{pp1}
\Delta\approx \frac{1}{3}\left(1-ft\overline{\varepsilon}^{a}\right)\exp\left(-t\overline{\varepsilon}^a\right),
\end{equation} as a function of 
$\ep=\varepsilon/\varepsilon_0$
to effectively account for the 17 NS instances (including the data of the newly announced PSR J0437-4715\,\cite{Choud24} by the navy diamond in the right panel of FIG.\,\ref{fig_Delta}), valid for $2.5\lesssim\overline{\varepsilon}\lesssim4.5$ (data available). Here $a\approx 2.24$ and $t\approx 0.03$ are two parameters, and $f\equiv \rm{L}^{-1}_W(-3^{-1}/\rm{e}\Delta_{\rm{GR}})\approx0.9539$ with $\rm{L}_W(x)$ being the Lambert-$W$ function defined as the solution of the equation $w\exp(w)=x$.
This parametrization has the following properties: (i) $\Delta\to0$ for $\overline{\varepsilon}\to\infty$\,\cite{Bjorken83}, and (ii) $\Delta\to1/3$ as $\overline{\varepsilon}\to0$; it also generates a minimum $\Delta_{\min}=\Delta_{\rm{GR}}\approx-0.041$ at $\overline{\varepsilon}_{\rm{GR}}=[(f+1)/ft]^{1/a}\approx6.5$ by construction (through the factor $f$).
Numerically, we find $\Delta$ in panel (a) drops most quickly at roughly about $\overline{\varepsilon}\approx3.5$ from being $\gtrsim0.1$ to $\lesssim0.1$.
This feature actually strongly connects with the possible peaked behavior of the squared sound speed\,\cite{Fuji22}.
The expected peak emerges at $\overline{\varepsilon}_{\rm{pk}}\approx4.3$ or $\varepsilon_{\rm{pk}}\approx645\,\rm{MeV}/\rm{fm}^3$ with $s^2\approx0.62$; this is because $\Delta\to \Delta_{\rm{GR}}$ somewhere and thus a peak emerges at $\overline{\varepsilon}\approx3.7$ in the derivative term ``$-\overline{\varepsilon}\d\Delta/\d\overline{\varepsilon}$''\,\cite{Fuji22}.
So the peak position in $s^2$ is larger than that in $s_{\rm{deriv}}^2$, see the analytical demonstration given in Subsection \ref{sub_decomTA} and Eq.\,(\ref{for3}).
More interestingly,  the GR bound $\Delta_{\min}=\Delta_{\rm{GR}}<0$ induces a valley in $s^2$ at $\overline{\varepsilon}_{\rm{vl}}\approx8.4$ with $s^2\approx0.25$, since $\Delta\to0>\Delta_{\rm{GR}}$ as $\varepsilon/\varepsilon_0\to\infty$ due to a valley appearing at about $\overline{\varepsilon}\approx8.2$ in the derivative part; however this density may exceed the one allowed in realistic NSs.
{\color{xll}In this case, the pQCD limit for $\Delta$ is the origin for the emergence of a valley in $s^2$ instead of a peak.}
We summarize the fitting properties of SSS at the peak and valley positions as well as its decomposition terms in TAB.\,\ref{tab_ppppp}, under the parametrization (\ref{pp1}).

\begin{table}[h!]
\renewcommand{\arraystretch}{1.5}
\centerline{\normalsize
\begin{tabular}{c|c||c|c||c|c|c|c||c|c|c|c} 
  \hline
$a$&$t$&$f$&$\overline{\varepsilon}_{\rm{GR}}$&$\overline{\varepsilon}_{\rm{pk}}$&$s_{\rm{pk}}^2$&$\overline{\varepsilon}_{\rm{vl}}$&$s_{\rm{vl}}^2$&$\overline{\varepsilon}_{\rm{deriv,pk}}$&$s_{\rm{deriv,pk}}^2$&$\overline{\varepsilon}_{\rm{deriv,vl}}$&$s_{\rm{deriv,vl}}^2$\\\hline\hline
2.24&0.03&0.9539&6.5&4.3&0.62&8.4&0.25&3.7&0.34&8.2&$-0.11$\\\hline
    \end{tabular}}
        \caption{Properties of the SSS (at the peak and valley positions) and its decomposition terms using the parametrization (\ref{pp1}); here $s^2_{\rm{pk}}\equiv s^2(\ep_{\rm{pk}})$,  $s^2_{\rm{vl}}\equiv s^2(\ep_{\rm{vl}})$, etc., $\ep=\varepsilon/\varepsilon_0$ with $\ep_0\approx150\,\rm{MeV}/\rm{fm}^3$.
        }\label{tab_ppppp}       
\end{table}

\begin{figure}[h!]
\centering
\includegraphics[width=9.cm]{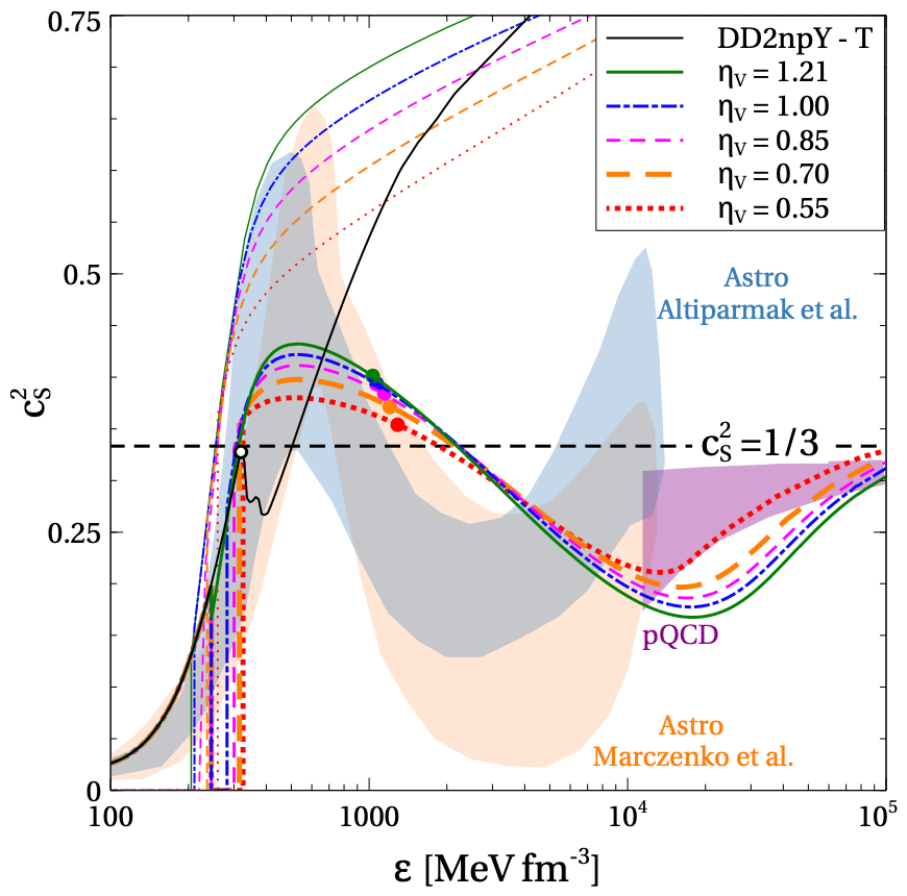}
\caption{(Color Online). The SSS obtained in a three-flavor nonlocal NJL model of quark matter with the scalar attractive, vector repulsive and diquark pairing interaction channels. Figure taken from Ref.\,\cite{Ivanytskyi2024}.
}\label{fig_IVK}
\end{figure}
What we learned above from parametrizing $\Delta$ with Eq.\,(\ref{pp1}) is consistent with a recent study on SSS by using a three-flavor nonlocal NJL model of quark matter with the scalar attractive, vector repulsive and diquark pairing interaction channels\,\cite{Ivanytskyi2024}, see FIG.\,\ref{fig_IVK}.
It also shows that the valley position $\ep_{\rm{vl}}$ in $s^2$ is slightly larger than that ($\ep_{\rm{vl,deriv}}$) in $s^2_{\rm{deriv}}$. This is consistent with the analytical analysis in Subsection \ref{sub_decomTA}, see Eq.\,(\ref{valley-k}).
In addition, whether including the newly announced PSR J04317-4715\,\cite{Choud24} with both mass and radius observations available under the parametrization (\ref{pp1})
may essentially impact qualitatively both the position and magnitude of the peak (valley), as shown in TAB.\,\ref{tab_navy}.

\begin{table}[h!]
\renewcommand{\arraystretch}{1.5}
\centerline{\normalsize
\begin{tabular}{c|c|c|c|c|c} 
  \hline
& $\ep_{\rm{GR}}$&$\ep_{\rm{pk}}$&$s^2_{\rm{pk}}$&$\ep_{\rm{vl}}$&$s^2_{\rm{vl}}$\\\hline\hline
with PSR J04317-4715&6.5&4.3&0.62&8.4&0.25\\\hline
without PSR J04317-4715&5.8&4.0&0.67&7.2&0.23\\\hline
    \end{tabular}}
        \caption{Influence of the mass and radius observational data of PSR J0437-4715\,\cite{Choud24} on the $s^2$ and related quantities using the parametrization (\ref{pp1}).}\label{tab_navy}     
\end{table}

\begin{figure}[h!]
\centering
\includegraphics[height=6.5cm]{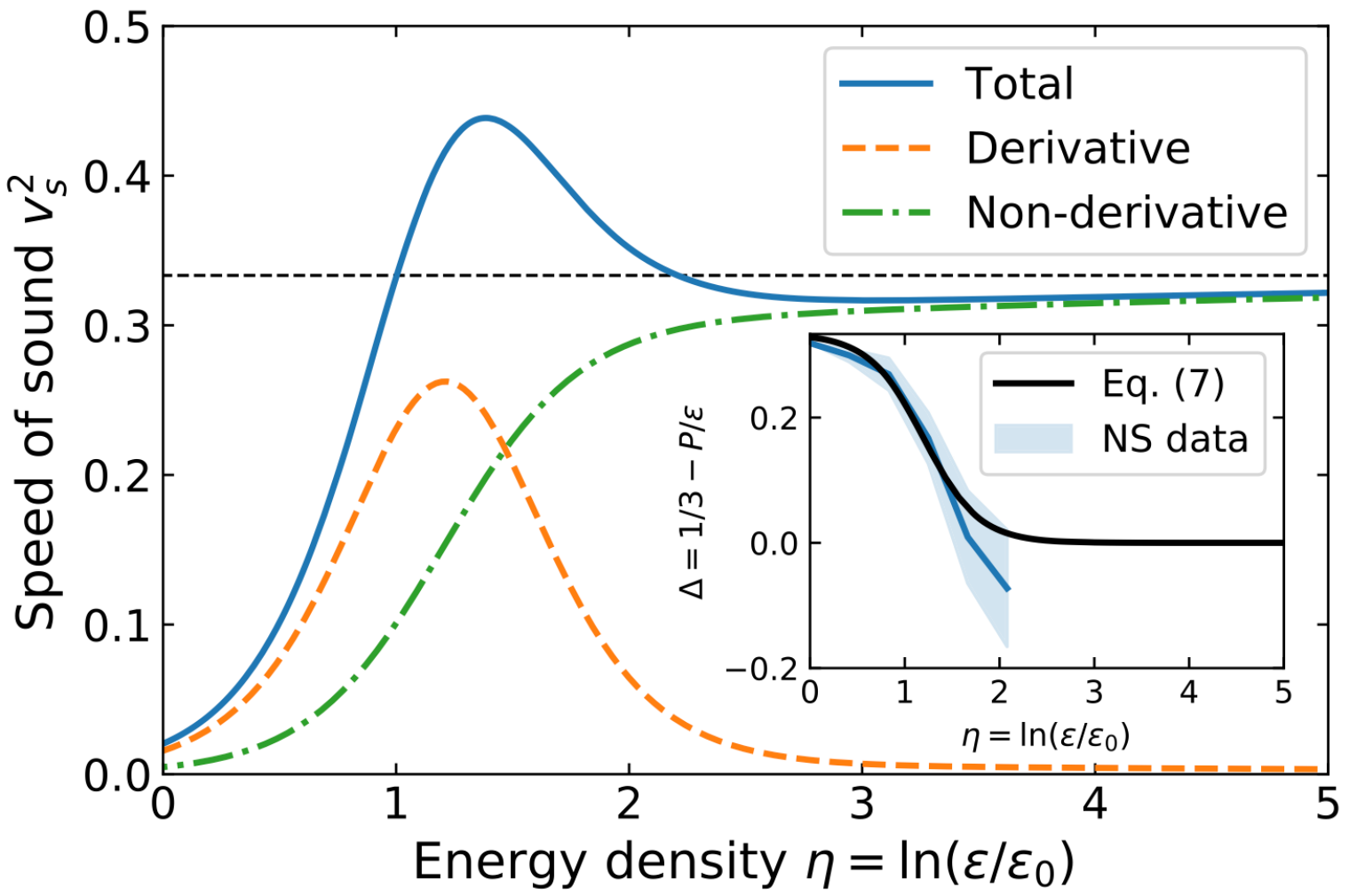}
\caption{(Color Online). Decomposition of $s^2$ into the derivative and non-derivative parts as in Eq.\,(\ref{for1}). The peak in $s^2$ is attributed to the peak appearing in the derivative part in the work of Ref.\,\cite{Fuji22}.
Figure taken from Ref.\,\cite{Fuji22}.
}\label{fig_Fuji22PRLs2Decom}
\end{figure}

\begin{figure}[h!]
\centering
\includegraphics[width=8.cm]{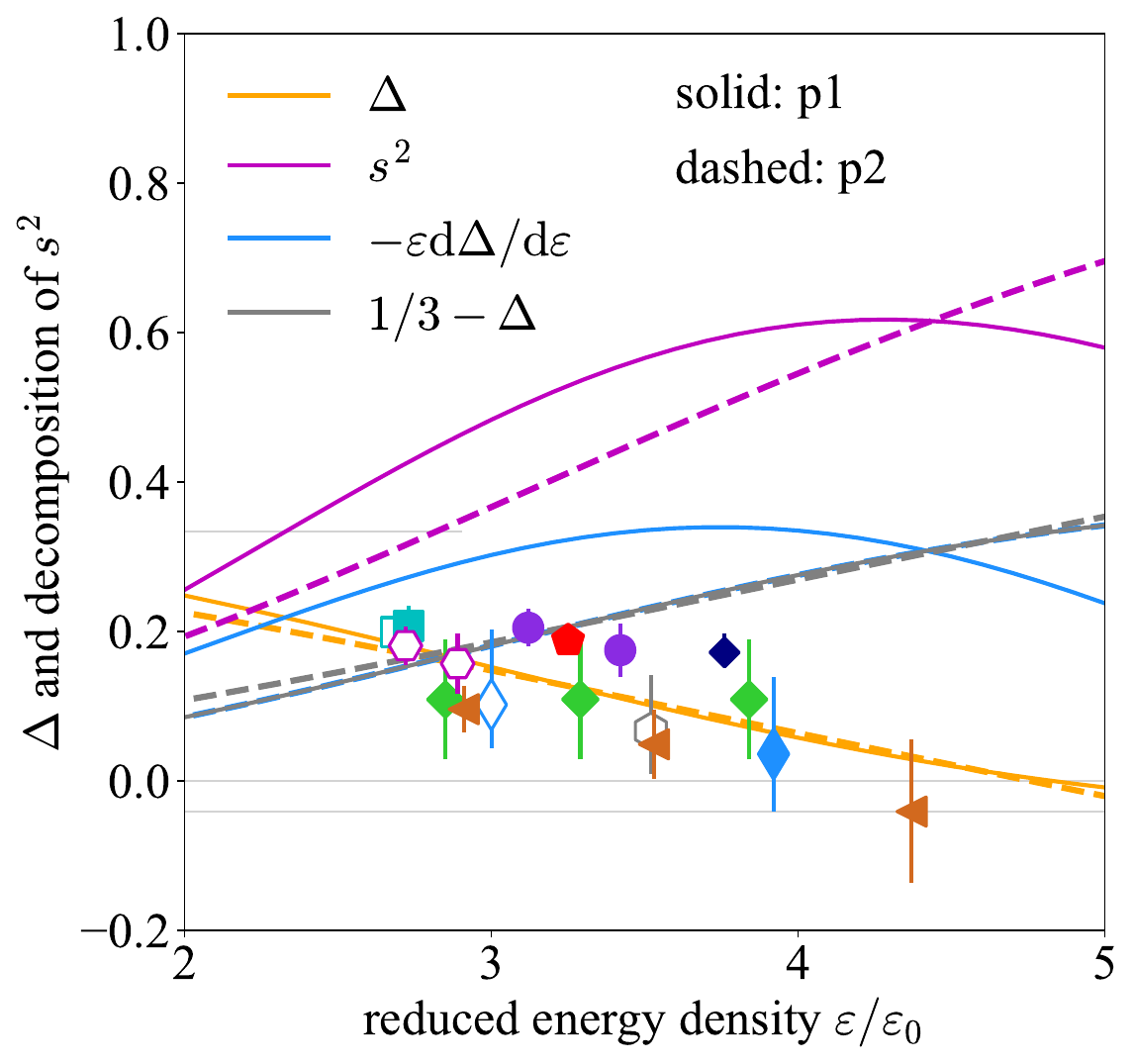}
\caption{(Color Online). Within the effective data region about $2.5\lesssim\overline{\varepsilon}\lesssim4.5$, parametrization (\ref{pp1}) and parametrization (\ref{pp2}) for $\Delta$ behaves very similarly; here p1 and p2 denote  parametrizations $\Delta\approx 3^{-1}(1-ft\overline{\varepsilon}^{a})\exp(-t\overline{\varepsilon}^a)$
and $3^{-1}\exp(-k_1\ep^2)-k_2\ep$, respectively.
Figure taken from Ref.\,\cite{CL24-b}.
}\label{fig_fffff}
\end{figure}

On the other hand, if we adopt another effective parametrization for $\Delta$ considering only the constraint $\lim_{\overline{\varepsilon}\to0}\Delta=3^{-1}$ (without its large $\overline{\varepsilon}$-limit) but can describe the same NS data approximately equally well as the previous one as shown in panel (b) of FIG.\,\ref{fig_s2ab}, e.g., 
\begin{equation}\label{pp2}
\Delta\approx \frac{1}{3}\exp\left(-k_1\overline{\varepsilon}^2\right)-k_2\overline{\varepsilon},
\end{equation} with two parameters $k_1\approx0.037$ and $k_2\approx0.032$, the resulting $s^2$ just monotonically increases with $\overline{\varepsilon}$ in spite of the fact that a broad peak exists in the derivative term ``$-\overline{\varepsilon}\d\Delta/\d\overline{\varepsilon}$''. 
Specifically, we find the peak in the derivative part $-\overline{\varepsilon}\d\Delta/\d\overline{\varepsilon}$ occurs at $\overline{\varepsilon}\approx6.2$ with the full $s^2\approx0.88$.
This is different from the analysis given in Ref.\,\cite{Fuji22} which indicates that a peak in the derivative part may induce a peaked behavior of the full $s^2$, as shown in FIG.\,\ref{fig_Fuji22PRLs2Decom}; see also Ref.\,\cite{Carl2024PRD} for the emergence of the peaked structure in the derivative part of $s^2$ using a non-local NJL model.
Within the effective data region in energy density, parametrizations (\ref{pp1}) and (\ref{pp2}) behave very similarly and they are approximately equally accurate, see FIG.\,\ref{fig_fffff}.
{\color{xll}It demonstrates that a peaked shape of $s^2$ is an implication considering some well-founded theoretical limits but not practically inevitable, and it is not a direct consequence of the currently available NS data themselves.}

As more observational data of mass, radius or red-shift becomes available (e.g., see the sketch in FIG.\,\ref{fig_DEL_s2peak}), our approach based on NS compactness and mass scalings can better constrain the trace anomaly $\Delta$ in a wide range of energy densities, locate precisely the possible peak positions of $s^2$ relying on NS data alone without using other physical inputs\,\cite{Zhou2024}, and restrict the NS EOS to a much narrower band.

\subsection{Brief summary of the existing causality boundaries for NS mass-radius relations}\label{sub_cau_BF}

Besides the Schwarzschild boundary,  a spherical star composed of uniform incompressible fluids is a special and simplest case\,\cite{Buch59}.
For such model, $\varepsilon(r)=\varepsilon_{\rm{c}}=\rm{const.}$, the TOV equations could be solved analytically to give
\begin{equation}\label{bb-1}
\x=\frac{P_{\rm{c}}}{\varepsilon_{\rm{c}}}=\frac{1-\sqrt{1-2\xi}}{3\sqrt{1-2\xi}-1}
\approx\frac{\xi}{2}\left(1+2\xi+\frac{17}{4}\xi^2+\frac{37}{4}\xi^3+\cdots\right),~~\xi=M_{\rm{NS}}/R.
\end{equation}
To avoid the singularity in the denominator of (\ref{bb-1}), one needs $R>9M_{\rm{NS}}/4$, giving the {\color{xll}Buchdahl causality boundary},
\begin{equation}\label{bb-Buchlim}
\boxed{
\mbox{Buchdahl causality boundary:}~~
R_{\rm{Buch}}/\rm{km}\gtrsim3.32M_{\rm{NS}}/M_{\odot}.}
\end{equation}
In addition, if we put the energy dominant condition (or equivalently the causality condition) on (\ref{bb-1}), namely $\x\leq1$, then (\ref{bb-1}) gives $R>8M_{\rm{NS}}/3$.
Therefore the {\color{xll}Buchdahl limit with energy dominant condition (EDC)} is 
\begin{equation}\label{bb-Buchlim-EDC}
\boxed{
\mbox{Buchdahl causality boundary with EDC:}~~
R_{\rm{Buch}}^{\rm{EDC}}/\rm{km}\gtrsim3.94M_{\rm{NS}}/M_{\odot}.}
\end{equation}
Denoting the upper limit for $\x$ generally as $\x_+$, i.e., $\x\leq\x_+$,  (\ref{bb-1}) gives then
\begin{empheq}[box=\fbox]{align}\label{bb-Buchlim-X}
&\mbox{Buchdahl causality boundary with $\x_+$:~~}\notag\\
&\hspace{2cm}
R_{\rm{Buch}}^{(+)}\gtrsim \frac{(1+3\x_+)^2}{2\x_+(1+2\x_+)}M_{\rm{NS}}\leftrightarrow\frac{R_{\rm{Buch}}^{(+)}}{\rm{km}}\gtrsim1.477\frac{(1+3\x_+)^2}{2\x_+(1+2\x_+)}\left(\frac{M_{\rm{NS}}}{M_{\odot}}\right),
\end{empheq}
here the superscript ``+'' in the radius reminds us that this is obtained under the condition $\x\leq\x_+$.
Using our upper limit $\x_+\approx0.374$ (of Eq.\,(\ref{Xupper})) extracted directly from the TOV equations, we then have
\begin{equation}\label{bb-Buchlim-374}
\boxed{
\mbox{Buchdahl causality boundary with $\x_+\approx0.374$:~~}
R_{\rm{Buch}}^{\x_+\approx0.374}/\rm{km}\gtrsim5.01M_{\rm{NS}}/M_{\odot}.}
\end{equation}

Later, Buchdahl considered an improved EOS\,\cite{Buch67} by generalizing the constant $\varepsilon_{\rm{c}}$ assumption to $\varepsilon(P)=12\sqrt{P_{\ast}P}-5P$ with $P_{\ast}$ a constant. In this case {\color{xll}(the Buchdahl fluid)}, one can work out the following boundary\,\cite{Buch67,Saes2024}
\begin{equation}\label{bb-2}
\boxed{
\x=\frac{\xi}{2-5\xi}\approx\frac{\xi}{2}\left(1+\frac{5}{2}\xi+\frac{25}{4}\xi^2+\frac{125}{8}\xi^3+\cdots\right),~~\mbox{or}~~\frac{R_{\rm{BF}}^{(+)}}{\rm{km}}\gtrsim1.477\left(\frac{5}{2}+\frac{1}{2\x_+}\right)\left(\frac{M_{\rm{NS}}}{M_{\odot}}\right).}
\end{equation}
Paralleling to inequalities of (\ref{bb-Buchlim}), (\ref{bb-Buchlim-EDC}) and (\ref{bb-Buchlim-374}), we should similarly obtain
\begin{equation}\label{bb-Buch-fluid}
\boxed{
R_{\rm{BF}}/\rm{km}\gtrsim3.69M_{\rm{NS}}/M_{\odot};~~R_{\rm{BF}}^{\rm{EDC}}/\rm{km}\gtrsim4.43M_{\rm{NS}}/M_{\odot};~~R_{\rm{BF}}^{\x_+\approx0.374}/\rm{km}\gtrsim5.67M_{\rm{NS}}/M_{\odot};}
\end{equation}
by taking $\x_+\to\infty$ (avoiding the  singularity in $\x$), $\x_+=1$ (energy dominant condition) and $\x_+\approx0.374$ (upper limit predicted by TOV equations) in (\ref{bb-2}), respectively;
here ``BF'' is abbreviated for ``Buchdahl fluid''.

\begin{figure}[h!]
\centering
\includegraphics[height=10.cm]{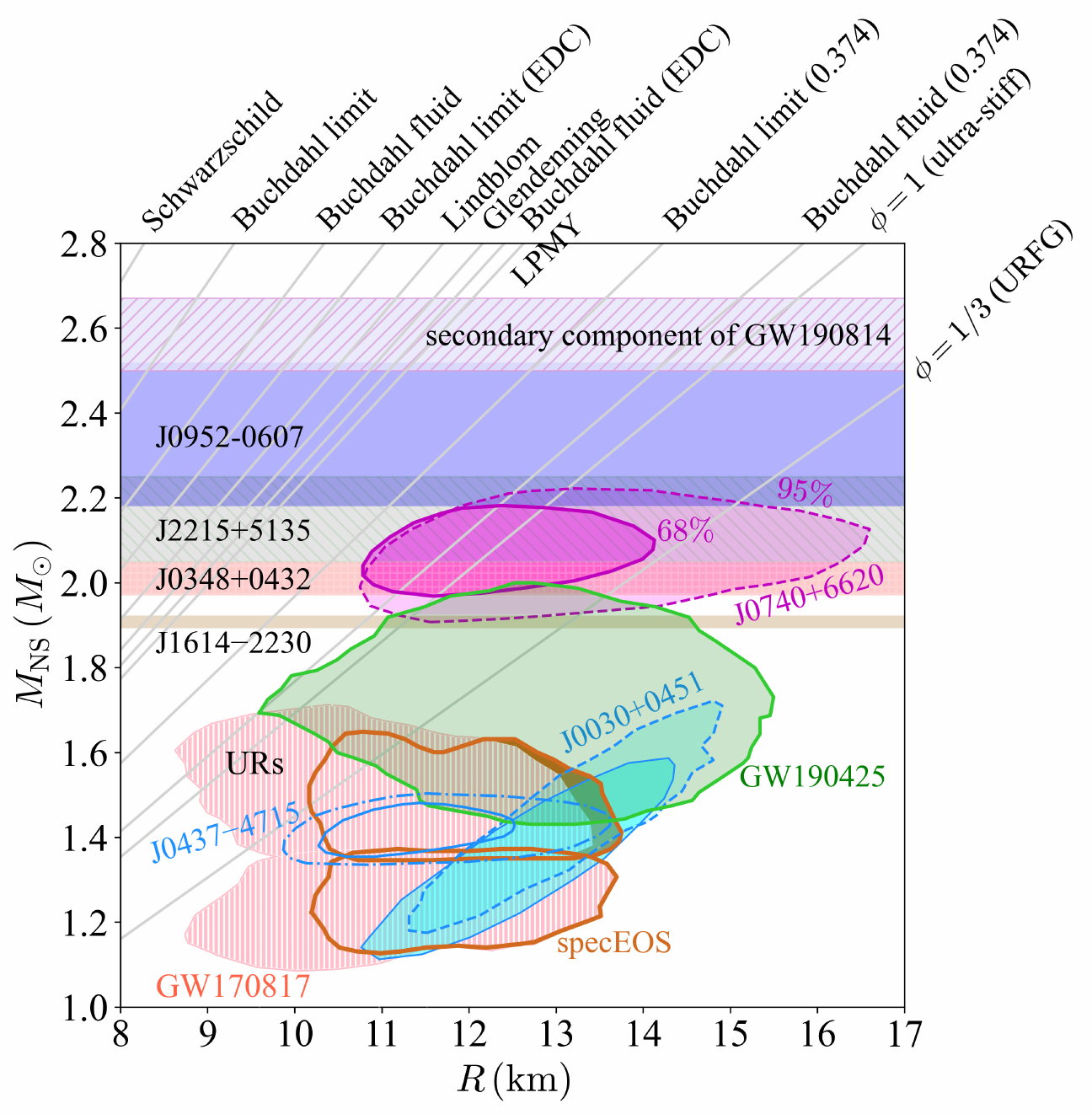}
\caption{(Color Online). Same as FIG.\,\ref{fig_NSMR-REV} but with a few causality boundaries (solid lightgrey lines) added: Schwarzschild relation (black hole constraint) $R=2M_{\rm{NS}}$; the Buchdahl limit $R=9M_{\rm{NS}}/4$, the Buchdahl limit with dominant energy condition $R=8M_{\rm{NS}}/3$ and the condition $\x\lesssim0.374$ as $R\approx3.44M_{\rm{NS}}$; the Buchdahl fluid $R=5M_{\rm{NS}}/2$ as well as that with the dominant energy condition $R=7M_{\rm{NS}}/2$ and the condition $\x\lesssim0.374$ as $R\approx3.83M_{\rm{NS}}$; and $R\approx2.83M_{\rm{NS}}$ from Ref.\,\cite{Lind84},  $R\approx 3.05M_{\rm{NS}}$ from Ref.\,\cite{Lattimer1990} and $R\approx2.94M_{\rm{NS}}$ from Ref.\,\cite{Glendenning1992}; the $R=4M_{\rm{NS}}$ from using an ``ultra-stiff'' EOS $P=\varepsilon$ (or $\phi=1$) and $R=14M_{\rm{NS}}/3$ using an URFG EOS.
}\label{fig_NSMR-REV-cau}
\end{figure}

Ref.\,\cite{Lind84} found that the red-shift $z$ has a maximum value about $0.863$, which inversely gives the causality boundary as\,\cite{Lind84,ZhangLi2019EPJA}
\begin{equation}\label{bb-Lind}
\boxed{
\mbox{Lindblom:}~~
R/\rm{km}\gtrsim4.18M_{\rm{NS}}/M_{\odot}.}
\end{equation}
This is perhaps one of the two most frequently used boundaries in the NS literature.
The other one was given in Ref.\,\cite{Lattimer1990} by further improving the NS EOS modeling\,\cite{Lattimer1990,ZhangLi2019EPJA} using the prescription $s^2\to1$ above a fiducial energy density,
\begin{equation}\label{bb-LPMY}
\boxed{
\mbox{Lattimer et al. (LPMY):}~~
R/\rm{km}\gtrsim4.51M_{\rm{NS}}/M_{\odot}.}
\end{equation}
In addition, Ref.\,\cite{Glendenning1992} constrained the boundary to be 
\begin{equation}
\boxed{
\mbox{Glendenning:}~~
R/\rm{km}\gtrsim4.34M_{\rm{NS}}/M_{\odot},}
\end{equation}
by considering first-order phase transitions with more than one conserved charge in NSs.
In Ref.\,\cite{Link1999}, the boundary is established by restricting the fractional moment of inertia of NS crust to $\Delta I/I\gtrsim1.4\%$,
\begin{equation}\label{Link99}
\boxed{
\mbox{Link et al.:}~~
R/\rm{km}\gtrsim3.6+3.9M_{\rm{NS}}/M_{\odot}.}
\end{equation}
Stellar models that are compatible with the lower bound on $\Delta I$ must fall below this line.
Furthermore, the causality boundaries predicted by the linear EOS $P=\zeta\varepsilon$ with $\zeta=\rm{const.}$ are $R/\rm{km}\gtrsim5.91M_{\rm{NS}}/M_{\odot}$ for an ``ultra-stiff'' EOS with $\phi=1$ and 
$R/\rm{km}\gtrsim6.89M_{\rm{NS}}/M_{\odot}$ for an URFG with $\phi=1/3$; see Eq.\,(\ref{MR-UST}) and (\ref{MR-URFG}).
For comparisons and ease of further discussions, we summarize the above causality boundaries on the M-R plot with several NS observational data in FIG.\,\ref{fig_NSMR-REV-cau} (see also FIG.\,\ref{fig_NSMR-REV}).

\subsection{Upper bound for the compactness of NSs at TOV configuration}\label{sub_UpperCompt}

\begin{figure}[h!]
\centering
\includegraphics[height=7cm]{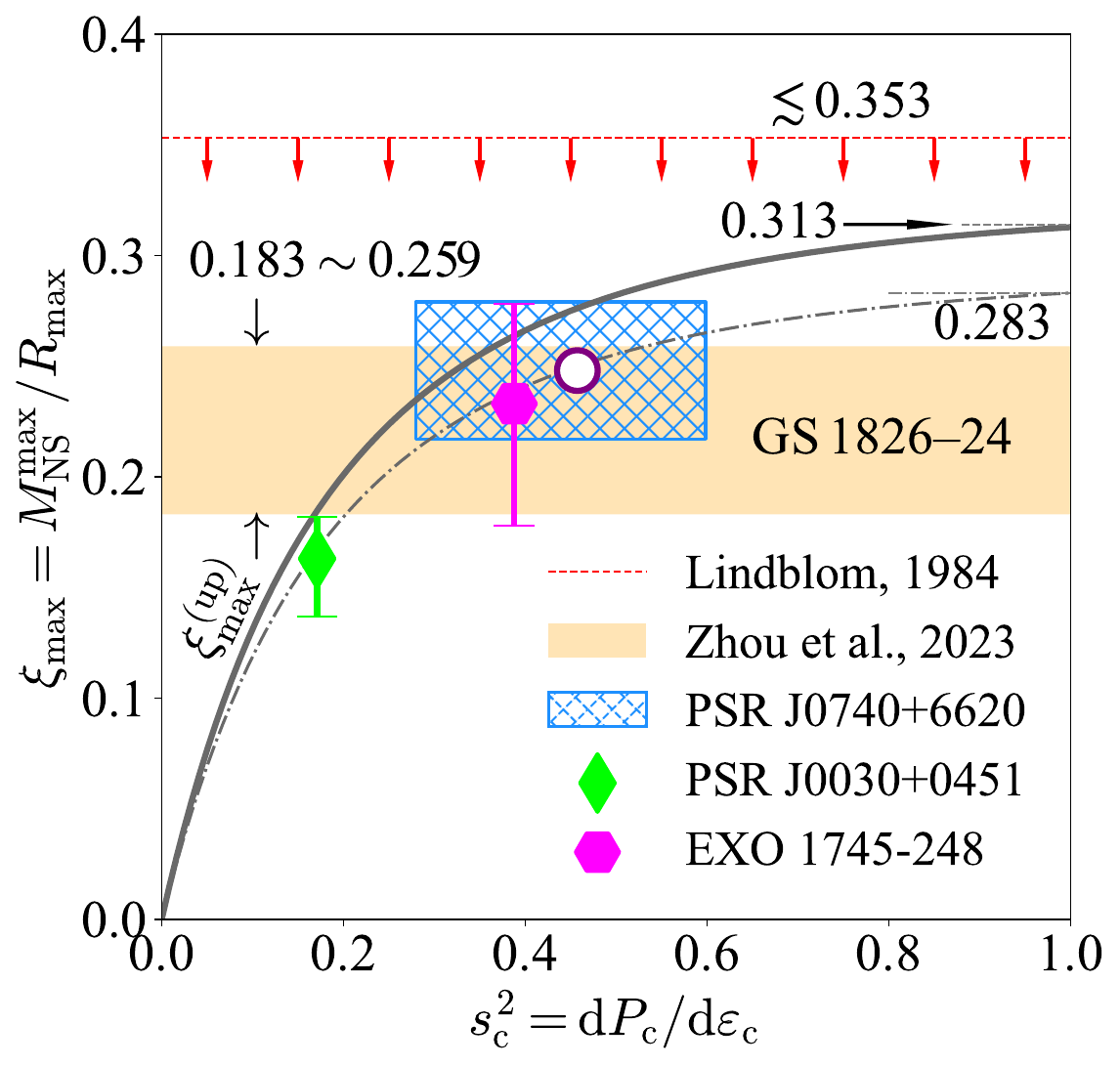}
\caption{(Color Online). Compactness parameter $\xi_{\max}$ for the maximum-mass configuration as a function of central SSS $s_{\rm{c}}^2$.  Figure taken from Ref.\,\cite{CLZ23-b} with a correction made on the compactness of the NS in GS 1826-24.}\label{fig_Compt}
\end{figure}

In this subsection, we study the upper limit for NS compactness, which is closely related to the lower (upper) bound on $\Delta$ (on $\phi$).
For the maximum-mass configuration on the NS M-R curve, one obtains the upper bound for $\xi_{\max}={M_{\rm{NS}}^{\max}}/{R_{\max}}$ by neglecting the intercepts $0.106$ and $0.64$ in Eq.\,(\ref{Mmax-G}) and Eq.\,(\ref{Rmax-n}) respectively, as
\begin{equation}
\xi_{\max}\equiv\frac{M_{\rm{NS}}^{\max}}{R_{\max}}
=\frac{M_{\rm{NS}}^{\max}/M_{\odot}}{R_{\max}/\rm{km}}
\left(\frac{M_{\odot}}{\rm{km}}\right)
<\frac{1.73\times10^3\Gamma_{\rm{c}}}{1.05\times 10^3\nu_{\rm{c}}}\left(\frac{M_{\odot}}{\rm{km}}\right)\approx\frac{2.44{\x}}{1+3{\x}^2+4{\x}}
\equiv \xi_{\max}^{\rm{(up)}}.
\label{C-ud}
\end{equation}
Therefore, $\xi_{\max}\lesssim0.313_{-0.01}^{+0.01}$ by taking ${\x}\approx0.374$, which is about $30\%$ smaller than the Buchdahl limit $4/9\approx0.444$.
In FIG.\,\ref{fig_Compt}, we plot the compactness parameter $\xi_{\max}$ of Eq.\,(\ref{C-ud}) as a function of the central SSS $s_{\rm{c}}^2$ which depends on the ${\x}$ via Eq.\,(\ref{sc2-TOV}), where the compactness for PSR J0740-6620 about $0.217\mbox{$\sim$}0.279$ (with the central value about 0.248) is shown by the lightblue hatched band (and the purple shallow circle) directly from its observational mass $\approx2.08_{-0.07}^{+0.07}M_{\odot}$\,\cite{Fon21} and radius $\approx12.39_{-0.98}^{+1.30}\,\rm{km}$\,\cite{Riley21}.
The compactness about $0.163_{-0.026}^{+0.019}$ for PSR J0030+0451\,\cite{Miller19} is shown by the light-green diamond (the SSS is obtained from its mass and radius using Eqs.\,(\ref{Mmax-G}), (\ref{Rmax-n}) and (\ref{sc2-TOV})).
A similar result for NS EXO 1745-248\,\cite{Ozel2016} is shown by the magenta hexagon.

We notice that neglecting the intercepts $0.106$ and $0.64$ in Eq.\,(\ref{Mmax-G}) and Eq.\,(\ref{Rmax-n}) slightly overestimates the compactness. Actually, by rewriting the form of $\xi_{\max}=M_{\rm{NS}}^{\max}/R_{\max}$, we can obtain 
\begin{equation}
\xi_{\max}
\lesssim
\Pi_{\rm{c}}(0.374)
\left(\frac{M_{\odot}}{\rm{km}}\right)
\frac{A_{\rm{M}}^{\max}}{A_{\rm{R}}^{\max}}\left[1+\frac{\rm{km}}{R_{\max}}
\left(\frac{B_{\rm{M}}^{\max}A_{\rm{R}}^{\max}}{A_{\rm{M}}\Pi_{\rm{c}}(0.374)}-B_{\rm{R}}^{\max}\right)\right]
\approx 0.313_{-0.01}^{+0.01}\cdot\left(1-\frac{1.14_{-0.3}^{+0.3}\,\rm{km}}{R_{\max}}\right),\end{equation} where $A_{\rm{M}}^{\max}$, $B_{\rm{M}}^{\max}$, $A_{\rm{R}}^{\max}$ and $B_{\rm{R}}^{\max}$ are defined in Eqs.\,(\ref{Mmax-G}) and (\ref{Rmax-n}),  and $1.14_{-0.3}^{+0.3}/R_{\max}\ll1$ could be treated as a correction to Eq.\,(\ref{C-ud}).  Considering a typical NS radius $R_{\max}\approx12.5\pm1\,\rm{km}$ (with a reduction about 10\% on $\xi_{\max}$),  it then leads to 
\begin{equation}\label{upp-xi}
\boxed{
\xi_{\max}\equiv \xi_{\rm{TOV}}\lesssim0.283_{-0.014}^{+0.014},~~\mbox{for NSs at the TOV configuration.}}
\end{equation}
shown by the grey dash-dotted line in FIG.\,\ref{fig_Compt}.
In this sense, the $\xi_{\max}\lesssim0.313$ from Eq.\,(\ref{C-ud}) provides an upper limit for the NS compactness.
An early constraint on the compactness $\xi$ about $\xi\lesssim0.353$\,\cite{Lind84} is indicated by the dashed red line (with a row of arrows).
Furthermore, the constraint on NS compactness can also be obtained from the gravitational red-shift $z$ defined by 
\begin{equation}\label{RF}
1+z=\left(1-\frac{2M_{\rm{NS}}}{R_{\rm{GR}}}\right)^{-1}
=\left(
1-\frac{2\xi}{\sqrt{1+z}}
\right)^{-1},~~R_{\rm{GR}}=\sqrt{1+z}R,~~
\xi=M_{\rm{NS}}/R.
\end{equation}
 Interestingly, a new constraint on $z$ was derived very recently from comparing X-ray burst model simulations with observational data for GS 1826-24 using newly measured atomic masses around the rp-process waiting-point nucleus $^{64}\rm{Ge}$\,\cite{Zhou23}. The resulting NS compactness is about $0.183\mbox{$\sim$}0.259$ with the most probable value 0.221 at 95\% confidence level (shown by the tan band). Its impact on both the EOS of supra-dense matter and the nature of GW190814's second component was studied very recently in Ref.\,\cite{Xie:2024mxu}.
 It was found that the EOS of high-density symmetric nuclear matter has to be softened appreciably while the symmetry energy at supersaturation densities stiffened compared to our prior knowledge. 
 
\begin{figure}[h!]
\centering
\includegraphics[height=6.5cm]{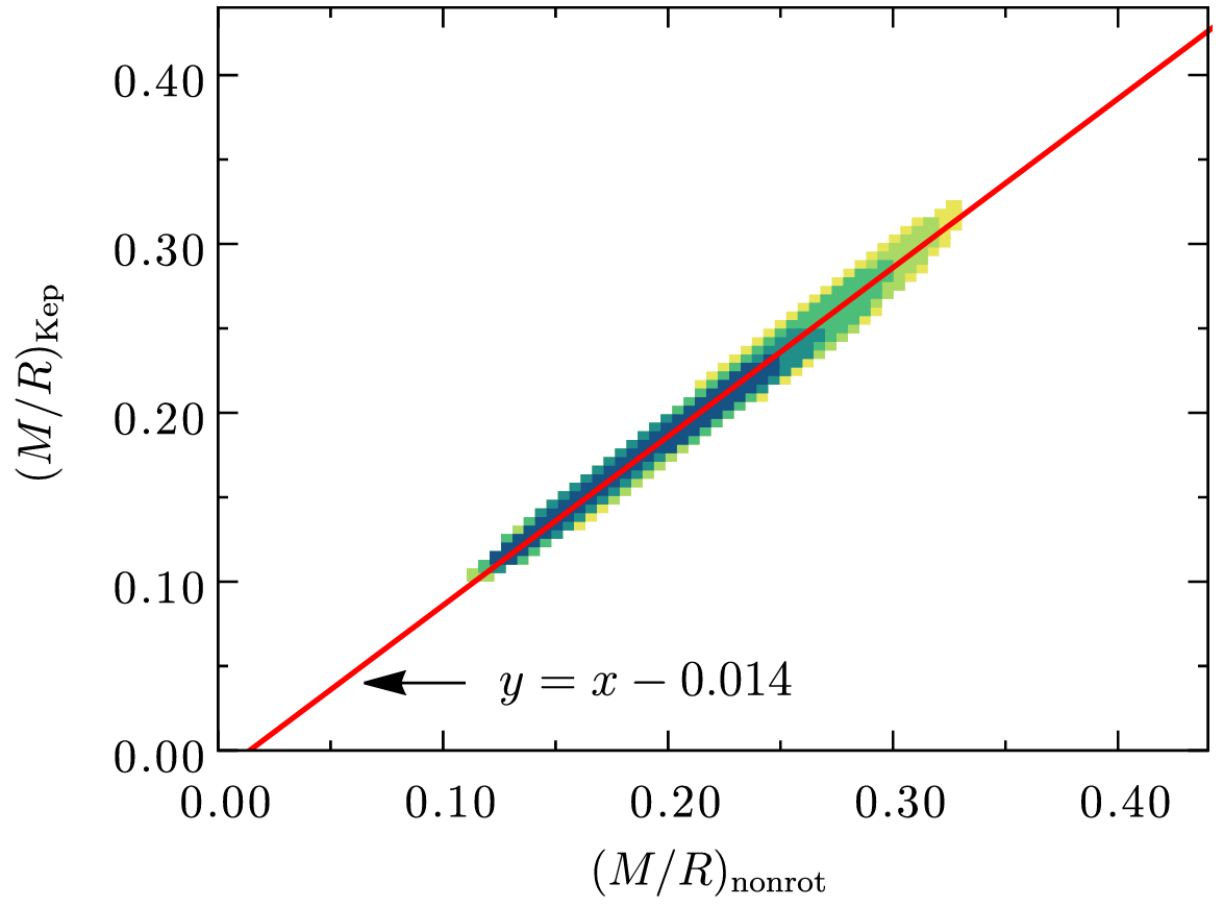}
\caption{(Color Online).  Universal relation between non-rotating and maximally rigidly rotating (Kepler) NS compactnesses under the restriction conditions $M_{\rm{NS}}^{\max}\gtrsim2M_{\odot}$ and the dimensionless tidal deformability of GW170817 being smaller than 720. Figure taken from Ref.\,\cite{Ann22}.
}
\label{fig_Ann22}
\end{figure}

In a recent study\,\cite{Ann22}, the universal relation between non-rotating and maximally rigidly rotating (Kepler) NS compactnesses  was explored under the conditions $M_{\rm{NS}}^{\max}\gtrsim2M_{\odot}$ and the dimensionless tidal deformability of GW170817 being smaller than 720, see FIG.\,\ref{fig_Ann22}.
They found that the compactness is upper limited as $\lesssim0.33$\,\cite{Ann22}.
{\color{xll}Compared to all of the above constrains available on NS compactness, ours is probably the most stringent one so far.}

We notice by passing that the compactness parameter is closely related to some bulk properties of NSs such as the moment of inertial $I$ and/or the NS binding energy $E_{\rm{b}}$\,\cite{Lattimer2001,Lattimer2004,Lattimer2007PR,XuJ,Bejger2003,
Steiner2016,LS05,Breu2016}. Therefore, the constraint on $\xi$ may naturally lead to some constraints on these quantities. For example, via the empirical expression for the moment of inertial as $I/M_{\rm{NS}}R^2\approx0.237(1+2.84\xi^2+18.9\xi^4)$ given by Ref.\,\cite{LS05}, one may obtain that $I/M_{\rm{NS}}R^2\lesssim0.49$ using $\xi\lesssim0.313$ or $I/M_{\rm{NS}}R^2\lesssim0.46$ using $\xi\lesssim0.283$.
Similarly, Ref.\,\cite{Lattimer2001} approximated the $E_{\rm{b}}$ as $E_{\rm{b}}/M_{\rm{NS}}\approx 0.6\xi/(1-\xi/2)$, and therefore $E_{\rm{b}}/M_{\rm{NS}}\lesssim0.22$ and $E_{\rm{b}}/M_{\rm{NS}}\lesssim0.20$ using $\xi\lesssim0.313$ and $\xi\lesssim0.283$, respectively.
In addition, the crustal fraction of the NS moment of inertia is approximately given by\,\cite{Lattimer2001,XuJ},
\begin{equation}
\frac{\Delta I}{I}\approx\frac{28\pi P_{\rm{t}}R^3}{3M_{\rm{NS}}}\frac{1.67\xi-0.6\xi^2}{\xi}\left[1+\frac{2P_{\rm{t}}(1+5\xi-14\xi^2)}{\rho_{\rm{t}}M_{\rm{N}}\xi^2}\right]^{-1},
\end{equation}
where $\rho_{\rm{t}}$ and $P_{\rm{t}}$ are the core-crust transition density and the pressure there, respectively.
We will not discuss these quantities further in this review since they could be treated as direct corollaries of the constraints on $\xi$.

\subsection{Can a massive NS have a radius smaller than about 10\,km?}\label{sub_10km}

The mass scaling of Eq.\,(\ref{Mmax-G}) and radius scaling of Eq.\,(\ref{Rmax-n}) together enable us to obtain a causality boundary, since they describe NSs at the TOV configuration.
Since the factor $(1+3{\x}^2+4{\x})/{\x}$ takes its minimum $\approx7.8$ at ${\x}\approx0.374$, we have from Eqs.\,(\ref{Mmax-G}) and (\ref{Rmax-n}) or equivalently Eq.\,(\ref{rel-1}) that\,\cite{CLZ23-b},
\begin{equation}\label{rel-2}
\boxed{
\mbox{intrinsic analysis from scaled TOV equations (this work):}~~
{R_{\max}}/{\rm{km}}\gtrsim 4.73{M_{\rm{NS}}^{\max}}/{M_{\odot}}+1.14.}
\end{equation}

\begin{figure}[h!]
\centering
\includegraphics[width=9.cm]{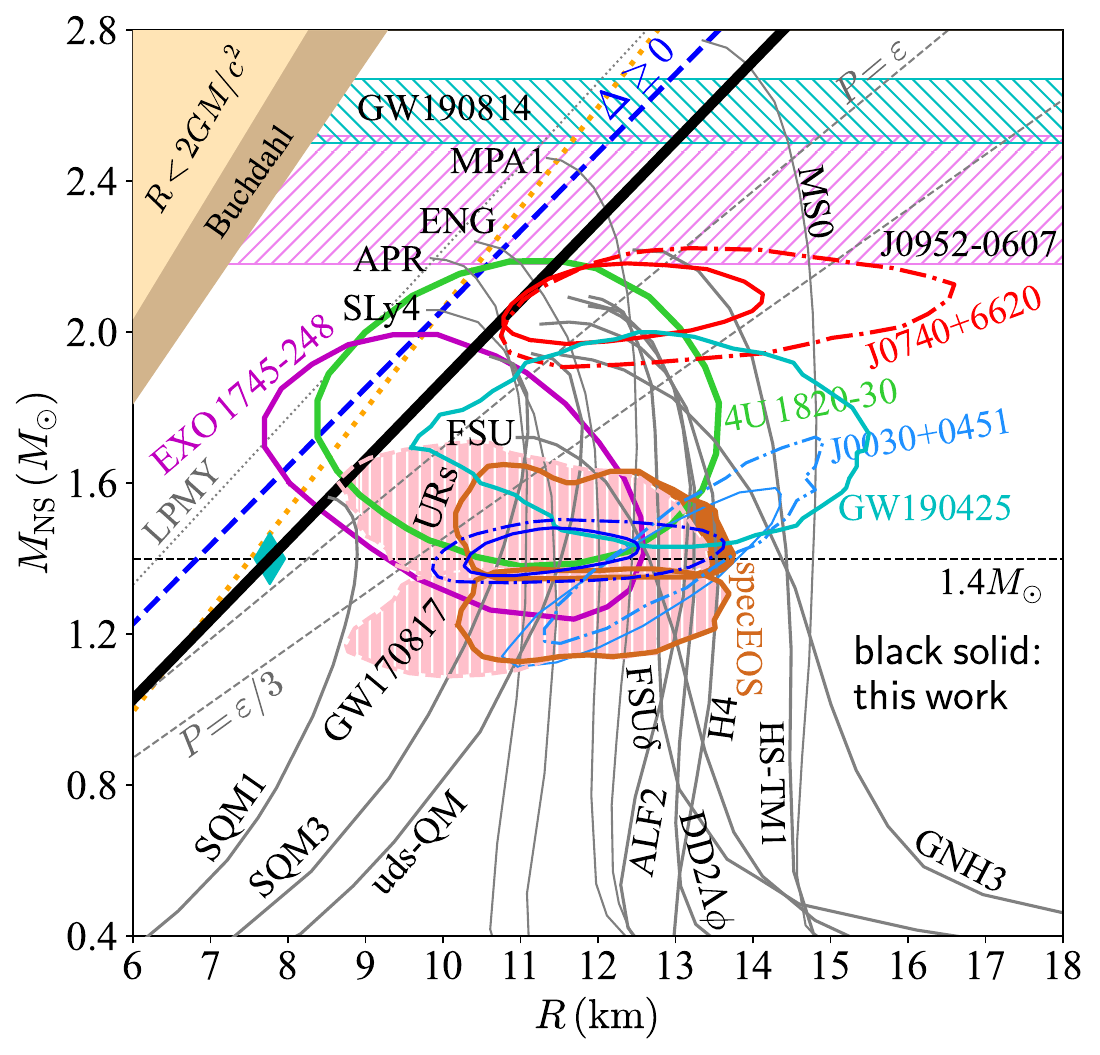}
\caption{(Color Online). Causality boundaries for NS M-R curve, here the black solid line is based on inequality (\ref{rel-2}) corresponding to the black line in the right panel of FIG.\,\ref{fig_MmaxS}.  The prediction using the trace anomaly $\Delta\equiv1/3-P/\varepsilon\geq0$ (blue dashed line) as well as the previous constraint $R_{\max}/\rm{km}\gtrsim  4.51M_{\rm{NS}}^{\max} /M_{\odot}$ (grey dotted line) are also shown for comparisons. The M-R constraints for several NSs (magenta/green/lightblue/red contours), M-R curves from a few typical empirical dense matter EOSs (grey solid curves), and the constraints for the M-R relation based on events GW170817 (chocolate/pink bands)\,\cite{Abbott2017,Abbott2018} and GW190425 (cyan solid contour)\,\cite{Abbott2020-a} are also shown.
See text for details.
Figure taken from Ref.\,\cite{CLZ23-b} with modifications by adding the observed mass and radius of PSR J0437-4715\,\cite{Choud24} (shown as the blue contours without nearby captions) and the boundary given in Ref.\,\cite{Ofeng2023} (orange dotted line).
}\label{fig_MR-C}
\end{figure}

\begin{figure}[h!]
\centering
\includegraphics[height=6.15cm]{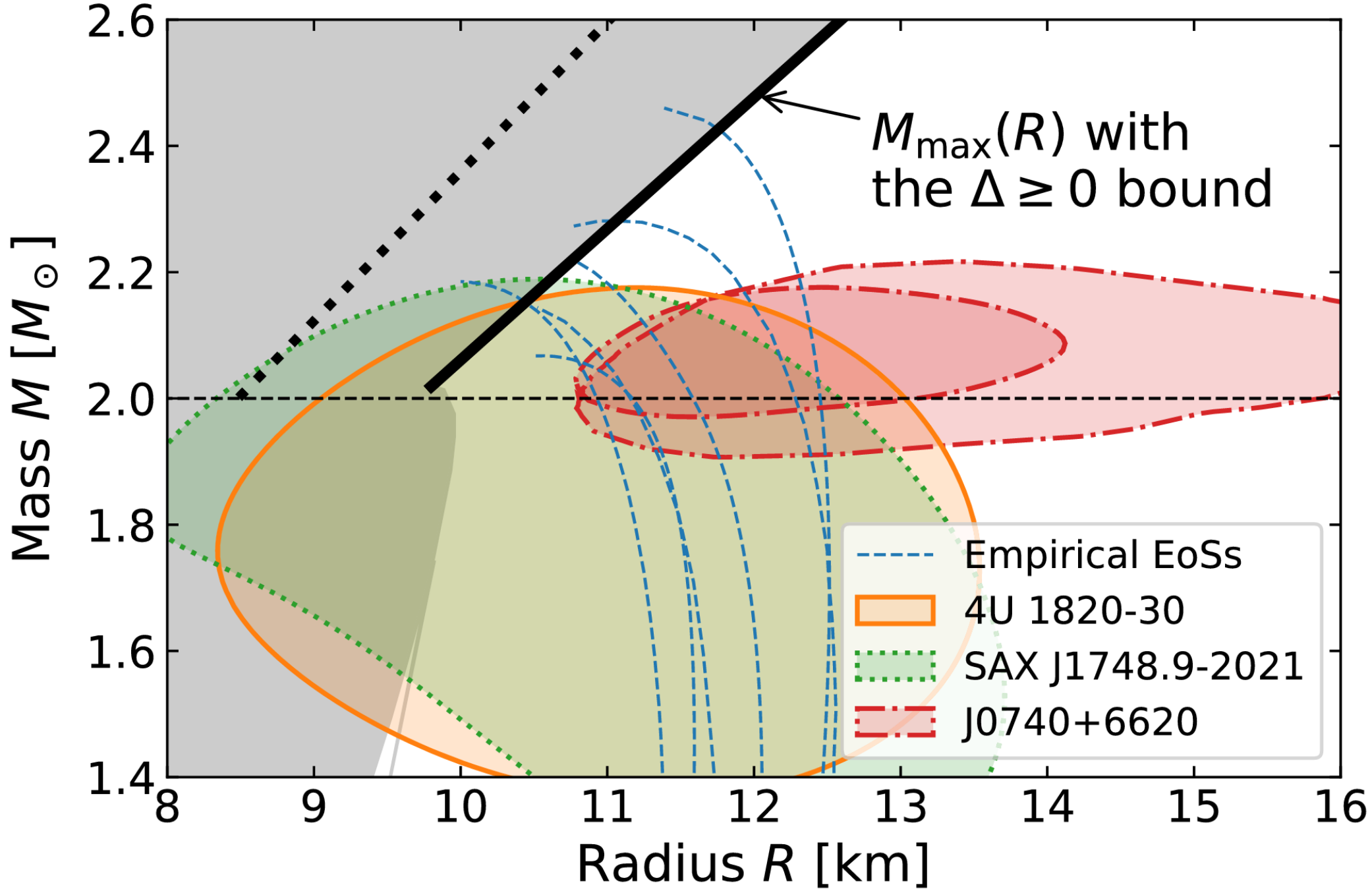}\qquad
\includegraphics[height=6.25cm]{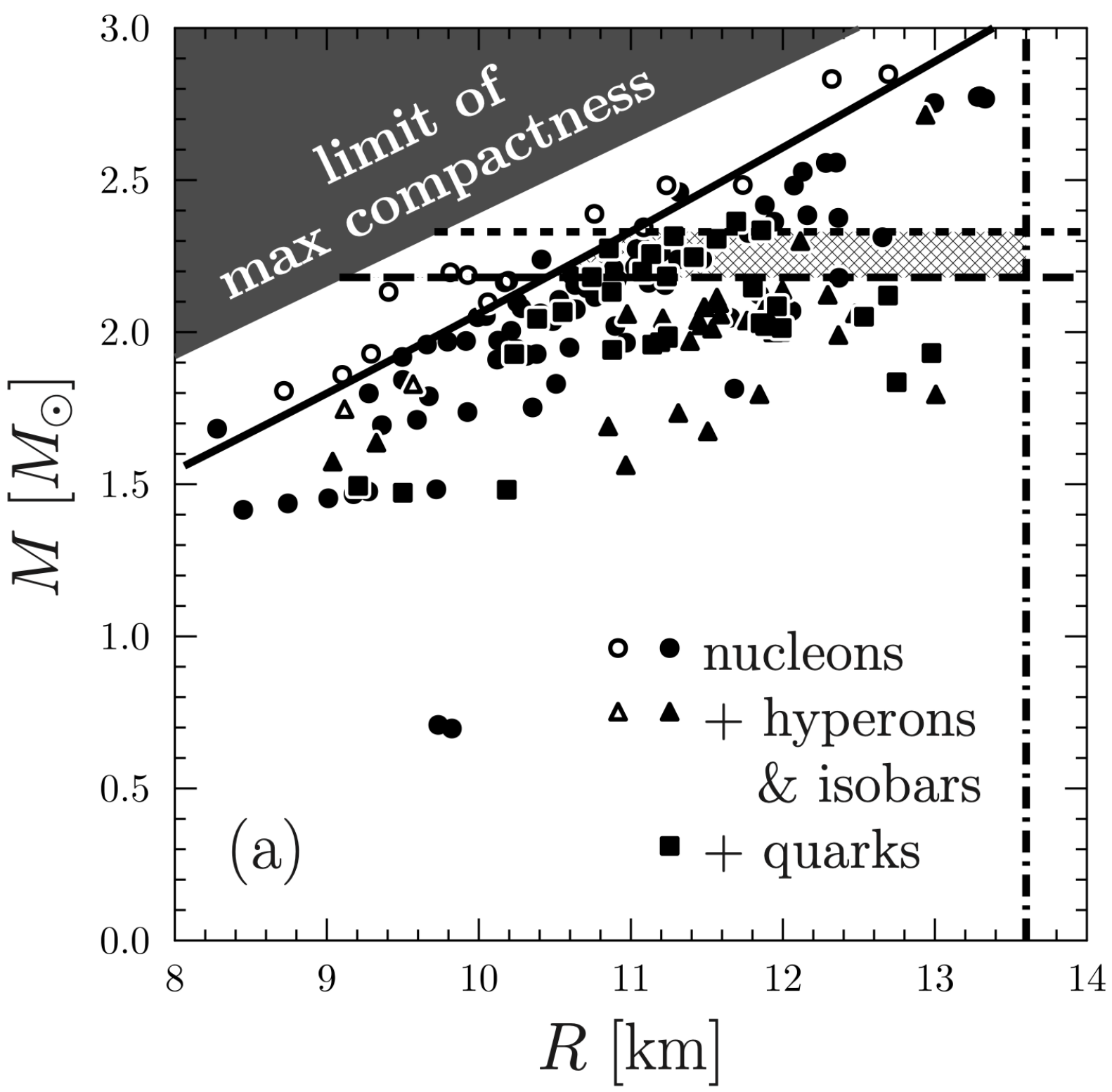}
\caption{(Color Online).  Left panel: NS M-R curve with the causality boundary via $\Delta\geq0$ is shown by the thick solid black line. Figure taken from Ref.\,\cite{Fuji22}.
Right panel: NS M-R curve with a recent empirical analysis on the causality boundary shown by the solid black line using phenomenological EOSs with/without considering exotic degree of freedoms such as quarks and/or hyperons. Figure taken from Ref.\,\cite{Ofeng2023}.
}\label{fig_Fuji-C}
\end{figure}

In FIG.\,\ref{fig_MR-C}, we show the causality boundary of (\ref{rel-2}) (black solid line) together with the M-R constraints for a few NSs including the PSR J0740+6620, PSR J0030+0451 and PSR J0437-0475\,\cite{Fon21,Riley21,Miller21,Salmi22,Riley19,Miller19,Choud24}, the 4U 1820-30 and EXO 1745-248\,\cite{Ozel16-a}, the recently reported black widow PSR J0952-0607 with a mass about 2.35$M_{\odot}$\,\cite{Romani22},  GW190814’s secondary component with a mass of $2.59^{+0.08}_{-0.09}M_{\odot}$\,\cite{Abbott2020} and the M-R curves from several empirical dense matter EOSs\,\cite{Muther87,Engvik96,Muller96,Akmal1998,Glendenning85,Douchin01,Lackey06, AFL2,Prakash95,Hempel10,Chu14,Banik14,FSU05,LiF2022} (grey solid curves).
The violet and the chocolate bands are obtained from the GW170817 event using the universal relations (URs) and the spectral EOS (specEOS) approaches\,\cite{Abbott2018}, respectively, while the cyan solid contour is based on the GW190425 event for a compact binary coalescence with a total mass about 3.4$M_{\odot}$\,\cite{Abbott2020-a}. Additionally, the grey dotted line marked as ``LPMY'' was empirically given by Ref.\,\cite{Lattimer1990} as in the inequality (\ref{bb-LPMY}).
The causality boundary for NS M-R curve was recently localized using an analysis of trace anomaly\,\cite{Fuji22} via the condition $\Delta\geq0$. It can be written as (blue dashed line in FIG.\,\ref{fig_Fuji-C})\,\cite{Fuji22}
\begin{equation}\label{FujiCB}
\boxed{
\mbox{Fujimoto et al.:}~~
R_{\max}/\rm{km}\gtrsim4.83M_{\rm{NS}}^{\max}/M_{\odot}+0.04.}
\end{equation}
See the left panel of FIG.\,\ref{fig_Fuji-C}.
In addition, a recent empirical analysis on the causality boundary using phenomenological EOSs with/without considering exotic degree of freedoms such as quarks and hyperons/isobars found that\,\cite{Ofeng2023}, 
\begin{equation}\label{Ofeng23CB}
\boxed{
\mbox{Ofengeim et al.:}~~
R_{\max}/\rm{km}\gtrsim3.75M_{\rm{NS}}^{\max}/M_{\odot}+2.27,}
\end{equation}
see the right panel of FIG.\,\ref{fig_Fuji-C}.

We can see from FIG.\,\ref{fig_MR-C} that even an ``ultra-stiff'' EOS $P=\varepsilon$ may introduce some tension with the observational data of NSs,  by comparing the upper grey dashed line (marked by ``$P=\varepsilon$'') and the red solid contour for PSR J0740+6620 (at a 68\% confidence level).
This is because the dense matter in NS cores (especially near the centers) could not be described by the linear EOS $P=\phi\varepsilon$; consequently the SSS is nontrivial and nonlinear depending on the $\x$.
Furthermore, compared with the observational data, the URFG EOS (with $P=\varepsilon/3$) leads to quite an inconsistent causality boundary (shown as the lower grey dashed line marked by ``$P=\varepsilon/3$''), see Eq.\,(\ref{MR-URFG}).

\begin{figure}[h!]
\centering
\includegraphics[width=14.cm]{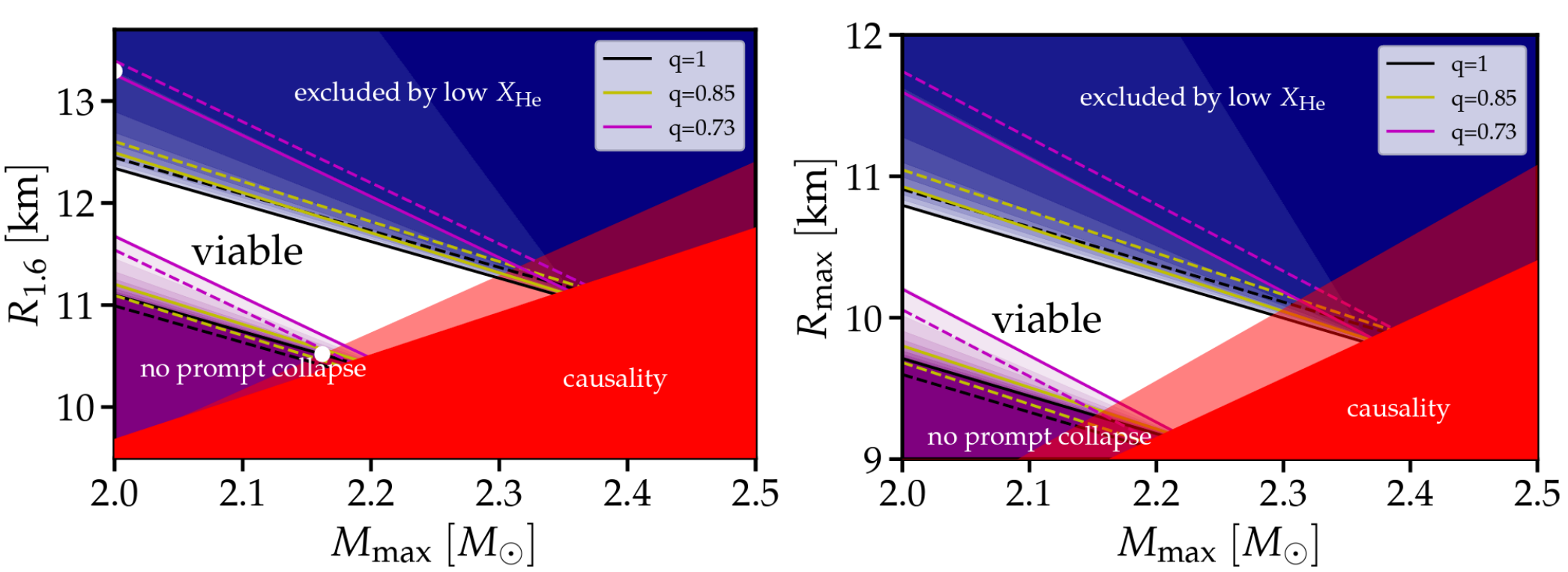}
\caption{(Color Online). Constraints on the radius of a $1.6M_{\odot}$ NS (left) and that of a NS at the TOV configurations using the threshold mass for prompt black-hole formation from the absence of strong helium features in the kilonova spectrum and the signals from GW 170817;
viable regions are obtained considering multiple constraints. For more details see Ref.\,\cite{Sneppen2024} where this figures is taken from.
}\label{fig_SneppenF1}
\end{figure}

Interestingly, our result (\ref{rel-2}) puts a more stringent constraint on the M-R curve than the frequently-used boundaries of Ref.\,\cite{Lind84} and Ref.\,\cite{Lattimer1990}.
More importantly,  it is very consistent with the observational data which basically excludes several celebrated empirical dense matter EOSs.
In particular, the observational boundary of M-R data for PSR J0740+6620\,\cite{Fon21,Riley21,Miller21,Salmi22} and those from the events GW170817\,\cite{Abbott2018} and GW190425\,\cite{Abbott2020-a} are highly consistent with our relation (\ref{rel-2}).
Therefore, it is important to check the new causality boundary (\ref{rel-2}) as more observational data on NS masses/radii or GW events become available (see similar discussions given in Refs.\,\cite{Fuji22,Ofeng2023} and the plots in FIG.\,\ref{fig_Fuji-C}). 
We would like to emphasize {\color{xll}that the causality boundary (\ref{rel-2}) is a direct and intrinsic consequence of the TOV equations,  without relying on any input EOS models. Therefore, it should not necessarily be consistent with microscopic nuclear EOSs;
and its correctness could only be checked by observational NS data similar as the upper bound for $\x$ about 0.374. Inversely, incorporating the boundary (\ref{rel-2}) may further help us construct more reliable nuclear EOSs.}
As an illustration of applying the new causal boundary on the M-R curve, we give a lower limit for the radius of the black widow PSR J0952-0607 as\label{Romani22}
\begin{equation}
R_{\max}\gtrsim12.25\,\rm{km}, \mbox{  for  PSR J0952-0607}.
\end{equation}
On the other hand, for a canonical NS, its radius is bounded from below to about 7.77\,km (shown as the cyan solid diamond in FIG.\,\ref{fig_MR-C}).
For massive NSs, we have
\begin{equation}
    \boxed{R_{\max}\gtrsim10.6\,\rm{km},~~\mbox{for}~~M_{\rm{NS}}^{\max}/M_{\odot}\gtrsim2,
    }
\end{equation}
this means massive NSs can hardly have radii smaller than about 10\,km.
On the opposite side, if the radii of NSs were known (constrained), e.g., $R_{\max}\approx12.39\,\rm{km}$ (for PSR J0740+6620\,\cite{Riley21} from NICER), using the causal limit ${\x}\lesssim0.374$ gives an upper limit $M_{\rm{NS}}^{\max}\lesssim2.38M_{\odot}$. Similarly if $R_{\max}\approx11.41\,\rm{km}$ or $13.69\,\rm{km}$ (the lower/upper constraints for the radius of PSR J0740+6620\,\cite{Riley21}) is adopted, one has $M_{\rm{NS}}^{\max}\lesssim2.17M_{\odot}$ or $M_{\rm{NS}}^{\max}\lesssim2.65M_{\odot}$, respectively.
These features were discussed around FIG.\,\ref{fig_Ypm_sk}. Interestingly, in a very recent study\,\cite{Sneppen2024}, the radius $R_{1.6}$ of a $1.6M_{\odot}$ NS and $R_{\max}$ of a NS at the TOV configuration are constrained by using the astrophysical helium spectrum combined with the GW170817 signals. They found that $R_{1.6}\approx11.4\,\rm{km}$ for $M_{\rm{TOV}}/M_{\odot}\approx2.3$ (shown in the left panel of FIG.\,\ref{fig_SneppenF1}). Moreover, $R_{\max}\gtrsim10.2\,\rm{km}$ is found if $M_{\rm{TOV}}/M_{\odot}\approx2$ (right panel of FIG.\,\ref{fig_SneppenF1}). Their finding is very close to our limit on $R_{\max}$ given previously.
For $M_{\rm{TOV}}/M_{\odot}\approx2.3$, the $R_{\max}$ is found to fall within a very narrow region around 10\,km, indicating that for even larger $M_{\rm{TOV}}/M_{\odot}\gtrsim2.3$ no parameter space for $R_{\max}$ exists\,\cite{Sneppen2024}.
The decreasing upper limit (based on helium spectrum) on radius (for both $R_{1.6}$ and $R_{\max}$) with $M_{\rm{TOV}}$ from  Ref.\,\cite{Sneppen2024} contrasts with predictions of many EOS models, where large radii are typically accompanied by large values of $M_{\rm{TOV}}$. This finding may put more stringent constraints on modeling NS EOS.

\begin{figure}[h!]
\centering
\includegraphics[height=7.5cm]{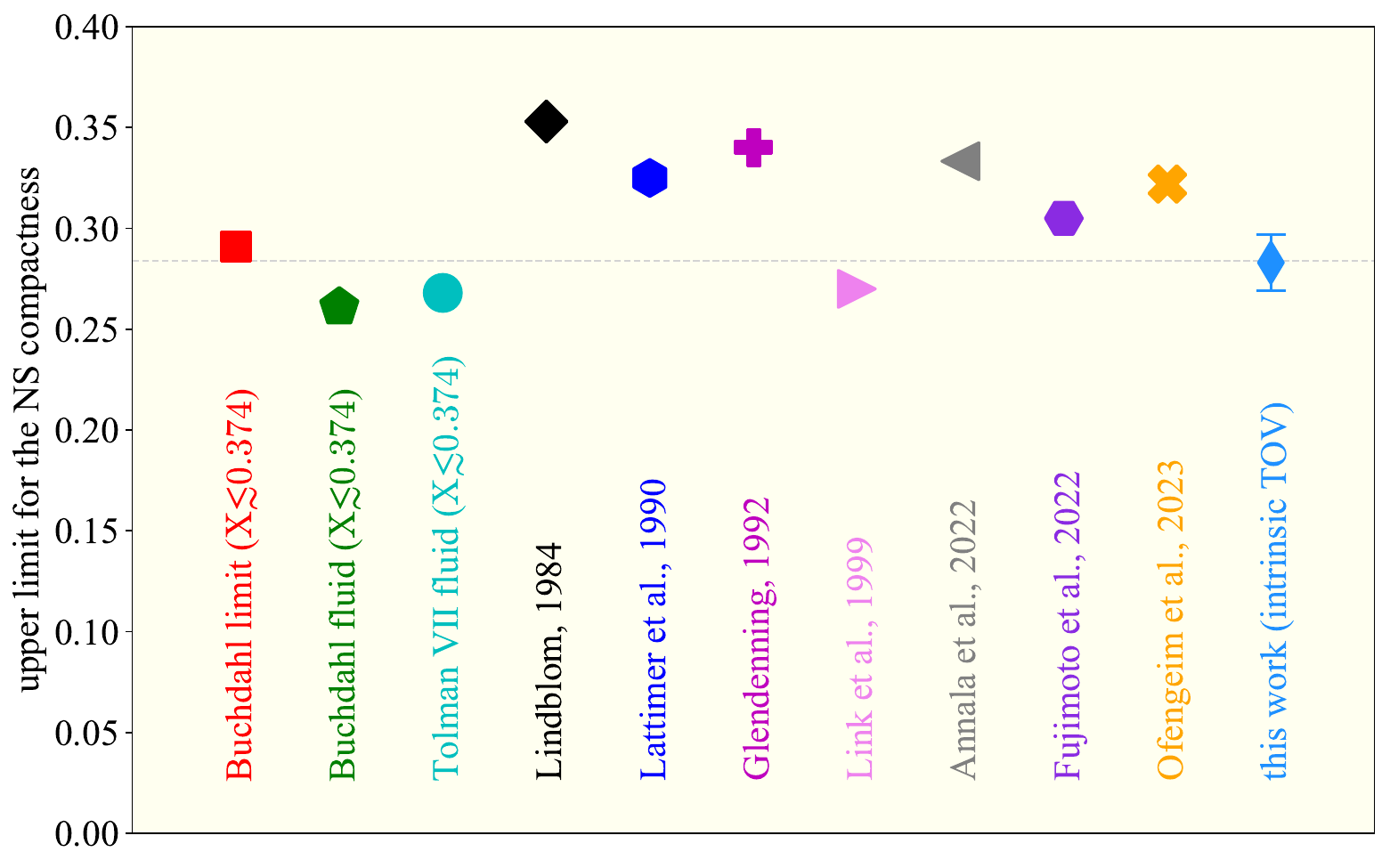}
\caption{(Color Online).  A partial list of the upper limits for the NS compactness, see the text for details.
}\label{fig_CompBuch}
\end{figure}

One feature of FIG.\,\ref{fig_NSMR-REV-cau} is that the causality boundaries given by the uniform incompressible fluid model (Buchdahl limit)\,\cite{Buch59} and the Buchdahl fluid\,\cite{Buch67} combined with the upper limit on $\x$ revealed by our analysis about 0.374 (inequality (\ref{Xupper})) are quite consistent with (\ref{rel-2}) as well as the inequalities (\ref{FujiCB}) and (\ref{Ofeng23CB}).
This can be more clearly demonstrated using the compactness limits. Using the uniform incompressible fluid model with $\x\lesssim0.374$, the compactness is upper bounded as $\xi\lesssim0.291$ since $R\gtrsim3.44M_{\rm{NS}}$ and the Buchdahl fluid model gives $\xi\lesssim0.261$ under the same condition $\x\lesssim0.374$. Our boundary of (\ref{rel-2}) then gives $\xi\lesssim0.283$ as shown in Eq.\,(\ref{upp-xi}) by adopting $R\approx12.5\,\rm{km}$.
Under the same radius condition, inequalities (\ref{FujiCB}) and (\ref{Ofeng23CB}) restrict the compactness as $\xi\lesssim0.305$ and $\xi\lesssim0.322$, respectively; while (\ref{Link99}) gives $\xi\lesssim0.270$.
As a further example, an analytical expression for $\x$ is available for the Tolman VII fluid, where the energy density profile takes the form of $\varepsilon(r)=\varepsilon_{\rm{c}}[1-(r/R)^2]$\,\cite{TOV39-1}.
The result is\,\cite{Saes2024}
\begin{align}\label{bb-3}
\x=&\frac{2}{15}\left\{\sqrt{\frac{3}{\xi}}\tan\left[
\arctan\sqrt{\frac{1}{3}\frac{\xi}{1-2\xi}}+\frac{1}{2}\ln\left(\frac{1}{6}+\sqrt{\frac{1-2\xi}{3\xi}}\right)
-\frac{1}{2}\ln\left(\frac{1}{\sqrt{3\xi}}-\frac{5}{6}\right)
\right]-\frac{5}{2}\right\}\notag\\
\approx&\frac{\xi}{2}\left(1+\frac{133}{60}\xi+\frac{599}{112}\xi^2+\frac{17915}{1344}\xi^3+\cdots\right)
.\end{align}
Requiring $\x\lesssim0.374$ gives $\xi\lesssim0.268$.

We summarize these results in FIG.\,\ref{fig_CompBuch}. It is seen that the constraints on the upper limit for $\xi$ are consistent.
{\color{xll}Although the uniform incompressible fluid (Buchdahl limit), the Buchdahl fluid and the Tolman VII fluid are simplified models (on the dense matter EOS),  the expressions for $\x$ in these models in Eqs.\,(\ref{bb-1}), (\ref{bb-2}) and (\ref{bb-3}) contain important information especially near the NS centers, therefore they could give reasonable prediction on the $\xi$.}
Moreover, the leading-order terms in Eqs.\,(\ref{bb-1}), (\ref{bb-2}) and (\ref{bb-3}) for $\x$ are all $\xi/2$ while the higher-order terms in $\xi$ differ among them.

\subsection{Ultimate energy density allowed in generally stable NSs}\label{sub_ultimate}

Actually, the correlation $M_{\rm{NS}}^{\max}$-$\Gamma_{\rm{c}}$ alone is already useful for inferring the ultimate (maximum) energy density $\varepsilon_{\rm{ult}}$ as well as the ultimate pressure $P_{\rm{ult}}$ allowed in NSs before they collapse into BHs.
By rewriting  Eq.\,(\ref{Mmax-G}) and considering the casual limit $\x\lesssim0.374$, we obtain
\begin{equation}\label{eps_ult}
\boxed{
\varepsilon_{\rm{c}}\leq\varepsilon_{\rm{ult}}\equiv 6.32\left(\frac{M_{\rm{NS}}^{\max}}{M_{\odot}}+0.106\right)^{-2}\,\rm{GeV}/\rm{fm}^3.}
\end{equation}
Its scaling with mass is $\varepsilon_{\rm{ult}}\sim M_{\rm{NS}}^{\max,-2}\cdot[1+\cdots]$, see Eq.\,(\ref{ci-1}) for a similar relation holding for generally stable NSs away from the TOV configuration.
Since (\ref{eps_ult}) is based on the mass scaling for NSs at the TOV configuration, this gives the absolute upper limit for the energy density (the meaning of ``ultimate'').
Using the constraint ${\x}\lesssim0.374$ once again gives,
\begin{equation}
\boxed{
\label{P_ult}
P_{\rm{c}}\leq P_{\rm{ult}}\equiv 2.36\left(\frac{M_{\rm{NS}}^{\max}}{M_{\odot}}+0.106\right)^{-2}\,\rm{GeV}/\rm{fm}^3.}
\end{equation}
We note that the correlation between $R_{\max}$ and $\nu_{\rm{c}}$ of Eq.\,(\ref{Rmax-n}) is not used in establishing these limits.

\begin{figure}[h!]
\centering
\includegraphics[width=9.cm]{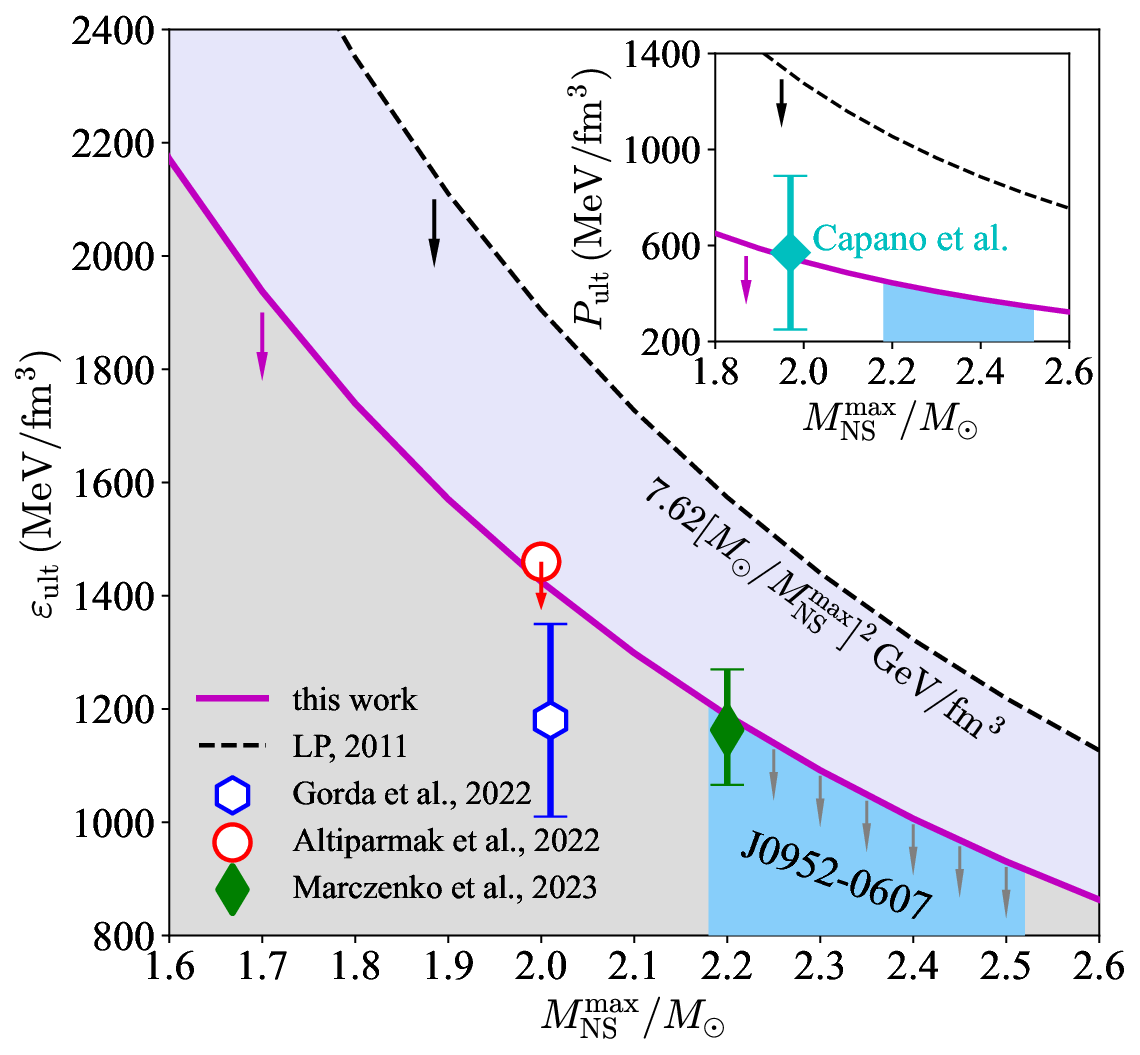}
\caption{(Color Online). The ultimate energy density $\varepsilon_{\rm{ult}}$ and the pressure $P_{\rm{ult}}$ (inset) allowed in NS cores as functions of the maximum-mass $M_{\rm{NS}}^{\max}$ of NS M-R relation.
The previous predictions on $\varepsilon_{\rm{ult}}$ and $P_{\rm{ult}}$ by Lattimer and Prakash in Ref.\,\cite{Lattimer10} are also shown for comparison (black dashed line).  Figure taken from Ref.\,\cite{CLZ23-b}.
}\label{fig_ULTeps}
\end{figure}

\begin{figure}[h!]
\centering
\includegraphics[height=8.cm]{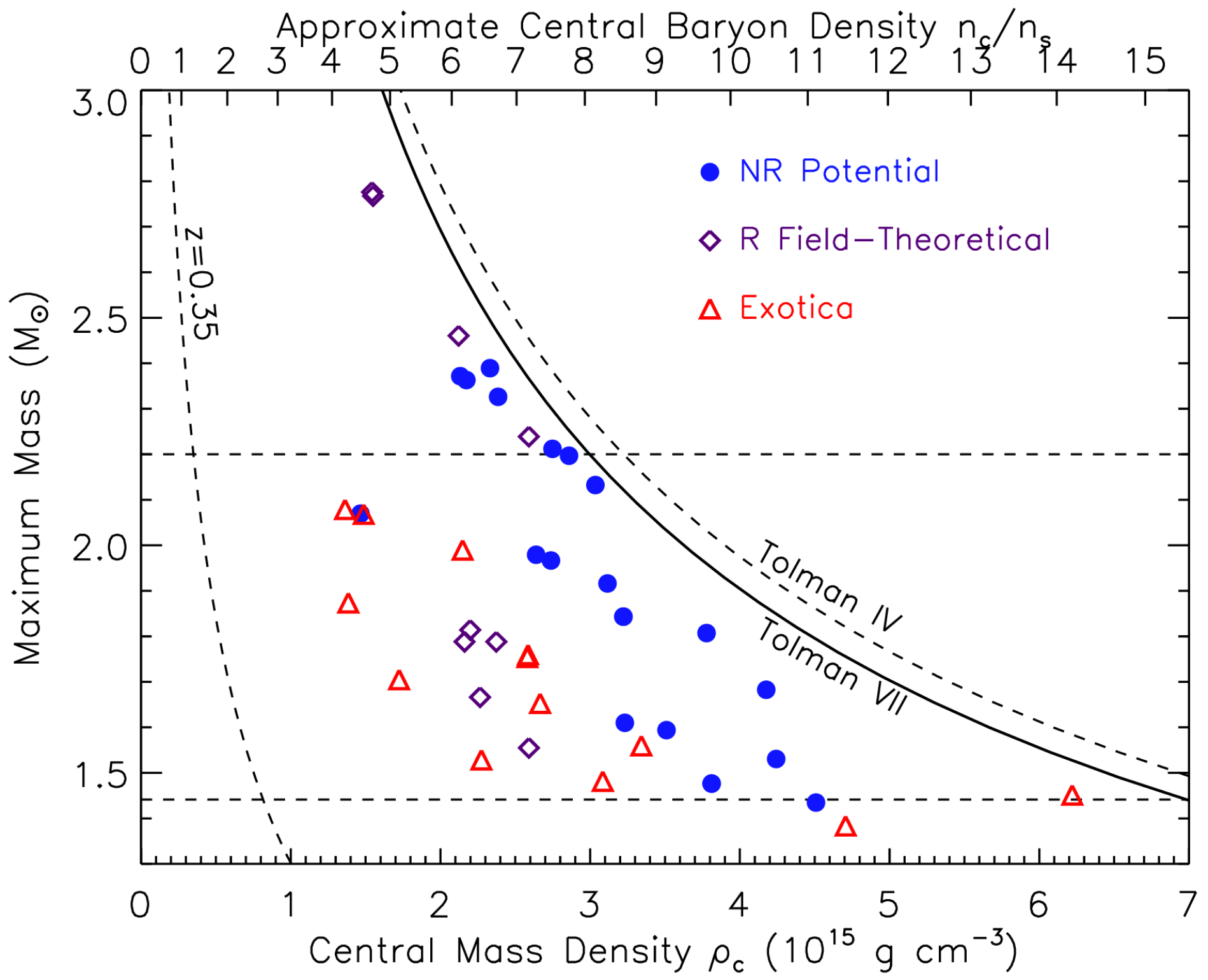}
\caption{(Color Online). Dependence of $M_{\rm{NS}}^{\max}$ on the central energy density obtained in Ref.\,\cite{Lattimer05}, see also an updated discussion given in Ref.\,\cite{Lattimer10}. Figure taken from Ref.\,\cite{Lattimer05}.
}\label{fig_LP05FIG}
\end{figure}

Based on Eq.\,(\ref{eps_ult}) for $\varepsilon_{\rm{ult}}$ and Eq.\,(\ref{P_ult}) for $P_{\rm{ult}}$,  we find that the existence of a $2.08M_{\odot}$ ($1.97M_{\odot}$) NS leads to $\varepsilon_{\rm{c}}\lesssim1.32\,\rm{GeV}/\rm{fm}^3$ ($1.47\,\rm{GeV}/\rm{fm}^3$) and $P_{\rm{c}}\lesssim494\,\rm{MeV}/\rm{fm}^3$ ($548\,\rm{MeV}/\rm{fm}^3$), respectively.
It is necessary to point out that the upper limit $P_{\rm{c}}\lesssim548\,\rm{MeV}/\rm{fm}^3$ obtained from assuming $M_{\rm{NS}}^{\max}\gtrsim1.97M_{\odot}$\,\cite{Dem10} is quite consistent with $P_{\max}\lesssim570\,\rm{MeV}/\rm{fm}^3$ shown by the cyan diamond with error bars in the inset of FIG.\,\ref{fig_ULTeps} obtained from a recent inference in Ref.\,\cite{Capano2020}.
The general $M_{\rm{NS}}^{\max}$-dependence of $\varepsilon_{\rm{ult}}$ and that of $P_{\rm{ult}}$ are shown by the magenta lines (together with the grey band). Also shown are the predictions on $\varepsilon_{\rm{ult}}$ and $P_{\rm{ult}}$ (black dashed lines with the lavender band) by Lattimer and Prakash in Ref.\,\cite{Lattimer10}, 
\begin{equation}\label{LP11ult}
\boxed{
\varepsilon_{\rm{ult}}\approx7.62\left(\frac{M_{\odot}}{M_{\rm{NS}}^{\max}}\right)^2\,\rm{GeV}/\rm{fm}^3,~~
 P_{\rm{ult}}\approx5.12\left(\frac{M_{\odot}}{M_{\rm{NS}}^{\max}}\right)^2\,\rm{GeV}/\rm{fm}^3,}
 \end{equation}
 which are cited as ``LP, 2011'' in FIG.\,\ref{fig_ULTeps}. Shown in FIG.\,\ref{fig_LP05FIG} is the original plot of the central energy density dependence of the mass $M_{\rm{NS}}^{\max}$ given in Refs.\,\cite{Lattimer05,Lattimer10}.
Considering $M_{\rm{NS}}^{\max}/M_{\odot}\approx2.08$, the above relations (\ref{LP11ult}) lead to the estimates $\varepsilon_{\rm{ult}}\lesssim1.76\,\rm{GeV}/\rm{fm}^3$ and $P_{\rm{ult}}\lesssim1.18\,\rm{GeV}/\rm{fm}^3$, which are about 33\% and 139\% larger than those from Eq.\,(\ref{eps_ult}) and Eq.\,(\ref{P_ult}), respectively. These differences are easy to understand since Refs.\,\cite{Lattimer05,Lattimer10} used simplified EOS models like the Tolman VII fluid and the Buchdahl fluid. Their limits on the ultimate energy density (as well as the pressure) are thus not tight enough, compared with our largely EOS model independent predictions of Eq.\,(\ref{eps_ult}) and Eq.\,(\ref{P_ult})
as we have analytically demonstrated and numerically verified\,\cite{CLZ23-a}.
In addition, we find that the constraint $\varepsilon_{\rm{c}}\lesssim1.41\,\rm{GeV}/\rm{fm}^3$ (by using $M_{\rm{NS}}^{\max}/M_{\odot}\approx2.01$\,\cite{Ant13} in Eq.\,(\ref{eps_ult})) is close to $\varepsilon_{\rm{c}}\lesssim1.18_{-0.17}^{+0.17}\,\rm{GeV}/\rm{fm}^3$ from recent ab-initio QCD calculations\,\cite{Gorda2023}, which also assumed the NS masses are greater than $2.01M_{\odot}$ as shown by the blue hexagon in  FIG.\,\ref{fig_ULTeps}. Similarly, we have the limit $\varepsilon_{\rm{c}}\lesssim1.42\,\rm{GeV}/\rm{fm}^3$ (using $M_{\rm{NS}}^{\max}/M_{\odot}\approx2$ in Eq.\,(\ref{eps_ult})), which is close to the limit $1.46\,\rm{GeV}/\rm{fm}^3$ of Ref.\,\cite{Altiparmak2022} where the algorithm rejects masses $\leq 2M_{\odot}$ (red circle).
Furthermore, the maximum central energy density $\lesssim1.16_{-0.10}^{+0.11}\,\rm{GeV}/\rm{fm}^3$ from Ref.\,\cite{Mar23} is also shown using the green diamond, in which the maximum mass $M_{\rm{NS}}^{\max}$ is greater than about 2.2$M_{\odot}$.
Using our estimate of Eq.\,(\ref{eps_ult}), for such NSs with $M_{\rm{NS}}^{\max}\approx2.2M_{\odot}$ the central energy density is found to be about $1.19\,\rm{GeV}/\rm{fm}^3$.
We summarize in TAB.\,\ref{tab_ult} these constraints on $\varepsilon_{\rm{ult}}$ and $P_{\rm{ult}}$.
\begin{table}[h!]
\renewcommand{\arraystretch}{1.5}
\centerline{\normalsize
\begin{tabular}{c|c|c|c||c|c} 
  \hline
$M_{\rm{NS}}^{\max}/M_{\odot}$&$\varepsilon_{\rm{ult}}$ ($\rm{GeV}/\rm{fm}^3$)&$P_{\rm{ult}}$ ($\rm{GeV}/\rm{fm}^3$)&Ref.&Eq.\,(\ref{eps_ult})&Eq.\,(\ref{P_ult})\\\hline\hline
       $\gtrsim1.97$&/&$570_{-320}^{+320}$&\,\cite{Capano2020}&1.47&548\\\hline
       $\gtrsim2.0$&1.46&/&\,\cite{Altiparmak2022}&1.42&533\\\hline
       $\gtrsim2.01$&$1.18_{-0.17}^{+0.17}$&/&\,\cite{Gorda2023}&1.41&528\\\hline
       $\gtrsim2.2$&$1.16_{-0.10}^{+0.11}$&/&\,\cite{Mar23}&1.19&444\\\hline
       \end{tabular}}
        \caption{Constraints on $\varepsilon_{\rm{ult}}$ and $P_{\rm{ult}}$ existing in the literature and from Eqs.\,(\ref{eps_ult}) and (\ref{P_ult}). Table taken from Ref.\,\cite{CLZ23-b}.}\label{tab_ult} 
\end{table}

\begin{figure}[h!]
\centering
\includegraphics[width=7.5cm]{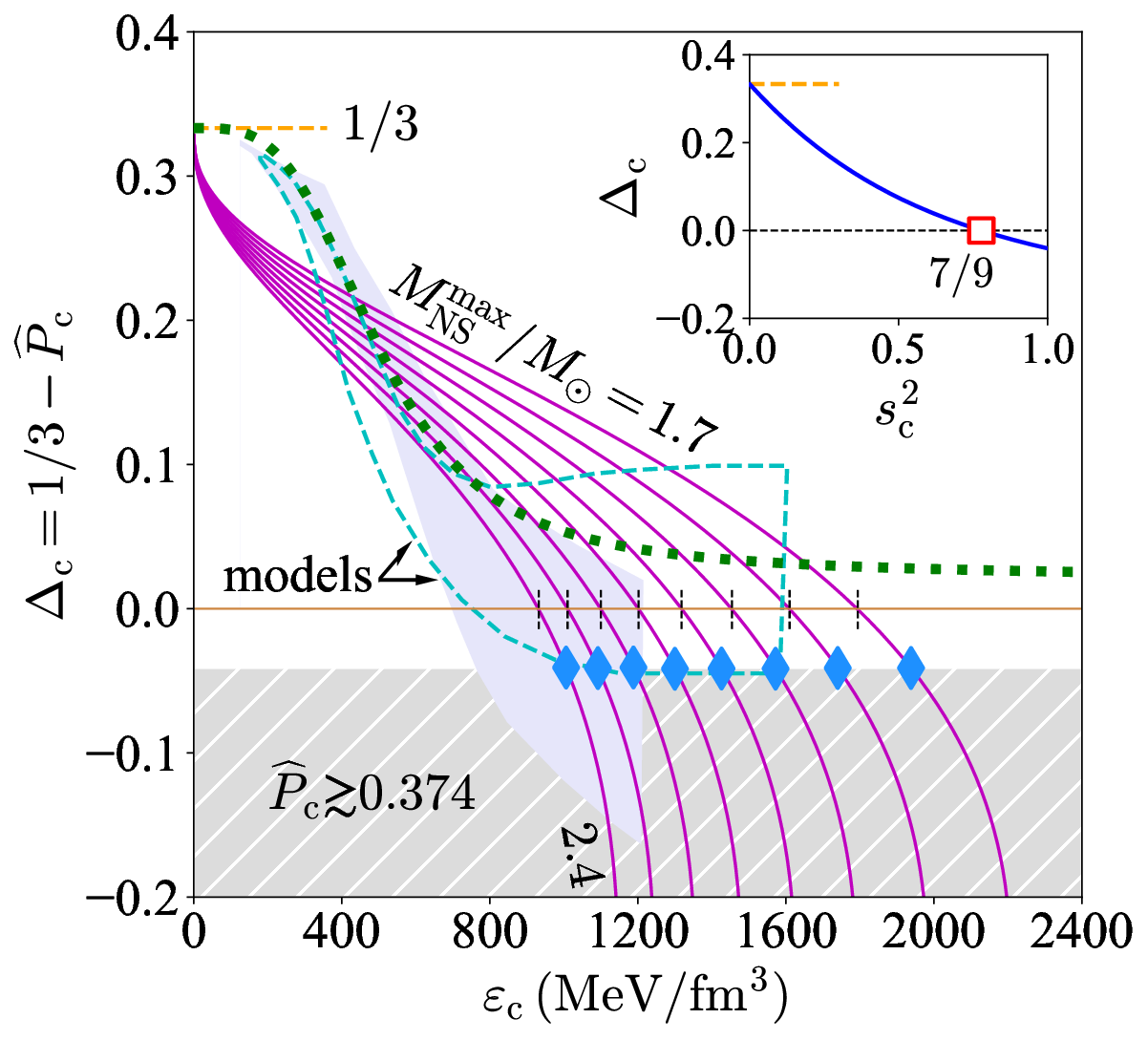}
\caption{(Color Online).  Central trace anomaly $\Delta_{\rm{c}}=1/3-{\x}$ as a function of $\varepsilon_{\rm{c}}$ for different $M_{\rm{NS}}^{\max}/M_{\odot}$ (from 1.7 to 2.4), where lightblue solid diamonds mark the points set by ${\x}\lesssim0.374$ while the vertical black dashed lines are for the vanishing points of $\Delta_{\rm{c}}$ for each $M_{\rm{NS}}^{\max}$.
Two {model calculations} (lavender/cyan)\,\cite{Fuji22,Gorda2023} and a parametrization for $\Delta$ (green dotted)\,\cite{Fuji22} are also shown for comparisons. Figure taken from Ref.\,\cite{CLZ23-b}.
}\label{fig_TAc}
\end{figure}
Finally, by considering the black widow PSR J0952-0607\,\cite{Romani22} with its mass about 2.35$M_{\odot}$, the ultimate energy density $\varepsilon_{\rm{ult}}$ and pressure $P_{\rm{ult}}$ {\color{xll}for all NSs existing in the Universe} are estimated to be smaller than about $1.05\,\rm{GeV}/\rm{fm}^3$ and $392\,\rm{MeV}/\rm{fm}^3$ (lightblue bands in FIG.\,\ref{fig_ULTeps}), respectively.
The analyses of very massive NSs are important for extracting the trace anomaly especially its high-density limit as demonstrated in FIG.\,\ref{fig_Delta}. We show in FIG.\,\ref{fig_TAc} the central trace anomaly $\Delta_{\rm{c}}=1/3-{\x}$ as a function of $\varepsilon_{\rm{c}}$ for different values of $M_{\rm{NS}}^{\max}$ (varying from 1.7$M_{\odot}$ to $2.4M_{\odot}$) using the mass scaling of Eq.\,(\ref{Rmax-n}). The parametrization for $\Delta$ suggested by Ref.\,\cite{Fuji22} is also shown (green dotted line), which is positive-definite by construction.
In addition, two {model predictions for $\Delta$ are represented by lavender\,\cite{Fuji22} and cyan dashed\,\cite{Gorda2023} bands using machine-learning algorithms and ab-initio QCD calculations, respectively.}
For relatively light NSs, the condition $\Delta_{\rm{c}}\geq0$ generally holds since the central energy density $\varepsilon_{\rm{c}}$ is relatively low, e.g., for $M_{\rm{NS}}^{\max}/M_{\odot}=1.7$ the condition $\varepsilon_{\rm{c}}\lesssim1.9\,\rm{GeV}/\rm{fm}^3$ is safely satisfied. On the other hand,  the $\varepsilon_{\rm{c}}$ may exceed the value set by $\Delta_{\rm{c}}\approx0$ as $M_{\rm{NS}}^{\max}$ increases; so FIG.\,\ref{fig_TAc} explains why the criterion $\Delta\geq0$ tends to {break down} for massive NSs with increasing $M_{\rm{NS}}^{\max}$\,\cite{Ecker23,Mus24,Pang24,Cao23,Tak23,Ann23,Mar24,Brandes2023-a}, instead of for light NSs.
In this sense, in order to observe a pattern of $\Delta$ on energy density indicated by the magenta line in FIG.\,\ref{fig_Dp_sk}, observations of massive and compact NSs are relevant/necessary.

\subsection{Implications of the conjecture of a positive trace anomaly on the peaked structure of SSS}\label{sub_conjecture}

Finally, we discuss the implications of a positive $\Delta$ on the possible peaked behavior of $s^2$\,\cite{Mar24aa}.
The relation between $\Delta$ and the peak position in $s^2$ was recently analyzed by using the average speed of sound\,\cite{Mar24aa}:
\begin{equation}\label{def-AVERS2}
    \langle s^2(\varepsilon)\rangle\equiv\frac{1}{\varepsilon}\int_0^{\varepsilon}\d\varepsilon's^2(\varepsilon')=\frac{P}{\varepsilon}=\phi.
\end{equation}
Under the assumptions (1) that the trace anomaly vanishes at some reference energy density $\varepsilon_{\rm{rf}}$ (so the average SSS averages to $\langle s^2_{\rm{rf}}\rangle\equiv\langle s^2(\varepsilon_{\rm{rf}})\rangle\to1/3$) and (2) $\Delta\geq0$ at all densities, the SSS may develop a local maximum above 1/3 at an energy density $\varepsilon\leq\varepsilon_{\rm{rf}}$.
This can be analyzed as follows.

\begin{figure}[h!]
\centering
\hspace{1.cm}\includegraphics[width=11.cm]{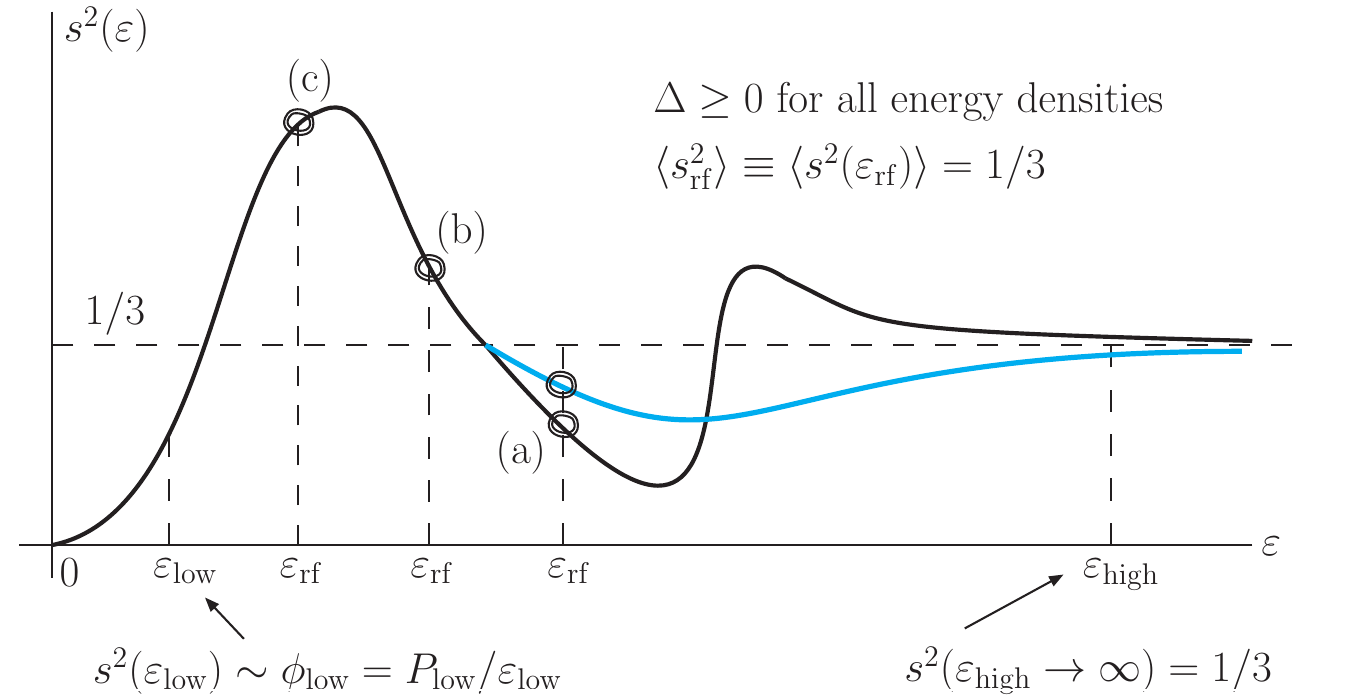}\\[1.cm]
\includegraphics[width=9.5cm]{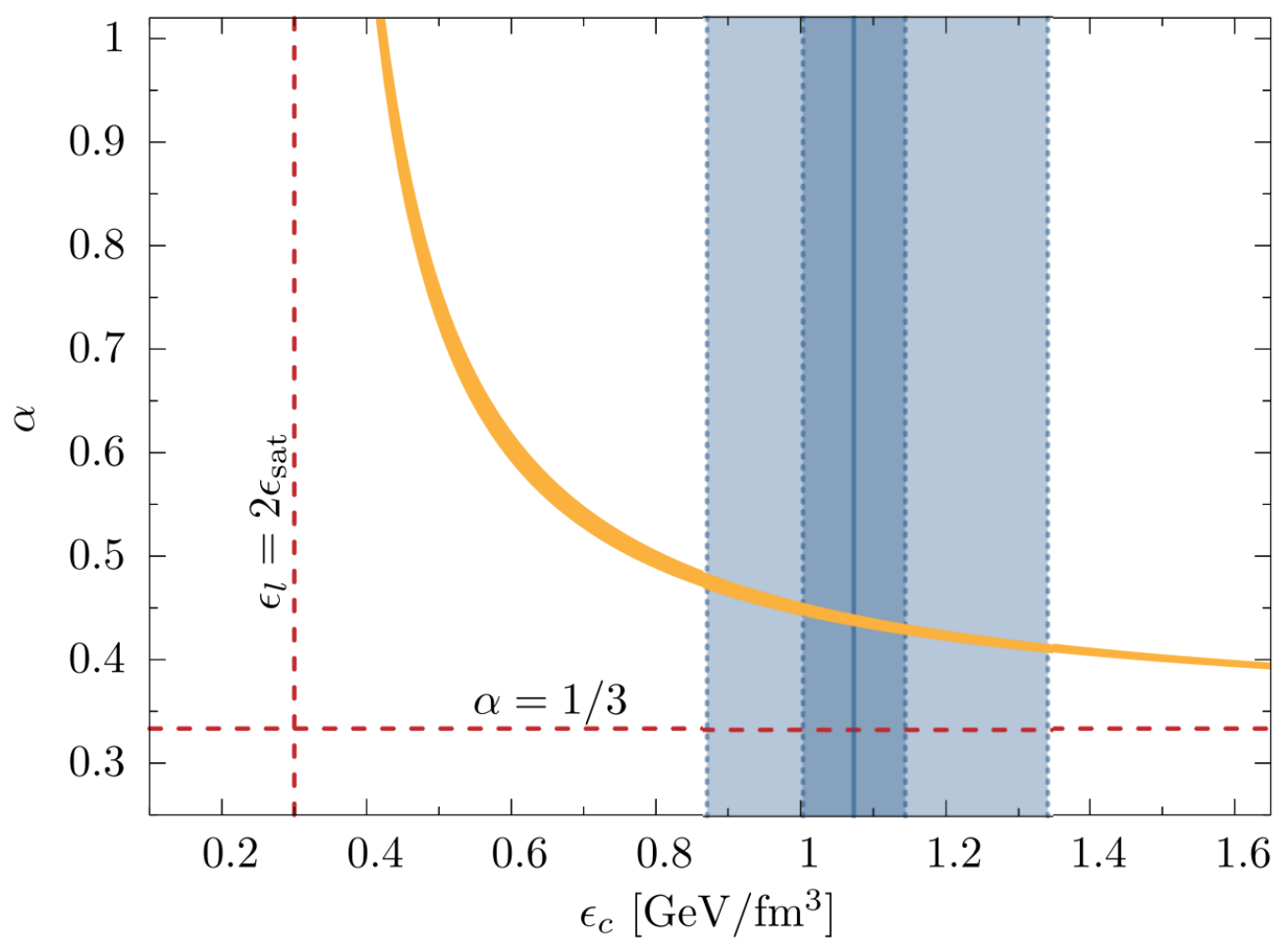}
\caption{(Color Online). 
Upper panel: a sketch of $s^2(\varepsilon)$ to help understand the peak position in $s^2(\varepsilon)$ considering trace anomaly constraints; here $\varepsilon_{\rm{low}}<\varepsilon_{\rm{rf}}<\varepsilon_{\rm{high}}$.
Lower panel: the $\varepsilon_{\rm{rf}}$-dependence of parameter $\alpha$ defined in Ref.\,\cite{Mar24aa}. Figure taken from Ref.\,\cite{Mar24aa}, here $\varepsilon_{\rm{c}}\leftrightarrow\varepsilon_{\rm{rf}}$ under our notations.
}\label{fig_Mar24aa-alpha} 
\end{figure}

The first assumption gives
\begin{equation}
    \langle s_{\rm{rf}}^2\rangle\equiv\frac{1}{\varepsilon_{\rm{rf}}}\int_0^{\varepsilon_{\rm{rf}}}\d\varepsilon s^2(\varepsilon)=\frac{1}{3}\leftrightarrow \int_0^{\varepsilon_{\rm{low}}}\d\varepsilon s^2(\varepsilon)+\int_{\varepsilon_{\rm{low}}}^{\varepsilon_{\rm{rf}}}\d\varepsilon s^2(\varepsilon)=\frac{1}{3}\varepsilon_{\rm{low}}+\frac{1}{3}\left(\varepsilon_{\rm{rf}}-\varepsilon_{\rm{low}}\right).
\end{equation}
Applying the second assumption $\Delta\geq0$ to the density region of $0\leq\varepsilon\leq\varepsilon_{\rm{low}}$, we have
\begin{equation}
    \langle s_{\rm{low}}^2\rangle\equiv\frac{1}{\varepsilon_{\rm{low}}}\int_0^{\varepsilon_{\rm{low}}}\d\varepsilon s^2(\varepsilon)\leq\frac{1}{3}\leftrightarrow \int_0^{\varepsilon_{\rm{low}}}\d\varepsilon s^2(\varepsilon)\leq\frac{1}{3}\varepsilon_{\rm{low}}.
\end{equation}
Combining these two results leads to the following average SSS ($\alpha$) over $\varepsilon_{\rm{low}}\leq\varepsilon\leq\varepsilon_{\rm{rf}}$
\begin{equation}\label{lk-1}
\boxed{
    \alpha\equiv \langle s^2(\varepsilon_{\rm{low}}\to\varepsilon_{\rm{rf}})
    \rangle\equiv\frac{1}{\varepsilon_{\rm{rf}}-\varepsilon_{\rm{low}}}\int_{\varepsilon_{\rm{low}}}^{\varepsilon_{\rm{rf}}}\d\varepsilon s^2(\varepsilon)\geq\frac{1}{3}.}
    \end{equation}
This relation shows that the average SSS in the region $\varepsilon_{\rm{low}}\leq\varepsilon\leq\varepsilon_{\rm{rf}}$ is equal to or greater than 1/3.
Assume that the EOS at low densities $\varepsilon\lesssim\varepsilon_{\rm{low}}$ is well known (e.g., from CEFT), one can further obtain $\langle s^2_{\rm{low}}\rangle<1/3$; then 
\begin{equation}
    \alpha=\frac{1}{3}+\frac{3^{-1}-\langle s_{\rm{low}}^2\rangle}{1-\varepsilon_{\rm{low}}/\varepsilon_{\rm{rf}}}\frac{\varepsilon_{\rm{low}}}{\varepsilon_{\rm{rf}}}>\frac{1}{3}.
\end{equation}
Consequently, there must be some density smaller than $\varepsilon_{\rm{rf}}$ at which the SSS is greater than 1/3.
Using the parameter $\alpha$, we can rewrite the average SSS $\langle s_{\rm{rf}}^2\rangle$ as $\langle s_{\rm{rf}}^2\rangle=\langle s_{\rm{low}}^2\rangle(\varepsilon_{\rm{low}}/\varepsilon_{\rm{rf}})+\alpha(1-\varepsilon_{\rm{low}}/\varepsilon_{\rm{rf}})$.

Next, applying the assumption $\Delta\geq0$ to the even larger density region $0\leq\varepsilon\leq\varepsilon_{\rm{high}}$, we may obtain
\begin{align}
    &\langle s_{\rm{high}}^2\rangle\equiv
    \frac{1}{\varepsilon_{\rm{high}}}\int_0^{\varepsilon_{\rm{high}}}\d\varepsilon s^2(\varepsilon)\leq\frac{1}{3}\notag\\
\leftrightarrow    &\frac{1}{\varepsilon_{\rm{high}}}\left[\int_0^{\varepsilon_{\rm{low}}}\d\varepsilon s^2(\varepsilon)+\int_{\varepsilon_{\rm{low}}}^{\varepsilon_{\rm{rf}}}\d\varepsilon s^2(\varepsilon)+\int_{\varepsilon_{\rm{rf}}}^{\varepsilon_{\rm{high}}}\d\varepsilon s^2(\varepsilon)\right]\leq\frac{1}{3}\notag\\
\leftrightarrow&\int_{\varepsilon_{\rm{rf}}}^{\varepsilon_{\rm{high}}}\d\varepsilon s^2(\varepsilon)<\frac{1}{3}\varepsilon_{\rm{low}}-\int_0^{\varepsilon_{\rm{low}}}\d\varepsilon s^2(\varepsilon)+\frac{1}{3}\left(\varepsilon_{\rm{high}}-\varepsilon_{\rm{rf}}\right),~~\mbox{using}~~\int_{\varepsilon_{\rm{low}}}^{\varepsilon_{\rm{rf}}}\d\varepsilon s^2(\varepsilon)>\frac{1}{3}\left(\varepsilon_{\rm{rf}}-\varepsilon_{\rm{low}}\right)
\notag\\
    \leftrightarrow& \langle s^2(\varepsilon_{\rm{rf}}\to\varepsilon_{\rm{high}})\rangle\equiv\frac{1}{\varepsilon_{\rm{high}}-\varepsilon_{\rm{rf}}}\int_{\varepsilon_{\rm{rf}}}^{\varepsilon_{\rm{high}}}\d\varepsilon s^2(\varepsilon)<\frac{1}{3}+\frac{3^{-1}\varepsilon_{\rm{low}}-\int_0^{\varepsilon_{\rm{low}}}\d\varepsilon s^2(\varepsilon)}{\varepsilon_{\rm{high}}-\varepsilon_{\rm{rf}}}.
    \end{align}
Taking $\varepsilon_{\rm{high}}\to\infty$ gives
\begin{equation}
\boxed{
    \langle s^2(\varepsilon_{\rm{rf}}\to\varepsilon_{\rm{high}})\rangle<\frac{1}{3},}
\end{equation}
this means the average SSS over density $\varepsilon\geq\varepsilon_{\rm{rf}}$ is smaller than 1/3.
Equivalently, there must exist some density region over which the SSS is smaller than 1/3.
If the SSS approaches 1/3 at extremely high density from above then the pattern for $s^2$ takes the form of the black line in the upper panel of FIG.\,\ref{fig_Mar24aa-alpha}; and it may have the light-blue pattern if $s^2$ approaches 1/3 from below.
In case (a) or (b) in the upper panel of FIG.\,\ref{fig_Mar24aa-alpha}, where $s_{\rm{rf}}^2<1/3$ or $s_{\rm{rf}}^2>1/3$, the peak in $s^2$ emerges somewhere $\leq\varepsilon_{\rm{rf}}$.
In case (c), the peak in $s^2$ emerges at $\varepsilon\geq\varepsilon_{\rm{rf}}$.
In all three cases, the $s^2$ develops a maximum above 1/3 at energy densities $\leq\varepsilon_{\rm{rf}}$\,\cite{Mar24aa}.
Numerically, using the knowledge of the EOS with $\varepsilon_{\rm{low}}\approx2\varepsilon_0$ and requiring that $\varepsilon_{\rm{rf}}$
has to reproduce the estimated central energy density of maximally massive NSs $\varepsilon_{\rm{TOV}}$ about $1\,\rm{GeV}/\rm{fm}^3$, Ref.\,\cite{Mar24aa} estimated that $\alpha\approx0.41\mbox{-}0.49$, showing that the SSS exceeds the conformal value.
Taking the basic requirement $\alpha\geq3^{-1}$ and the same CEFT constraint on NS EOS with $\varepsilon\lesssim\varepsilon_{\rm{low}}$, Ref.\,\cite{Mar24aa} also obtained $\langle s^2_{\rm{rf}}\rangle\gtrsim0.26$ with $\varepsilon_{\rm{rf}}\approx1\,\rm{GeV}/\rm{fm}^3$.
The lower panel of FIG.\,\ref{fig_Mar24aa-alpha} shows the $\varepsilon_{\rm{rf}}$-dependence of the parameter $\alpha$ defined and used in Ref.\,\cite{Mar24aa}; an upper bound on $\alpha$ as $\alpha\leq1$ implies that $\varepsilon_{\rm{rf}}\gtrsim0.42\,\rm{GeV}/\rm{fm}^3$.

At this point, we would like to point out that the reference density $\varepsilon_{\rm{rf}}$ introduced in Ref.\,\cite{Mar24aa} is a relevant quantity based on which certain/special assumptions are made, e.g., $\langle s^2_{\rm{rf}}\rangle\approx\langle s^2_{\rm{TOV}}\rangle\equiv\langle s^2(\varepsilon_{\rm{TOV}})\rangle\to1/3$.
However, the central ratio of pressure over energy density $\phi_{\rm{c}}=\x=\langle s^2_{\rm{c}}\rangle$ for NSs at the TOV configuration may either be smaller or larger than 1/3. It may thus partially reduce the predicting power of the approach.
For example, by letting generally $\langle s^2_{\rm{low}}\rangle\lesssim\x_{\rm{low}}$, $\langle s^2_{\rm{rf}}\rangle\approx \langle s^2_{\rm{TOV}}\rangle\lesssim\x_{\rm{rf}}$ and $\Delta\gtrsim\Delta_{\rm{GR}}=3^{-1}-\x_+$, where $\x_{\rm{low}}$ is smaller than $\x_{\rm{rf}}$ and $\x_+\approx0.374$, we may similarly obtain
\begin{equation}
    \langle s^2(\varepsilon_{\rm{low}}\to\varepsilon_{\rm{rf}})\rangle\gtrsim\x_{\rm{rf}}+\frac{\x_{\rm{rf}}-\x_{\rm{low}}}{\varepsilon_{\rm{rf}}/\varepsilon_{\rm{low}}-1}\approx\frac{(\varepsilon_{\rm{rf}}/\varepsilon_{\rm{low}})\x_{\rm{rf}}-\x_{\rm{low}}}{\varepsilon_{\rm{rf}}/\varepsilon_{\rm{low}}-1}.
\end{equation}
In addition, we have $\langle s^2(\varepsilon_{\rm{rf}}\to\varepsilon_{\rm{high}})\rangle\lesssim\x_+$ by taking $\varepsilon_{\rm{high}}\to\infty$.
Adopting artificially $\x_{\rm{rf}}\approx\x_{\rm{low}}\approx\x_+\approx1/3$ recovers the analysis in Ref.\,\cite{Mar24aa}.
The peak position analysis in such case depends on the ratio $\varepsilon_{\rm{rf}}/\varepsilon_{\rm{low}}$ and the three $\x$'s, which is more involved.

\begin{figure}[ht!]
\centering
\includegraphics[width=16.5cm]{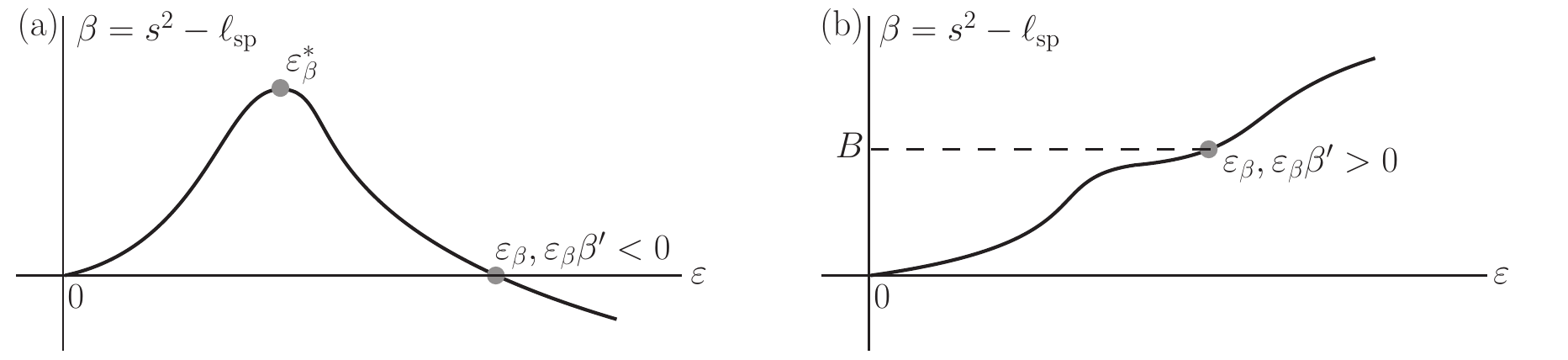}
\caption{Two types of the $\varepsilon$-dependence of the curvature $\beta(\varepsilon)$: the $\beta$ function increases with $\varepsilon$ at small $\varepsilon$, then it reaches the maximum point $\varepsilon_{\beta}^{\ast}$ and the zero point $\varepsilon_\beta$ at which $\beta(\varepsilon_{\beta})=0$ (left), or it monotonically increases with $\varepsilon$ (right).
}\label{fig_beta_function}
\end{figure}

In a recent work of Ref.\,\cite{Marc24Curvature}, a peaked $s^2$ in NSs was attributed to the vanishing of the curvature $\beta$ of energy per particle. This can be analyzed as follows.
Using the thermodynamic relation $\d\varepsilon=\mu\d\rho$ and $P=\rho^2\d(\varepsilon/\rho)/\d\rho$ with $\mu$ the baryon chemical potential and $\varepsilon/\rho$ the energy per particle, we rewrite the SSS as:
\begin{align}
    s^2=&\frac{\d P}{\d\varepsilon}=\frac{1}{\mu}\frac{\d }{\d\rho}
    \left(\rho^2\frac{\d(\varepsilon/\rho)}{\d\rho}\right)   =\frac{2\rho}{\mu}\frac{\d(\varepsilon/\rho)}{\d\rho}+\frac{\rho^2}{\mu}\frac{\d^2(\varepsilon/\rho)}{\d\rho^2},
\end{align}
the first term is the slope of the energy per particle and the second term is the curvature. We denote them as $\ell_{\rm{sp}}$ and $\beta$, respectively.
The slope could be further expressed as\,\cite{Marc24Curvature}
\begin{align}\label{def_CC_alpha}
    \ell_{\rm{sp}}=&\frac{2\rho}{\mu}\frac{\d(\varepsilon/\rho)}{\d\rho}
    =2\left(1-\frac{\varepsilon}{\rho}\frac{\d\rho}{\d\varepsilon}\right)=2\left(1-\frac{\varepsilon}{\rho}\frac{\rho}{P+\varepsilon}\right)=\frac{2P}{P+\varepsilon}=\frac{2}{1+\phi^{-1}}=2\frac{\Delta-1/3}{\Delta-4/3},
\end{align}
where the relation $P=\rho\d\varepsilon/\d\rho-\varepsilon=\rho\mu-\varepsilon$ is used.
Consequently, we have
\begin{equation}\label{def_CC_beta}
    \beta=s^2-\ell_{\rm{sp}}=-\varepsilon{\d\Delta}/{\d\varepsilon}+({3}^{-1}-\Delta)\left[1+{2}/({\Delta-4/3})\right].
\end{equation}
Since $\phi$ is a monotonically increasing function of $\varepsilon$, $\ell_{\rm{sp}}$ also increases with $\varepsilon$.
Consequently, if $s^2$ has a peak somewhere, it should be attributed to the curvature term $\beta$.
Moreover, $\beta(\varepsilon)\approx2\Delta'^2(0)\varepsilon^2+2\Delta'^3(0)\varepsilon^3$ increases with $\varepsilon$, where $\Delta'(0)<0$ is the first-order derivative of $\Delta(\varepsilon)$ with respect to $\varepsilon$ at $\varepsilon=0$, i.e., $\Delta(\varepsilon)\approx 3^{-1}+\Delta'(0)\varepsilon$.
So $\beta(\varepsilon)\geq0$ for small $\varepsilon$ and it develops a local maximum $\varepsilon_{\beta}^{\ast}$ and then a zero point $\varepsilon_{\beta}$, here $\varepsilon_{\beta}>\varepsilon_{\beta}^{\ast}$.
See the left panel of FIG.\,\ref{fig_beta_function} for the $\varepsilon$-dependence of such curvature function.
The curvature $\beta$ of energy per particle is closely related to the incompressibility coefficient $K_{\rm{NM}}$ in nuclear matter\,\cite{Ivan22}
\begin{equation}
    K_{\rm{NM}}(\rho)\equiv9\rho^2\frac{\d^2(\varepsilon/\rho)}{\d\rho^2}=9\mu\left(s^2-\frac{2P}{P+\varepsilon}\right)=9\mu\left(s^2-\frac{2}{1+\phi^{-1}}\right),~~\phi=P/\varepsilon,
\end{equation}
consequently $\beta=K_{\rm{NM}}/9\mu$.
If $K_{\rm{NM}}(\rho)$ is defined as $9\d P/\d\rho$, then the relation becomes $\beta=K_{\rm{NM}}/9\mu-2/(1+\phi^{-1})$.
Connecting the $\beta$ with the nuclear matter incompressibility coefficient provides a useful perspective to investigate the conformal properties of NS matter\,\cite{Ivan22}.

Expanding the curvature $\beta$ around the zero point $\varepsilon_{\beta}$ as
\begin{equation}\label{beta_func_exp1}
    \beta(\varepsilon)\approx\beta'\left(\varepsilon-\varepsilon_\beta\right)+\frac{1}{2}\beta''\left(\varepsilon-\varepsilon_{\beta}\right)^2,~~\beta'\equiv\left.\frac{\d\beta}{\d\varepsilon}\right|_{\varepsilon=\varepsilon_{\beta}}<0,~~\beta''\equiv\left.\frac{\d^2\beta}{\d\varepsilon^2}\right|_{\varepsilon=\varepsilon_{\beta}},
\end{equation}
from which we obtain $\varepsilon_{\beta}^{\ast}-\varepsilon_\beta=-\beta'/\beta''$; therefore $\beta''<0$ since $\varepsilon_{\beta}^{\ast}-\varepsilon_\beta<0$. Notice that if we include $6^{-1}\beta'''(\varepsilon-\varepsilon_\beta)^3$ in the expansion of $\beta(\varepsilon)$, then $\beta''$ could be positive as long as $\beta'''>0$ since now $\varepsilon_\beta^{\ast}-\varepsilon_\beta=-[\beta''+\sqrt{\beta''^2-2\beta'\beta'''}]/\beta'''<0$; for our purpose the expansion (\ref{beta_func_exp1}) to order $\beta''$ is enough. 
Similarly, we expand the trace anomaly $\Delta$ around $\varepsilon_\beta$ as:
\begin{equation}\label{Delta_func_exp}
\Delta(\varepsilon)\approx\Delta_\beta+\Delta'_\beta\left(\varepsilon-\varepsilon_\beta\right)+2^{-1}\Delta_\beta''\left(\varepsilon-\varepsilon_\beta\right)^2.
\end{equation}
Putting it into $\beta(\varepsilon)$ of (\ref{def_CC_beta}) and keeping terms to order $(\varepsilon-\varepsilon_\beta)^2$ gives
\begin{align}\label{beta_func_exp2}
    \beta(\varepsilon)\approx&-\varepsilon_\beta\Delta_\beta'-\frac{1}{3}\frac{2-3\Delta_\beta-9\Delta_\beta^2}{4-3\Delta_\beta}\notag\\
    &-\frac{1}{(4-3\Delta_\beta)^2}\left[
    2\left(7-24\Delta_\beta+9\Delta_\beta^2\right)\Delta_\beta'
    +\left(16-24\Delta_\beta+9\Delta_\beta^2\right)\varepsilon_\beta\Delta_\beta''
    \right]\left(\varepsilon-\varepsilon_\beta\right)\notag\\
    &+\frac{3}{2}\frac{1}{(4-3\Delta_\beta)^3}\left[
    36\Delta_\beta'^2-\left(40-126\Delta_\beta+108\Delta_\beta^2-27\Delta_\beta^3\right)\Delta_\beta''
    \right]\left(\varepsilon-\varepsilon_\beta\right)^2.
\end{align}
The vanishing of the first line (there is no constant term in $\beta(\varepsilon)$ according to (\ref{beta_func_exp1})) leads to the expression for $\Delta_\beta'$:
\begin{equation}
    \Delta_\beta'=-\frac{1}{3\varepsilon_\beta}\frac{(2+3\Delta_\beta)(1-3\Delta_\beta)}{4-3\Delta_\beta},
\end{equation}
which is semi-negative definite for $-2/3\leq\Delta_\beta\leq1/3$.
Equaling the coefficient of the second line of (\ref{beta_func_exp2}) with $\beta'$ and putting $\Delta_\beta'$ into it gives the expression for $\Delta_\beta''$:
\begin{equation}
    \Delta_\beta''=-\frac{\beta'}{\varepsilon_\beta}+\frac{2}{3\varepsilon_\beta^2}\frac{(2+3\Delta_\beta)(7-3\Delta_\beta)(1-3\Delta_\beta)^2}{(4-3\Delta_\beta)^3},
\end{equation}
which is semi-positive definite for $-2/3\leq\Delta_\beta\leq1/3$.
Now, we have ($\delta\varepsilon\equiv\varepsilon-\varepsilon_\beta$):
\begin{equation}
\boxed{
    \Delta(\varepsilon)\approx\Delta_\beta-\frac{\delta\varepsilon}{3\varepsilon_\beta}\frac{(2+3\Delta_\beta)(1-3\Delta_\beta)}{4-3\Delta_\beta}-\frac{\delta\varepsilon^2}{2\varepsilon_\beta^2}\left[\varepsilon_\beta\beta'-\frac{2}{3}\frac{(2+3\Delta_\beta)(7-3\Delta_\beta)(1-3\Delta_\beta)^2}{(4-3\Delta_\beta)^3}\right].}
\end{equation}
Consequently, a local minimum develops at $\varepsilon_\Delta>\varepsilon_\beta$ for $\Delta(\varepsilon)$: 
\begin{equation}
    \varepsilon_\Delta/\varepsilon_\beta=1-\frac{\Delta_\beta'}{\Delta_\beta''\varepsilon_\beta}
    =\frac{3(2+3\Delta_\beta)(1-3\Delta_\beta)(10-24\Delta_\beta+9\Delta_\beta^2)-3\varepsilon_\beta\beta'(4-3\Delta_\beta)^3}{2(2+3\Delta_\beta)(7-3\Delta_\beta)(1-3\Delta_\beta)^2-3\varepsilon_\beta\beta'(4-3\Delta_\beta)^3},
\end{equation}
with the minimum $\Delta_{\min}\equiv    \Delta(\varepsilon_\Delta)$ given by:
\begin{equation}
\Delta_{\min}=\Delta_\beta-\frac{1}{2}\frac{\Delta_\beta'^2}{\Delta_\beta''}
    =-\frac{(2+3\Delta_\beta)(1-3\Delta_\beta)^2(8-78\Delta_\beta+27\Delta_\beta^2)+18\varepsilon_\beta\beta'\Delta_\beta(4-3\Delta_\beta)^3}{12(2+3\Delta_\beta)(7-3\Delta_\beta)(1-3\Delta_\beta)^2-18\varepsilon_\beta\beta'(4-3\Delta_\beta)^3}.
\end{equation}
For $\Delta_\beta\approx0$, the above relations become
\begin{equation}\label{rlll}
\Delta(\varepsilon)\approx-\frac{1}{6}\frac{\delta\varepsilon}{\varepsilon_\beta}+\frac{1}{2}\left(\frac{7}{48}-\varepsilon_\beta\beta'\right)\frac{\delta\varepsilon^2}{\varepsilon_\beta^2},~~ \varepsilon_\Delta/\varepsilon_\beta\approx\frac{15-48\varepsilon_\beta\beta'}{7-48\varepsilon_\beta\beta'},~~\Delta_{\min}\approx\frac{2}{3}\frac{1}{48\varepsilon_\beta\beta'-7}.
\end{equation}
A small negative $\Delta_{\min}\gtrsim-2/21\approx-0.1$ (corresponding to $\varepsilon_\beta\beta'\approx0$) and $\varepsilon_{\Delta}/\varepsilon_\beta\gtrsim1$ would be obtained, considering that $0\lesssim-\varepsilon_\beta\beta'\approx\mathcal{O}(1)$; e.g., we have $\varepsilon_{\Delta}/\varepsilon_\beta\approx1.26$ and $\Delta_{\min}\approx-0.02$ by taking $\varepsilon_\beta\beta'\approx-0.5$.

\begin{figure}[ht!]
\centering
\includegraphics[width=8.1cm]{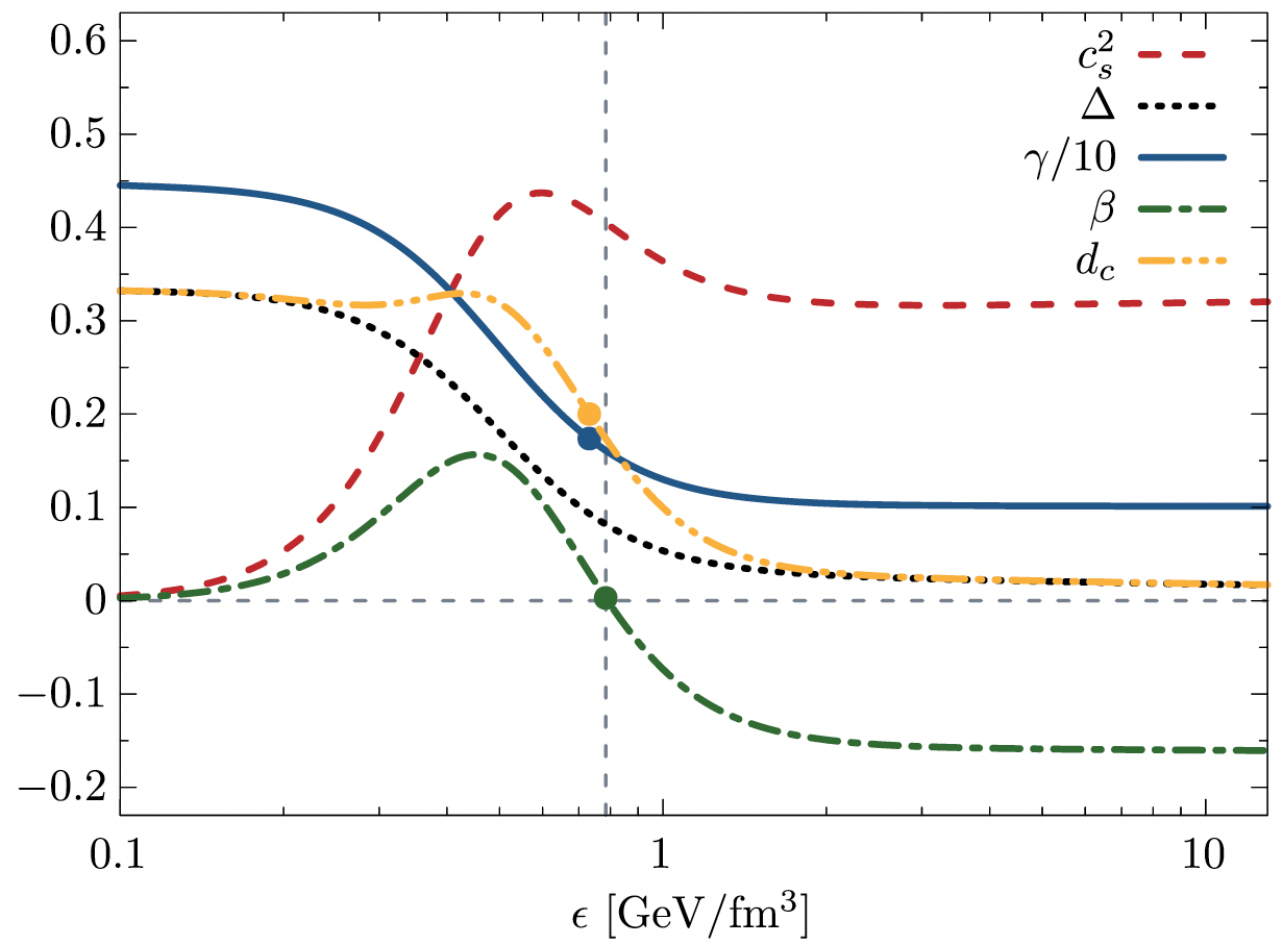}\qquad
\includegraphics[width=8.cm]{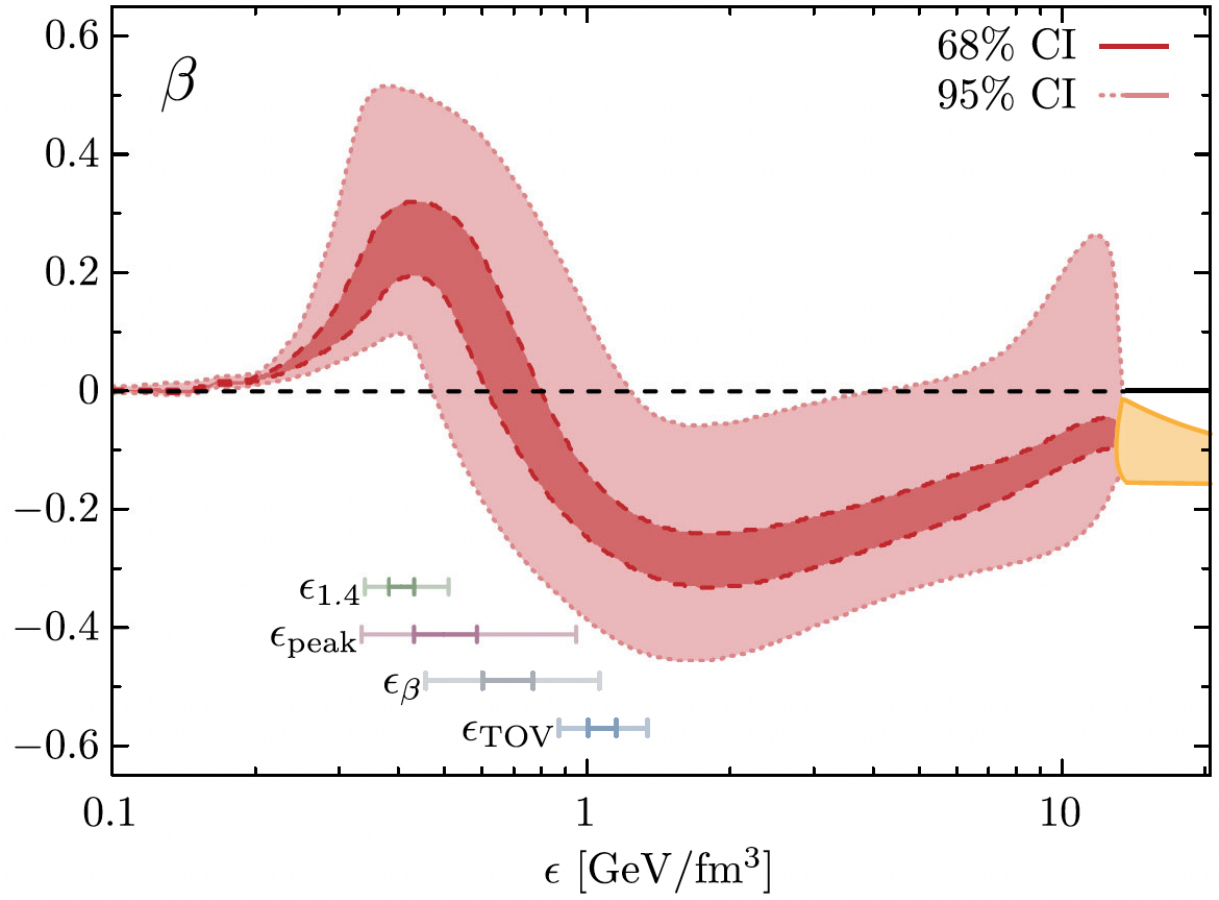}
\caption{(Color Online). Left panel: the curvature $\beta$ 
and several other related quantities obtained under a specific parametrization for $\Delta(\varepsilon)$.
Right panel: the energy density dependence of the curvature $\beta$ obtained using constraints from astrophysical observations, nuclear-many calculations and from pQCD at high densities. Figures taken from Ref.\,\cite{Marc24Curvature}.
}\label{fig_MarcCC}
\end{figure}

After obtaining the expressions for $\Delta_\beta'$ and $\Delta_\beta''$, the third line of (\ref{beta_func_exp2}) gives an equation for determining $\Delta_\beta$ (in terms of $\beta'$ and $\beta''$), which is a complicated function of $\Delta_\beta$; the approximation for $\Delta_\beta$ is:
\begin{equation}
    \Delta_\beta\approx\frac{2}{3}\frac{29-240\varepsilon_\beta\beta'+128\varepsilon_\beta^2\beta''}{159-744\varepsilon_\beta\beta'+320\varepsilon_\beta^2\beta''},
\end{equation}
then $\Delta_\beta\approx0.16$ by taking $\varepsilon_\beta\beta'\approx\varepsilon_\beta^2\beta''\approx-1$.
For our purpose, we may not analyze this equation for $\Delta_\beta$ further and remember that $\Delta_\beta$ is generally small $\lesssim1/3$ at the zero point $\varepsilon_\beta$ of $\beta(\varepsilon)$.
Next, by letting $\d s^2(\varepsilon)/\d\varepsilon=0$ and using the $\Delta(\varepsilon)$ of Eq.\,(\ref{Delta_func_exp}), we obtain the peak position $\varepsilon_{\rm{pk}}$ in $s^2$ as:
\begin{equation}\label{rtpk}
\boxed{
    \frac{\varepsilon_{\rm{pk}}}{\varepsilon_\beta}\approx
    \frac{10-32\varepsilon_\beta\beta'}{7-48\varepsilon_\beta\beta'}\left[1
    +12\Delta_\beta\cdot\frac{3+8\varepsilon_\beta\beta'}{(5-16\varepsilon_\beta\beta')(7-48\varepsilon_\beta\beta')}
    \right],}
\end{equation}
therefore:
\begin{equation}\label{rtpk1}
s_{\rm{pk}}^2\equiv s^2(\varepsilon_{\rm{pk}})
\approx\frac{1}{32}\frac{121-672\varepsilon_\beta\beta'+256\varepsilon_\beta^2\beta'^2}{7-48\varepsilon_\beta\beta'}\left[1-\frac{3}{4}\frac{2001-34464\varepsilon_\beta\beta'+108800\varepsilon_\beta^2\beta'^2}{(7-48\varepsilon_\beta\beta')(121-672\varepsilon_\beta\beta'+256\varepsilon_\beta^2\beta'^2)}\cdot\Delta_\beta\right],
\end{equation}
and
\begin{equation}
\left.\frac{\d^2s^2}{\d\varepsilon^2}\right|_{\varepsilon=\varepsilon_{\rm{pk}}}\approx-\frac{28-192\varepsilon_\beta\beta'-75\Delta_\beta}{64\varepsilon_\beta^2}<0.
\end{equation}
The negativeness of the second-order derivative of $s^2$ at $\varepsilon_{\rm{pk}}$ implies it is a local maximum point (we can see that $28-75\Delta_\beta-192\varepsilon_\beta\beta'>3$ since $-2/3\leq\Delta_\beta\leq1/3$ and $\varepsilon_\beta\beta'<0$).
Assuming $-\varepsilon_\beta\beta'\approx\mathcal{O}(1)$, we then have $\varepsilon_{\rm{pk}}/\varepsilon_\beta\approx\mathcal{O}(1)$ (generally $2/3\lesssim\varepsilon_{\rm{pk}}/\varepsilon_\beta\lesssim10/7$), e.g., $\varepsilon_{\rm{pk}}/\varepsilon_\beta\approx0.75$ (or $\approx1$) by adopting $\varepsilon_\beta\beta'\approx-1$ (or $\varepsilon_\beta\beta'\approx-3/16$) and $\Delta_\beta\approx0$; so the turning-point of the curvature $\beta(\varepsilon)$ from being positive to negative is strongly correlated with the peak position in $s^2$\,\cite{Marc24Curvature}; moreover we have  $s_{\rm{pk}}^2\approx0.60$ by adopting $\Delta_\beta\approx0$ and $\varepsilon_\beta\beta'\approx-1$  (a minimum of $s_{\rm{pk}}^2=1/2$ occurs at $\varepsilon_\beta\beta'=-3/16$ according to (\ref{rtpk1})).
Therefore, the maximum SSS ($s_{\rm{pk}}^2$) in NSs may slightly be larger than 1/2\,\cite{Ecker2022,Cao2022PRD}.
In addition, magnitude of the second term in the bracket of (\ref{rtpk}) is smaller than about $1.8\%$ (compared with 1) by using $\varepsilon_\beta\beta'\approx-1$ and the maximum $\Delta_\beta\approx1/3$.
The $\varepsilon$-dependence of the curvature $\beta$ from a very specific parametrization for $\Delta(\varepsilon)$, i.e., $\Delta(\varepsilon)\approx3^{-1}[1-[1+\exp(-3.45(\eta-1.2))]^{-1}][1-2/(20+\eta^2)]$\,\cite{Fuji22} with $\eta=\ln(\varepsilon/\varepsilon_0)$, is shown in the left panel of FIG.\,\ref{fig_MarcCC}\,\cite{Marc24Curvature}, from which one can obtain three peak positions $\varepsilon_\beta\approx790\,\rm{MeV}/\rm{fm}^3$, $\varepsilon_{\rm{pk}}\approx597\,\rm{MeV}/\rm{fm}^3$ and $\varepsilon_\beta^{\ast}\approx453\,\rm{MeV}/\rm{fm}^3$, therefore $\varepsilon_{\rm{pk}}/\varepsilon_\beta\approx0.76$, $\varepsilon_\beta^{\ast}/\varepsilon_\beta\approx0.57$ and $\varepsilon_{\rm{pk}}/\varepsilon_\beta^{\ast}\approx1.31$.
Furthermore, the $\Delta$ curve in the left panel also indicates $\Delta_\beta''>0$ (black dotted line).
Similarly, the peak position $\varepsilon_{\rm{deriv,pk}}$ in the derivative part of SSS is
\begin{equation}
\frac{\varepsilon_{\rm{deriv,pk}}}{\varepsilon_\beta}\approx\frac{3}{4}\frac{10-32\varepsilon_\beta\beta'}{7-48\varepsilon_\beta\beta'}\left[1
+12\Delta_\beta\cdot\frac{3+8\varepsilon_\beta\beta'}{(5-16\varepsilon_\beta\beta')(7-48\varepsilon_\beta\beta')}
    \right];
\end{equation}
so $\varepsilon_{\rm{pk}}>\varepsilon_{\rm{deriv,pk}}$.
The relation between $\varepsilon_{\rm{pk}}$ and $\varepsilon_\beta^{\ast}$ could be obtained by using Eq.\,(\ref{rtpk}) and $\varepsilon_\beta^{\ast}\approx\varepsilon_\beta-\beta'/\beta''$:
\begin{equation}
    \frac{\varepsilon_{\rm{pk}}-\varepsilon_\beta^{\ast}}{\varepsilon_\beta}\approx\frac{3+16\varepsilon_\beta\beta'}{7-48\varepsilon_\beta\beta'}+\frac{\beta'}{\varepsilon_\beta\beta''}.
    \end{equation}
As long as $\beta'/\varepsilon_\beta\beta''>1/3$ (a quasi-linear shape near $\varepsilon_\beta$) or equivalently $\d\ln\beta'/\d\ln\varepsilon_\beta>3$, we always have $\varepsilon_{\rm{pk}}>\varepsilon_\beta^\ast$, i.e., the peak position in $s^2$ is on the right side of that in the curvature $\beta$.

In the right panel of FIG.\,\ref{fig_MarcCC}, the $\varepsilon$-dependence of the curvature obtained from Ref.\,\cite{Marc24Curvature} is shown. 
They found that $\varepsilon_\beta\approx678\,\rm{MeV}/\rm{fm}^3$ and $\varepsilon_{\rm{pk}}\approx500\,\rm{MeV}/\rm{fm}^3$, so $\varepsilon_{\rm{pk}}/\varepsilon_\beta\approx0.74$.
We can also find from the figure that $\varepsilon_\beta^\ast\approx445\,\rm{MeV}$ and thus $\varepsilon_{\rm{pk}}/\varepsilon_\beta^\ast\approx0.89$.
Under the assumption $\Delta\geq0$, Ref.\,\cite{Marc24Curvature} further found $s_{\beta}^2\equiv s^2(\varepsilon_\beta)\leq1/2$ and $\gamma_\beta\equiv\gamma(\varepsilon_\beta)\geq3/2$, based on the relation  $s_\beta^2+\gamma_\beta=2$ and the expressions for $\Delta_\beta$\,\cite{Marc24Curvature}:
\begin{equation}\label{yp-1}
    \Delta_\beta=\frac{4}{3}\frac{s_\beta^2-1/2}{s_\beta^2-2}
    =\frac{4}{3}\frac{\gamma_\beta-3/2}{\gamma_\beta},~~\mbox{or}~~s_\beta^2=\frac{2-6\Delta_\beta}{4-3\Delta_\beta},~~\gamma_\beta=\frac{6}{4-3\Delta_\beta}.
\end{equation}
If these inequalities are violated, then a negative $\Delta$ is implied\,\cite{Marc24Curvature}.
The polytropic index $\gamma$ around the transition point $\varepsilon_\beta$ could be obtained as:
\begin{equation}
    \gamma(\varepsilon)\approx\frac{6}{4-3\Delta_\beta}\left[1
    -\frac{\delta\varepsilon}{\varepsilon_\beta}\frac{(2-9\Delta_\beta)-8\varepsilon_\beta\beta'(4-9\Delta_\beta)}{(1-3\Delta_\beta)(4-3\Delta_\beta)^2}
    \right],~~\beta'<0.
\end{equation}
The second term in the square bracket is negative for $\varepsilon>\varepsilon_\beta$, indicating the $\gamma$ index decreases with $\varepsilon$ above $\varepsilon_\beta$, which may eventually approach the conformal limit.
Taking $\Delta_\beta\approx0$ and $-\varepsilon_\beta\beta\approx-1$, we have $
    \gamma\approx(3/2)[1-(17/8)\delta\varepsilon/\varepsilon_\beta]$.
The approaching of $\gamma$ to the conformal limit is a direct consequence of the assumption that $\beta$ experiences a transition from positive to negative (so $\varepsilon_\beta$ exists and $\beta'<0$).
Similarly, the SSS $s^2$ for $\varepsilon$ near $\varepsilon_\beta$ is:
\begin{equation}
    s^2/s_{\beta}^2\approx1+
    \frac{2}{s_\beta^2}\frac{\delta\varepsilon}{\varepsilon_\beta}\frac{3(2-3\Delta_\beta)+8\varepsilon_\beta\beta'(4-9\Delta_\beta)}{(4-3\Delta_\beta)^3},~~\beta'<0,
\end{equation}
so $s^2/s_\beta^2<1$ for $\varepsilon>\varepsilon_\beta$ as $3(2-3\Delta_\beta)+8\varepsilon_\beta\beta'(4-9\Delta_\beta)$ may probably be negative for $-\varepsilon_\beta\beta'\approx\mathcal{O}(1)$ and $\Delta_\beta\approx0$.

However, we notice that the curvature $\beta$ in realistic NSs may have no turning-point from positive to negative (so the zero point $\varepsilon_\beta$ is not well defined).
For example, Ref.\,\cite{Perego2022} found that $K_{\rm{NM}}^{\rm{TOV}}\equiv [9\d P/\d\rho]_{\rm{TOV}}\approx10\,\rm{GeV}$ using $\xi_{\max}\approx0.28$ (see their FIG.\,4), taking $\rho_{\rm{TOV}}\approx6\rho_0$, $P_{\rm{TOV}}\approx400\,\rm{MeV}/\rm{fm}^3$, $\varepsilon_{\rm{TOV}}\approx1\,\rm{GeV}/\rm{fm}^3$, one obtains $\mu_{\rm{TOV}}\approx(\varepsilon_{\rm{TOV}}+P_{\rm{TOV}})/\rho_{\rm{TOV}}\approx1.5\,\rm{GeV}$ and $\phi_{\rm{TOV}}=P_{\rm{TOV}}/\varepsilon_{\rm{TOV}}\approx0.4$.
Consequently $\beta_{\rm{TOV}}=K_{\rm{NM}}^{\rm{TOV}}/9\mu_{\rm{TOV}}-2/(1+\phi^{-1}_{\rm{TOV}})\approx0.19$.
Without requiring $\beta'<0$ (used in the above discussion) and in the mean while introducing a constant term $B$ in the expansion of $\beta$ over $\varepsilon$ as $\beta\approx B+\beta'\delta\varepsilon+2^{-1}\beta''\delta\varepsilon^2$, we can similarly work out 
\begin{align}
 \Delta(\varepsilon)\approx&\Delta_\beta-\frac{\delta\varepsilon}{3\varepsilon_\beta}\frac{2+12B-(3+9B)\Delta_\beta-9\Delta_\beta^2}{4-3\Delta_\beta}\notag\\
 &-\frac{\delta\varepsilon^2}{2\varepsilon_\beta^2}\left[\varepsilon_\beta\beta'-\frac{2}{3}\frac{(7-3\Delta_\beta)(1-3\Delta_\beta)[2+12B-(3+9B)\Delta_\beta-9\Delta_\beta^2]}{(4-3\Delta_\beta)^3}\right],
 \end{align}
 and\begin{align}
 \gamma(\varepsilon)\approx&\frac{6}{4-3\Delta_\beta}\left[1
    -\frac{2(1+6B)-9(1+5B)\Delta_\beta-8\varepsilon_\beta\beta'(4-9\Delta_\beta)}{(1-3\Delta_\beta)(4-3\Delta_\beta)^2}\frac{\delta\varepsilon}{\varepsilon_\beta}
    \right],\label{yp-2}\\
    s^2/s_{\beta}^2\approx&1-
    \frac{1}{s_\beta^2}\frac{18(1+3B)\Delta_\beta-12(1+6B)-16\varepsilon_\beta\beta'(4-9\Delta_\beta)}{(4-3\Delta_\beta)^3}\frac{\delta\varepsilon}{\varepsilon_\beta},~~s_\beta^2=B+\frac{2-6\Delta_\beta}{4-3\Delta_\beta}.\label{yp-3}
\end{align}
Then $\Delta(\varepsilon)$ develops the local minimum $\Delta_{\min}(B)=\Delta(\varepsilon_{\Delta})\approx-2(1+6B)^2/[21(1+6B)-144\varepsilon_\beta\beta']$ at $\varepsilon_{\Delta}/\varepsilon_\beta\approx[15(1+6B)-48\varepsilon_\beta\beta']/[7(1+6B)-48\varepsilon_\beta\beta']$ with $\Delta_\beta\approx0$; the dependence $B(\varepsilon_\beta\beta')$ can then be obtained via, e.g., $\Delta_{\rm{min}}(B)\approx\Delta_{\rm{GR}}\approx-0.041$.
By comparing the relations of (\ref{rlll}) we find $\Delta_{\min}(B)<\Delta_{\min}(0)<0$ for $B>0$ and $\varepsilon_\beta\beta'<0$.
Moreover, we have $B\lesssim s_\beta^2\lesssim B+2^{-1}$ for $1/3\gtrsim\Delta_\beta\gtrsim0$; 
the basic condition $0\leq s_\beta^2\leq1$ requires
\begin{equation}
    \frac{6\Delta_\beta-2}{4-3\Delta_\beta}\leq B\leq\frac{2+3\Delta_\beta}{4-3\Delta_\beta}.
\end{equation}
According to (\ref{yp-2}), we find that as long as
\begin{equation}\label{yp-cond}
    2(1+6B)-9(1+5B)\Delta_\beta-8\varepsilon_\beta\beta'(4-9\Delta_\beta)<0,
\end{equation}
the correction in the square bracket of (\ref{yp-2}) would be greater than zero for $\delta\varepsilon/\varepsilon_\beta>0$, implying $\gamma$ increases after $\varepsilon_\beta$.
Specifically, by rewriting (\ref{yp-cond}) we have $\gamma/\gamma_\beta>1$ for $\delta\varepsilon/\varepsilon_\beta>0$ if $B$ satisfies:
\begin{equation}\label{yp-cond-aa}
    0<B<\min\left[
   \frac{8\varepsilon_\beta\beta'(4-9\Delta_\beta)-(2-9\Delta_\beta)}{12-45\Delta_\beta},\frac{2+3\Delta_\beta}{4-3\Delta_\beta}
    \right],~~\mbox{where}~~\varepsilon_\beta\beta'>0.
\end{equation}
The $\varepsilon$-dependence of such $\beta$ function is sketched in the right panel of FIG.\,\ref{fig_beta_function}.
Under the condition (\ref{yp-cond}):
\begin{equation}
    s^2/s_\beta^2\gtrsim1+\frac{\delta\varepsilon}{\varepsilon_\beta}
    \frac{1}{s_\beta^2}
    \frac{16(1+6B)-36(1+4B)\Delta_\beta}{(4-3\Delta_\beta)^3},
\end{equation}
which may be greater than 1, indicating the $s^2$ is an increasing function of $\varepsilon$ around $\varepsilon_\beta$.
For example, if we take $\Delta_\beta\approx0$ for $\varepsilon_\beta\approx1\,\rm{GeV}/\rm{fm}^3$ and $\varepsilon_\beta\beta'\approx0.2$, then $B\approx0.1$ satisfies the condition (\ref{yp-cond-aa}), so $[16(1+6B)-36(1+4B)\Delta_\beta]/(4-3\Delta_\beta)^3\approx0.4$, $s_{\beta}^2\approx2^{-1}+B\approx0.6$ and $s^2/s_\beta^2\gtrsim1+2\delta\varepsilon/3\varepsilon_\beta$.

These discussions imply that $B\approx0$ and $\beta'<0$ (as shown in the left panel of FIG.\,\ref{fig_beta_function}) 
construct two basic conditions (assumptions) for generating a $\beta$ function with zero point and forcing the $\gamma$ parameter to approach its conformal limit at energy densities above this zero point.

\setcounter{equation}{0}
\section{Conclusions, Caveats and Future Perspectives}\label{SEC_8}

We have studied the EOS of supra-denser matter in NS cores and many related issues listed in the INTRODUCTION by analyzing perturbatively the scaled TOV equations and using only NS observational data in a largely nuclear EOS-model independent manner. Our main conclusions are summarized below (corresponding to the questions listed in the INTRODUCTION).

\begin{enumerate}[label=(\alph*)]
\item By perturbatively analyzing the dimensionless TOV equations (\ref{TOV-ds}), we established the scaling between NS mass $M_{\rm{NS}}$ and $\Gamma_{\rm{c}}=\varepsilon_{\rm{c}}^{-1/2}\Pi_{\rm{c}}^{3/2}$ and that between NS radius $R$ and $\nu_{\rm{c}}=\varepsilon_{\rm{c}}^{-1/2}\Pi_{\rm{c}}^{1/2}$, here $\varepsilon_{\rm{c}}$ is the central energy density and $\Pi_{\rm{c}}=\x/(1+3\x^2+4\x)$ with $\x=P_{\rm{c}}/\varepsilon_{\rm{c}}$ and $P_{\rm{c}}\equiv P(\varepsilon_{\rm{c}})$ is the central EOS, see the left panel of FIG.\,\ref{fig_MmaxS}, the  right panel of FIG.\,\ref{fig_Phi_c} and FIG.\,\ref{fig_Rnc}.
The scalings are verified quantitatively by using more than a hundred microscopic/phenomenological nuclear EOS models and about $10^5$ randomly generated meta-model nuclear EOSs satisfying all contemporary terrestrial and astrophysical constraints.
The mass and radius scalings can be effectively mapped onto power-law forms, from which we deduce that $P_{\rm{c}}\sim M_{\rm{NS}}^{\max,3}/R_{\max}^5$ and $\x\sim(M_{\rm{NS}}^{\max}/R_{\max})^2$ for TOV NSs.
On the other hand, for Newtonian stars, $\Pi_{\rm{c}}\approx\x$ and so $M_{\rm{NS}}\sim P_{\rm{c}}^{3/2}\varepsilon_{\rm{c}}^{-2}$ and $R\sim P_{\rm{c}}^{1/2}\varepsilon_{\rm{c}}^{-1}$, as well as $\xi\approx2\x$; inversely one obtains $P_{\rm{c}}\sim M_{\rm{NS}}^2/R^4$.

\item Centers of NSs at the maximum-mass configuration (TOV configuration) on the M-R curve contain the stable densest visible matter existing in our Universe. If either the mass $M_{\rm{NS}}^{\max}\equiv M_{\rm{TOV}}$ or the radius $R_{\max}\equiv R_{\rm{TOV}}$
is available/observed, one can extract the EOS of the densest visible matter using either the $M_{\rm{NS}}$-$\Gamma_{\rm{c}}$ or $R$-$\nu_{\rm{c}}$ scaling at the TOV configuration. If both the mass and radius of a NS are measured we can determine individually the $P_{\rm{c}}$ and $\varepsilon_{\rm{c}}$ at the core of this NS, see FIG.\,\ref{fig_eP-PSR740+6620}.
In particular, for PSR J0740+6620 we found that $P_{\rm{c}}\equiv P(\varepsilon_{\rm{c}})\approx0.012\varepsilon_{\rm{c}}^{4/3}\cdot(1+0.047\varepsilon_{\rm{c}}^{1/3}+0.0026\varepsilon_{\rm{c}}^{2/3}+0.00016\varepsilon_{\rm{c}}+\cdots)$ for roughly $
640\,\rm{MeV}/\rm{fm}^3\lesssim\varepsilon_{\rm{c}}\lesssim1210\,\rm{MeV}/\rm{fm}^3$.
Our predictions are independent of any input nuclear EOS.

\item Because the General Relativity for strong-field gravity is a nonlinear theory and the TOV equations inherit this nonlinearity, the EOS inferred from the latter is also essentially nonlinear\,\cite{CL24-c}.
This means a linear EOS of the form $P=\zeta\varepsilon+\rm{const.}$ with $\zeta$ being a constant is fundamentally inconsistent with the TOV equations.
Specifically, using such linear EOS leads  to singularities for the pressure and energy density at NS centers. In addition, a polytropic index $\gamma$ of 1 of such linear EOS is basically in conflict with the fact that $\gamma\geq4/3$ by analyzing directly the dimensionless TOV equations without using any input nuclear EOS, see Eq.\,(\ref{sc2-TOV}) or Eq.\,(\ref{sc2-GG}).
Equivalently, an EOS model with a constant speed of sound (CSS) can not consistently describe the dense matter in NS cores; i.e., the widely used CSS EOS model in the literature is an approximation for describing core NS matter; one may take caution when making inference on the core NS EOS using such model.

\item The NS compactness $\xi=M_{\rm{NS}}/R$ is found to scale with $\x=\phi_{\rm{c}}={\x}=P_{\rm{c}}/\varepsilon_{\rm{c}}$ according to $\xi\approx2\x/(1+3\x^2+4\x)=2\Pi_{\rm{c}}$; this scaling is verified quantitatively in the left panel of FIG.\,\ref{fig_Phi_c} using $10^5$ randomly generated meta-model EOSs.
The $\xi$-$\Pi_{\rm{c}}$ scaling enables us to extract directly the central ratio $\x$ and thus the trace anomaly $\Delta$ from NS observational data independent of any input EOS model.

\item The SSS $s^2=\d P/\d\varepsilon$ is closely related to the compactness $\xi$ via  $s^2=g(\xi)$ with $g$ being a dimensionless function. In particular, the central SSS can be written as $s_{\rm{c}}^2\approx 3^{-1}\tau^{-1}(4+\Psi)\xi+(4/3)\tau^{-2}(5+2\Psi)\xi^2+\cdots$, where $\tau\approx2$ and $\Psi\equiv 2\d\ln M_{\rm{NS}}/\d\ln\varepsilon_{\rm{c}}>0$, see Eq.\,(\ref{EQ-1}). This means the central stiffness (characterized by $s_{\rm{c}}^2$) also characterizes the physics of compactness $\xi$ of NSs, see Eq.\,(\ref{EQ-CS}), or 
\begin{empheq}[box=\fbox]{align}
\mbox{NS compactness }\xi={M_{\rm{NS}}}/{R}
\leftrightarrow&
\mbox{central ratio }\x=\phi_{\rm{c}}=\widehat{P}_{\rm{c}}=P_{\rm{c}}/\varepsilon_{\rm{c}}\notag\\
\leftrightarrow&\mbox{average SSS from surface to center }\x=\langle s_{\rm{c}}^2\rangle=\frac{1}{\varepsilon_{\rm{c}}}\int_0^{\varepsilon_{\rm{c}}}\d\varepsilon' s^2(\varepsilon')\notag\\
\leftrightarrow
&\mbox{central stiffness }s_{\rm{c}}^2=\x\left(1+\frac{1+\Psi}{3}\frac{1+3\x^2+4\x}{1-3\x^2}\right).
\end{empheq}
However, it does not mean $s^2$ is physically equivalently to $\xi$ since $s^2$ is not a monotonic function of the radial distance from the NS center,  i.e., there is no one-to-one correspondence between the stiffness at places away from NS centers (characterized by $s^2$ instead of $s_{\rm{c}}^2$) and the compactness.
Similarly, although $\x$ characterizes the same physics of $\xi$ when describing the compactness, i.e., a larger $\x$ corresponds to a larger $\xi$, the ratio $\phi$ is not generally equivalent to $\xi$.
This is because a NS with larger $\x$ (and therefore a larger $\xi$) may have a smaller $\phi$ at finite $\hr$, see the relevant discussion given at the end of Subsection \ref{sub_s2_1st}.

\item Due to the nonlinearity of the EOS in NS cores and the fact that the equivalence between $s^2\leq1$ (causality principle) and $\phi\equiv P/\varepsilon\leq1$ only holds for the linear EOS $P\sim\varepsilon$, the gravitational upper limit for $\phi$ is expected to be smaller than 1.
Using the expression for the central SSS as $s_{\rm{c}}^2=\x[1+3^{-1}(1+3\x^2+4\x)/(1-3\x^2)]$ and $s_{\rm{c}}^2\leq1$, we found that $\phi\leq\x\lesssim0.374$, see Eq.\,(\ref{sc2-TOV}).
Though the Principle of Special Relativity requires $P\leq \varepsilon$, the strong-field gravity in General Relativity and the induced nonlinearity of the EOS render the upper bound for the ratio $\phi$ to be much smaller than 1\,\cite{CL24-c}.
As a direct consequence, the dimensionless trace anomaly $\Delta=1/3-\phi=1/3-P/\varepsilon$ is lower bounded by GR as $\Delta\gtrsim\Delta_{\rm{GR}}\approx-0.041$.
The GR effects significantly reduce the factor $\Pi_{\rm{c}}^{\rm{N}}=\x$ with $\x\leq3/4$ in Newtonian case being upper bounded as $\Pi_{\rm{c}}^{\rm{N}}\lesssim0.75$ to $\Pi_{\rm{c}}=\x/(1+3\x^2+4\x)\lesssim0.128$; and it has apparent influence (since $\Pi_{\rm{c}}^3/\Pi_{\rm{c}}^{\rm{N},3}\approx200$) on estimating the upper bound on $\varepsilon_{\rm{c}}$ (and that on $M_{\rm{TOV}}$) which involves $\Pi_{\rm{c}}^3$, see Eq.\,(\ref{cda-1}).
In addition, the dependence of $s_{\rm{c}}^2$ on $\x$ can be effectively mapped onto a power-law form of $\x^2$, therefore $s_{\rm{c}}^2\sim (M_{\rm{NS}}^{\max}/R_{\max})^4$, namely the central SSS for TOV NSs roughly scales as the NS compactness to the fourth.
Similarly for Newtonian stars, $s_{\rm{c}}^2\sim (M_{\rm{NS}}/R)^1$ so $s_{\rm{c}}\sim (M_{\rm{NS}}/R)^{1/2}$.

\item Using the upper bound for $\x$ as $\x\lesssim0.374$ as well as the mass and radius scalings of Eq.\,(\ref{Mmax-G}) and (\ref{Rmax-n}),  we can obtain an inequality which lower bounds the radius $R_{\max}$ for NS at TOV configuration as $R_{\max}/\rm{km}\gtrsim4.73M_{\rm{NS}}^{\max}/M_{\odot}+1.14$, see (\ref{rel-2}). Because NSs at the maximum-mass configuration are the densest, this gives a new causality boundary for NSs. It is shown to be consistent with observational data of several NSs including those in GW170817\,\cite{Abbott2017,Abbott2018} and the PSR J0740+6620\,\cite{Riley21} albeit in certain tension
with predictions using a few empirical EOS for NS matter in the literature.  As a direct corollary, the radius of PSR J0952-0607\,\cite{Romani22} with a mass about $2.35M_{\odot}$ is found to be $\gtrsim12.25\,\rm{km}$, see FIG.\,\ref{fig_MR-C}.
Similarly, for massive NSs ($M_{\rm{NS}}^{\max}\gtrsim2M_{\odot}$), the radii should be greater than about 10.6\,km.

\item Rewriting the causality boundary $R_{\max}/\rm{km}\gtrsim4.73M_{\rm{NS}}^{\max}/M_{\odot}+1.14$ gives $\xi_{\rm{TOV}}\equiv \xi_{\max}\equiv M_{\rm{NS}}^{\max}/R_{\max}\lesssim0.313(1-1.14\,\rm{km}/R_{\max})$;
therefore an absolute upper bound for $\xi_{\max}$ about $0.313$ is obtained. Considering $R_{\max}\approx12.5\pm1\,\rm{km}$ further limits the $\xi_{\max}$ to be about $\xi_{\max}\lesssim0.283\pm0.014$, see (\ref{upp-xi}) and FIG.\,\ref{fig_CompBuch} for the summary on existing upper bounds on the compactness.
This bound for $\xi$ is consistent with all observed NSs and is much tighter than existing bounds for NS compactness (the right eight points of FIG.\,\ref{fig_CompBuch}).
Similarly,  using the mass scaling of Eq.\,(\ref{Mmax-G}) alone allows us to obtain an inequality for the ultimate central energy density as $\varepsilon_{\rm{c}}\lesssim\varepsilon_{\rm{ult}}\approx6.32(M_{\rm{NS}}^{\max}/M_{\odot}+0.106)^{-2}\,\rm{GeV}/\rm{fm}^3$; therefore the observations of massive NSs may essentially upper limit the ultimate energy density existing in our Universe, see Eq.\,(\ref{eps_ult}).
For instance, taking $M_{\rm{NS}}^{\max}/M_{\odot}\approx2.3$ gives us $\varepsilon_{\rm{ult}}\approx1.1\,\rm{GeV}/\rm{fm}^3$.
By using the bound $\x\lesssim0.374$ once again, we obtain $P_{\rm{c}}\lesssim P_{\rm{utl}}\approx2.36(M_{\rm{NS}}^{\max}/M_{\odot}+0.106)^{-2}\,\rm{GeV}/\rm{fm}^3$, see Eq.\,(\ref{P_ult}), so $M_{\rm{NS}}^{\max}/M_{\odot}\approx2.3$ induces $P_{\rm{ult}}\approx408\,\rm{MeV}/\rm{fm}^3$.
In establishing these bounds, only the mass scaling (\ref{Mmax-G}) is used.

\item Using the radius scaling of (\ref{Rmax-n}), we obtain an estimate on the central baryon density of NSs at the TOV configuration as $\rho_{\rm{c}}/\rho_{\rm{sat}}\approx7350\x/(1+3\x^2+4\x)\cdot(R_{\max}/\rm{km}-0.64)^{-2}$,  see Eq.\,(\ref{Rrhoc}). Taking $R_{\max}\approx12.5\,\rm{km}$ leads to $\rho_{\rm{c}}/\rho_{\rm{sat}}\lesssim6.7$. This relation is useful for estimating the highest central baryon density in NSs when future radius observation/measurement become eventually more accurate.

\item The conformal bound $s^2\leq1/3$ is likely violated in NSs at the TOV configuration with $M_{\rm{NS}}^{\max}\gtrsim1.9M_{\odot}$, see FIG.\,\ref{fig_sckk}.
It is obvious from Eq.\,(\ref{sc2-TOV}) that $s_{\rm{c}}^2\to1/3$ occurs much earlier than $\x\to1/3$, implying the violation of $s^2\leq1/3$ occurs at a mild $\x$ being smaller than $1/3$.
In fact, if one lets $\x\leq1/3$, then Eq.\,(\ref{sc2-TOV}) predicts that $s_{\rm{c}}^2\leq7/9\approx0.778$.
For a canonical NS with $M_{\rm{NS}}\approx1.4M_{\odot}$ and $R\approx12\pm1\,\rm{km}$,  the $\x$ is found to be about $\x\approx0.15\pm0.02$ via the compactness-$\Pi_{\rm{c}}$ scaling (see TAB.\,\ref{tab_cEOS}) and the factor $\Psi\equiv2\d\ln M_{\rm{NS}}/\d\ln\varepsilon_{\rm{c}}\approx2.85\pm0.29$ via the $\Psi$-$M_{\rm{NS}}$ relation of FIG.\,\ref{fig_k-fac}; therefore $s_{\rm{c}}^2\approx0.47\pm0.09$ using the formula (\ref{sc2-GG}), see TAB.\,\ref{tab_M14_sc2}. Therefore, the conformal bound for $s^2$ breaks down even for light NSs. Moreover, $\Psi\approx2.85$ implies that the upper limit $\phi\leq\x\lesssim0.24$ for canonical NSs via the SSS formula (\ref{sc2-GG}).
The relation between $\Psi$ and the polytropic index $n$ appearing in the EOS $P\sim \varepsilon^{1+1/n}$ is established as $n=3/(1+\Psi)$, see Eq.\,(\ref{n-Psi}); which provides an alternative viewpoint on $\gamma=1+n^{-1}\geq4/3$.

\item Consider some recent studies showing that $s^2$ is perhaps upper bounded to a lower value different from 1\,\cite{Hipp24}, the reduction on the stiffness $s_{\rm{c}}^2$ correspondingly reduces the compactness $\xi$ as we established in the above; therefore the radius $R$ for a given NS mass is expected to increase (in order to reduce $\xi$).
By adopting $s^2\lesssim0.781$\,\cite{Hipp24} from a new development of transport theories for nuclear matter, in particular, we find that the radius $R$ for a $2M_{\odot}$ NS may increase by about 0.8\,km, see the upper panel of FIG.\,\ref{fig_Psi_on_R}; future astrophysical observations on $R$ have the potential to tell whether such lower bound on $s^2$ is reasonable.
Similarly, if the curvature term $\beta=-\varepsilon\d\Delta/\d\varepsilon+(3^{-1}-\Delta)[1+2/(\Delta-4/3)]$ of the energy per particle develops a zero point $\varepsilon_\beta\approx1\,\rm{GeV}$ (defined by $\beta(\varepsilon_\beta)=0$) with $-\varepsilon_\beta\beta'\approx\mathcal{O}(1)$ and $\Delta_\beta\lesssim0.1$, the maximum SSS $s_{\rm{pk}}^2$ is found to be smaller than about $0.5\mbox{-}0.6$, see (\ref{rtpk1}), i.e., $s_{\rm{pk}}^2\lesssim0.5\mbox{-}0.6$.

\item Since $s_{\rm{c}}^2>0$ as implied by Eqs.\,(\ref{sc2-TOV}) and (\ref{sc2-GG}), a sharp phase transition (PT) signaled by the vanishing of $s^2$ is basically excluded; while a continuous crossover characterized by a smooth reduction of $s^2$ is allowed in cores of massive NSs.
This means massive NSs may have soft cores ($s^2(r)>s_{\rm{c}}^2\equiv s^2(r=0)$).

\item The strong-field gravity in GR compresses NS matter and thus makes $\phi\sim\x=P_{\rm{c}}/\varepsilon_{\rm{c}}$ sizable ($\sim\mathcal{O}(0.1)$), while simultaneously extrudes a possible peak in the density/radius profile of $s^2$ whence $\x\gtrsim0.1$, see FIG.\,\ref{fig_s2_GRNewt}. The peak eventually reduces as $\x$ increases and finally disappears in the GR causal limit $\x\to0.374$.
The position of the peak $r_{\rm{pk}}$ is at about $(30\%$-$40\%)R$ from the NS center and the relative enhancement $\Delta s^2=s_{\max}^2/s_{\rm{c}}^2-1$ is generally smaller than about 10\%, here $s_{\max}^2\equiv s^2(r_{\rm{pk}})$, see the left panel of FIG.\,\ref{fig_s2peak} and FIG.\,\ref{fig_s2_peak_stat-af}.
Moreover, using the compactness scaling one can extract the $\x$ directly from observed NS masses/radii; the dimensionless trace anomaly $\Delta$ as well as its energy density dependence could then be evaluated. As a result, we find using the trace anomaly decomposition of SSS that the currently available NS data (masses and radii) can not invariably generate a peaked $s^2$ profile, 
see FIG.\,\ref{fig_s2ab} and FIG.\,\ref{fig_fffff}.

\item For Newtonian stars, the ratio $\phi$ or $\x$ is generally small, and we have in this case $s^2\approx s_{\rm{c}}^2-4\widehat{r}^2/15+\hr^4/{60s_{\rm{c}}^2}$, here $\hr$ is the dimensionless distance from the star center. It monotonically decreases when going outward from the center, see Eq.\,(\ref{for_s2Newt-1}).
In particular, it vanishes at $\hr^2=6s_{\rm{c}}^2$ and $\d s^2/\d\hr\approx-(8\hr/15)(1-\hr^2/8s_{\rm{c}}^2)$ is always negative for $0\leq \hr^2\leq 6s_{\rm{c}}^2$, see FIG.\,\ref{fig_s2New_r4}.
In addition, we have generally for Newtonian stars that $s^2\approx2\phi+\cdots$ around $\phi\approx0$, see (\ref{s2_on_phi}), therefore Newtonian stars have $\gamma_{\rm{N}}\approx2$.

\item The dense matter in NS cores could fundamentally not be conformal under the basic definition $s^2/\phi\to1$ since a linear EOS of the form $P=\zeta\varepsilon+\rm{const.}$ (with $\zeta$ being constant) is excluded by the TOV equations.
Using some empirical criterion such that $\Theta\lesssim0.2$ with $\Theta=[(3^{-1}-\phi)^2+(s^2-\phi)^2]^{1/2}$,  the core matter maybe conformal however the likelihood is very low, see the right panel of FIG.\,\ref{fig_tc}, TAB.\,\ref{sstab} and Eqs.\,(\ref{gr-1}), (\ref{Tr-1}), (\ref{gm-1}) and (\ref{Tm-1}).
In particular, we find the central $\Theta_{\rm{c}}$ for $\hr=0$ is probably greater than 0.19 and the $\Theta$ for finite $\hr$ is even larger.
The $\Theta_{\rm{c}}$ parameter for a canonical NS with $R\approx12\pm1\,\rm{km}$ is found to be about $0.38\pm0.05$.
Similarly, the polytropic index at NS centers is obtained as $\gamma_{\rm{c}}\approx4/3+0.55\xi_{\max}$, see Eq.\,(\ref{gamma-xi}), therefore $\gamma_{\rm{c}}>4/3$; for a NS at the TOV configuration with radius about 12\,km to have $\gamma_{\rm{c}}\lesssim1.75$ its mass should be smaller than about $1.88M_{\odot}$, see FIG.\,\ref{fig_gamma-xi-rela}, implying massive NSs could hardly have a $\gamma_{\rm{c}}\lesssim1.75$.
Moreover, Eq.\,(\ref{zf-3}) gives $s^2/\phi=(4/3+4\x/3+32\mu/15)\x/\phi$, which is generally greater than 4/3 near NS centers, here $\mu=\heps-1$ is small negative in NS cores.
Furthermore, we have $s^2$ as a function of $\phi$ near NS centers as $s^2(\phi)/\x\approx4/3+(32/5)(1-19\x/4)(\phi/\x-1)-(876/25)(1-3439\x/219)(\phi/\x-1)^2$, which is non-monotonic (see Eq.\,(\ref{DDDk})), implying $s^2$ and $\phi$ are basically two different attributes of NSs.
For Newtonian stars, $s^2\approx2\phi$ (see (\ref{s2_on_phi})) and so $\Theta_{\rm{N}}\approx3^{-1}-\phi+3\phi^2/2\gtrsim0.2$ considering $\phi\approx0$.

\item The mass scaling of Eq.\,(\ref{gk-mass}) implies that the upper bound for NS central energy density $\varepsilon_{\rm{c}}$ decreases with increasing $M_{\rm{NS}}$ and roughly $\varepsilon_{\rm{c}}\sim M_{\rm{NS}}^{-2}$, reflecting the self-gravitating nature of NSs.
On the other hand, the stability condition $\d M_{\rm{NS}}/\d\varepsilon_{\rm{c}}>0$ implies that the $\varepsilon_{\rm{c}}$ increases with $M_{\rm{NS}}$. 
Therefore, a maximum mass for stable NSs unavoidably exists by considering these two different tendencies in varying the $\varepsilon_{\rm{c}}$, see FIG.\,\ref{fig_Ypm_sk} for the relevant sketch.
Using the specific mass, radius and compactness scalilngs given in this review, the maximum mass $M_{\rm{NS}}^{\max}\equiv M_{\rm{TOV}}$ is estimated to be about $M_{\rm{TOV}}\approx2.28\pm0.28M_{\odot}$, see FIG.\,\ref{fig_X-MR} and FIG.\,\ref{fig_MTOV-STAT} for the summary of the status on $M_{\rm{TOV}}$.

\item The derivative $\d R/\d M_{\rm{NS}}$ depends essentially on $\Psi$ as $\d R/\d M_{\rm{NS}}\sim (1-2\Psi^{-1})M_{\rm{NS}}^{-(2/3)(1+\Psi^{-1})}$ or $\d R/\d M_{\rm{NS}}\sim (1-2\Psi^{-1})R^{-2(1+\Psi^{-1})/(1-2\Psi^{-1})}\sim\Psi-2$, see Eq.\,(\ref{fgk-2}) and Eq.\,(\ref{fgk-8}). Therefore, a weak dependence of $R$ on $M_{\rm{NS}}$ or the empirical evidence of a ``vertical'' shape on the M-R curve could be traced back to the quasi-linear growth of $M_{\rm{NS}}$ with $\varepsilon_{\rm{c}}$ around $M_{\rm{NS}}/M_{\odot}\approx1.4\mbox{-}2.2$, i.e., $M_{\rm{NS}}\sim\varepsilon_{\rm{c}}^{\Psi/2}\approx\varepsilon_{\rm{c}}$ for $\Psi\approx2$.
This approximation has been verified quantitatively using $10^5$ samples of our meta-model EOSs, see FIG.\,\ref{fig_k-fac}.
For a super-liner growth rate for low-mass NSs with $\Psi>2$, we have $\d R/\d M_{\rm{NS}}>0$ or equivalently $\d M_{\rm{NS}}/\d R>0$ (from the lower-left corner to the upper-right corner on the NS M-R curve).

\item By considering generally $\Delta(\overline{\varepsilon})\approx\Delta_\ell+2^{-1}\Delta_{\ell}''(\overline{\varepsilon}-\overline{\varepsilon}_{\ell})^2+6^{-1}\Delta_{\ell}'''(\overline{\varepsilon}-\overline{\varepsilon}_{\ell})^3$ where $\Delta_{\ell}\equiv\Delta(\overline{\varepsilon}_{\ell})$, $\Delta''_{\ell}=\Delta''(\overline{\varepsilon}_{\ell})$ and $\Delta'''_{\ell}=\Delta'''(\overline{\varepsilon}_{\ell})$, $\overline{\varepsilon}_{\ell}$ is the peak position in $\Delta(\overline{\varepsilon})$ so that $\Delta_{\ell}''$ is negative, we can work out that $\overline{\varepsilon}_{\ell}^{\ast}/\overline{\varepsilon}_{\ell}\equiv\rm{H}(k)\equiv(3/4)(1-k^{-1})+(4k)^{-1}\sqrt{k^2-2k+9}$ with $k\equiv  \overline{\varepsilon}_{\ell}\Delta_{\ell}'''/\Delta_{\ell}''=\d\ln\Delta_\ell''/\d\ln\ep_{\ell}$, see Eq.\,(\ref{ratk}), $\overline{\varepsilon}_{\ell}^{\ast}$ is the peak position in the SSS $s^2$.
Therefore $1/2\lesssim\overline{\varepsilon}_{\ell}^{\ast}/\overline{\varepsilon}_{\ell}\lesssim1$ for all possible $k$, see Eq.\,(\ref{epast}) and FIG.\,\ref{fig_func_Wk}.
A ratio $\overline{\varepsilon}_{\ell}^{\ast}/\overline{\varepsilon}_{\ell}$ deviating from 1/2 or 1 indicates a non-negligible kurtosis coefficient $\Delta_{\ell}''''\equiv\Delta_\ell^{(4)}$.
\end{enumerate}
We summarize schematically in panel (a) of FIG.\,\ref{fig_SumNS} the most important results and the main steps used to obtain them in this review, using the IPAD-TOV approach.
In panel (b) of FIG.\,\ref{fig_SumNS}, we give an example on the similarity between microscopic and macroscopic quantities involving the dimensionless quantities characterizing NS properties.
While the ratio $\phi$ of the pressure over energy density of Eq.\,(\ref{def-phi}) is monotonic (near $\hr\approx0$), the SSS $s^2$ of Eq.\,(\ref{def-s2}) may not be monotonic; similarly, although the compactness $\xi$ of Eq.\,(\ref{def-compactness}) is positive-definite, the slope $\d M_{\rm{NS}}/\d R$ of the NS M-R curve may either be positive or negative depending on the magnitude of $\Psi$.

\begin{figure}[h!]
\centering
\includegraphics[width=16.5cm]{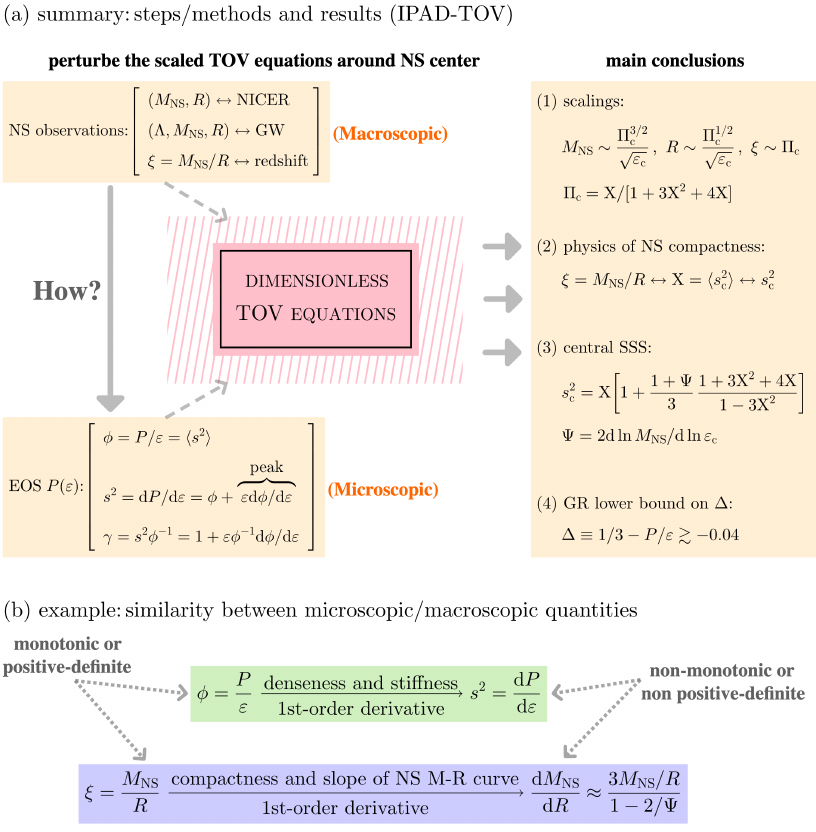}
\caption{(Color Online). Upper panel: a schematic summary of the main steps and most important results of the IPAD-TOV approach.
Lower panel: an example of the similarity between microscopic and macroscopic quantities characterizing NS properties.
}\label{fig_SumNS}
\end{figure}

There are certainly caveats in our analyses. As in solving any differential equations using polynomials, one has to truncate the series of expansions somewhere. In our case studied so far, 
the scalings for NS mass, radius and compactness are obtained by truncating the perturbative expansion of $P$ and $\varepsilon$ to low orders in reduced radius $\hr$. While the results on the EOS and related quantities are quite consistent with existing constraints from state-of-the-art simulations/inferences, their refinement by including even higher-order $\hr$ terms would be important for studying more accurately the radius profile of $\hP$, $\heps$, $\Delta$ and $s^2$.
In the Appendix of Ref.\,\cite{CL24-c}, an effective correction to the mass scaling is investigated, where $\Gamma_{\rm{c}}$ is replaced by $\Gamma_{\rm{c}}(1+18\x/25)$, see also Ref.\,\cite{CL24-b} for some related discussions and implications of this correction term.
Such correction is natural from the viewpoint of the perturbative expansion based on $\x$.
Generally, expanding to even high-order terms inevitably make the analytical analyses more involved, and the determinations of the high-order coefficients will need more observational data beyond just the NS masses and radii. Nevertheless, solving the TOV equations using polynomials has a clear step-by-step procedure as we have demonstrated in this review. 
Moreover, more accurate data of different observables beyond the masses and radii are expected as more progress is being made in this era of multi-messenger astronomy.

In addition, to check the accuracy of our expansions and/or determine the minimum truncation order, we used some information about the NS EOS predicted by nuclear theories or constraints from nuclear experiments. Thus, most of the novel scaling properties of NS properties revealed in our analyses are largely instead of absolutely independent of nuclear EOS models. 

Given the caveats mentioned above, the quantitatively verified novel scalings of NS properties and the gained new insights into several critical issues are clearly very useful. They enabled us to reveal fundamental physics directly from NS observations with little biases due to the still very uncertain predictions of nuclear many-body theories on supra-dense matter EOS. Since solving differential equations using polynomials is a standard method in mathematics, the fruitful analyses presented in this review justify further applications of this approach in solving more unresolved fundamental issues in NS physics. Looking forward, as examples we list below a few future perspectives in this regard.

1.\;The (dimensionless) tidal deformability $
\Lambda=2k_2/3\xi^5$\,\cite{Hinderer2008} provides very useful constraint on NS EOS\,\cite{Abbott2017,Abbott2018}.
Here the quadrupolar tidal Love number $k_2$ is determined by ($\xi=M_{\rm{NS}}/R$ is the NS compactness)
\begin{empheq}[box=\fbox]{align}
k_2=&\frac{8}{5}\xi^5(1-2\xi)^5\left[2-y_R+2\xi\left(y_R-1\right)\right]
\times
\big\{6\xi\left[2-y_R+\xi\left(5y_R-8\right)\right]\notag\\
&\hspace*{1cm}+4\xi^3\left[13-11y_R+\xi\left(3y_R-2\right)+2\xi^2\left(1+y_R\right)\right]\notag\\
&\hspace*{1cm}+3\left(1-2\xi\right)^2\ln\left(1-2\xi\right)
\left[2-y_R+2\xi\left(y_R-1\right)\right]\big\},
\end{empheq}
and the function $y(r)$ satisfies the differential equation (prime is taken with respect to $r$):
\begin{equation}\label{ode_y}
\boxed{
\nu y'+y^2+ye^{\lambda}\left[1+4\pi r^2(P-\varepsilon)\right]+r^2Q=0,~~
Q=4\pi e^{\lambda}\left(5\varepsilon+9P+\frac{\varepsilon+P}{\d P/\d\varepsilon}\right)
-\frac{6e^{\lambda}}{r^2}-\nu'^2.}
\end{equation}
In this differential equation, $\nu(r)$ and $\lambda(r)$ are the time and space components of the spacetime metric; the function
$y(r)$ satisfies the boundary condition $y(0)=2$ and $y_R$ is its value at the radius $R$.
It would be interesting to investigate if the Love number $k_2$ scales with some combination of NS internal properties, thus building possibly new routes towards understanding NS EOS, compared to what we have already learned from the mass and radius scalings. 

In addition, the frequencies and damping times of various oscillation modes of isolated NSs or the remnant of their mergers are also controlled by the NS EOS in differential equations, see, a recent review in Ref.\,\cite{Li2019}. Some of these observables, e.g., the frequencies of f and g modes will be measured by future GW detectors. Solving the differential equations determining them using polynomials may reveal also largely nuclear EOS-model independent information about the nature and EOS of NSs.

\begin{figure}[h!]
\centering
\includegraphics[width=13.cm]{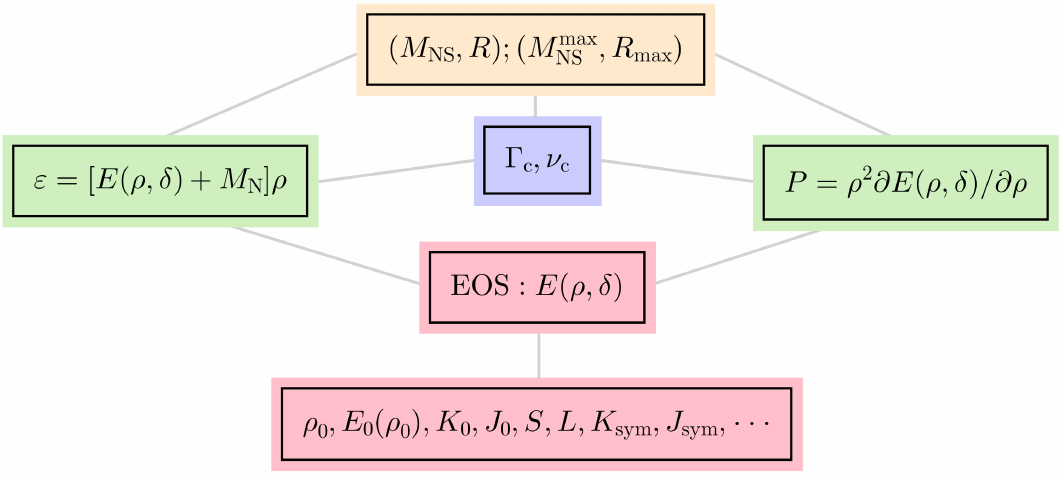}
\caption{(Color Online).  Connections among nuclear characteristic quantities $\rho_0$, $E_0(\rho_0)$, $K_0$, $J_0$, $S$,  $L$, $K_{\rm{sym}}$ and $J_{\rm{sym}}$, etc., defined in Eqs.\,(\ref{fE0}) and (\ref{fEsym}), and the observational masses and radii of NSs.
}\label{fig_EPeps}
\end{figure}

2.\;Since the TOV equations couple only the system pressure $P$ and energy density $\varepsilon$, and in order to investigate how certain nuclear properties\,\cite{LCK08}, e.g., the nuclear symmetry energy and the nucleon effective mass\,\cite{Li2018PPNP}, or quantities characterizing quark matter\,\cite{Baym2018},  may correlate with the NS observations (e.g., masses, radii, tidal polarizabilities and oscillation frequencies), the energy per nucleon/baryon expressed as a function of baryon density $\rho$ and isospin asymmetry $\delta$, namely $E(\rho,\delta)$ is necessary to be introduced directly into the dimensionless TOV equations. It is worth studying whether (nearly) EOS-model independent predictions may emerge, e.g., how does $M_{\rm{NS}}$ or $M_{\rm{NS}}^{\max}$ depend on the skewness parameter $J_0$ of symmetric nuclear matter of Eq.\,(\ref{fE0})? Can we reveal such dependence similar as (\ref{gk-mass}) or (\ref{gk-radius})? 
See FIG.\,\ref{fig_EPeps} for a sketch on the connections among the relevant quantities. Such investigation may extend the analyses in this review. In particular, we used in the main text the approximation $\varepsilon(\rho)\approx M_{\rm{N}}\rho$ to estimate the central nucleon density $\rho_{\rm{c}}$ using the $\heps_{\rm{c}}$ obtained from the mass and radius scalings, and some improvements along this line are needed.

3.\;The slow rotation of NSs could be treated as a type of perturbations\,\cite{Hartle1967,Hartle1968,Hartle1969,Friedman2013rotating}; and the related stellar equations become the Hartle--Thorne (HT) equations. We can similarly analyze perturbative structures of the HT equations and find the connection between certain characteristic quantities and the NS EOS.
Obviously, the HT equations are much more involved than the TOV equations.
A related quantity is the moment of inertia $I$ (defined in Eq.\,(\ref{def_MI})); the crustal fraction $\Delta I/I$ is strongly correlated with the core-crust transition density $\rho_{\rm{t}}$ and the pressure $P_{\rm{t}}$ there. They may have important impact on certain NS properties\,\cite{Lattimer2007PR}. 

\section*{Acknowledgments}
This work was supported in part by the U.S. Department of Energy, Office of Science, under Award Number DE-SC0013702, the CUSTIPEN (China-U.S. Theory Institute for Physics with Exotic Nuclei) under the US Department of Energy Grant No. DE-SC0009971.
\begin{appendices}


\end{appendices}

\renewcommand*{\bibfont}{\footnotesize}
{\begin{spacing}{0.96}
{
\bibliography{main-REF}}
\end{spacing}
}


\end{document}